\documentclass[pdflatex,sn-mathphys,iicol]{sn-jnl}

\usepackage{
graphicx,
amsfonts,
textcomp,
program,
amssymb,
listings,
rotating,
booktabs,
algorithmicx,
mathrsfs,
url,
appendix,
etoolbox,
algpseudocode,
multirow,
algorithm,
amsmath,
hyperref,
manyfoot,
hhline,
pifont,
tabularx,
natbib}

\usepackage[export]{adjustbox}

\newcommand\shline{\specialrule{0.85pt}{0pt}{0pt}}

\renewcommand{\x}{{\mathbf{x}}}
\newcommand{\xhat}{{\hat{\mathbf{x}}}}
\newcommand{\Xhat}{{\hat{\mathbf{X}}}}
\newcommand{\Rbb}{{\mathbb{R}}}
\newcommand{\Ebb}{{\mathbb{E}}}
\renewcommand{\y}{{\mathbf{y}}}
\renewcommand{\z}{{\mathbf{z}}}
\renewcommand{\Z}{{\mathbf{Z}}}

\renewcommand{\A}{{\mathbf{A}}}
\newcommand{\IM}{{\mathbf{I}_M}}
\newcommand{\n}{{\mathbf{n}}}
\newcommand{\M}{{\mathbf{M}}}
\newcommand{\Set}{{\mathbf{S}}}
\newcommand{\F}{{\mathcal{F}}}
\renewcommand{\T}{{\mathbf{T}}}
\newcommand{\Loss}{{\mathcal{L}}}
\newcommand{\Th}{{\mathbf{\Theta}}}
\newcommand{\Appx}{{\textcolor{blue}{\textit{\textbf{Appendix}}}}}
\newcommand{\MP}{{\textit{\textbf{\textcolor{blue}{main paper}}}}}

\jyear{2023}
\theoremstyle{thmstyleone}

\theoremstyle{thmstyletwo}

\theoremstyle{thmstylethree}

\begin{document}

\title[Article Title]{Self-Supervised Scalable Deep Compressed Sensing}

\author[1]{\fnm{Bin} \sur{Chen}}
\author[1]{\fnm{Xuanyu} \sur{Zhang}}
\author[2]{\fnm{Shuai} \sur{Liu}}
\author*[3]{\fnm{Yongbing} \sur{Zhang}}\email{ybzhang08@hit.edu.cn}
\author*[1]{\fnm{Jian} \sur{Zhang}}\email{zhangjian.sz@pku.edu.cn}
\affil[1]{School of Electronic and Computer Engineering, Peking University}
\affil[2]{Shenzhen International Graduate School, Tsinghua University}
\affil[3]{School of Computer Science and Technology, Harbin Institute of Technology (Shenzhen)}

\abstract{Compressed sensing (CS) is a promising tool for reducing sampling costs. Current deep neural network (NN)-based CS approaches face the challenges of collecting labeled measurement-ground truth (GT) data and generalizing to real applications. This paper proposes a novel \textbf{S}elf-supervised s\textbf{C}alable deep CS method, comprising a deep \textbf{L}earning scheme called \textbf{SCL} and a family of \textbf{Net}works named \textbf{SCNet}, which does not require GT and can handle arbitrary sampling ratios and matrices once trained on an incomplete measurement set. Our SCL contains a dual-domain loss and a four-stage recovery strategy. The former encourages a cross-consistency on two measurement parts and a sampling-reconstruction cycle-consistency regarding arbitrary ratios and matrices to maximize data utilization. The latter can progressively leverage the common signal prior in external measurements and internal characteristics of test samples and learned NNs to improve accuracy. SCNet combines both the explicit guidance from optimization algorithms and the implicit regularization from advanced NN blocks to learn a collaborative signal representation. Our theoretical analyses and experiments on simulated and real captured data, covering 1-/2-/3-D natural and scientific signals, demonstrate the effectiveness, superior performance, flexibility, and generalization ability of our method over existing self-supervised methods and its significant potential in competing against many state-of-the-art supervised methods. Code is available at \url{https://github.com/Guaishou74851/SCNet}.}

\keywords{Compressed sensing, inverse imaging problems, self-supervised learning, and algorithm unrolling.}

\maketitle

\section{Introduction}
Compressed sensing (CS) \cite{donoho2006compressed,candes2006compressive} is a pioneering paradigm that effectively breaks through the limit of Nyquist-Shannon theorem \cite{shannon1949communication} for signal acquisition and has derived various practical applications like single-pixel imaging (SPI) \cite{duarte2008single,li2022dual,mur2022deep}, accelerated magnetic resonance imaging (MRI) \cite{lustig2008compressed,sun2016deep}, sparse-view computational tomography (CT) \cite{szczykutowicz2010dual}, and snapshot compressive imaging (SCI) \cite{yuan2021snapshot,fu2021coded,zhang2023progressive}. The fundamental goal of image CS reconstruction is to predict the high-dimensional image $\x \in \Rbb^N$ from its low-dimensional degraded measurement $\y =\A \x + \n\in\Rbb^M$, where the linear projection is achieved by a sampling matrix $\A\in\Rbb^{M\times N}$, and $\n\sim q_\n$ is commonly assumed to be the additive white Gaussian noise (AWGN). In practice, extremely low values of CS ratio (or sampling rate) defined as $\gamma =M/N$ with $M\ll N$ bring not only the benefits of sampling reduction (\textit{e.g.}, energy saving \cite{duarte2008single}, acceleration \cite{sun2016deep}, and suppression of X-ray radiation \cite{szczykutowicz2010dual}), but also the difficulty of inferring $\x$ from only $\y$ and $\A$ as it becomes ill-posed. Early attempts, represented by the iterative optimization algorithms \cite{aharon2006k,gleichman2011blind,li2013efficient,candes2012exact}, exploit image priors (\textit{e.g.} sparsity \cite{zhang2014group} and low-rankness \cite{dong2014compressive}) to impose regularization with theoretical guarantees \cite{cai2014data,aghagolzadeh2015new,chun2019convolutional}.

Deep neural networks (NN) have achieved significant success in CS community over the past decade \cite{sun2016deep,mousavi2015deep,kulkarni2016reconnet,gilton2019neumann,shi2019scalable,chen2020learning,sun2020dual,you2021ista,gilton2021deep,you2021coast,zhang2021amp,chen2022content,chen2022fsoinet,li2022d3c2,zhang2023deep}. Typically, the traditional \textit{supervised} methods fit a reconstruction NN $\F_\Th$ with a parameter set $\Th$ to the latent target mapping $\y \mapsto \x$ by solving $\min_\Th \sum_{i=1}^l \lVert \F_\Th(\y_i)-\x_i \rVert_2^2$ on a labeled dataset $\{(\y_i,\x_i)\}$, consisting of $l$ pairs of measurement $\y$ and ground truth (GT) $\x$. However, registered $(\y,\x)$ pairs can be prohibitively expensive and even physically impossible to obtain in many applications like medical imaging, ghost imaging \cite{wang2022far}, biomicroscopic imaging \cite{wu2021imaging}, and astronomical imaging \cite{sun2021deep}. In addition, insufficient training data, domain gaps between natural and scientific images, or simulated and real measurements \cite{lyu2017deep,higham2018deep}, and low-quality GTs can introduce bias and overfitting to NNs, and result in their poor generalization with large performance drops to the real deployments of CS systems \cite{quan2022dual}.

Recent progress in \textit{unsupervised} methods provides viable solutions to alleviate data requirements. Untrained NN priors \cite{qayyum2022untrained} can leverage the implicit regularization of deep networks (\textit{e.g.}, convolutional \cite{ulyanov2018deep,heckel2020compressive} and Bayesian \cite{pang2020self}) with early stopping and dropout \cite{quan2020self2self} techniques to recover a faithful estimation from single measurement. Building on the remarkable advancements brought by Noise2Noise (N2N) \cite{lehtinen2018noise2noise} and Stein's unbiased risk estimator (SURE) \cite{stein1981estimation,eldar2008generalized}-based methods \cite{soltanayev2018training,zhussip2019extending} for image denoising, \cite{xia2019training,yaman2020self,liu2020rare} train recovery and artifact-removal networks on measurement pairs. Some works \cite{metzler2018unsupervised,zhussip2019training} combine SURE and the learned denoising-based approximate message passing (LDAMP) \cite{metzler2017learned} framework for single measurements. Equivariance constraints are introduced in \cite{chen2021equivariant,chen2022robust,chen2023imaging} to learn image components in the non-trivial nullspace of $\A$. Further advancements in signal model learning from incomplete measurements sensed by multiple operators are achieved by recent research \cite{tachella2022unsupervised,tachella2022sensing}, where the necessary and sufficient conditions are both presented. Inspired by the recorrupted-to-recorrupted (R2R) \cite{pang2021recorrupted} denoising approach, \cite{quan2022dual} develops a double-head noise injection-based method and a fast adaption scheme. In a related work \cite{wang2022self}, an adaptive stochastic gradient Langevin dynamics-based sampling is employed to approximate the Bayesian estimator of GTs. These methods break free from the dependence on large amounts of labeled data. Moreover, there is also a line of works \cite{bora2017compressed,bora2018ambientgan,kabkab2018task,wu2019deep,raj2019gan,wang2022optimal} using generative adversarial networks (GAN) to reconstruct without requirements on $(\y,\x)$ pairs. Some plug-and-play (PnP) \cite{venkatakrishnan2013plug,sun2019online,ryu2019plug,kadkhodaie2021stochastic,kamilov2023plug}, regularization by denoising (RED) \cite{romano2017little,liu2022online}, and diffusion model-based \cite{kawar2022denoising,wang2023zero,chung2023diffusion,feng2023score} methods employ deep denoiser priors \cite{zhang2022plug,liu2022recovery} in a zero-shot manner, but they require GTs for modeling the clean image distribution $q_\x$.

\textbf{Research Scope and Task Setting:} This work investigates \textit{self-supervised} deep NN-based CS reconstruction, where some external measurements $\{\y^{train}_i\}$ sensed by a single fixed sampling matrix $\A^{train}\sim q_\A$ of CS ratio $\gamma^{train}\in(0,1]$ can be available for the offline training of NN $\F_\Th$ \cite{quan2022dual}. Our aim is to obtain the GT estimation $\{\xhat_i\}$ from test samples $\{\y^{test}_i\}$, matrix $\A^{test}$ of ratio $\gamma^{test}$, and $\F_\Th$ without access to GT $\x \sim q_\x$. We focus on image CS reconstruction and its SPI applications \cite{duarte2008single,li2022dual,mur2022deep} with orthonormalized Gaussian matrices \cite{donoho2009message} satisfying $\A \A^\top=\IM$ and Bernoulli matrices \cite{zhang2010compressed}. Our work can be extended to other settings \cite{tachella2022unsupervised} and data types \cite{zhao2016video}.

\textbf{Weaknesses of Existing Self-Supervised CS Methods:} \underline{\textcolor{blue}{(1) Unsatisfactory performance.}} Current self-supervised methods \cite{ulyanov2018deep,heckel2020compressive,pang2020self,xia2019training,yaman2020self,liu2020rare,metzler2018unsupervised,zhussip2019training,chen2021equivariant,chen2022robust,tachella2022unsupervised,quan2022dual,quan2022learning,wang2022self} may not compete against state-of-the-art supervised ones \cite{kulkarni2016reconnet,zhang2018ista,sun2020dual,chen2020learning,you2021ista,you2021coast}. We attribute this fact to two factors. Firstly, their insufficient utilization of data and shared knowledge among different CS tasks leads to weak regularization and the high risk of overfitting to incomplete measurements. Secondly, most approaches solely focus on one part of NN architecture and learning scheme designs (\textit{e.g.} loss function), resulting in unbalanced developments and underfitting due to their dated components and training techniques. A comprehensive guidance from advanced NN frameworks like neural representation \cite{chen2021learning,liu2022recovery,lee2022local,li2023busifusion}, algorithm unrolling \cite{gregor2010learning,monga2021algorithm,zhang2023physics}, and its augmentations \cite{chen2020learning,ning2020accurate,song2023deep,cui2022fast,song2023dynamic} can alleviate this challenge. \underline{\textcolor{blue}{(2) Lack of flexibility and scalability.}} Many self-supervised deep NN-based methods \cite{ulyanov2018deep,pang2020self,xia2019training,chen2021equivariant,chen2022robust,tachella2022unsupervised,quan2022dual,wang2022self} have been developed for image CS reconstruction. However, their flexibility is generally limited. This is because, although these methods can address CS reconstruction problems with various sampling matrices or CS ratios by replacing the input and training data, they consider the problems as independent tasks. As a result, to address unseen CS tasks of different sampling matrices or CS ratios, they require additional measurement samplings to create new training datasets, alongside extensive NN re-trainings or considerable model adaptations for new conditions. Such a process not only increases the demand on sampling, model number, and training time but also potentially brings detrimental effects to the object being captured (\textit{e.g.} human body) and will consume considerable storage and computational overheads, making them impractical for deployments \cite{shi2019scalable}. For instance, in wireless broadcast \cite{li2013new,yin2016compressive}, terminal users can receive images of different ratios based on their channel conditions. In medical and biomicroscopic imaging, additional measurements can be acquired to enhance tissue and cell details. Suppose there are $d$ systems equipped with $d$ complete matrices, current methods \cite{xia2019training,chen2021equivariant,chen2022robust,tachella2022unsupervised,quan2022dual} may need up to $dN$\footnote{There are at most $N$ discrete CS ratios in practice, \textit{i.e.}, $\gamma$ can only be selected from $\{1/N,2/N,\cdots,N/N\}$. In our experiments, $d$ and $N$ can reach $2$ and $1089$ (or $16384$), respectively. They can be even much larger in real CS applications.} dataset constructions, NN trainings, or adaptions to address all possible requirements. Thus, a ratio-scalable and matrix-adaptive reconstruction method \cite{boyce2015overview,shi2019scalable,you2021coast,zhong2022scalable,chen2022content} is preferred for real applications, which is not adequately addressed by existing self-supervised deep CS methods.

This paper proposes a novel \textbf{S}elf-supervised s\textbf{C}alable deep NN-based CS method, consisting of a \textbf{L}earning scheme (\textbf{SCL}) and a \textbf{Net}work family (\textbf{SCNet}). Inspired by the impressive success of N2N \cite{lehtinen2018noise2noise} and controllable NNs \cite{zhang2018ffdnet,he2019modulating,he2020interactive,cai2021toward,you2021ista,you2021coast,chen2022content,zhang2022plug,mou2022metric}, SCL imposes a cross-supervision mechanism \cite{yaman2020self} between two divided parts of each CS measurement with random lengths and a cycle-consistency constraint on sampling-reconstruction pipeline regarding arbitrary CS ratios and matrices \cite{you2021coast} to improve task diversity and NN robustness. SCNet combines the concepts of algorithm unrolling \cite{monga2021algorithm} and implicit neural representation \cite{liu2022recovery} to develop a new recovery NN that achieves a better trade-off among performance, flexibility, scalability, complexity, and interpretability compared to existing approaches. Our method can be used for arbitrary scenarios $\x\sim q_\x$, CS ratios $\gamma=(M/N) \in (0,1]$, and sampling matrices $\A\sim q_\A$ once trained on the measurement set of a single matrix with fixed ratio (\textit{e.g.} 10\%), and can flexibly adapt to test samples for further improvements if needed. By resolving the above-mentioned two limitations \textcolor{blue}{(1)} and \textcolor{blue}{(2)}, it paves a practical way for high-quality and scalable CS for data-limited and requirement-changeable real environments.

\begin{figure*}[!t]
\centering
\includegraphics[width=1.0\textwidth]{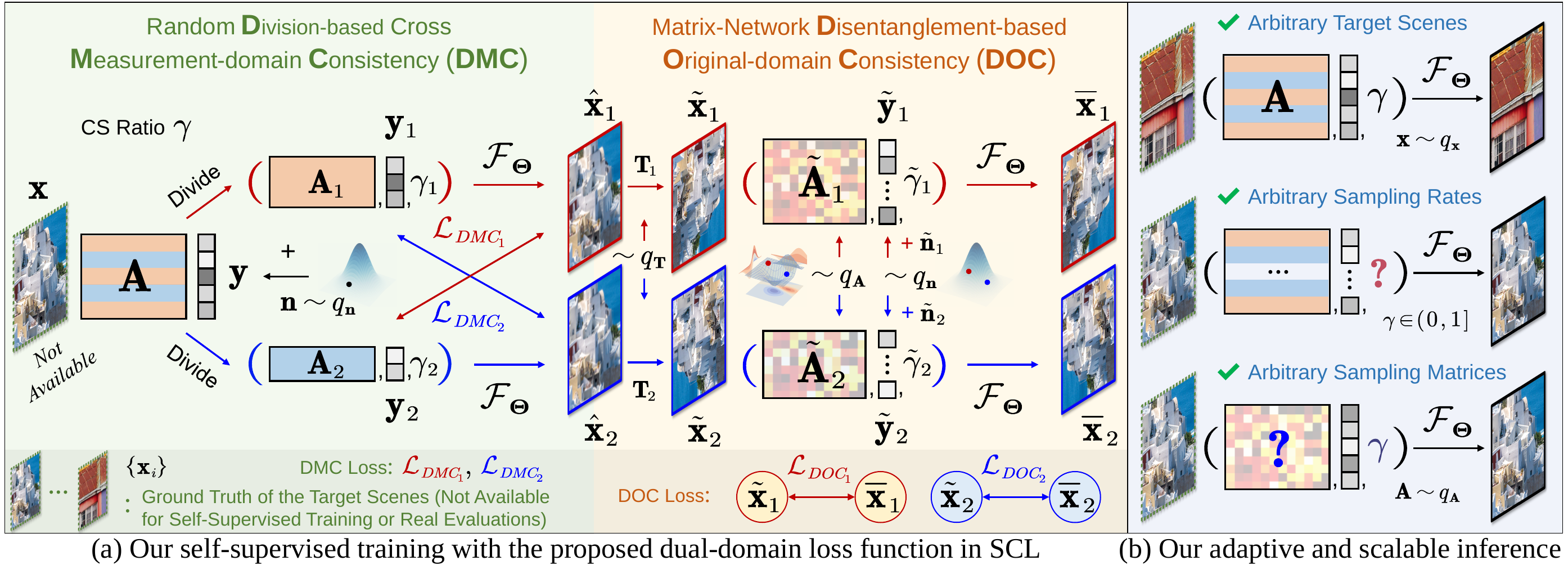}
\caption{Illustration of our self-supervised dual-domain loss. \textcolor{blue}{(a)} We propose to encourage both the DMC and DOC for any sampling matrices, CS ratios, and observation noises, aiming to learn the common priors of latent signals and the entire CS domain without overfitting to specific ones. \textcolor{blue}{(b)} Once the recovery NN is learned, it can be generalized to arbitrary in-distribution target scenes, ratios, and matrices not seen in the training set, to efficiently adapt to changeable requirements.}
\label{fig:loss}
\end{figure*}

\textbf{Contributions:} \textbf{(1)} A novel self-supervised scalable CS learning scheme named SCL, which includes a dual-domain loss function and a four-stage deep progressive reconstruction strategy. \textbf{(2)} A new NN family dubbed SCNet, which combines explicit guidance from traditional proximal gradient descent (PGD) algorithm \cite{parikh2014proximal} and implicit regularization\footnote{Implicit regularization enables NNs to learn meaningful image patterns inherently, reducing overfitting and aligning results with actual image distributions without explicit terms or guidance. Building upon existing self-supervised image reconstruction research \cite{ulyanov2018deep,quan2022dual,qin2023ground}, our method utilizes its effect, as detailed in Sec.~\ref{subsec:SCNet}. We substantiate the effectiveness and rationale of NN's implicit regularization through experiments documented in both the \MP~and \Appx.} from advanced NN components. \textbf{(3)} Extensive experiments conducted on simulated and real data of 1D, 2D, and 3D signals exhibit the significant superiority and effectiveness of our method in improving both recovery quality and generalization ability. A real SPI optical system is established for performance validation on fluorescence microscopy \cite{lichtman2005fluorescence}. Corresponding analyses and discussions are also provided to try to offer an understanding of our method's working principle.

Note that this work introduces three novel contributions that have not been extensively explored previously, despite existing literature presenting similar methodologies in dual-domain loss, four-stage strategy, and unrolled NNs. While these concepts may not be entirely new, our work extends these ideas, demonstrating unique insights and notable performance in a specific context of image CS. For a comprehensive discussion of how our contributions differ from related work, please refer to Sec.~\ref{sec:relationships_to_other_works}.

\section{Proposed Method}
This section introduces our SCL and SCNet. To be concrete, Sec.~\ref{subsec:SCL} presents SCL, comprising a dual-domain loss and a four-stage recovery strategy. The dual-domain loss includes (1) a Division-based cross Measurement-domain Consistency (DMC) constraint, which uses a maximized fine-grained data augmentation (by up to approximately $2^M \times$ expansion in the size of measurement set), as well as (2) a matrix-network Disentanglement-based Original-domain Consistency (DOC) constraint, that leverages the available information of distributions $q_\x$, $q_\n$, and $q_\A$ to impose consistency on the sampling-reconstruction cycle. The four-stage strategy first learns the common signal prior among external measurements and then exploits the internal statistics of test samples in a coarse-to-fine manner. Sec.~\ref{subsec:SCNet} presents SCNet, a family of simple yet efficient recovery NNs that can work synergistically and achieve mutual support with SCL.

\subsection{Self-Supervised Scalable Reconstruction Learning (SCL)}
\label{subsec:SCL}
\textbf{(1) Dual-Domain Loss Function.} Given NN architecture $\F_\Th$, measurement $\y$, matrix $\A$, and ratio $\gamma$, a very widely adopted and straightforward practice of deep self-supervised CS reconstruction is to enforce the measurement consistency (MC) on $\F_\Th$ using a loss function of form like $\Loss_{MC}=\lVert \A\xhat - \y \rVert_2^2$ with $\xhat=\F_\Th(\y)$, which, however, easily brings overfitting to inharmonious results, faced with two critical problems \cite{pang2020self,chen2022robust,quan2022dual}: (i) The presence of noise. $\n$ is not suppressed and will propagate from $\y$ to $\xhat$. (ii) The solution ambiguity. $\xhat$ converges freely to $\A^\dagger\y+\mathcal{H}(\y)$ satisfying $\A\xhat=\A\A^\dagger\y+\A\mathcal{H}(\y)=\y$ with $\mathcal{H}(\y)\in \text{null}(\A)$, where $\A^\dagger$ is the right pseudo-inverse of $\A$ and $\text{null}(\A)=\{\mathbf{w}\in\Rbb^N:\A\mathbf{w}=\mathbf{0}_M\}$ is the $(N-M)$-dimensional nullspace of $\A$. This causes unstable outputs without meeting $\xhat = \x$.

To mitigate these issues and make $\F_\Th$ adaptive to the changes in CS ratio and sampling matrix, we develop a new Division-based Measurement-domain Consistency (DMC) loss, which employs the combinations of measurement elements from $\y$ to construct thousands of new tasks with different matrices and ratios for augmentation on data diversity. As illustrated in Fig.~\ref{fig:loss} (a, left), we randomly divide the original task $(\y,\A,\gamma)$ into two complementary parts $(\y_1,\A_1,\gamma_1)$ and $(\y_2,\A_2,\gamma_2)$ satisfying the conditions of $\Set_{\y_1}\cup~\Set_{\y_2}=\Set_{\y}$, $\Set_{\A_1}\cup~\Set_{\A_2}=\Set_{\A}$,
$\Set_{\y_1}\cap~\Set_{\y_2}=\Set_{\A_1}\cap~\Set_{\A_2}=\varnothing$, and $\gamma_1 + \gamma_2 = \gamma$. Here, $\mathbf{S}_\y=\{(i,\y[i])\}_{i=1}^M$ and $\mathbf{S}_\A=\{(i,\A[i])\}_{i=1}^M$ are the set representations of $\y$ and $\A$. $\y[i]$ and $\A[i]$ denote the measurement element and matrix row at index $i$, while $M_1$ and $M_2$ represent the dimensions of $\y_1$ and $\y_2$ with $M_1+M_2=M$, respectively. Our goal is to reconstruct $\xhat_1 =\F_\Th(\y_1,\A_1,\gamma_1)$ that takes matrix and ratio as conditional inputs and can conform to the following two constraints:
\begin{gather}
\textit{Self-MC Constraint:}\quad\A_1\xhat_1 = \A_1\x,~~
\label{eq:self_mc}
\end{gather}
\begin{gather}
\textit{Cross-MC Constraint:}\quad\A_2\xhat_1 = \A_2\x.
\label{eq:cross_mc}
\end{gather}
While the constraint in Eq.~(\ref{eq:self_mc}) can be partially imposed by a self-MC loss $\Loss_{self_1}=\lVert \A_1\xhat_1 - \y_1 \rVert_2^2$ \cite{xia2019training} and a SURE-based loss \cite{chen2022robust}, here we focus on the encouragement of Eq.~(\ref{eq:cross_mc}) by a cross-MC loss $\Loss_{cross_1}=\lVert \A_2\xhat_1 - \y_2 \rVert_2^2$, which is noise-resistant under an assumption as shown below:

\vspace{5pt}
\textbf{Assumption (Distribution of the observation noise).} \textit{The observation noise $\n$ is additive white Gaussian noise (AWGN) of zero mean and standard deviation (or noise level) $\sigma$, \textit{i.e.}, $\n \sim q_\n=\mathcal{N}(\mathbf{0}_M,\sigma^2\mathbf{I}_M)$. The noise components $\n_1$ and $\n_2$ in the divided measurements $\y_1=\A_1 \x+\n_1$ and $\y_2=\A_2 \x+\n_2$ are mutually independent.}

\vspace{5pt}
Based on the above assumption, here we show a property of $\Loss_{cross_1}$ by the following theorem:

\vspace{5pt}
\textbf{Theorem (Equivalence of the cross-MC loss to its noise-free counterpart).} \textit{The cross-MC loss is equivalent in terms of expectation to noise-free one for imposing Eq.~(\ref{eq:cross_mc}), i.e., there is:}
\begin{gather*}
J= \Ebb_{\y_1,\y_2} \{ \lVert \A_2 \F_\Th (\y_1,\A_1,\gamma_1) - \y_2 \rVert_2^2 \}\\
= \Ebb_{\x,\n_1} \{ \lVert \A_2 \left[ \F_\Th (\A_1 \x+\n_1,\A_1,\gamma_1) - \x \right] \rVert_2^2 \} + M_2\sigma^2.
\end{gather*}
\begin{proof}
The expectation $J$ can be expanded as:
\begin{align*}
J=~&\Ebb_{\y_1,\y_2}\left\{\lVert \A_2 \F(\y_1,\A_1,\gamma_1) - \y_2 \rVert_2^2\right\}\\
=~&\Ebb_{\x,\n_1,\n_2}\{\lVert \A_2 \F(\A_1\x+\n_1,\A_1,\gamma_1) \\
& \quad \quad \quad \quad \quad - (\A_2\x+\n_2) \rVert_2^2 \} \\
=~&\Ebb_{\x,\n_1,\n_2}\left\{\lVert \A_2\left[ \F(\A_1\x+\n_1,\A_1,\gamma_1) - \x \right] -\n_2 \rVert_2^2\right\}\\
=~&\Ebb_{\x,\n_1,\n_2}\left\{\lVert \A_2\left[ \F(\A_1\x+\n_1,\A_1,\gamma_1) - \x \right] \rVert_2^2 \right\}\\
&-\Ebb_{\x,\n_1,\n_2}\left\{2\n_2^\top\A_2\left[ \F(\A_1\x+\n_1,\A_1,\gamma_1) - \x \right]\right\}\\
&+\Ebb\left\{\n_2^\top \n_2 \right\}.
\end{align*}
For the second term, since $\n_2$ has zero mean and is independent to $\n_1$ and $\x$, we have:
\begin{align*}
&\Ebb_{\x,\n_1,\n_2}\left\{2\n_2^\top\A_2\left[ \F(\A_1\x+\n_1,\A_1,\gamma_1) - \x \right]\right\}\\
=~&2{\underbrace{\left(\Ebb\left\{\n_2\right\}\right)}_{\mathbf{0}_{M_2}}}^\top\Ebb_{\x,\n_1}\left\{\A_2\left[ \F(\A_1\x+\n_1,\A_1,\gamma_1) - \x \right]\right\}=0.
\end{align*}
For the last term, since different elements of $\n_2$ are mutually independent, with variance $\sigma^2$, we have:
\begin{align*}
&\Ebb\left\{\n_2^\top \n_2 \right\}=\sum_{i=1}^{M_2}\Ebb\left\{(\n_2[i])^2\right\}\\
=~&\sum_{i=1}^{M_2}~[\sigma^2-{\underbrace{\left(\Ebb\left\{\n_2[i]\right\}\right)}_{0}}^2]=M_2\sigma^2.
\end{align*}
Combining the above three equations, we have:
\begin{gather*}
J=\Ebb_{\y_1,\y_2}\left\{\lVert \A_2 \F(\y_1,\A_1,\gamma_1) - \y_2 \rVert_2^2\right\}\\
=\Ebb_{\x,\n_1,\n_2}\left\{\lVert \A_2\left[ \F(\A_1\x+\n_1,\A_1,\gamma_1) - \x \right] \rVert_2^2 \right\}+M_2\sigma^2.
\end{gather*}
The proof is done.
\end{proof}

The above theorem can be regarded as an extension of the theoretical results presented in N2N \cite{lehtinen2018noise2noise} and Self2Self (S2S) \cite{quan2020self2self}. N2N demonstrates that training an NN on noisy image pairs with independent, zero-mean Gaussian noise distributions is equivalent to training on noisy-clean pairs. S2S shows that training on pairs of a Bernoulli sampled image and its complementary one from a single noisy image is equivalent to training with a Bernoulli sampled image and the ground truth image. While both methods are designed for denoising under the special setting of $\A = \mathbf{I}_N$, our theorem extends their scope. It validates that denoising in the measurement domain through random division is viable for the general cases of $\A \neq \mathbf{I}_N$. It is worth noting that the theorem is built upon the expectation calculation. It implicitly assumes that the training dataset is sufficiently large to generate many pairs of split measurements, $\y_1$ and $\y_2$, with noise components $\n_1$ and $\n_2$. The measurements $\{\y\}$ correspond to a sufficiently large number of different latent clean images $\{\x\}$ that often share similarities, such as similar patterns and contents. For example, in our experiments on real data, we collected thousands of measurements from scenes of similar biological samples for training. In such cases, the trained network becomes a general reconstructor capable of successfully recovering images from measurements of unseen images. Conversely, if the training measurements are too few, for instance, as few as one, the network might overfit to the measurements and become ineffective in recovering new, unseen images.

\begin{table*}[!t]
\caption{Details of the developed four-stage self-supervised progressive CS reconstruction strategy in our SCL method.}
\label{tab:four_stage_strategy}
\centering
\resizebox{\textwidth}{!}{\begin{tabular}{>{\columncolor[HTML]{EFEFEF}}ll}
\shline
(Stage-1) Training-time offline external learning: & Fit $\F_\Th$ on $\{\y_i^{train}\}$ of size $l^{train}$ and $\A^{train}$ by our $\Loss$ \\
(Stage-2) Test-time cross-image internal learning: & Fit $\F_\Th$ on $\{\y_i^{test}\}$ of size $l^{test}$ and $\A^{test}$ by our $\Loss$ \\
(Stage-3) Test-time single-image internal learning: & Fit $\F_\Th$ on the specific task $(\y^{test},\A^{test},\gamma^{test})$ by our $\Loss$ \\
(Stage-4) Test-time single-image self-ensemble: & Generate estimation by $\xhat^* = \Ebb_{\M,\T}\left\{\T^{-1}\F_{\Th}(\y,\M\A\T,\gamma)\right\}$ \\
\shline
\end{tabular}}
\end{table*}

\vspace{5pt}
Moving beyond the cross-MC loss, we propose to generalize the $\ell_2$ error form in $\Loss_{cross_1}$ from $\lVert \cdot \rVert_2^2$ to the $\ell_p$ form $\lVert \cdot \rVert_p^p$ and set $p=1$ by default. The reason is that the original $\ell_2$ loss can lead to undesirable biased learning of $\F_\Th$ toward low-ratio tasks, since their loss values may become almost 1-3 orders of magnitude larger than those of high-ratio ones due to their less input information and higher recovery difficulty, while our $\ell_p$ loss can significantly alleviate this issue in practice. Another loss term $\lVert \A_1\xhat_2 - \y_1 \rVert_p^p$ symmetric to $\Loss_{cross_1}$ with $\xhat_2=\F_\Th(\y_2,\A_2,\gamma_2)$ is introduced to make the CS ratios balanced and data utilization adequate in the training process, thus improving final performance of NNs. We do not enforce the self-MC in Eq.~(\ref{eq:self_mc}) as we observe that the cross-MC encouragement in Eq.~(\ref{eq:cross_mc}) is dominant and competent for assurance of reconstruction accuracy. To summarize, our DMC loss is defined as follows:
\begin{gather}
\Loss_{DMC}=\Loss_{DMC_1}+\Loss_{DMC_2},\nonumber\\
\Loss_{DMC_1}=\lVert \A_2\xhat_1 - \y_2 \rVert_p^p,\nonumber \\
\Loss_{DMC_2}=\lVert \A_1\xhat_2 - \y_1 \rVert_p^p.
\label{eq:dmc_loss}
\end{gather}
To liberate the recovery NN $\F_\Th$ from the dependence on specific internal information of $(\y,\A,\gamma)$ instances and enable its learning beyond the measurement $\y$, matrix $\A$, and CS ratio range $(0,\gamma]$, while generalizing across the entire matrix space and range $(0,1]$, here we introduce a new matrix-network Disentanglement-based Original-domain Consistency (DOC) loss. As Fig.~\ref{fig:loss} (a, right) shows, it leverages estimation $\xhat$ from $\Loss_{DMC}$ calculation and jointly exploits the information of $q_\x$, $q_\n$, and $q_\A$ to establish a connection between the measurement and original (image) domains, enforcing a cycle-consistency of sampling-reconstruction for arbitrary in-distribution CS tasks:
\begin{gather}
\textit{DOC Constraint:}\nonumber \\
\forall \T,~\forall \Tilde{\A},~\forall \Tilde{\n},~\forall \Tilde{M}\in\{1,\cdots,N\},\nonumber \\
\bar{\x}=\F_\Th(\Tilde{\y},\Tilde{\A}, \Tilde{\gamma})=\Tilde{\x}.
\label{eq:doc_cons}
\end{gather}
Here, $\Tilde{\y}=\Tilde{\A}\Tilde{\x}+\Tilde{\n}$ is a new simulated observation, $\Tilde{\x}=\T\xhat$ is a diversity-augmented version of $\xhat$ using a geometric transformation matrix $\T\in\Rbb^{N\times N}$ \cite{chen2023imaging}, while the ratio is given by $\Tilde{\gamma} =\Tilde{M}/N$. Based on Eq.~(\ref{eq:doc_cons}), our DOC loss is defined as a sum of two terms for both estimations $\xhat_1$ and $\xhat_2$:
\begin{gather}
\Loss_{DOC}=\Loss_{DOC_1}+\Loss_{DOC_2},\nonumber \\
\Loss_{DOC_1}=\lVert \bar{\x}_1 - \Tilde{\x}_1 \rVert_p^p, \quad
\Loss_{DOC_2}=\lVert \bar{\x}_2 - \Tilde{\x}_2 \rVert_p^p, \nonumber \\
\forall i\in\{1,2\}, \quad \bar{\x}_i=\F_\Th(\Tilde{\y}_i,\Tilde{\A}_i,\Tilde{\gamma}_i), \nonumber \\
\Tilde{\y}_i=\Tilde{\A}_i \Tilde{\x}_i +\Tilde{\n}_i, \quad \Tilde{\x}_i=\T_i \xhat_i, \quad \Tilde{\gamma}_i=\Tilde{M}_i/N.
\label{eq:doc_loss}
\end{gather}
For each estimation $\xhat_i$ in the training batch, $\T_i\sim q_\T$ is first randomly drawn from eight classic group actions on the natural image set \cite{chen2021equivariant,chen2022robust}, including rotations, flippings, and their combinations \cite{chen2022content}. $\Tilde{M}_i \sim \text{Uniform}(\{1,\cdots,N\})$, $\Tilde{\A}_i \sim q_\A$, and $\Tilde{\n}_i \sim q_\n$ are then sampled to create CS task $(\Tilde{\y}_i,\Tilde{\A}_i,\Tilde{\gamma}_i)$ for DOC encouragement\footnote{In this paper, we do not strictly distinguish distributions of different data dimensions corresponding to the same data type (\textit{e.g.}, observation noise) using separate notations (\textit{e.g.}, $q_{\n,\Tilde{M}_1}$ and $q_{\n,\Tilde{M}_2}$) for conciseness without raising confusion.}. Finally, we combine the two loss parts $\Loss_{DMC}$ and $\Loss_{DOC}$ in Eqs.~(\ref{eq:dmc_loss}) and (\ref{eq:doc_loss}) using a weighting factor $\alpha$ to obtain our dual-domain self-supervised loss $\Loss =\Loss_{DMC} + \alpha \Loss_{DOC}$.

\begin{figure*}[!t]
\centering
\includegraphics[width=1.0\textwidth]{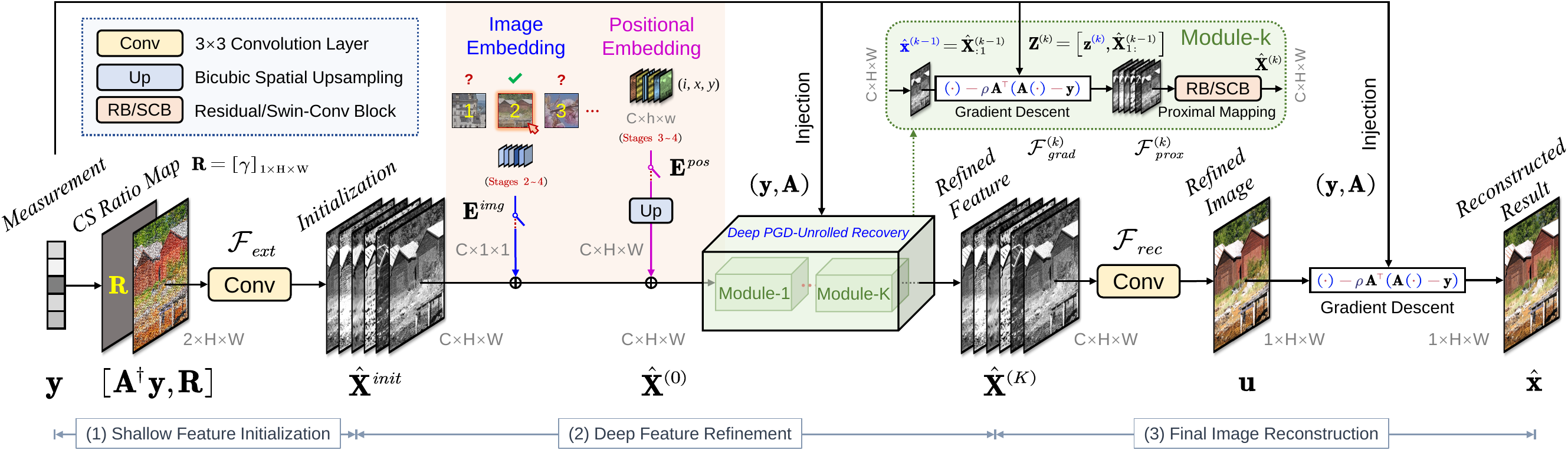}
\caption{Illustration of our developed SCNet, which first extracts a shallow feature, injects two image-specific and position-aware learnable embeddings, and then refines the feature by $K$ generalized algorithm-unrolled modules. It combines the explicit guidance of proximal gradient descent (PGD) algorithm and implicit regularization of NN blocks, yielding a collaborative representation. The final reconstruction result is obtained by a convolution with a gradient descent (GD) update.}
\label{fig:net}
\end{figure*}

\textbf{(2) Four-Stage Progressive Reconstruction Strategy.} As Fig.~\ref{fig:loss} (b) exhibits, our self-supervised loss $\Loss$ learns the latent common signal prior from external incomplete measurements in an offline manner \cite{quan2022learning}, making $\F_\Th$ adaptive for arbitrary CS tasks once training completed. Nevertheless, the success of dataset-free approaches \cite{ulyanov2018deep,pang2020self,wang2022self} shows the potential of fitting NNs to specific test $\y$ samples, enabling an online reconstruction and further improvement. Based on the insights in \cite{fu2021coded,zhang2023deep,quan2022dual}, we design a strategy that can take advantage of the GT-free nature of $\Loss$ for both training and test to alleviate the risks posed by external data bias and out-of-distribution $\y$ samples, and further enhance final performance. To be concrete, as shown in Tab.~\ref{tab:four_stage_strategy}, it consists of four stages for CS reconstruction learning.

Considering that the measurements in a single test set can share common knowledge of the same scene, such as similar structures and textures contained in their latent GTs, our test-time internal learning (stages 2 and 3) first fits $\F_\Th$ to the entire test set and then adapts it to each specific measurement in a coarse-to-fine manner. To exploit the generalization ability of $\F_\Th$ to arbitrary sampling matrices, we propose to use a self-ensemble inference in stage-4, which averages $D$ recovered images from $\F_\Th$ with $D$ randomly perturbed $\A$ inputs to approximate the final estimation $\xhat^*$ via Monte Carlo sampling \cite{gal2016dropout}:

\begin{gather}
\textit{Self-Ensemble:}\nonumber \\
\xhat^*=\Ebb_{\M,\T}\left\{\T^{-1}\F_{\Th}(\y,\M\A\T,\gamma)\right\}\nonumber \\
\approx \frac{1}{D} \sum_{i=1}^{D} \T_i^{-1}\F_\Th(\y,\M_i\A\T_i,\gamma),
\label{eq:self_ensemble}
\end{gather}
where $\M\in\{0,1\}^{M\times M}$ is a diagonal matrix whose elements have a probability of $r$ for being $0$ and $(1-r)$ for being $1$. This self-ensemble stems from our observation that the geometric transformations in $\Loss_{DOC}$ and random maskings on $\A$ with a small $r\in[0,1]$ can preserve the low-frequency smooth image areas with high confidence, and generate predictions for high-frequency textures, which exhibit certain degree of independence and can be averaged to suppress artifacts and yield better results \cite{quan2022learning}. It can be closely related to our unrolled NN design that is described in Sec.~\ref{subsec:SCNet}. In practice, several customizations on the strategy can be flexibly implemented based on specific requirements, like the fully-activated one denoted by $(1\rightarrow 2\rightarrow 3 \rightarrow 4)$ for best recovery quality, $(1\rightarrow 4)$ for training resource-constrained devices, and $(3 \rightarrow 4)$ for dataset-free deployments. The NN parameters in $\Th$ are randomly initialized for the first stage of each configuration and saved to the next ones once learning completed stage-by-stage.

\subsection{Collaborative Representation-based Reconstruction Network (SCNet)}
\label{subsec:SCNet}
CS recovery from $(\y,\A,\gamma)$ can be formulated as a regularized optimization problem: $\min_\x~\frac{1}{2}\lVert \A\x-\y \rVert_2^2 + \lambda\mathcal{R}(\x)$, which can be solved by the proximal gradient descent (PGD) algorithm \cite{parikh2014proximal} as follows:
\begin{gather}
\textit{Gradient Descent (GD):}\nonumber \\ \z^{(k)}=\xhat^{(k-1)}-\rho \A^\top(\A \xhat^{(k-1)}-\y)\in \Rbb^N,
\label{eq:pgd_grad_original}
\end{gather}
\begin{gather}
\textit{Proximal Mapping (PM):}\nonumber \\ \xhat^{(k)}=\mathtt{prox}_{\lambda \mathcal{R}}(\z^{(k)})\nonumber \\
=\arg\underset{\x}{\min}~\frac{1}{2}\lVert \x-\z^{(k)} \rVert_2^2 + \lambda \mathcal{R}(\x)\in \Rbb^N,
\label{eq:pgd_prox_original}
\end{gather}
where $k$ and $\rho$ denote the iteration index and step size, respectively. Following \cite{gregor2010learning,monga2021algorithm} and their augmentations \cite{zhang2023physics,ning2020accurate,song2023deep,cui2022fast,song2023dynamic}, we develop SCNet to implement $\F_\Th$ that is inspired by PGD and can adapt to different tasks once trained with SCL. As Fig.~\ref{fig:net} shows, its recovery consists of three sub-processes:

\textbf{(1) Shallow Feature Initialization.} The measurement $\y$ is first transformed to the image domain by $\A^\dagger \y$ and concatenated with a CS ratio map $\mathbf{R}$, which has the same shape as $\x\in\Rbb^{1\times H\times W}$ with all elements being $\gamma$. The shallow feature of data concatenation $[\A^\dagger \y,\mathbf{R}]\in\Rbb^{2\times H\times W}$ is then extracted by a convolution layer $\F_{ext}$ for initialization as $\Xhat^{init}=\F_{ext}([\A^\dagger \y,\mathbf{R}])\in\Rbb^{C\times H\times W}$.

\textbf{(2) Deep Feature Refinement.} We extend deep PGD unrolling \cite{monga2021algorithm} from traditional iterative image-level optimization to a high-throughput feature-level recovery with a maximized information flow passing through the whole network trunk \cite{zhang2023physics}. Specifically, given an initial starting feature $\Xhat^{(0)}\in\Rbb^{C\times H\times W}$, we generalize the original PGD steps in Eqs.~(\ref{eq:pgd_grad_original}) and (\ref{eq:pgd_prox_original}) to feature-level unrolling, denoted by $\F_{grad}^{(k)}$ and $\F_{prox}^{(k)}$, as shown below:
\begin{gather}
\textit{Generalized GD from }(\ref{eq:pgd_grad_original})\textit{:}\nonumber \\
\Z^{(k)}=\F_{grad}^{(k)}(\Xhat^{(k-1)};\y,\A)\nonumber \\
=\left[\z^{(k)},\Xhat_{1:}^{(k-1)}\right]\in\Rbb^{C\times H\times W},
\label{eq:pgd_grad_generalized}
\end{gather}
\begin{gather}
\textit{Generalized PM from }(\ref{eq:pgd_prox_original})\textit{:}\nonumber \\
\Xhat^{(k)}=\F_{prox}^{(k)}(\Z^{(k)})\in\Rbb^{C\times H\times W}\nonumber \\
\text{(}\F_{prox}^{(k)}\text{ is an RB or SCB)}.
\label{eq:pgd_prox_generalized}
\end{gather}

Here, $\Z^{(k)}$ is obtained by analytically updating only the first channel of $\Xhat^{(k-1)}$ using Eq.~(\ref{eq:pgd_grad_original}) with $\xhat^{(k-1)}=\Xhat_{:1}^{(k-1)}$, while leaving the other feature channels $\Xhat_{1:}^{(k-1)}$ unchanged. The second update step of $\Xhat^{(k)}$ is implemented by a single residual block (RB) \cite{he2016deep} for the convolutional SCNet version called \textbf{SC-CNN}, or an advanced Swin-Conv block (SCB) \cite{zhang2022practical} for the Transformer-based SCNet version named \textbf{SCT}. We unroll a total of $K\in\mathbb{Z}^{+}$ generalized PGD iterations with each one being converted into an NN module that sequentially conducts the two steps in Eqs.~(\ref{eq:pgd_grad_generalized}) and (\ref{eq:pgd_prox_generalized}). We set $\rho\equiv 1$ for $\sigma=0$ cases and re-parameterize it as $\rho=\text{Sigmoid}(\tau)$ to make it data-driven and limited in the range of $(0,1]$ with a learnable parameter $\tau \in\mathbb{R}$ for $\sigma >0$ cases. We have empirically observed that these straightforward $\rho$ settings can be competent for facilitating our self-supervised ``SCL+SCNet'' training.

To enhance flexibility and the balance between performance and interpretability, SCNet not only constrains results by explicit PGD framework and implicit regularizations of locality and translation invariance brought by convolutions, RBs, and SCBs, but also introduces two learnable embeddings to increase optimization freedom and prevent underfitting. The first is an image embedding (IE) $\mathbf{E}^{img}\in\Rbb^{C\times 1\times 1}$ activated only for stages 2-4 and generated for each test image to help $\F_\Th$ distinguish each specific measurement ($\y$) instance from the entire test set. The second is a positional embedding (PE) $\mathbf{E}^{pos}\in\Rbb^{C\times h\times w}$ activated only for stages 3-4 and first upscaled by a bicubic interpolation to match the feature size. It can have various values for different areas and make $\F_\Th$ position-aware and spatially-variant. Both $\mathbf{E}^{img}$ and $\mathbf{E}^{pos}$ are zero-initialized and added to $\Xhat^{init}$ channel-/element-wise to generate feature $\Xhat^{(0)}$ before PGD-unrolled recovery. Output is the refined feature $\Xhat^{(K)}$ from the $K$-th deep module.

\textbf{(3) Final Image Reconstruction.} A refined image is first generated from deep feature by $\mathbf{u}=\F_{rec}(\Xhat^{(K)})\in\Rbb^{1\times H\times W}$ using a convolution layer $\F_{rec}$. The final recovered result is then obtained by an extra GD step similar to Eq.~(\ref{eq:pgd_grad_original}) as follows:
\begin{gather}
\textit{Final GD:}\quad\xhat=\mathbf{u}-\rho\A^\top(\A\mathbf{u}-\mathbf{y})\nonumber \\
=\left[\rho\A^\dagger\y+(\mathbf{I}_N-\rho\A^\dagger\A)\mathbf{u}\right]\in\Rbb^N\nonumber \\
(\text{with~}\A^\top =\A^\dagger).
\label{eq:final_gd}
\end{gather}
It should be highlighted that when $\sigma=0$, our setting $\rho\equiv 1$ ensures that $\xhat$ strictly conforms to the self-MC in Eq.~(\ref{eq:self_mc}), \textit{i.e.}, $\A\xhat =\y=\A\x$, with self-loss always being zero if calculated. In the presence of noise $(\sigma>0)$, the measurement of $\xhat$ becomes a weighted sum given by $\A\xhat=\left[\rho \y + (1-\rho)\A\mathbf{u}\right]\in\Rbb^M$, while the nullspace component \cite{chen2020deep} of $\xhat$ will remain the same as that of $\mathbf{u}$, \textit{i.e.}, $(\mathbf{I}_N-\A^\dagger\A)\xhat=(\mathbf{I}_N-\A^\dagger\A)\mathbf{u}\in\Rbb^N$. This final GD step is motivated by the success of residual learning \cite{zhang2017beyond} for measurement denoising and has been demonstrated to be effective in facilitating our ``SCL+SCNet'' training\footnote{While our method is primarily developed and analyzed within the context of using orthonormalized Gaussian matrices, where $\A^\top = \A^\dagger$ holds true, we note a deviation in this assumption when applying Bernoulli matrices in our experiments on real data. In these instances, the equality $\A^\top = \A^\dagger$ typically does not hold. Consequently, the final GD step cannot ensure self-MC. To address this issue, we adapt our methodology by redefining the step size $\rho$ in a more flexible manner: $\rho = \text{Sigmoid}(\tau)$, where $\tau$ is a learnable parameter, allowing the model to adjust this value effectively during training.}.

Formally, all the learnable parameters of our SCNet can be expressed as $\Th=\{\F_{ext},\F_{rec},\tau\}\cup \{\mathbf{E}_i^{img},\mathbf{E}_i^{pos}\}_{i=1}^{l^{test}}\cup \{\F_{prox}^{(k)}\}_{k=1}^K$, where $l^{test}$ is the size of test set. Each unrolled module of SCNet has its own parameters in $\F_{prox}^{(k)}$ implemented by an RB or SCB for SC-CNN or SCT. We define the SCNets trained by stage-1 only as the standard versions, while their enhanced ones are obtained by incorporating the complete SCL stages 1-4, denoted as \textbf{SC-CNN$^+$} and \textbf{SCT$^+$}, respectively.

\subsection{Relationship to Other Works}
\label{sec:relationships_to_other_works}
The utilization of NNs for self-supervised image reconstruction tasks has gained increasing interest, particularly in the context of image denoising, which can be a critical component for inverse imaging problems \cite{pang2020self}. Approaches such as deep image prior (DIP) \cite{ulyanov2018deep}, Noise2Noise (N2N) \cite{lehtinen2018noise2noise}, Stein's unbiased risk estimator (SURE) \cite{stein1981estimation}-based methods \cite{soltanayev2018training,zhussip2019extending}, and others \cite{krull2019noise2void,batson2019noise2self,quan2020self2self,pang2021recorrupted} show promising capabilities in measurement noise suppression. However, they mainly solve the special setting of $\A=\mathbf{I}_N$.

Many self-supervised inverse imaging methods are devoted to tackling the challenges of eliminating solution ambiguity. In this discussion, we focus on the works most relevant to our own research. Untrained NN priors \cite{ulyanov2018deep,pang2020self} and random sampling-based methods \cite{wang2022self} fit an NN to each test $\y$ sample, but suffer from low efficiency and underutilization of available datasets. Xia \textit{et al.} \cite{xia2019training} require paired measurements of the same scene, obtained from two different sensors, which may not be feasible in resource-limited environments. Similar ideas of cross-supervision are explored in previous works \cite{yaman2020self,liu2020rare,zhou2022dual,zhou2023dsformer}. Other methods \cite{metzler2018unsupervised,zhussip2019training} learn SURE-based denoisers from single measurements. They still face challenges related to unbalanced development, weak regularization, and limited data exploitation.

Two recent studies, equivariant imaging (EI) \cite{chen2021equivariant} and multiple operator imaging (MOI) \cite{tachella2022unsupervised}, address the problem of self-supervised reconstruction by encouraging equivariances to geometric transformations (such as shifts and rotations) and ensuring sampling-reconstruction consistency across different sampling operators beyond those used for sampling the training measurement set. Our dual-domain loss function shares similarities with these approaches, incorporating transformations $\T$ and employing numerous random sampling matrices during training. However, our method advances beyond these studies by leveraging the internal combinations of measurement data to foster cross-consistency and by fully utilizing the information of the observation noise and sampling matrix distributions. The proposed approach not only bolsters the performance of NNs through the maximized leverage of data but also facilitates \textit{ratio-scalable} and \textit{matrix-adaptive} CS recoveries. Such characteristics hold significant appeal for deployments.

Drawing on the concepts introduced by Noiser2Noise \cite{moran2020noisier2noise} and Self-Supervised learning via Data Undersampling (SSDU) \cite{yaman2020self}, a recent study \cite{millard2023theoretical} introduces a variable density Noiser2Noise approach, which exhibits parallels to our method of fostering cross-MC through the division of the matrix $\A$ into $\A_1$ and $\A_2$. A key difference between our approach and this recent study lies in the foundational assumption of their work: they assume that the undersampling mask is a random variable whose expectation is a square, diagonal, and full-rank matrix. In contrast, our method is predicated on a fixed sampling matrix $\A$ for training measurement set, meaning that our approach does not sample the nullspace component of images with respect to a single $\A$ during training. Hence, our research offers a complementary perspective to this recent work in this specific context.

In particular, we note that the recent PARCEL \cite{wang2022parcel} method is based on the concept of measurement re-undersampling, similar to our DMC loss. However, PARCEL uses predetermined and fixed undersampling masks/rates for specific reconstruction tasks. While recent approaches like DDSSL \cite{quan2022dual} and MetaCS \cite{qin2023ground}, inspired by the R2R method \cite{pang2021recorrupted}, introduce a dual-domain loss function and meta-learning respectively, akin to our ``DMC+DOC'' loss combination, they rely on noise injection and fixed sampling matrix information for learning. Our approach, in contrast, randomly divides measurement data, potentially creating up to $\sim 2^{M}$ new samples, and leverages the information of $q_\A$ and $q_\n$. As a result, our NNs are more flexible to handle different situations than those of PARCEL, DDSSL, and MetaCS. To be concrete, our SCNet can manage various CS tasks after training with a specific measurement set and fixed ratio. Moreover, our comprehensive four-stage strategy extends the model adaptation schemes of DDSSL and MetaCS by utilizing shared information across similar images (stage 2) and enhancing reconstruction through the generalization ability of our NNs (stage 4).

In addition, while recent studies \cite{zhou2022dual, wang2023hyperspectral} integrate deep unrolling with Transformers, our method differs in three main ways: (1) we extend deep PGD unrolling from traditional image optimization to generalized feature recovery; (2) we introduce image and positional embeddings for higher optimization freedom; and (3) we employ a final GD step for improved recovery. These enhancements are vital for self-supervised CS reconstruction.

In summary, our method offers a distinctive, comprehensive, and general solution. It enhances the current techniques and incorporates a complete design of loss functions, reconstruction strategies, and NN architectures, and can be applied to tasks with different matrix requirements and ratios once trained. Moreover, while most existing studies only conduct numerical simulations for evaluation, remaining gaps for real deployments, as the following section shows, our method has been validated on a real SPI optics system, paving the way for the applications of CS in scientific research.

\begin{figure*}[!t]
\centering
\begin{minipage}[c]{0.25\linewidth}
\centering
\scalebox{0.63}{
\begin{tabular}[c]{lcc}
\shline
\rowcolor[HTML]{EFEFEF} 
Method                       & (a)   & (b)   \\
$\A^\dagger \y$ & 4.17  & 13.76 \\
MC                           & 11.77 & 14.51 \\
DMC                          & 14.65 & 26.05 \\
MC+OC                        & 16.93 & 34.77 \\
MC+DOC                       & 28.74 & 36.60 \\
\textbf{DMC+DOC}                      & \textcolor{blue}{\underline{31.35}} & \textcolor{blue}{\underline{41.08}} \\
Sup                          & \textcolor{red}{\textbf{32.01}} & \textcolor{red}{\textbf{43.57}} \\ \shline
\end{tabular}}\\
\vspace{-1pt}\hspace{0.5pt}
\centering
\scalebox{0.85}{\includegraphics[width=\linewidth]{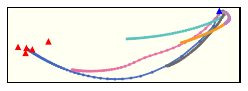}}
\end{minipage}
\hfill
\begin{minipage}[c]{0.26\linewidth}
\centering
\includegraphics[width=\linewidth]{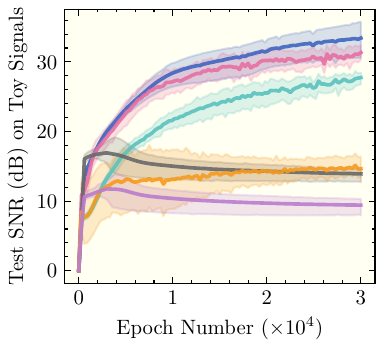}
\end{minipage}
\hfill
\begin{minipage}[c]{0.45\linewidth}
\centering
\includegraphics[width=\linewidth]{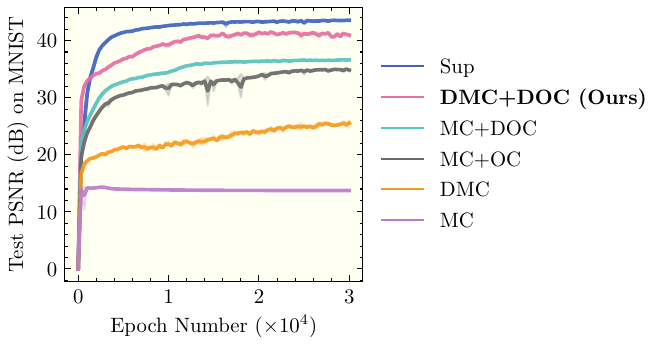}
\end{minipage}
\caption{Comparisons of the CS reconstruction for 1D toy signals and 2D MNIST digit images \cite{deng2012mnist} among $\A^\dagger \y$ and six SC-CNNs with scheme names (loss functions) in quotes (brackets): ``MC'' ($\Loss_{MC}$), ``DMC'' ($\Loss_{DMC}$), ``MC+OC'' ($\Loss_{MC}+\alpha \Loss_{OC}$), ``MC+DOC'' ($\Loss_{MC}+\alpha \Loss_{DOC}$), our ``DMC+DOC'' ($\Loss_{DMC}+\alpha \Loss_{DOC}$), and ``Sup'' ($\Loss_{sup}$). \textcolor{blue}{(Top-left)} The highest test signal-to-noise ratios (SNR, dB) on toy signals (a) and the peak SNRs (PSNR, dB) on MNIST (b) with CS ratio 50\%. \textcolor{blue}{(Bottom-left)} The optimization trajectories of NNs for MNIST, visualized by projecting the NN weights of training checkpoints to a 2D space with PCA \cite{pearson1901liii}. Blue and red triangles represent initial and final weights optimized by $\Loss_{sup}$ for ratio 50\%, and fine-tuned results for 10\%, 20\%, 30\%, and 40\%. \textcolor{blue}{(Middle/Right)} Test SNR/PSNR curves on toy signals/MNIST with ratio 50\%.}
\label{fig:toy_curve_and_tab}
\end{figure*}

\begin{table*}[!t]
\caption{Comparison of the average PSNR (dB) on four 2D natural image benchmarks \cite{kulkarni2016reconnet,martin2001database,huang2015single,agustsson2017ntire}. Throughout this paper, all the best and second-best results of each test case are highlighted in bold red and underlined blue, respectively. Unless we specifically mention otherwise, we test all methods in this paper using the same sampling matrix and CS ratio that are used during training.}
\label{tab:compare_sota_psnr_no_noise}
\centering
\resizebox{1.0\textwidth}{!}{
\begin{minipage}[c]{0.025\textwidth}
\resizebox{1.0\textwidth}{!}{
\begin{tabular}{c}
\multirow{6}{*}{\rotatebox[origin=c]{90}{Supervised Methods}} \\ \\ \\ \\ \\ \\ \\ \\ \\ \\ \\
\multirow{9}{*}{\rotatebox[origin=c]{90}{Self-Supervised Methods}} \\ \\ \\ \\ \\ \\ \\ \\
\end{tabular}}
\end{minipage}
\begin{minipage}[c]{0.965\textwidth}
\resizebox{1.0\textwidth}{!}{
\begin{tabular}{lc|ccc|ccc|ccc|ccc}
\shline
\rowcolor[HTML]{EFEFEF} 
\multicolumn{1}{l|}{\cellcolor[HTML]{EFEFEF}} &
  Test Set &
  \multicolumn{3}{c|}{\cellcolor[HTML]{EFEFEF}Set11} &
  \multicolumn{3}{c|}{\cellcolor[HTML]{EFEFEF}CBSD68} &
  \multicolumn{3}{c|}{\cellcolor[HTML]{EFEFEF}Urban100} &
  \multicolumn{3}{c}{\cellcolor[HTML]{EFEFEF}DIV2K} \\ \hhline{>{\arrayrulecolor[HTML]{EFEFEF}}->{\arrayrulecolor{black}}|-------------} 
\rowcolor[HTML]{EFEFEF} 
\multicolumn{1}{l|}{\multirow{-2}{*}{\cellcolor[HTML]{EFEFEF}Method}} &
  CS Ratio $\gamma$~~($\sigma =0$) &
  10\% &
  30\% &
  50\% &
  10\% &
  30\% &
  50\% &
  10\% &
  30\% &
  50\% &
  10\% &
  30\% &
  50\% \\ \hline \hline
\multicolumn{2}{l|}{ReconNet (CVPR 2016)}        & 24.08 & 29.46 & 32.76 & 23.92 & 27.97 & 30.79 & 20.71 & 25.15 & 28.15 & 24.41 & 29.09 & 32.15 \\
\multicolumn{2}{l|}{ISTA-Net$^+$ (CVPR 2018)}    & 26.49 & 33.70 & 38.07 & 25.14 & 30.24 & 33.94 & 22.81 & 29.83 & 34.33 & 26.30 & 32.65 & 36.88 \\
\multicolumn{2}{l|}{DPA-Net (TIP 2020)}         & 27.66 & 33.60 & -     & 25.33 & 29.58 & -     & 24.00 & 29.04 & -     & 27.09 & 32.37 & -     \\
\multicolumn{2}{l|}{MAC-Net (ECCV 2020)}         & 27.92 & 33.87 & 37.76 & 25.70 & 30.10 & 33.37 & 23.71 & 29.03 & 33.10 & 26.72 & 32.23 & 35.40 \\
\multicolumn{2}{l|}{ISTA-Net$^{++}$ (ICME 2021)} & 28.34 & 34.86 & 38.73 & 26.25 & 31.10 & 34.85 & 24.95 & 31.50 & 35.58 & 27.82 & 33.74 & 37.78 \\
\multicolumn{2}{l|}{COAST (TIP 2021)} & 28.78 & 35.10 & 38.90 & 26.28 & 31.08 & 34.72 & 25.32 & 31.90 & 35.89 & 27.98 & 33.85 & 37.87 \\\hline \hline
\multicolumn{2}{l|}{DIP (CVPR 2018)}   & 26.09 & 32.58 & 35.30 & 24.73 & 28.47 & 31.81 & 23.37 & 28.23 & 33.15 & 25.42 & 31.32 & 35.03 \\
\multicolumn{2}{l|}{BCNN (ECCV 2020)}   & 27.58 & 33.70 & 37.44 & 25.15 & 29.56 & 33.04 & 24.87 & 31.07 & 35.03 & 26.93 & 32.53 & 36.33        \\
\multicolumn{2}{l|}{EI (ICCV 2021)}   & 21.76 & 33.49 & 37.66 & 22.43 & 29.74 & 33.43 & 19.04 & 28.82 & 33.72 & 22.60 & 32.24 & 36.88        \\
\multicolumn{2}{l|}{ASGLD (CVPR 2022)}   & 27.81 & 34.17 & 37.46 & 25.23 & 29.58 & 32.37 & 24.54 & 31.09 & 34.91 & 27.01 & 32.67 & 36.14        \\
\multicolumn{2}{l|}{DDSSL (ECCV 2022)}   & 27.65 & 34.21 & 38.40 & 26.17 & 31.03 & 34.65 & 25.12 & 32.59 & 36.44 & 27.55 & 33.81 & 37.52        \\ \hline \hline
\multicolumn{2}{l|}{\textbf{SC-CNN (Ours)}}   & 27.93 & 34.95 & 39.07 & 25.81 & 31.03 & 34.84 & 23.75 & 31.52 & 36.13 & 27.33 & 33.86 & 38.13        \\
\multicolumn{2}{l|}{\textbf{SC-CNN$^+$ (Ours)}}   & \textcolor{red}{\textbf{29.42}} & \textcolor{red}{\textbf{36.12}} & \textcolor{blue}{\underline{40.16}} & \textcolor{red}{\textbf{26.27}} & \textcolor{blue}{\underline{31.51}} & \textcolor{blue}{\underline{35.38}} & \textcolor{blue}{\underline{27.73}} & \textcolor{blue}{\underline{34.56}} & \textcolor{blue}{\underline{38.47}} & \textcolor{blue}{\underline{28.47}} & \textcolor{blue}{\underline{34.77}} & \textcolor{blue}{\underline{39.01}} \\
\multicolumn{2}{l|}{\textbf{SCT (Ours)}}   & 28.21 & 35.32 & 39.50 & 25.99 & 31.20 & 35.06 & 24.04 & 31.80 & 36.54 & 27.51 & 34.03 & 38.39 \\
\multicolumn{2}{l|}{\textbf{SCT$^+$ (Ours)}}   & \textcolor{blue}{\underline{29.32}} & \textcolor{blue}{\underline{36.10}} & \textcolor{red}{\textbf{40.41}} & \textcolor{blue}{\underline{26.17}} & \textcolor{red}{\textbf{31.60}} & \textcolor{red}{\textbf{35.52}} & \textcolor{red}{\textbf{27.74}} & \textcolor{red}{\textbf{34.58}} & \textcolor{red}{\textbf{38.55}} & \textcolor{red}{\textbf{28.52}} & \textcolor{red}{\textbf{34.89}} & \textcolor{red}{\textbf{39.24}} \\
\shline
\end{tabular}}\end{minipage}}
\end{table*}

\section{Experiment}

In this section, we present the evaluations of our method on four types of data: (1) 1D synthesized sparse signals, (2) 2D handwritten digit images from the MNIST dataset, (3) 2D natural images, and (4) real captured measurements for SPI of 2D intensities and 3D cubes. All experiments are performed on an NVIDIA RTX 4090 GPU\footnote{Please refer to Sec. \ref{sec:more_experimental_details} in the \Appx~for more experimental details, results, analyses, and discussions.}.

\begin{figure*}[!t]
\centering
\begin{minipage}[c]{0.4\linewidth}
\centering
\includegraphics[width=\linewidth]{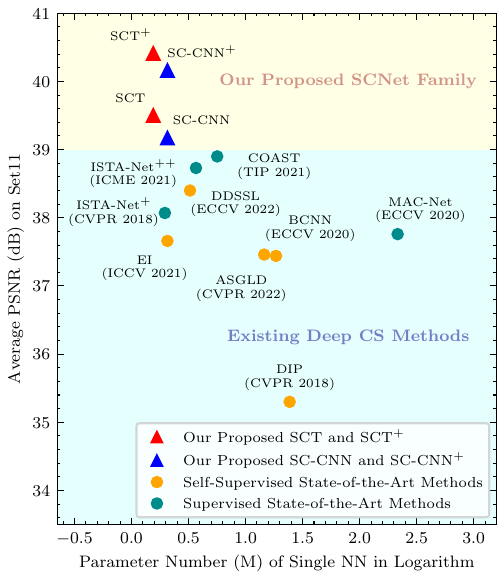}
\end{minipage}
\begin{minipage}[c]{0.58\linewidth}
\centering
\includegraphics[width=\linewidth]{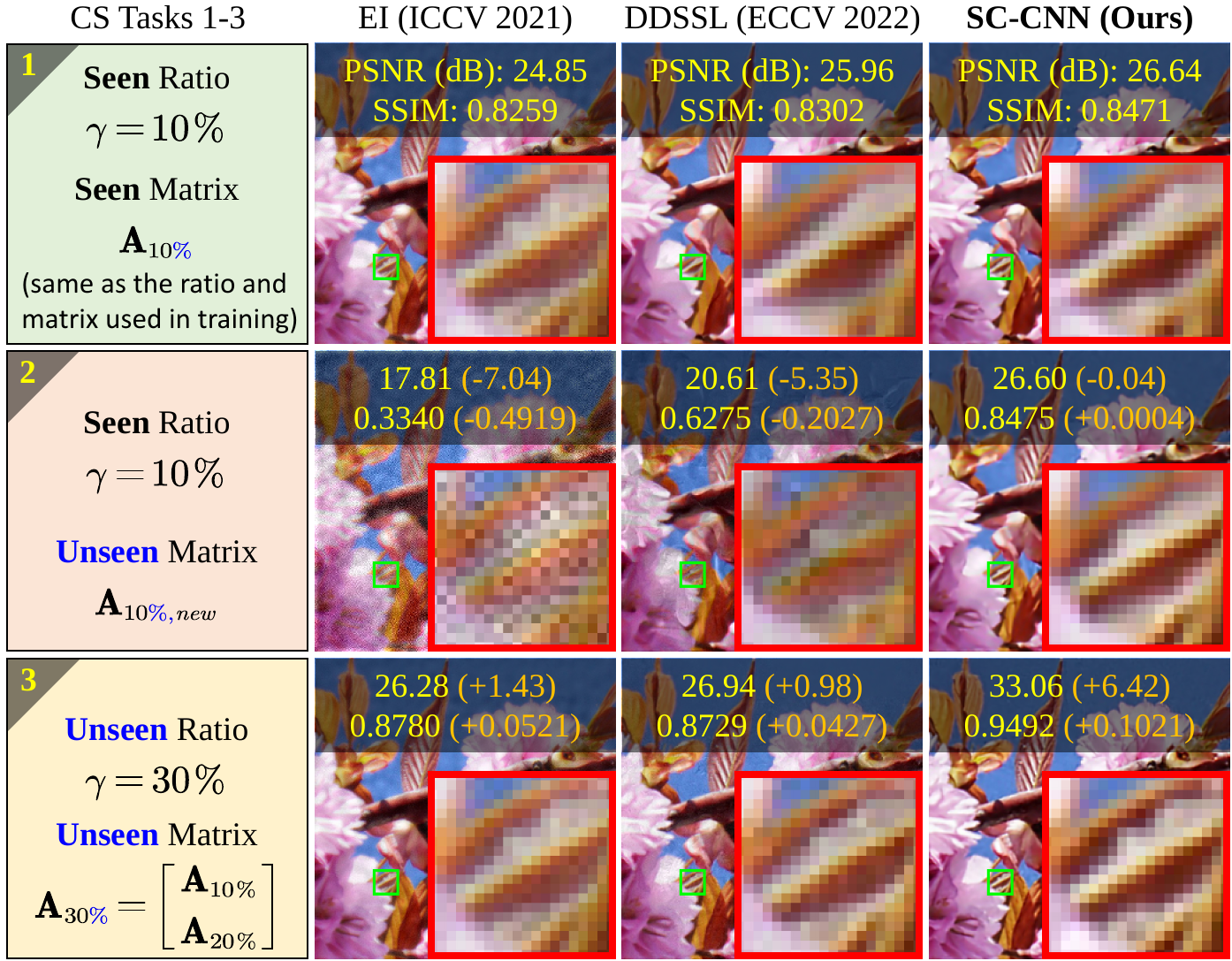}
\end{minipage}
\caption{\textcolor{blue}{(Left)} Comparison of PSNR (dB) and parameter number (M) among our four SCNets and nine existing methods with $\gamma =50\%$ and $\sigma =0$. \textcolor{blue}{(Right)} Visualization of the reconstructions of an image named ``0898'' from DIV2K \cite{agustsson2017ntire} produced by three self-supervised methods: EI \cite{chen2021equivariant}, DDSSL \cite{quan2022dual}, and our SC-CNN with only external learning (\textit{i.e.}, stage-1). \textcolor{blue}{(The 1st row)} NNs are first tested in the setting of ratio 10\% and a matrix consistent with the ones adopted during training. \textcolor{blue}{(The 2nd and 3rd rows)} They are then extended to new ratios and matrices that were not provided or seen in the training set (marked as ``Unseen''). Our SC-CNN remains robust to different settings once trained (with all matrices being random Gaussian and satisfying $\A \A^\top=\mathbf{I}_M$), while EI and DDSSL may fail and be sensitive to the changes of CS ratio or sampling matrix.}
\label{fig:compare_param_and_generalization_ability}
\end{figure*}

\subsection{CS Reconstruction for 1D Toy Signals and 2D MNIST Digit Images}
\textbf{Setup.} We first study the efficacy of our method on 1D 2-sparse toy signals and the MNIST \cite{deng2012mnist} dataset. The toy signals are generated by $\x=\mathbf{\Psi}\mathbf{s}$ of size 32, where $\mathbf{\Psi}$ denotes the DCT basis and $\mathbf{s}$ has its first 2 elements being sampled from a joint distribution, with the other 30 being zero. We employ 4 bivariate toy distributions \cite{sun2021deep}, including a Gaussian mixture, a uniform, \textit{etc.}, to construct 4 datasets. Each one contains 900 signals for training and 100 signals for test. The MNIST dataset consists of 60000 training samples and 10000 test samples with size $28\times 28$. We use six loss functions including the $\Loss_{MC}$ \cite{ulyanov2018deep},  the original-domain consistency (OC) loss $\Loss_{OC}=\lVert \xhat - \F_\Th(\A\xhat,\A,\gamma) \rVert_p^p$, our $\Loss_{DMC}$ and $\Loss_{DOC}$, the supervised loss $\Loss_{sup}=\lVert \x - \F_\Th(\A\x,\A,\gamma) \rVert_p^p$, and their various combinations to train six SC-CNNs with $K=3$, $C=16$, and without the final GD step for 30000 epochs. The measurements $\{\y_i\}$ are generated by a fixed Gaussian matrix of ratio $50\%$ from GTs.

\textbf{Results.} Fig.~\ref{fig:toy_curve_and_tab} compares the recovery accuracy among $\A^\dagger \y$ and SC-CNNs trained by the six losses. We observe that: (1) ``MC'' and ``MC+OC'' may overfit to undesirable results. This overfitting can be alleviated by our measurement divisions in ``DMC'', which exhibits a steady and continuous growth in test accuracy; (2) our DOC encouragement brings significant performance improvements (SNR $>$$10$dB on toy signals and PSNR $>$$1.8$dB on MNIST), and makes ``DMC+DOC'' competitive to the supervised counterpart, with distances $<$$2.5$dB. In Fig.~\ref{fig:toy_curve_and_tab} (bottom-left), we visualize the learning trajectories of NN weights for different methods. Remarkably, our DMC and DOC encouragements can guide the NNs to learn the common knowledge and signal priors among various CS tasks: the NN weights optimized with our $\Loss_{DMC}$ and $\Loss_{DOC}$ present an obvious and clear convergence tendency towards the weight ``cluster'' trained by $\Loss_{sup}$ across different sampling ratios, effectively avoiding some trivial solutions.

\begin{figure*}[!t]
\centering
\begin{minipage}[c]{0.39\linewidth}
\centering
\includegraphics[width=\linewidth]{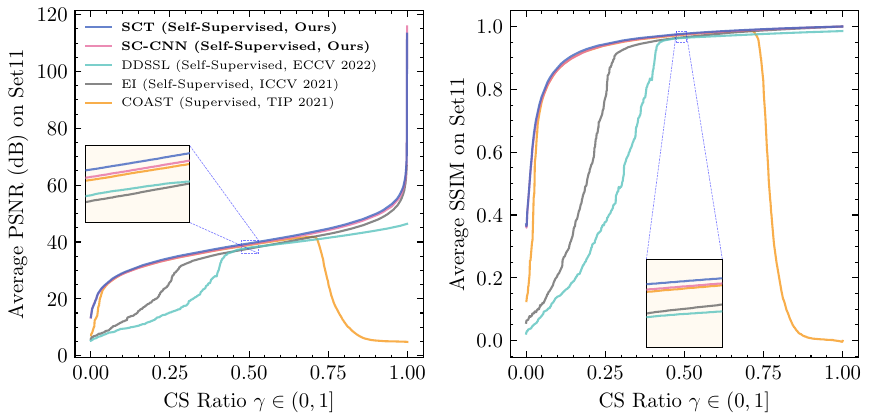}
\end{minipage}
\hfill
\begin{minipage}[c]{0.6\linewidth}
\centering
\includegraphics[width=\linewidth]{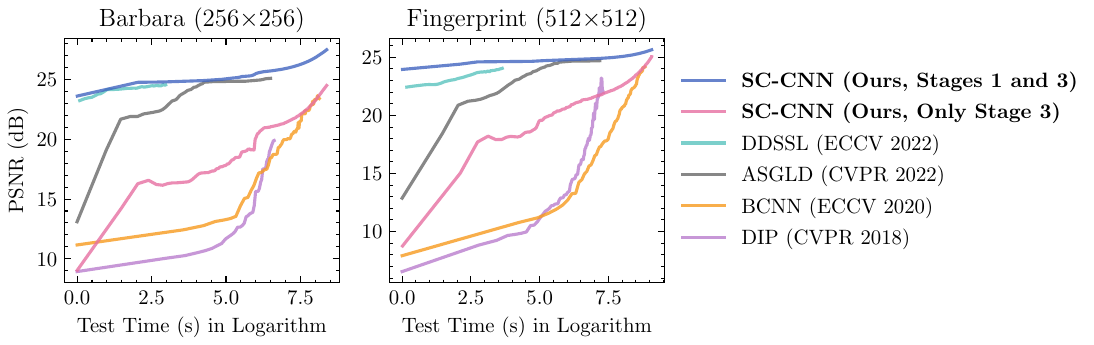}
\end{minipage}
\caption{\textcolor{blue}{(Left)} Comparison of the CS ratio-scalability among five CS NNs. The four self-supervised NNs are trained on a measurement set of ratio 50\%. The supervised COAST \cite{you2021coast} is trained on five ratios: 10\%, 20\%, 30\%, 40\%, and 50\%. SCNets generalizes well to the entire range $(0,1]$ once trained, with significant performance leading over others sensitive to ratio changes. \textcolor{blue}{(Right)} Comparison of six NNs that use test-time single-image internal learning (\textit{i.e.}, stage-3) on two images from Set11 \cite{kulkarni2016reconnet}. DIP \cite{ulyanov2018deep}, BCNN \cite{pang2020self}, ASGLD \cite{wang2022self}, and SC-CNN (only stage-3) are trained from scratch. DDSSL \cite{quan2022dual} and SC-CNN (stages 1 and 3) are initialized by the NN weights from their offline self-supervised learning on external measurements.}
\label{fig:compare_scalability_and_time_curves}
\end{figure*}

\subsection{CS Reconstruction for 2D Natural Images}
\textbf{Setup.} We simulate $\y=\A\x+\n$ on 88912 randomly cropped image blocks of size $N=33\times 33$ \cite{zhang2018ista} from the T91 dataset \cite{dong2014learning,kulkarni2016reconnet} to create a measurement set $\{\y_i^{train}\}$ using a fixed block-based (or block-diagonal) \cite{gan2007block,chun2017compressed} Gaussian matrix $\A$ for external learning. Six \textit{supervised} methods: ReconNet \cite{kulkarni2016reconnet}, ISTA-Net$^+$ \cite{zhang2018ista}, DPA-Net \cite{sun2020dual}, MAC-Net \cite{chen2020learning}, ISTA-Net$^{++}$ \cite{you2021ista}, and COAST \cite{you2021coast}, and six \textit{self-supervised} approaches: DIP \cite{ulyanov2018deep}, BCNN \cite{pang2020self}, EI \cite{chen2021equivariant}, REI \cite{chen2022robust}, ASGLD \cite{wang2022self}, and DDSSL \cite{quan2022dual} are incorporated for comparison. Following \cite{quan2022dual}, we implement EI by applying the EI loss to SC-CNN when $\sigma =0$, and replace it with its enhanced robust version REI when $\sigma >0$. For our SCNets, we set $K=20$ and $C=32$ by default. Four benchmarks: Set11 \cite{kulkarni2016reconnet}, CBSD68 \cite{martin2001database}, Urban100 \cite{huang2015single}, and DIV2K \cite{agustsson2017ntire} are employed. Test images from Urban100 and DIV2K are all $256\times 256$ center-cropped. Results are evaluated by PSNR and SSIM \cite{wang2004image} on the Y channel in YCrCb space.

\begin{figure*}[!t]
\centering
\begin{minipage}[c]{0.63\linewidth}
\centering
\includegraphics[width=1.0\textwidth]{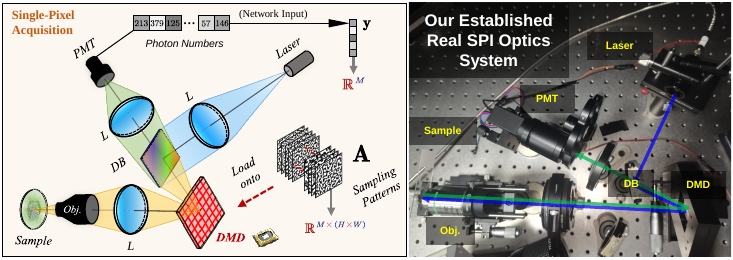}
\end{minipage}
\begin{minipage}[c]{0.3635\linewidth}
\centering
\resizebox{1.0\textwidth}{!}{
\tiny
\setlength{\tabcolsep}{0.5pt}
\begin{tabular}{cccccc}
\multirow{2}{*}{\rotatebox[origin=c]{90}{Nucleus~~~~~~}}~~~& $\A^\dagger \y$ & TVAL3 & MC & DMC & \textbf{DMC+DOC} \\
\multirow{2}{*}{\rotatebox[origin=c]{90}{FM~~~~~~~~}}~~~&
\includegraphics[width=0.225\textwidth]{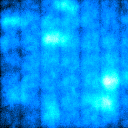} &
\includegraphics[width=0.225\textwidth]{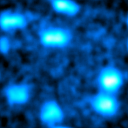} &
\includegraphics[width=0.225\textwidth]{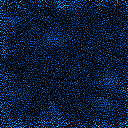} &
\includegraphics[width=0.225\textwidth]{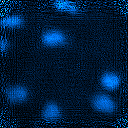} &
\includegraphics[width=0.225\textwidth]{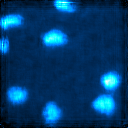}\\
\multirow{2}{*}{\rotatebox[origin=c]{90}{F-Actin~~~~~~}}~~~&
\includegraphics[width=0.225\textwidth]{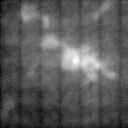} &
\includegraphics[width=0.225\textwidth]{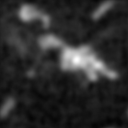} &
\includegraphics[width=0.225\textwidth]{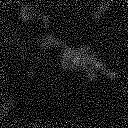} &
\includegraphics[width=0.225\textwidth]{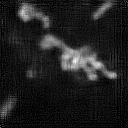} &
\includegraphics[width=0.225\textwidth]{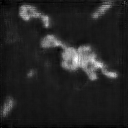}\\
&
\includegraphics[width=0.225\textwidth]{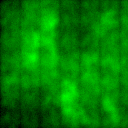} &
\includegraphics[width=0.225\textwidth]{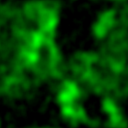} &
\includegraphics[width=0.225\textwidth]{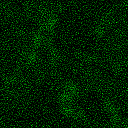} &
\includegraphics[width=0.225\textwidth]{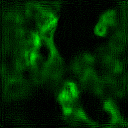} &
\includegraphics[width=0.225\textwidth]{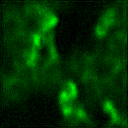}\\
\end{tabular}}
\end{minipage}
\caption{\textcolor{blue}{(Left)} Illustration of our established SPI system. A set of optical instruments and DMD modulation patterns are used to implement $\y= \A\x$. \textcolor{blue}{(Middle)} Layout of our SPI system on testbed. Blue arrows represent emission path of light from laser to sample, while green arrows correspond to excitation path of fluorescence from sample to PMT. \textcolor{blue}{Obj.}: objective; \textcolor{blue}{L}: lens; \textcolor{blue}{DMD}: digital micro-mirror device; \textcolor{blue}{DB}: dichroic beamsplitter; \textcolor{blue}{PMT}: photomultiplier. \textcolor{blue}{(Right)} Visual comparison among five methods. We provide the 1st 2D slice for 3D reconstruction of F-actin. Our method can yield best SPI qualities.}
\label{fig:real}
\end{figure*}

\textbf{Results.} Tab.~\ref{tab:compare_sota_psnr_no_noise} and Fig.~\ref{fig:compare_param_and_generalization_ability} (left) provide the comparisons among different CS recovery methods. We can observe that: (1) the standard SCNets are competitive with both the existing best supervised and self-supervised methods (\textit{e.g.}, COAST \cite{you2021coast} and DDSSL \cite{quan2022dual}), especially when $\gamma > 10\%$, while enjoying low parameter cost ($< 0.4$M) and real-time inference speeds ($> 30$ frames per second); (2) our internal learning and self-ensemble schemes in stages 2-4 of SC-CNN$^+$ and SCT$^+$ bring PSNR improvements of 0.18-3.98dB with a negligible additional parameter burden for single NN, demonstrating the effectiveness of our loss $\Loss$ and progressive reconstruction strategy. Furthermore, we find that integrating Transformer blocks into SCNet consistently improves recovery performance, as demonstrated by the improved PSNR observed for SCT and SCT$^+$ compared to SC-CNN and SC-CNN$^+$, which lack Transformer blocks, across the majority of cases. Figs.~\ref{fig:compare_param_and_generalization_ability} (right) and \ref{fig:compare_scalability_and_time_curves} (left) reveal the remarkable generalization ability of SCNets on the entire ratio range $(0,1]$ and arbitrary matrices once trained on an incomplete measurement set. In contrast, the competing methods EI \cite{chen2021equivariant} and DDSSL \cite{quan2022dual} suffer from unexpected performance drops and yield suboptimal results when the ratio and matrix change due to requirement variations and hardware malfunctions. While these methods may struggle to effectively adapt to such changeable settings in real-world deployments, our method not only achieves superior performance, but also exhibits favorable scalability and generalization ability given only fixed external measurements. In Fig.~\ref{fig:compare_scalability_and_time_curves} (right), we observe that our SC-CNN is also capable of providing recoveries by utilizing only the stage-3, with its NN capacity being much smaller than that of others, and can significantly benefit from the weight initialization of stage-1 training on external measurements, demonstrating its flexibility for different real environments.

\subsection{CS Reconstruction for 2D and 3D Fluorescence Microscopy with Real SPI Optics System}
\textbf{Setup.} To verify the effectiveness of our method in real scenarios, we establish an SPI system for fluorescence microscopy \cite{lichtman2005fluorescence}. The hardware setup is illustrated in Figs.~\ref{fig:real} (left) and (middle). More details are provided by Sec. \ref{sec:more_experimental_details_real} in \Appx. Our DMD sequences (in time) through $2M$ projections, implementing $M=5000$ rows of an $M\times N$ bipolar matrix $\A$, where each projection appears as an $N=128\times 128$ pattern, and all matrix elements are i.i.d. Bernoulli and has an equal probability of 0.5 for being +1 and -1. We use three types of samples: nucleus, fluorescent microsphere (FM), and filamentous actin (F-Actin). For the former two, we obtain 950 and 50 $\y$s of size $M$ for training and test, respectively. For F-actin, we capture 490 and 10 groups of training and test measurements, respectively. Each group contains 21 $\y$s corresponding to 21 2D slices ($\x$s) along the z-axis. We employ SC-CNN ($K=20$, $C=32$, and trained by only stage-1) without IE, PE, and final GD step. For 3D imaging of F-actin samples, we set $C=64$ and concatenate the 21 slices channel-wise to recover them jointly.

\textbf{Results.} Fig.~\ref{fig:real} (right) displays the reconstructed results of five methods at CS ratio $\gamma =(5000/16384)\approx 30.5\%$. We can observe that: (1) $\A^\dagger \y$, TVAL3 \cite{li2013efficient}, and ``MC'' \cite{ulyanov2018deep} methods can reconstruct some basic structures, but may also introduce undesirable artifacts and the loss of some important image features; (2) our ``DMC'' can well suppress noise and yield better results, while ``DMC+DOC'' can provide further improved preservation and balance of the primary features and details for all three sample types, producing enhanced and clearer informative nuclear speckles \cite{alexander2021p53} and FM/F-actin intensities. These findings confirm the efficacy of our method for real data, its performance superiority, and better generalization ability than existing methods, once trained on a single collected measurement set of fixed ratio and matrix. It is worth noting that the imaging results can be further enhanced by adopting our default SCT$^+$, equipped with the fully activated four-stage reconstruction $(1\rightarrow 2\rightarrow 3\rightarrow 4)$.

\section{Conclusion}
This work proposes a novel self-supervised scalable CS method, consisting of a learning scheme named SCL and a deep NN family called SCNet. Our SCL employs a dual-domain loss to learn a generalizable image-, ratio-, and matrix-adaptive mapping instead of specific ones from fixed incomplete measurements by augmenting and encouraging random cross-consistency and arbitrary sampling-reconstruction cycle consistency. A four-stage strategy further improves accuracy progressively. SCNet is designed based on PGD inspiration and is well-regularized by NN mechanisms. The combination and mutual promotion of SCL and SCNet effectively exploit available data and information to drive deep reconstruction toward valid results that can even surpass those predicted by supervised NNs. Experiments demonstrate the effectiveness and superiority of our method in achieving a better balance among imaging quality, flexibility, scalability, complexity, and interpretability than existing ones, thus paving the way for practical CS imaging applications. Our future work is to extend this method for other inverse imaging problems, including but not limited to inpainting, deconvolution \cite{quan2022learning}, MRI \cite{lustig2008compressed,sun2016deep}, CT \cite{szczykutowicz2010dual}, SCI \cite{yuan2021snapshot,fu2021coded}, and interferometric imaging \cite{sun2021deep,feng2023score}.

\section*{Acknowledgement}
This work was supported in part by the National Natural Science Foundation of China (No. 62331011) and the Shenzhen Research Project (No. JCYJ20220531093215035).

\clearpage

\appendix

\section*{Appendix}

\vspace{30pt}

\tableofcontents

\vspace{30pt}

\counterwithin{figure}{section}
\counterwithin{table}{section}

\section*{Overview of Appendix}
In this appendix, we present additional details for enhancing the clarity of our original presentations in \MP~and provide a deeper understanding of the working principles of our proposed method. Specifically, Sec.~\ref{sec:NN_blocks} delves into the structural designs of our adopted two basic NN blocks including RB and SCB. In Sec.~\ref{sec:more_experimental_details}, we provide further insights into the implementation details of our method and the SPI hardware setup utilized for our evaluations. Additionally, we present a comprehensive set of experimental results, including comparisons, ablation studies, and recovery analyses for different practical requirements, supported by extensive qualitative and quantitative analyses. Furthermore, Sec.~\ref{sec:discussions} offers discussions on the insights and explanations underlying our method, the limitations, and the broader impacts of our work.

\section{Structural Details of NN Blocks}
\label{sec:NN_blocks}

\begin{figure}
\centering
\includegraphics[width=0.48\textwidth]{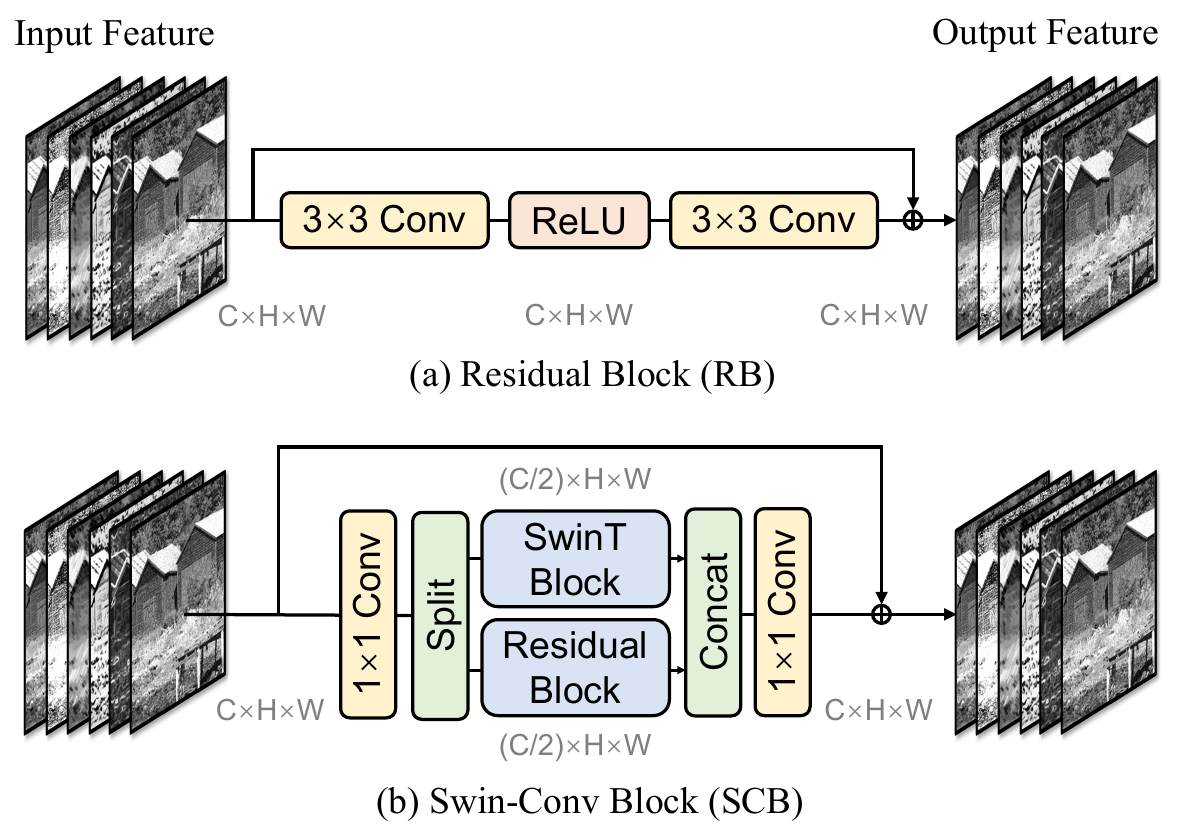}
\caption{Illustration of RB and SCB. These blocks allow SCNets to explore the short- and long-range dependencies in images, while regularizing NN optimization toward valid results via the inductive bias from local convolutions and sliding (or shifted) window-based attention modules.}
\label{fig:basic_blocks}
\end{figure}

Fig.~\ref{fig:basic_blocks} illustrates the structural design of the residual block (RB) \cite{he2016deep,lim2017enhanced} and the Swin-Conv block (SCB) \cite{zhang2022practical} employed in our proposed SCNet for feature refinement. In this section, we present the details of RB and SCB as follows.

\textbf{Residual Block (RB).} As shown in Fig.~\ref{fig:basic_blocks} (a), each RB consists of a $3\times 3$ convolution layer, a rectified linear unit (ReLU) \cite{nair2010rectified} activation, and another $3\times 3$ convolution layer. The final output is obtained by fusing the input and the residual features through an identity skip connection. All features have the same shape of $C\times H\times W$.

\begin{figure*}[!t]
\centering
\includegraphics[width=1.0\textwidth]{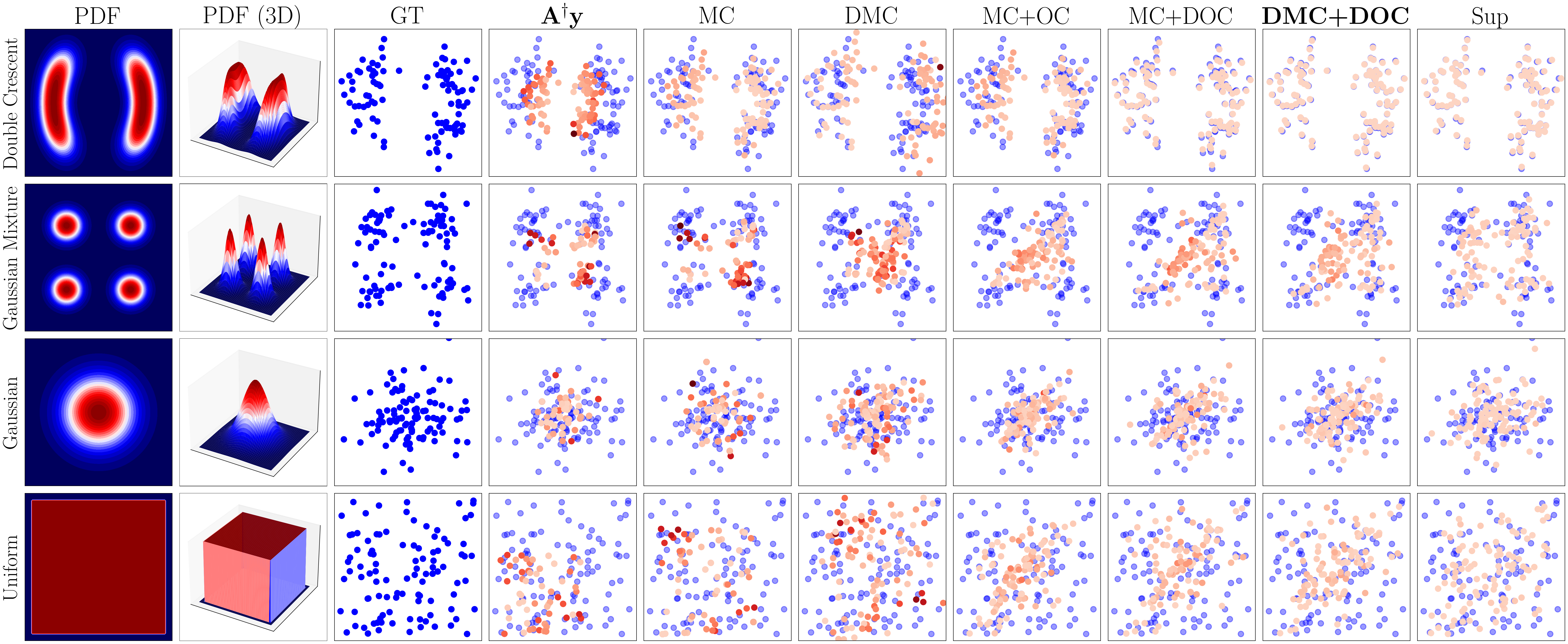}
\caption{Visualizations of the probability density functions (PDF) and recovered results corresponding to four toy signal types: double crescent (DC), Gaussian mixture (GM), Gaussian (G), and uniform (U). \textcolor{blue}{(The first two columns)} 2D ``heatmap'' and 3D ``surface'' visualizations of the PDFs of $q_{DC}$, $q_{GM}$, $q_{G}$, and $q_{U}$. \textcolor{blue}{(The last eight columns)} Visualizations of the CS reconstruction results from different methods. The 100 GT $\x$s and prediction $\xhat$s of each signal type are transformed using $\mathbf{s}=\mathbf{\Psi}^{-1}\x$ and $\hat{\mathbf{s}}=\mathbf{\Psi}^{-1}\xhat$, respectively. The first two DCT coefficients, corresponding to the most low-frequency components, are picked for 2D visualization. Blue points represent the 100 test GTs for reference, whereas red points correspond to the 100 recovered results from different methods. Lighter shades of red indicate lower prediction errors (\textit{i.e.}, smaller Euclidean distances to GTs).}
\label{fig:toy_vis_points}
\end{figure*}

\textbf{Swin-Conv Block (SCB) \cite{zhang2022practical}.} As illustrated in Fig.~\ref{fig:basic_blocks} (b), for input feature of shape $C\times H\times W$, it is first passed through an $1\times 1$ convolution layer and then divided into two features of shape $(C/2)\times H\times W$. One of the two features is fed into a Swin Transformer (SwinT) block \cite{liu2021swin,liang2021swinir} to explore the spatial correlations and non-local information. Additionally, the other of features is passed through an RB to capture the local dependency. Finally, the features are concatenated and transformed adaptively via an $1\times 1$ convolution. Similar to RB, the SCB also has a skip connection that fuses the input and residual features.

\section{More Experimental Details}
\label{sec:more_experimental_details}
\subsection{CS Reconstruction for 1D Toy Signals and 2D MNIST Digit Images}
\subsubsection{More Implementation Details}
\textbf{Four Bivariate Distributions for Generating Sparse Toy Signals.} Our experimental 1D 2-sparse toy signals are generated by simulating $\x=\mathbf{\Psi}\mathbf{s}\in\Rbb^{32}$, where $\mathbf{\Psi}\in\Rbb^{32\times 32}$ represents the discrete cosine transform (DCT) \cite{ahmed1974discrete} basis, while the coefficient vector $\mathbf{s}\in\Rbb^{32}$ has its first two elements randomly sampled from a joint distribution, corresponding to the first two most low-frequency base vectors of $\mathbf{\Psi}$. The remaining thirty elements are zero, \textit{i.e.}, $\mathbf{s}=[s_1,s_2,0,\cdots,0]^\top$. Following \cite{sun2021deep}, we construct and employ four bivariate joint distributions including a double crescent (DC) or bowtie-shaped $q_{DC}$, a Gaussian mixture (GM) $q_{GM}$, a Gaussian (G) $q_{G}$, and a uniform (U) $q_{U}$. The four probability density functions (PDF) of these distributions are visualized in the first two columns of Fig.~\ref{fig:toy_vis_points}. They can be mathematically formulated as follows (not strictly normalized):
\begin{align*}
& q_{DC}(s_1,s_2)\propto \left[e^{-\frac{25}{18}\left({4{s_1}-2}\right)^2} + e^{-\frac{25}{18}\left({4{s_1}+2}\right)^2}\right]\\
& \quad \quad \quad \quad \quad \quad \quad \times e^{-\frac{25}{18}\left({\sqrt{(4{s_1})^2+(4{s_2})^2}-2}\right)^2}, \\
& q_{GM}(s_1,s_2)\propto \left\{\mathcal{N}([1,1]^\top,\frac{\mathbf{I}_2}{10})+\mathcal{N}([-1,1]^\top,\frac{\mathbf{I}_2}{10})\right.\\
& \quad \quad \quad \left.+~\mathcal{N}([1,-1]^\top,\frac{\mathbf{I}_2}{10})+\mathcal{N}([-1,-1]^\top,\frac{\mathbf{I}_2}{10})\right\}, \\
& q_{G}(s_1,s_2)= \mathcal{N}(\mathbf{0}_2,\mathbf{I}_2), \\
& q_{U}(s_1,s_2)= \left\{ \begin{array}{l}
	1,\quad \left( s_1,s_2 \right) \in \left[ 0,1 \right] \times \left[ 0,1 \right],\\
	0,\quad \text{otherwise}.\\
\end{array} \right.
\end{align*}

\begin{table*}[!t]
\caption{Quantitative comparison of the best signal-to-noise ratio (SNR, dB) among seven different methods on four toy signal sets of ratio 50\% and size 100. Our ``DMC+DOC'' default scheme achieves the best self-supervised recovery results, which are competitive to those given by supervised learning.}
\label{tab:more_comp_toy}
\centering
\resizebox{0.9\textwidth}{!}{
\begin{tabular}{lccccccc}
\shline
\rowcolor[HTML]{EFEFEF}
Toy Signal Type & $\A^\dagger \y$ & MC    & DMC   & MC+OC & MC+DOC & \textbf{DMC+DOC} & Sup   \\ \hline \hline
Double Crescent (DC)  & 4.04            & 12.41 & 13.51 & 17.33 & 26.42  & \textcolor{red}{\textbf{31.78}}   & \textcolor{blue}{\underline{28.01}} \\
Gaussian Mixture (GM) & 4.12            & 9.46  & 13.40 & 14.65 & 28.68  & \textcolor{blue}{\underline{30.83}}   & \textcolor{red}{\textbf{32.41}} \\
Gaussian (G)  & 4.12            & 11.47 & 17.33 & 16.42 & 30.78  & \textcolor{red}{\textbf{32.85}}   & \textcolor{blue}{\underline{32.74}} \\
Uniform (U)           & 4.40            & 13.72 & 14.37 & 19.32 & 29.08  & \textcolor{blue}{\underline{29.92}}   & \textcolor{red}{\textbf{34.89}} \\ \hline \hline
Average               & 4.17            & 11.77 & 14.65 & 16.93 & 28.74  & \textcolor{blue}{\underline{31.35}}   & \textcolor{red}{\textbf{32.01}} \\ \shline
\end{tabular}}
\end{table*}

\textbf{More Network and Training Details.} We implement all the experiments on PyTorch \cite{paszke2019pytorch} framework and use Adam \cite{kingma2014adam} with momentum 0.9 and weight decay 0.999 for NN parameter optimization. For toy signals, we generate 1000 GTs for each distribution and randomly divide them to form a training set of size 900 and a test set of size 100. Note that only measurements sensed by a Gaussian matrix of ratio 50\% are available for our self-supervised learning. Training an SC-CNN with $K=3$, $C=16$, and all 1D convolution kernels of size 3 in stage-1 requires ten hours for 30000 epochs with batch size 900 and learning rate $1\times 10^{-4}$, where we set $p=1$ and $\alpha=0.1$ for stage-1. Regarding MNIST, SCL training with batch size 10000 and all convolutions being 2D $3\times 3$ takes approximately one day.

\begin{figure*}[!t]
\setlength{\tabcolsep}{0.5pt}
\resizebox{1.0\textwidth}{!}{
\tiny
\begin{tabular}{ccccccccccccccc}
\multirow{2}{*}{\rotatebox[origin=c]{90}{$\gamma=10\%$~~~~~}}~~~~~&$\A^\dagger \y$ & MC & DMC & MC+OC & MC+DOC & \textbf{DMC+DOC} & Sup~~~ & $\A^\dagger \y$ & MC & DMC & MC+OC & MC+DOC & \textbf{DMC+DOC} & Sup\\
&\includegraphics[width=0.07\textwidth]{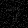} &
\includegraphics[width=0.07\textwidth]{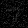} &
\includegraphics[width=0.07\textwidth]{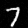} &
\includegraphics[width=0.07\textwidth]{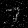} &
\includegraphics[width=0.07\textwidth]{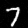} &
\includegraphics[width=0.07\textwidth]{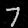} &
\includegraphics[width=0.07\textwidth]{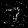}~~~ &
\includegraphics[width=0.07\textwidth]{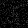} &
\includegraphics[width=0.07\textwidth]{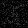} &
\includegraphics[width=0.07\textwidth]{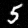} &
\includegraphics[width=0.07\textwidth]{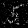} &
\includegraphics[width=0.07\textwidth]{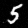} &
\includegraphics[width=0.07\textwidth]{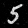} &
\includegraphics[width=0.07\textwidth]{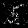}\\
\multirow{2}{*}{\rotatebox[origin=c]{90}{$\gamma=30\%$~~~~~}}~~~~~&11.84 & 12.00 & \textcolor{blue}{\underline{24.99}} & 12.97 & 24.73 & \textcolor{red}{\textbf{25.12}} & 13.09~~~ & 10.70 & 10.85 & \textcolor{blue}{\underline{23.86}} & 12.12 & 23.58 & \textcolor{red}{\textbf{24.38}} & 12.52\\
&\includegraphics[width=0.07\textwidth]{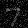} &
\includegraphics[width=0.07\textwidth]{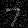} &
\includegraphics[width=0.07\textwidth]{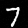} &
\includegraphics[width=0.07\textwidth]{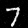} &
\includegraphics[width=0.07\textwidth]{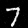} &
\includegraphics[width=0.07\textwidth]{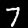} &
\includegraphics[width=0.07\textwidth]{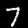}~~~ &
\includegraphics[width=0.07\textwidth]{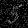} &
\includegraphics[width=0.07\textwidth]{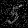} &
\includegraphics[width=0.07\textwidth]{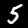} &
\includegraphics[width=0.07\textwidth]{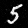} &
\includegraphics[width=0.07\textwidth]{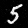} &
\includegraphics[width=0.07\textwidth]{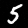} &
\includegraphics[width=0.07\textwidth]{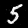}\\
\multirow{2}{*}{\rotatebox[origin=c]{90}{$\gamma=50\%$~~~~~}}~~~~~&13.28 & 13.72 & \textcolor{blue}{\underline{34.98}} & 27.00 & 27.83 & \textcolor{red}{\textbf{35.14}} & 29.39~~~ & 11.90 & 12.30 & \textcolor{blue}{\underline{35.21}} & 25.34 & 26.23 & \textcolor{red}{\textbf{36.42}} & 27.81\\
&\includegraphics[width=0.07\textwidth]{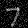} &
\includegraphics[width=0.07\textwidth]{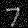} &
\includegraphics[width=0.07\textwidth]{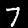} &
\includegraphics[width=0.07\textwidth]{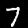} &
\includegraphics[width=0.07\textwidth]{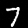} &
\includegraphics[width=0.07\textwidth]{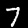} &
\includegraphics[width=0.07\textwidth]{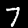}~~~ &
\includegraphics[width=0.07\textwidth]{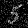} &
\includegraphics[width=0.07\textwidth]{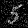} &
\includegraphics[width=0.07\textwidth]{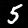} &
\includegraphics[width=0.07\textwidth]{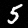} &
\includegraphics[width=0.07\textwidth]{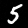} &
\includegraphics[width=0.07\textwidth]{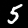} &
\includegraphics[width=0.07\textwidth]{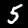}\\
&15.35 & 16.05 & 39.50 & 37.39 & 43.25 & \textcolor{red}{\textbf{45.48}} & \textcolor{blue}{\underline{44.96}}~~~ & 13.55 & 14.22 & 38.30 & 37.55 & 42.64 & \textcolor{red}{\textbf{46.18}} & \textcolor{blue}{\underline{45.96}}
\end{tabular}}
\caption{Visual comparison of different methods on two images of hand-written digits ``7'' \textcolor{blue}{(left)} and ``5'' \textcolor{blue}{(right)} from MNIST \cite{deng2012mnist} dataset with ratio $\gamma \in \{10\%,30\%,50\%\}$ and noise level $\sigma =0$.}
\label{fig:comparison_MNIST}
\end{figure*}

\subsubsection{More Performance Comparisons}

\textbf{Toy Signals.} The last eight columns of Fig.~\ref{fig:toy_vis_points} visualize the recovered results obtained from SC-CNNs trained by different loss functions in each case. Tab.~\ref{tab:more_comp_toy} presents a quantitative evaluation of best average test SNR on four synthesized toy signal sets. We observe that: (1) compared to the traditional measurement consistency (MC) constraint-based scheme, our random division-based MC (DMC) constraint is effective in mitigating overfitting and can achieve better results; (2) the introduction of original-domain consistency (OC) constraint to ``MC'' to form a dual-domain loss is effective for recovery enhancement, while our matrix-network disentanglement-based OC (DOC) constraint further improves accuracy of about 0.84-14.36dB on SNR, verifying that the utilization of $q_\A$ for sampling-reconstruction cycle-consistency encouragement can guide NNs to learn valid results and useful common knowledge of CS domain and signal priors without overfitting to specific settings; (3) our combination of DMC and DOC achieves the best self-supervised recovery performance and can even surpass supervised learning in ``DC'' and ``G'' cases. These results demonstrate the effectiveness and great potential of our method in deep reconstructions without access to GTs.

\textbf{MNIST Digit Images.} Fig.~\ref{fig:comparison_MNIST} provides the visual results of CS reconstruction among seven different methods on two MNIST \cite{deng2012mnist} images depicting hand-written digits ``7'' (left) and ``5'' (right). All NNs are trained on the same measurement set with ratio 50\%. We observe that: (1) our ``DMC'' and ``DOC'' schemes, and their combination produce high-quality recoveries, which can even compete against the results predicted by supervised learning-based NN; (2) the ``MC'', ``MC+OC'', and ``Sup'' schemes learned on the fixed ratio and matrix setting exhibit undesirable performance degradations with changes in the measurement size (or sampling ratio); (3) once trained on a pre-collected incomplete measurement set, our schemes can adapt to different ratios by a single NN model, ensuring scalability and flexibility without introducing extra burdens on the inference complexity and parameter number.

\begin{figure*}[!t]
\setlength{\tabcolsep}{0.5pt}
\hspace{-4pt}
\resizebox{1.0\textwidth}{!}{
\tiny
\begin{tabular}{cccccccccccccc}
    GT & ReconNet & ISTA-Net$^\text{+}$ & ISTA-Net$^\text{++}$ & COAST & DIP & BCNN & EI & ASGLD & DDSSL & \textbf{SC-CNN} & \textbf{SC-CNN$^\text{+}$} & \textbf{SCT} & \textbf{SCT$^\text{+}$}\\
    \includegraphics[width=0.08\textwidth]{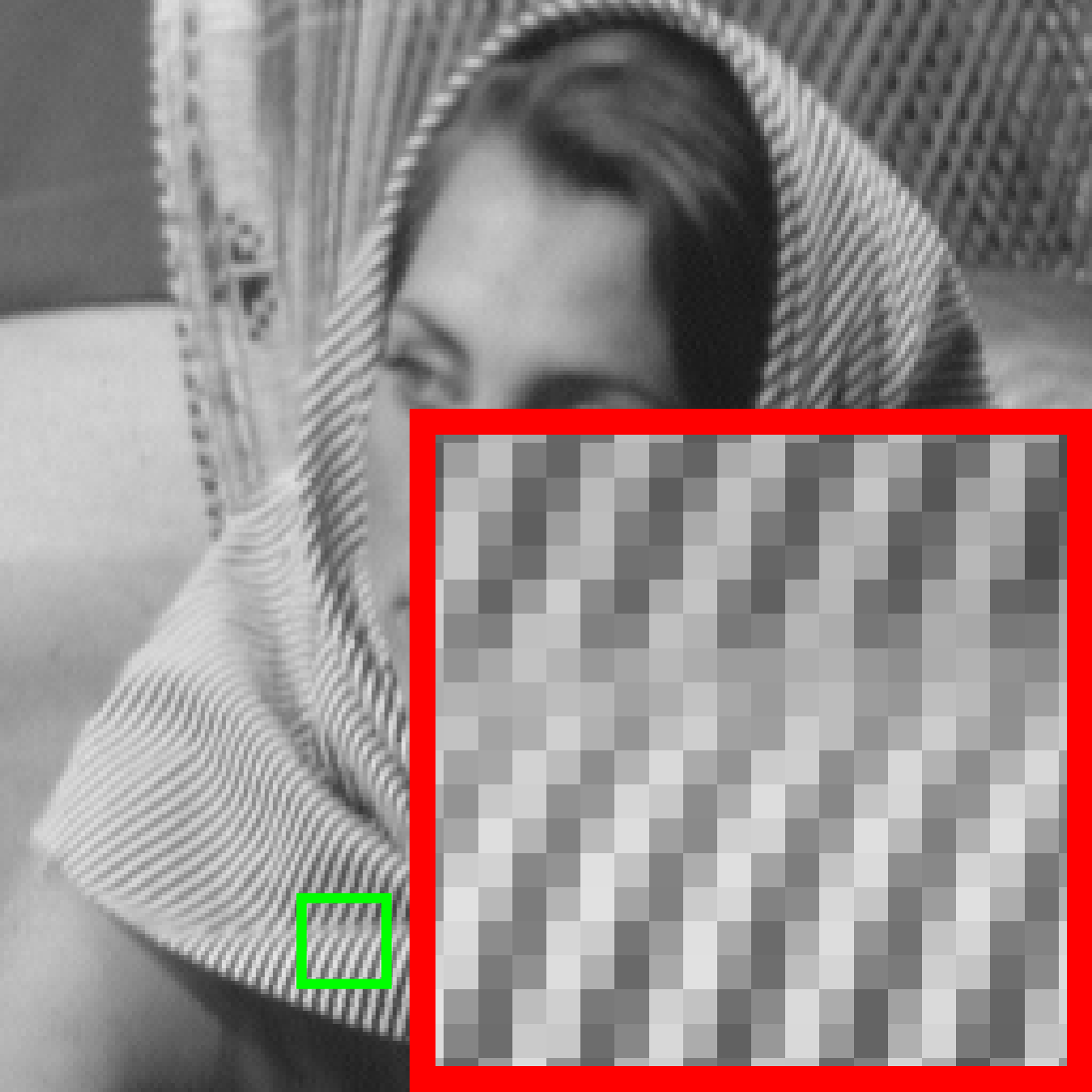}
    &\includegraphics[width=0.08\textwidth]{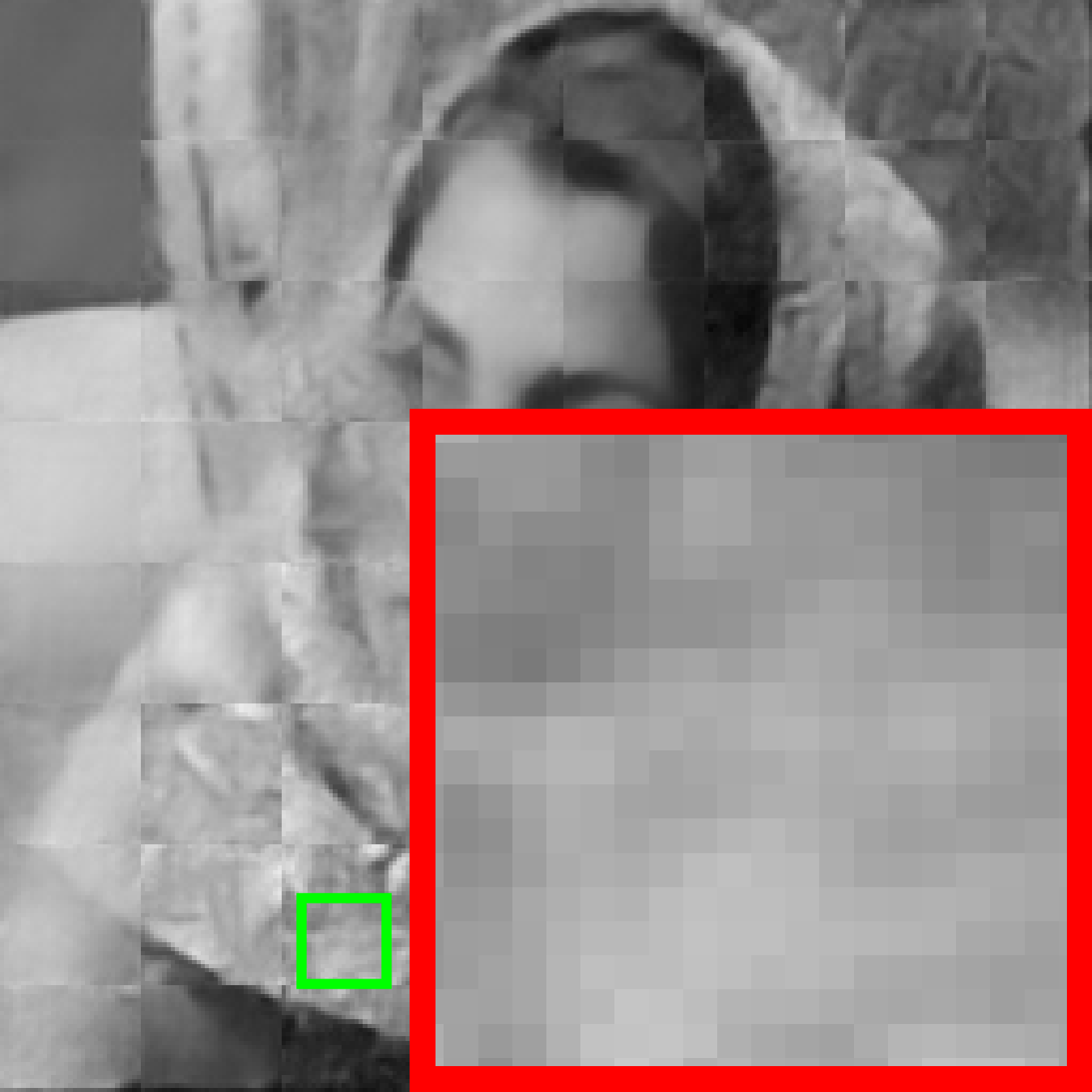}
    &\includegraphics[width=0.08\textwidth]{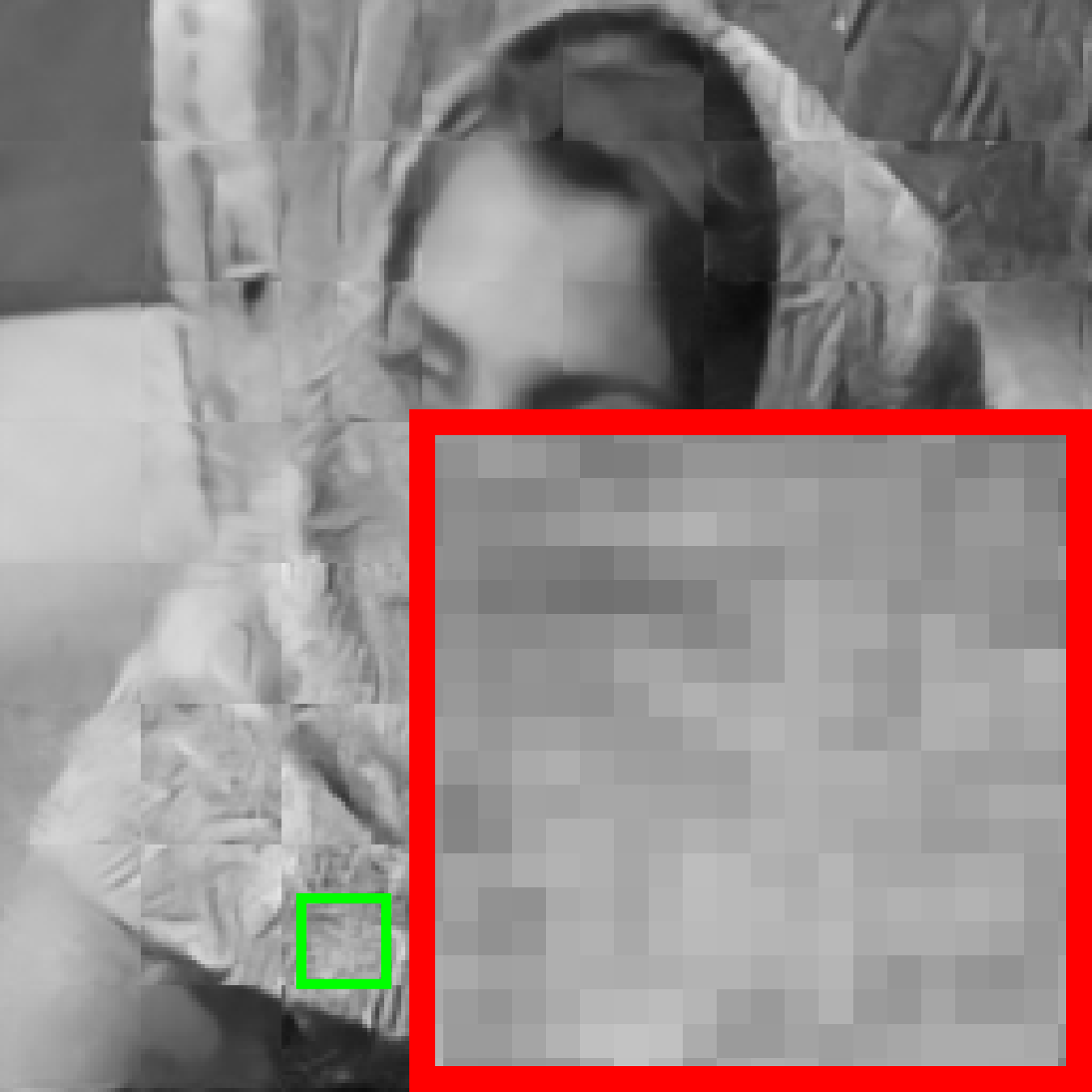}
    &\includegraphics[width=0.08\textwidth]{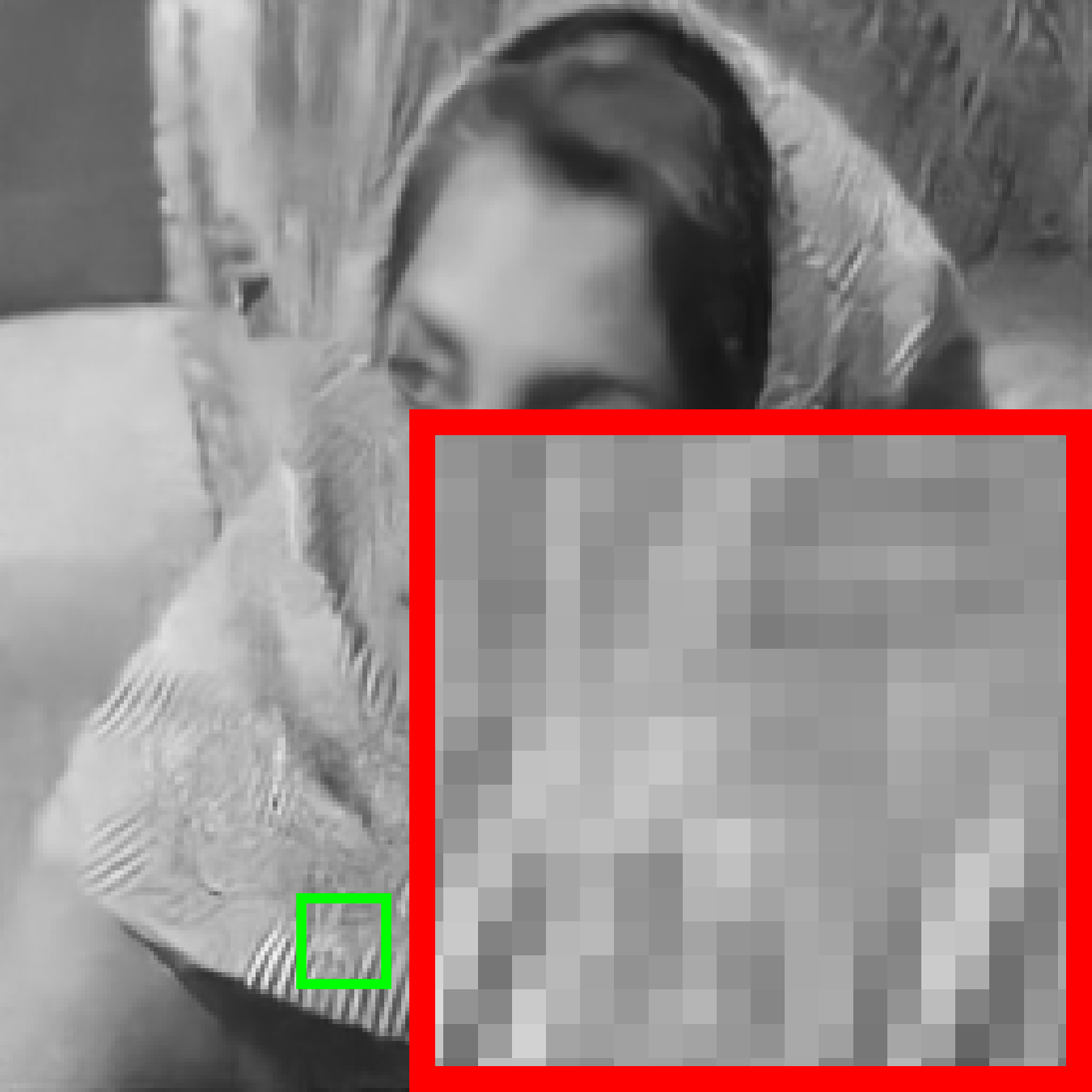}
    &\includegraphics[width=0.08\textwidth]{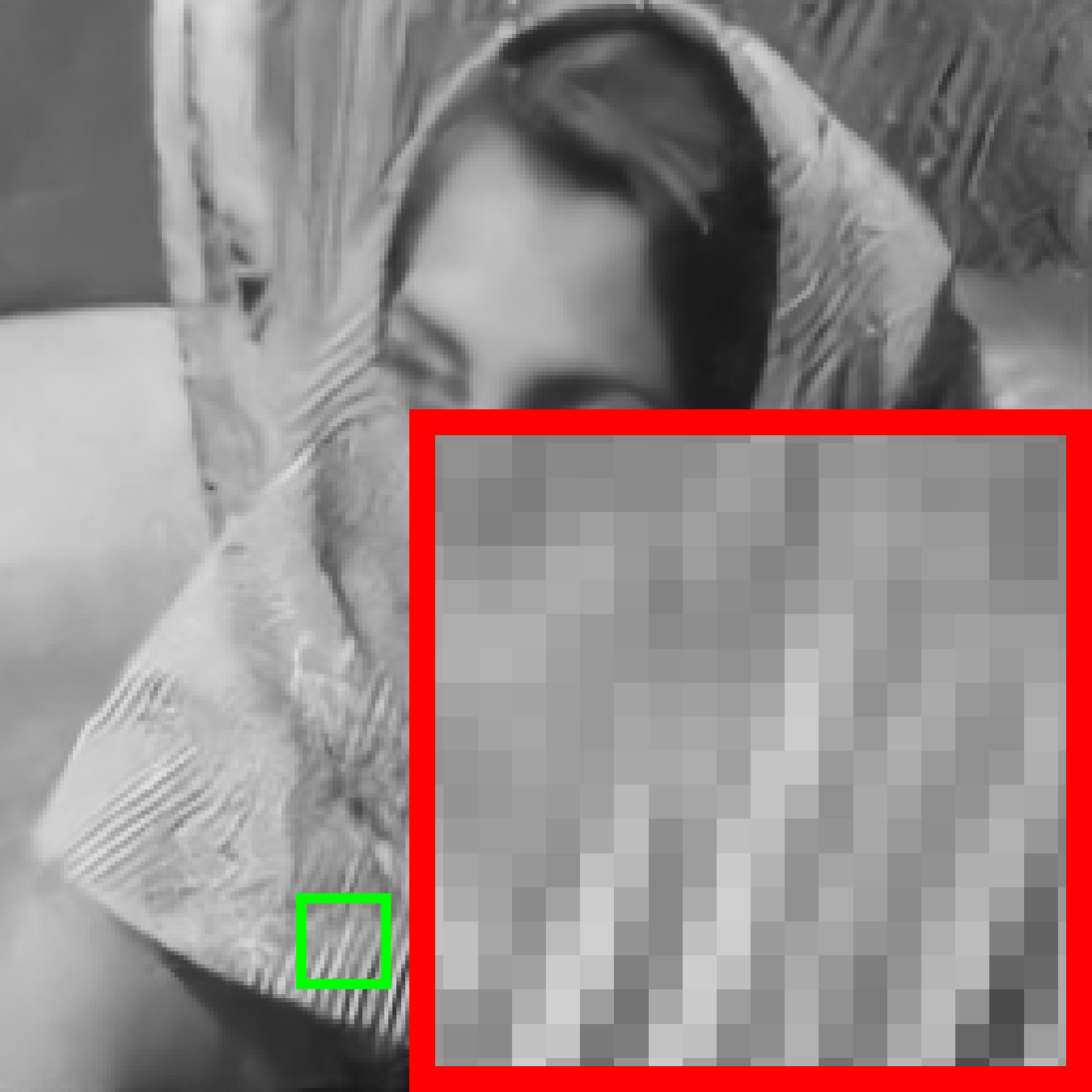}
    &\includegraphics[width=0.08\textwidth]{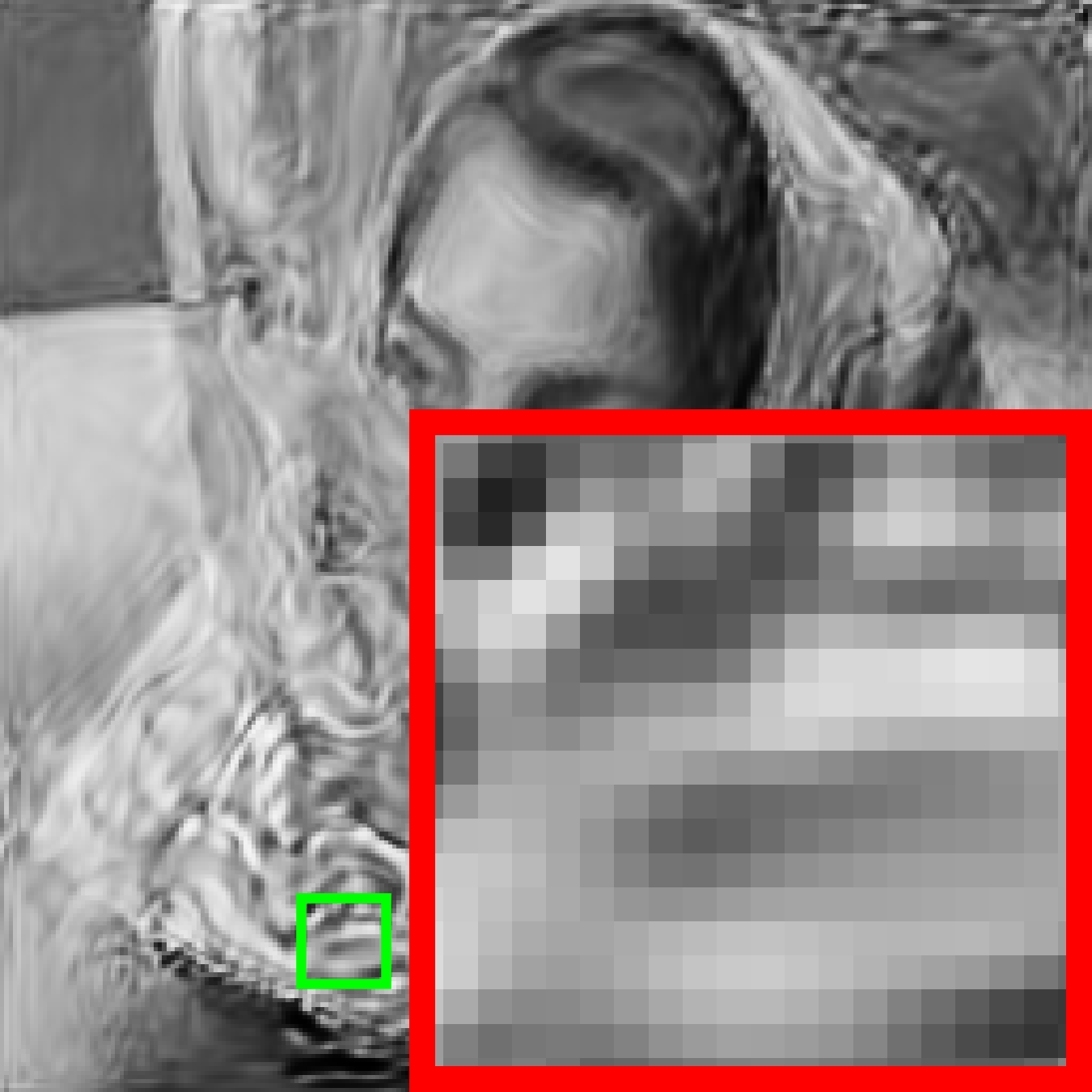}
    &\includegraphics[width=0.08\textwidth]{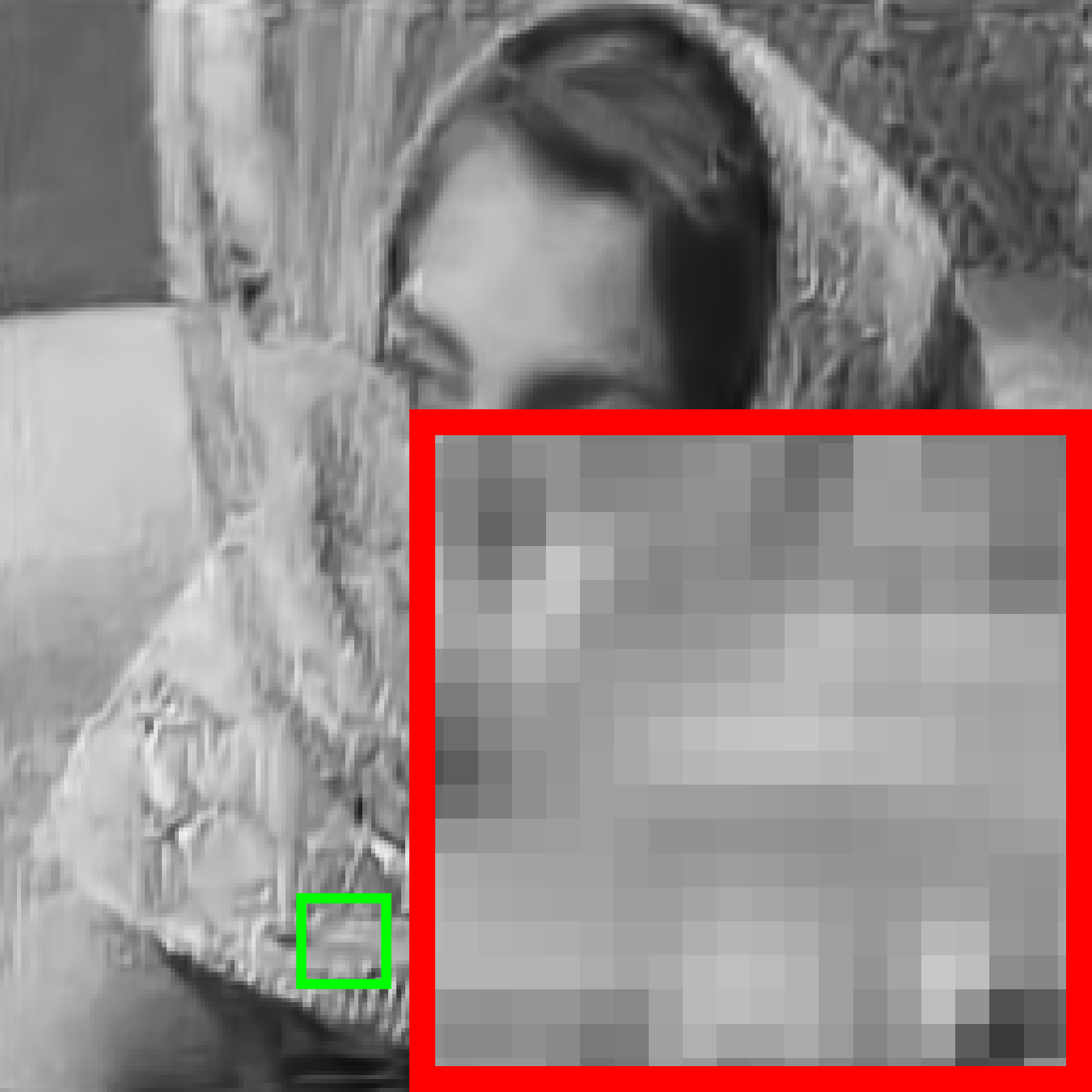}
    &\includegraphics[width=0.08\textwidth]{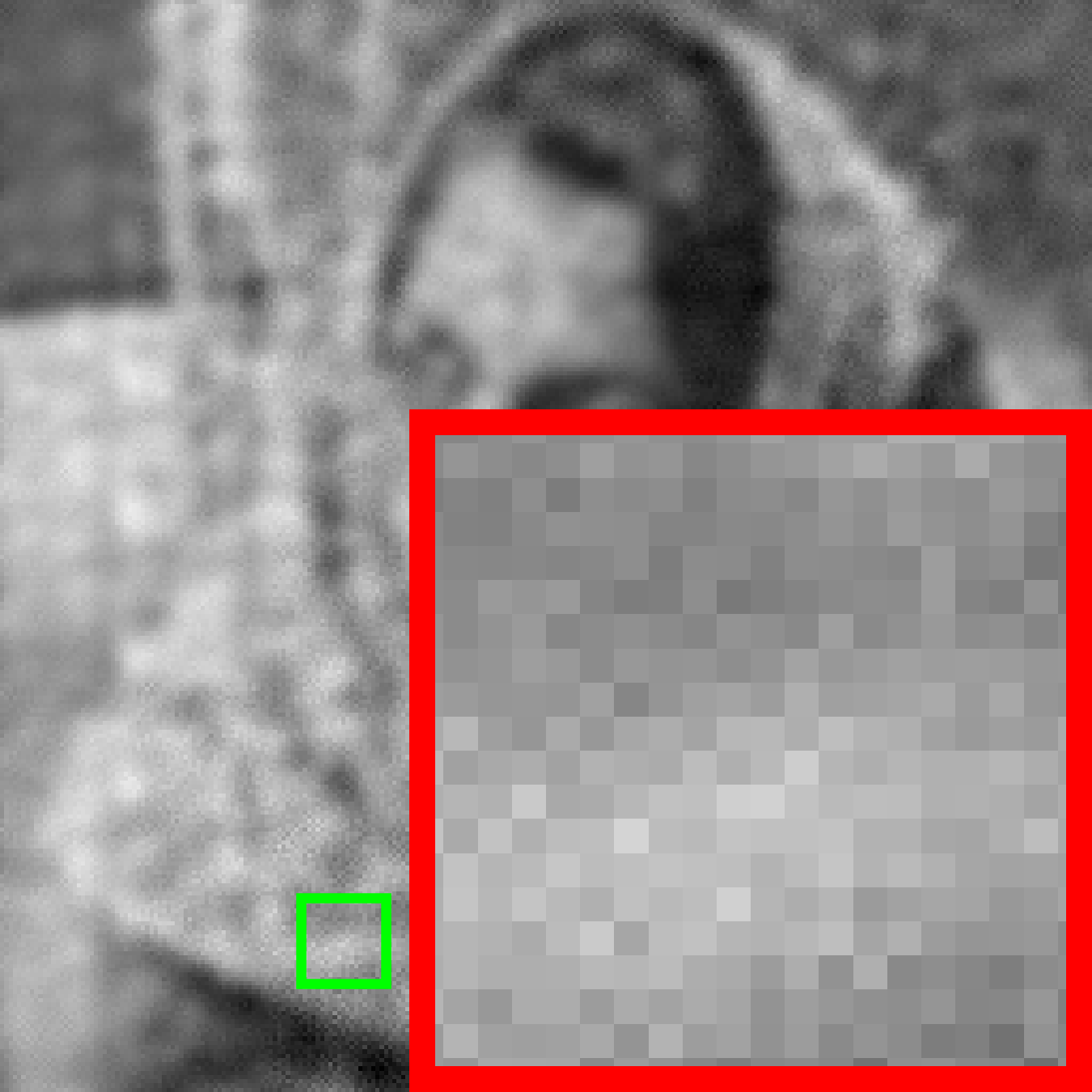}
    &\includegraphics[width=0.08\textwidth]{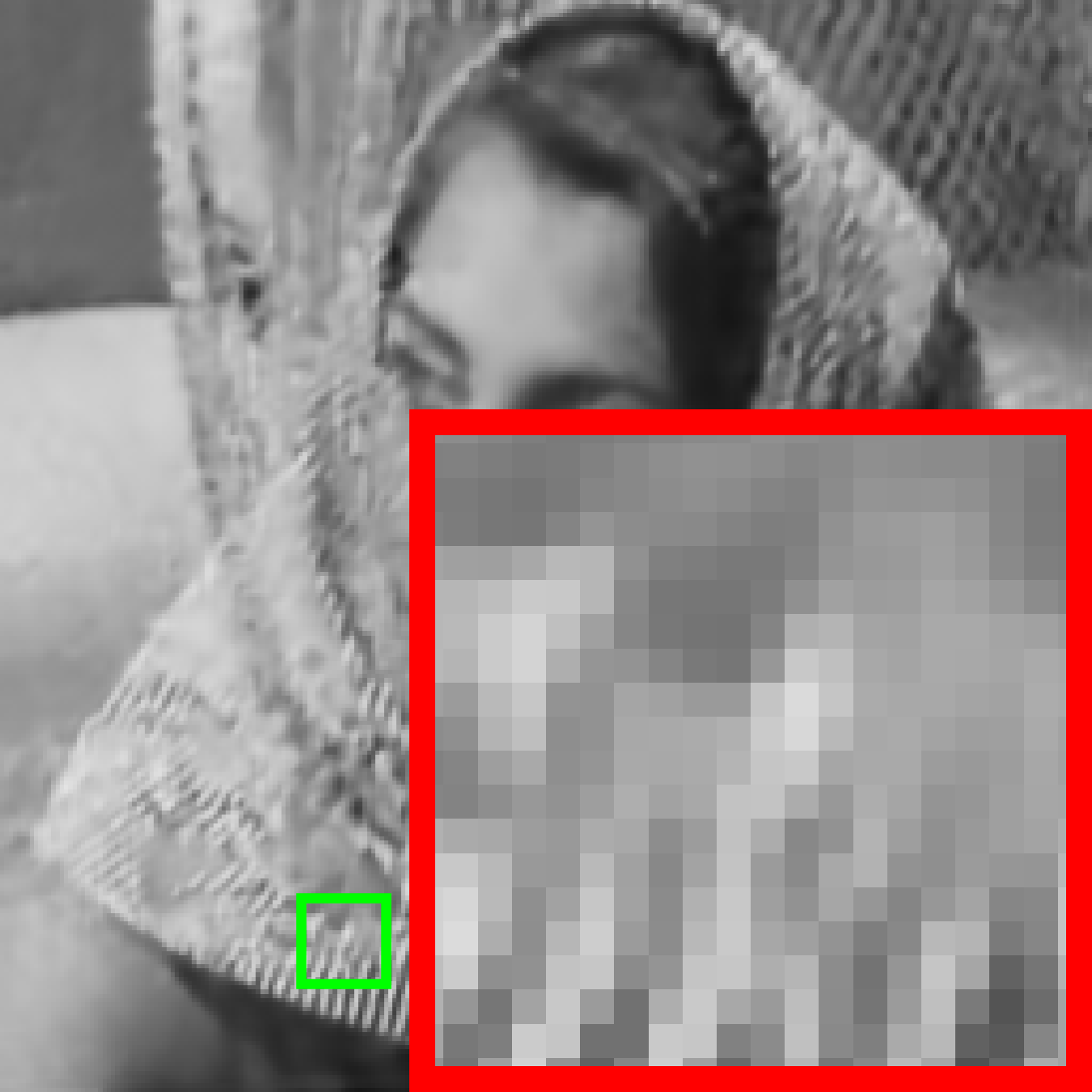}
    &\includegraphics[width=0.08\textwidth]{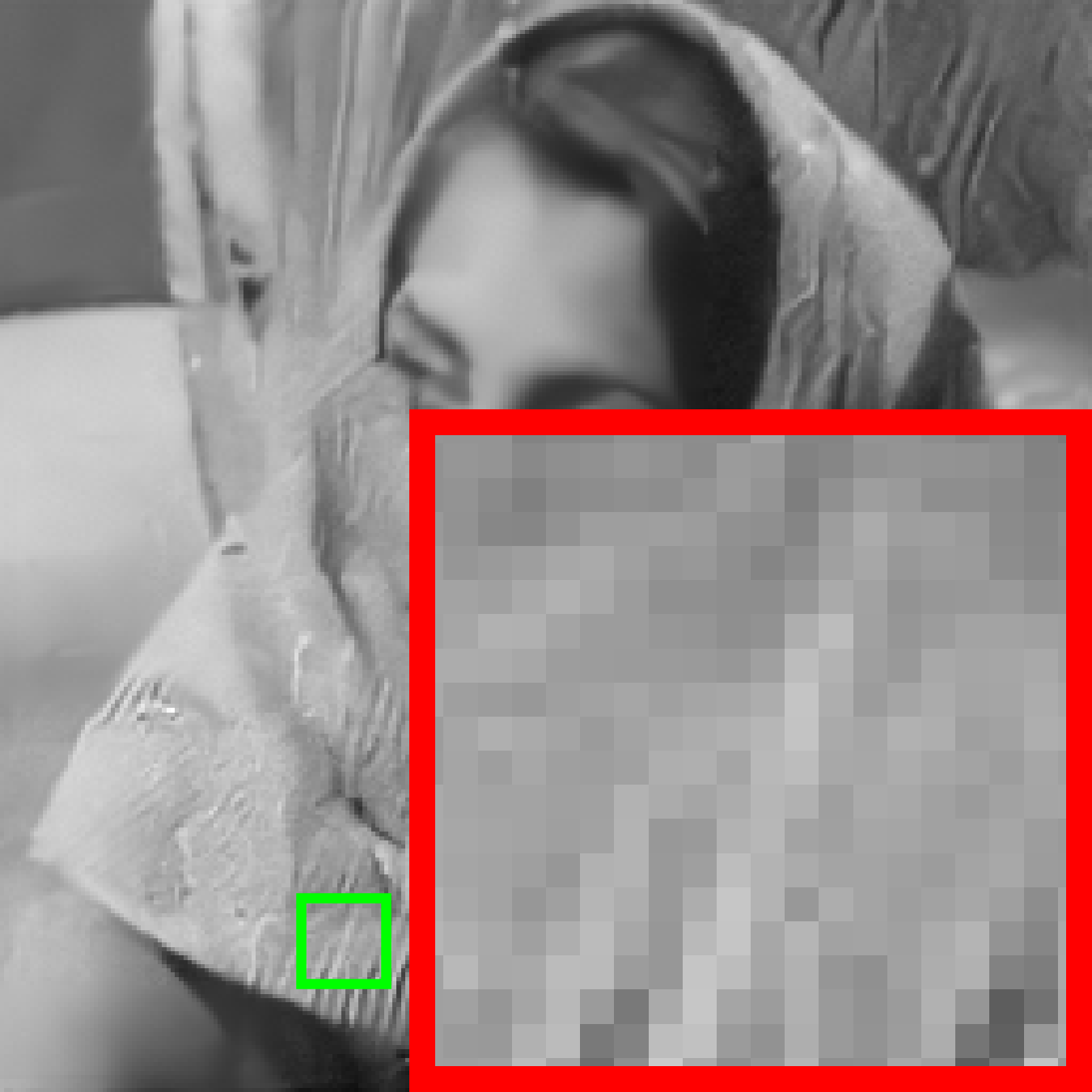}
    &\includegraphics[width=0.08\textwidth]{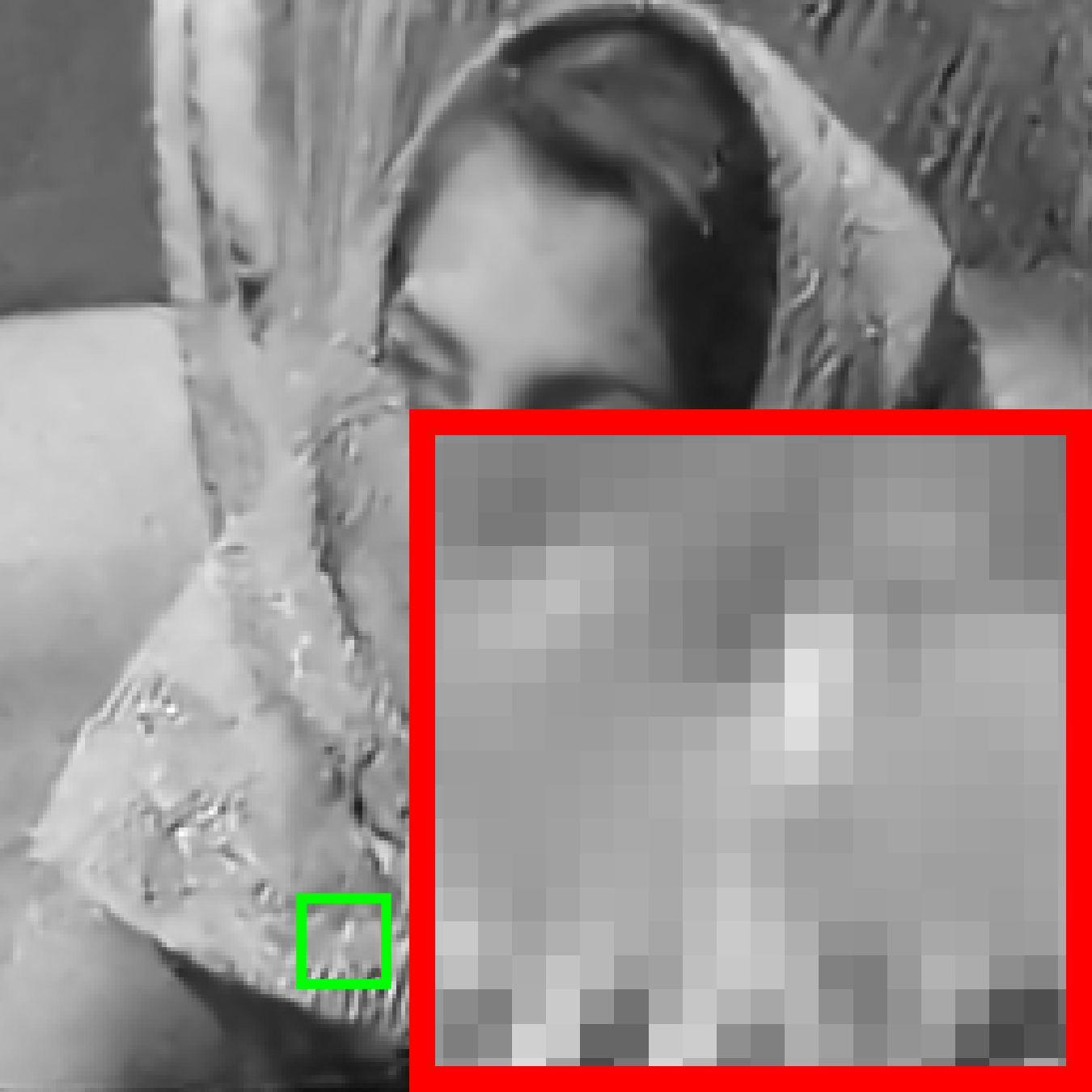}
    &\includegraphics[width=0.08\textwidth]{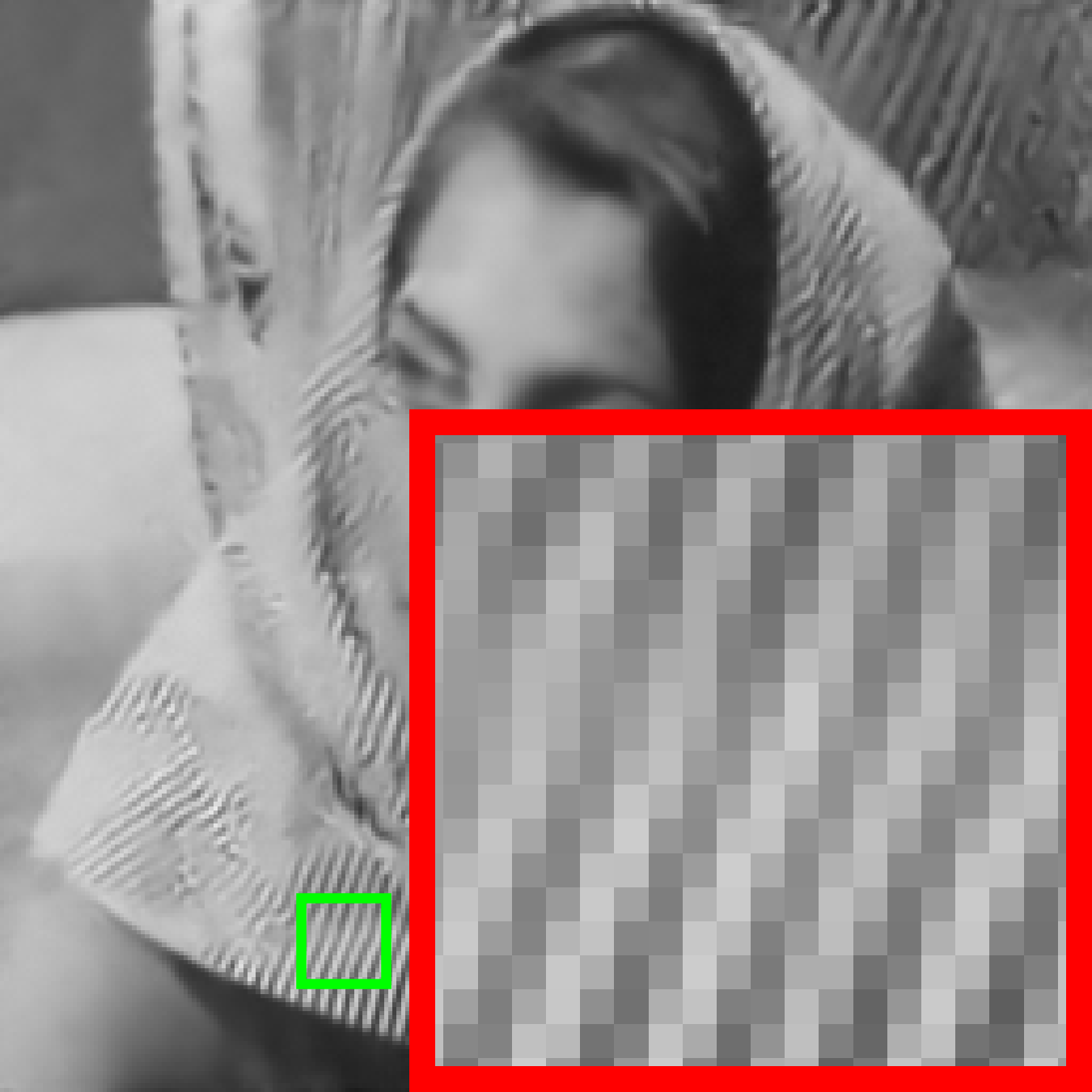}
    &\includegraphics[width=0.08\textwidth]{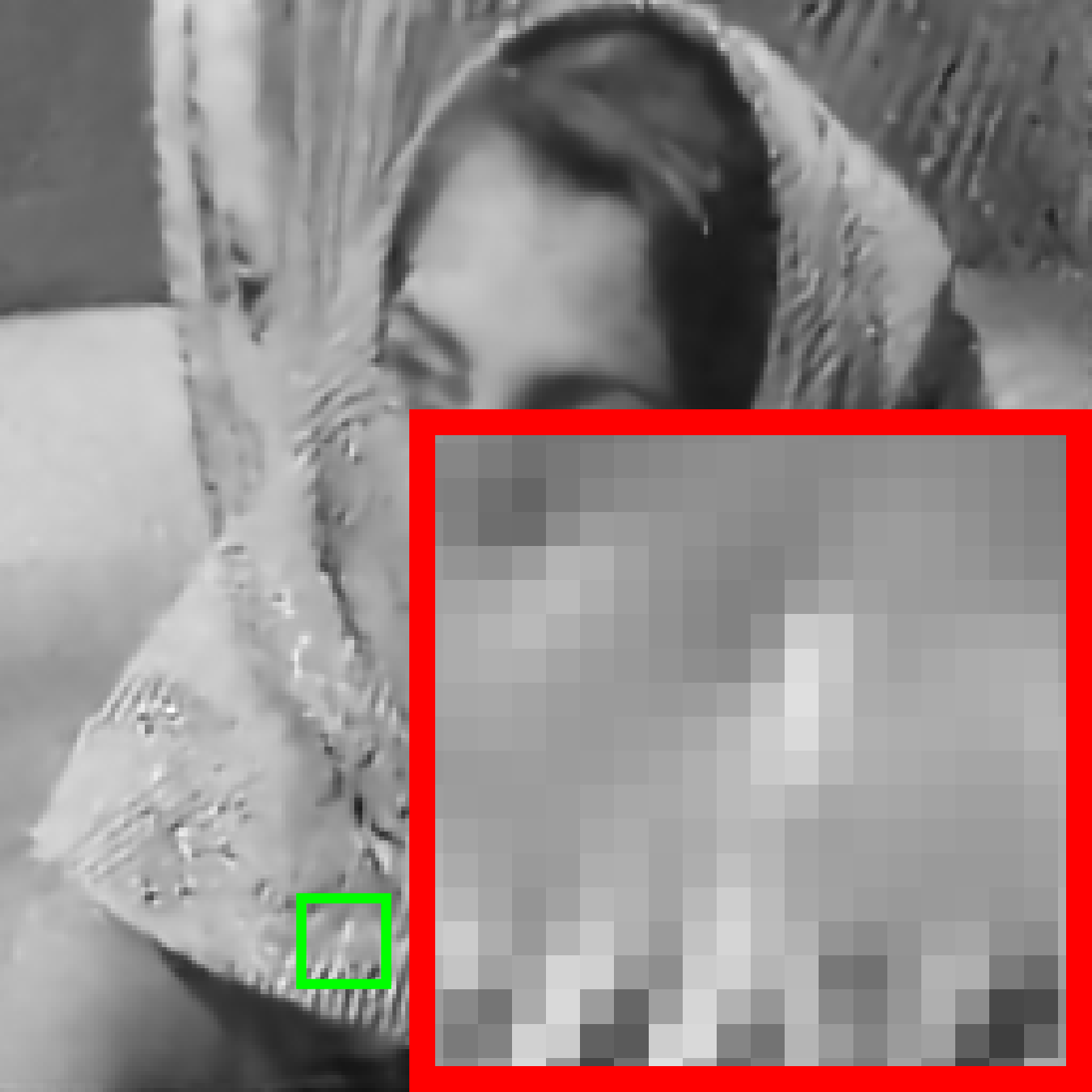}
    &\includegraphics[width=0.08\textwidth]{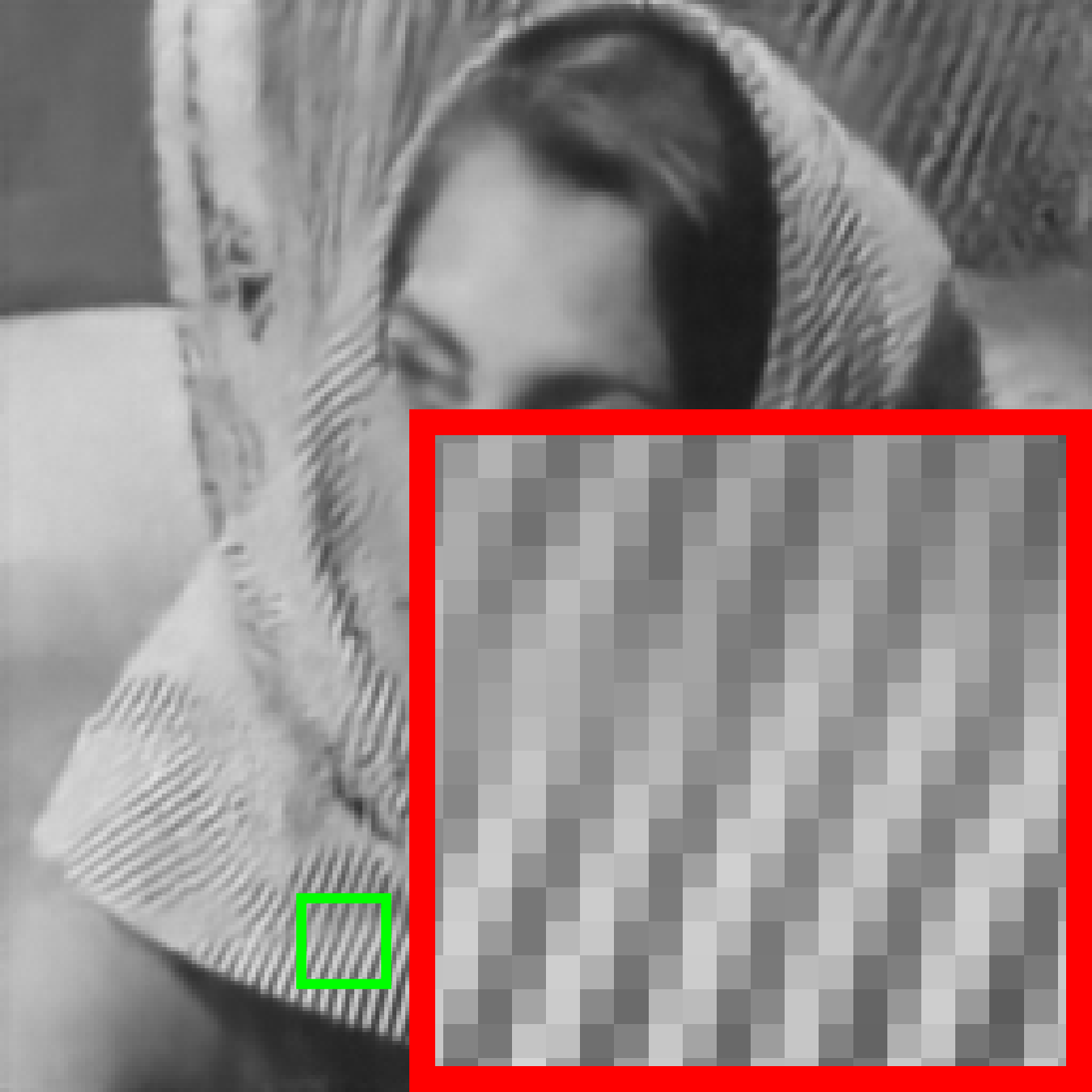}\\
    PSNR/SSIM & 22.56/0.63 & 23.52/0.69 & 25.07/0.76 & 25.22/0.78 & 19.90/0.56 & 23.41/0.71 & 21.94/0.55 & 25.13/0.77 & 24.51/0.75 & 23.62/0.73 & \underline{\textcolor{blue}{27.49}}/\underline{\textcolor{blue}{0.84}} & 24.09/0.75 & \textbf{\textcolor{red}{27.69}}/\textbf{\textcolor{red}{0.85}}\\
    \includegraphics[width=0.08\textwidth]{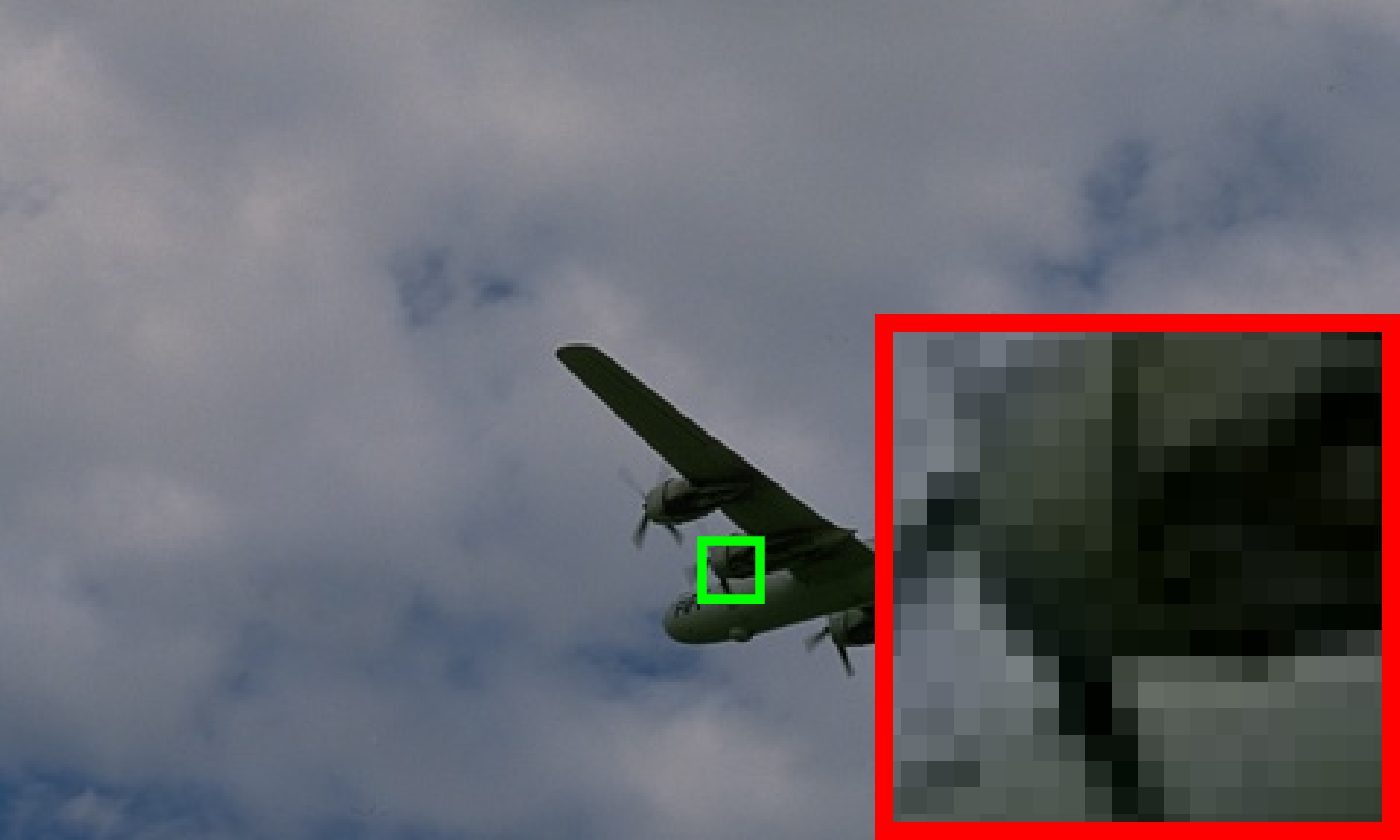}
    &\includegraphics[width=0.08\textwidth]{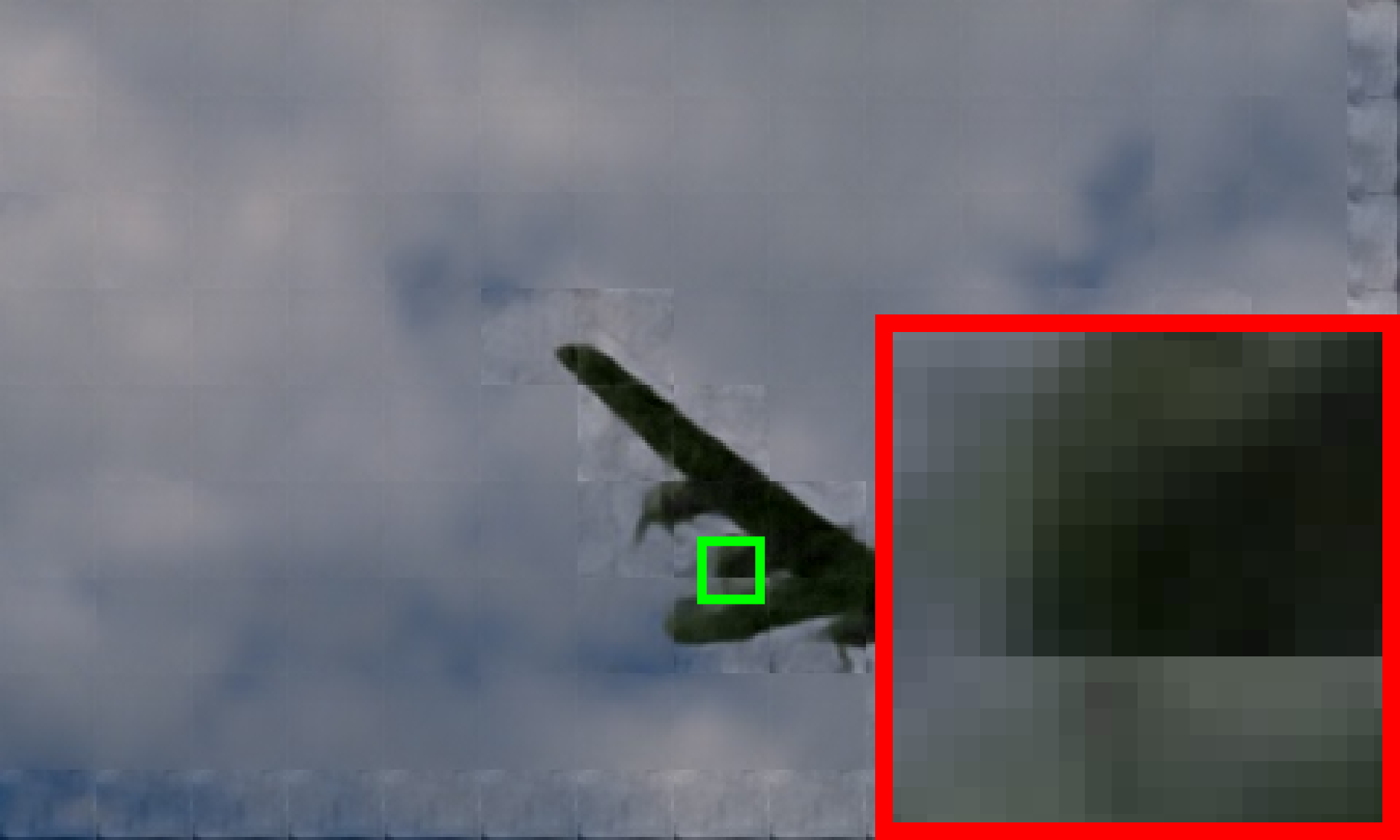}
    &\includegraphics[width=0.08\textwidth]{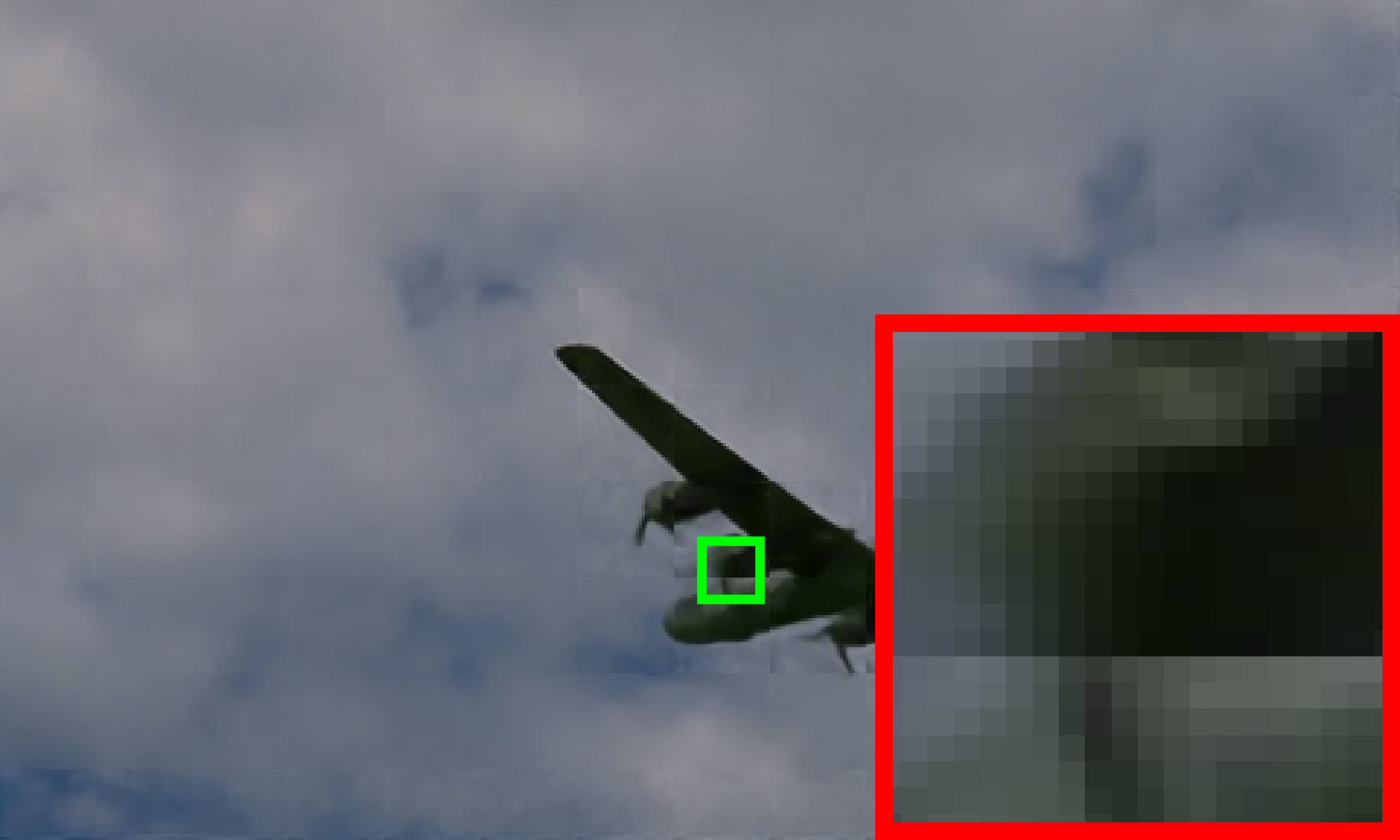}
    &\includegraphics[width=0.08\textwidth]{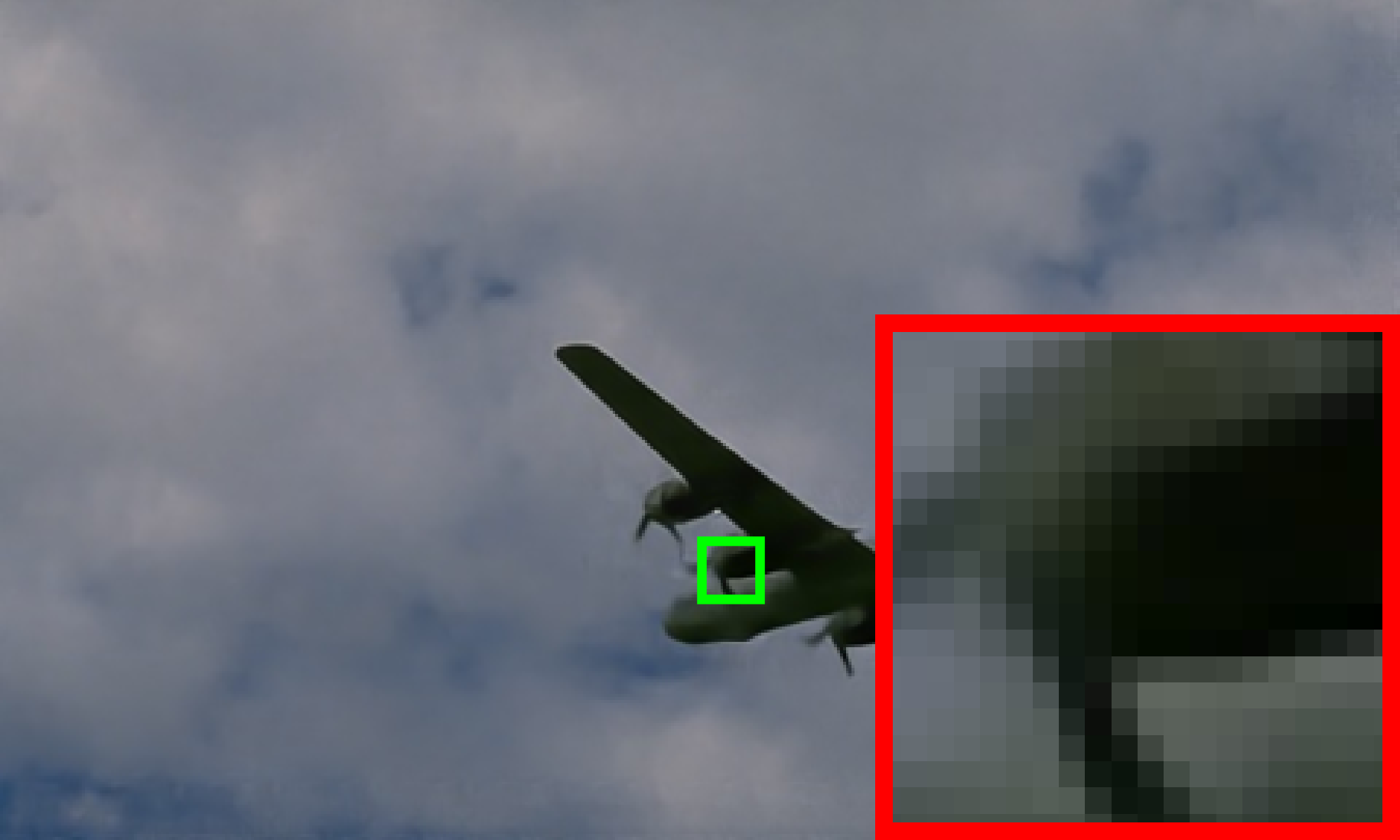}
    &\includegraphics[width=0.08\textwidth]{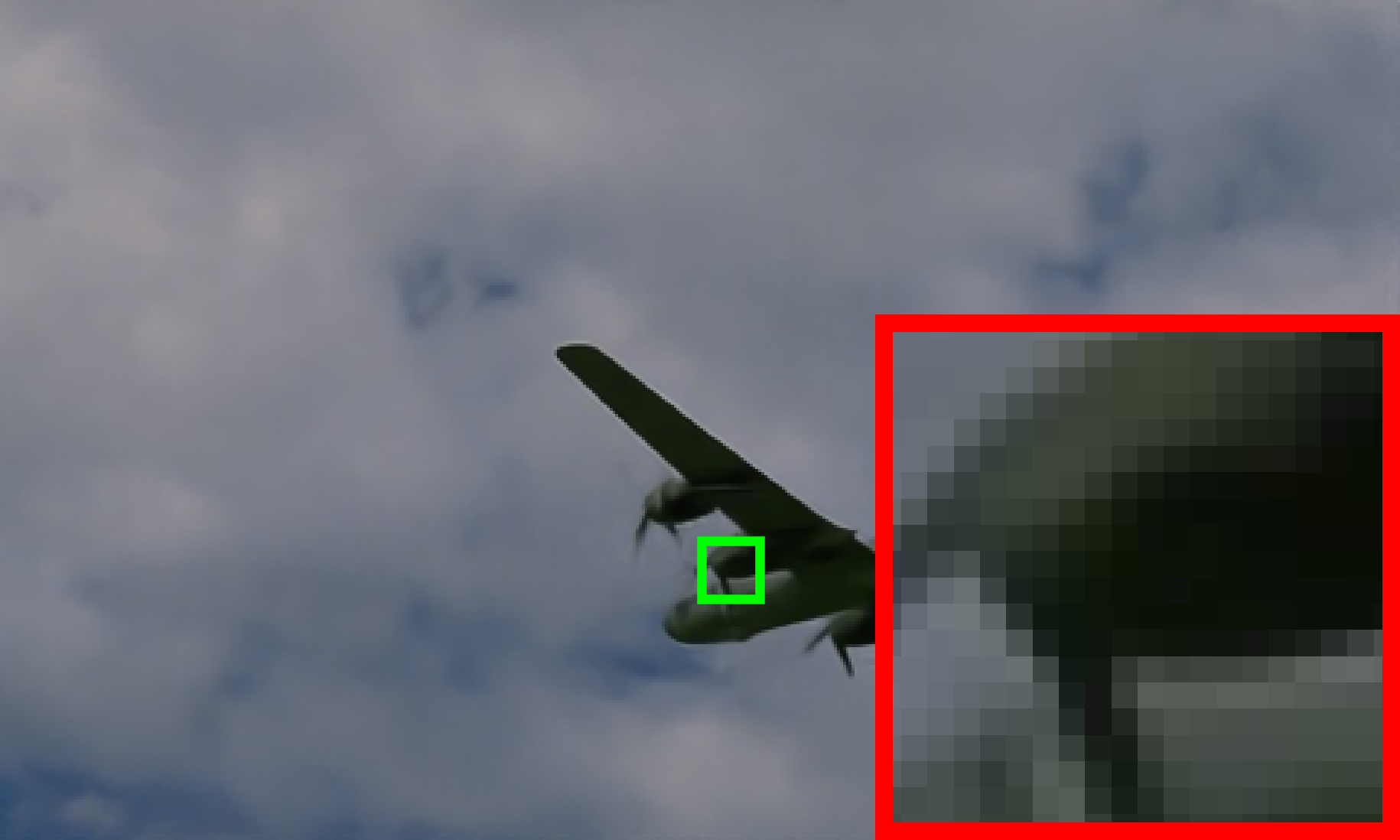}
    &\includegraphics[width=0.08\textwidth]{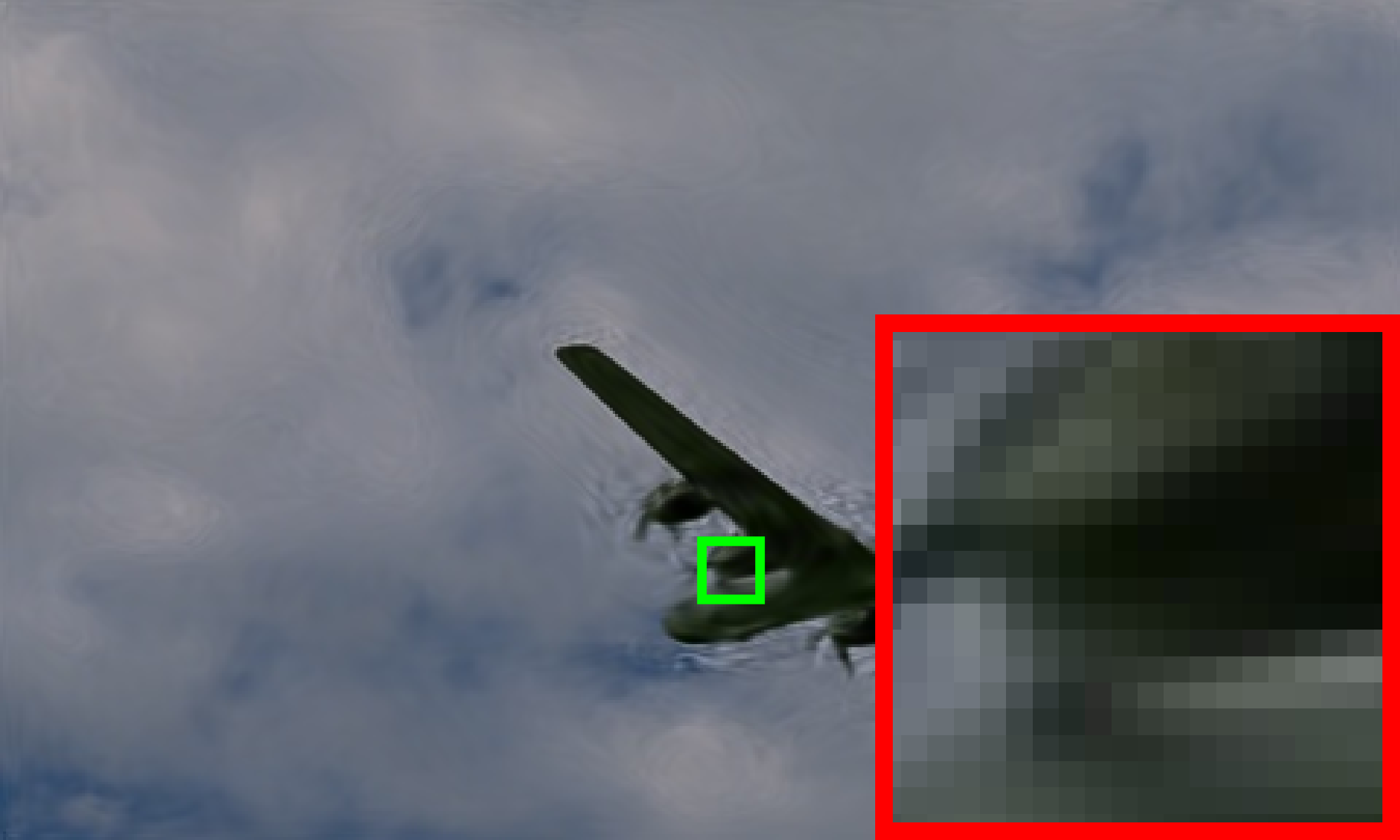}
    &\includegraphics[width=0.08\textwidth]{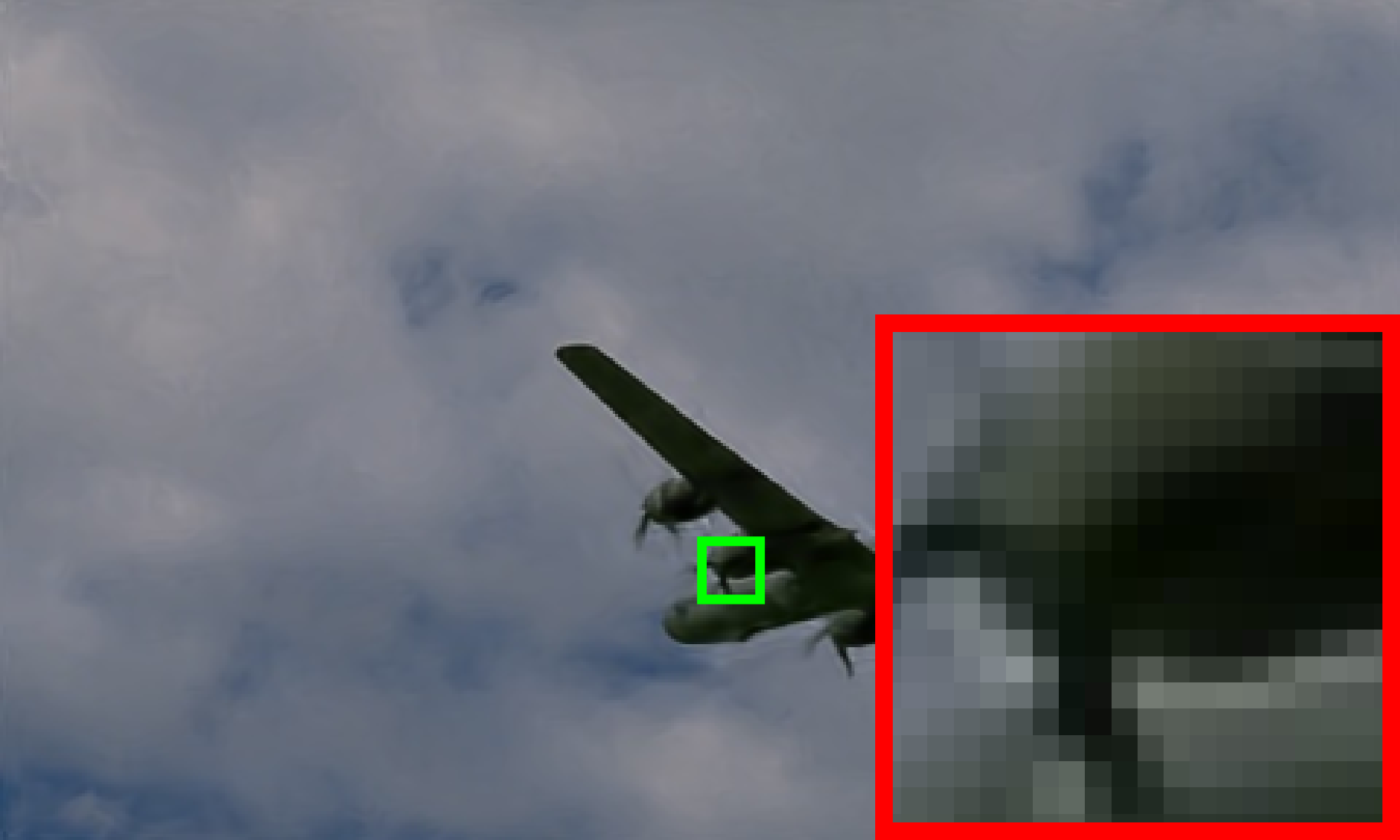}
    &\includegraphics[width=0.08\textwidth]{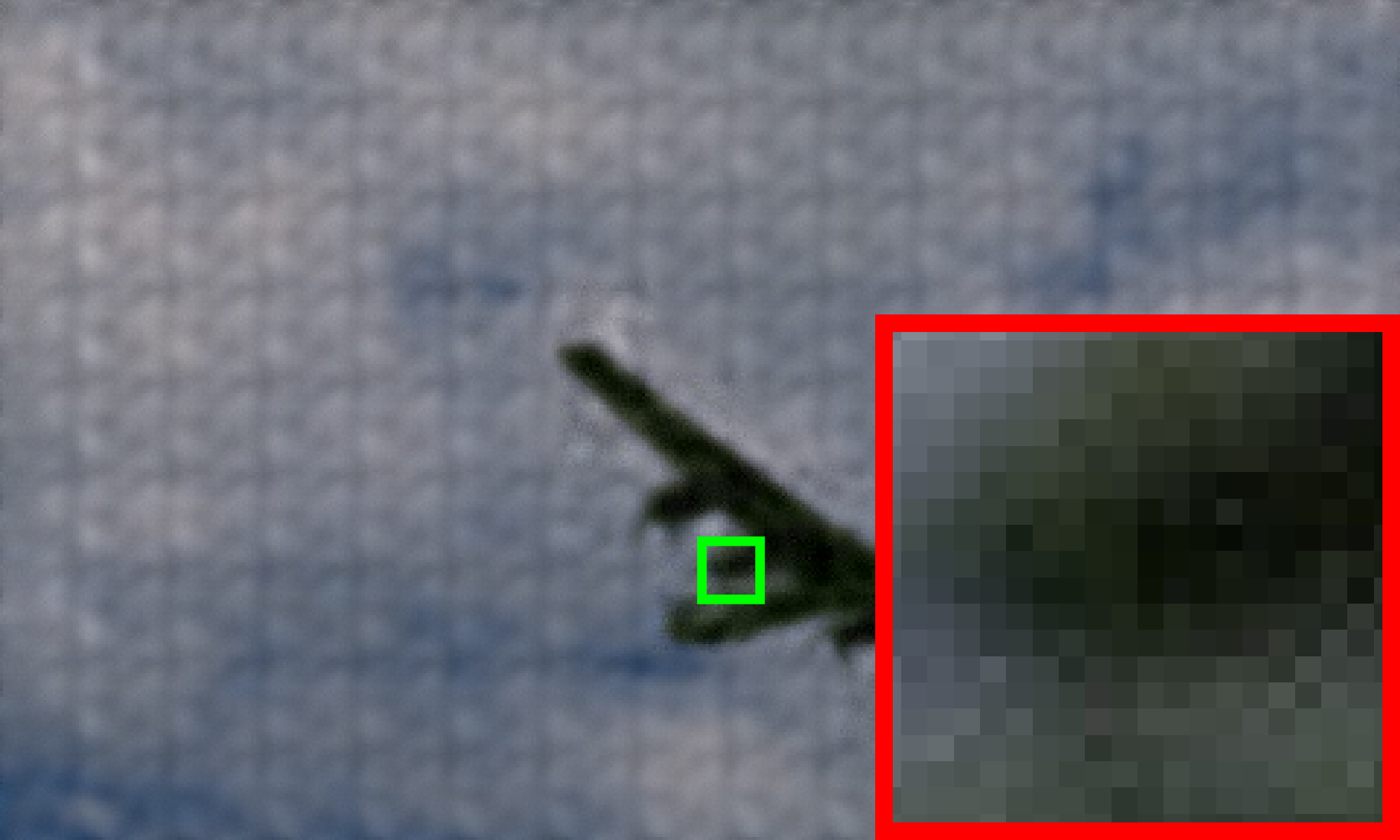}
    &\includegraphics[width=0.08\textwidth]{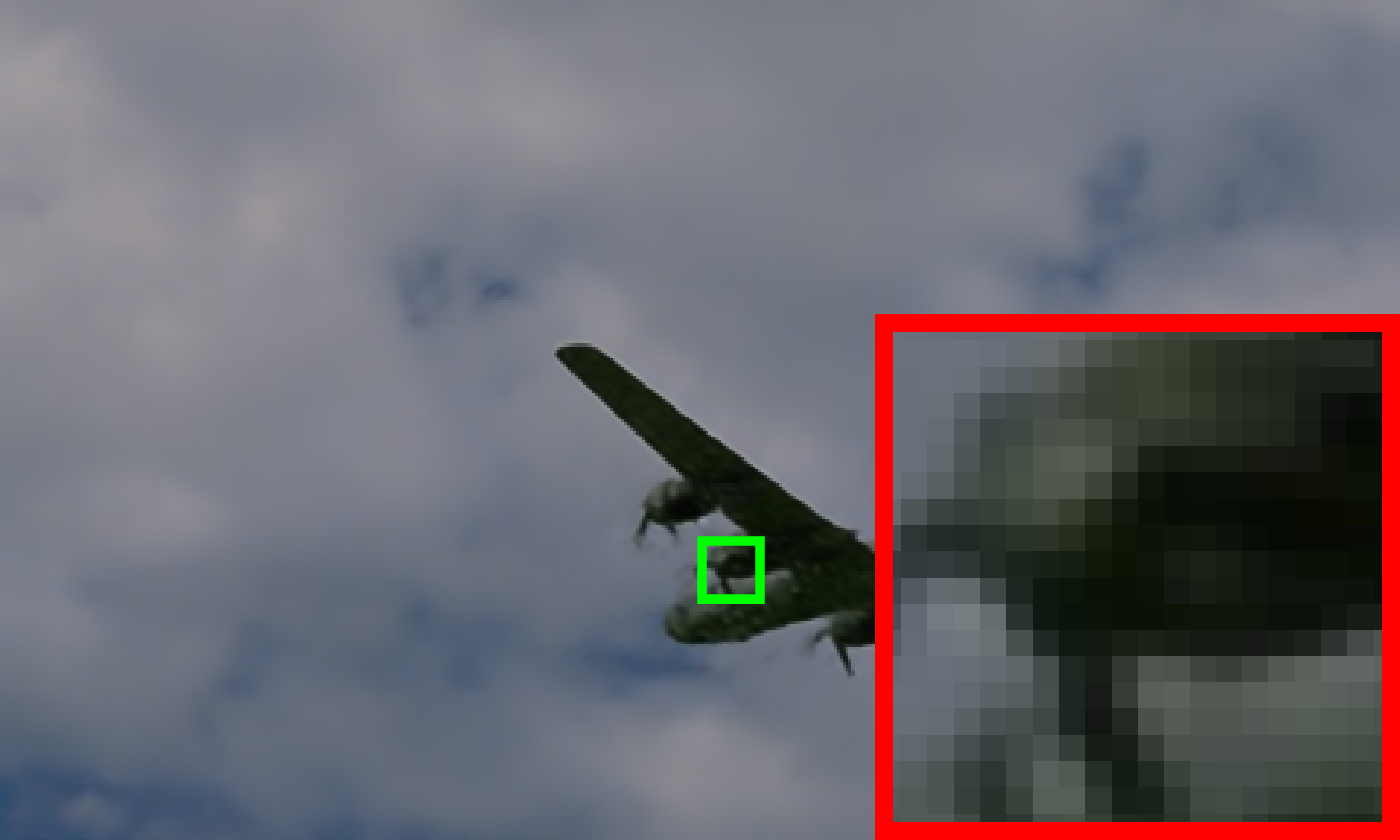}
    &\includegraphics[width=0.08\textwidth]{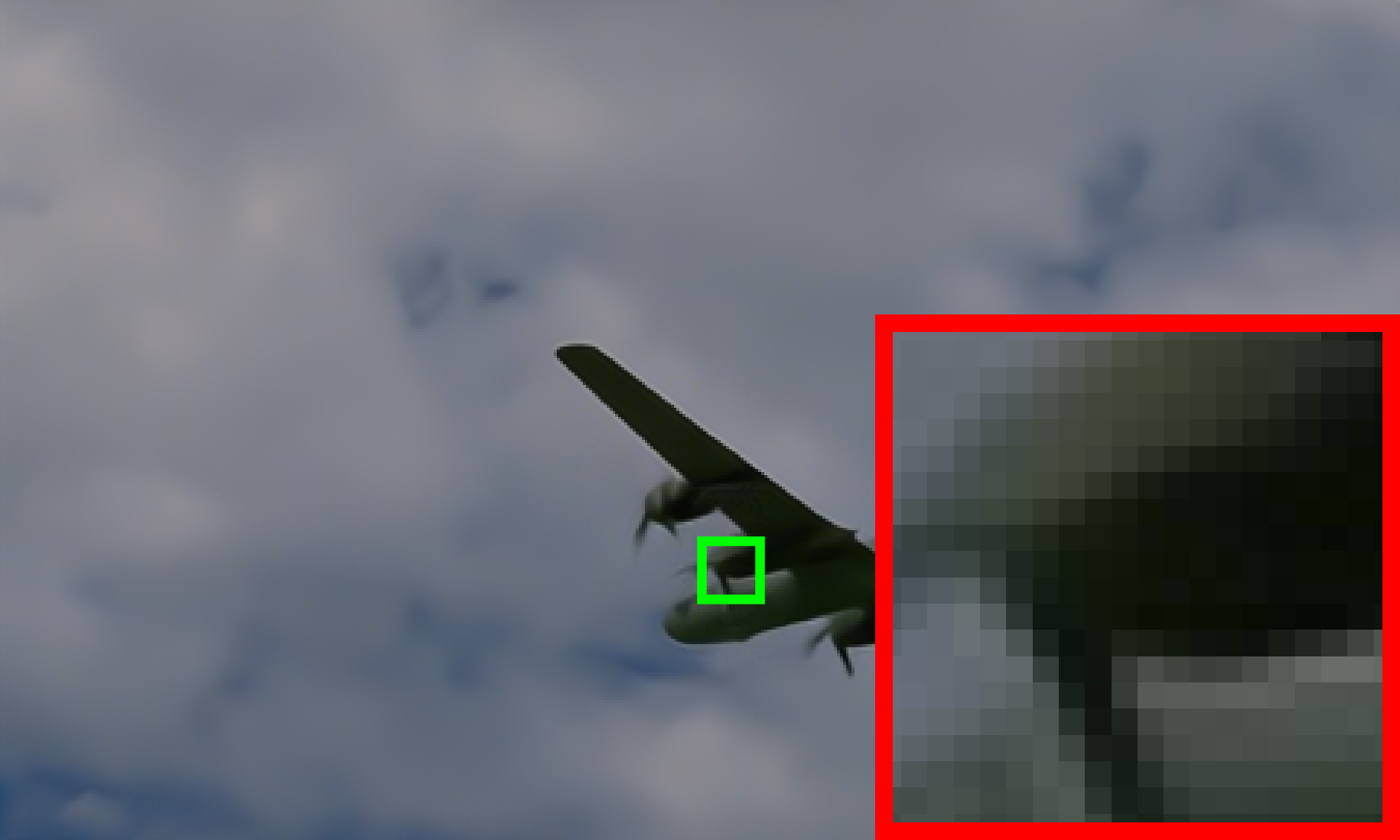}
    &\includegraphics[width=0.08\textwidth]{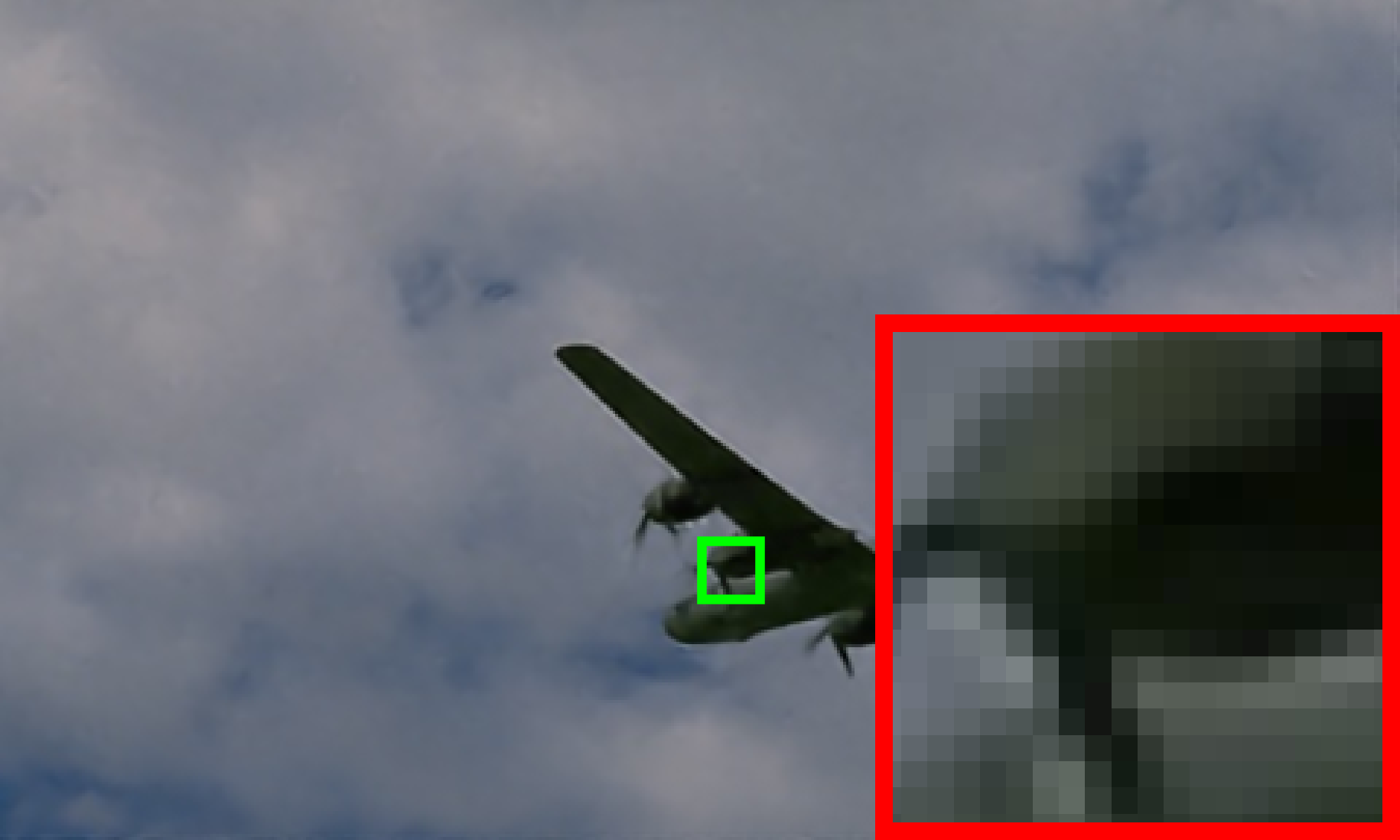}
    &\includegraphics[width=0.08\textwidth]{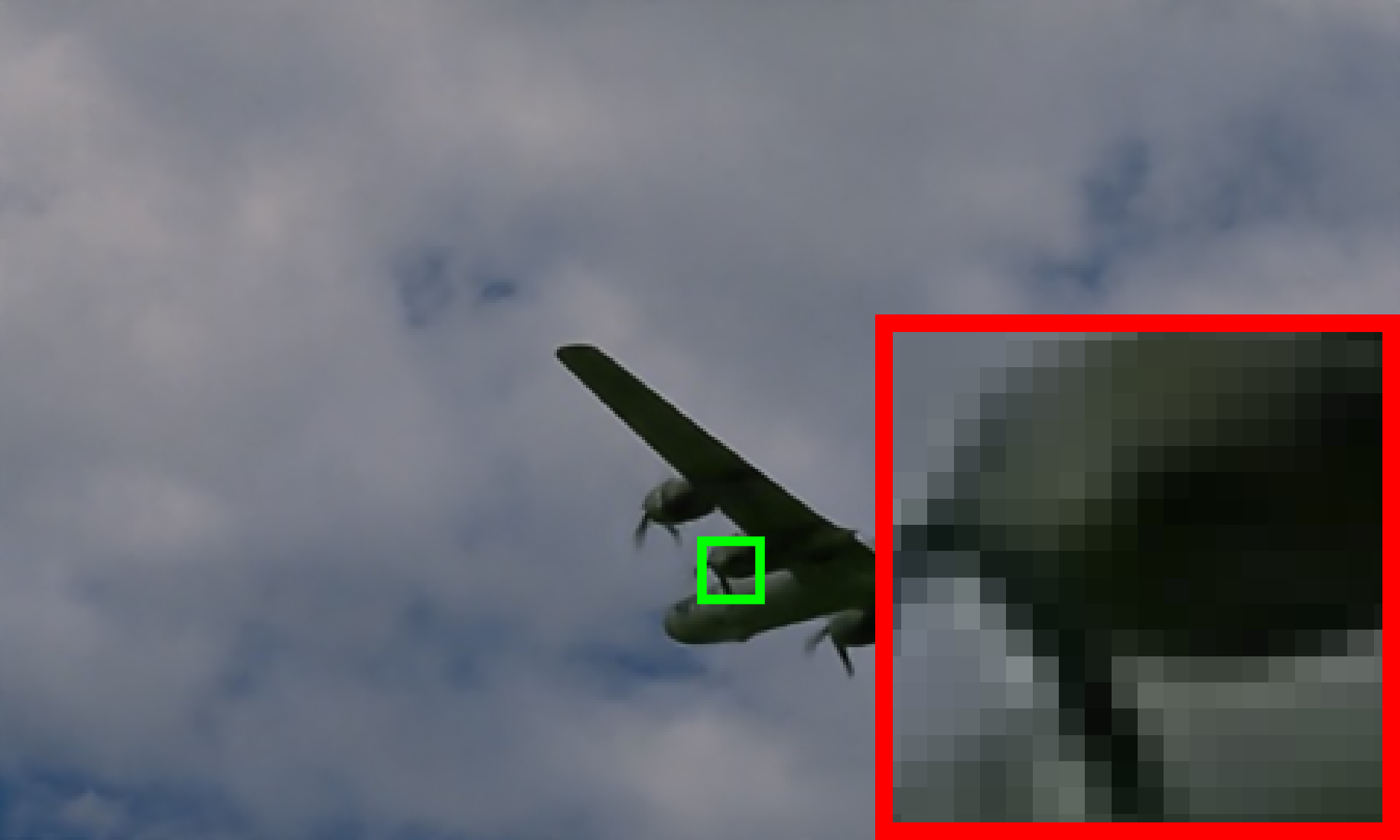}
    &\includegraphics[width=0.08\textwidth]{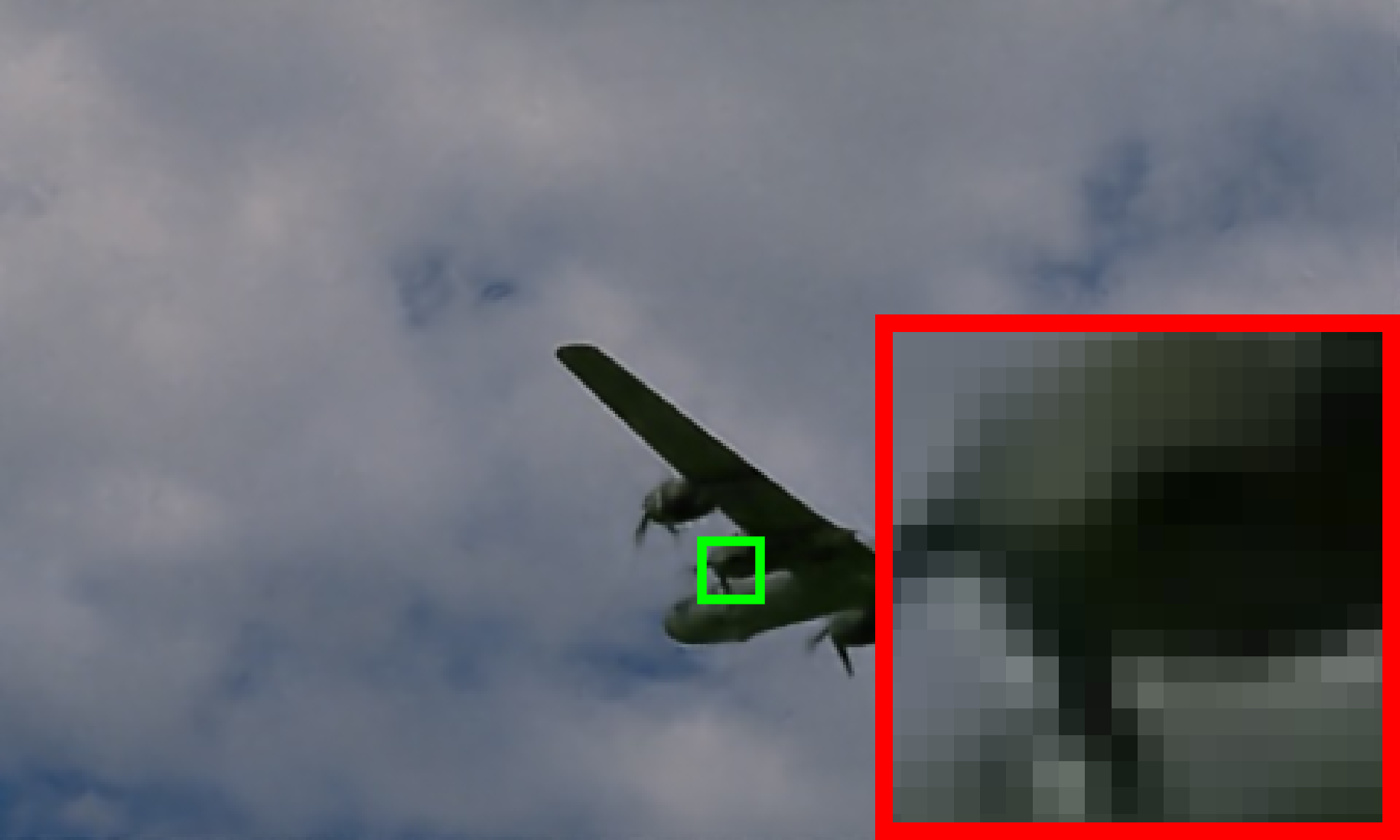}
    &\includegraphics[width=0.08\textwidth]{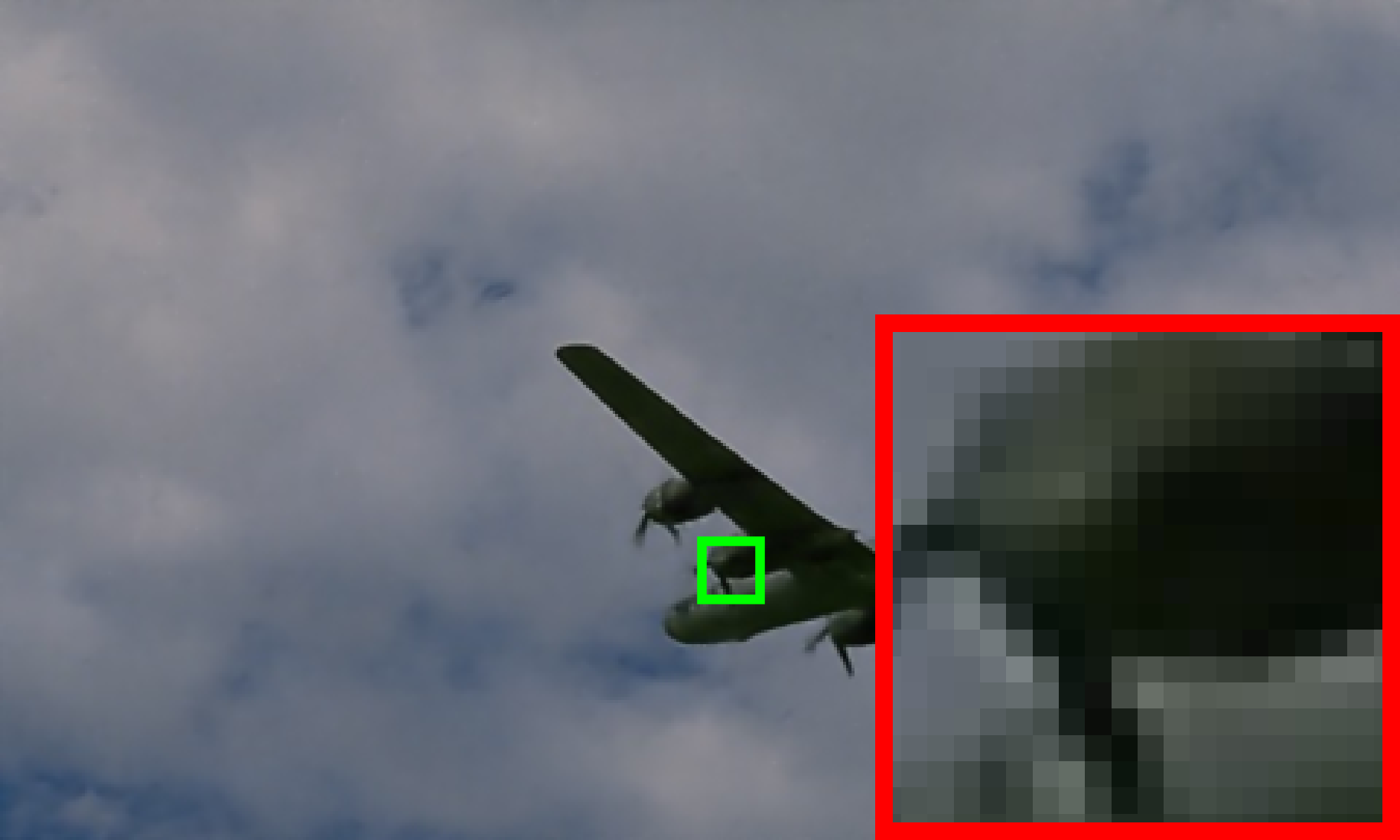}\\
    PSNR/SSIM & 31.98/0.91 & 37.07/0.96 & 39.89/\underline{\textcolor{blue}{0.97}} & 39.95/\underline{\textcolor{blue}{0.97}} & 34.98/0.94 & 38.71/\underline{\textcolor{blue}{0.97}} & 29.16/0.75 & 38.87/\underline{\textcolor{blue}{0.97}} & 39.26/\underline{\textcolor{blue}{0.97}} & 39.36/\underline{\textcolor{blue}{0.97}} & \textbf{\textcolor{red}{41.02}}/\textbf{\textcolor{red}{0.98}} & 39.71/\textbf{\textcolor{red}{0.98}} & \underline{\textcolor{blue}{40.53}}/\textbf{\textcolor{red}{0.98}}\\
    \includegraphics[width=0.08\textwidth]{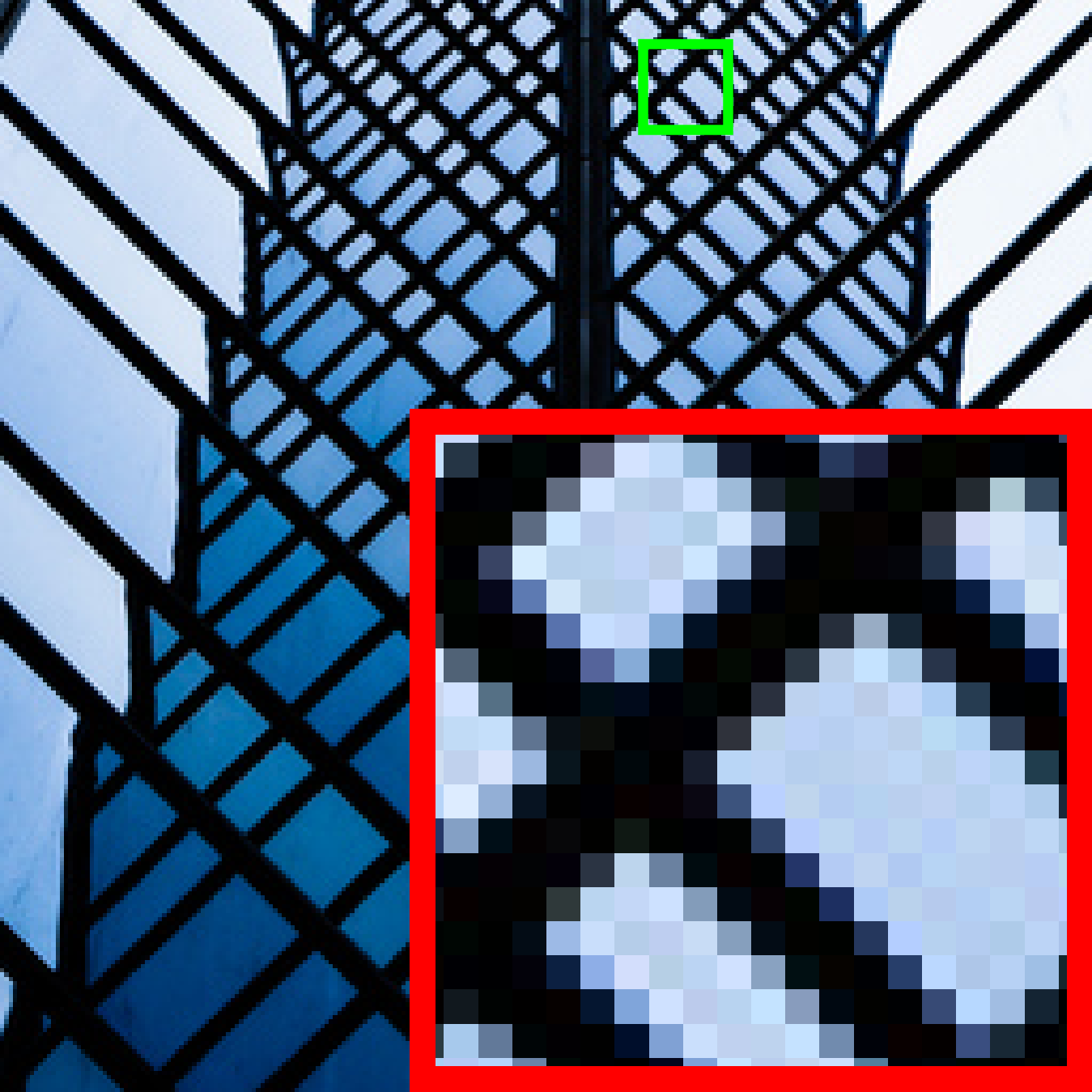}
    &\includegraphics[width=0.08\textwidth]{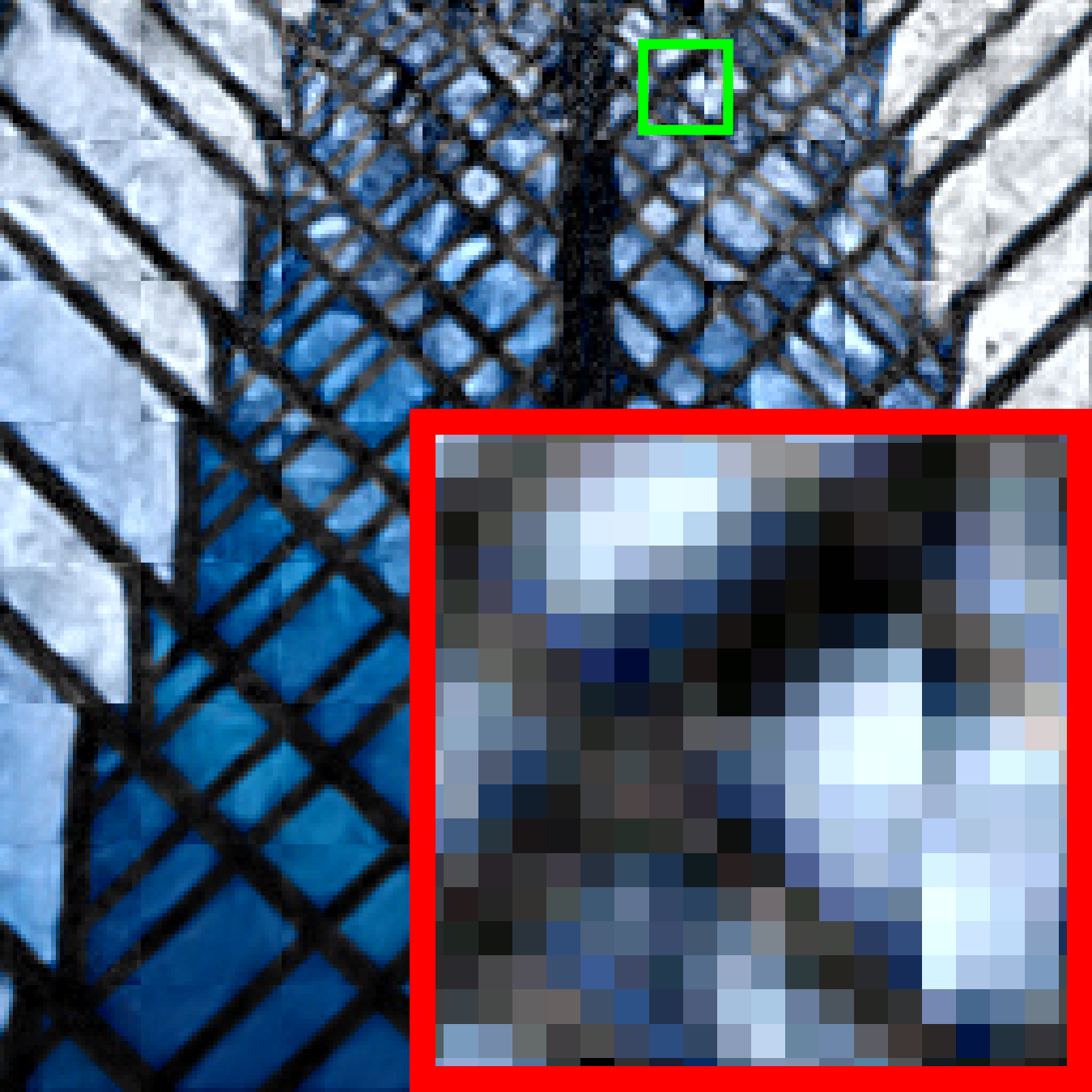}
    &\includegraphics[width=0.08\textwidth]{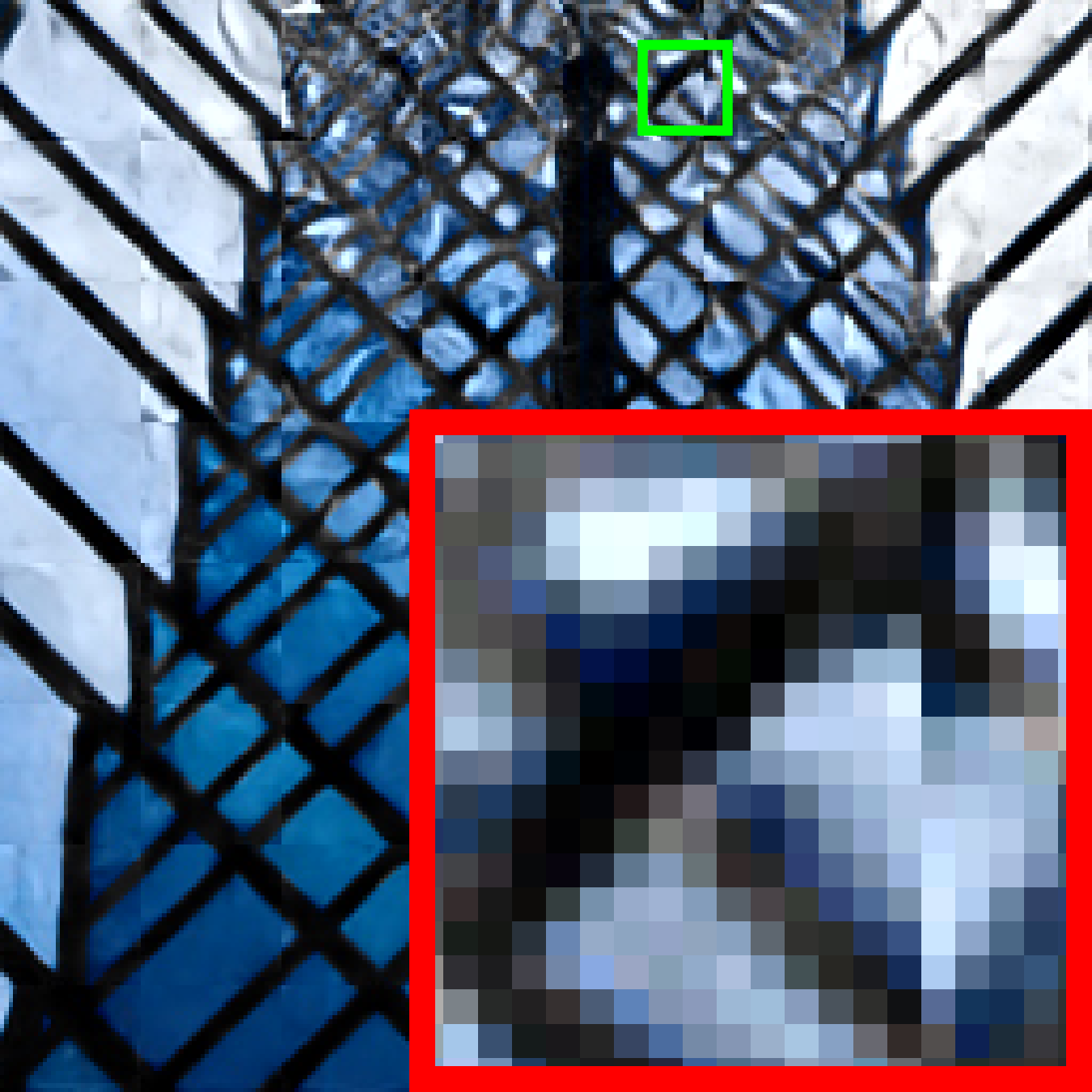}
    &\includegraphics[width=0.08\textwidth]{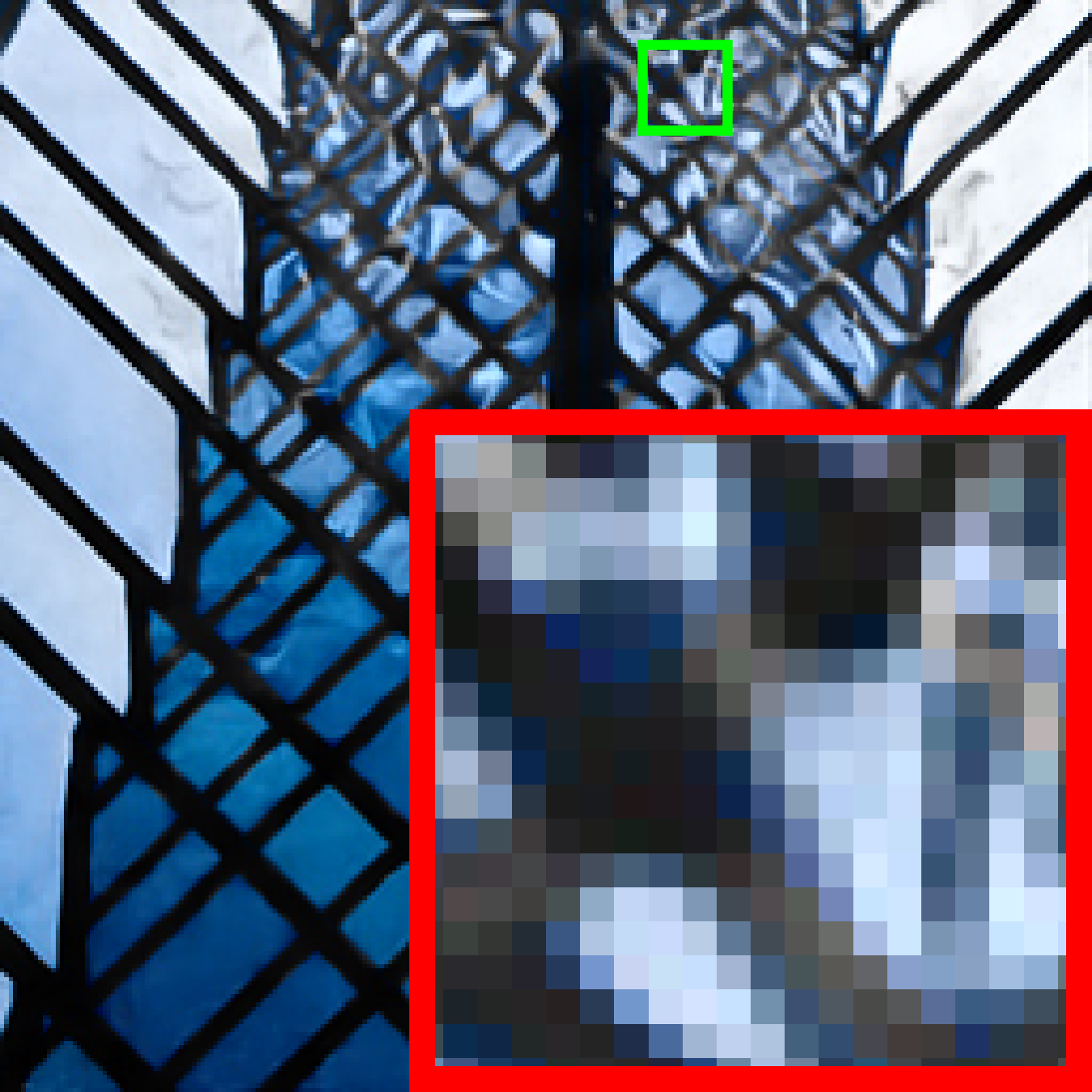}
    &\includegraphics[width=0.08\textwidth]{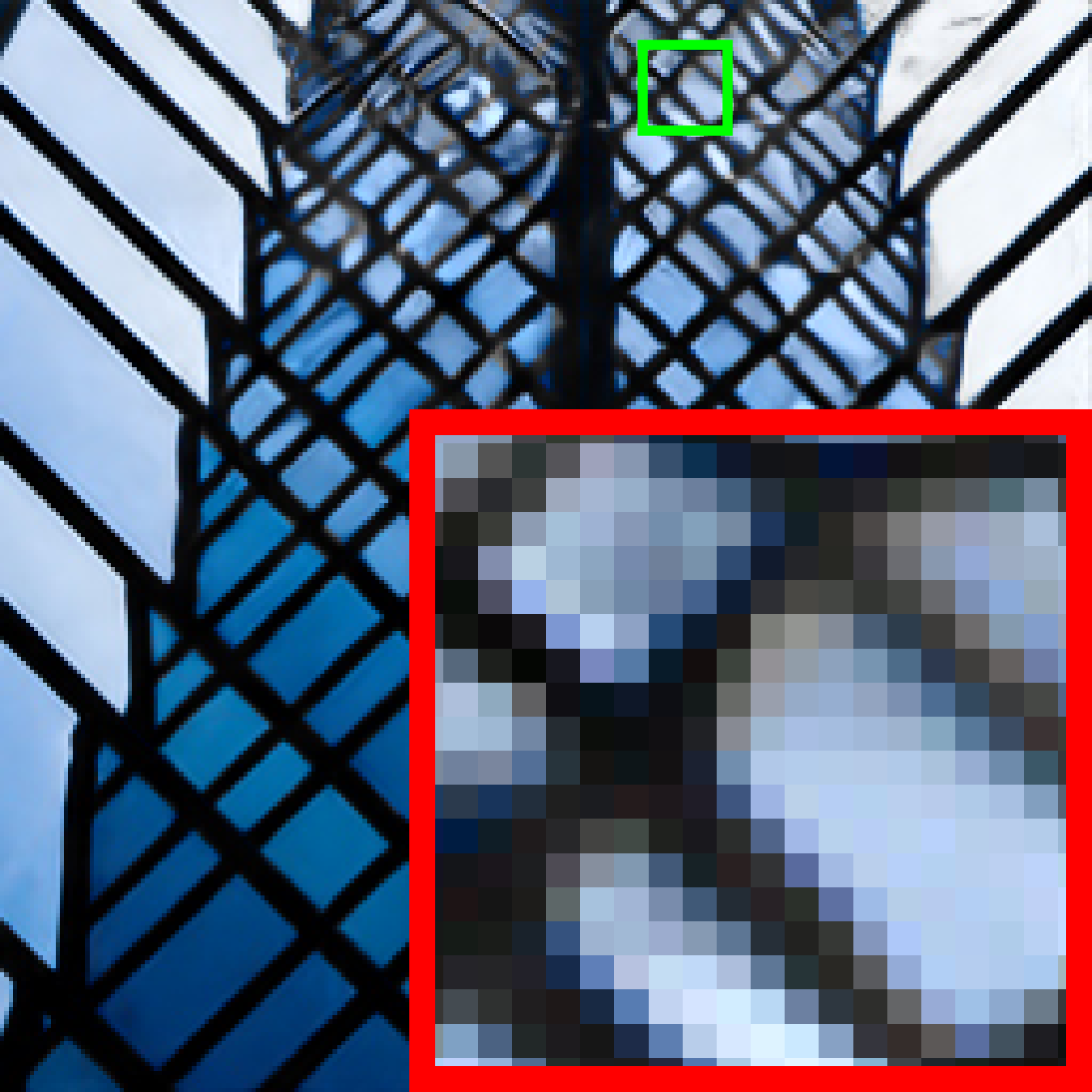}
    &\includegraphics[width=0.08\textwidth]{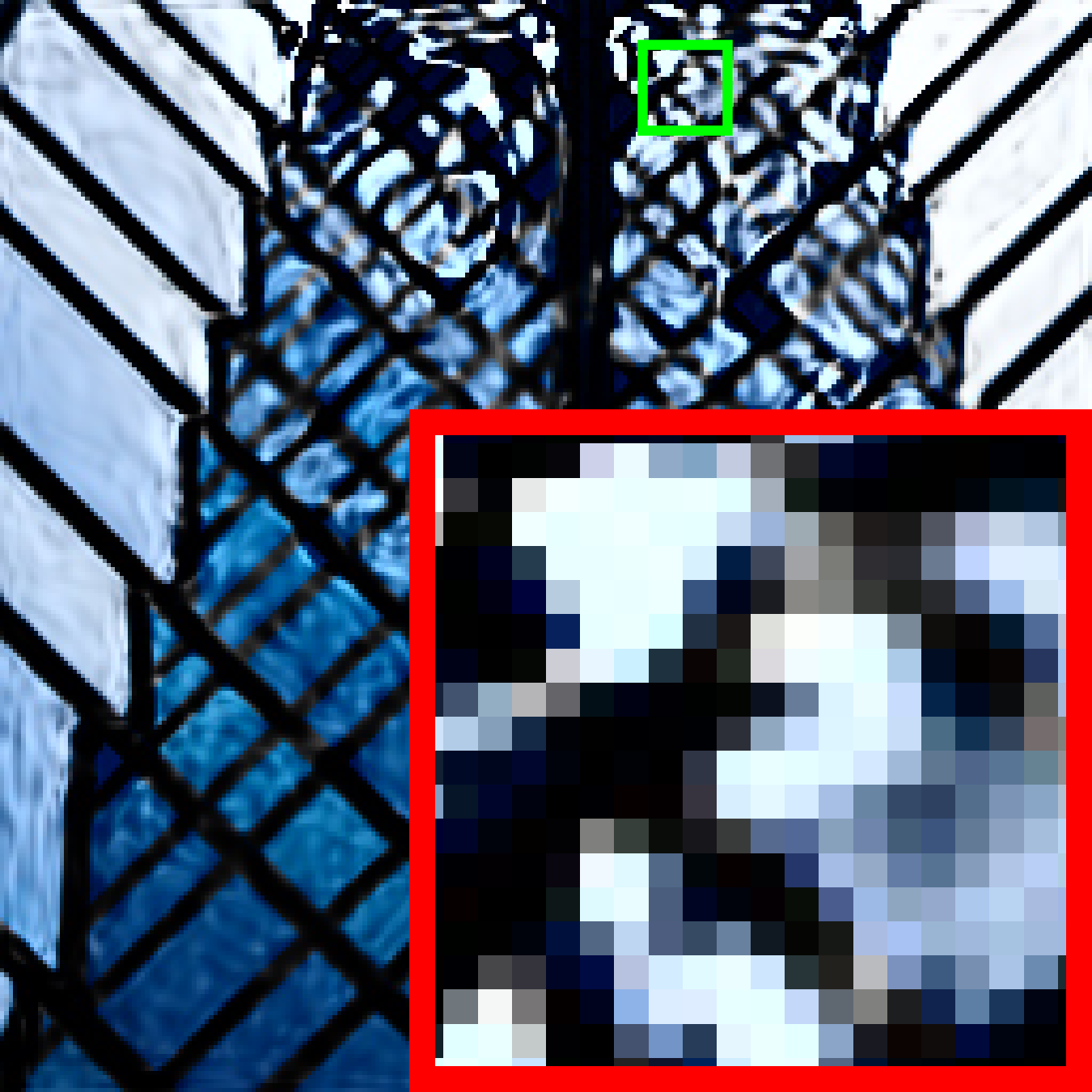}
    &\includegraphics[width=0.08\textwidth]{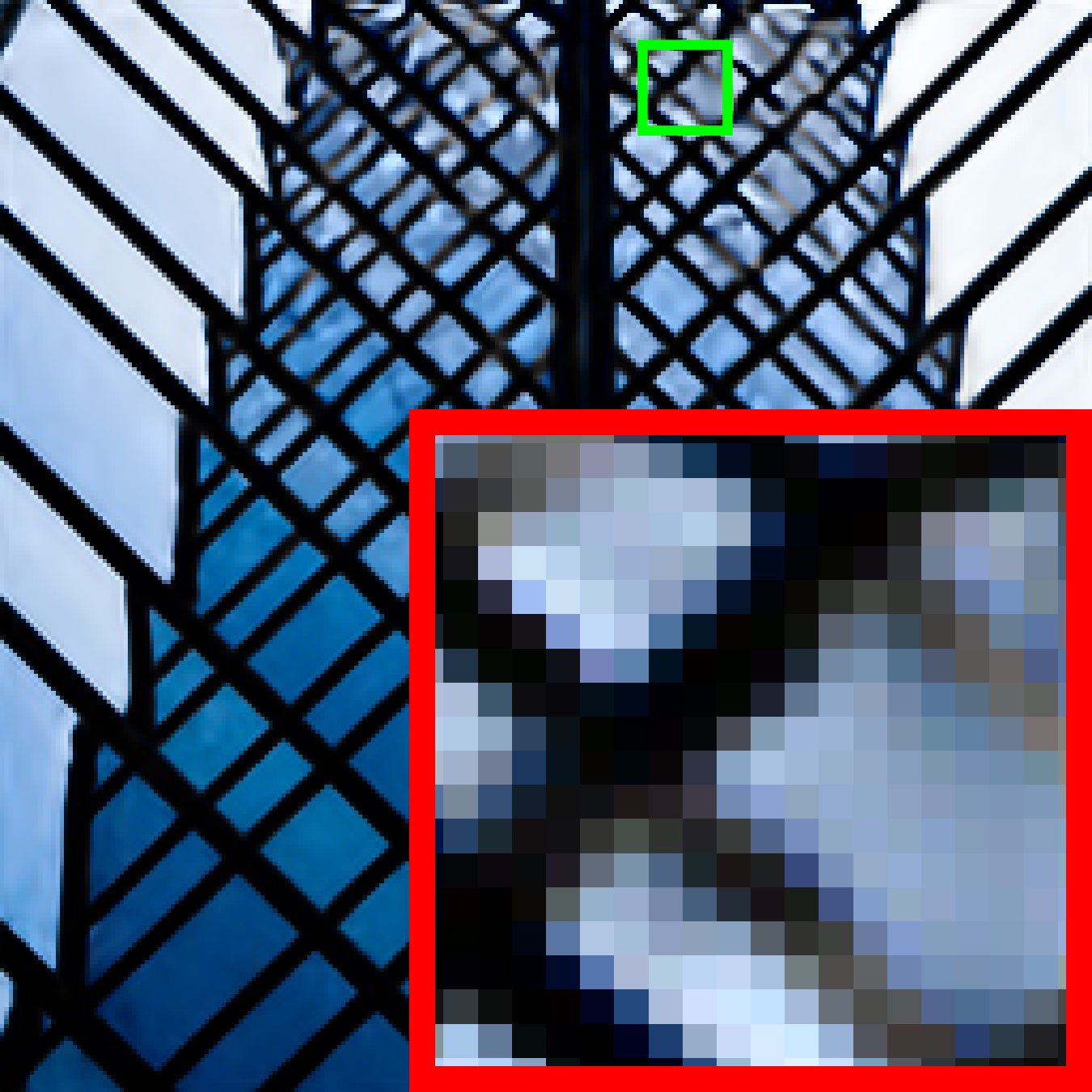}
    &\includegraphics[width=0.08\textwidth]{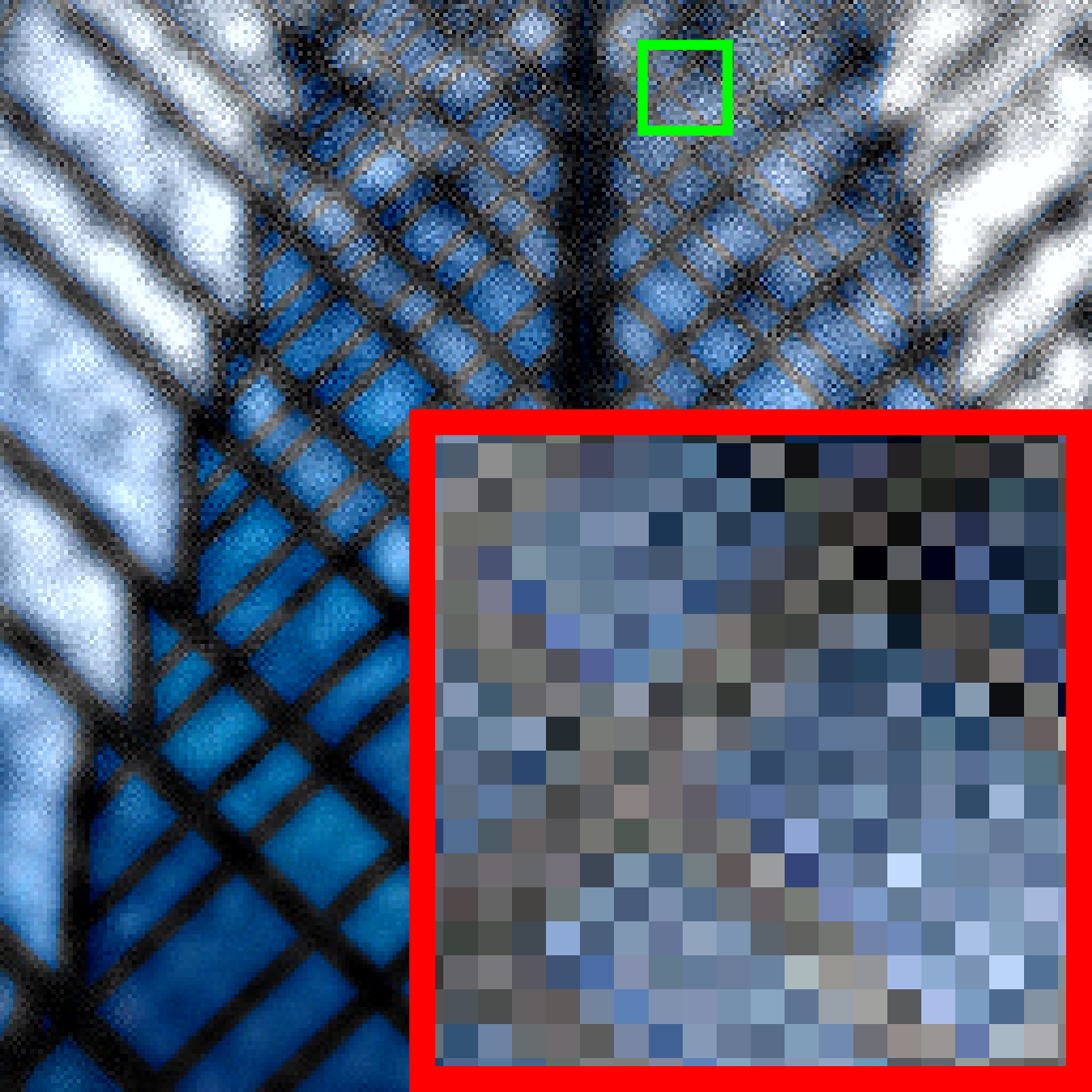}
    &\includegraphics[width=0.08\textwidth]{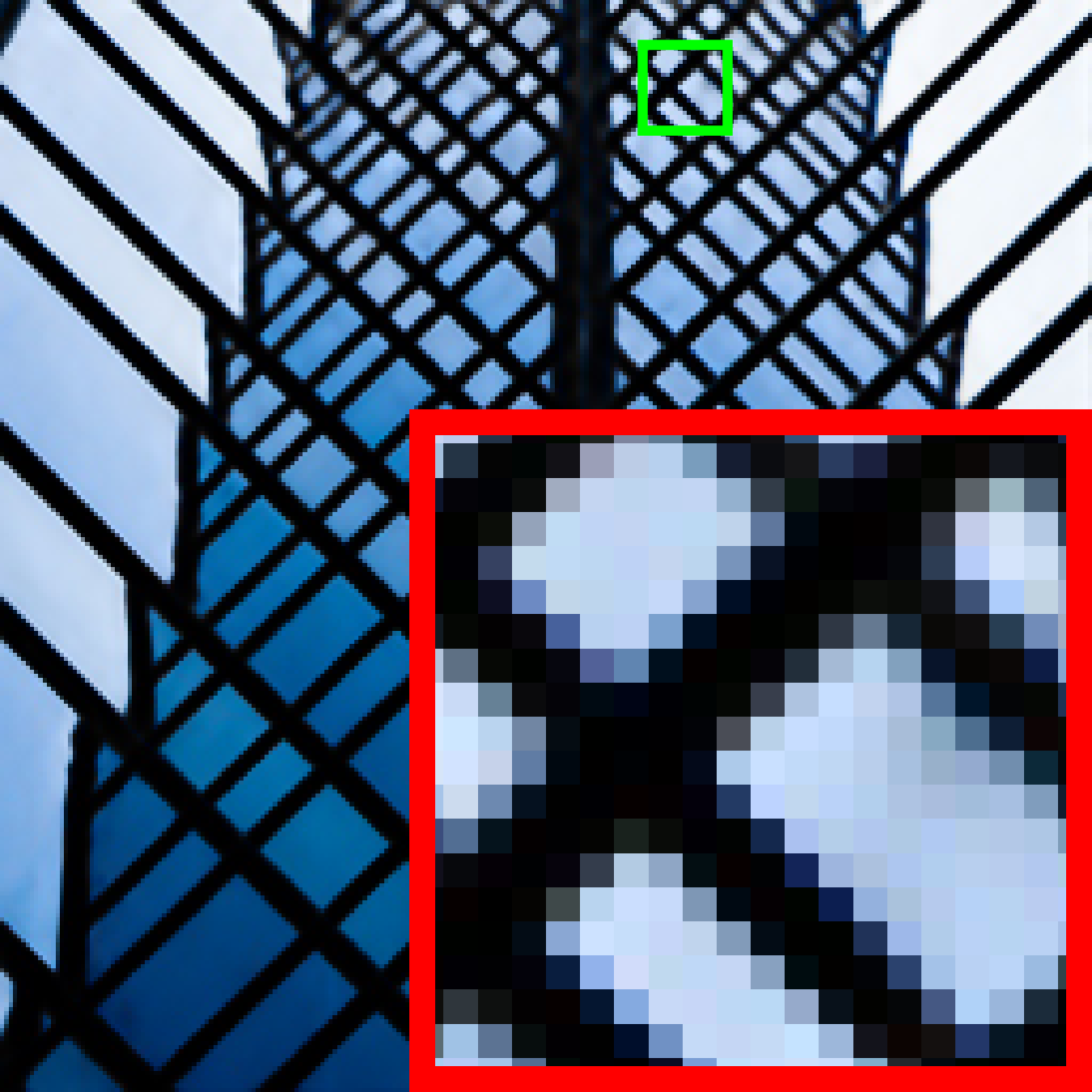}
    &\includegraphics[width=0.08\textwidth]{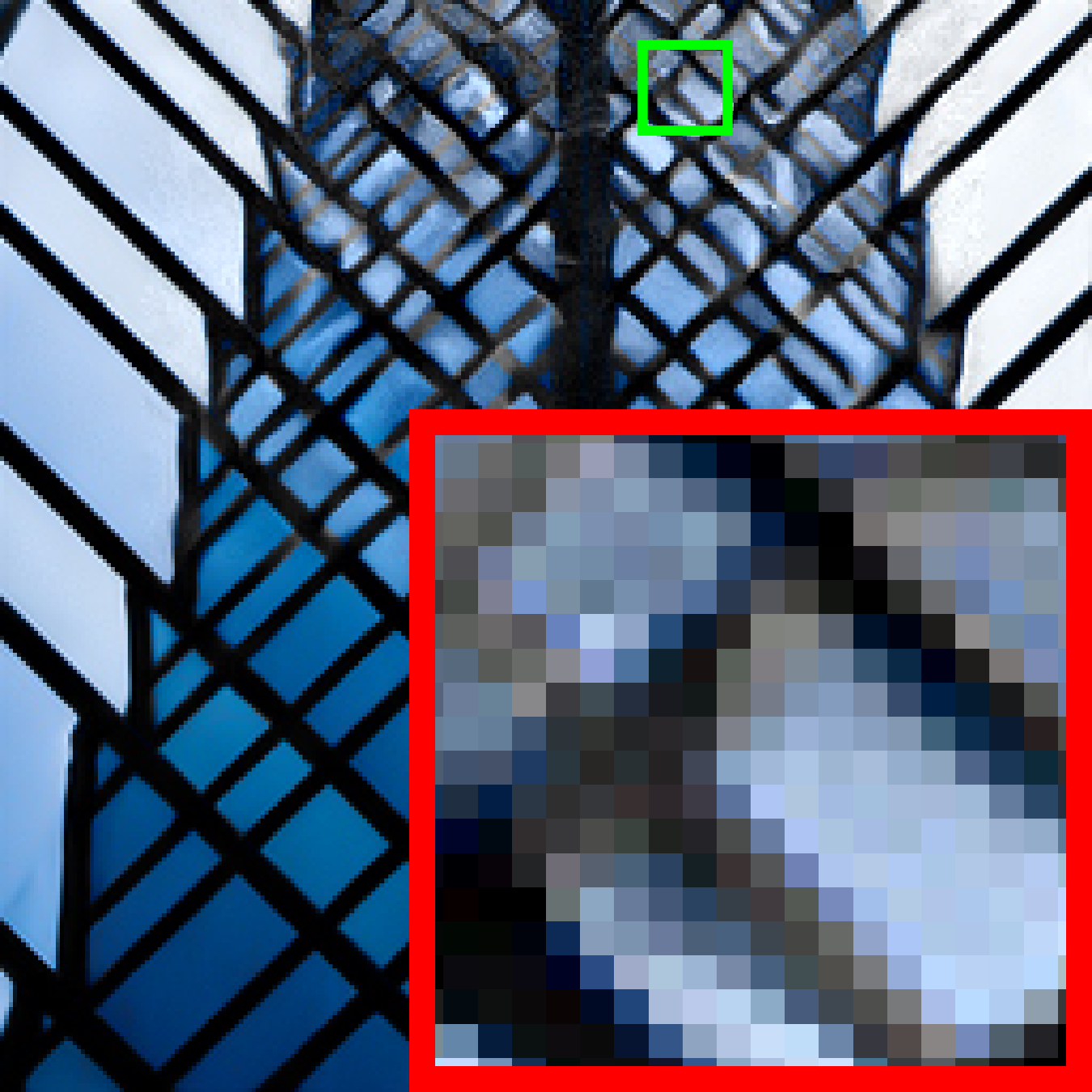}
    &\includegraphics[width=0.08\textwidth]{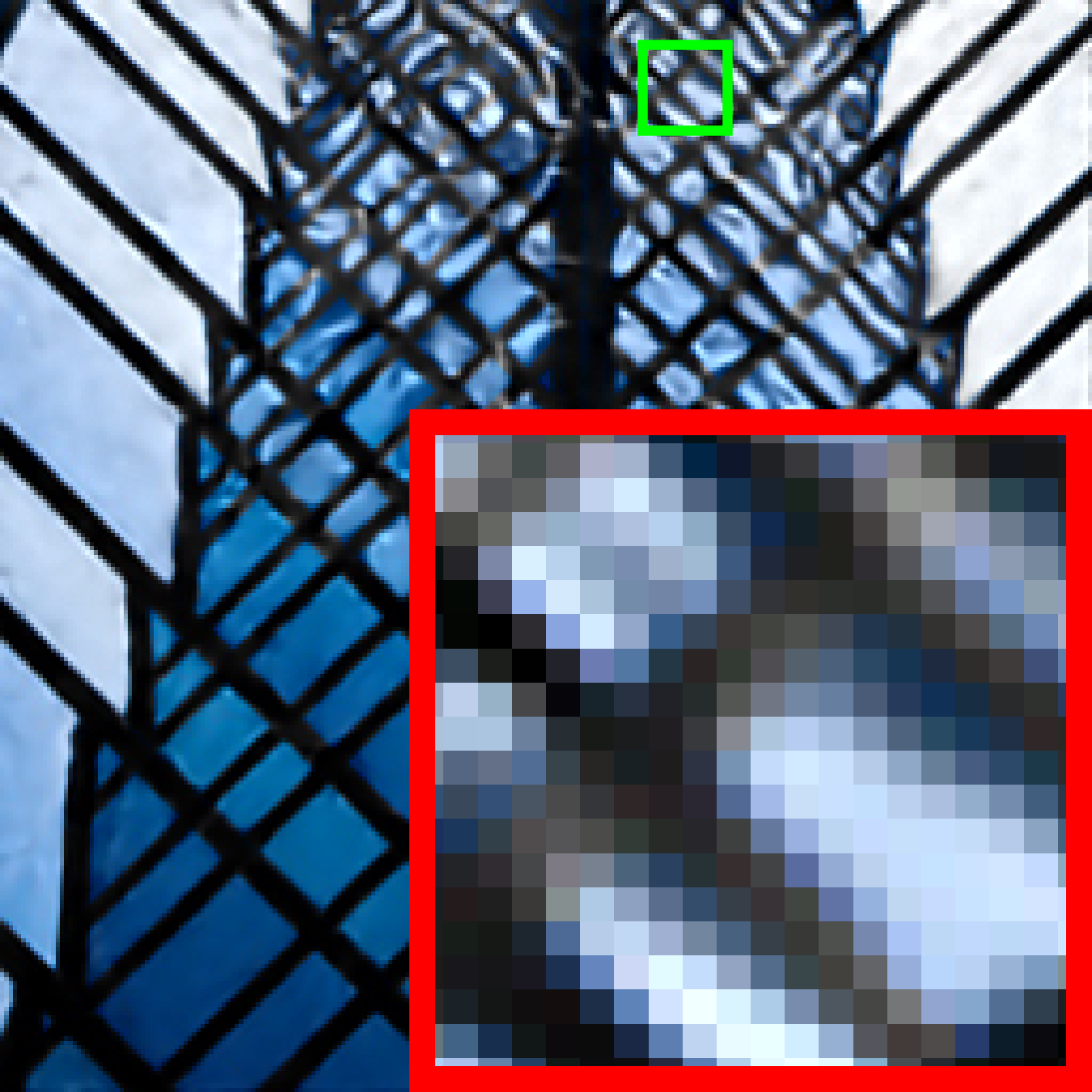}
    &\includegraphics[width=0.08\textwidth]{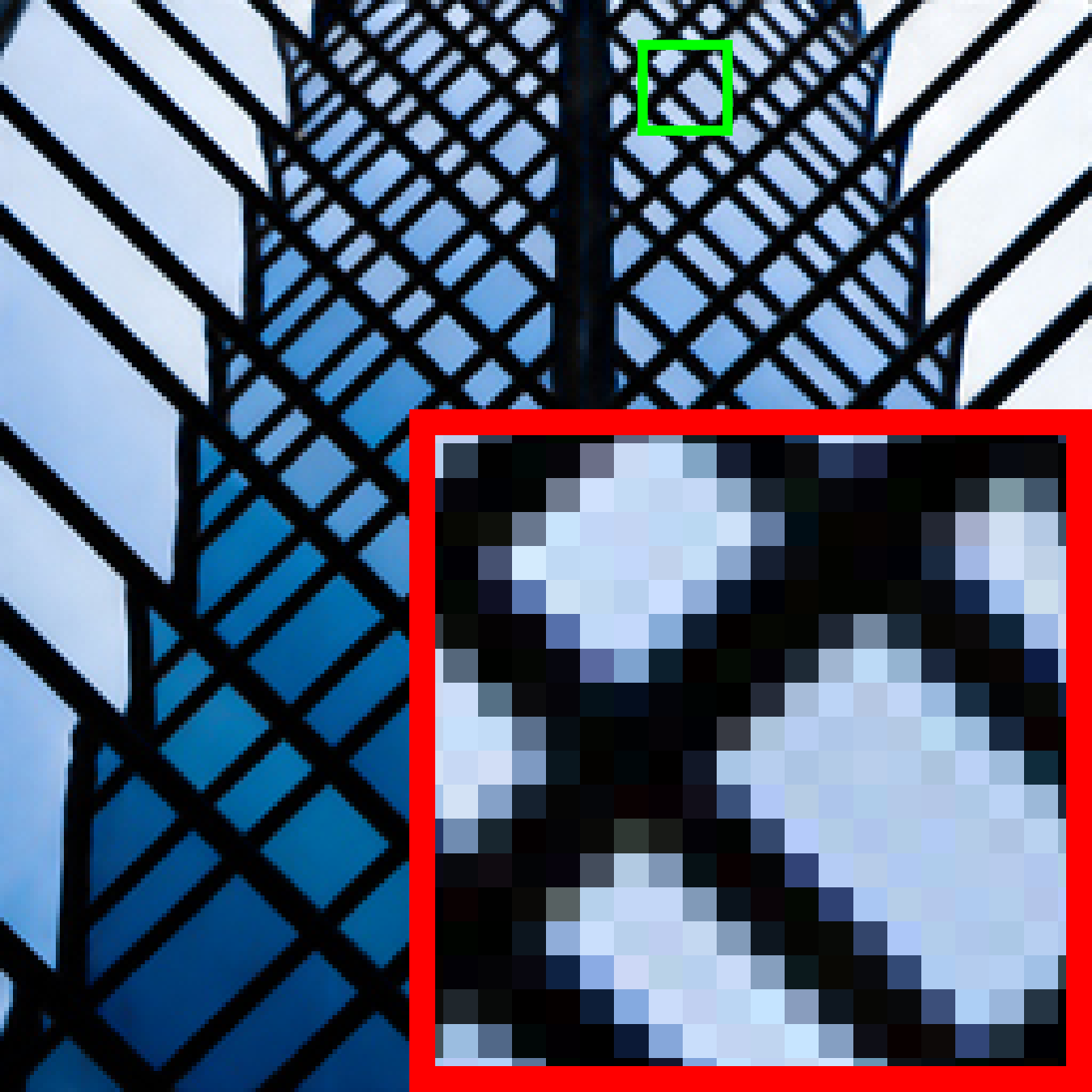}
    &\includegraphics[width=0.08\textwidth]{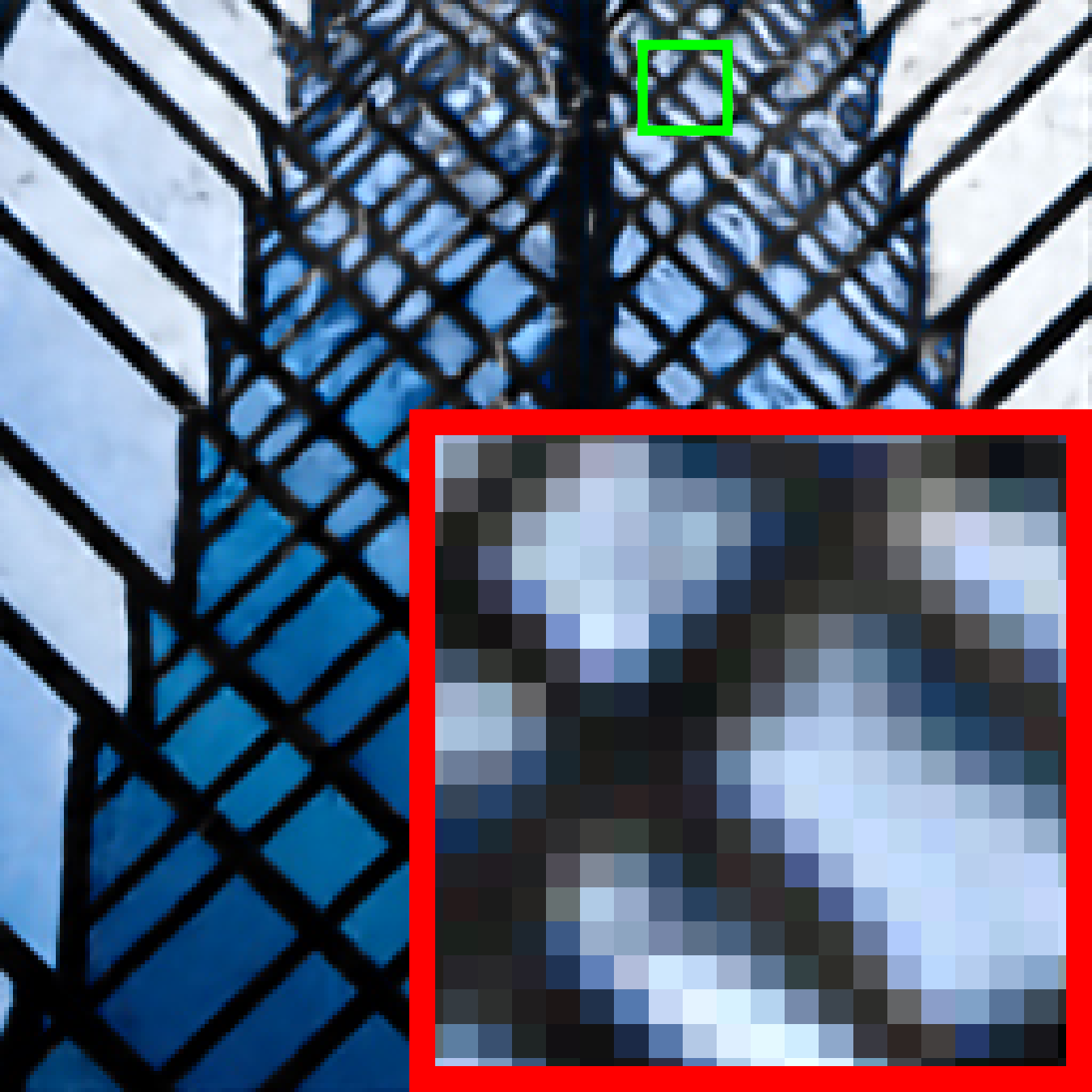}
    &\includegraphics[width=0.08\textwidth]{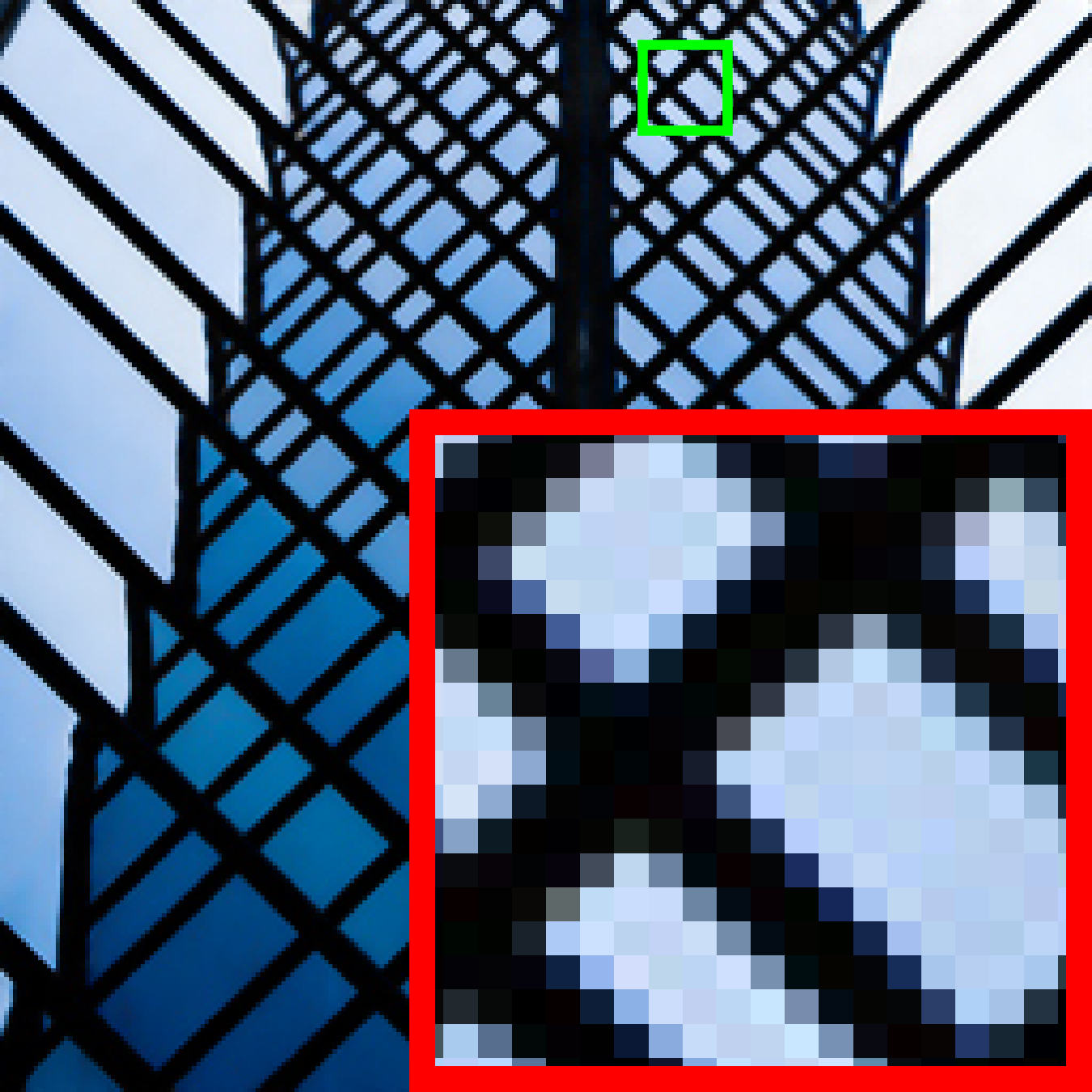}\\
    PSNR/SSIM & 16.45/0.63 & 17.66/0.75 & 18.18/0.80 & 20.85/0.90 & 15.63/0.67 & 22.33/\underline{\textcolor{blue}{0.93}} & 13.74/0.39 & 28.12/\textbf{\textcolor{red}{0.97}} & 19.12/0.86 & 18.89/0.82 & \underline{\textcolor{blue}{31.63}}/\textbf{\textcolor{red}{0.97}} & 19.87/0.86 & \textbf{\textcolor{red}{31.80}}/\textbf{\textcolor{red}{0.97}} \\
    \includegraphics[width=0.08\textwidth]{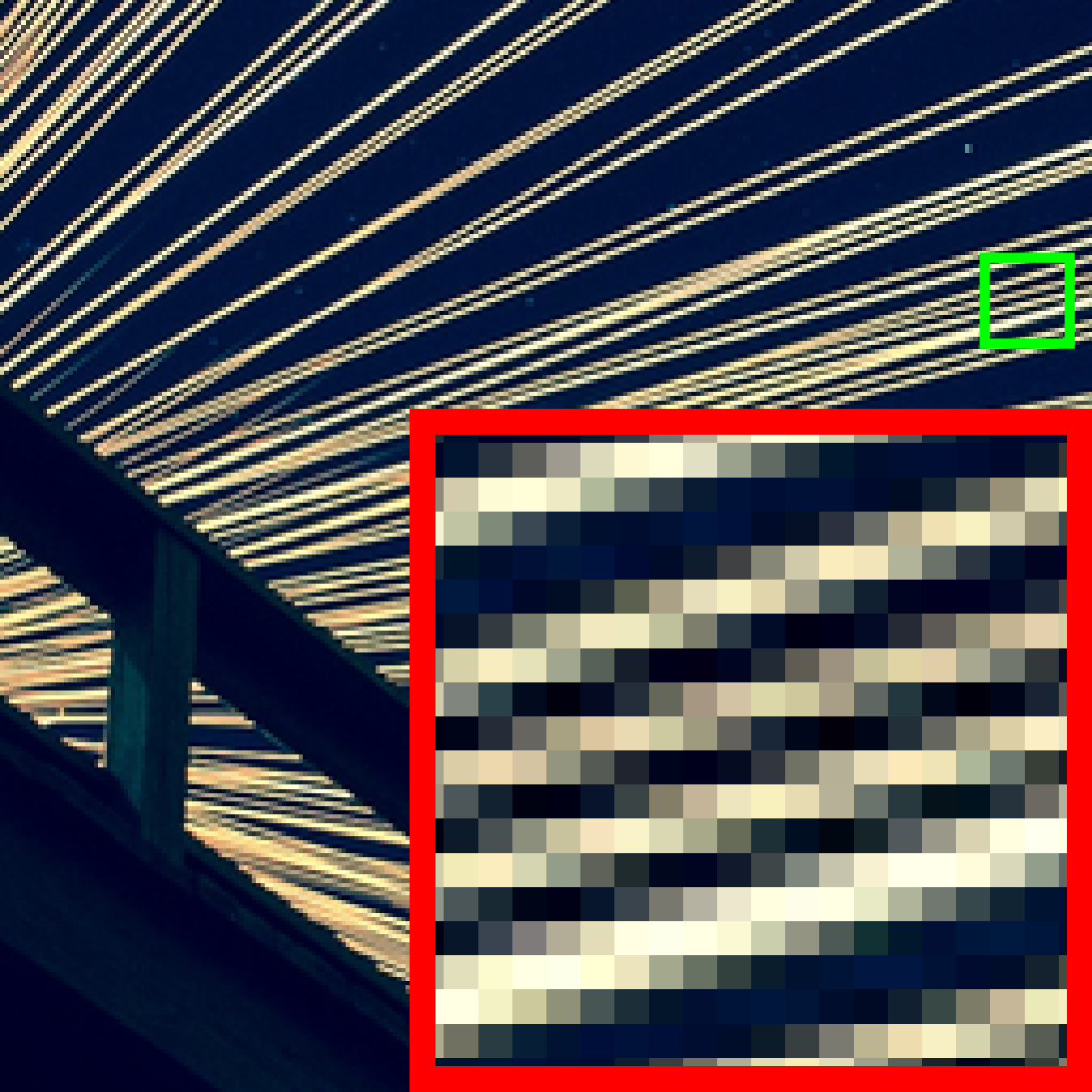}
    &\includegraphics[width=0.08\textwidth]{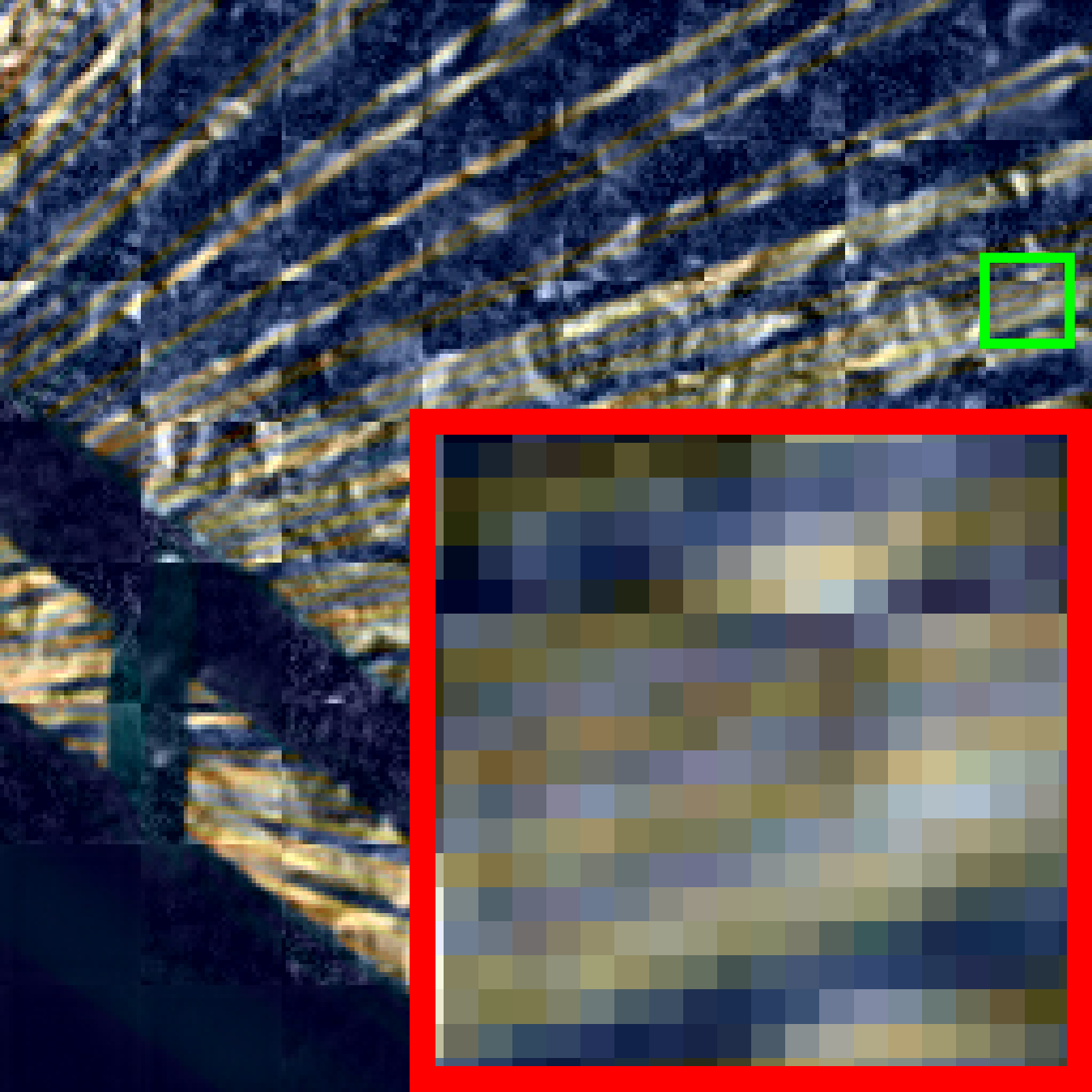}
    &\includegraphics[width=0.08\textwidth]{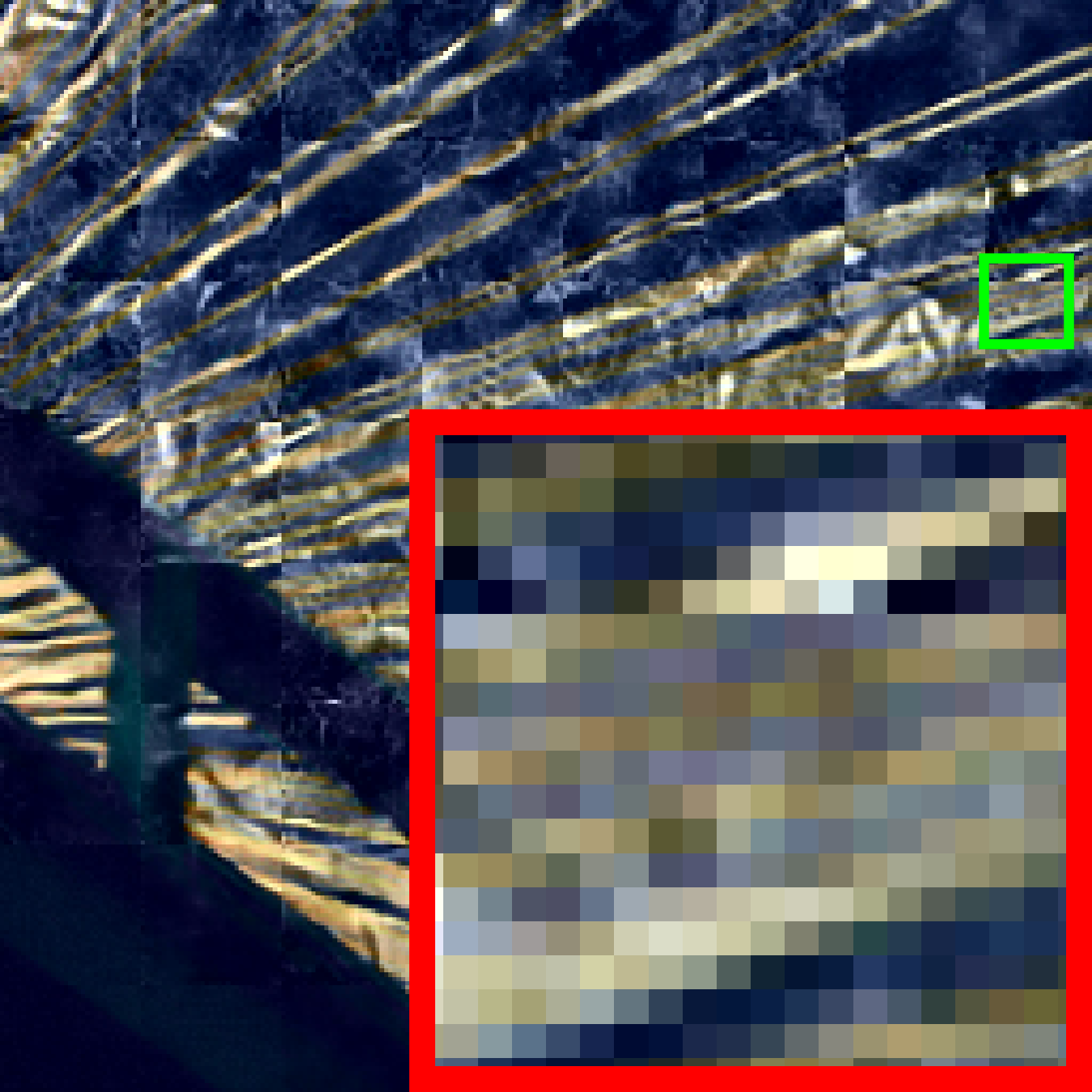}
    &\includegraphics[width=0.08\textwidth]{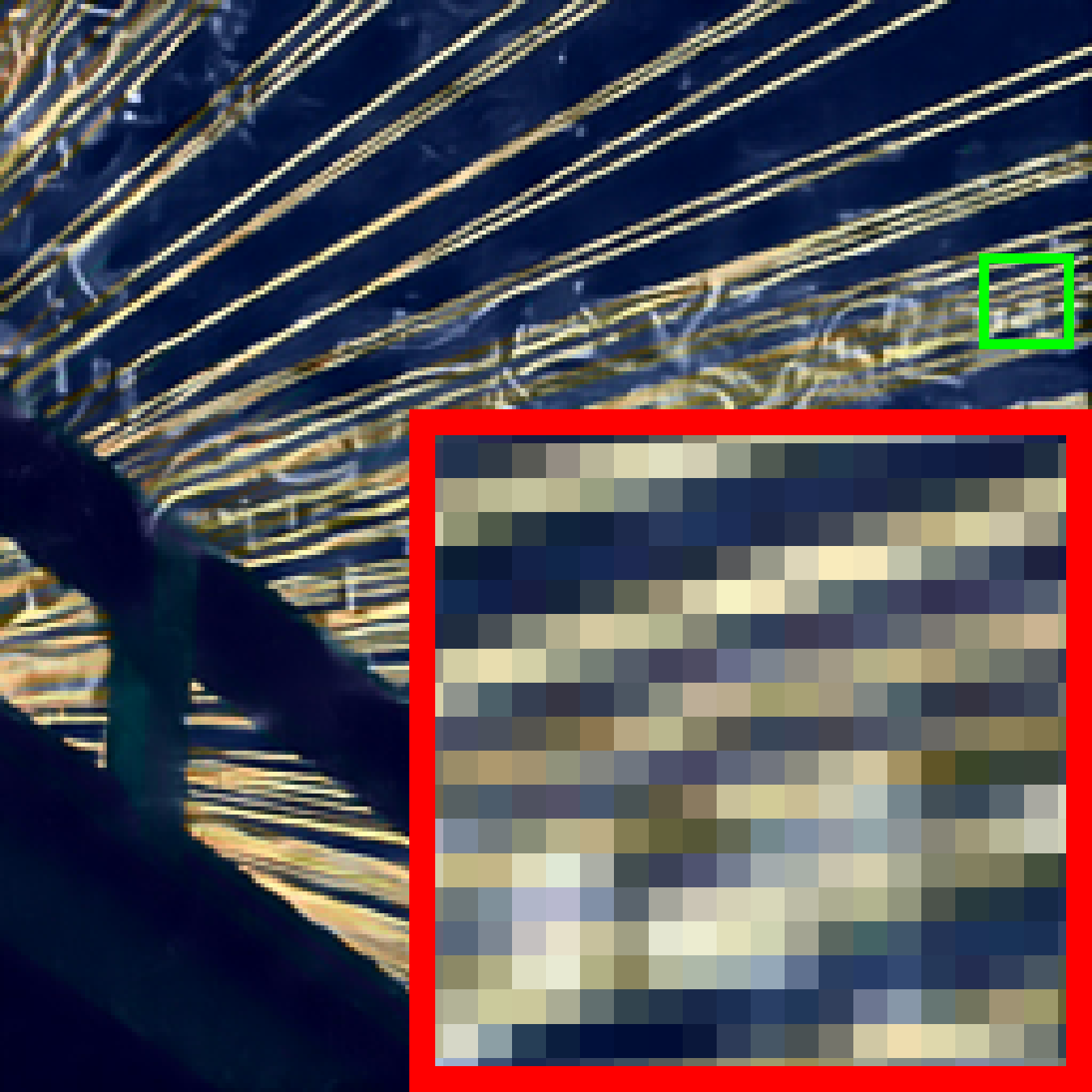}
    &\includegraphics[width=0.08\textwidth]{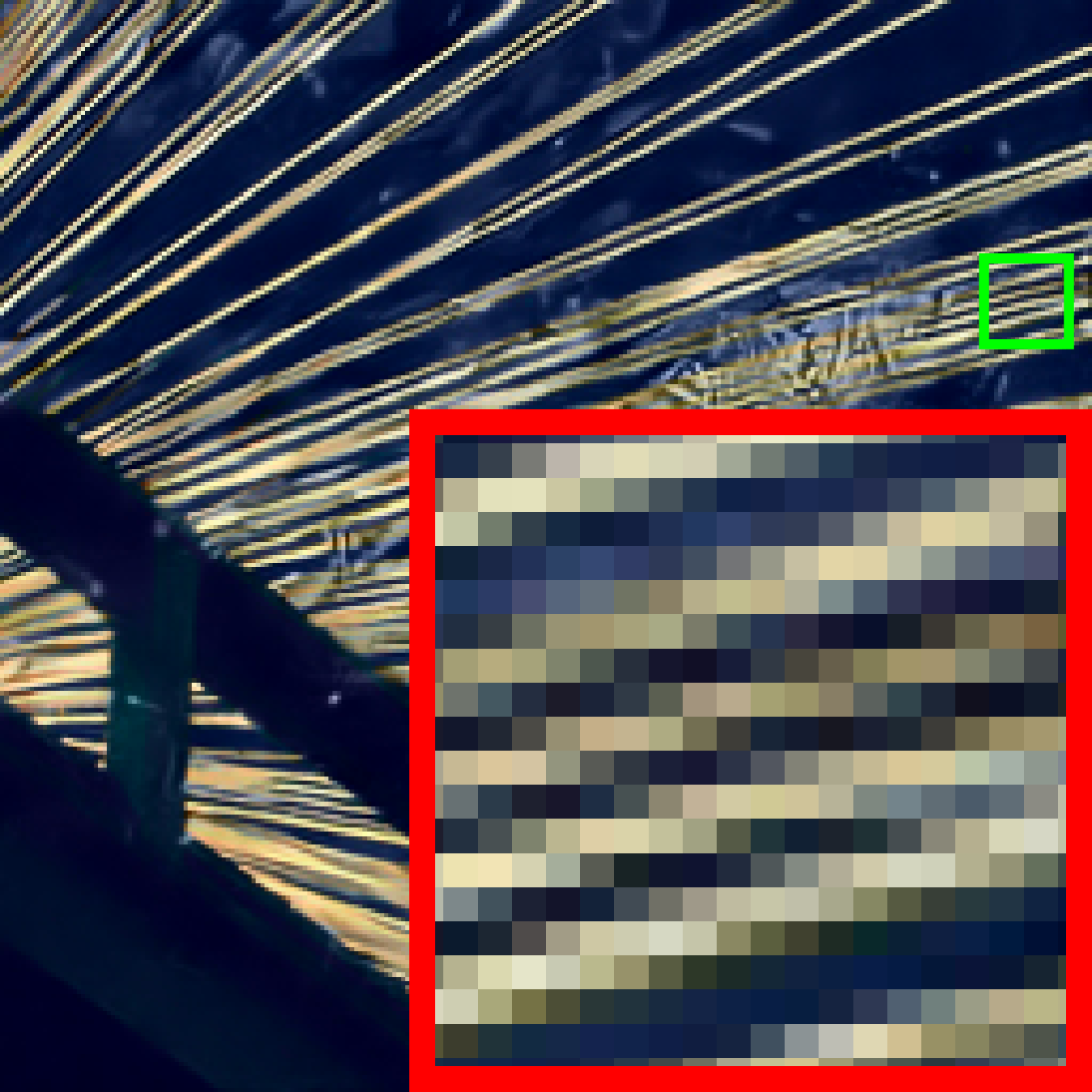}
    &\includegraphics[width=0.08\textwidth]{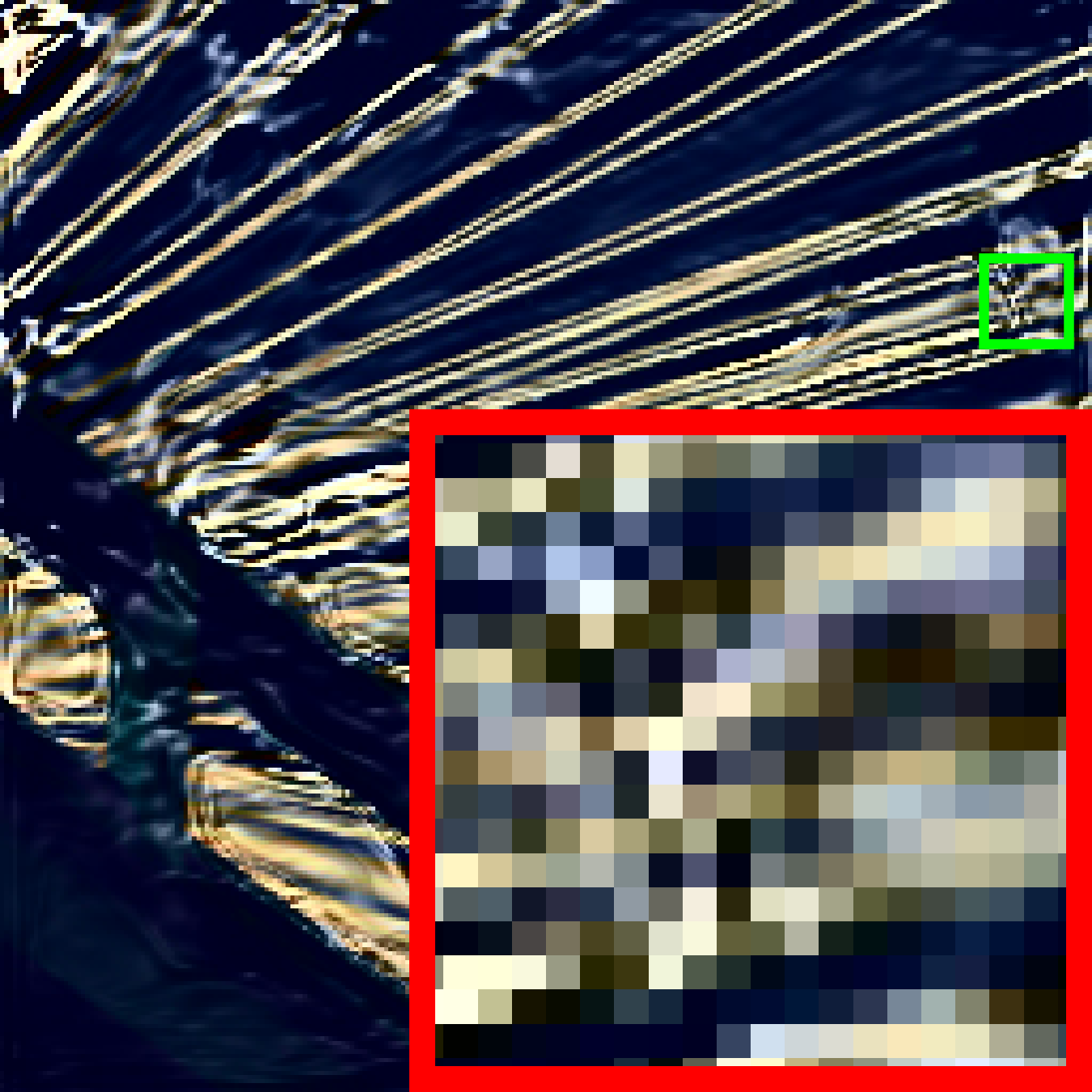}
    &\includegraphics[width=0.08\textwidth]{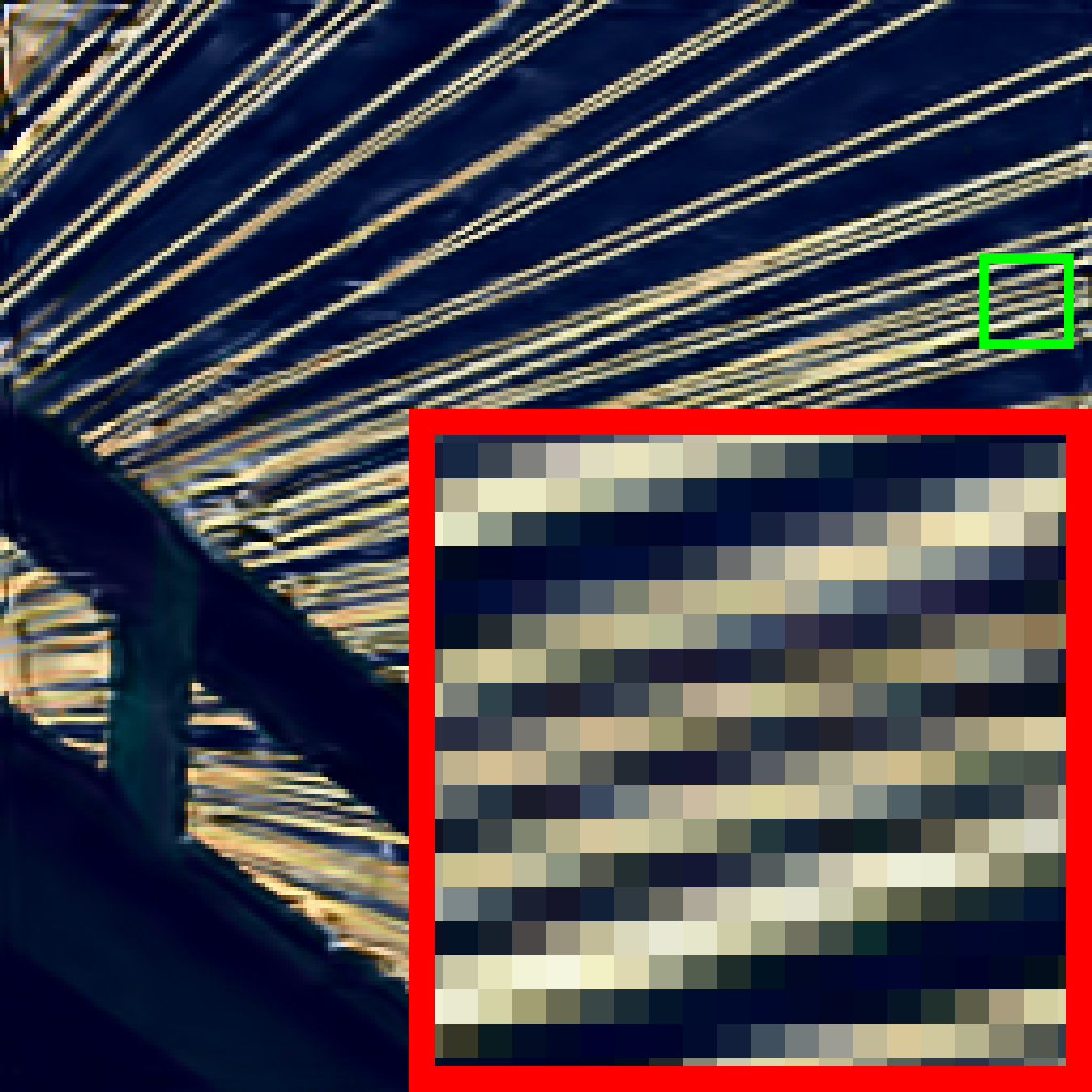}
    &\includegraphics[width=0.08\textwidth]{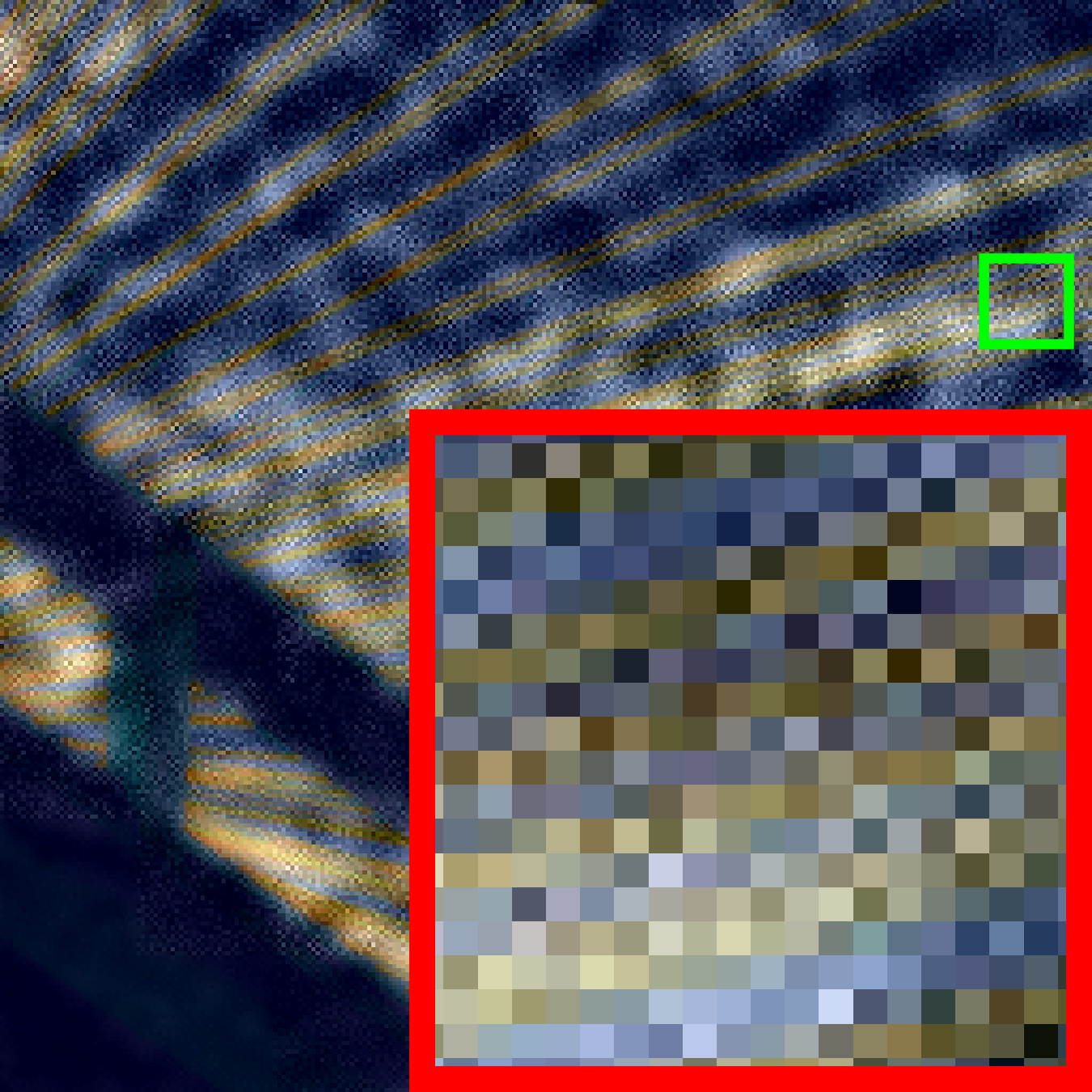}
    &\includegraphics[width=0.08\textwidth]{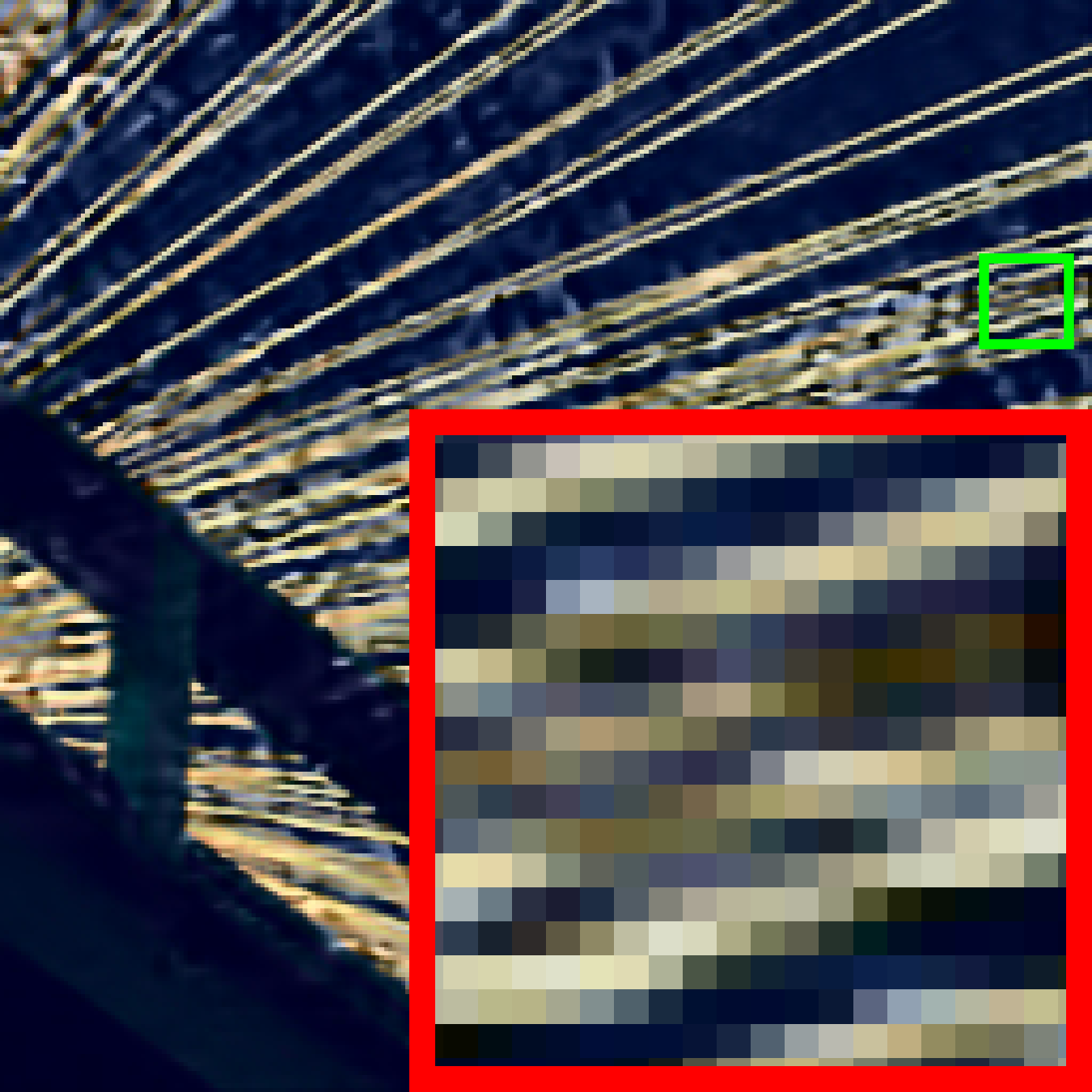}
    &\includegraphics[width=0.08\textwidth]{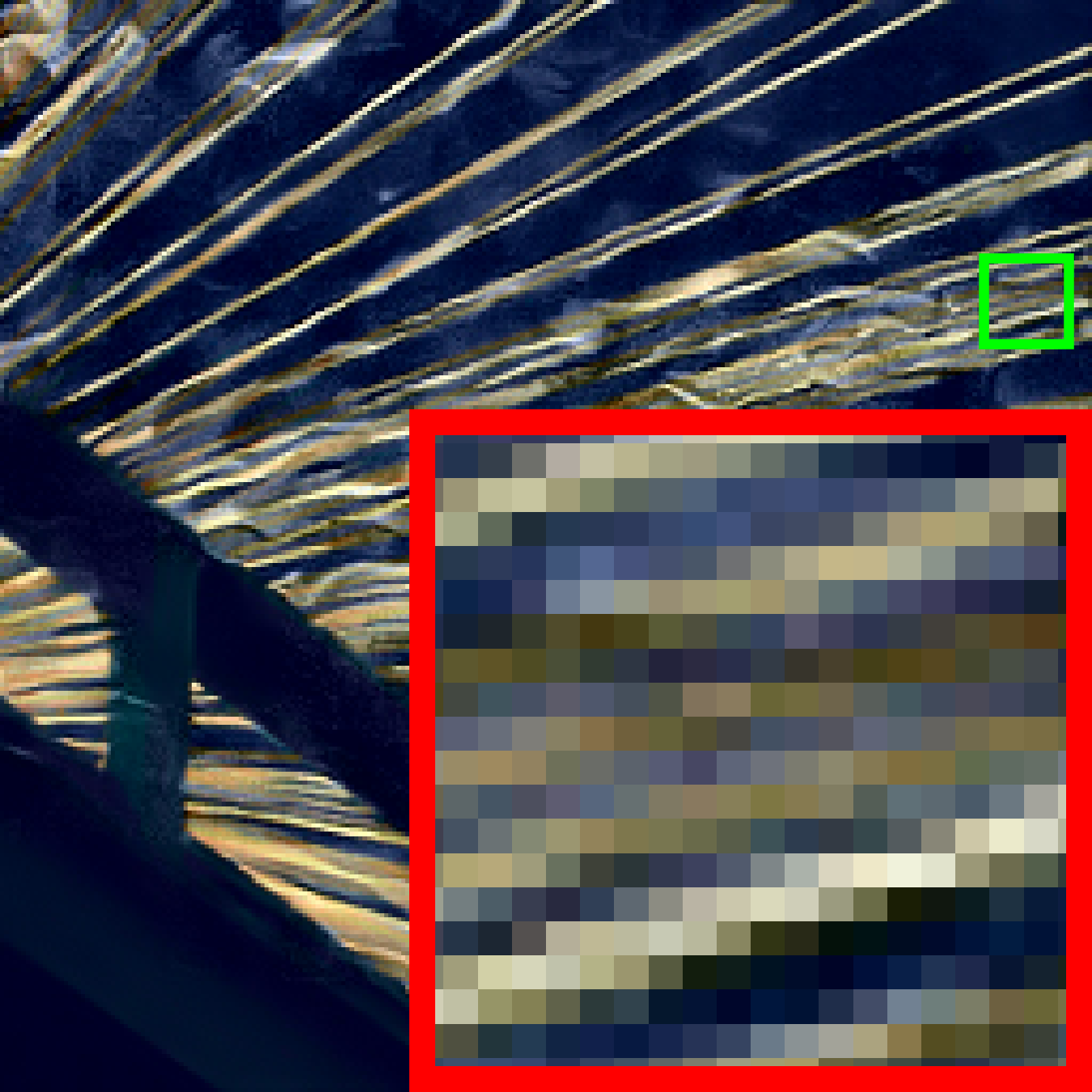}
    &\includegraphics[width=0.08\textwidth]{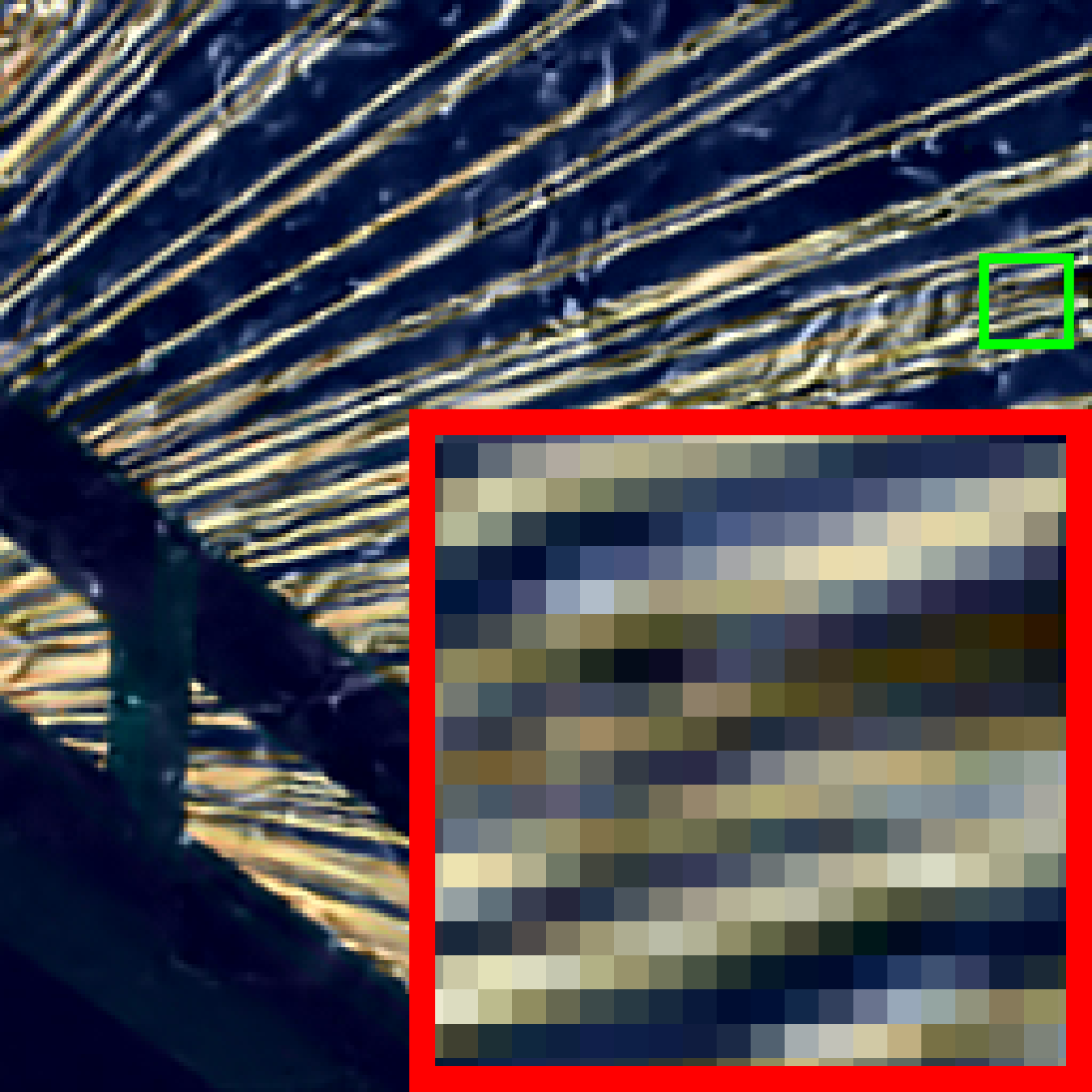}
    &\includegraphics[width=0.08\textwidth]{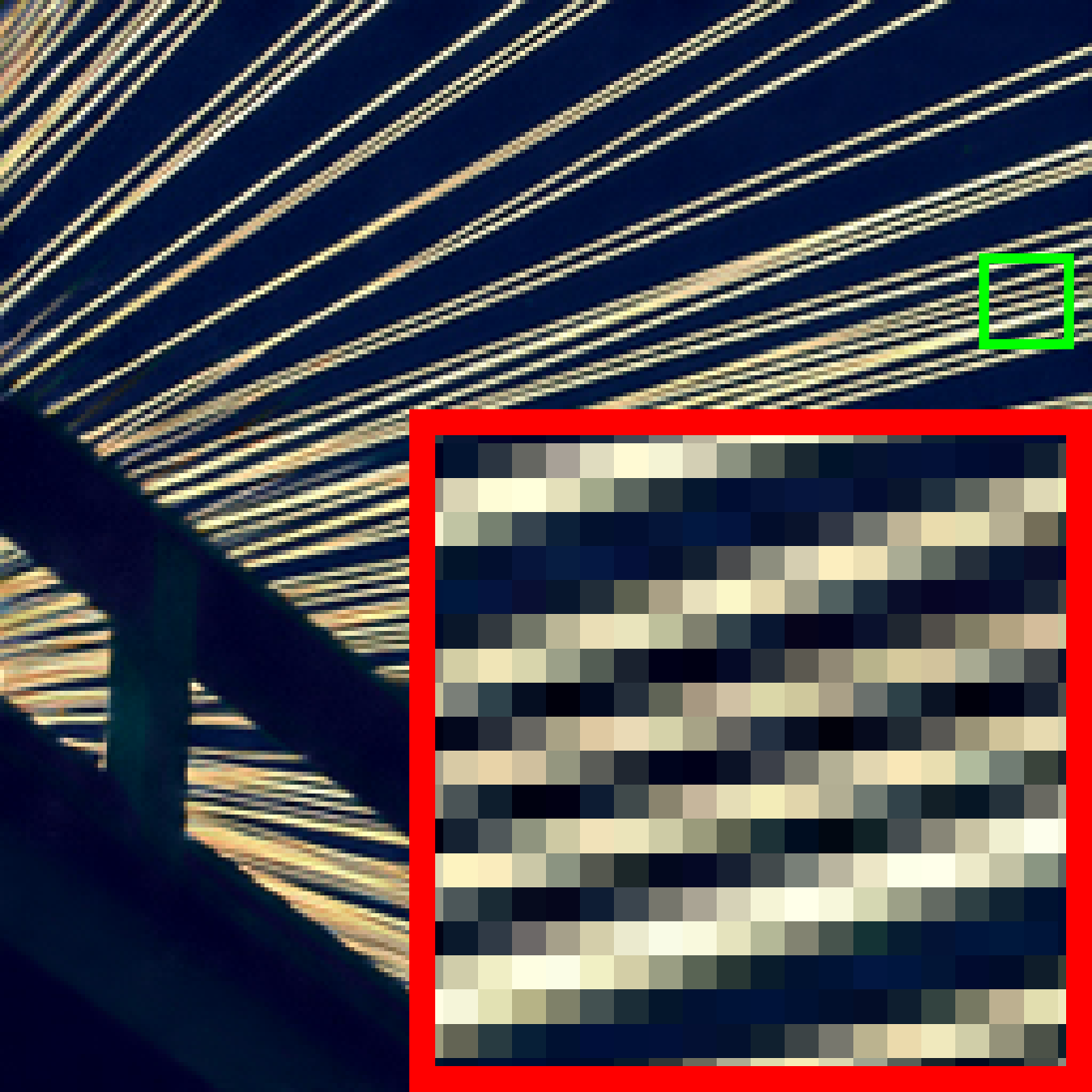}
    &\includegraphics[width=0.08\textwidth]{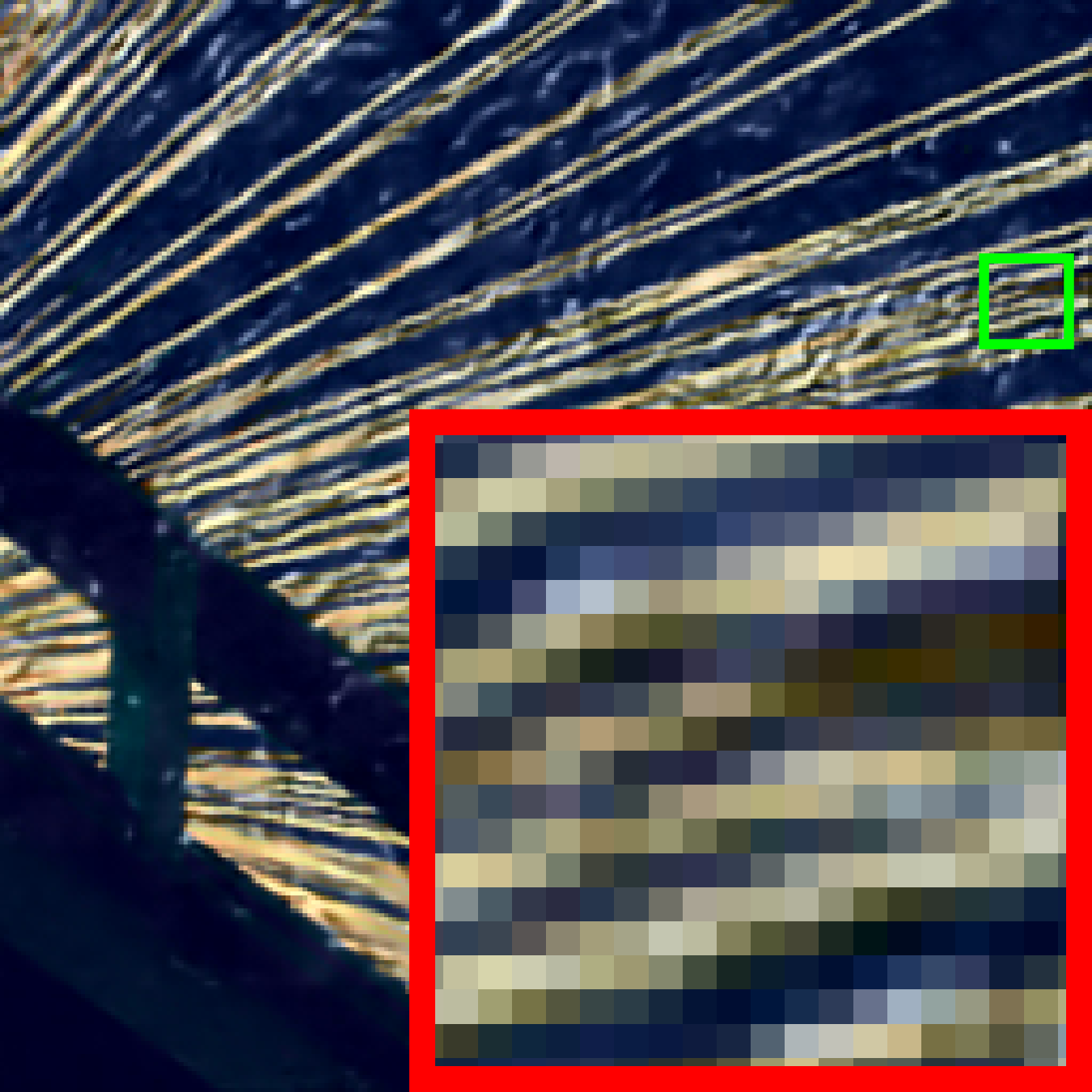}
    &\includegraphics[width=0.08\textwidth]{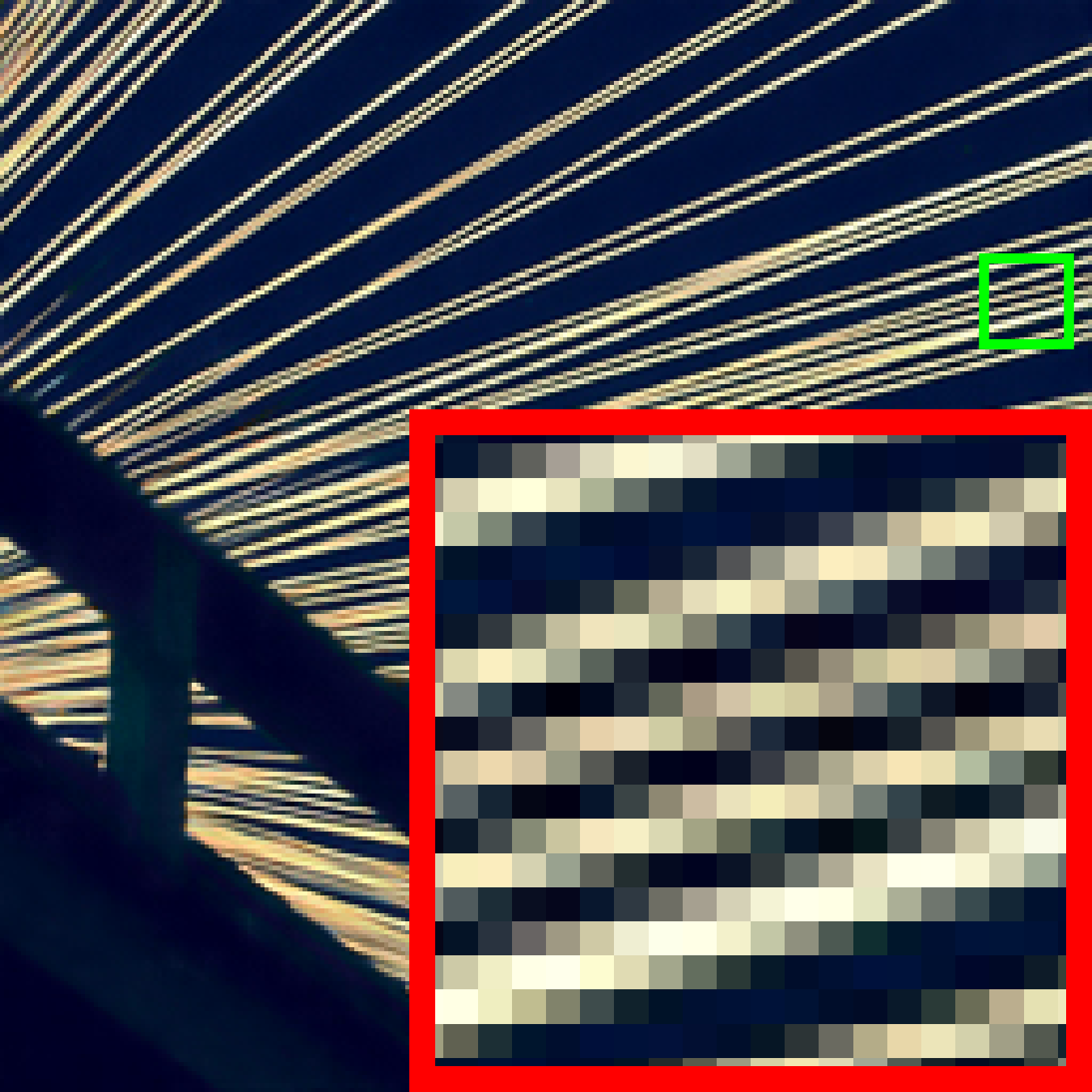}\\
    PSNR/SSIM & 12.28/0.34 & 13.32/0.49 & 16.90/0.79 & 17.76/0.81 & 14.62/0.65 & 19.30/\underline{\textcolor{blue}{0.85}} & 11.89/0.24 & 16.07/0.73 & 15.73/0.70 & 15.67/0.70 & \underline{\textcolor{blue}{26.86}}/\textbf{\textcolor{red}{0.96}} & 15.93/0.71 & \textbf{\textcolor{red}{27.59}}/\textbf{\textcolor{red}{0.96}}
\end{tabular}}
\caption{Visual comparison of different methods on four natural benchmark images named ``Barbara'', ``test\_66'', ``img\_067'' and ``0828'' from Set11~\cite{kulkarni2016reconnet} \textcolor{blue}{(top)}, CBSD68~\cite{martin2001database} \textcolor{blue}{(upper middle)}, Urban100~\cite{huang2015single} \textcolor{blue}{(lower middle)}, and DIV2K~\cite{agustsson2017ntire} \textcolor{blue}{(bottom)}, respectively, with $\gamma =10\%$ and $\sigma =0$.}
\label{fig:comparison_standard_natural_images_r10}
\end{figure*}

\begin{figure*}[!t]
\setlength{\tabcolsep}{0.5pt}
\hspace{-4pt}
\resizebox{1.0\textwidth}{!}{
\tiny
\begin{tabular}{cccccccccccccc}
    GT & ReconNet & ISTA-Net$^\text{+}$ & ISTA-Net$^\text{++}$ & COAST & DIP & BCNN & EI & ASGLD & DDSSL & \textbf{SC-CNN} & \textbf{SC-CNN$^\text{+}$} & \textbf{SCT} & \textbf{SCT$^\text{+}$}\\
    \includegraphics[width=0.08\textwidth]{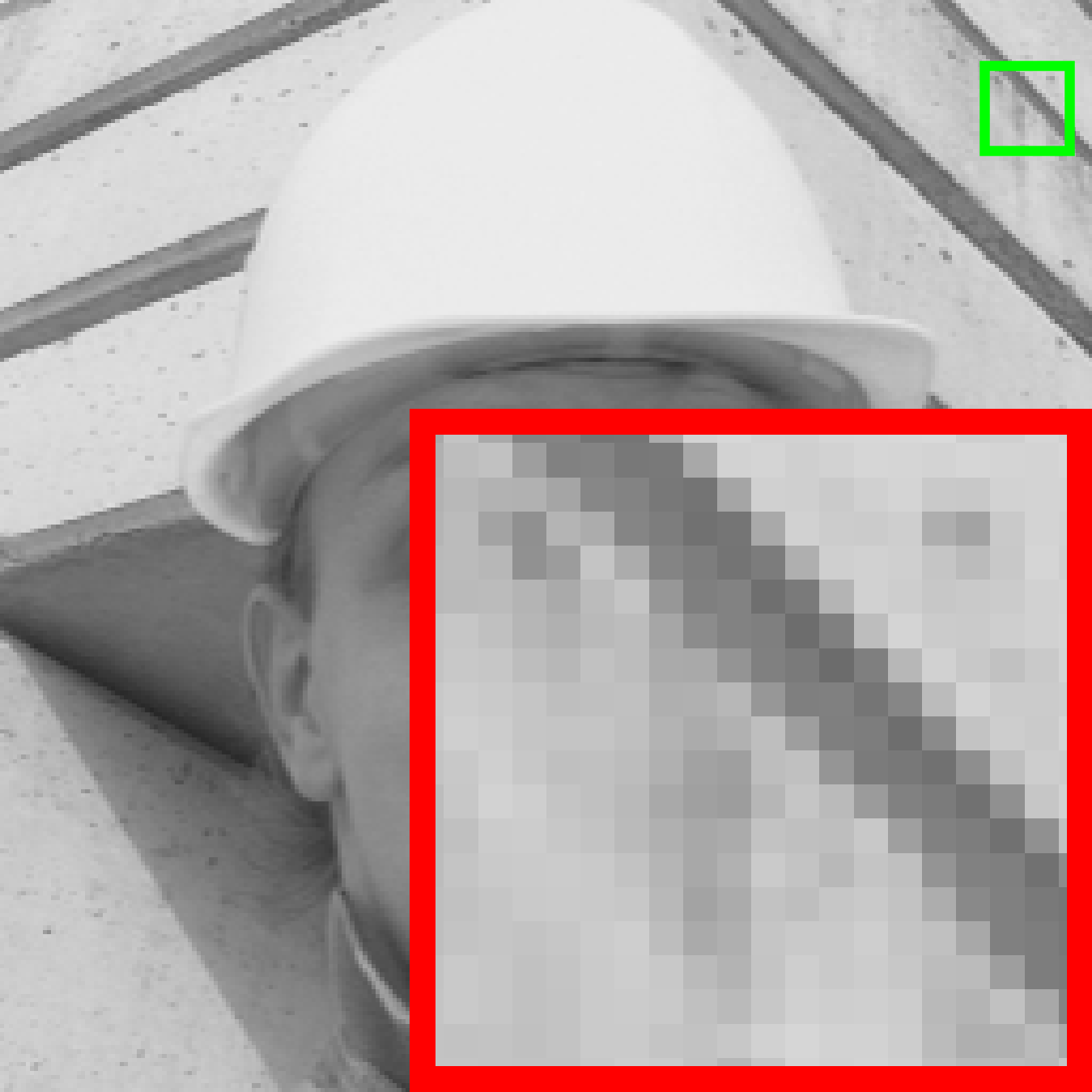}
    &\includegraphics[width=0.08\textwidth]{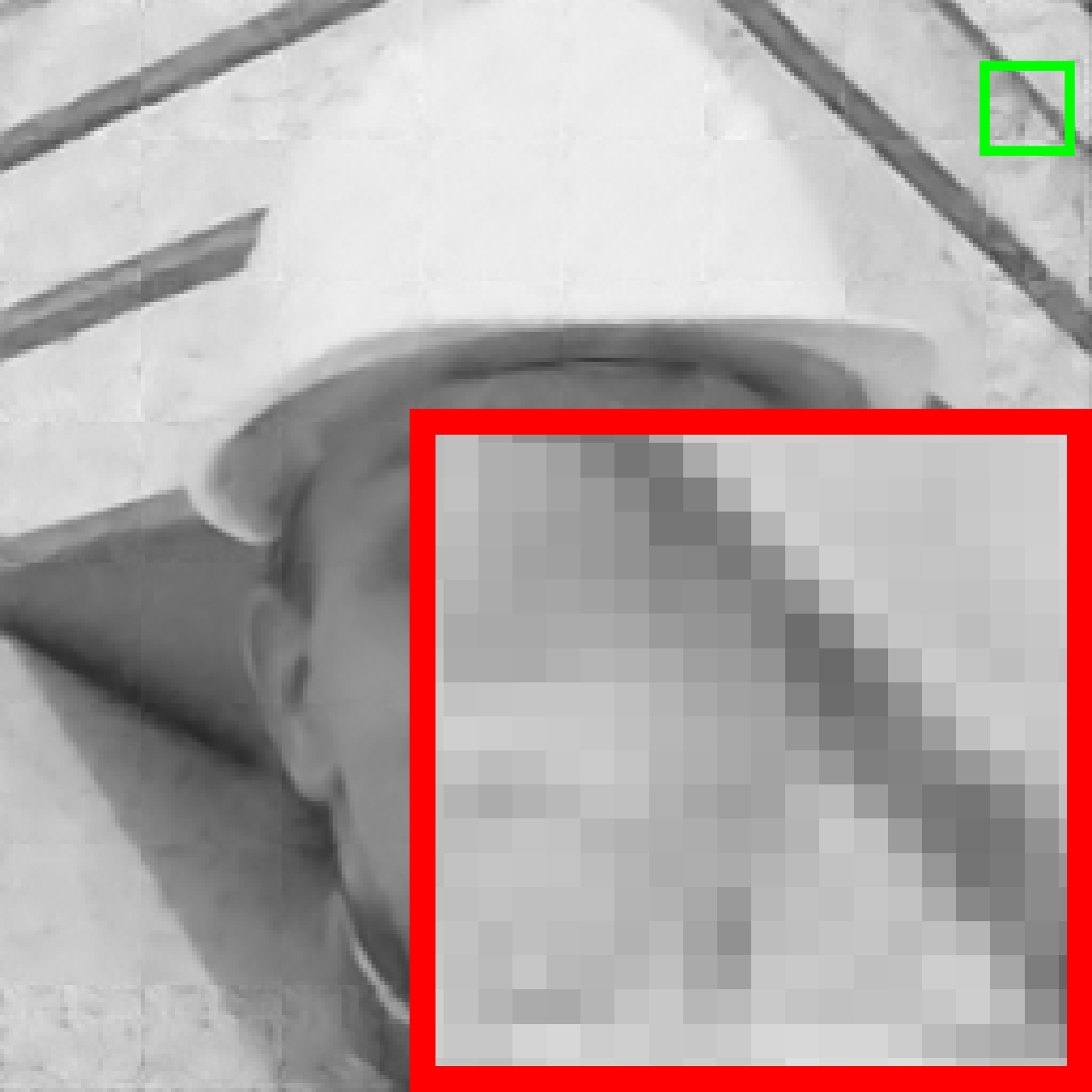}
    &\includegraphics[width=0.08\textwidth]{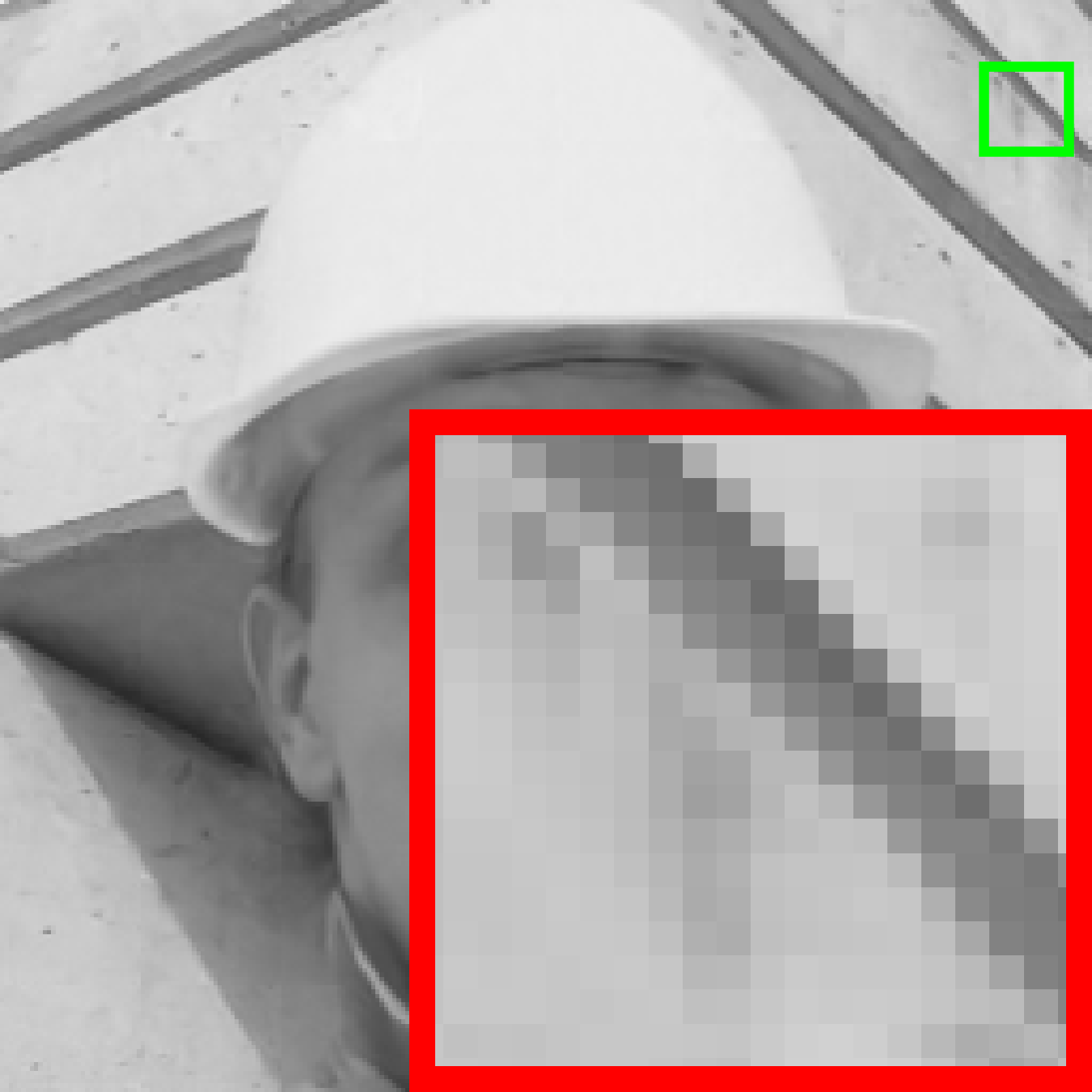}
    &\includegraphics[width=0.08\textwidth]{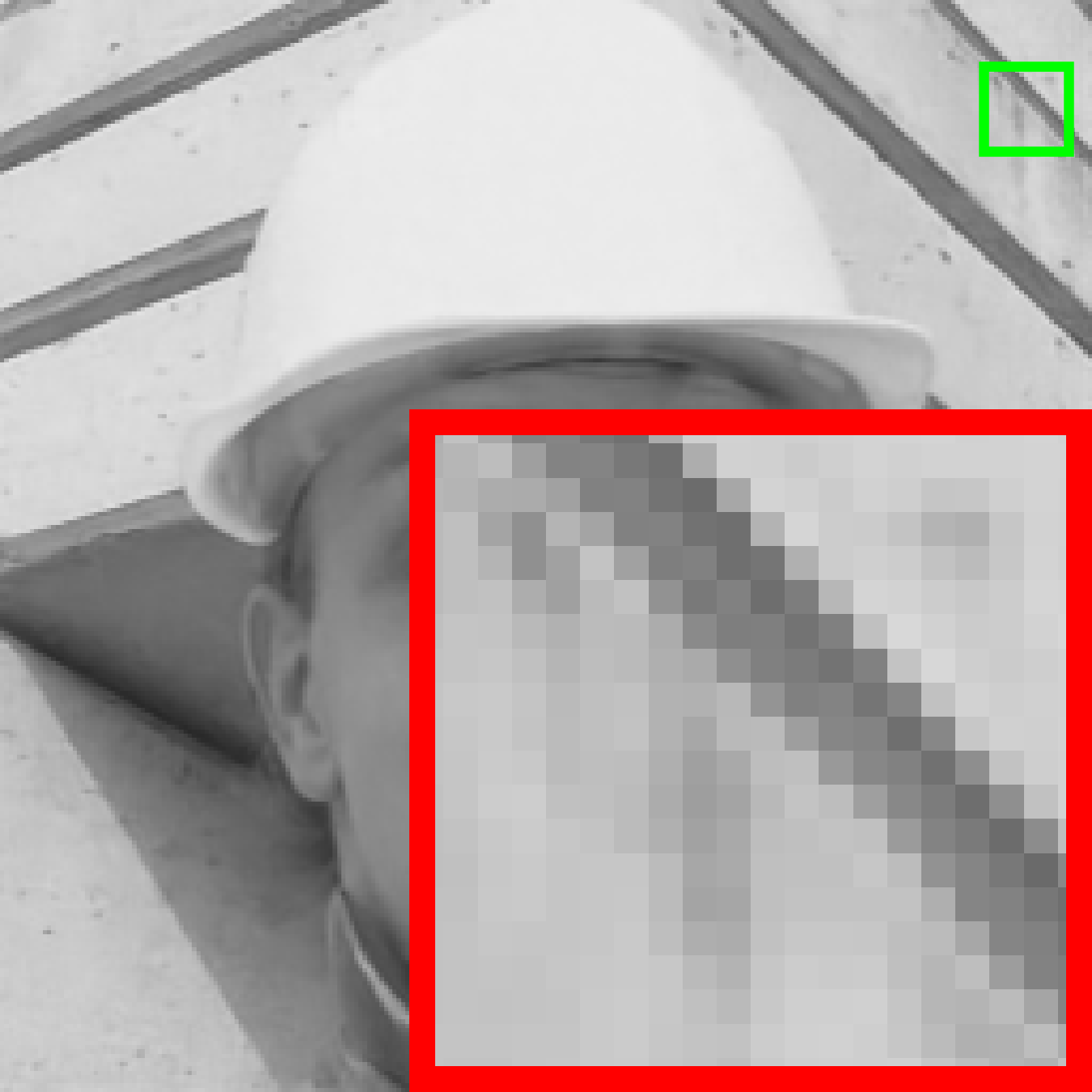}
    &\includegraphics[width=0.08\textwidth]{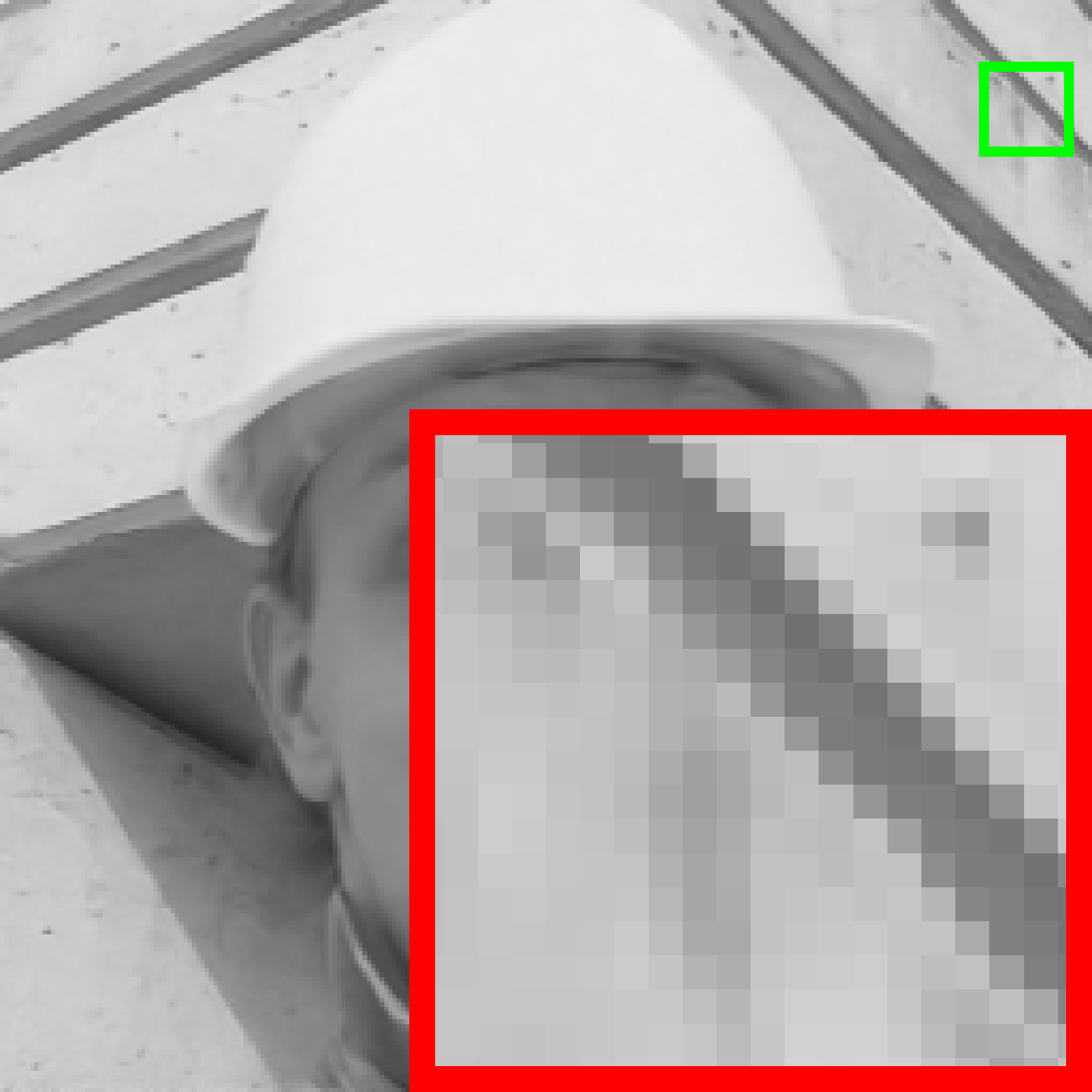}
    &\includegraphics[width=0.08\textwidth]{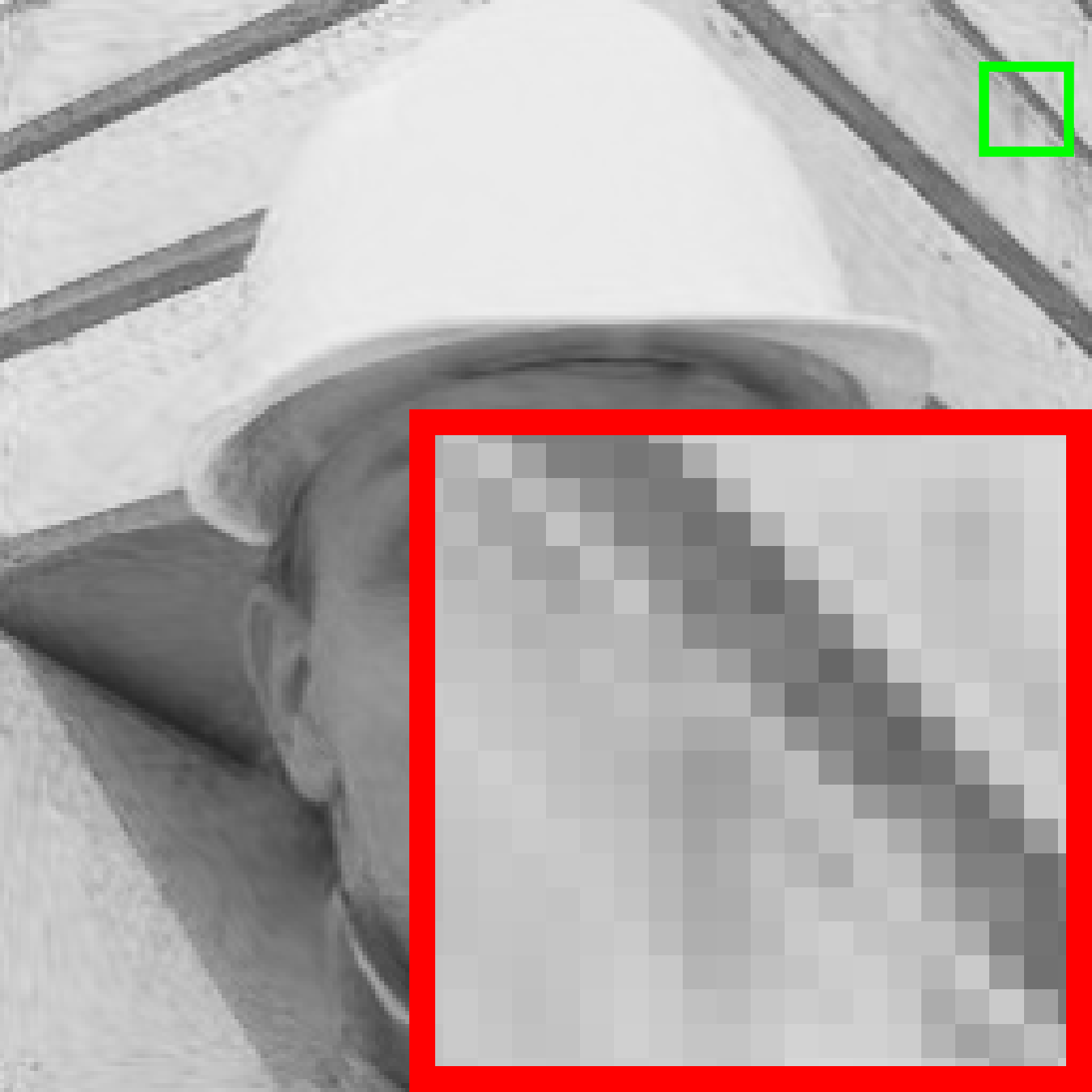}
    &\includegraphics[width=0.08\textwidth]{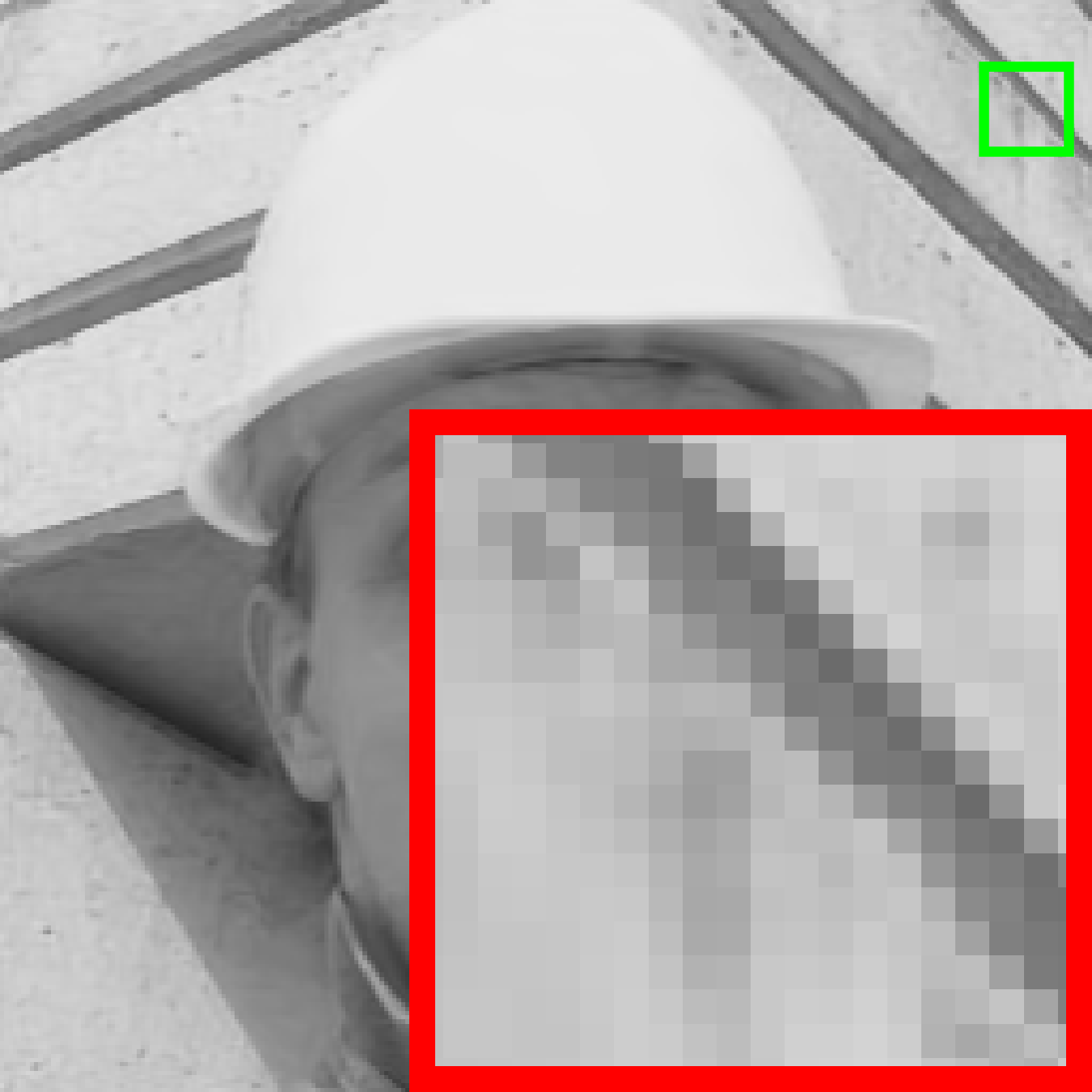}
    &\includegraphics[width=0.08\textwidth]{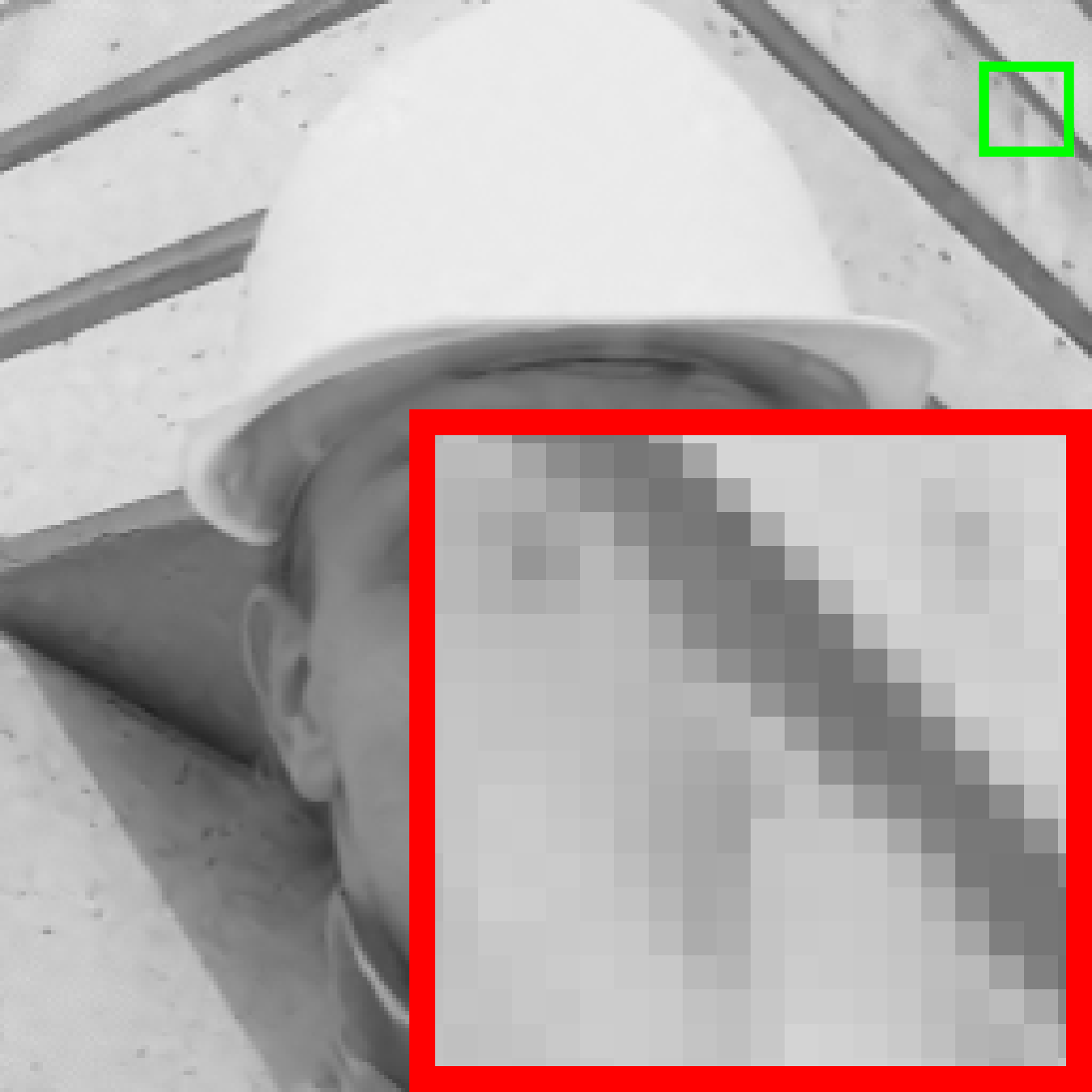}
    &\includegraphics[width=0.08\textwidth]{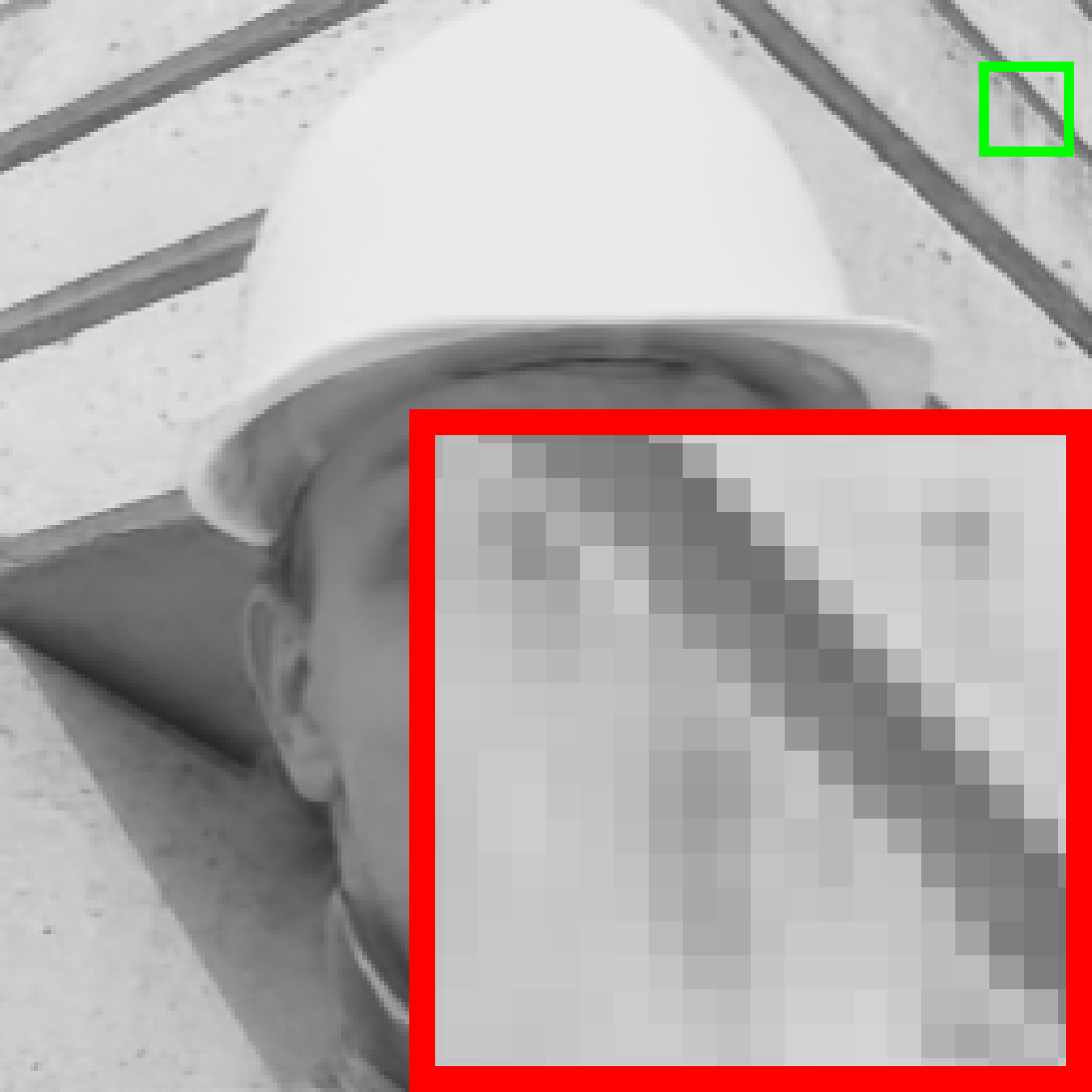}
    &\includegraphics[width=0.08\textwidth]{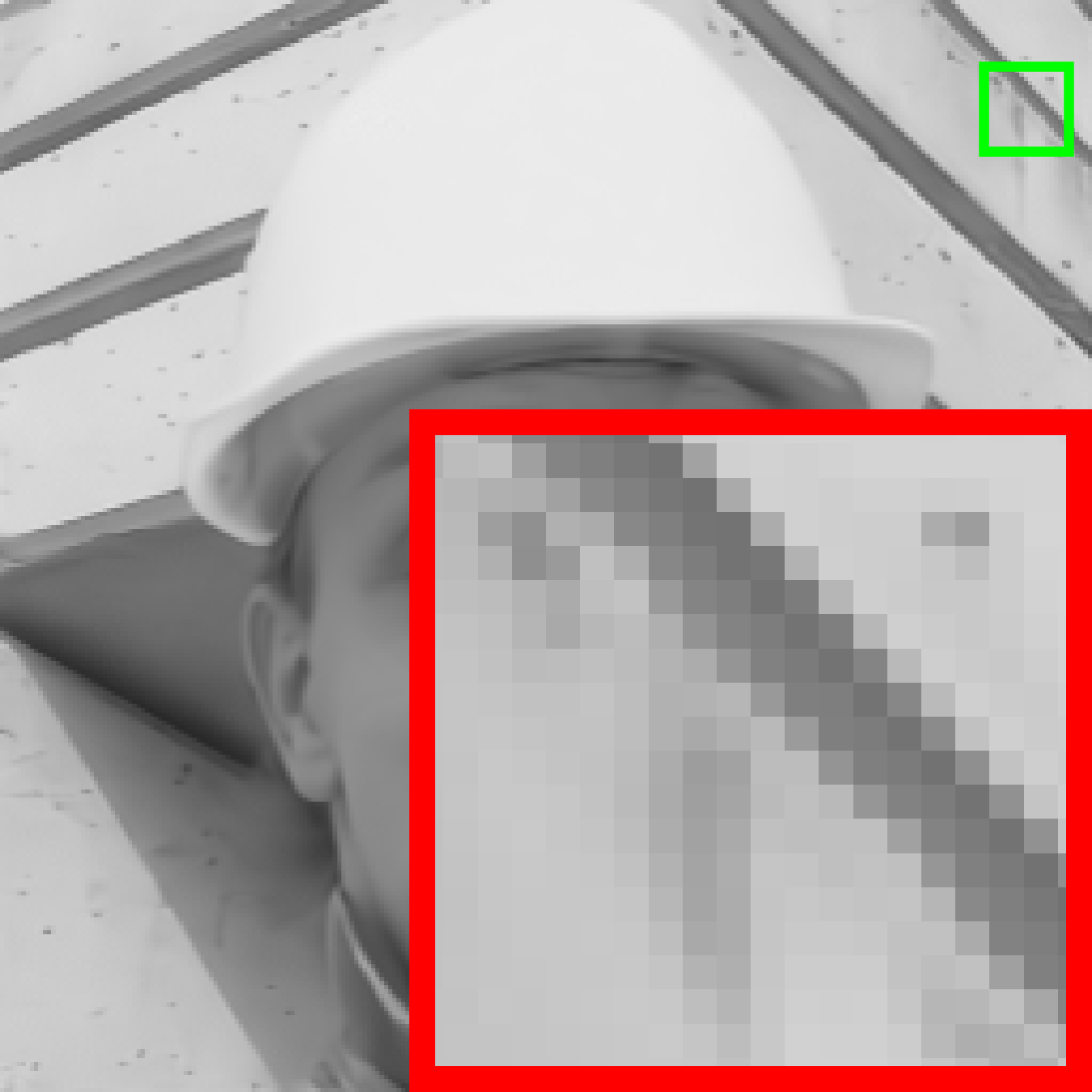}
    &\includegraphics[width=0.08\textwidth]{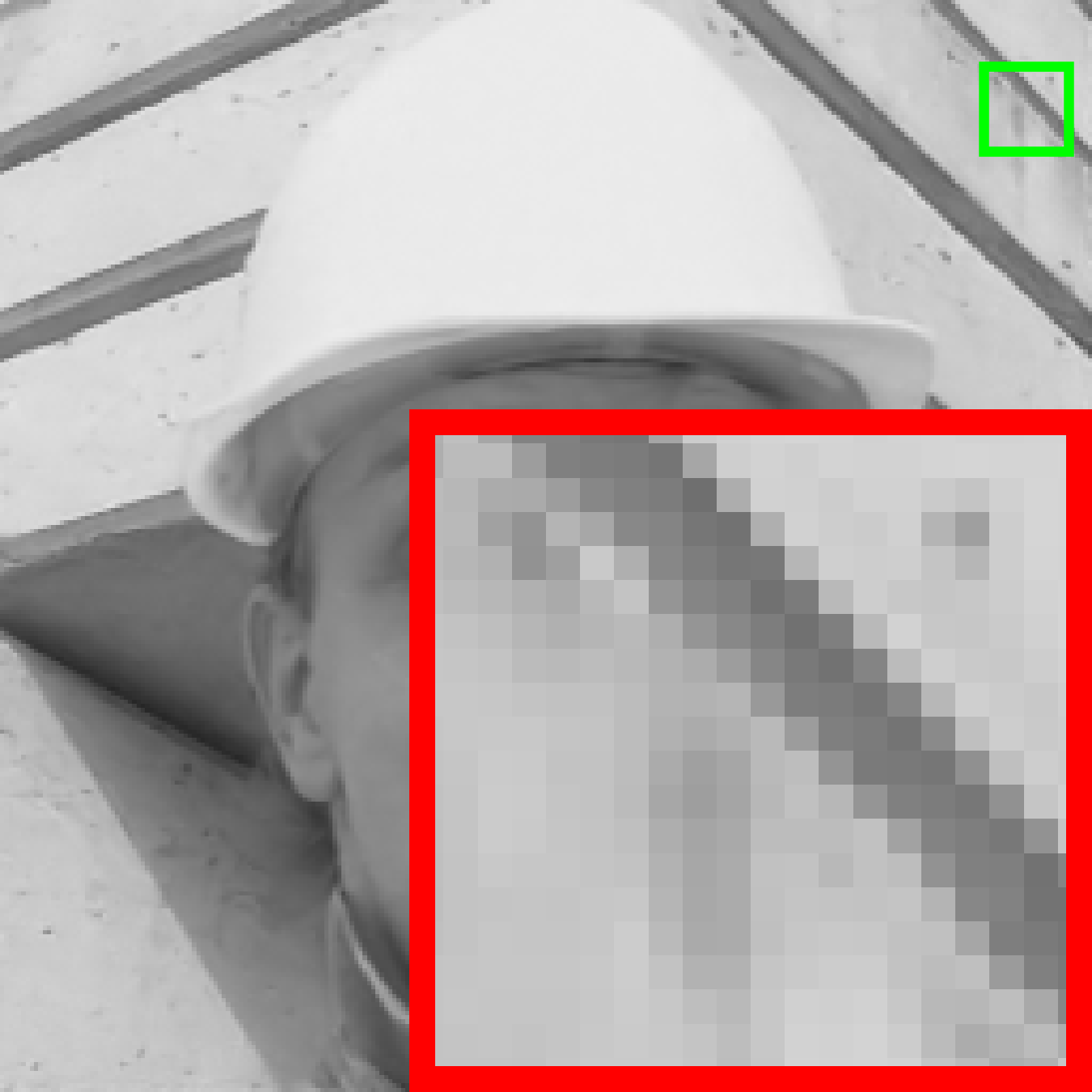}
    &\includegraphics[width=0.08\textwidth]{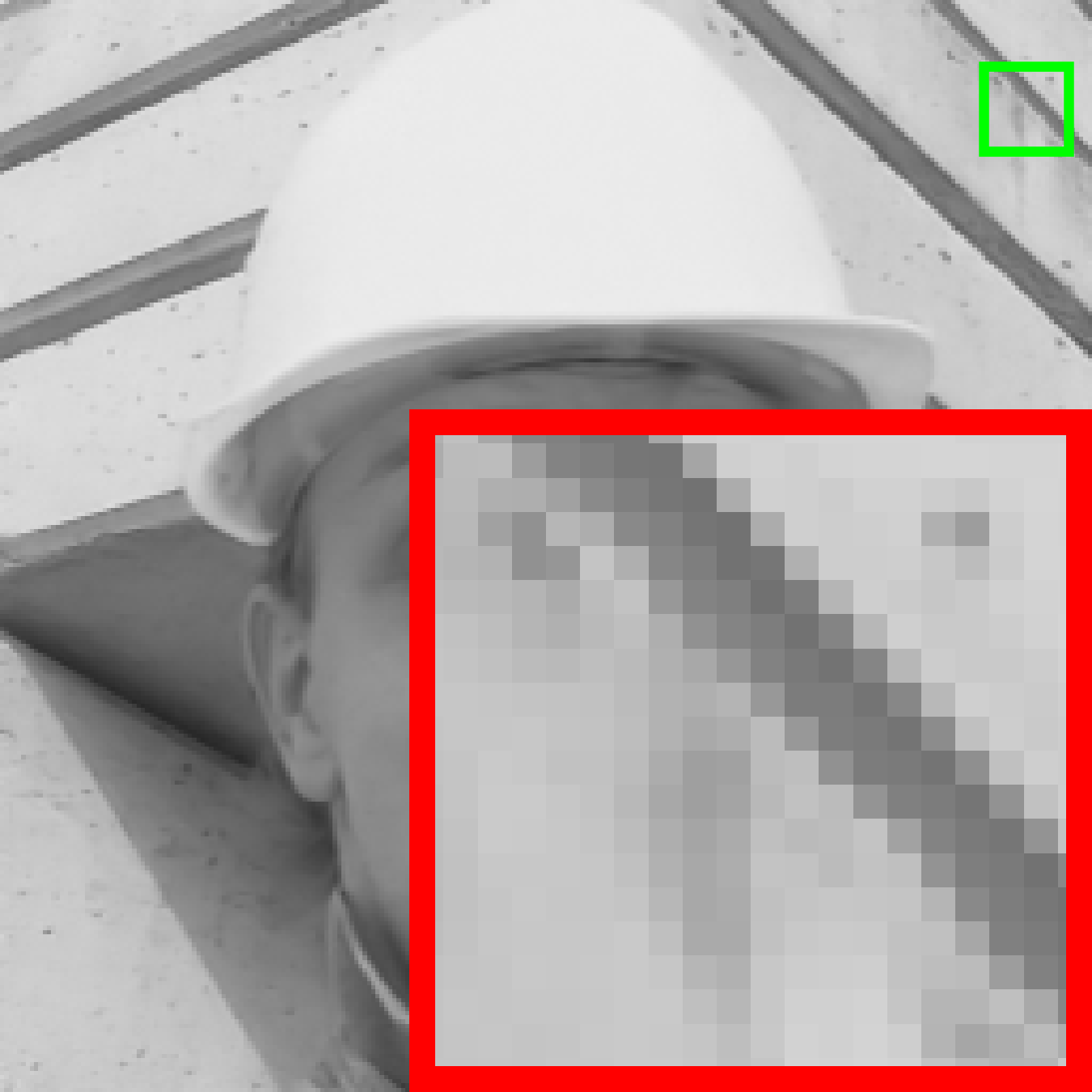}
    &\includegraphics[width=0.08\textwidth]{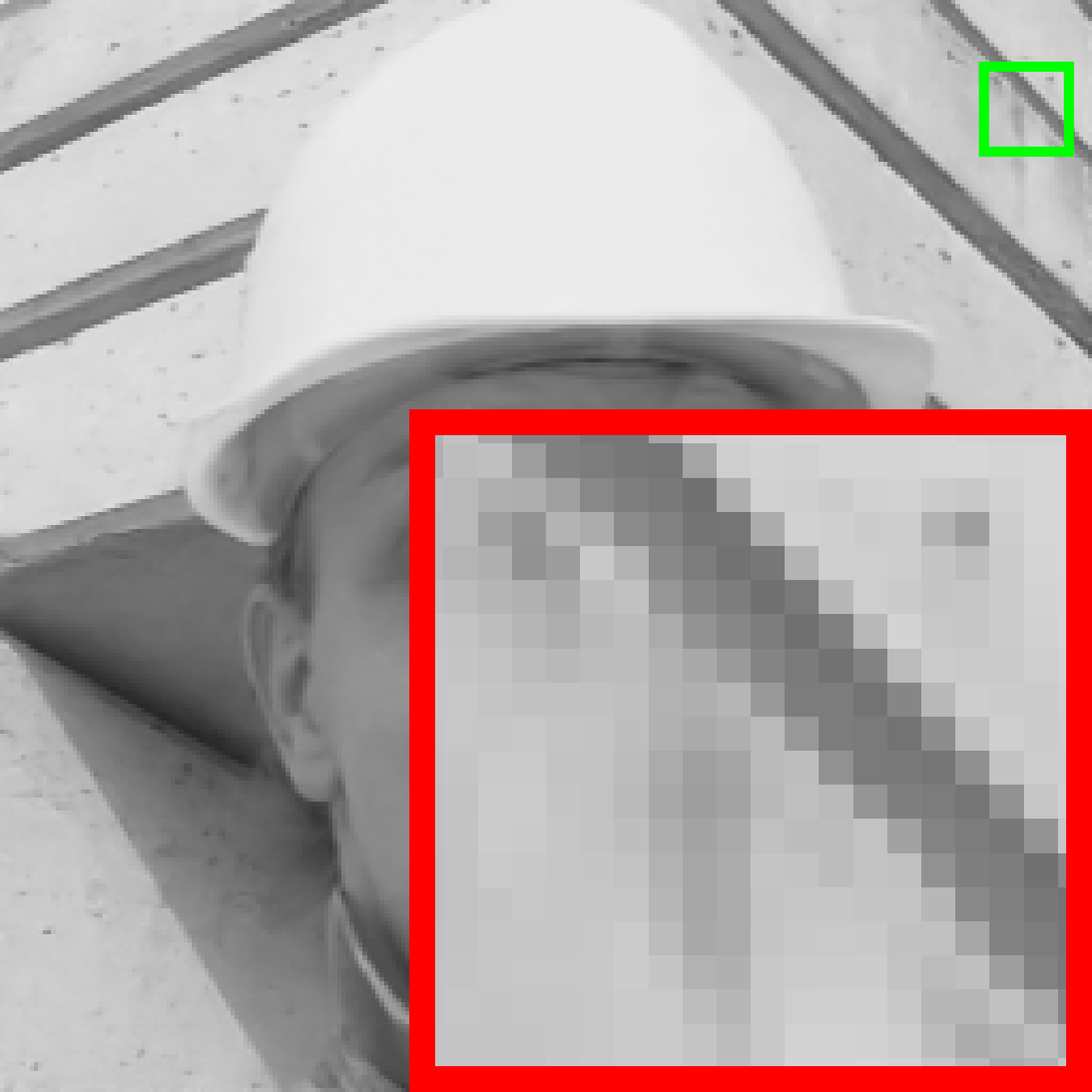}
    &\includegraphics[width=0.08\textwidth]{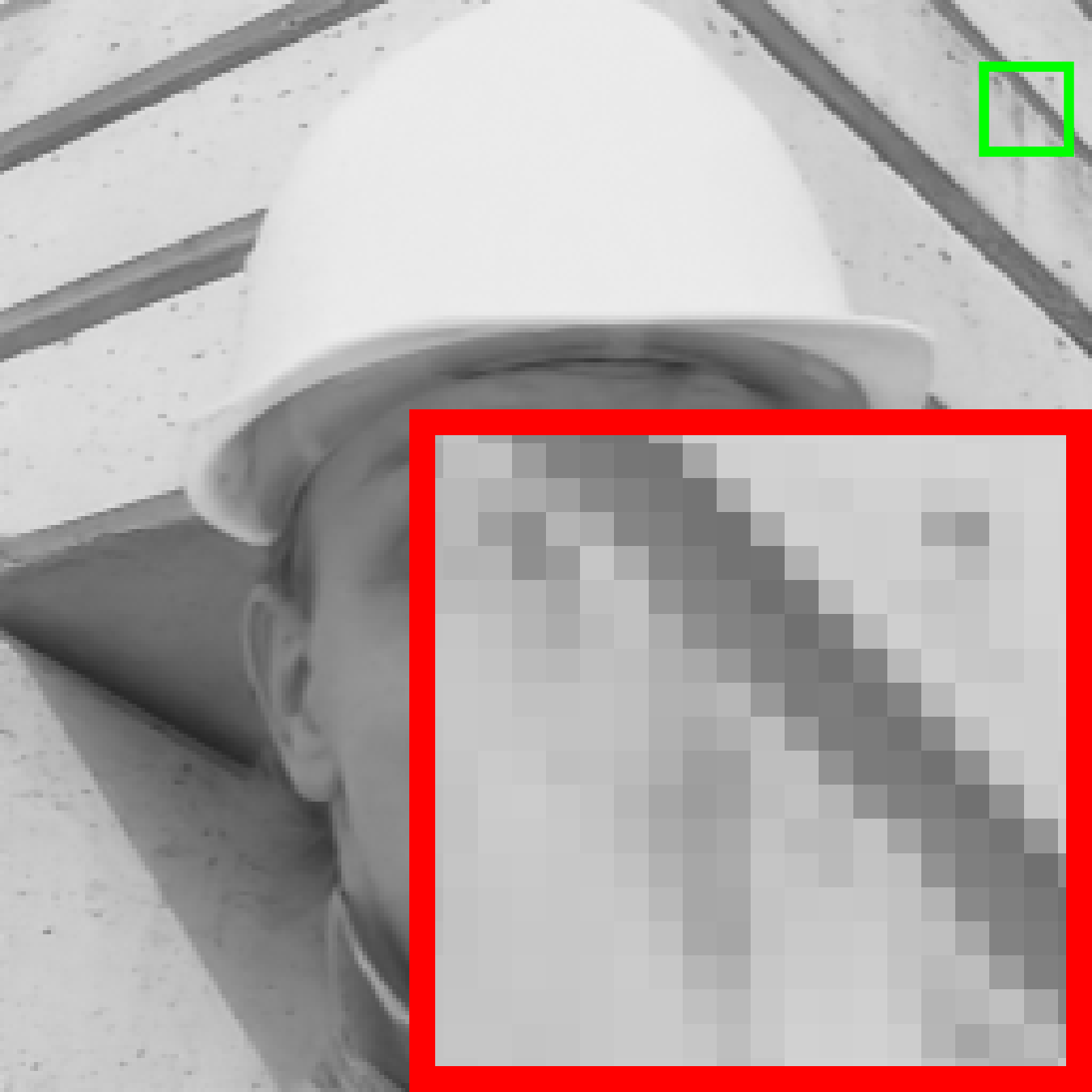}\\
    PSNR/SSIM & 34.16/0.91 & 40.22/\underline{\textcolor{blue}{0.97}} & 40.85/\underline{\textcolor{blue}{0.97}} & 41.22/\underline{\textcolor{blue}{0.97}} & 36.59/0.93 & 39.80/0.96 & 40.18/\underline{\textcolor{blue}{0.97}} & 40.48/\underline{\textcolor{blue}{0.97}} & 41.11/\underline{\textcolor{blue}{0.97}} & 41.52/\underline{\textcolor{blue}{0.97}} & \underline{\textcolor{blue}{42.41}}/\textbf{\textcolor{red}{0.98}} & 41.78/0.97 & \textbf{\textcolor{red}{42.67}}/\textbf{\textcolor{red}{0.98}}\\
    \includegraphics[width=0.08\textwidth]{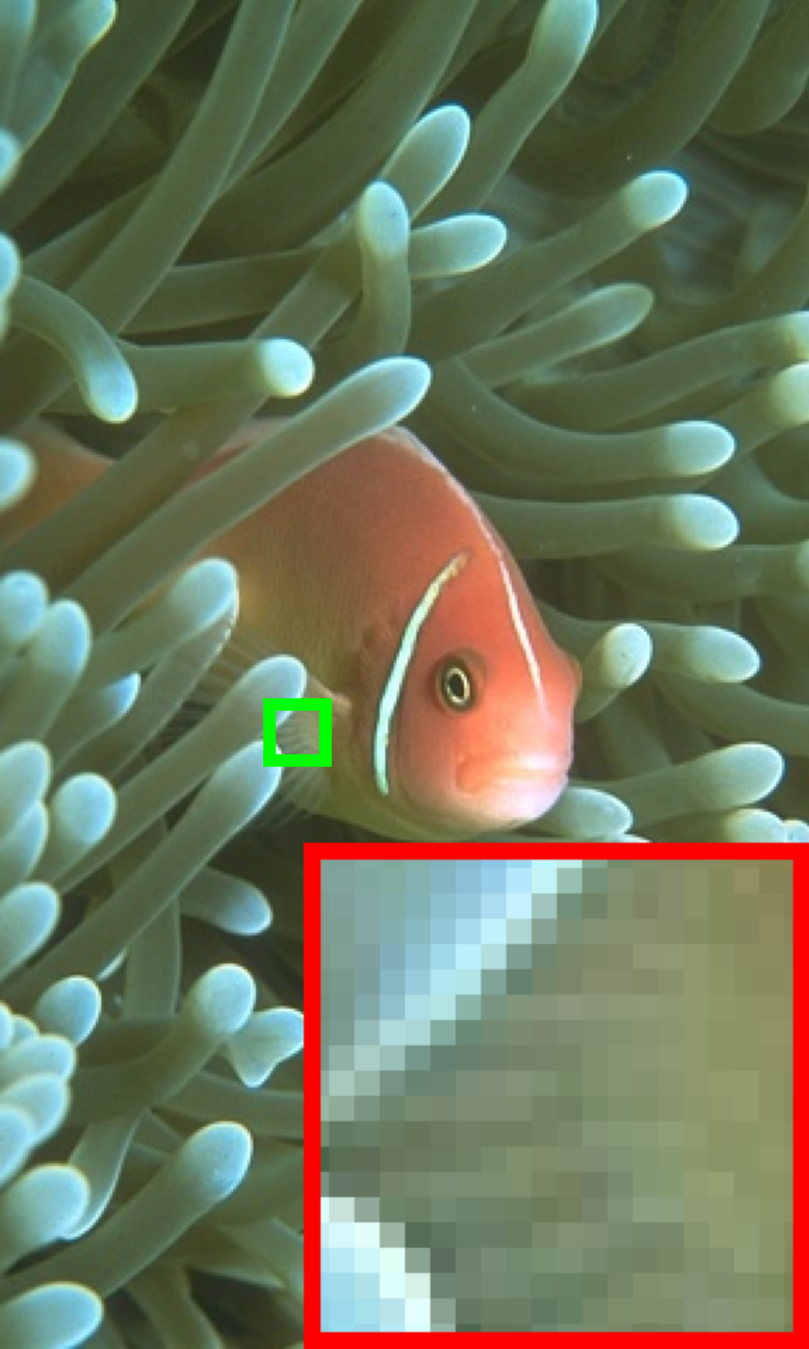}
    &\includegraphics[width=0.08\textwidth]{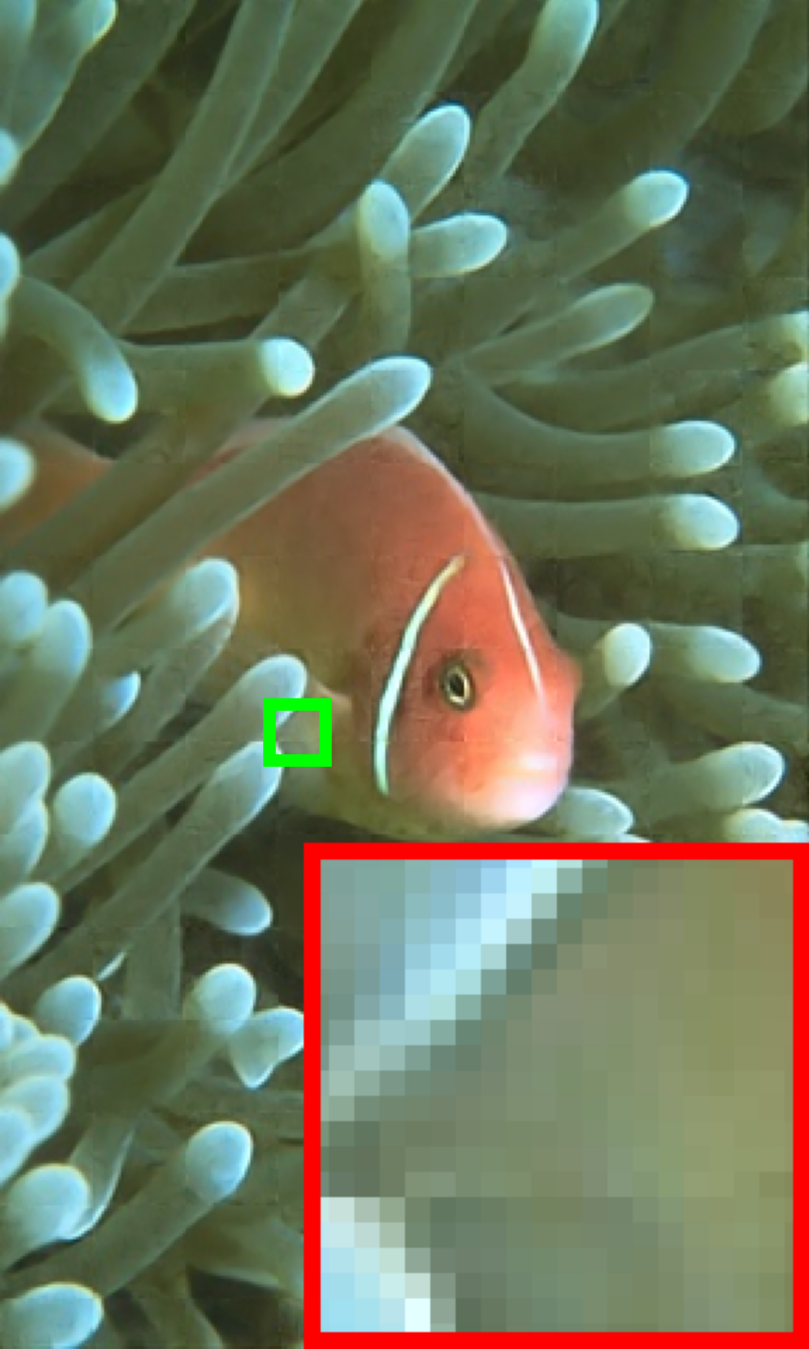}
    &\includegraphics[width=0.08\textwidth]{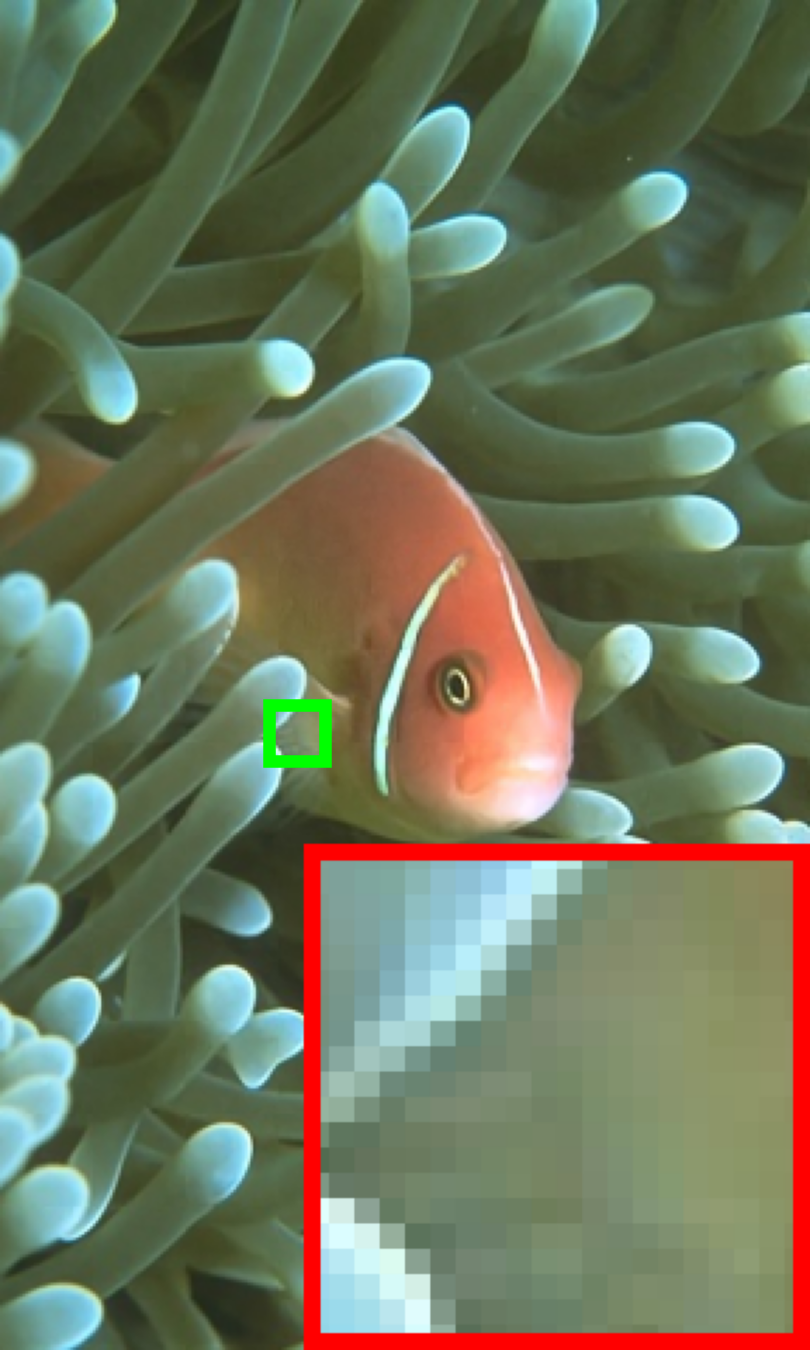}
    &\includegraphics[width=0.08\textwidth]{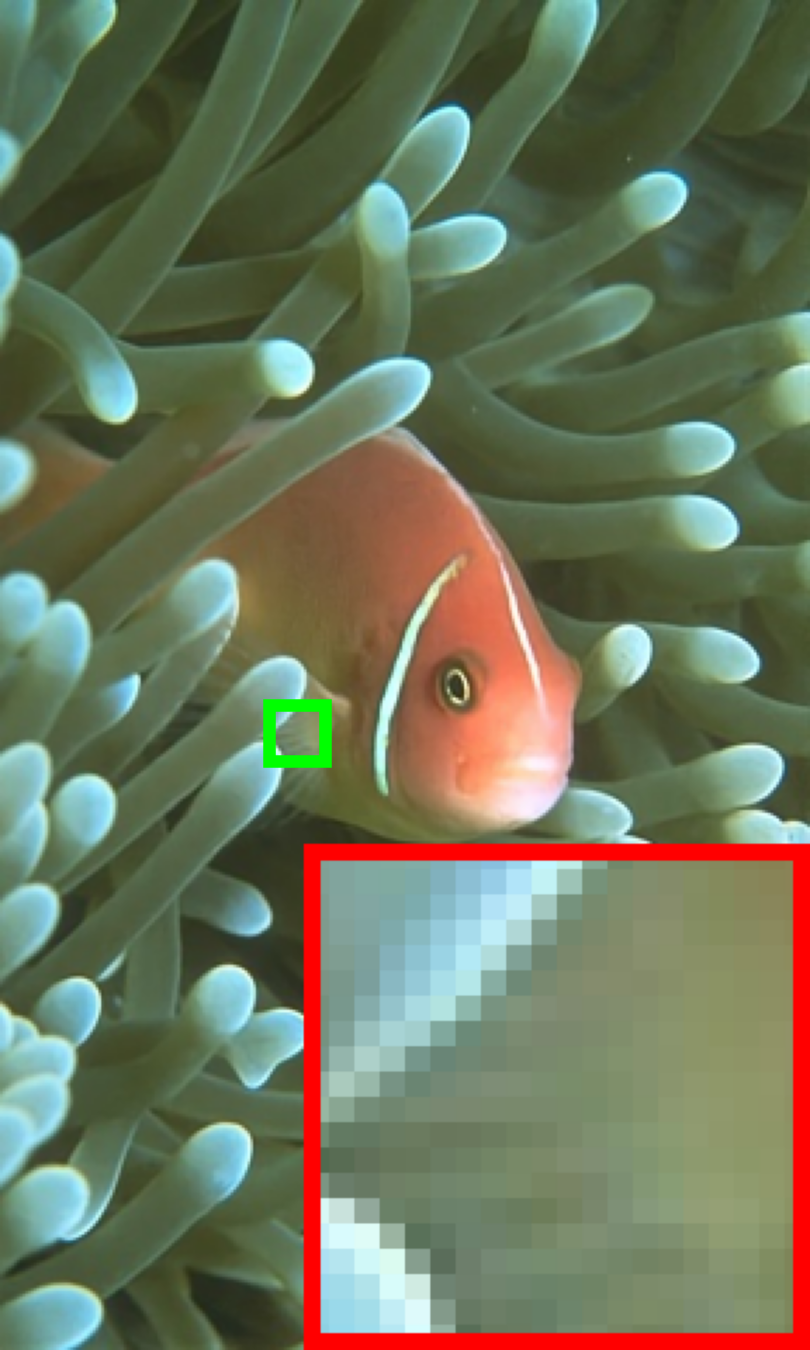}
    &\includegraphics[width=0.08\textwidth]{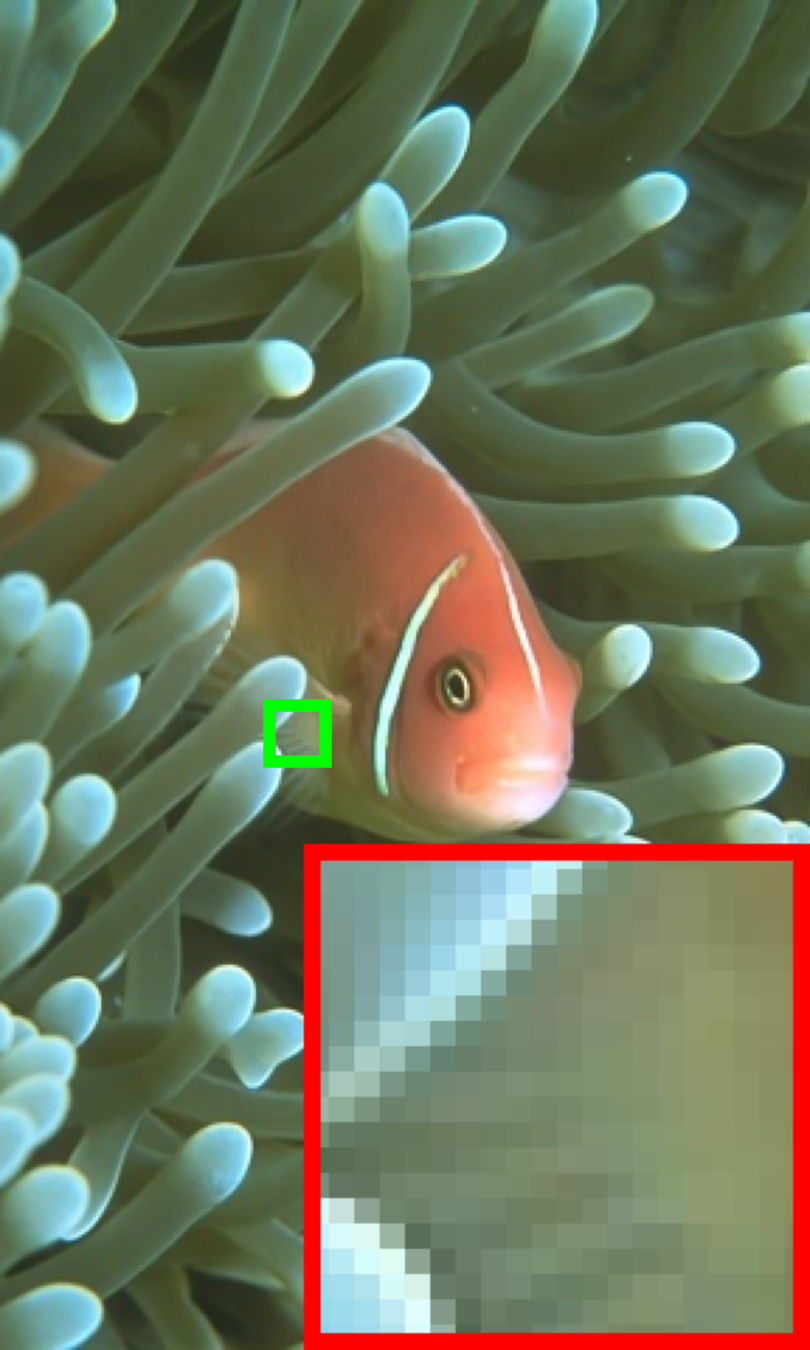}
    &\includegraphics[width=0.08\textwidth]{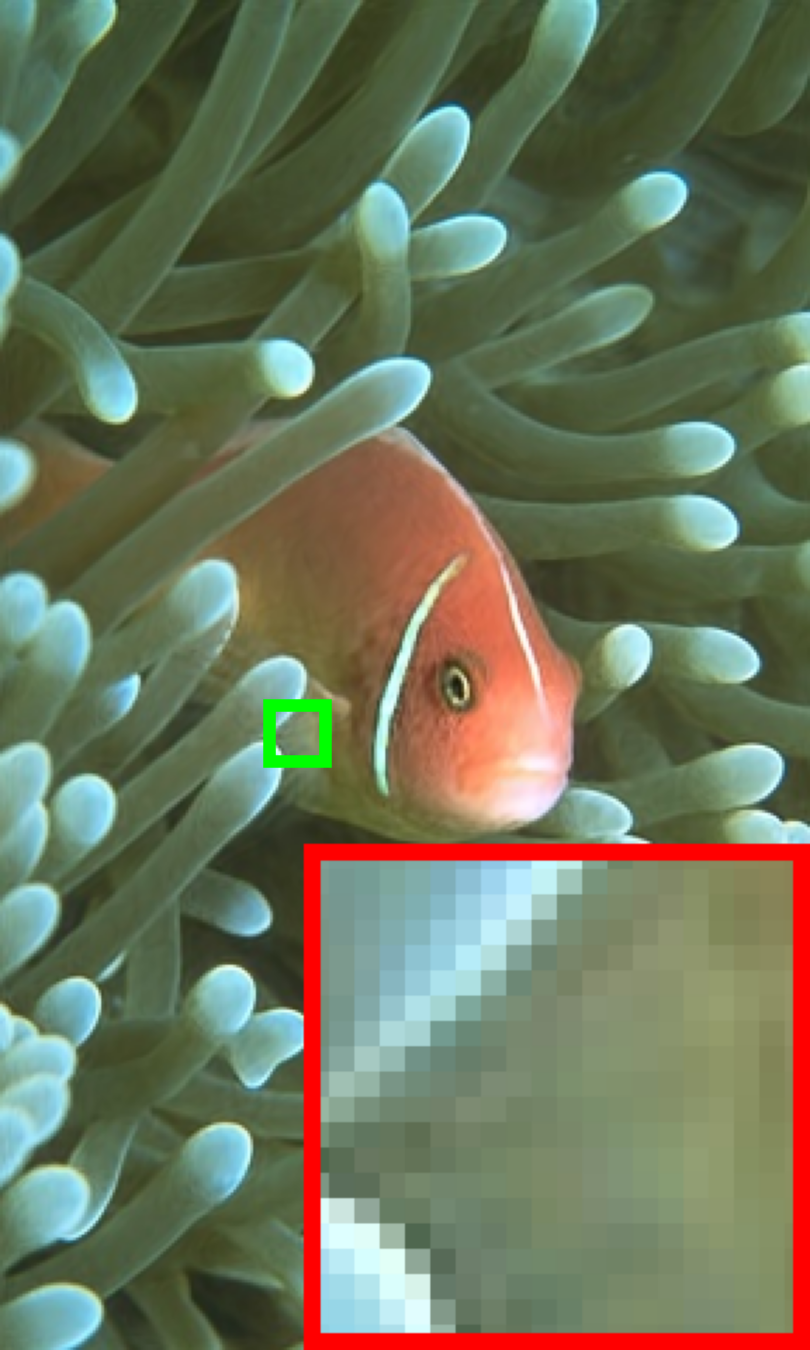}
    &\includegraphics[width=0.08\textwidth]{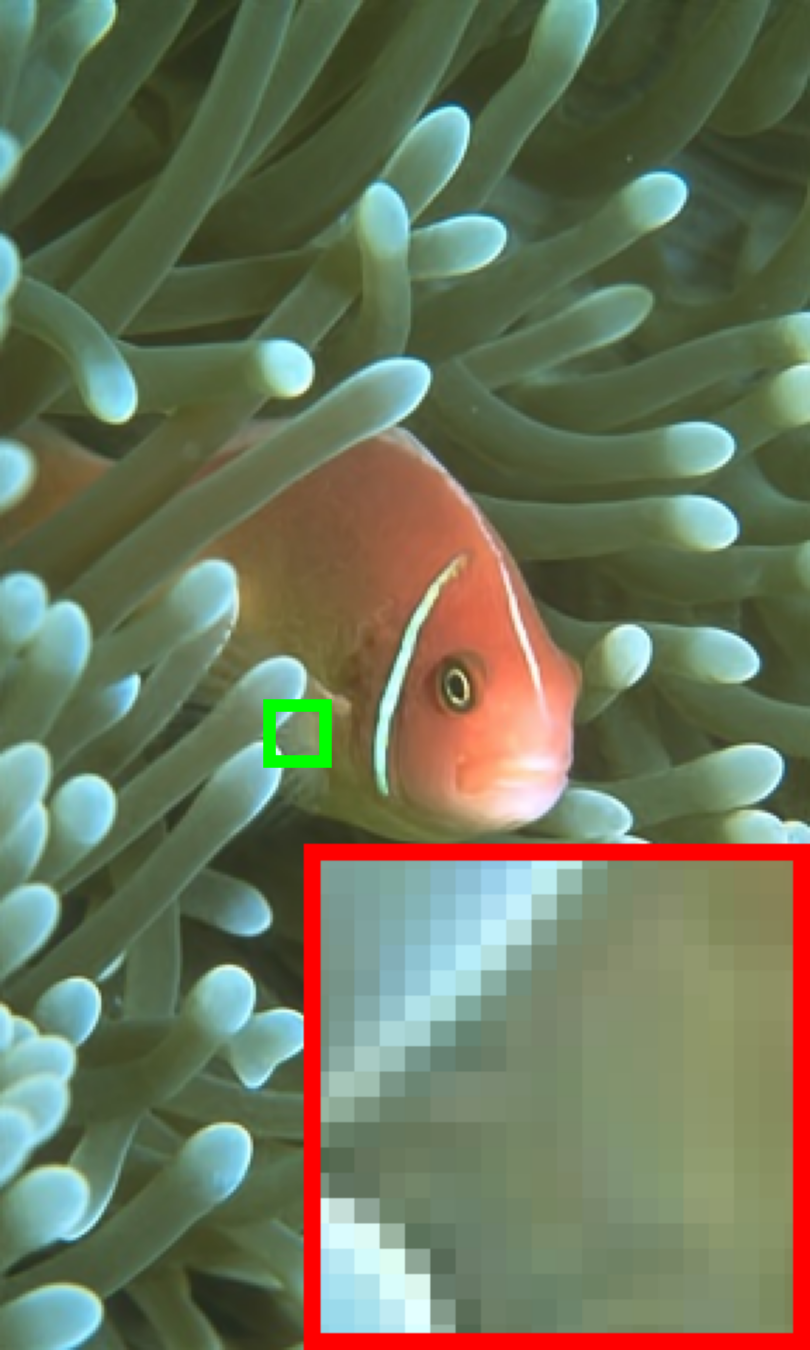}
    &\includegraphics[width=0.08\textwidth]{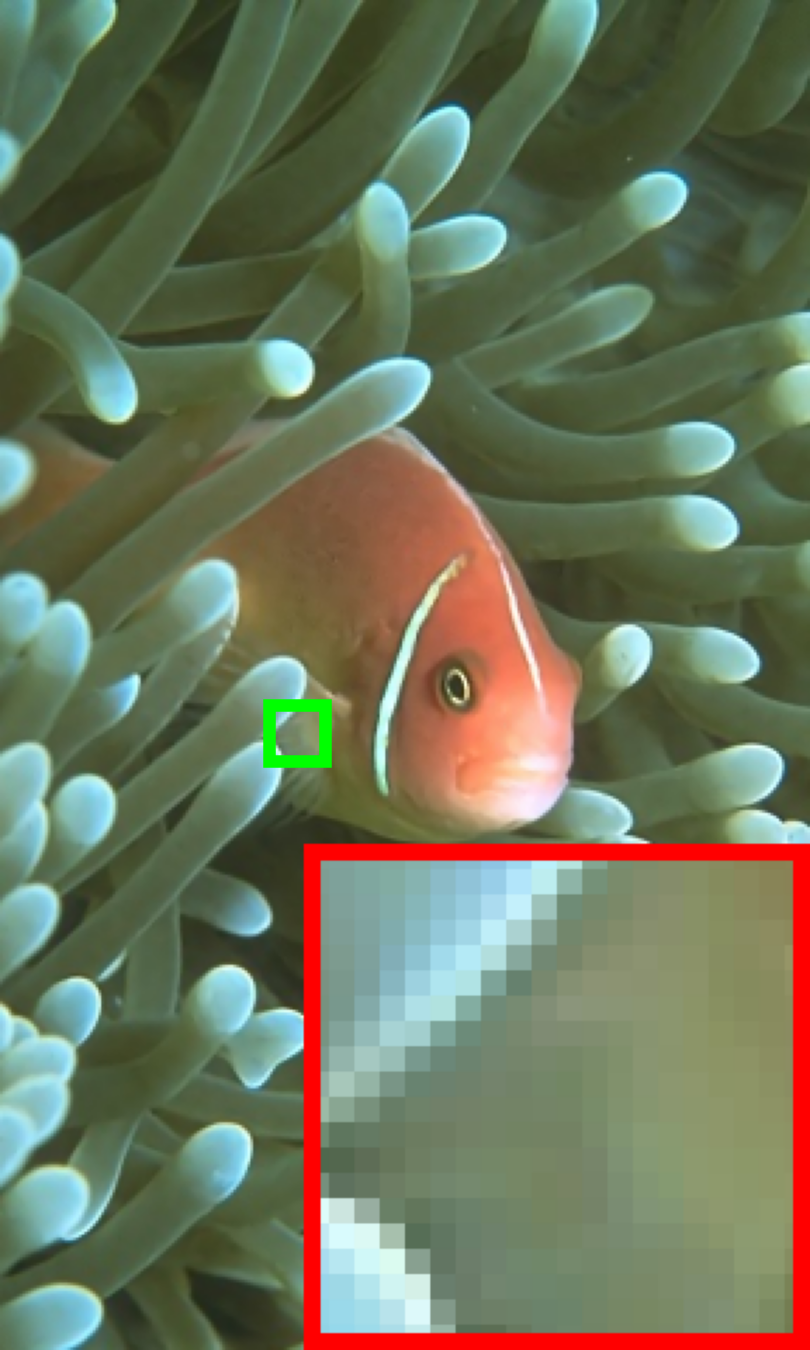}
    &\includegraphics[width=0.08\textwidth]{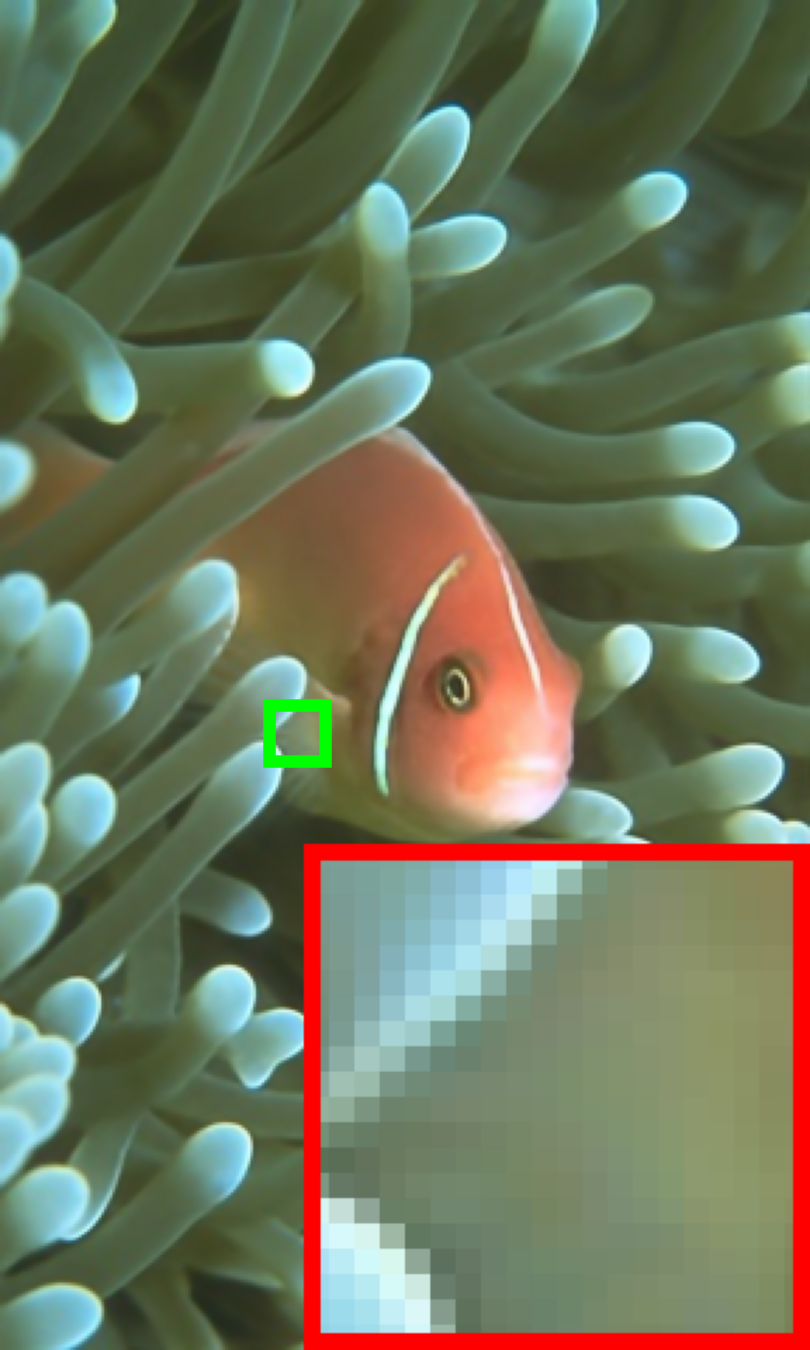}
    &\includegraphics[width=0.08\textwidth]{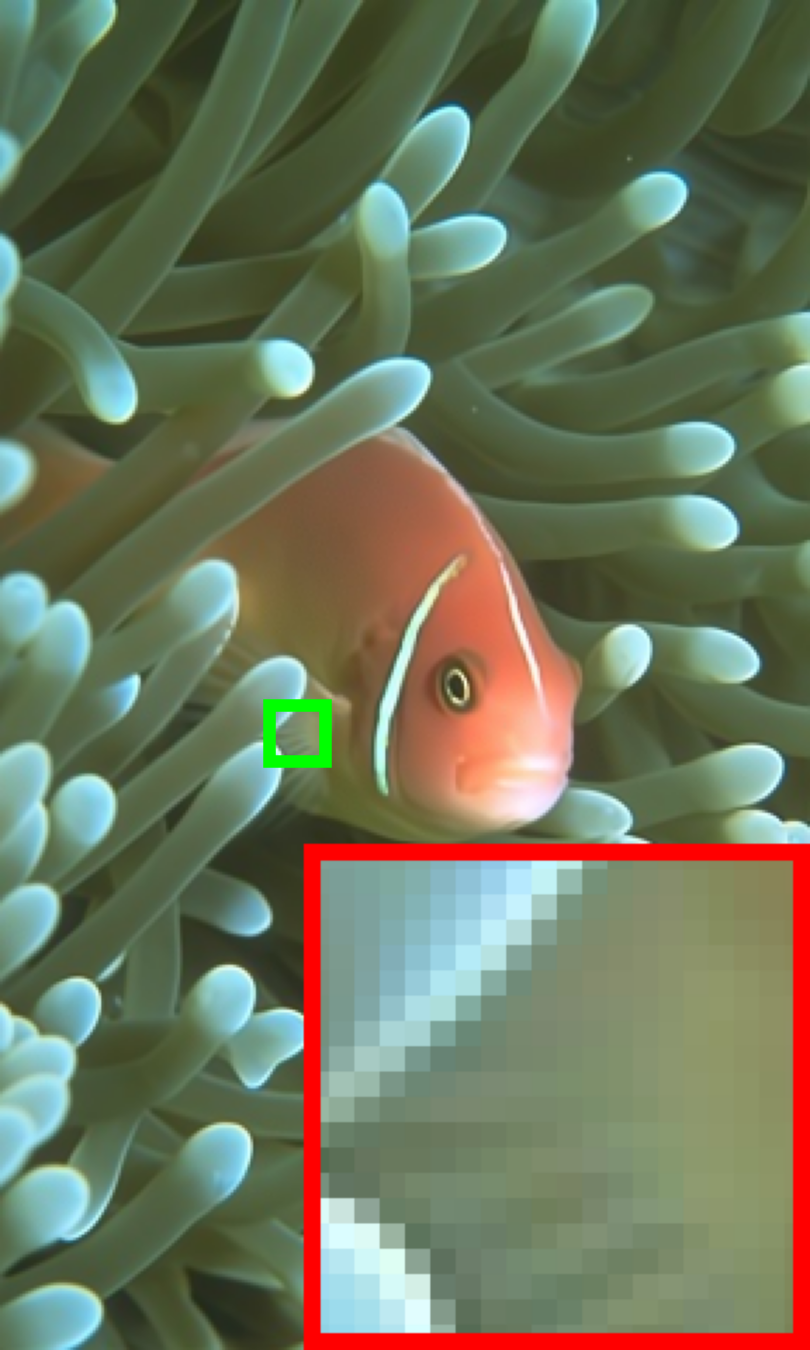}
    &\includegraphics[width=0.08\textwidth]{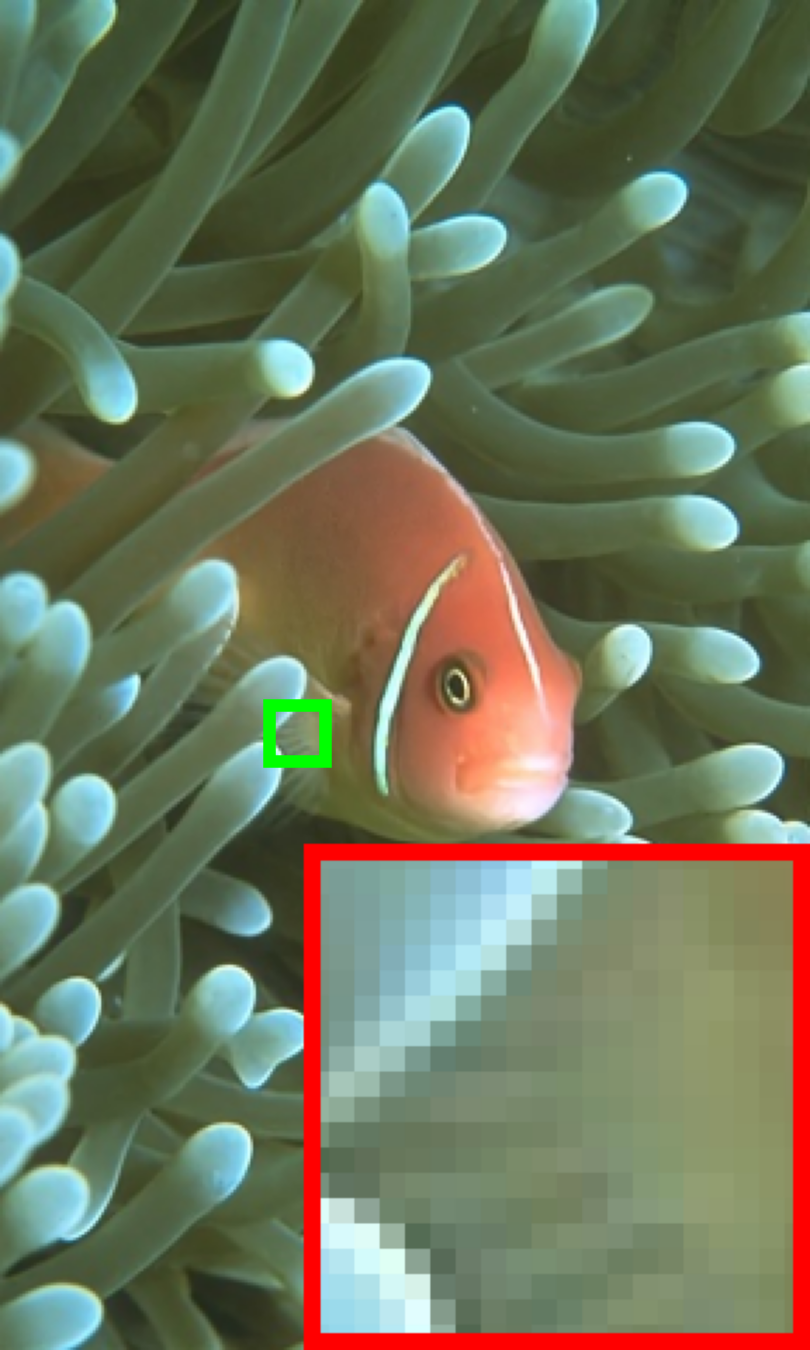}
    &\includegraphics[width=0.08\textwidth]{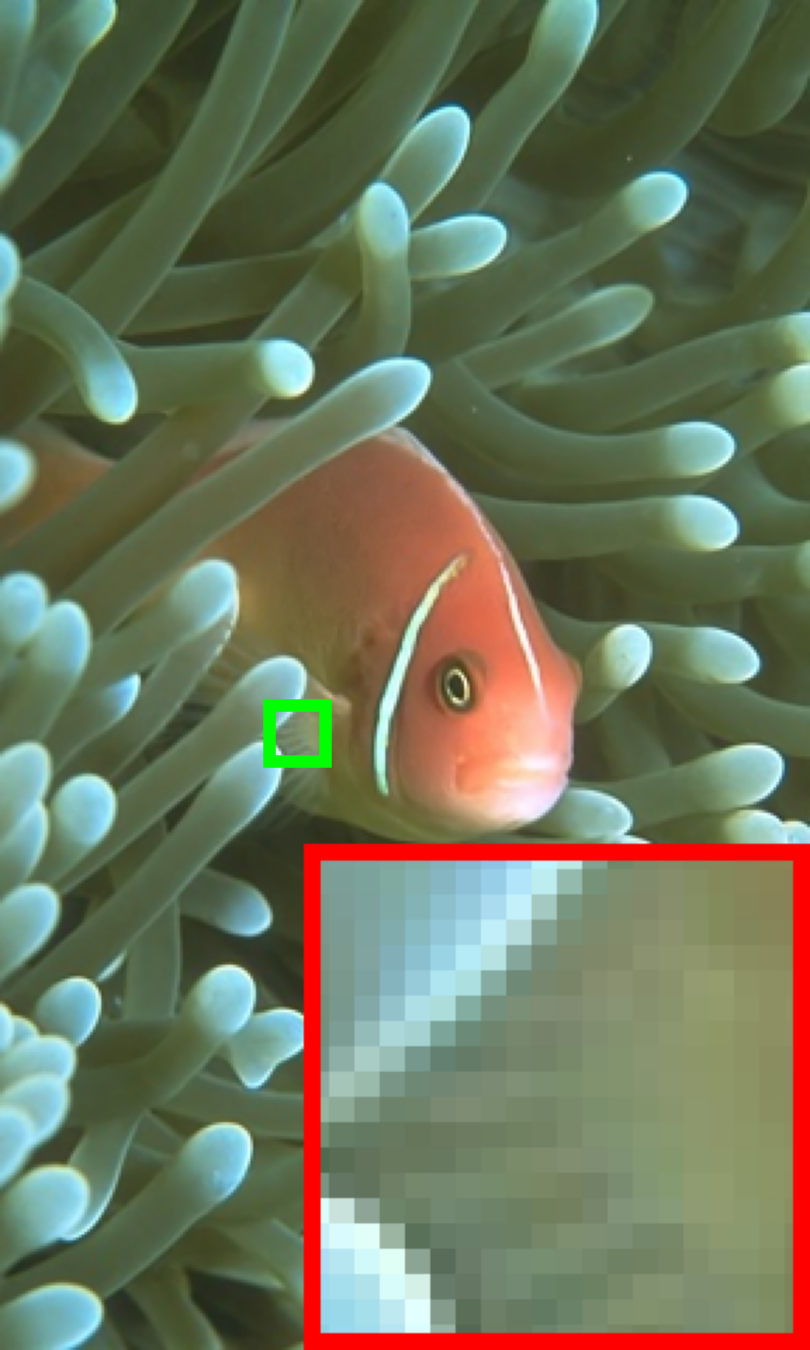}
    &\includegraphics[width=0.08\textwidth]{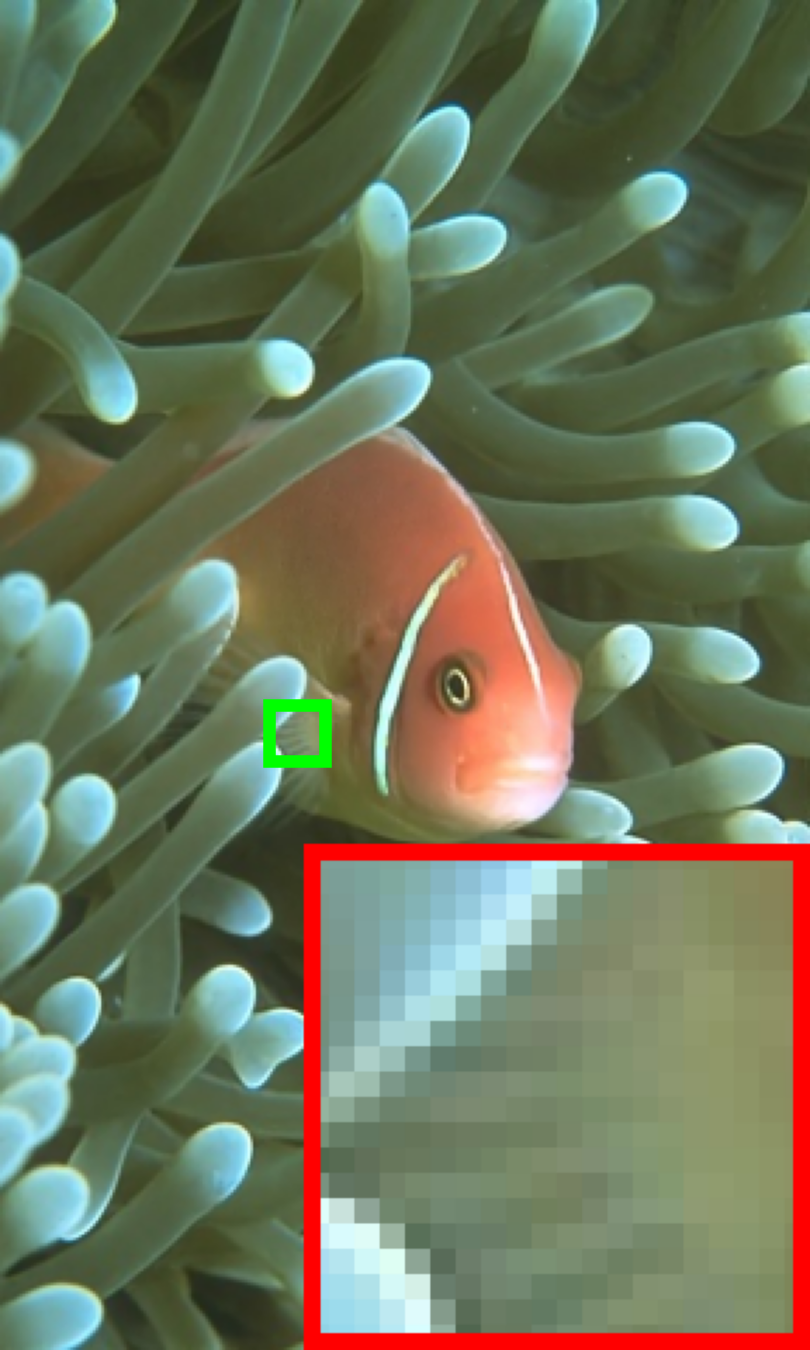}
    &\includegraphics[width=0.08\textwidth]{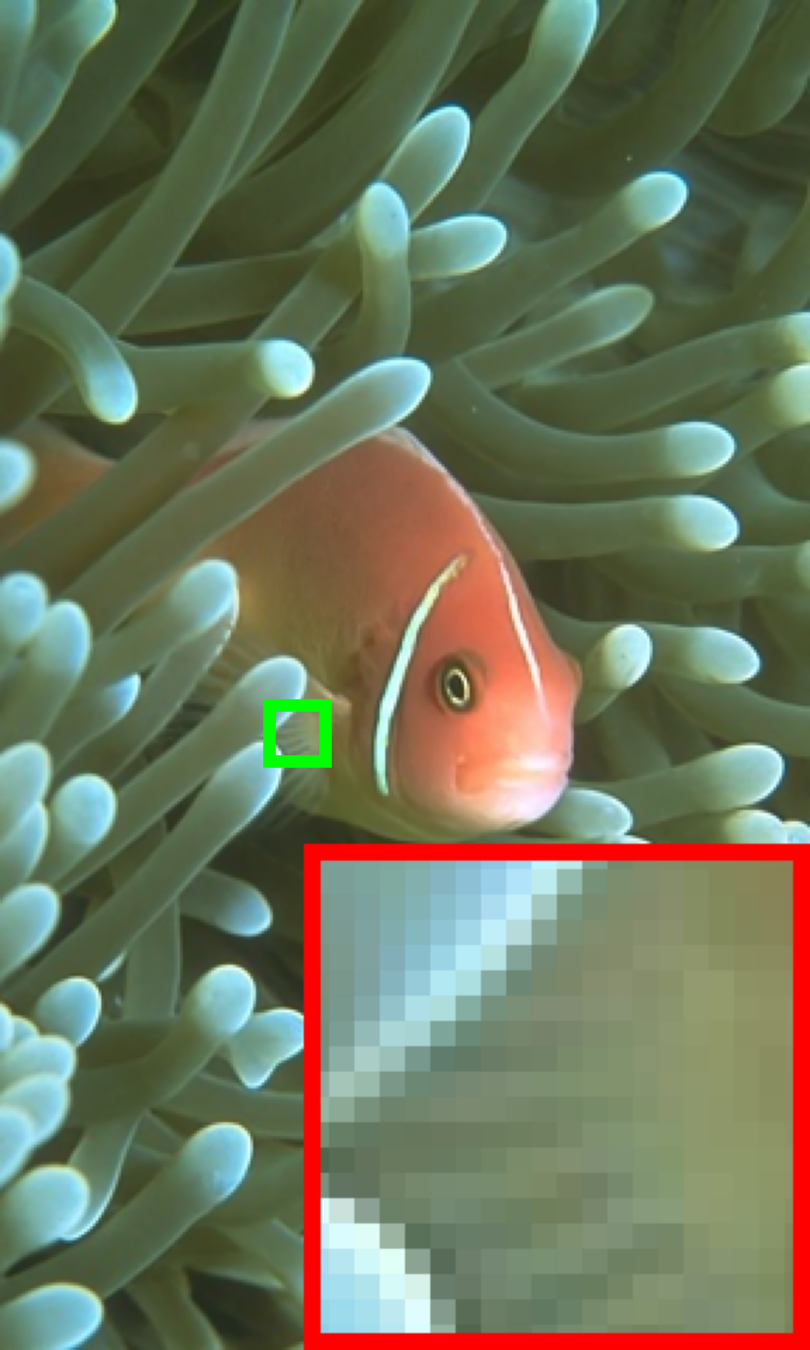}\\
    PSNR/SSIM & 34.47/0.93 & 41.75/\underline{\textcolor{blue}{0.98}} & 42.55/\underline{\textcolor{blue}{0.98}} & 43.08/\textbf{\textcolor{red}{0.99}} & 39.17/0.97 & 41.90/\underline{\textcolor{blue}{0.98}} & 42.27/\underline{\textcolor{blue}{0.98}} & 39.78/0.9722 & 42.53/\underline{\textcolor{blue}{0.98}} & 43.35/\textbf{\textcolor{red}{0.99}} & \underline{\textcolor{blue}{43.82}}/\textbf{\textcolor{red}{0.99}} & 43.42/\textbf{\textcolor{red}{0.99}} & \textbf{\textcolor{red}{43.88}}/\textbf{\textcolor{red}{0.99}} \\
    \includegraphics[width=0.08\textwidth]{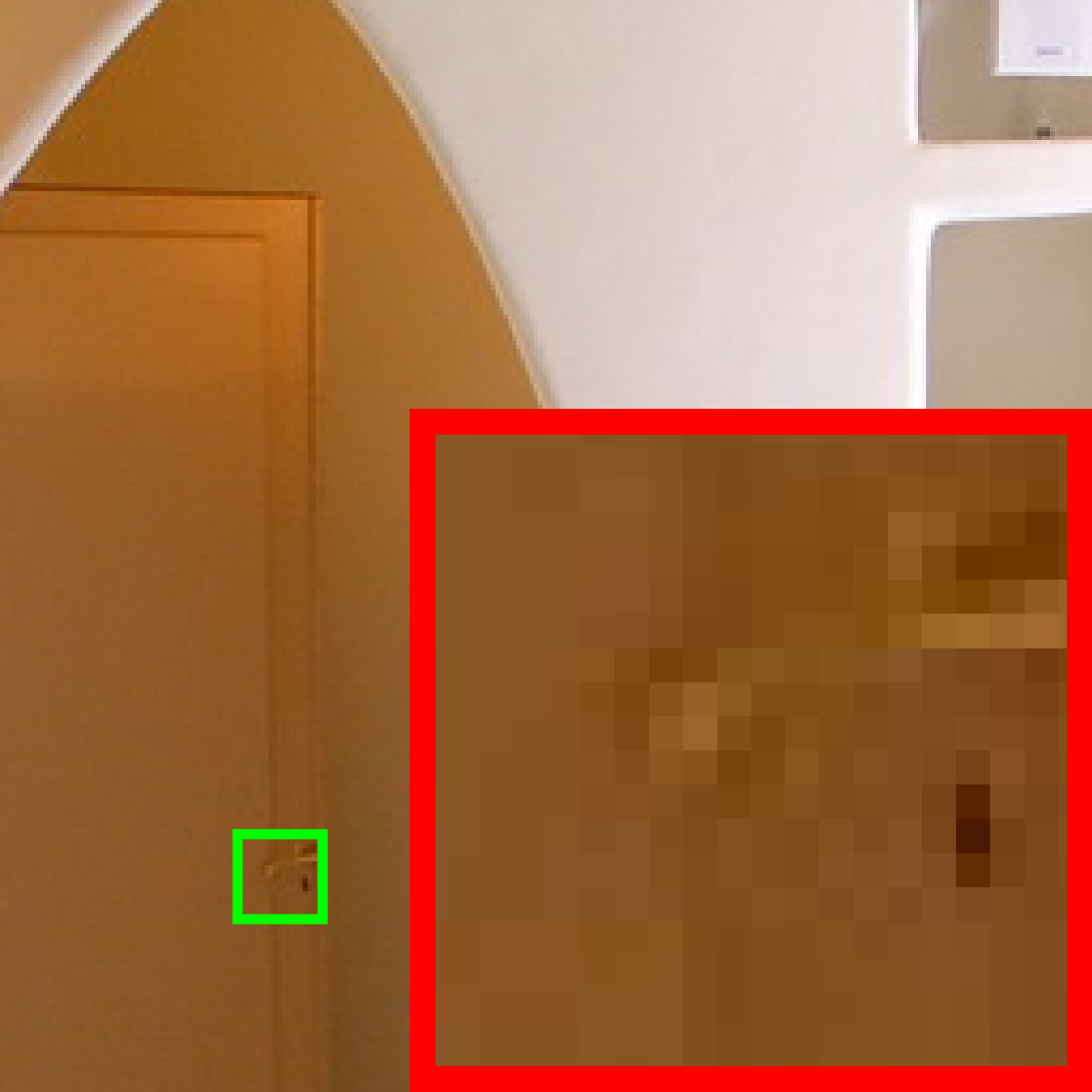}
    &\includegraphics[width=0.08\textwidth]{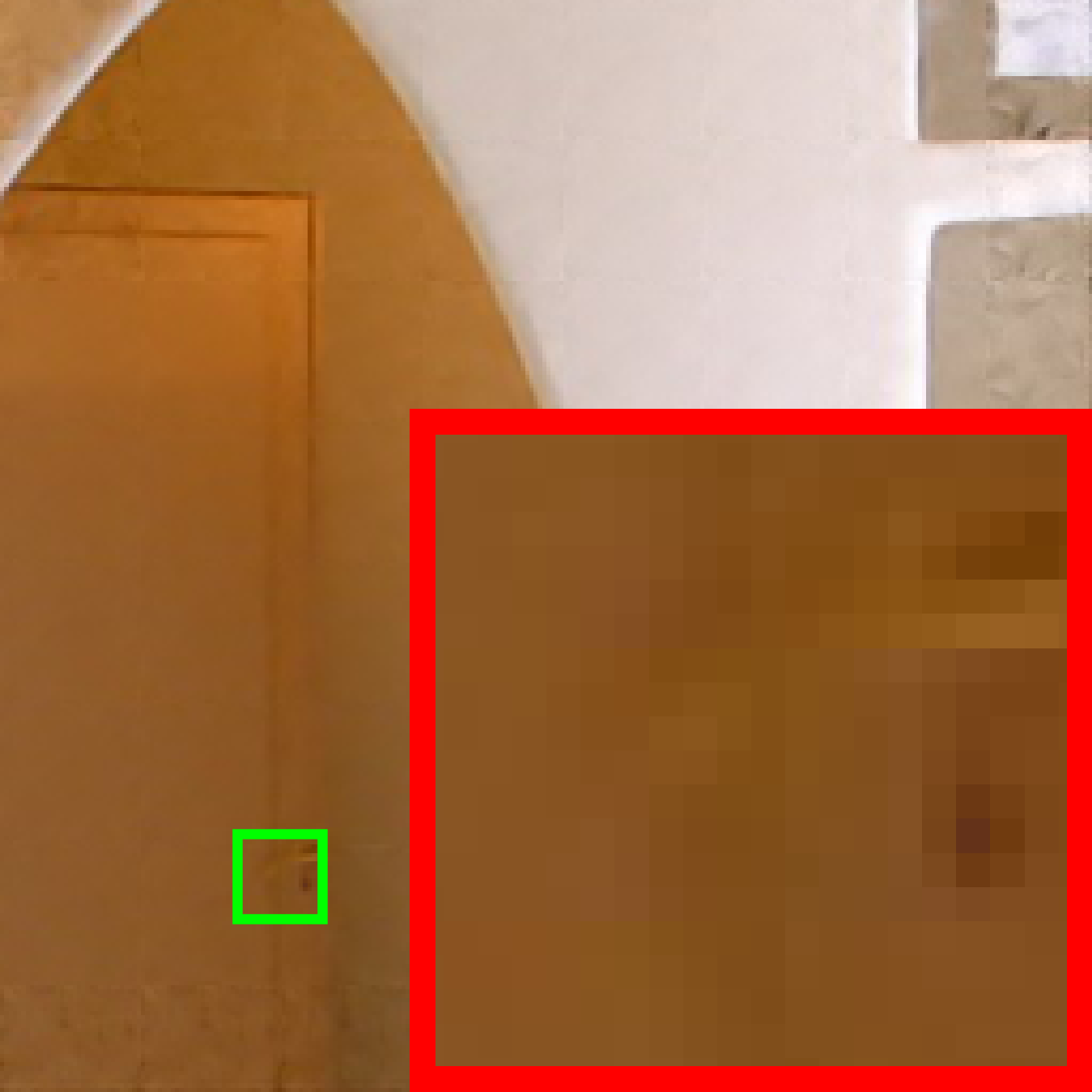}
    &\includegraphics[width=0.08\textwidth]{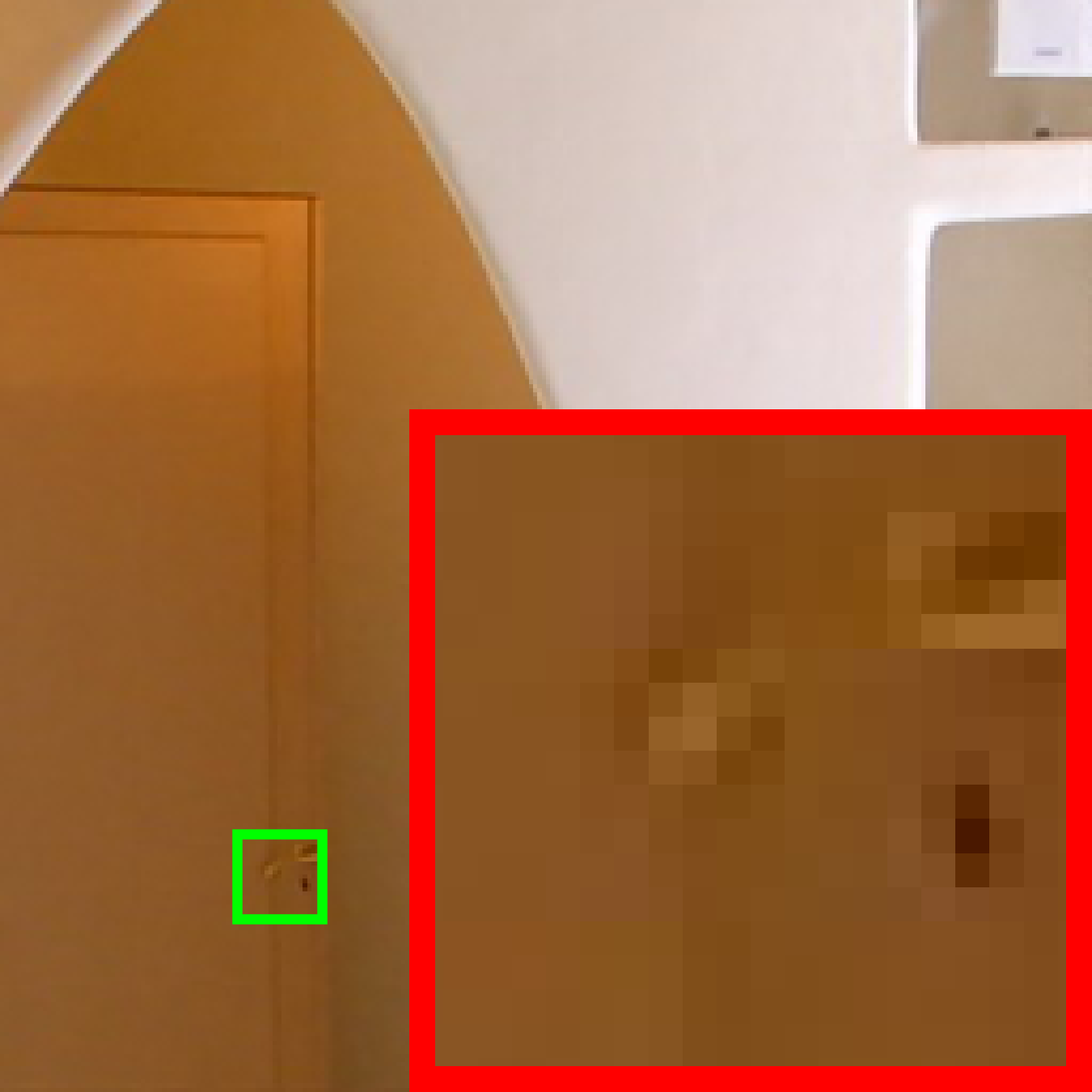}
    &\includegraphics[width=0.08\textwidth]{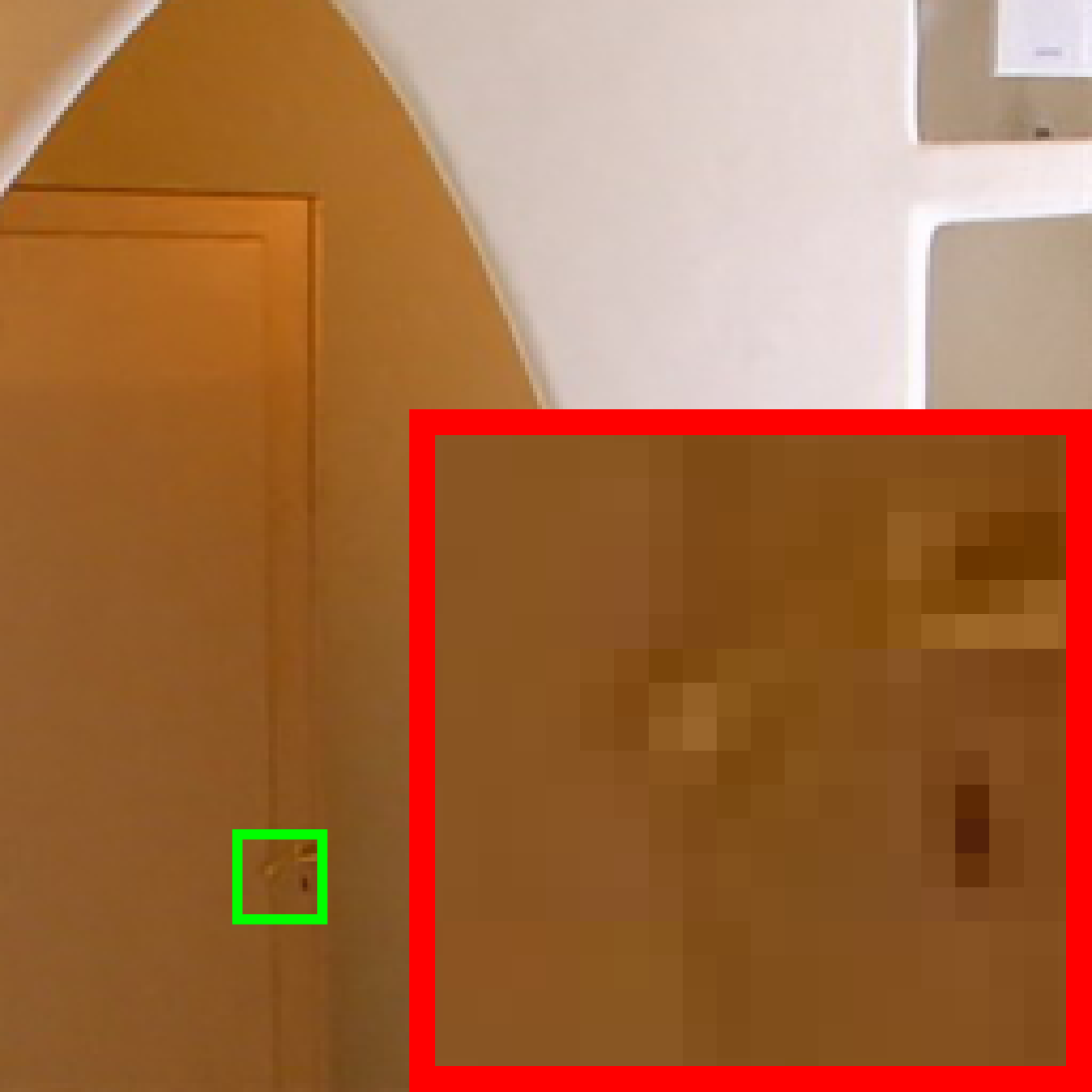}
    &\includegraphics[width=0.08\textwidth]{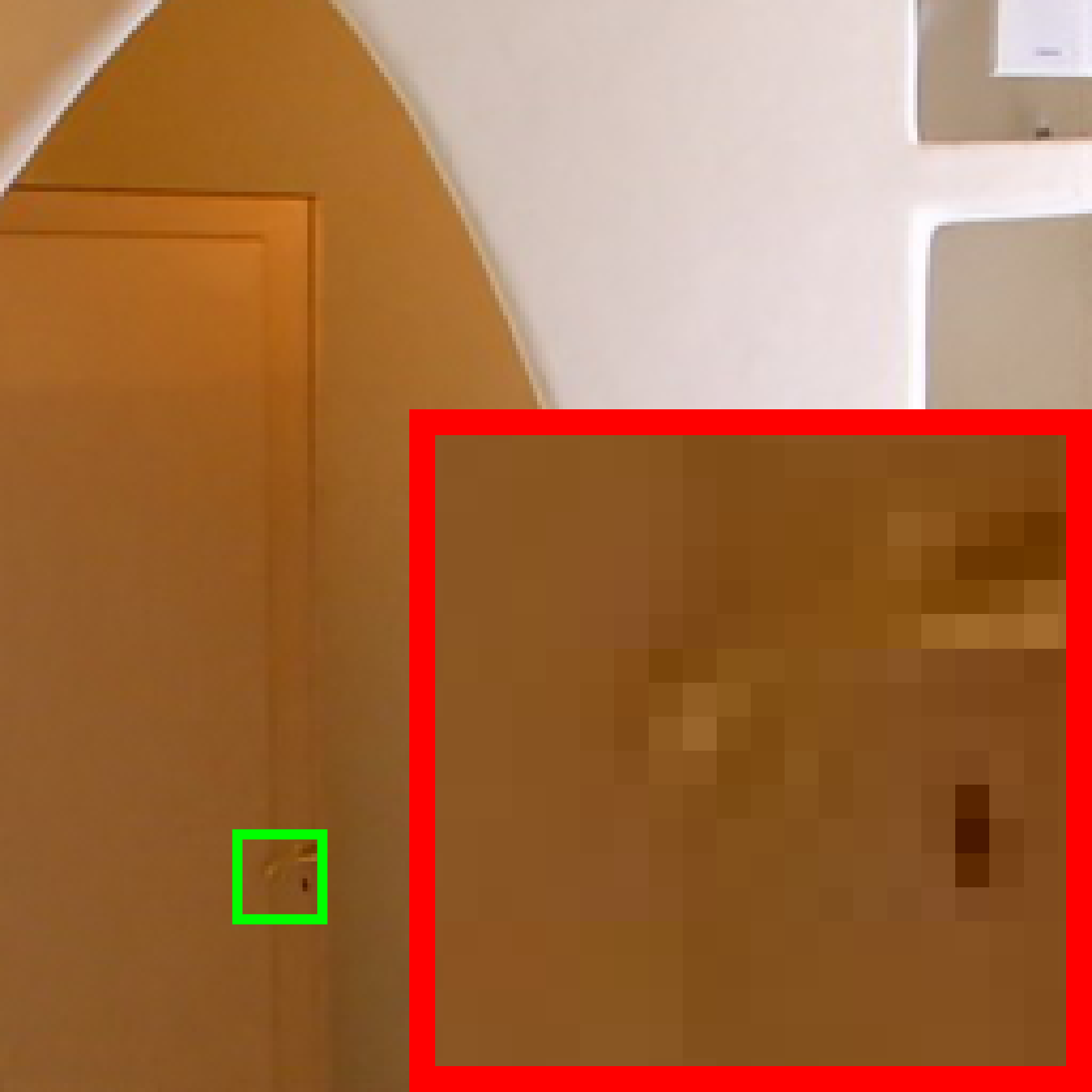}
    &\includegraphics[width=0.08\textwidth]{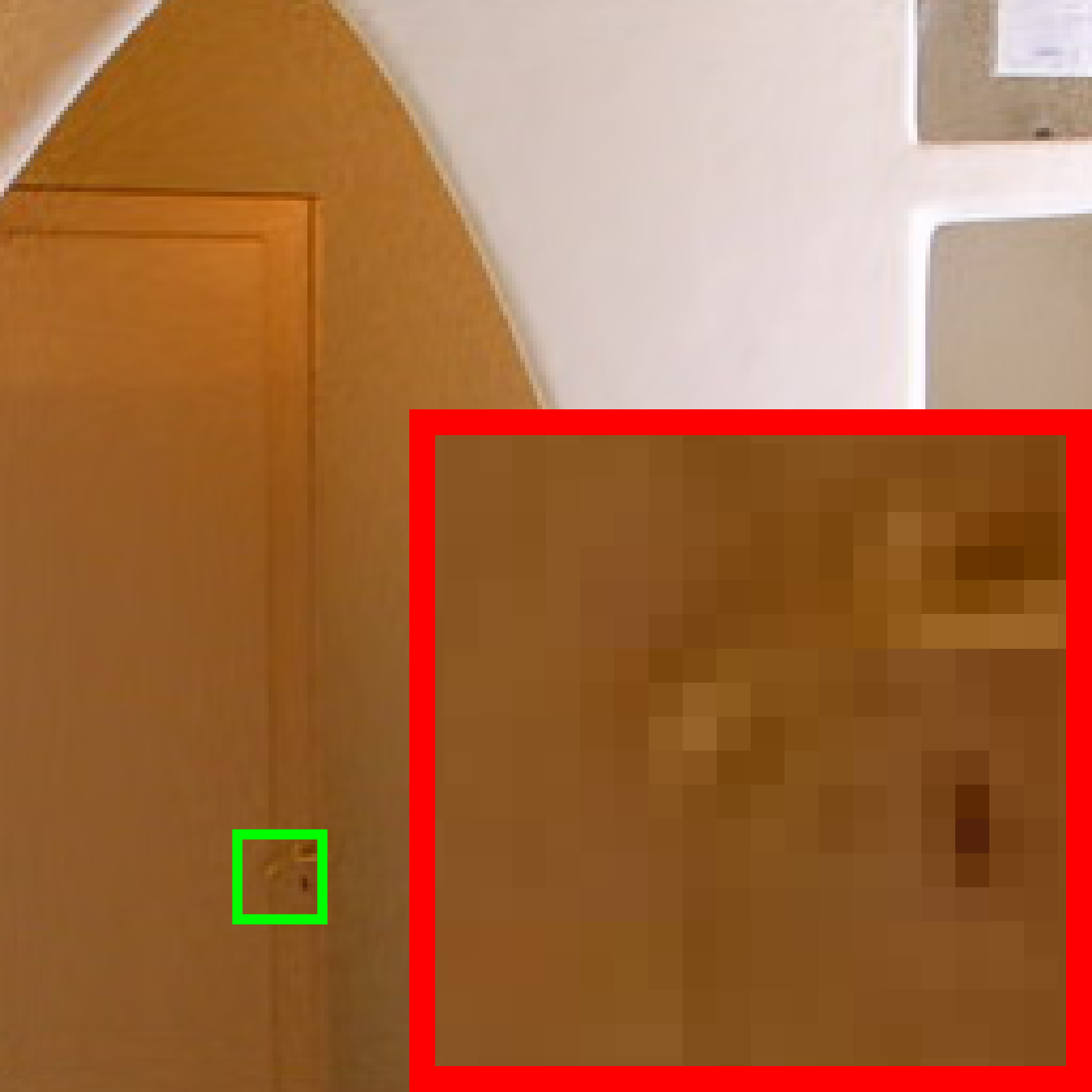}
    &\includegraphics[width=0.08\textwidth]{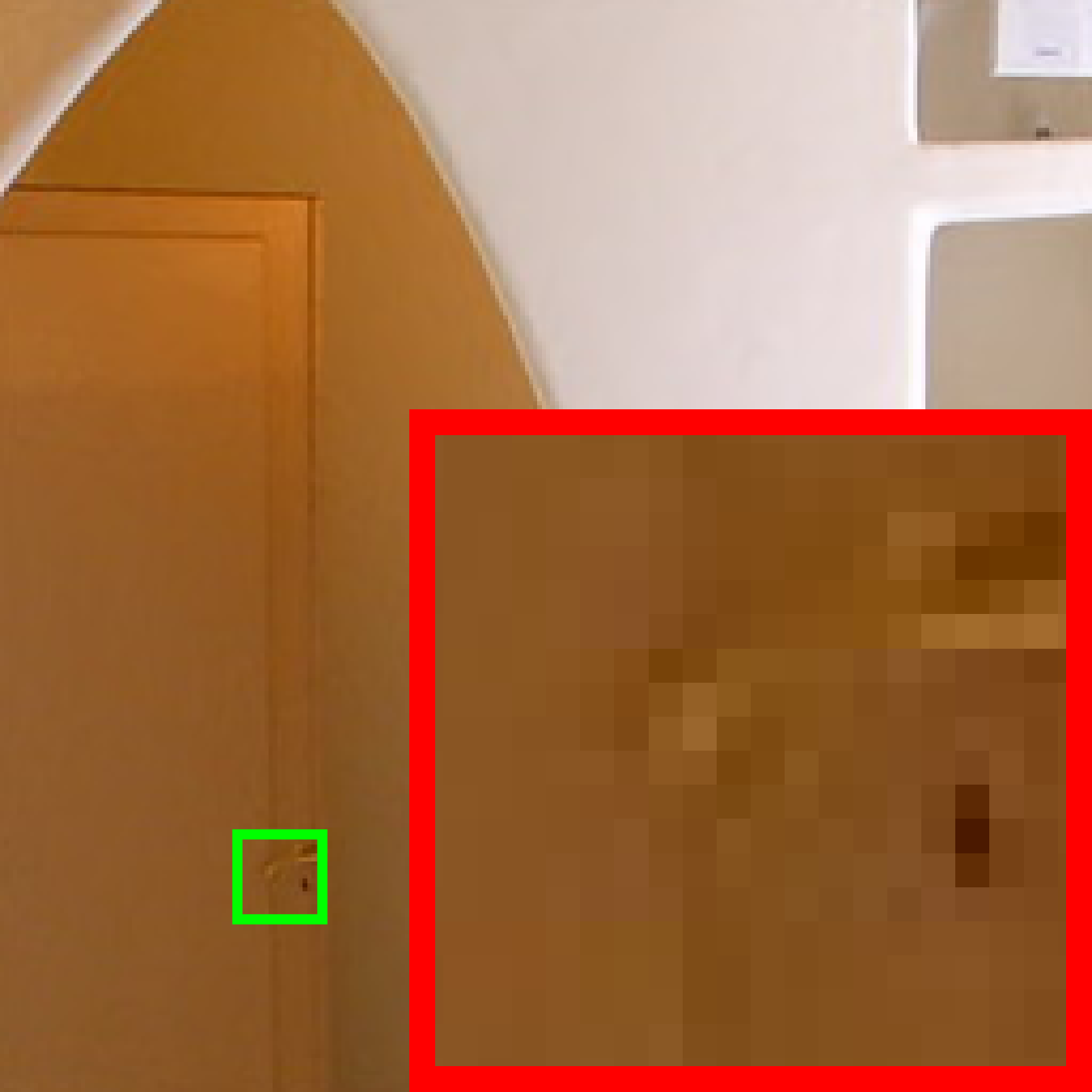}
    &\includegraphics[width=0.08\textwidth]{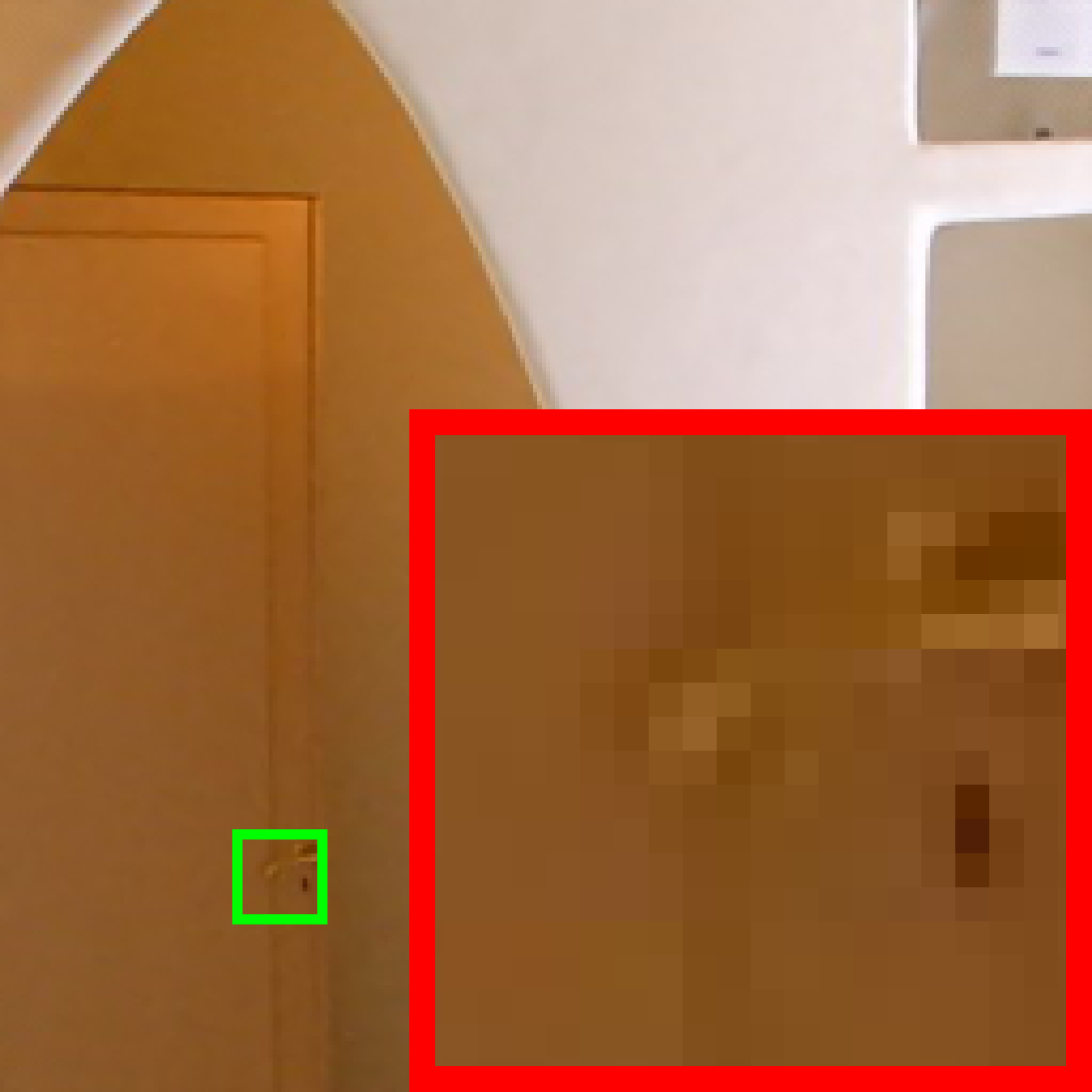}
    &\includegraphics[width=0.08\textwidth]{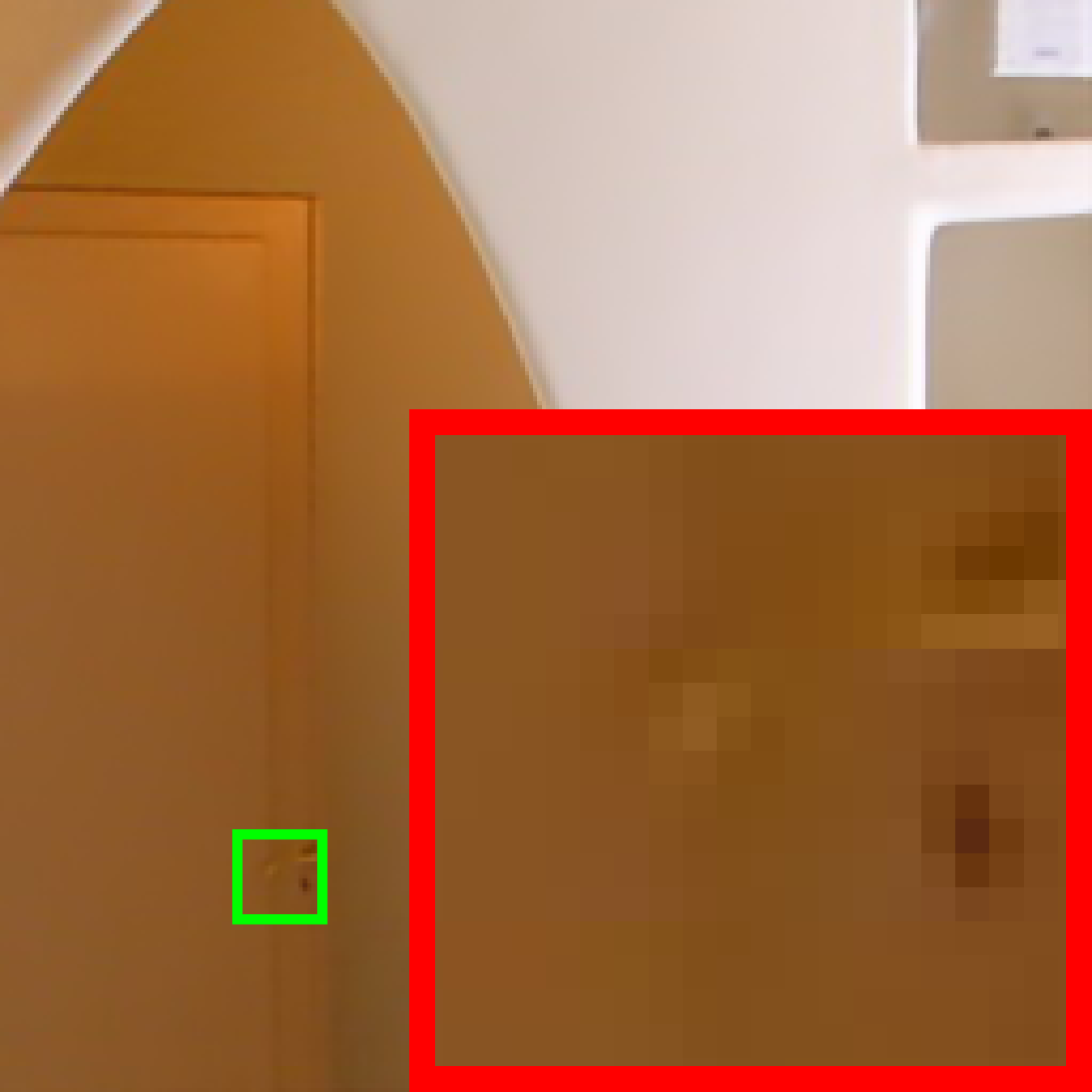}
    &\includegraphics[width=0.08\textwidth]{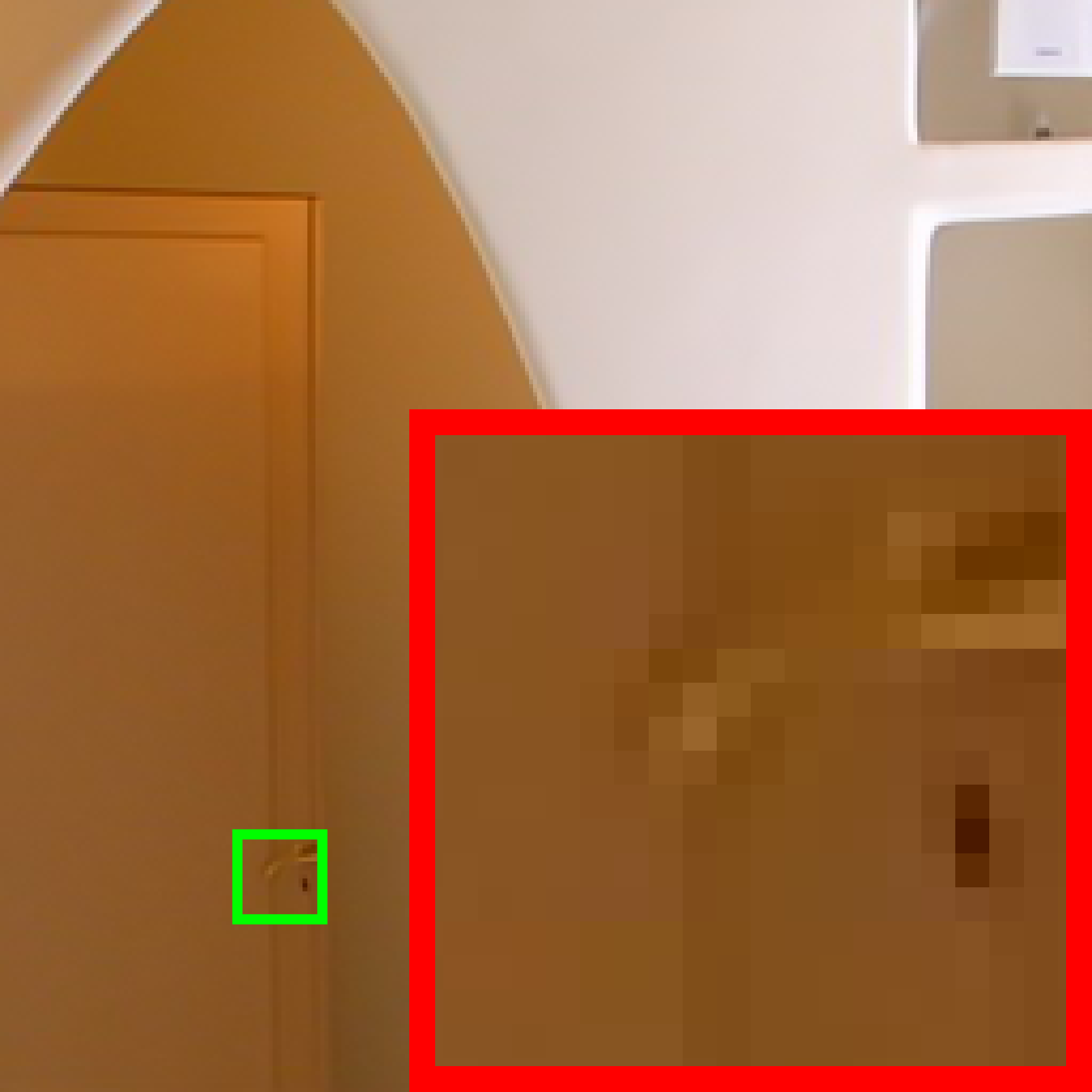}
    &\includegraphics[width=0.08\textwidth]{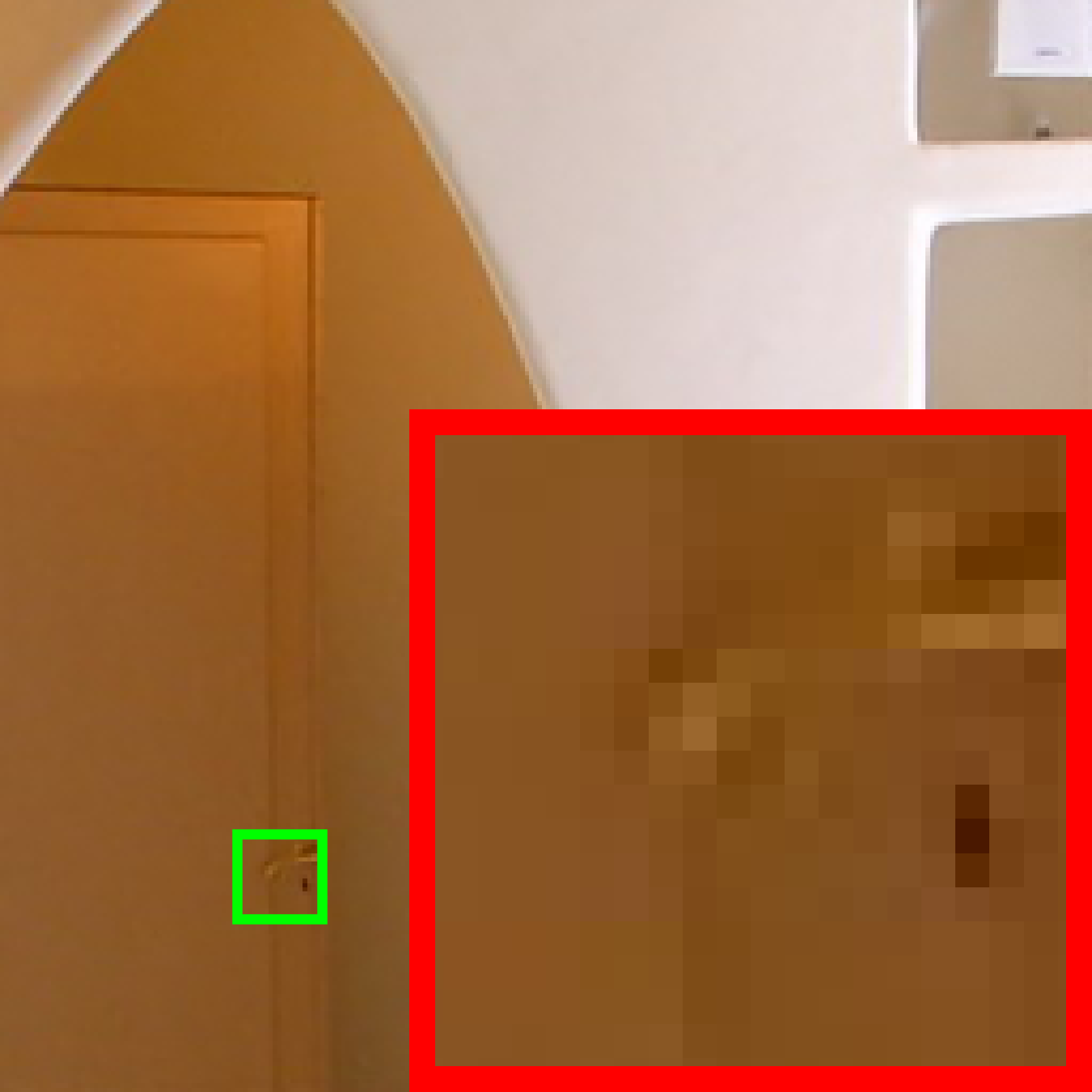}
    &\includegraphics[width=0.08\textwidth]{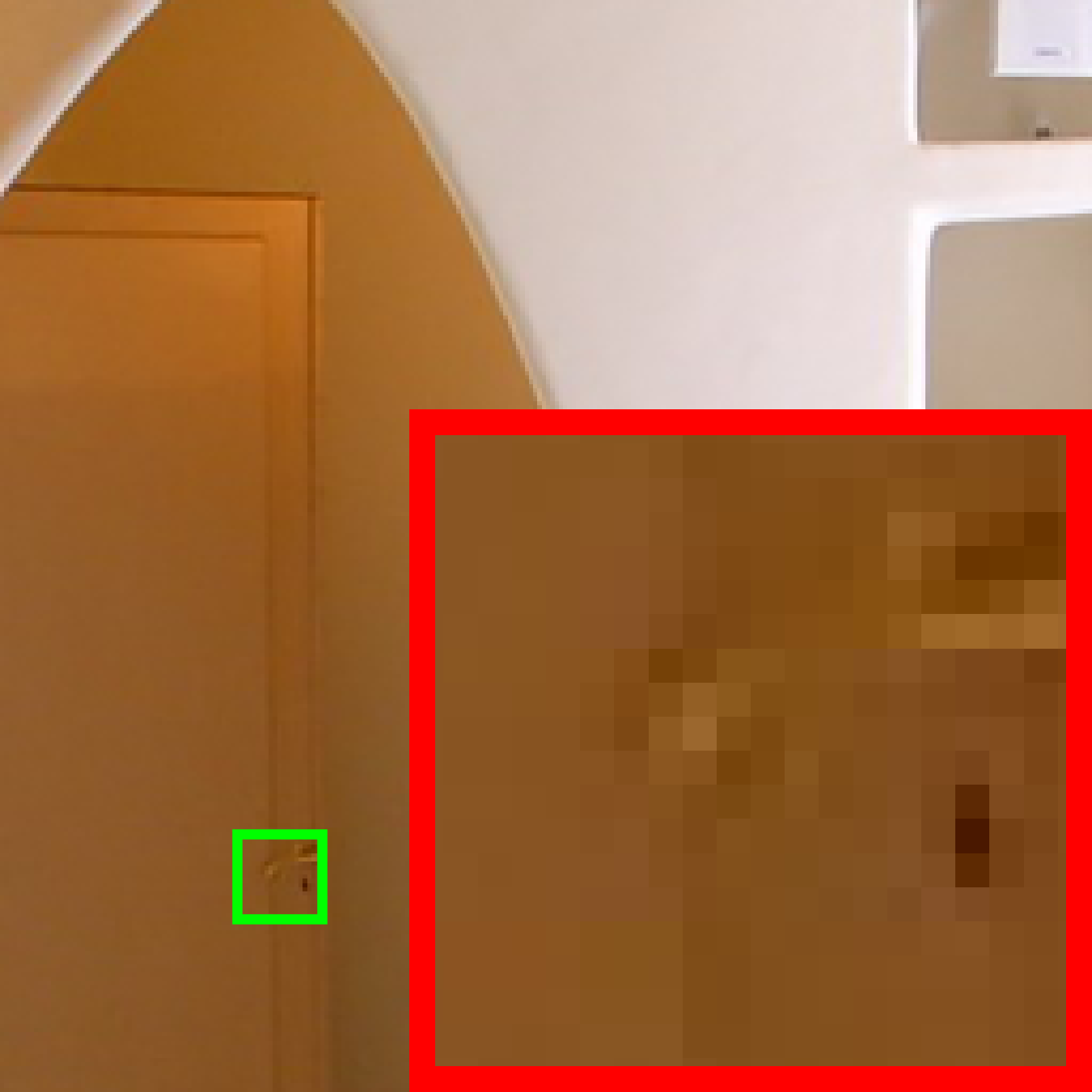}
    &\includegraphics[width=0.08\textwidth]{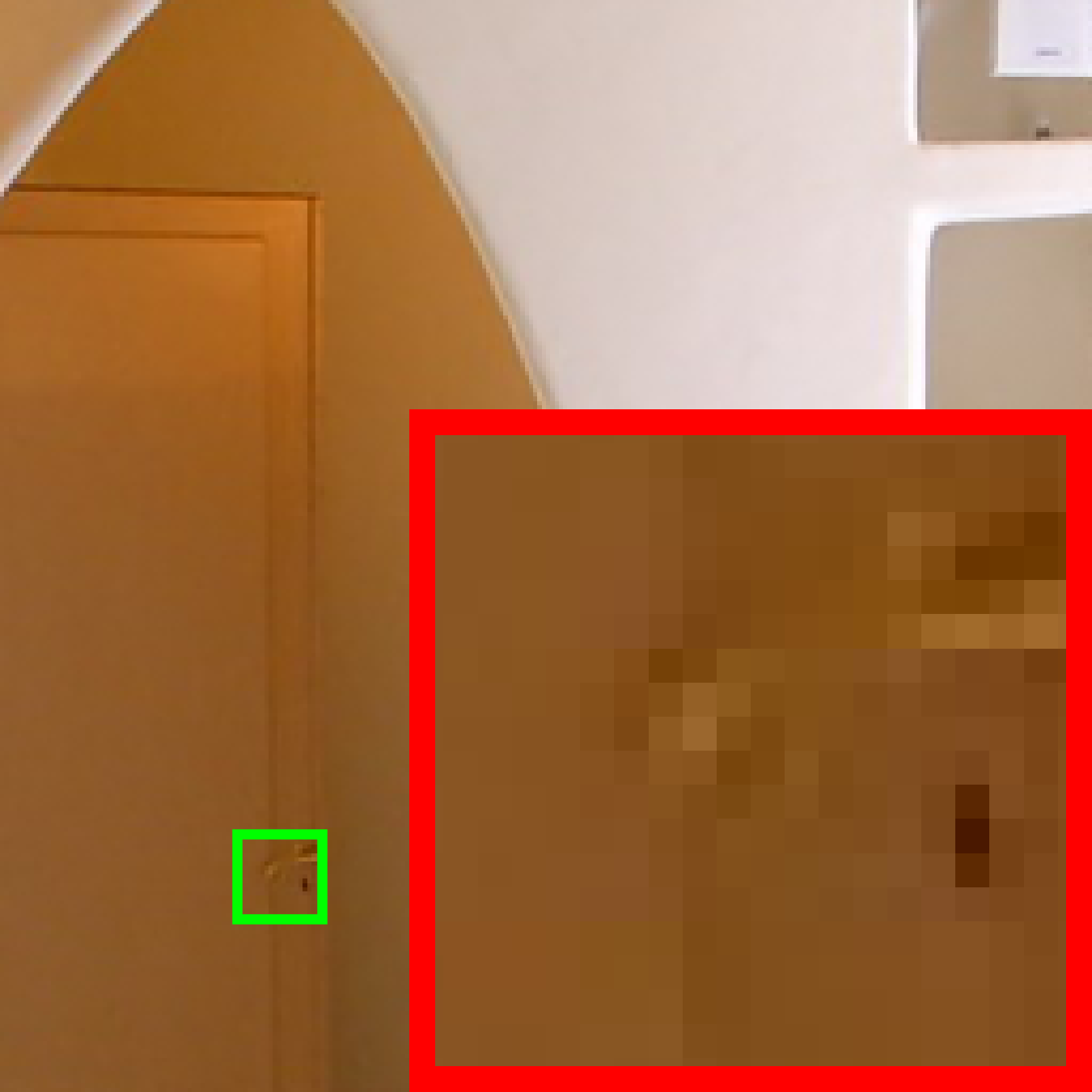}
    &\includegraphics[width=0.08\textwidth]{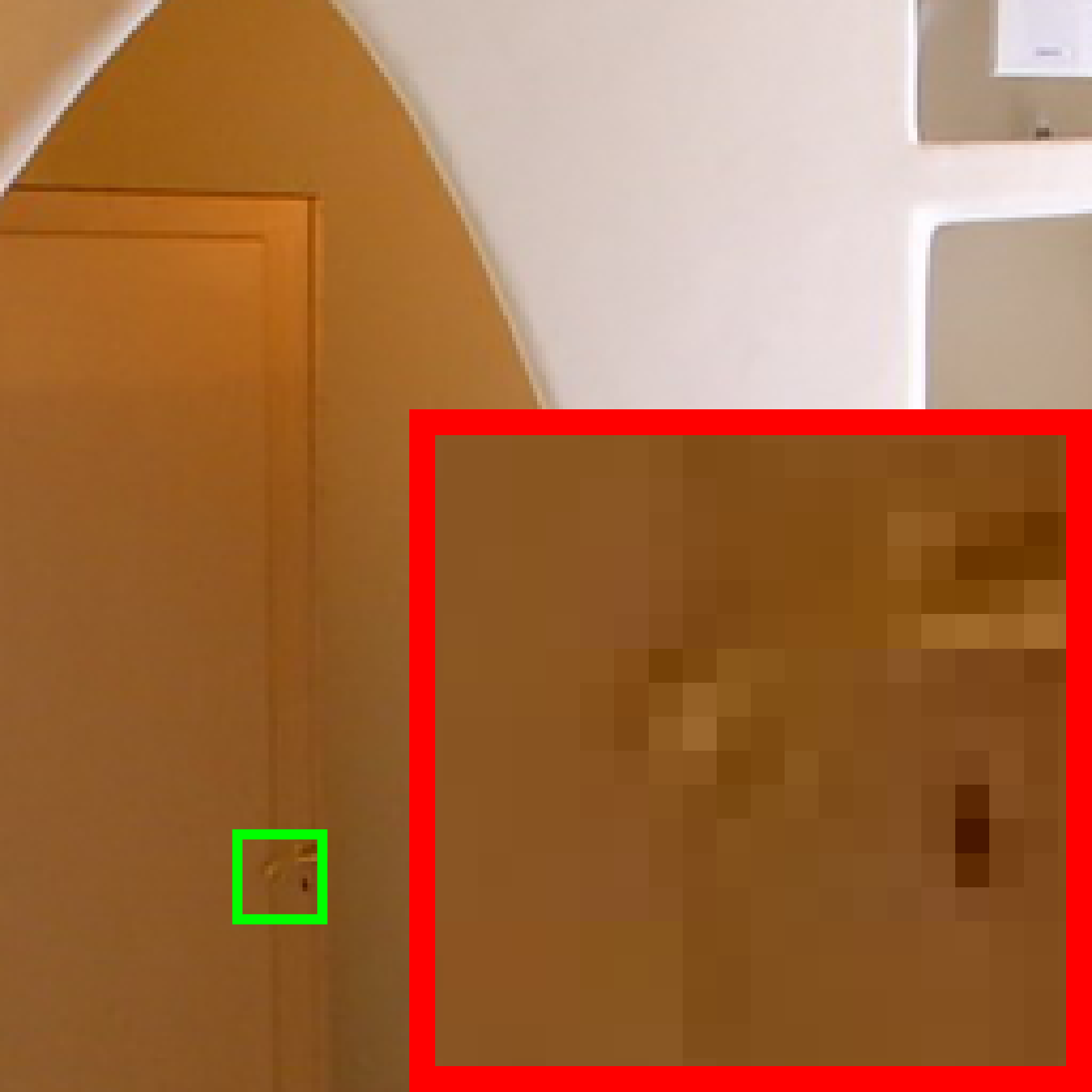}\\
    PSNR/SSIM & 37.15/0.94 & 47.23/\underline{\textcolor{blue}{0.98}} & 47.51/\textbf{\textcolor{red}{0.99}} & 47.85/\textbf{\textcolor{red}{0.99}} & 44.53/\underline{\textcolor{blue}{0.98}} & 46.80/\textbf{\textcolor{red}{0.99}} & 47.32/\textbf{\textcolor{red}{0.99}} & 45.40/\textbf{\textcolor{red}{0.99}} & 47.40/\textbf{\textcolor{red}{0.99}} & 48.94/\textbf{\textcolor{red}{0.99}} & \textbf{\textcolor{red}{49.56}}/\textbf{\textcolor{red}{0.99}} & 49.02/\textbf{\textcolor{red}{0.99}} & \underline{\textcolor{blue}{49.54}}/\textbf{\textcolor{red}{0.99}} \\
    \includegraphics[width=0.08\textwidth]{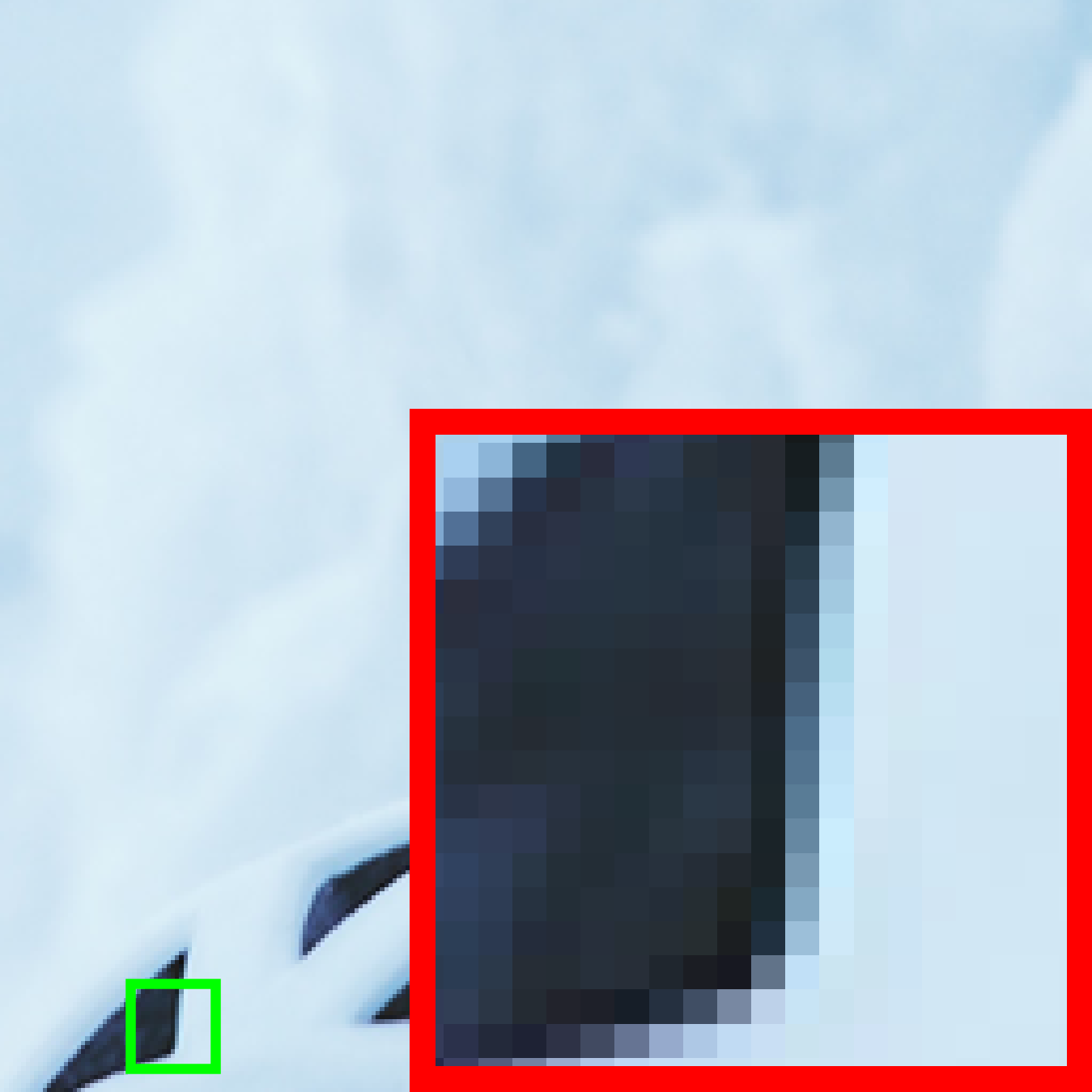}
    &\includegraphics[width=0.08\textwidth]{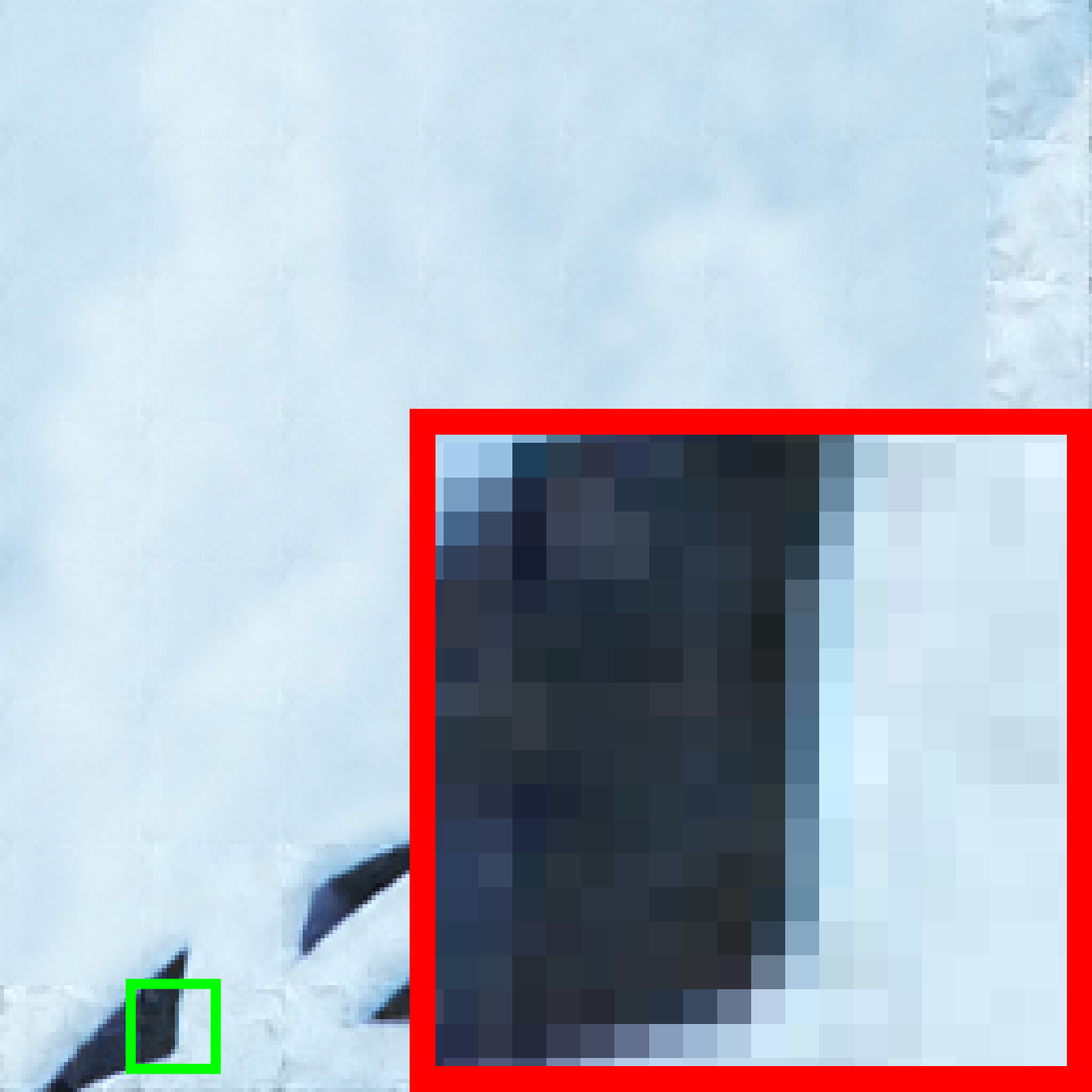}
    &\includegraphics[width=0.08\textwidth]{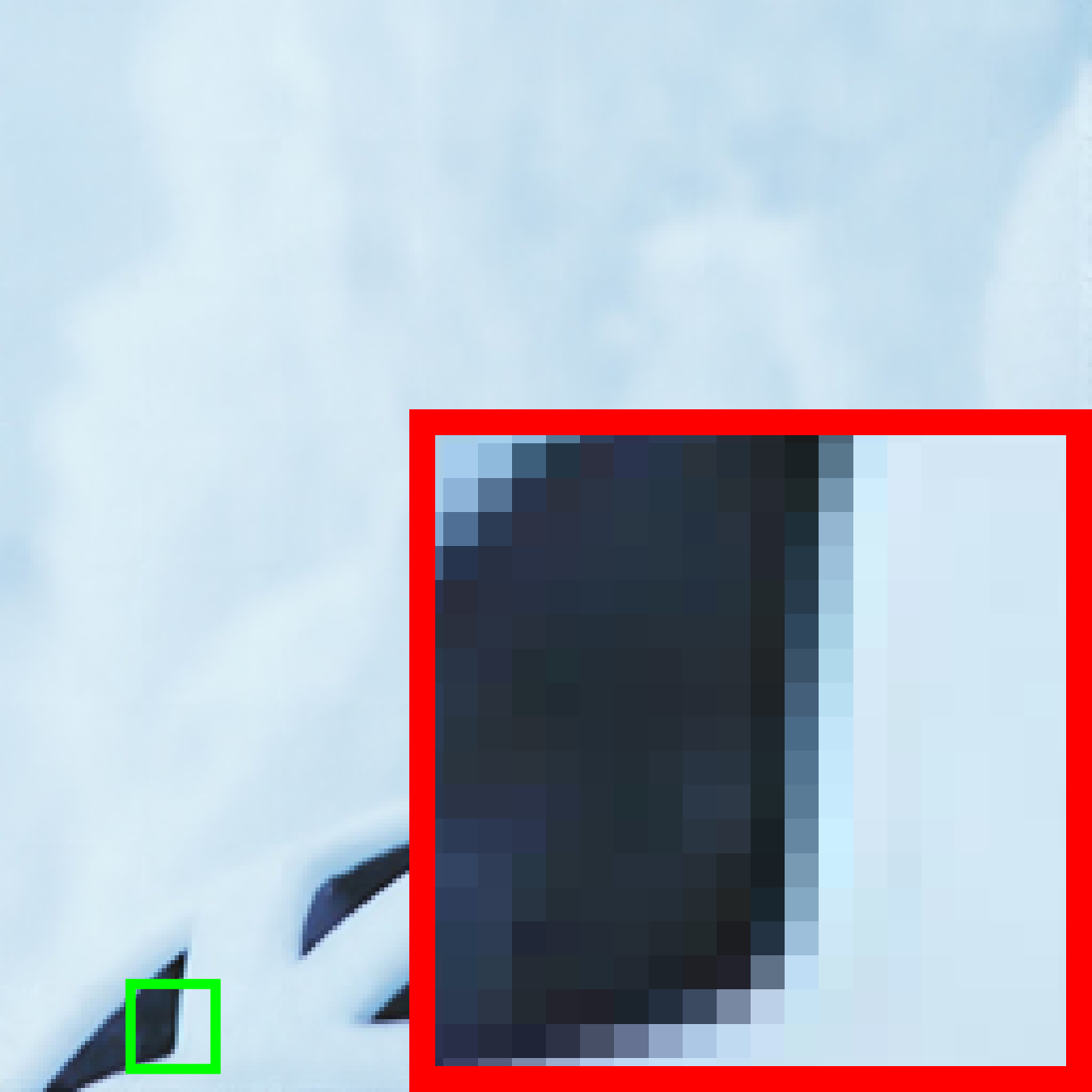}
    &\includegraphics[width=0.08\textwidth]{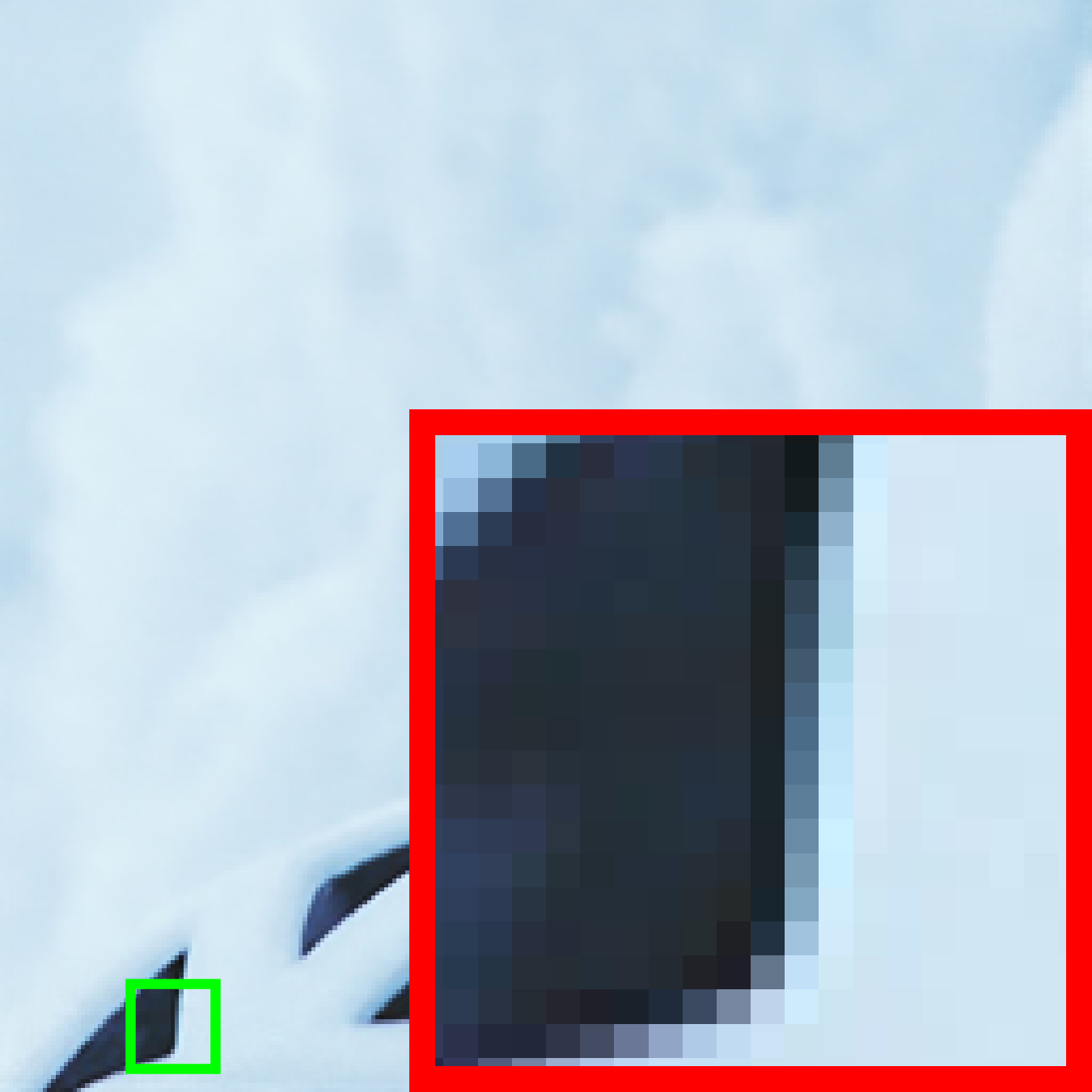}
    &\includegraphics[width=0.08\textwidth]{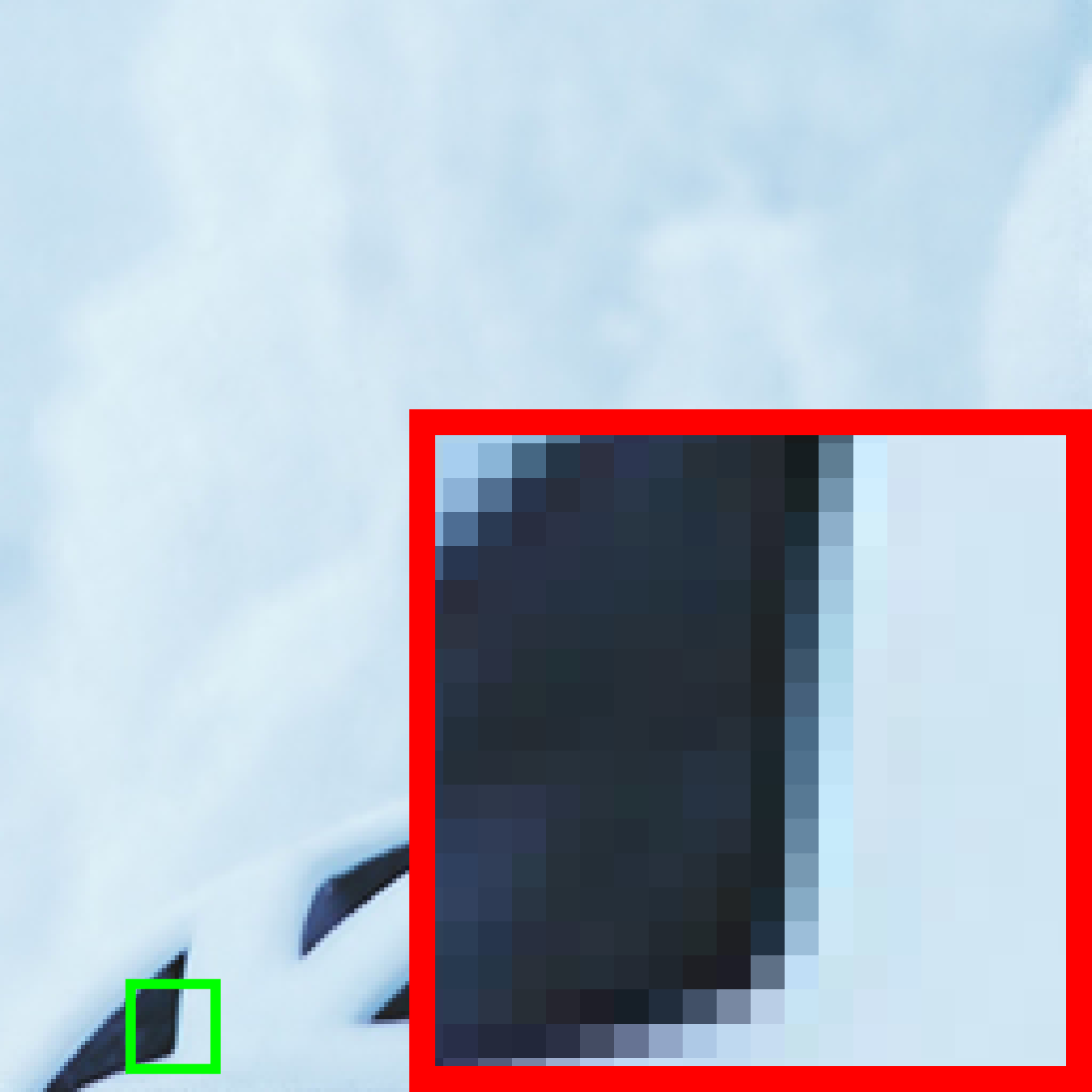}
    &\includegraphics[width=0.08\textwidth]{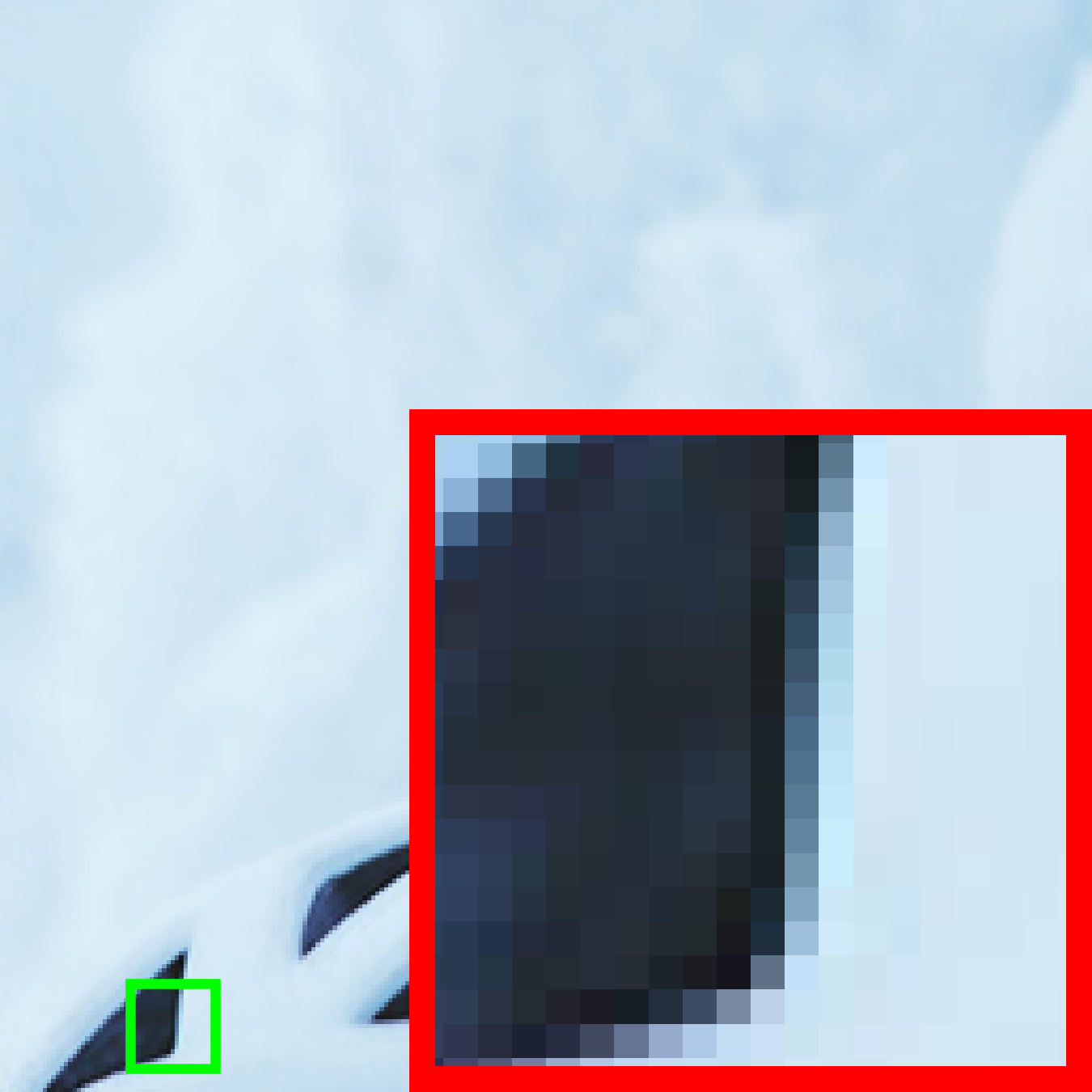}
    &\includegraphics[width=0.08\textwidth]{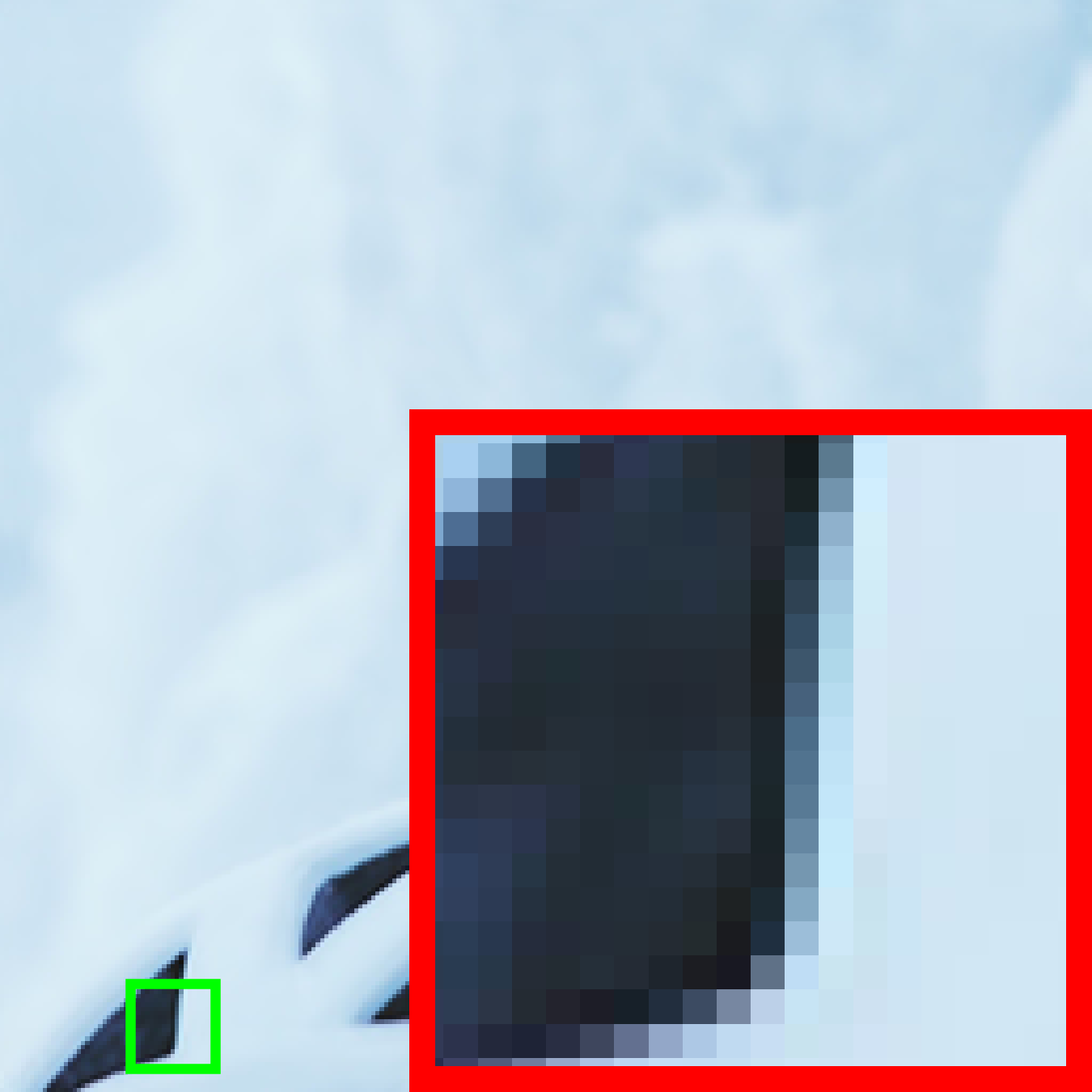}
    &\includegraphics[width=0.08\textwidth]{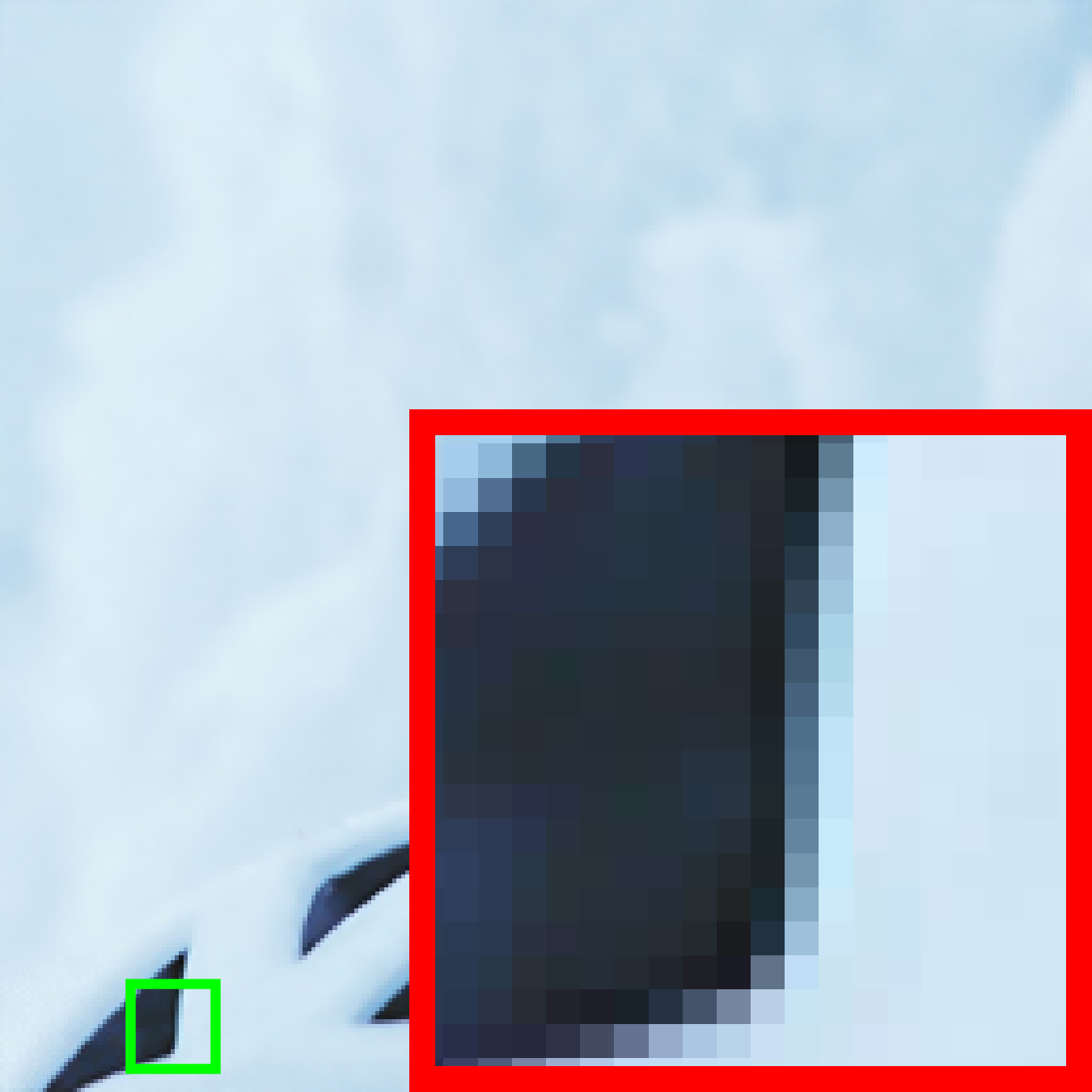}
    &\includegraphics[width=0.08\textwidth]{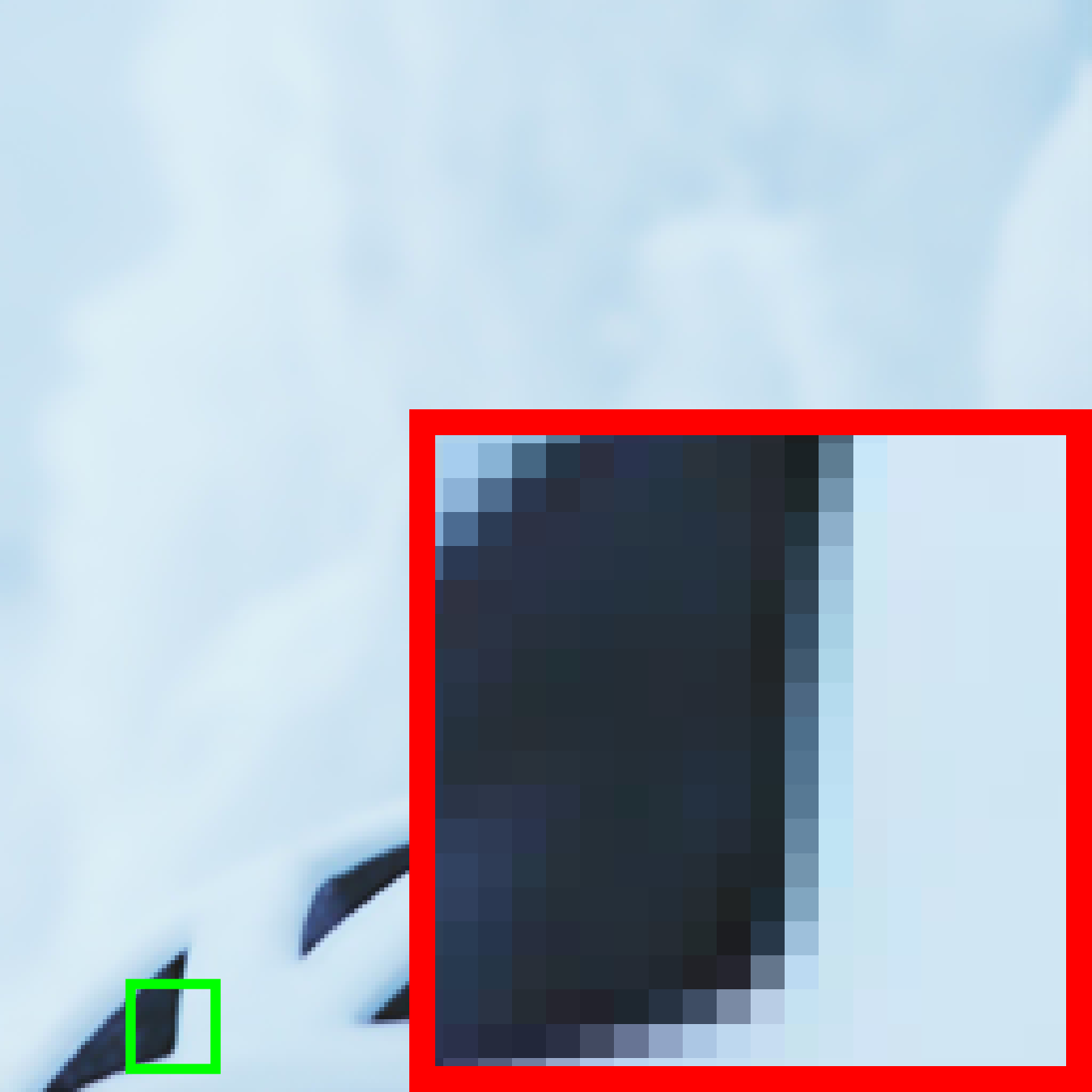}
    &\includegraphics[width=0.08\textwidth]{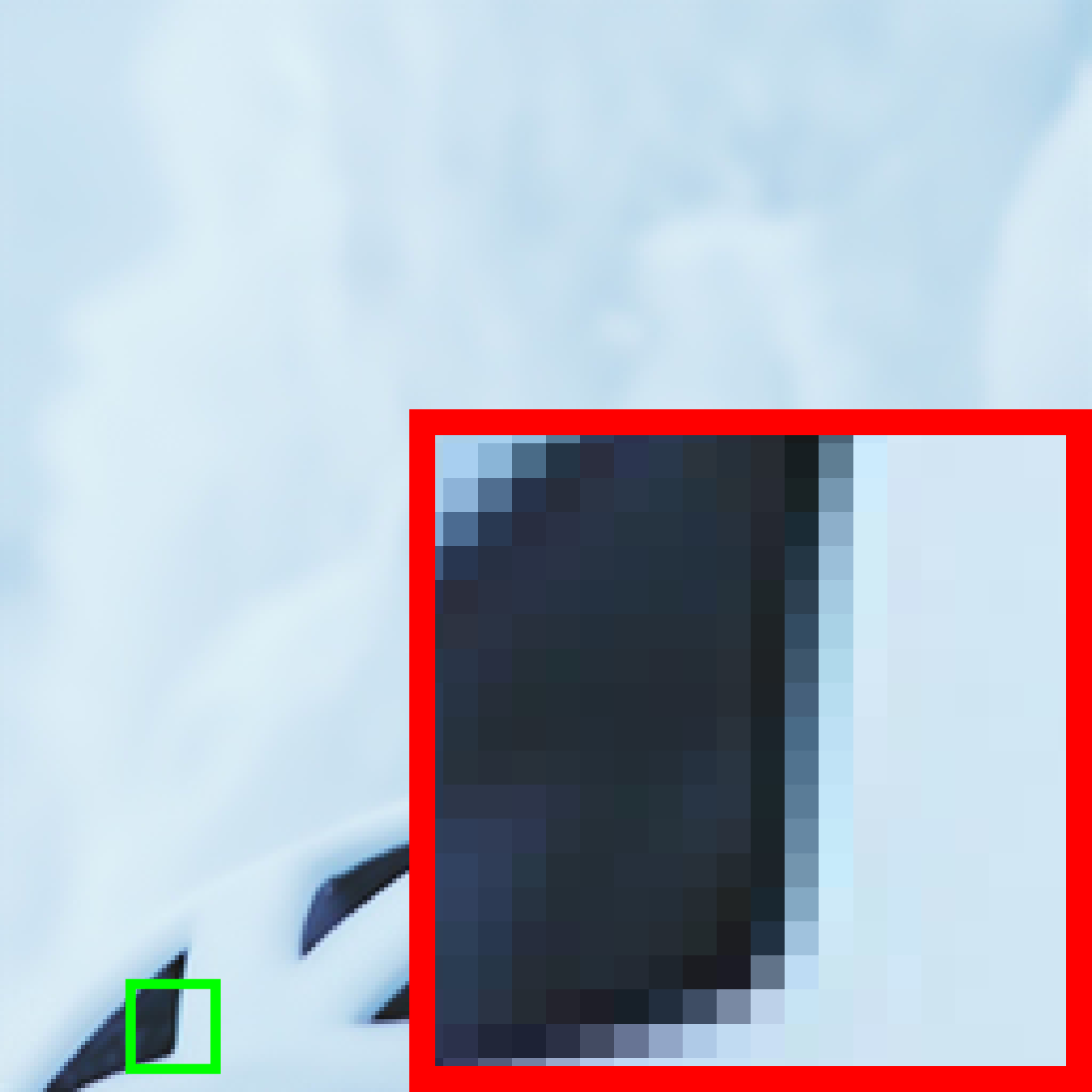}
    &\includegraphics[width=0.08\textwidth]{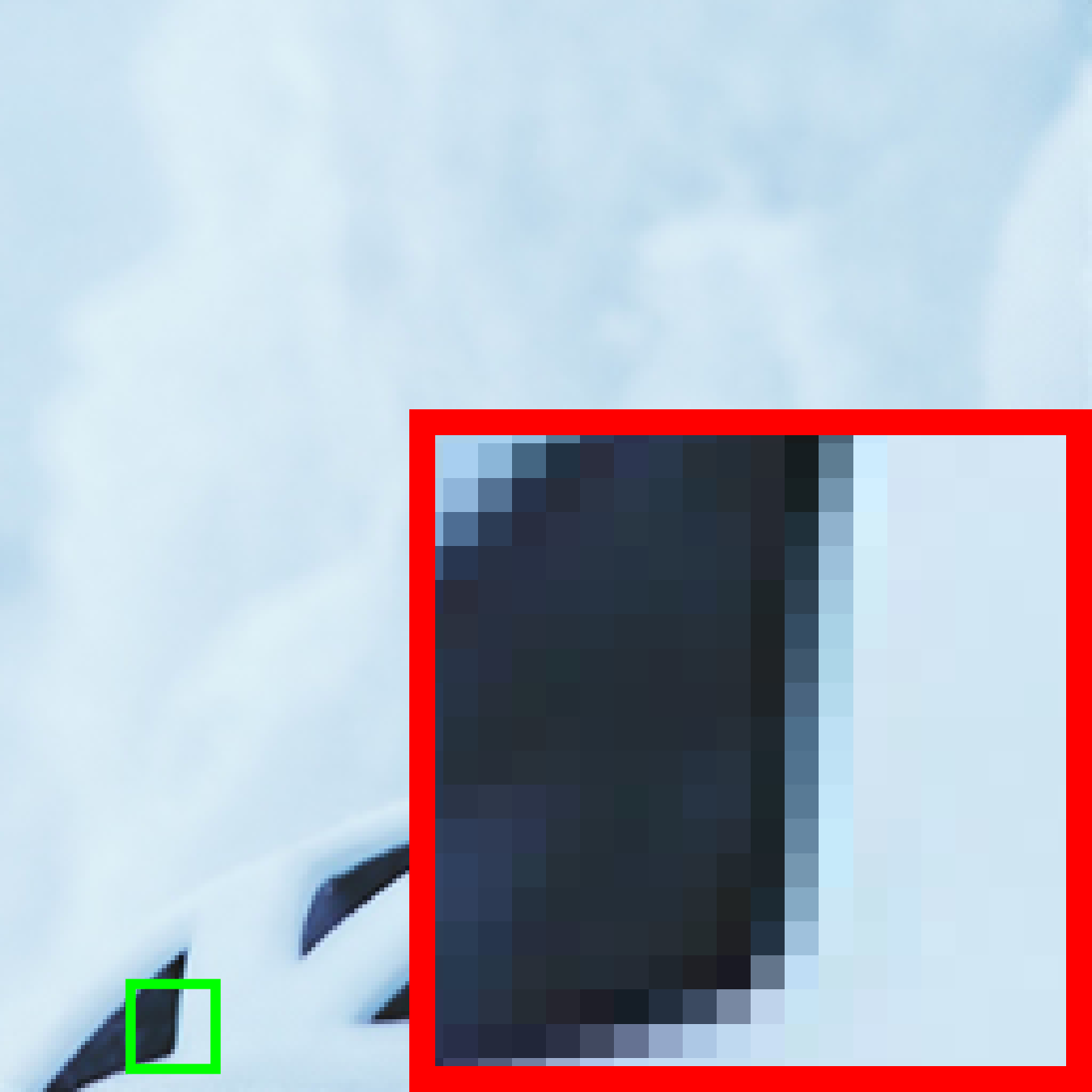}
    &\includegraphics[width=0.08\textwidth]{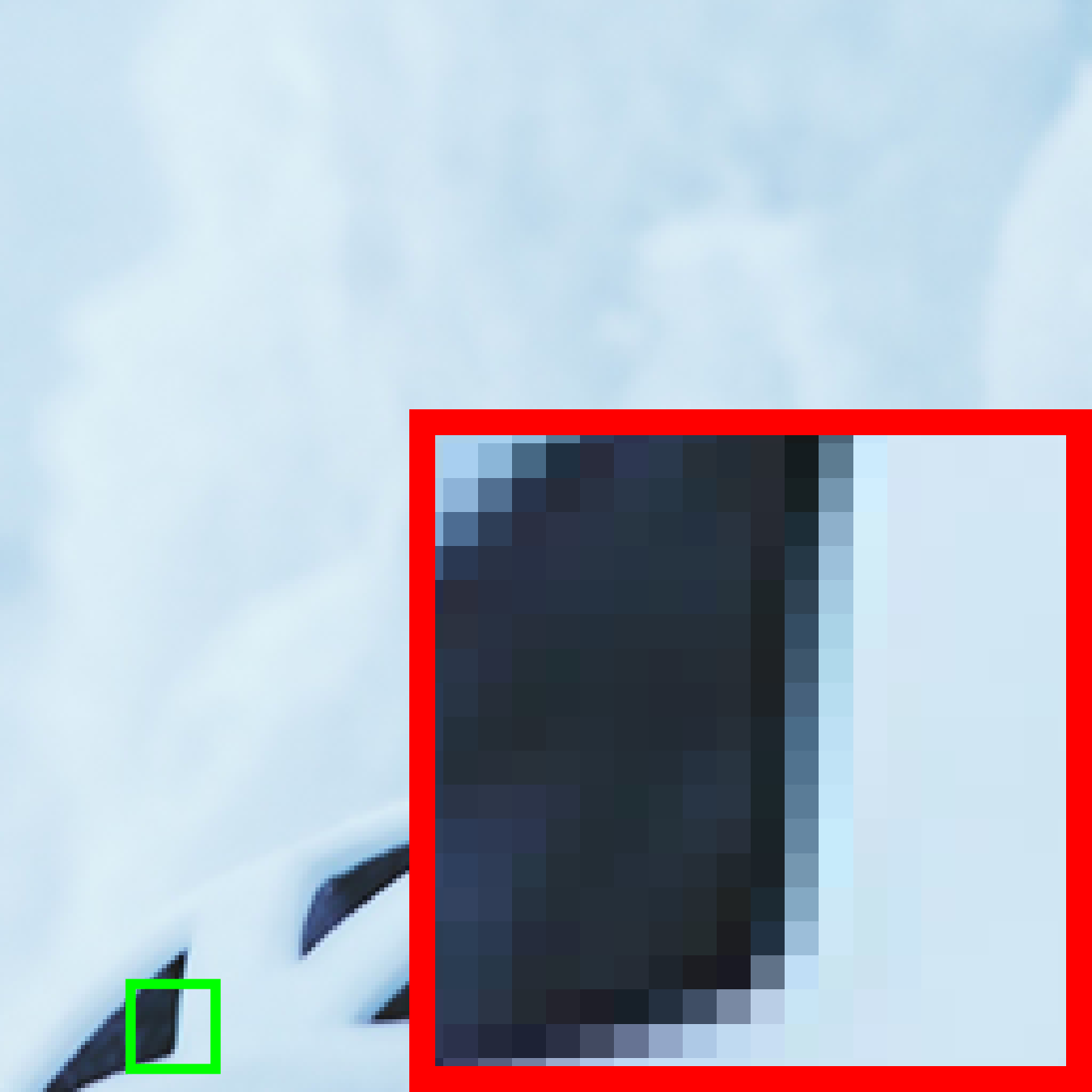}
    &\includegraphics[width=0.08\textwidth]{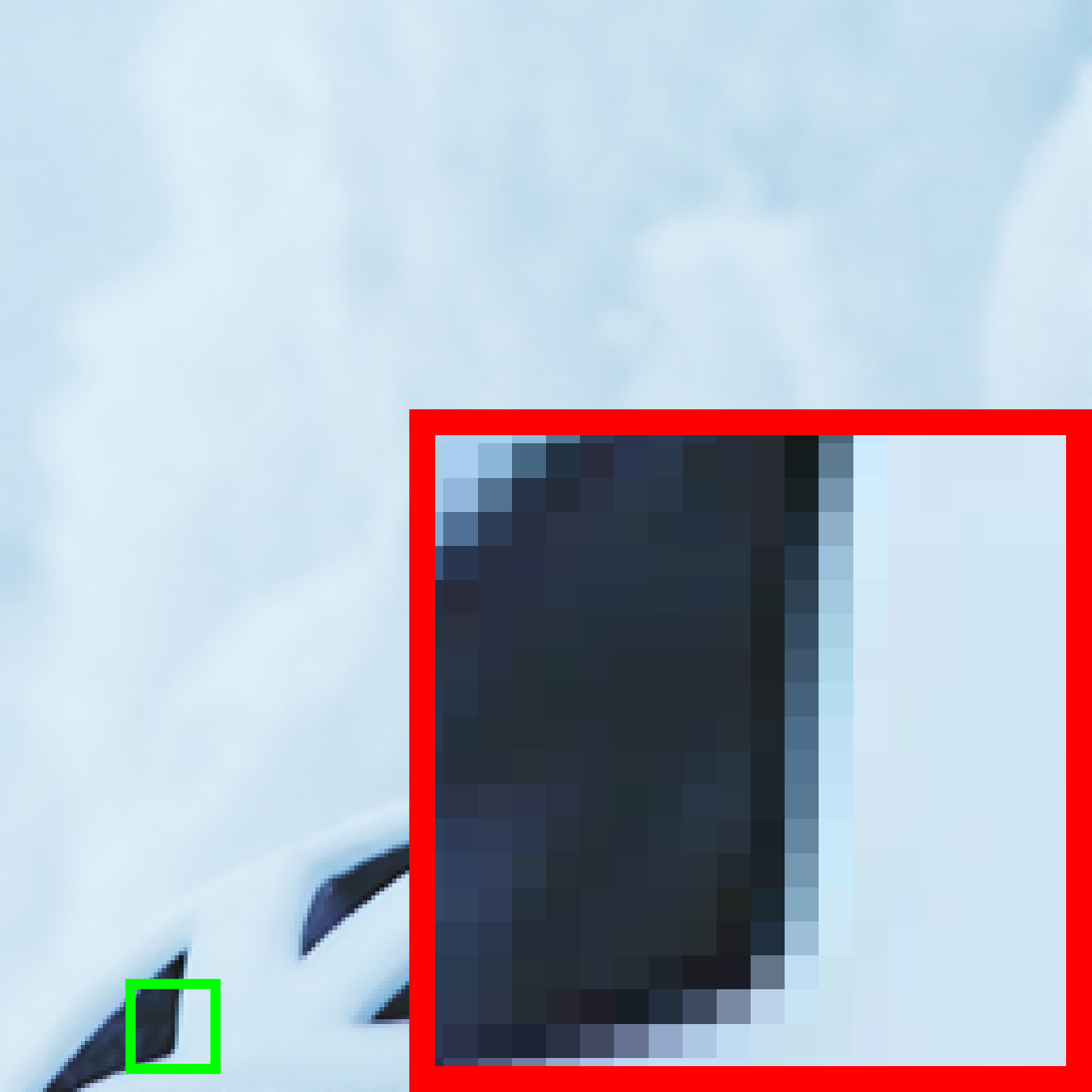}
    &\includegraphics[width=0.08\textwidth]{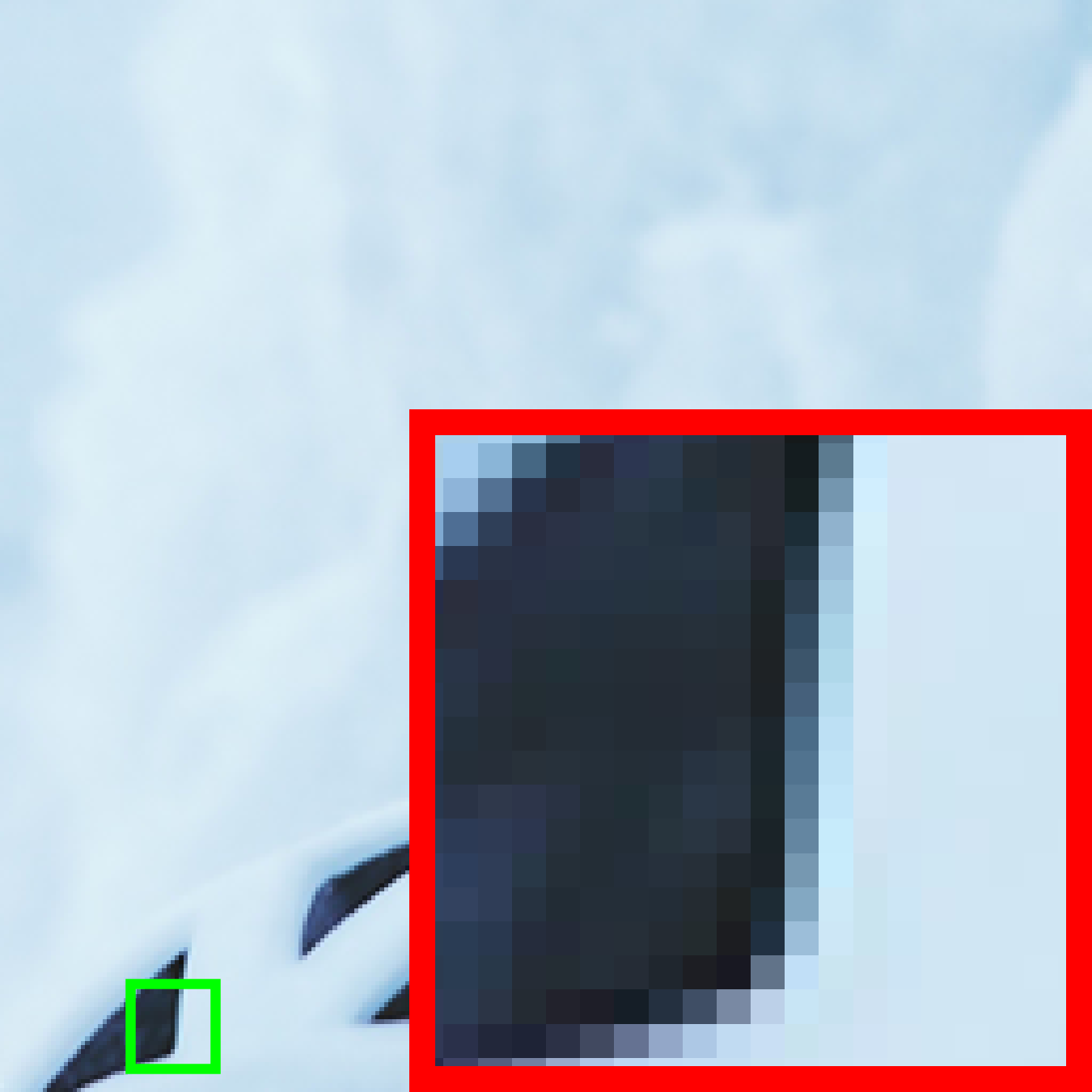}\\
    PSNR/SSIM & 35.91/0.93 & 48.02/\underline{\textcolor{blue}{0.99}} & 48.59/\underline{\textcolor{blue}{0.99}} & 49.39/\underline{\textcolor{blue}{0.99}} & 49.04/\underline{\textcolor{blue}{0.99}} & 48.56/\textbf{\textcolor{red}{1.00}} & 49.93/\textbf{\textcolor{red}{1.00}} & 47.54/\underline{\textcolor{blue}{0.99}} & 49.75/\underline{\textcolor{blue}{0.99}} & 51.19/\textbf{\textcolor{red}{1.00}} & \textbf{\textcolor{red}{53.05}}/\textbf{\textcolor{red}{1.00}} & \underline{\textcolor{blue}{52.02}}/\textbf{\textcolor{red}{1.00}} & \underline{\textcolor{blue}{52.02}}/\textbf{\textcolor{red}{1.00}}
\end{tabular}}
\caption{Visual comparison of different methods on four natural benchmark images named ``Foreman'', ``test\_39'', ``img\_080'' and ``0844'' from Set11~\cite{kulkarni2016reconnet} \textcolor{blue}{(top)}, CBSD68~\cite{martin2001database} \textcolor{blue}{(upper middle)}, Urban100~\cite{huang2015single} \textcolor{blue}{(lower middle)}, and DIV2K~\cite{agustsson2017ntire} \textcolor{blue}{(bottom)}, respectively, with $\gamma =30\%$ and $\sigma =0$.}
\label{fig:comparison_standard_natural_images_r30}
\end{figure*}

\begin{figure*}[!t]
\setlength{\tabcolsep}{0.5pt}
\hspace{-4pt}
\resizebox{1.0\textwidth}{!}{
\tiny
\begin{tabular}{cccccccccccccc}
    GT & ReconNet & ISTA-Net$^\text{+}$ & ISTA-Net$^\text{++}$ & COAST & DIP & BCNN & EI & ASGLD & DDSSL & \textbf{SC-CNN} & \textbf{SC-CNN$^\text{+}$} & \textbf{SCT} & \textbf{SCT$^\text{+}$}\\
    \includegraphics[width=0.08\textwidth]{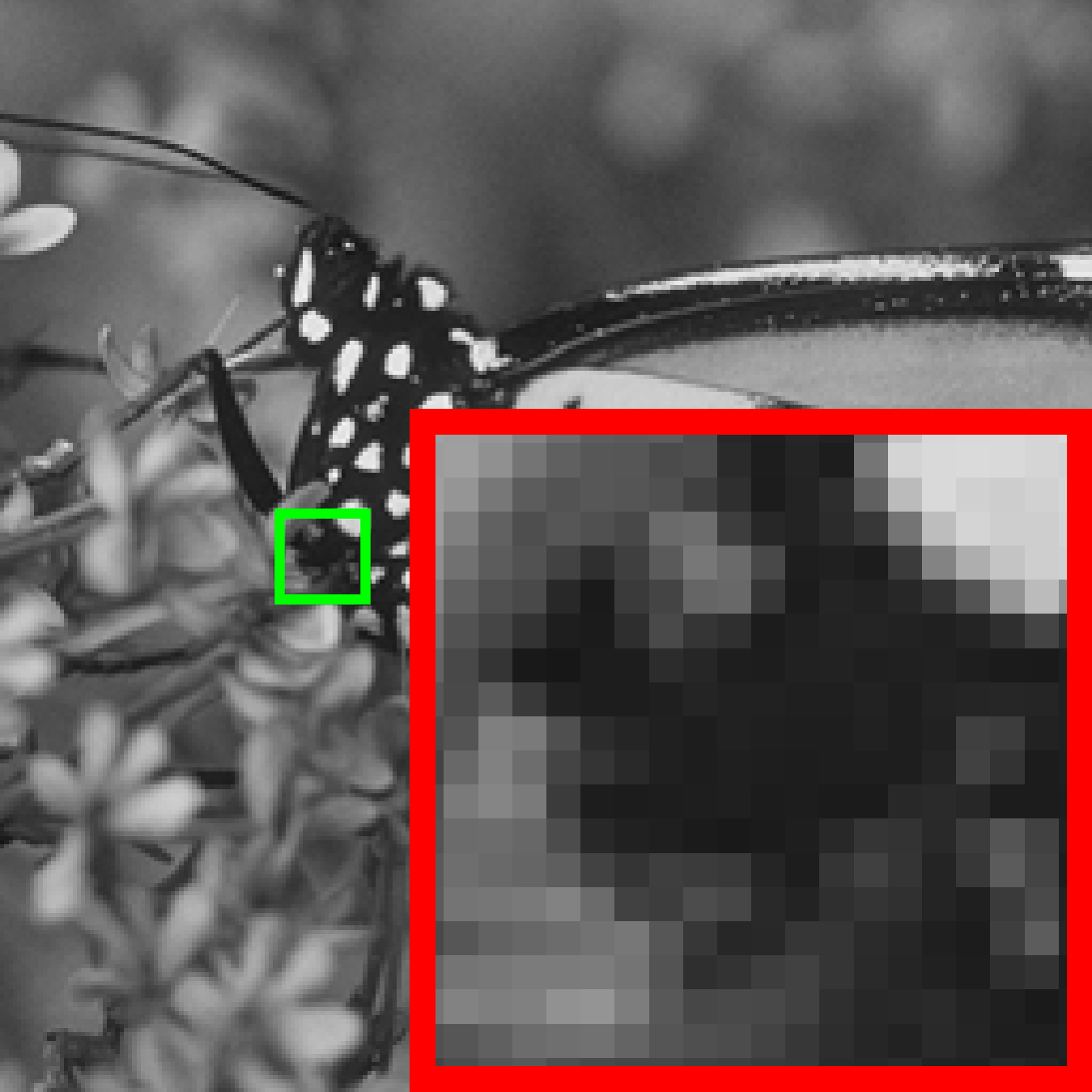}
    &\includegraphics[width=0.08\textwidth]{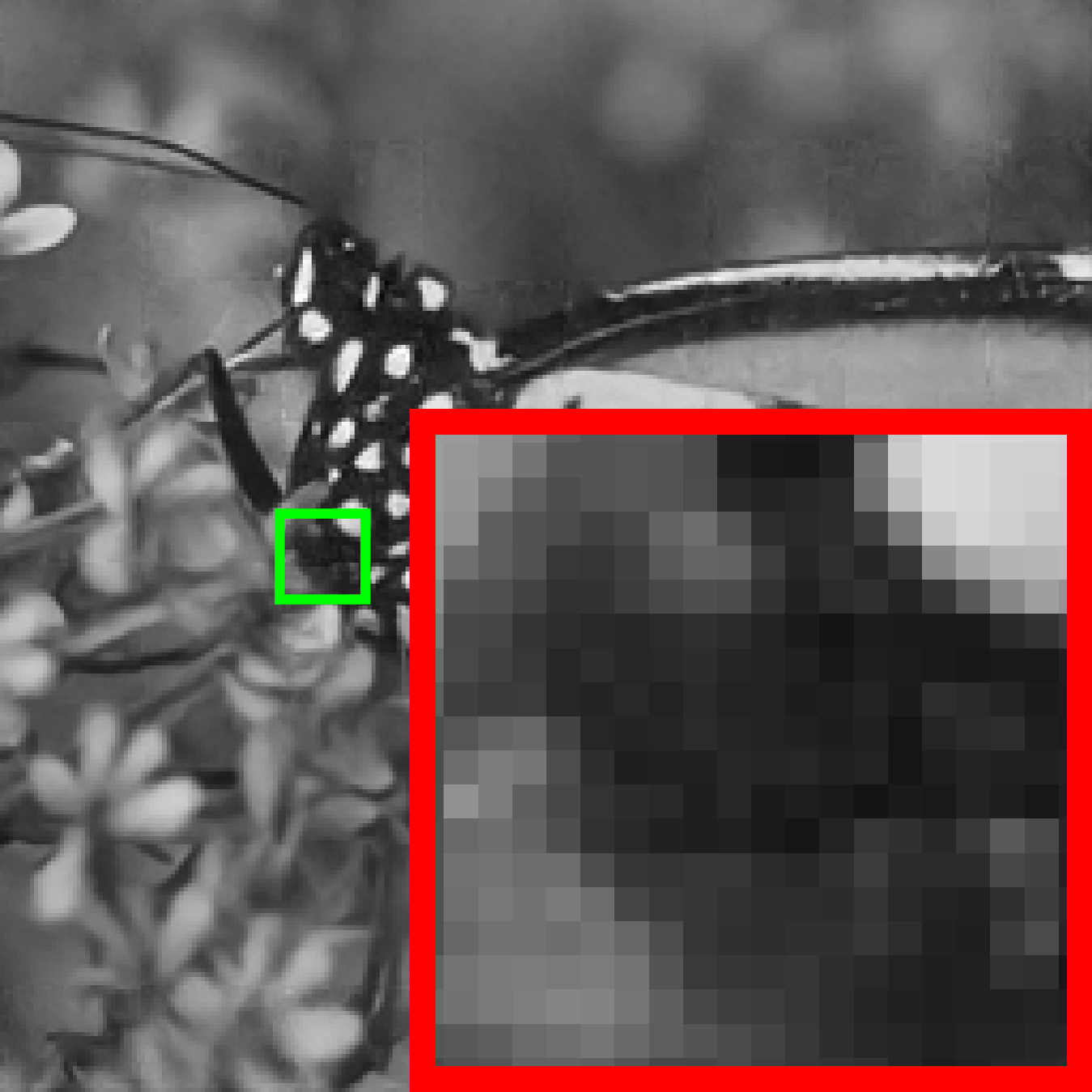}
    &\includegraphics[width=0.08\textwidth]{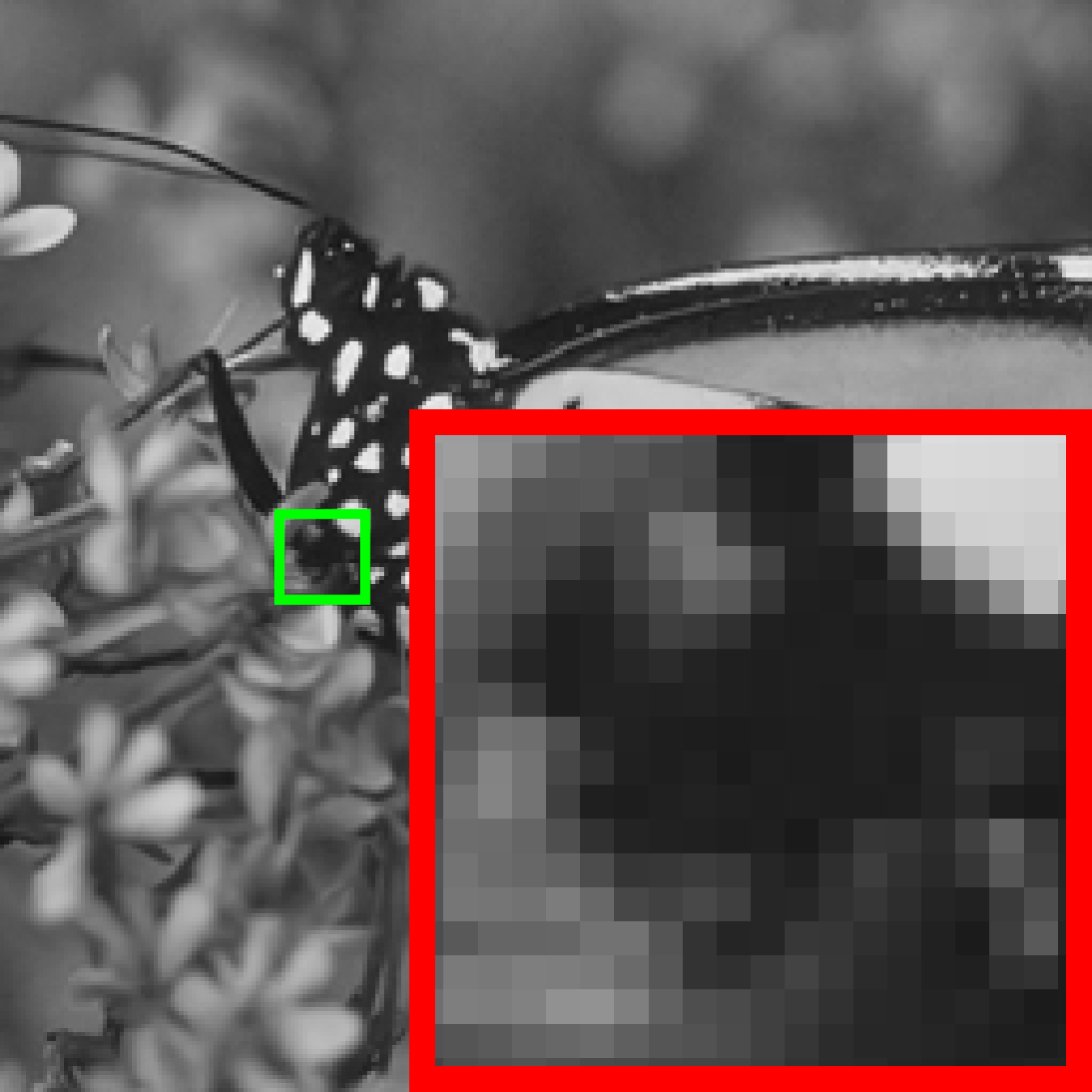}
    &\includegraphics[width=0.08\textwidth]{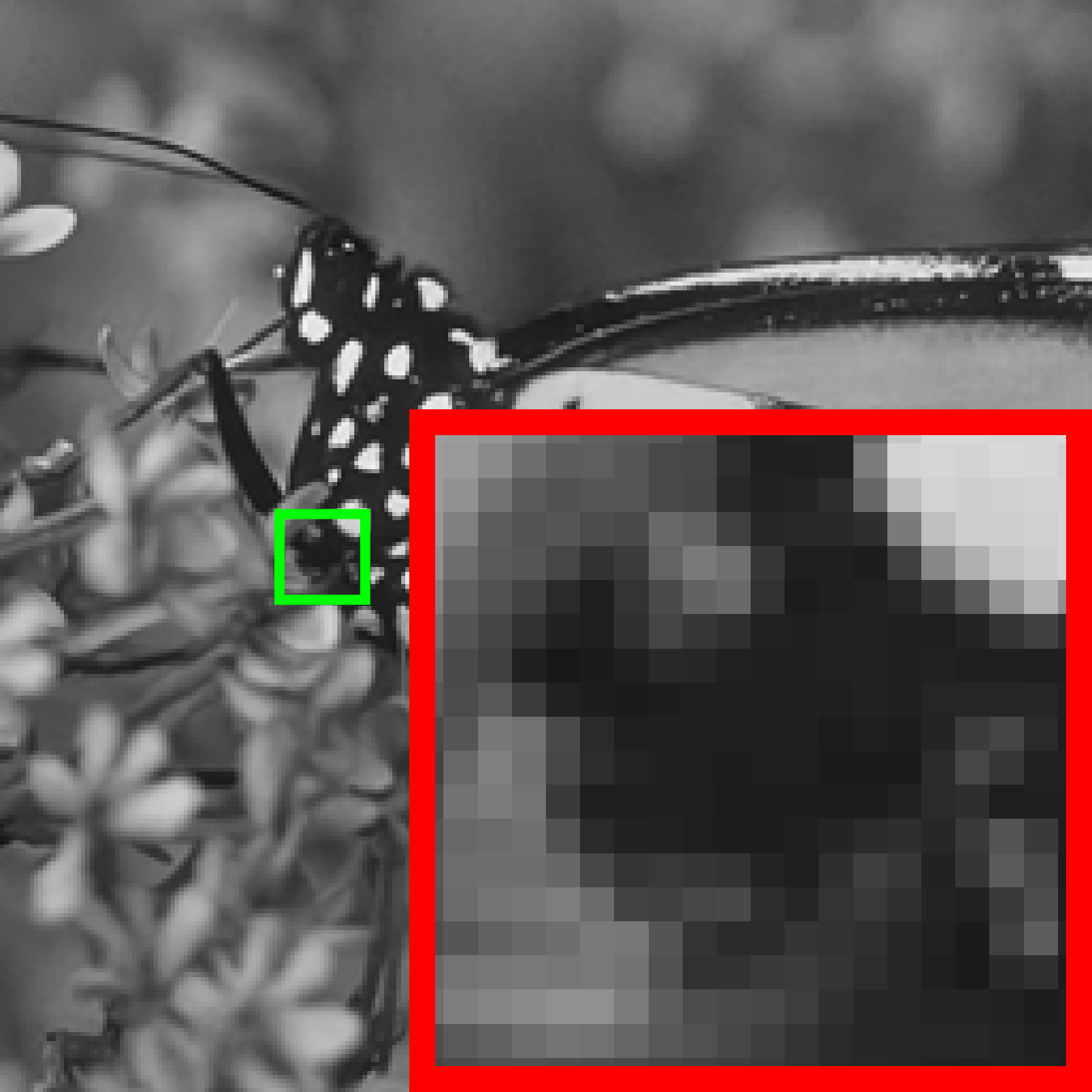}
    &\includegraphics[width=0.08\textwidth]{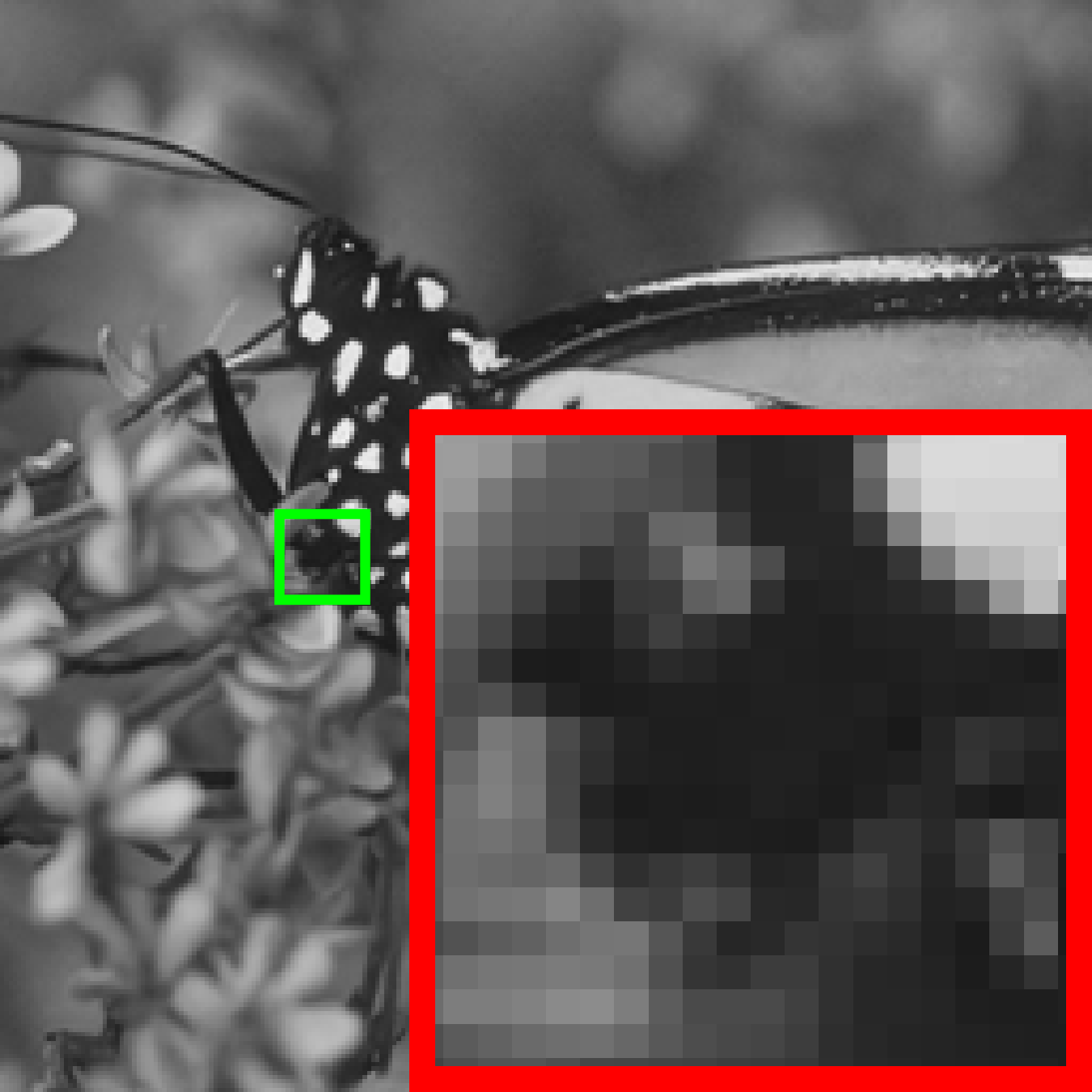}
    &\includegraphics[width=0.08\textwidth]{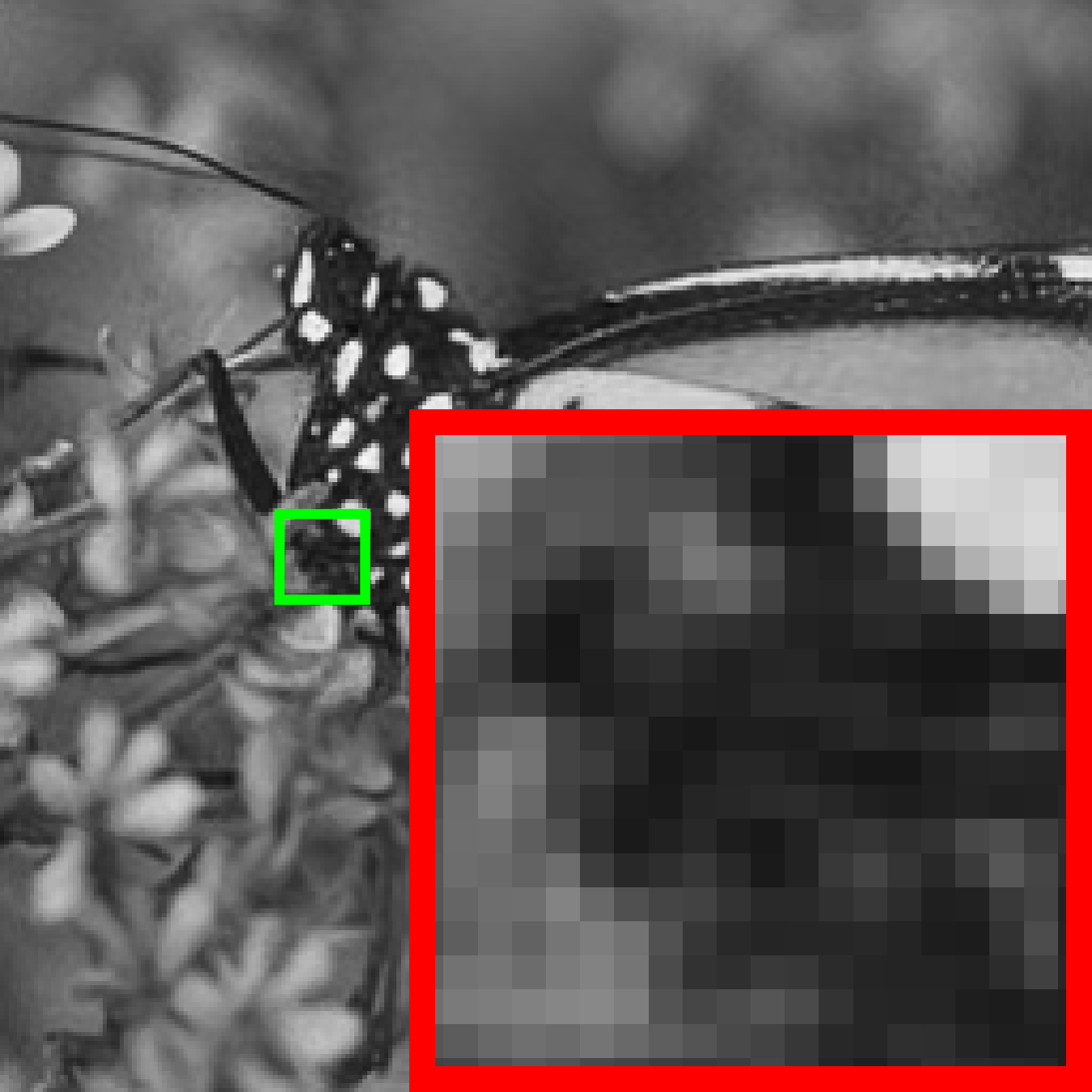}
    &\includegraphics[width=0.08\textwidth]{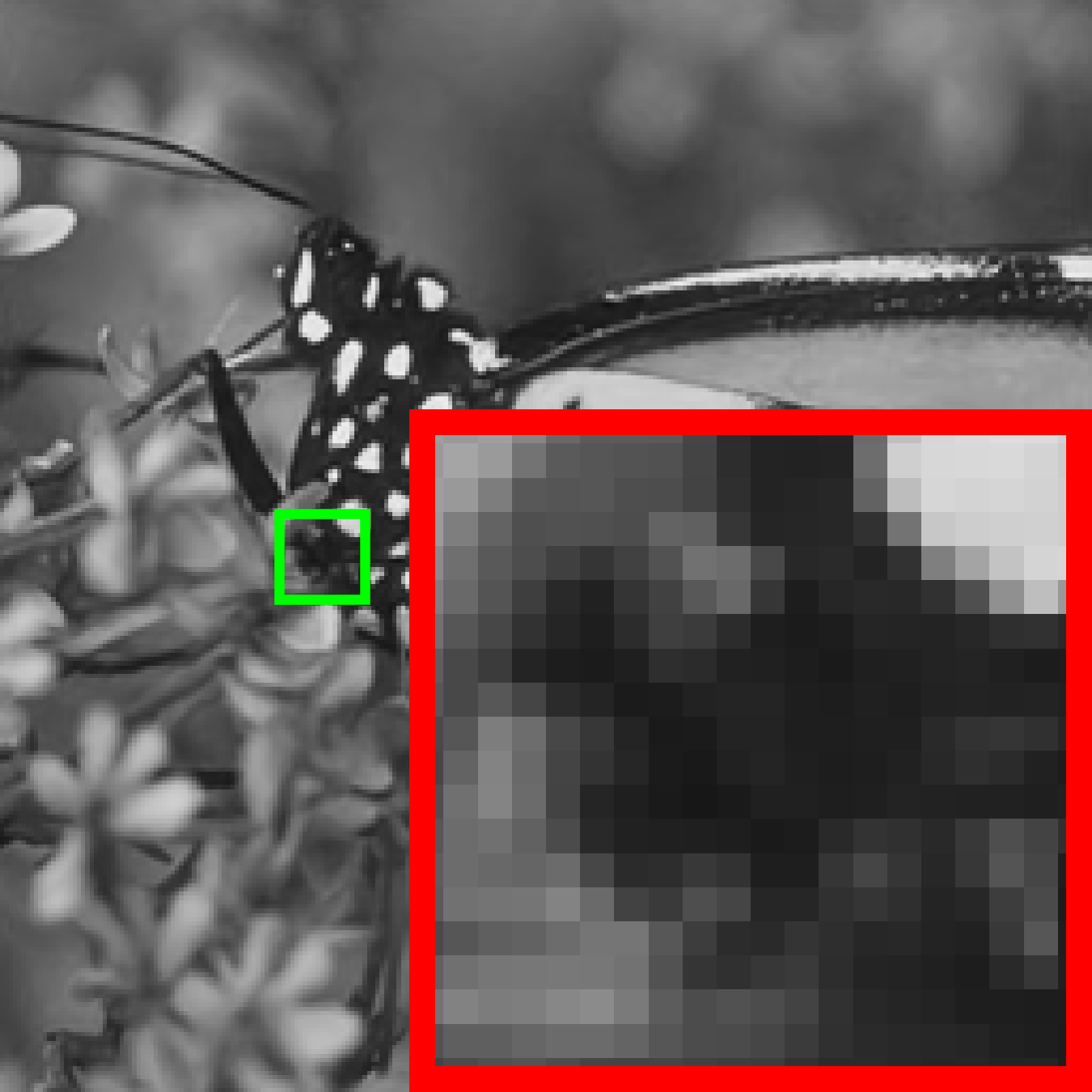}
    &\includegraphics[width=0.08\textwidth]{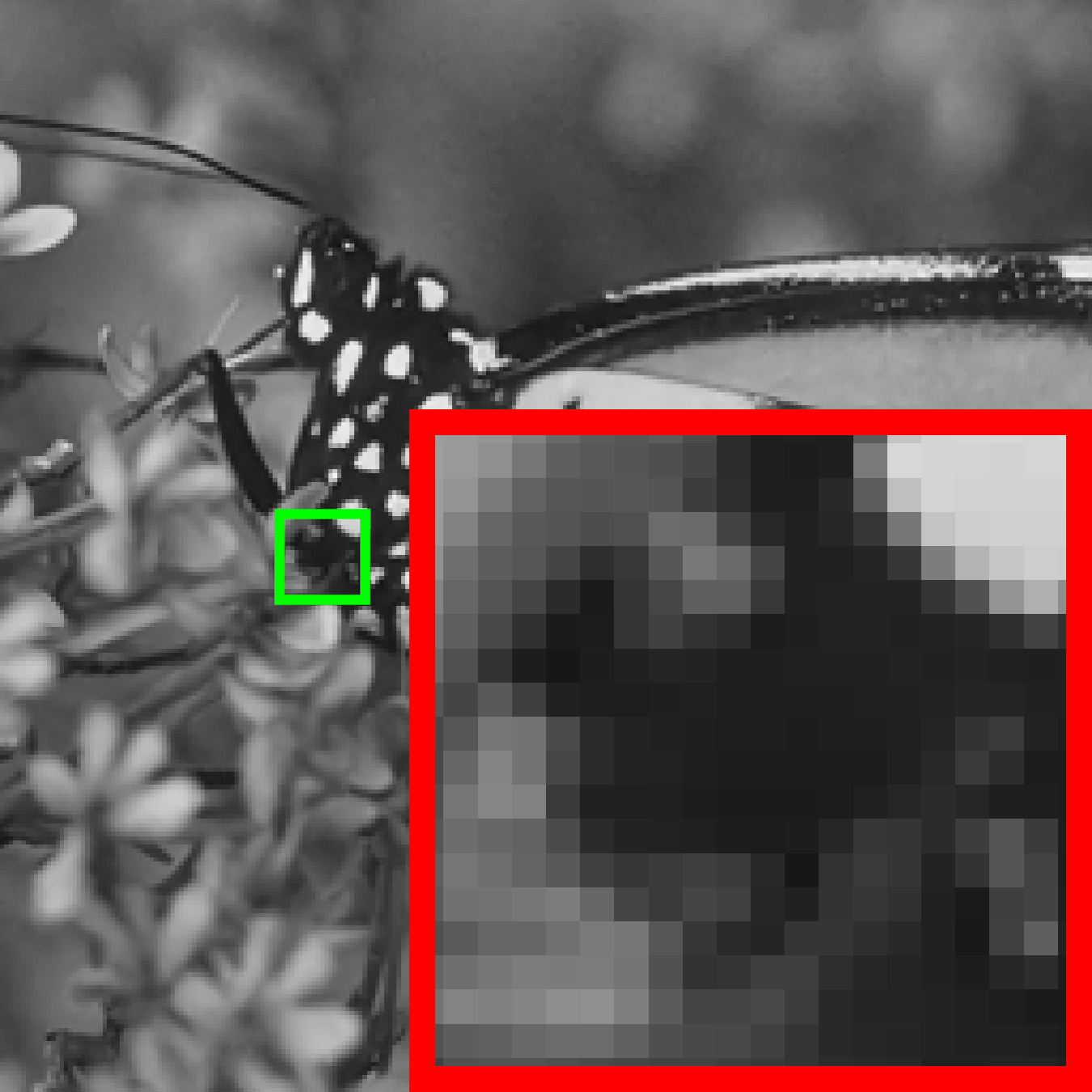}
    &\includegraphics[width=0.08\textwidth]{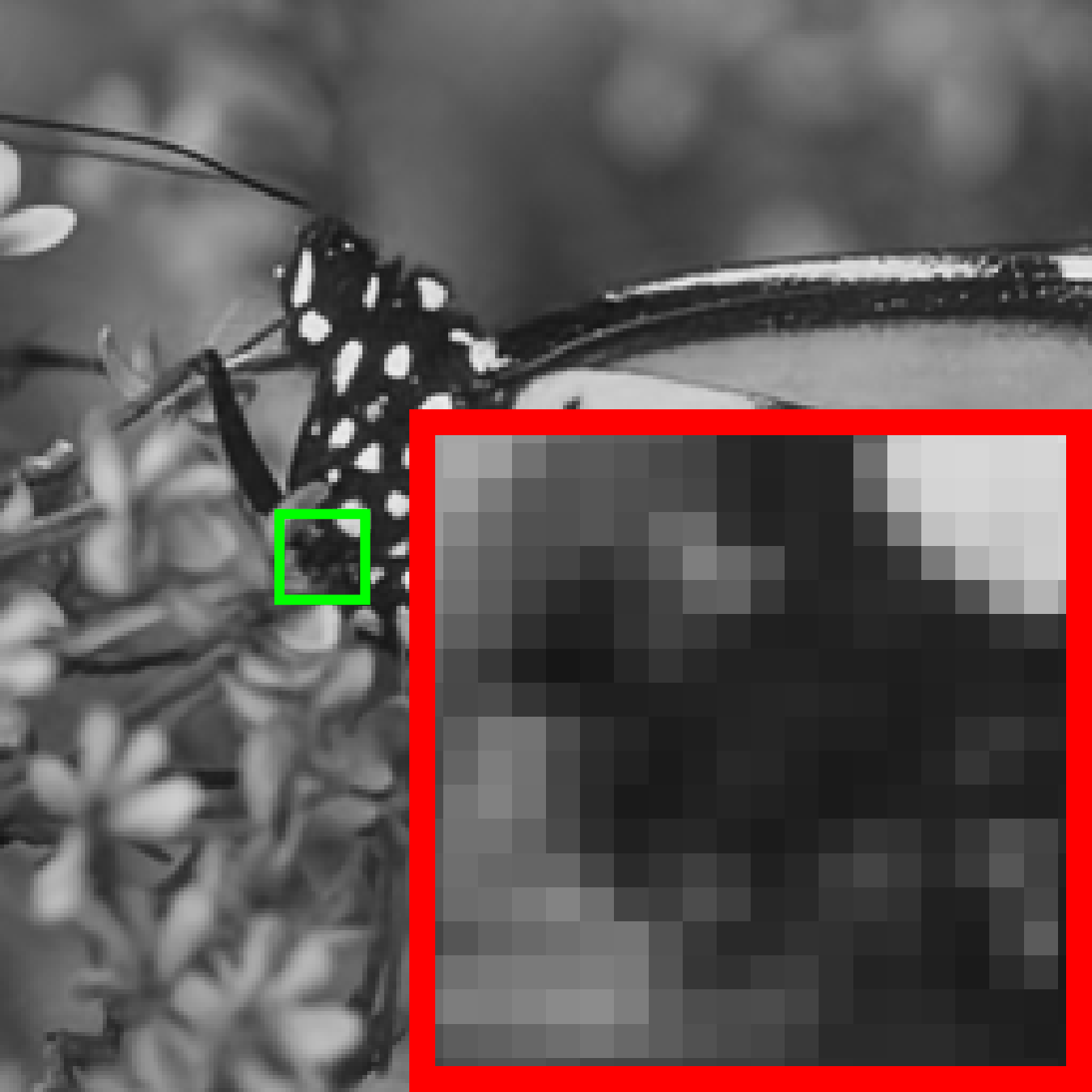}
    &\includegraphics[width=0.08\textwidth]{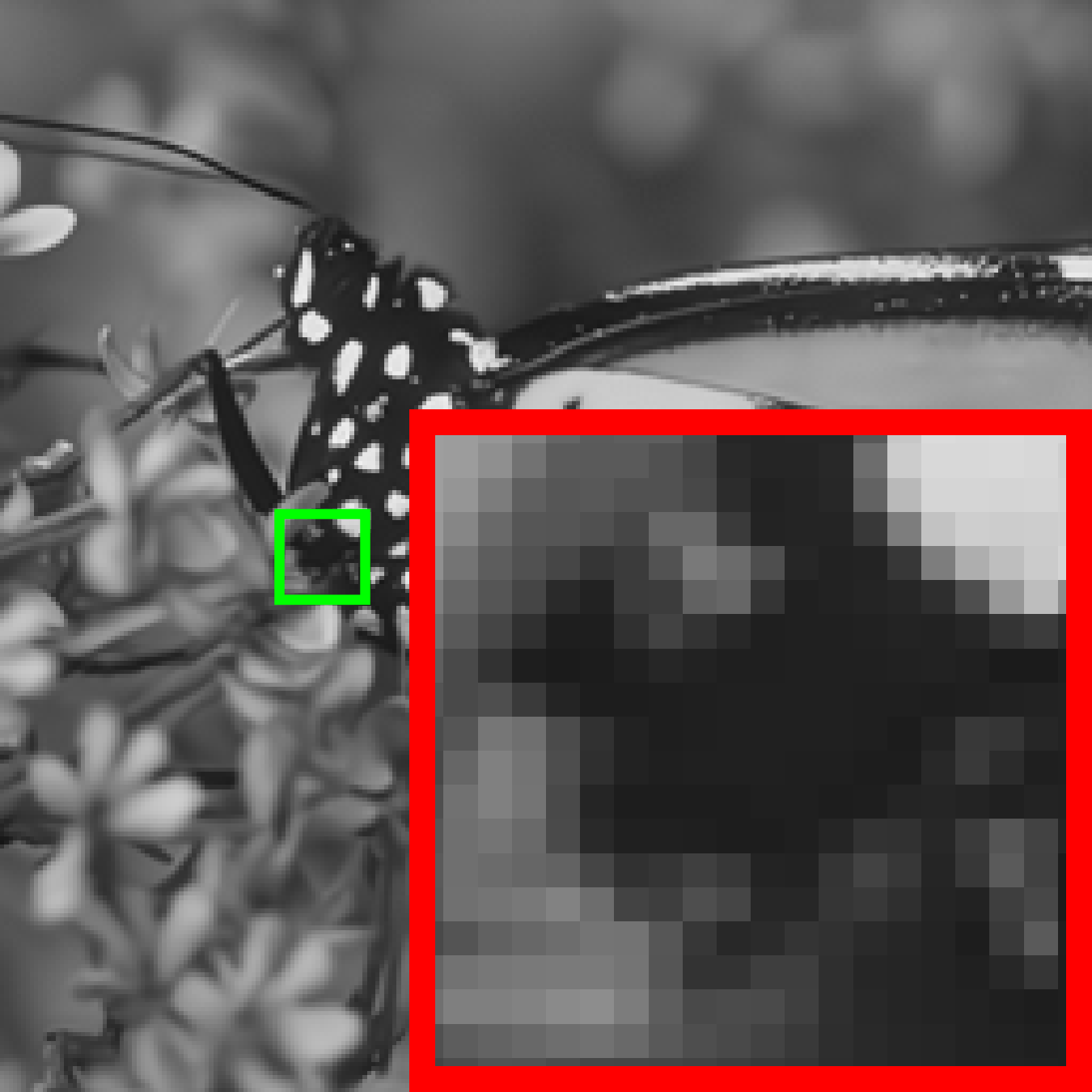}
    &\includegraphics[width=0.08\textwidth]{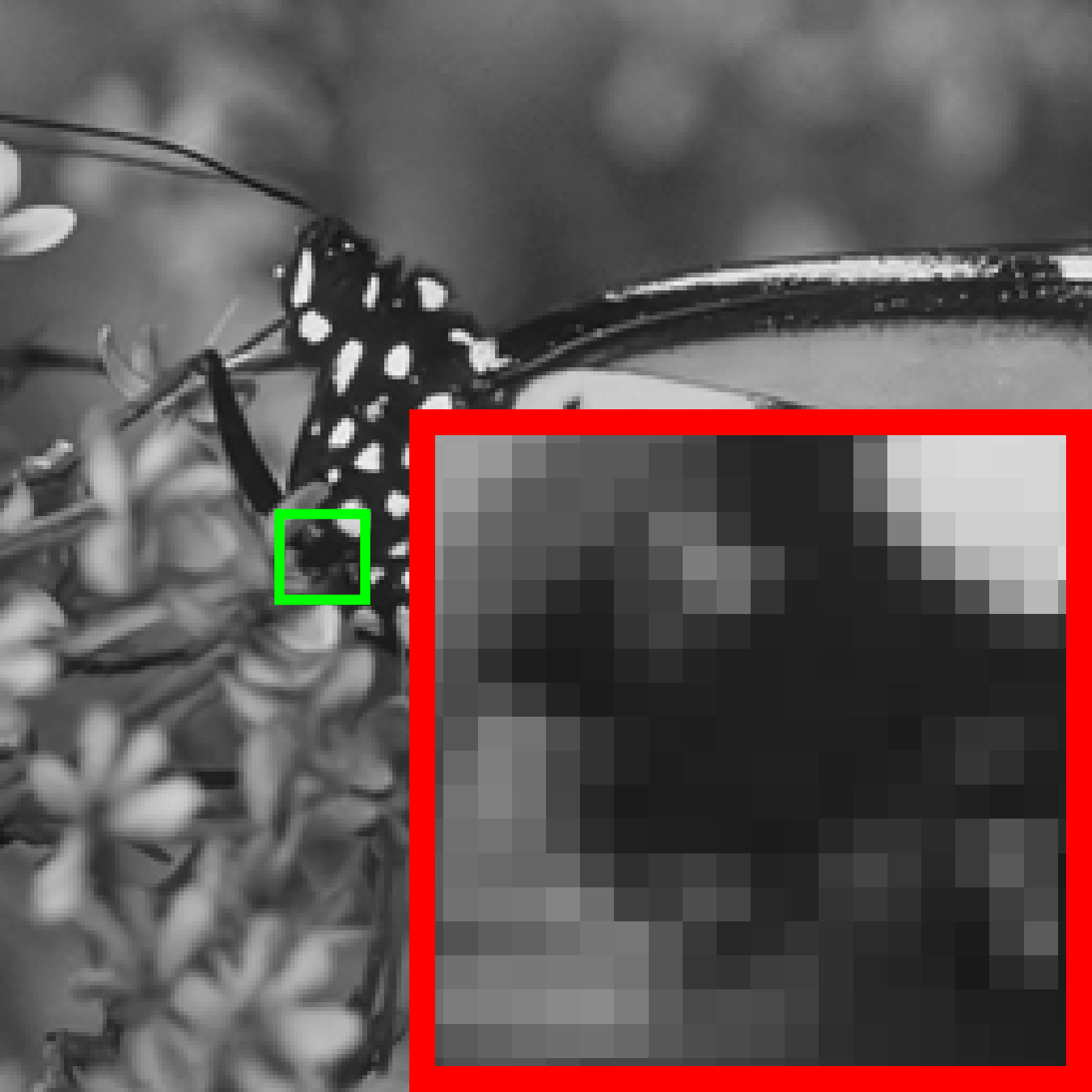}
    &\includegraphics[width=0.08\textwidth]{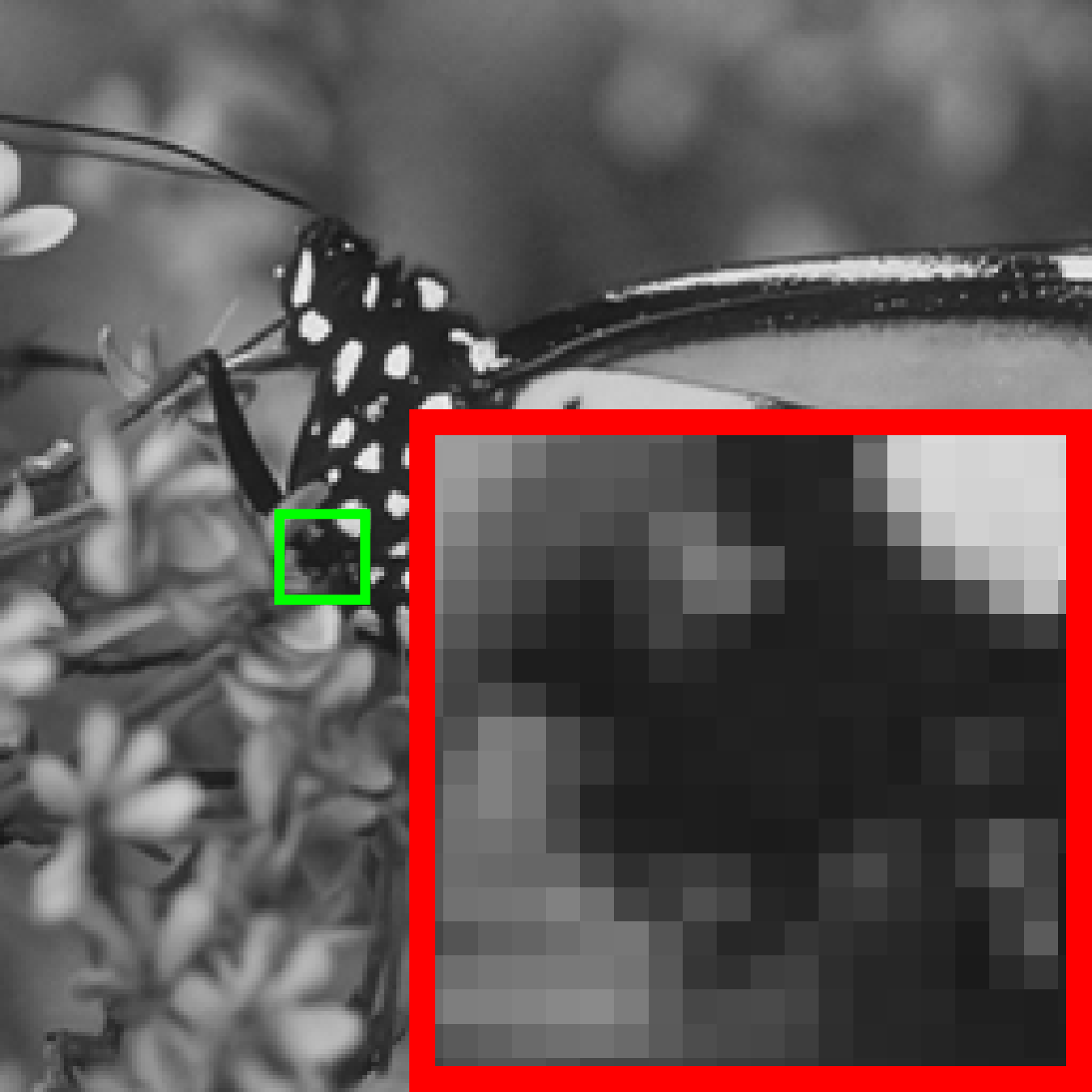}
    &\includegraphics[width=0.08\textwidth]{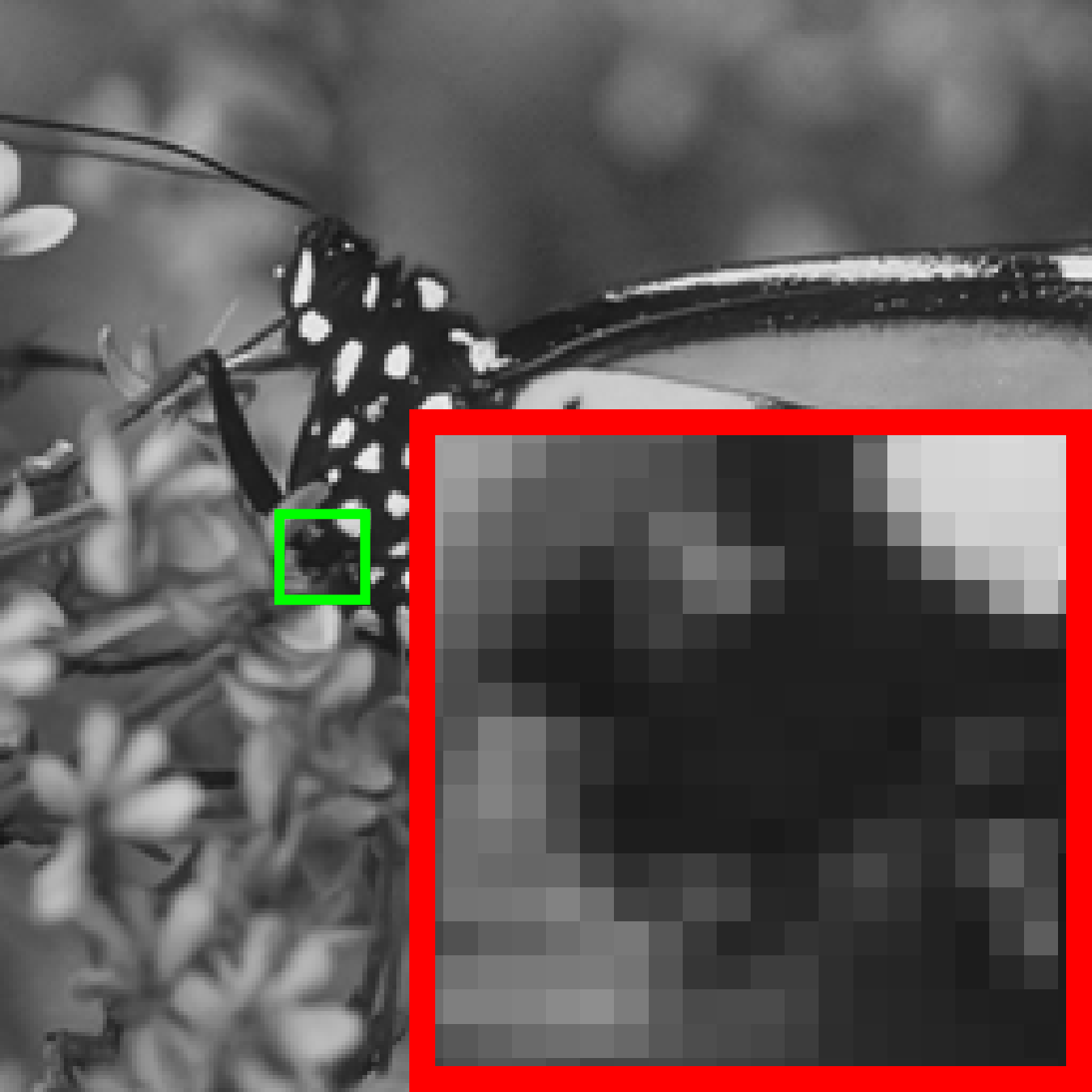}
    &\includegraphics[width=0.08\textwidth]{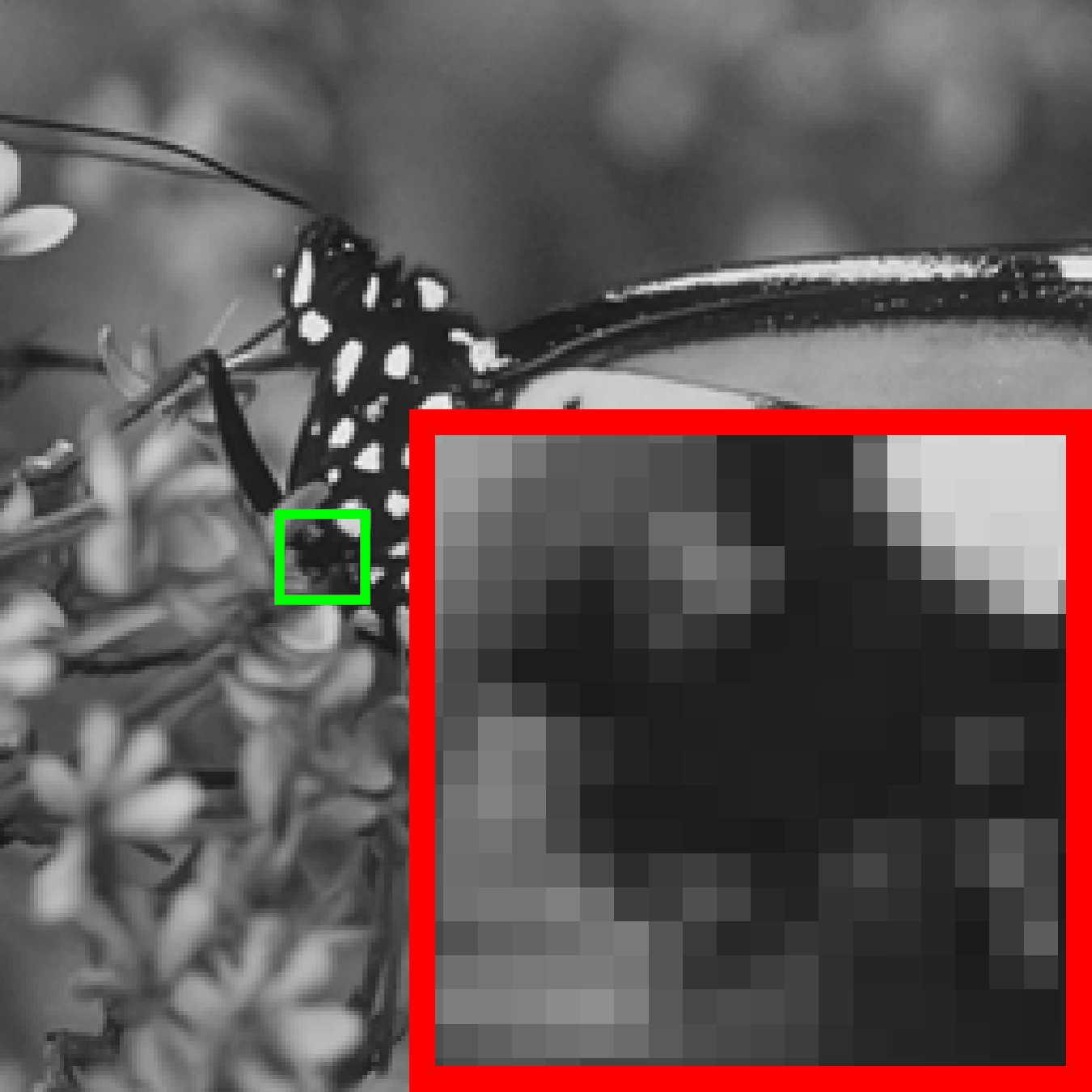}\\
    PSNR/SSIM & 32.30/0.94 & 40.22/\textbf{\textcolor{red}{0.99}} & 40.77/\textbf{\textcolor{red}{0.99}} & 41.05/\textbf{\textcolor{red}{0.99}} & 34.88/0.96 & 39.22/\underline{\textcolor{blue}{0.98}} & 40.24/\underline{\textcolor{blue}{0.98}} & 39.04/\underline{\textcolor{blue}{0.98}} & 40.58/\underline{\textcolor{blue}{0.98}} & 41.35/\textbf{\textcolor{red}{0.99}} & \underline{\textcolor{blue}{42.16}}/\textbf{\textcolor{red}{0.99}} & 41.62/\textbf{\textcolor{red}{0.99}} & \textbf{\textcolor{red}{42.60}}/\textbf{\textcolor{red}{0.99}}\\
    \includegraphics[width=0.08\textwidth]{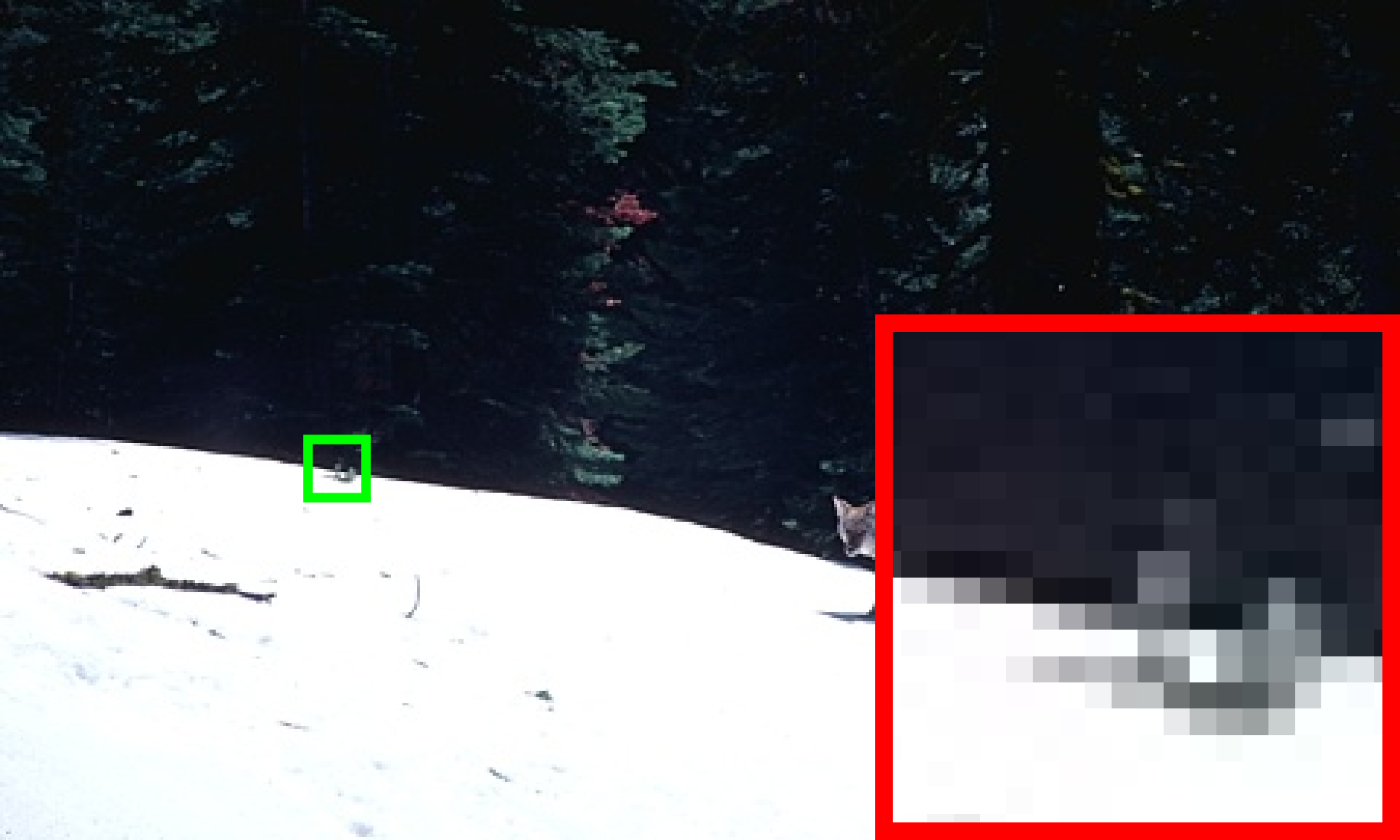}
    &\includegraphics[width=0.08\textwidth]{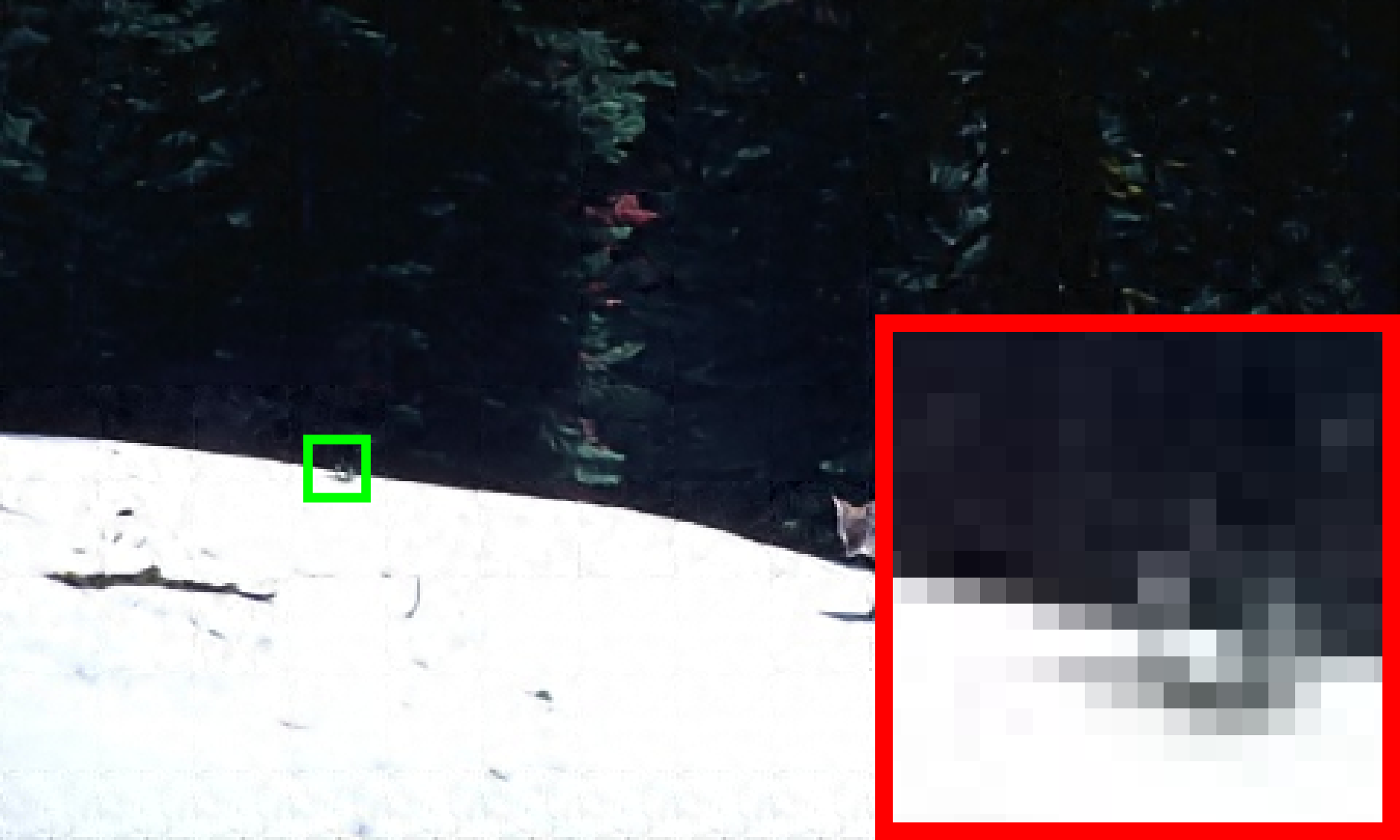}
    &\includegraphics[width=0.08\textwidth]{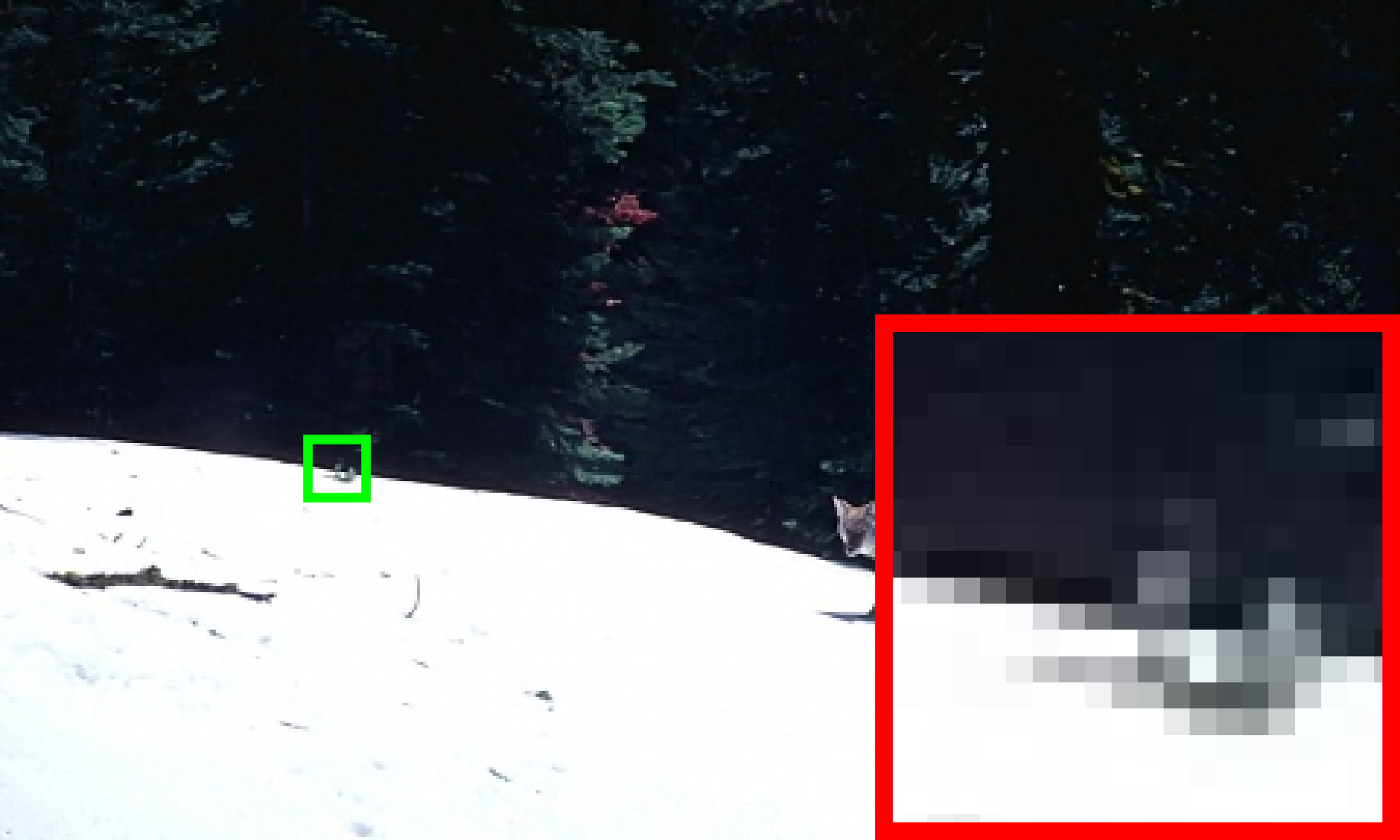}
    &\includegraphics[width=0.08\textwidth]{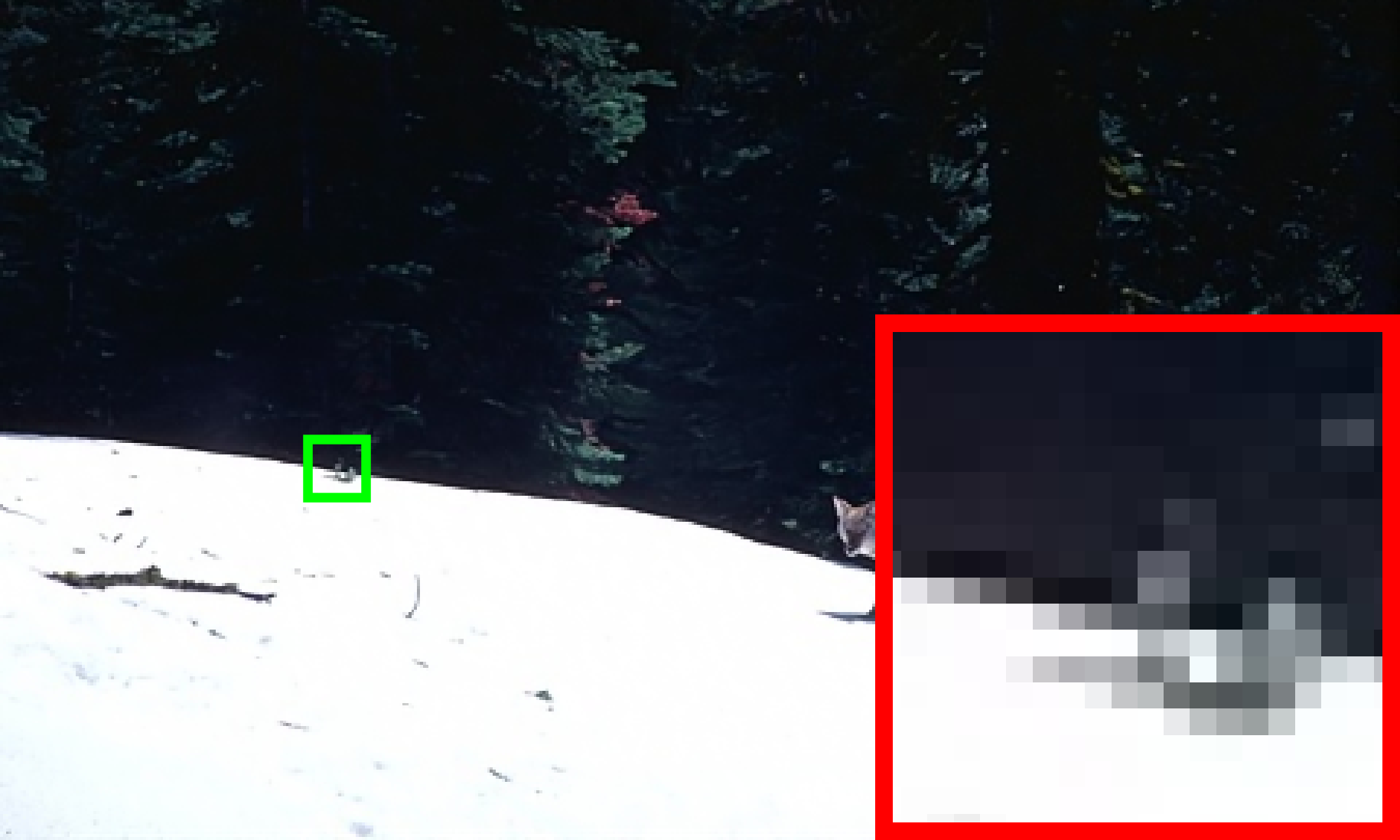}
    &\includegraphics[width=0.08\textwidth]{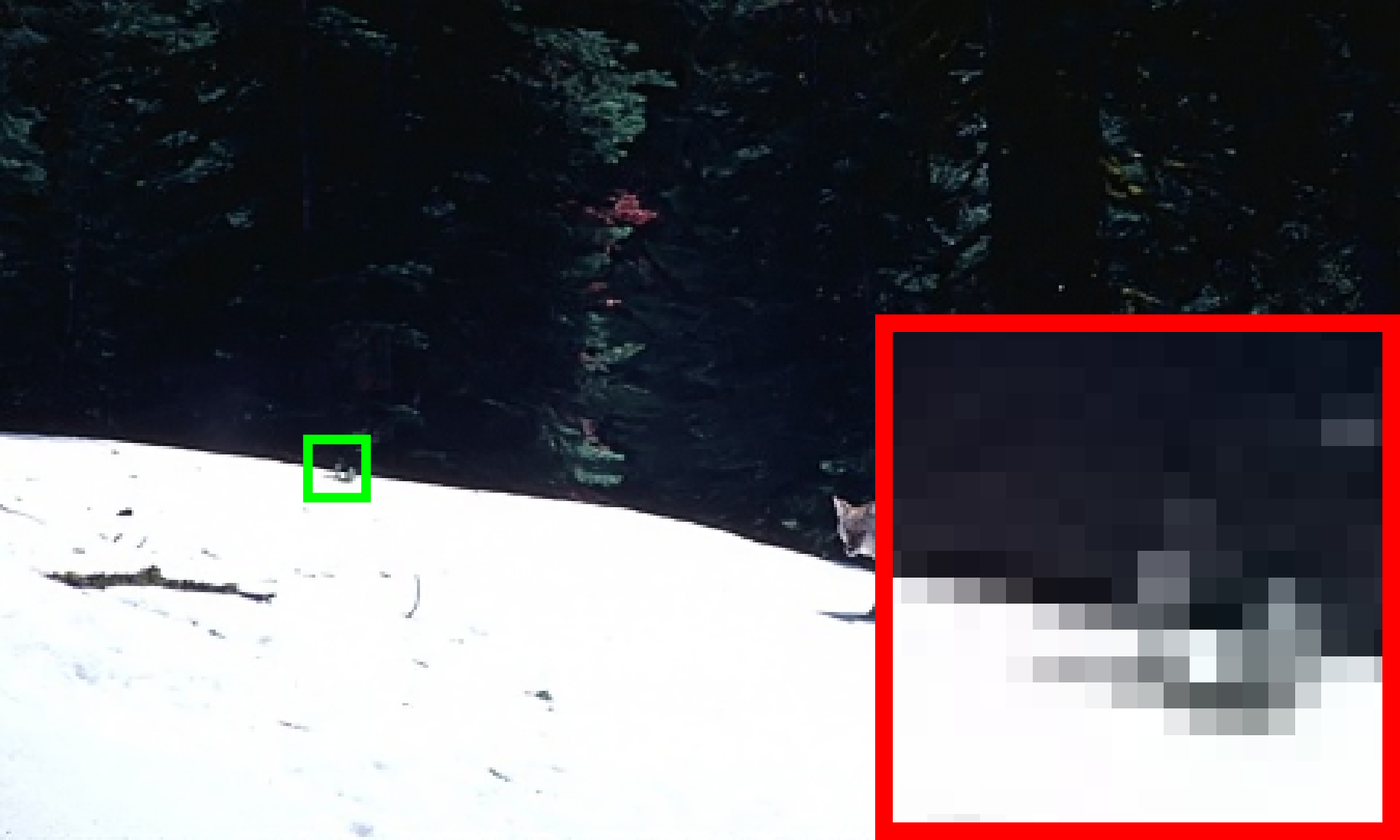}
    &\includegraphics[width=0.08\textwidth]{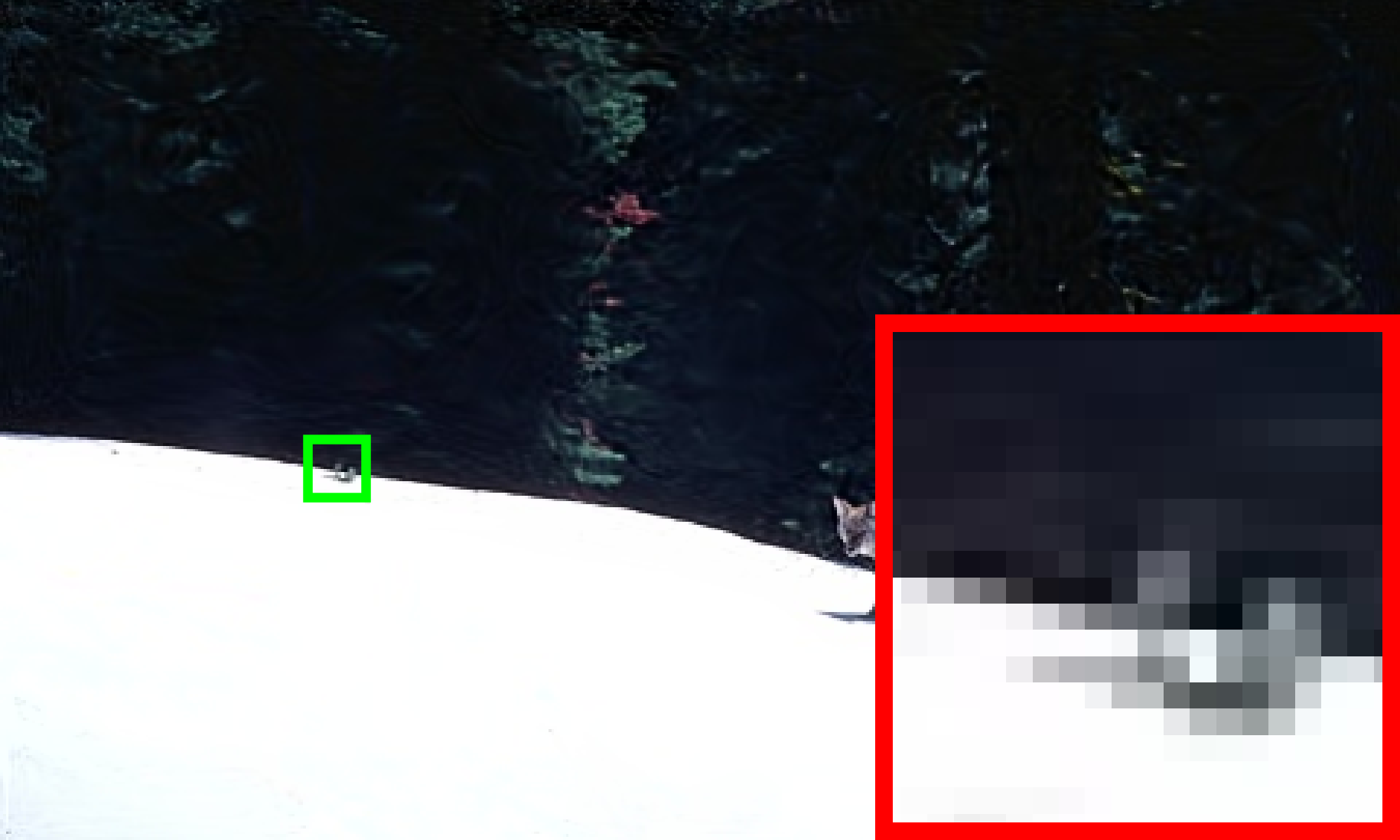}
    &\includegraphics[width=0.08\textwidth]{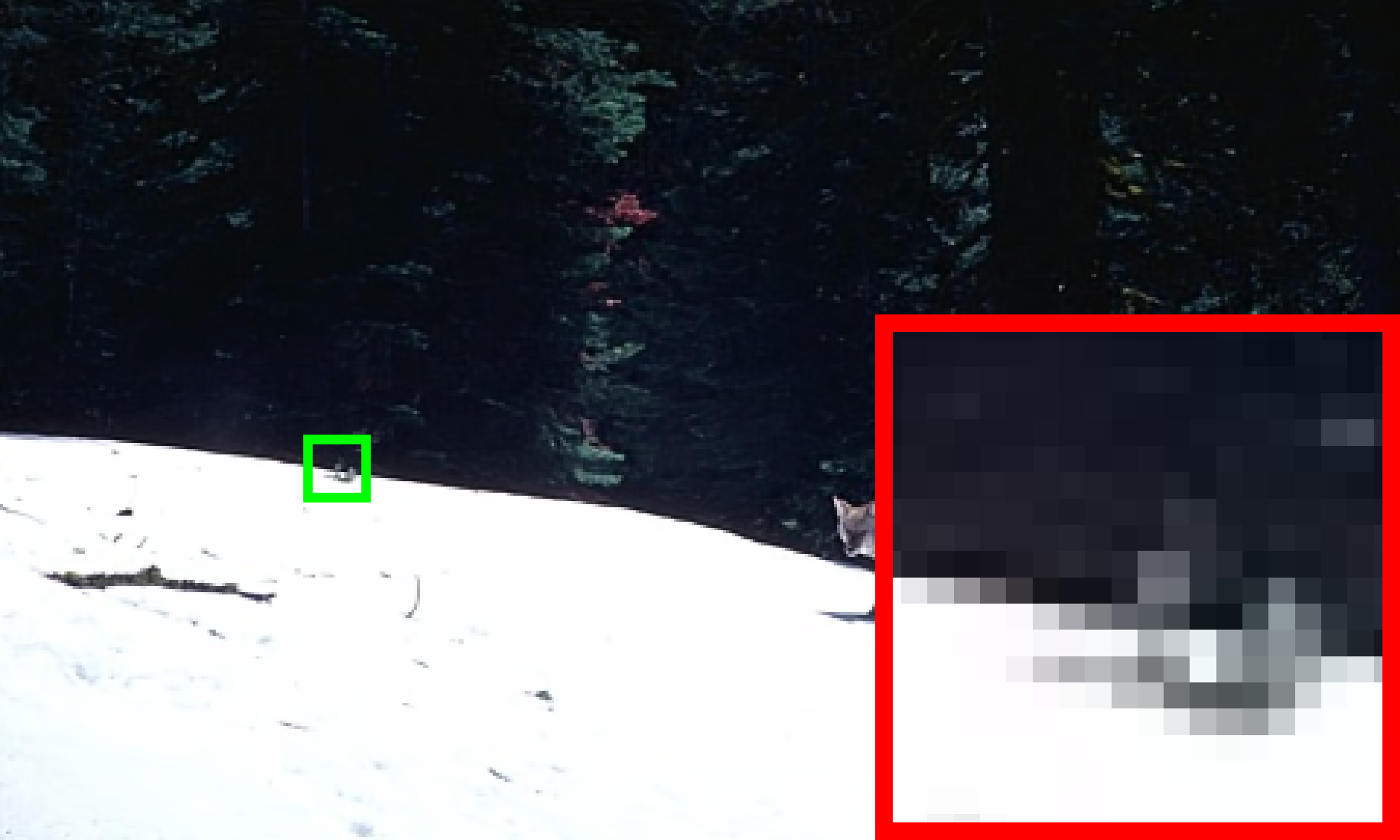}
    &\includegraphics[width=0.08\textwidth]{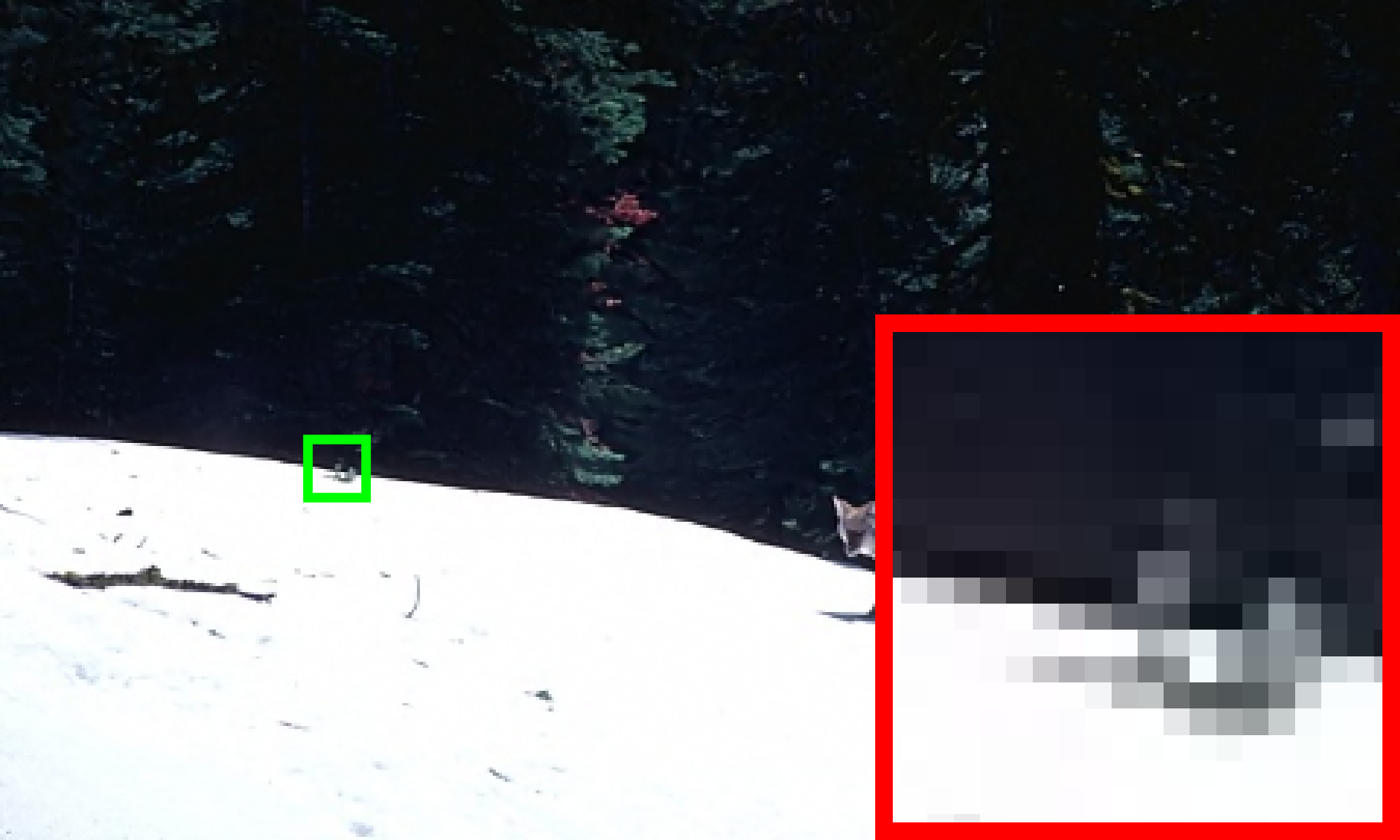}
    &\includegraphics[width=0.08\textwidth]{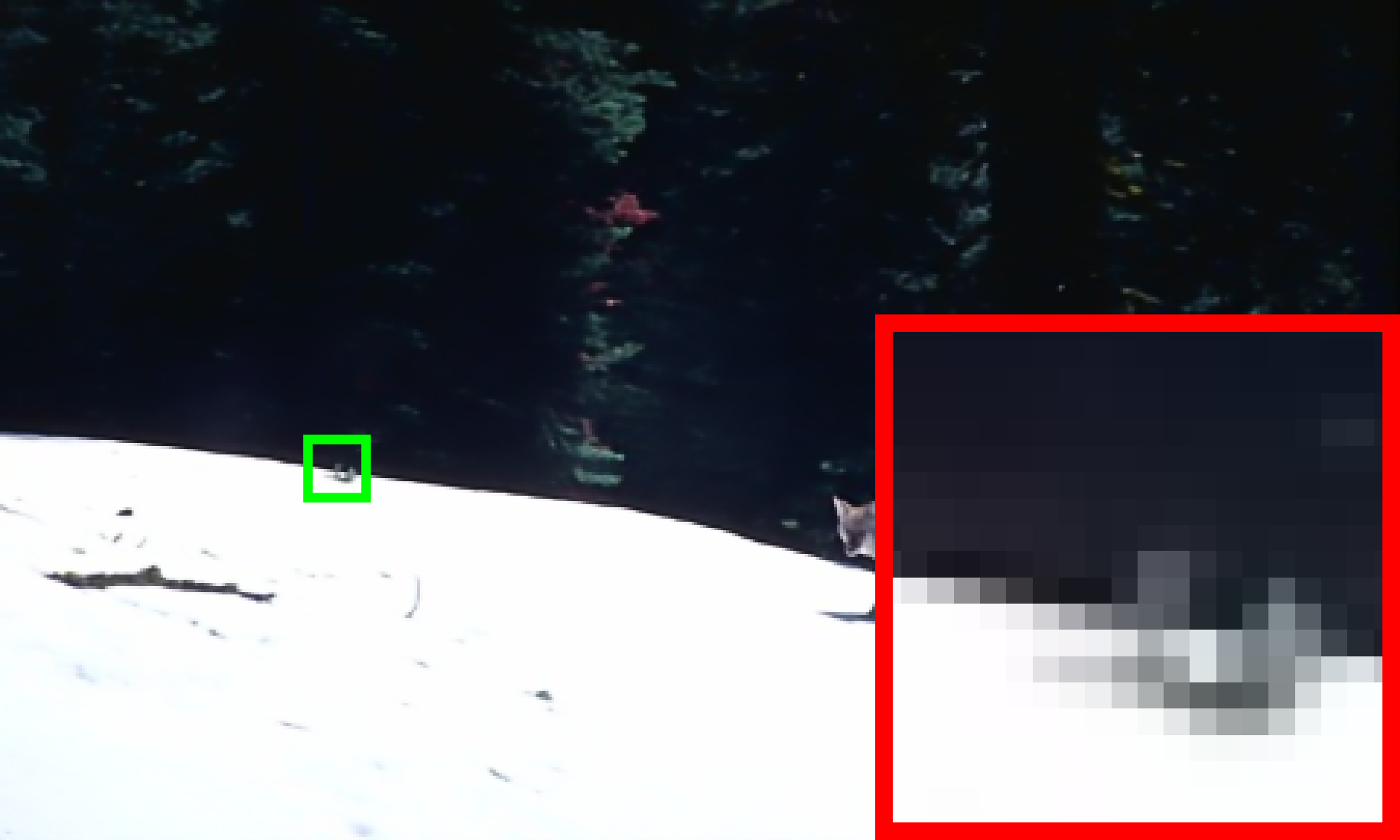}
    &\includegraphics[width=0.08\textwidth]{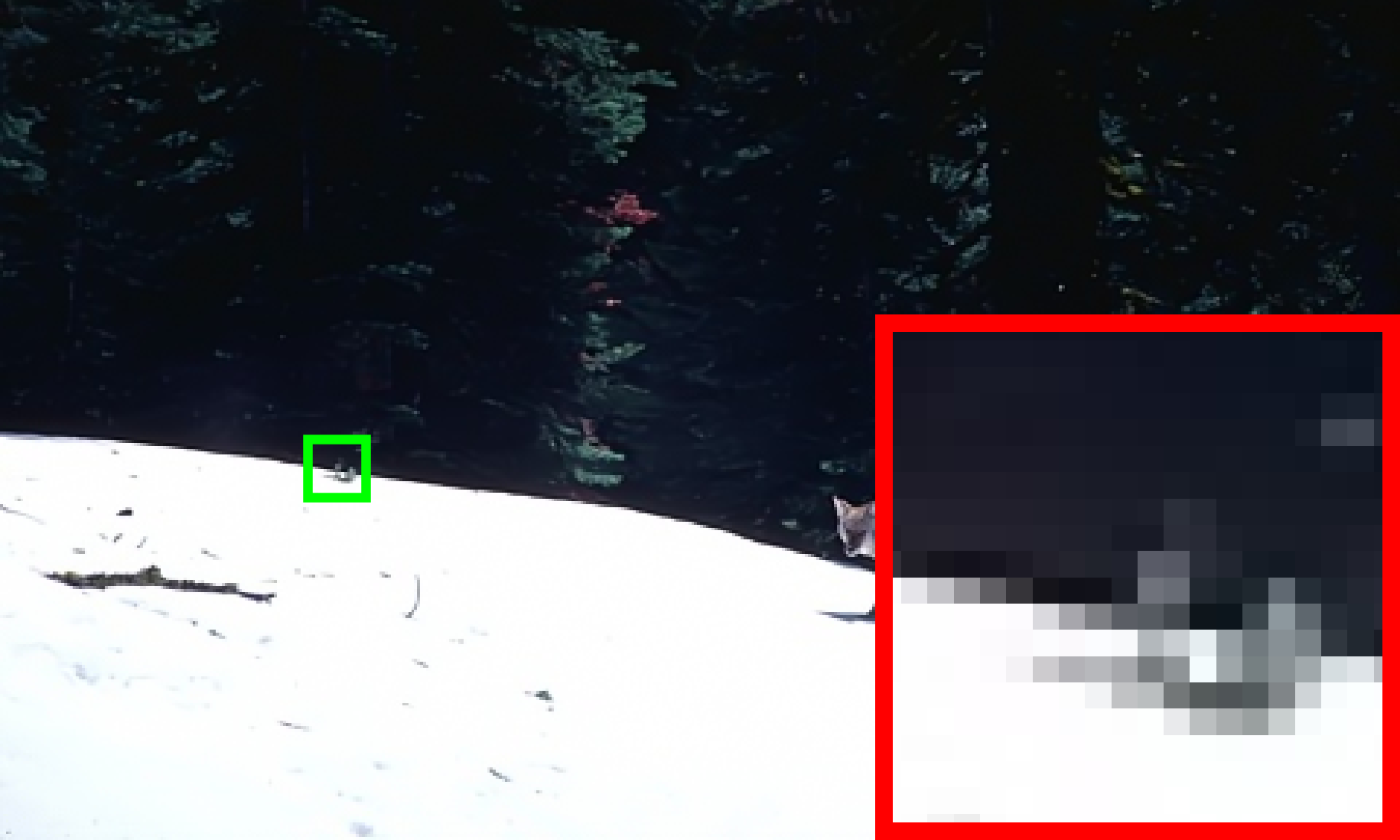}
    &\includegraphics[width=0.08\textwidth]{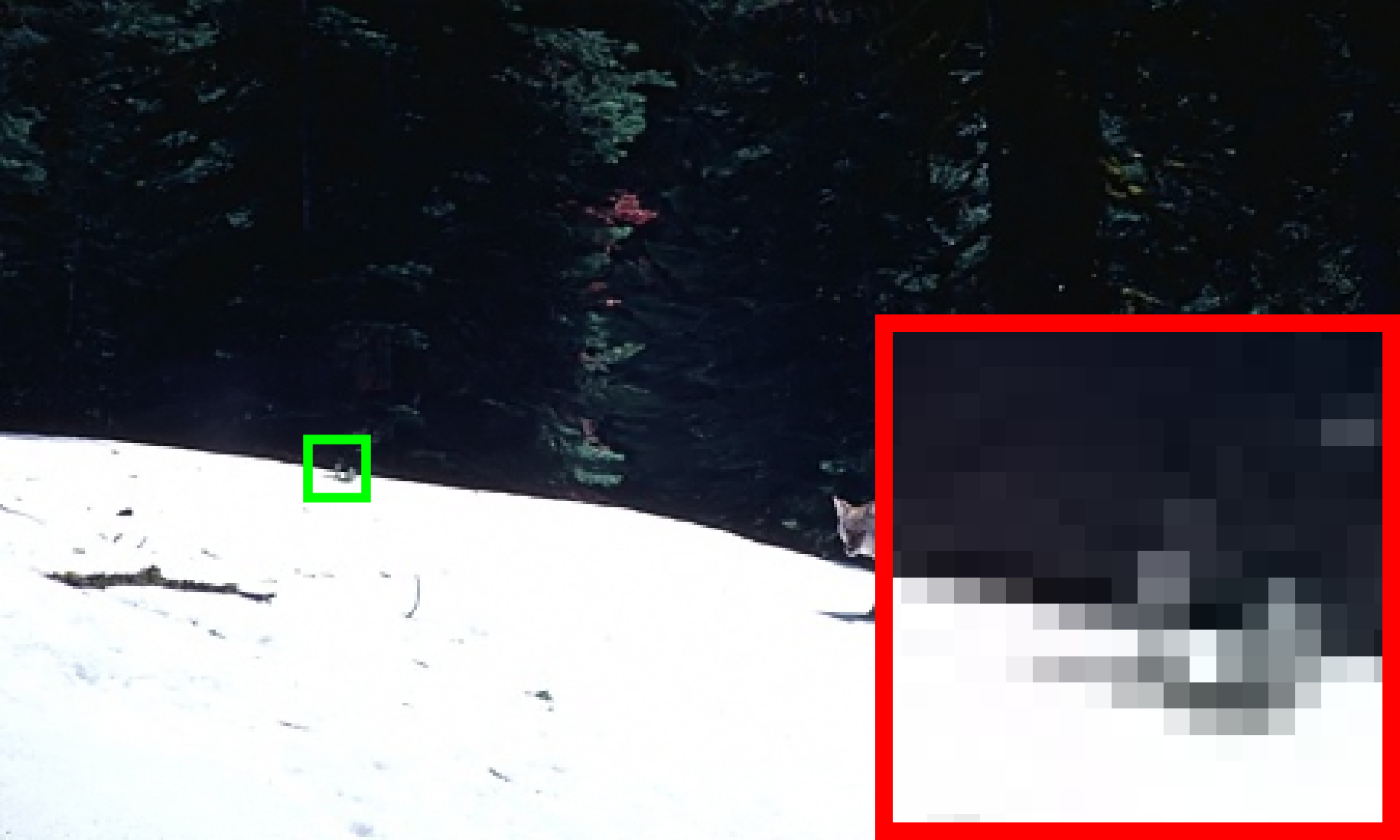}
    &\includegraphics[width=0.08\textwidth]{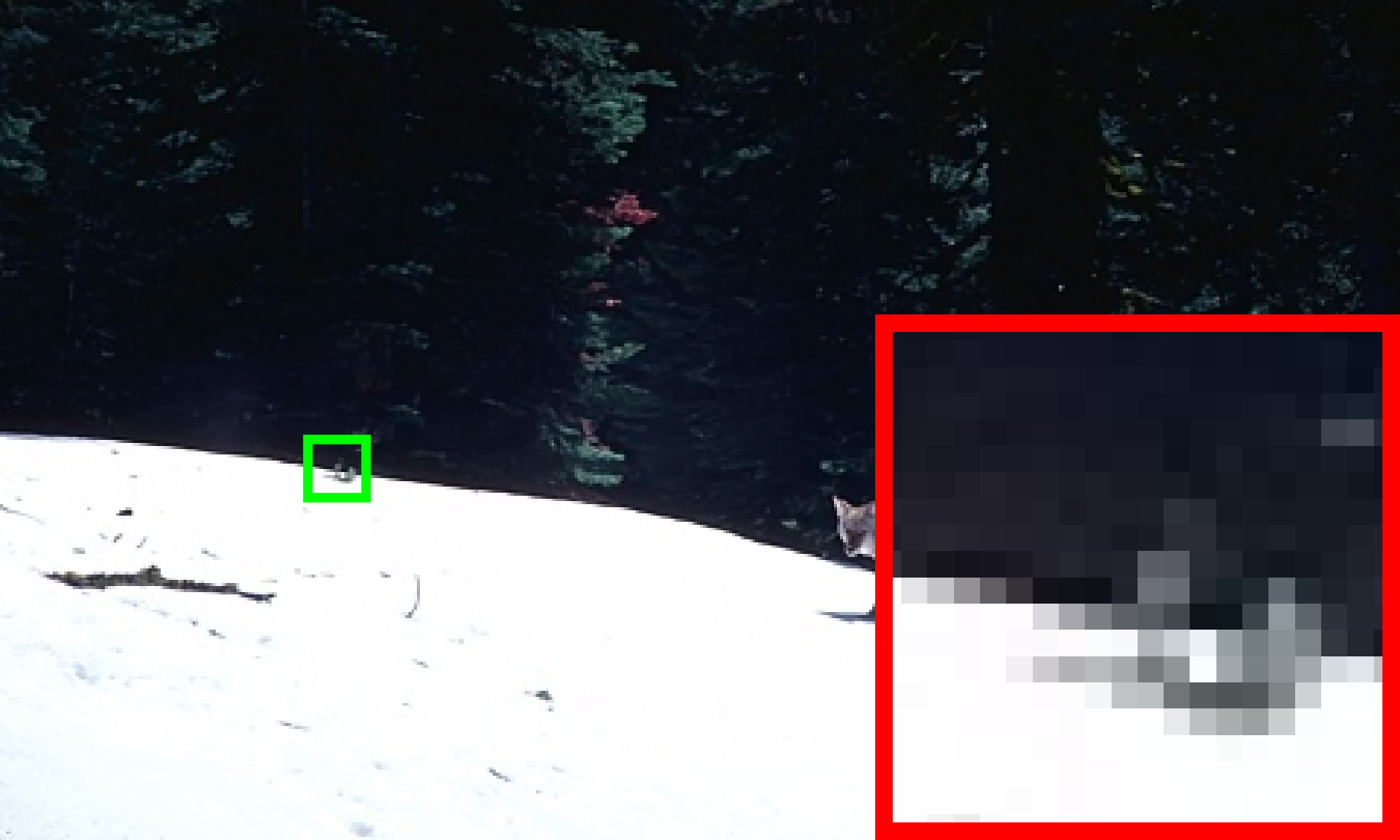}
    &\includegraphics[width=0.08\textwidth]{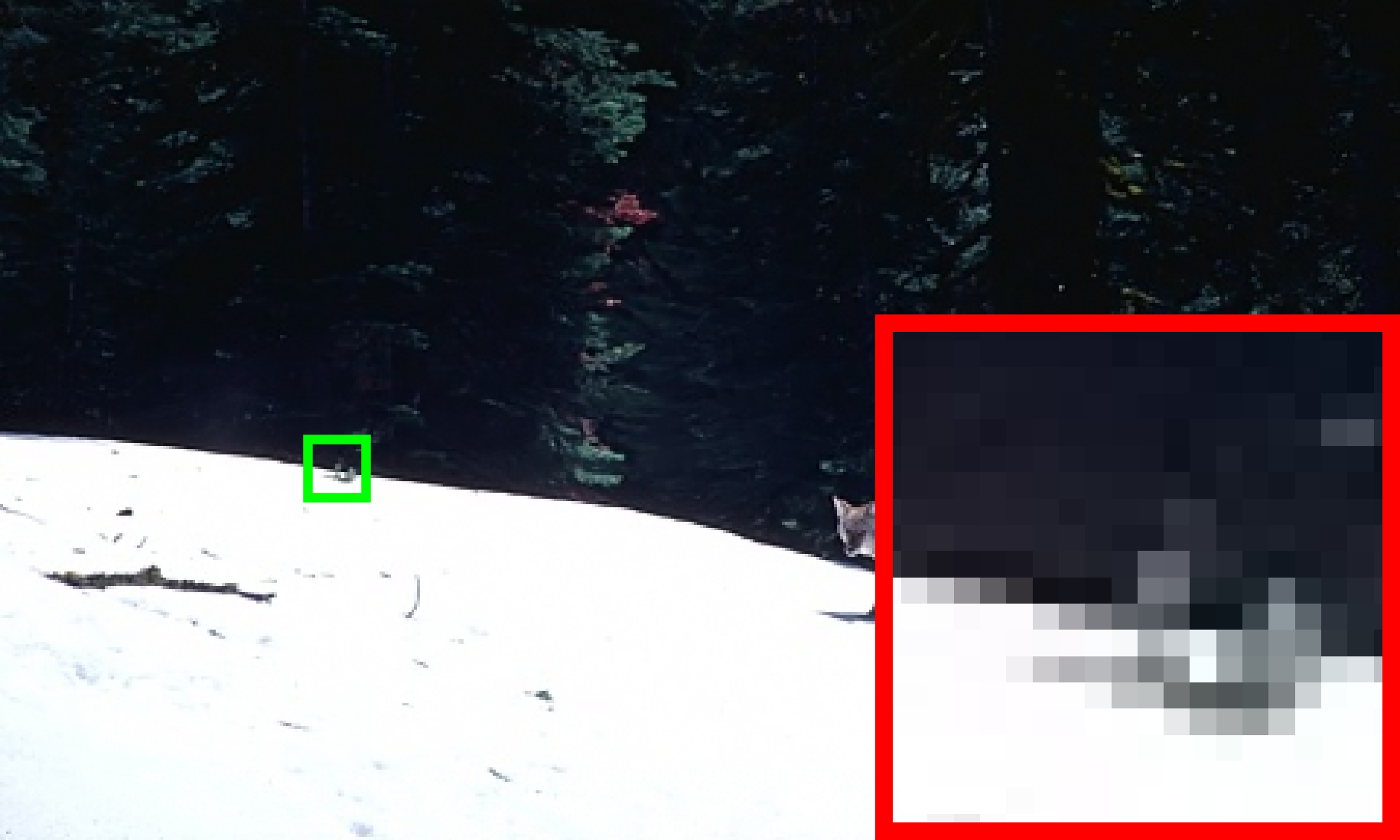}
    &\includegraphics[width=0.08\textwidth]{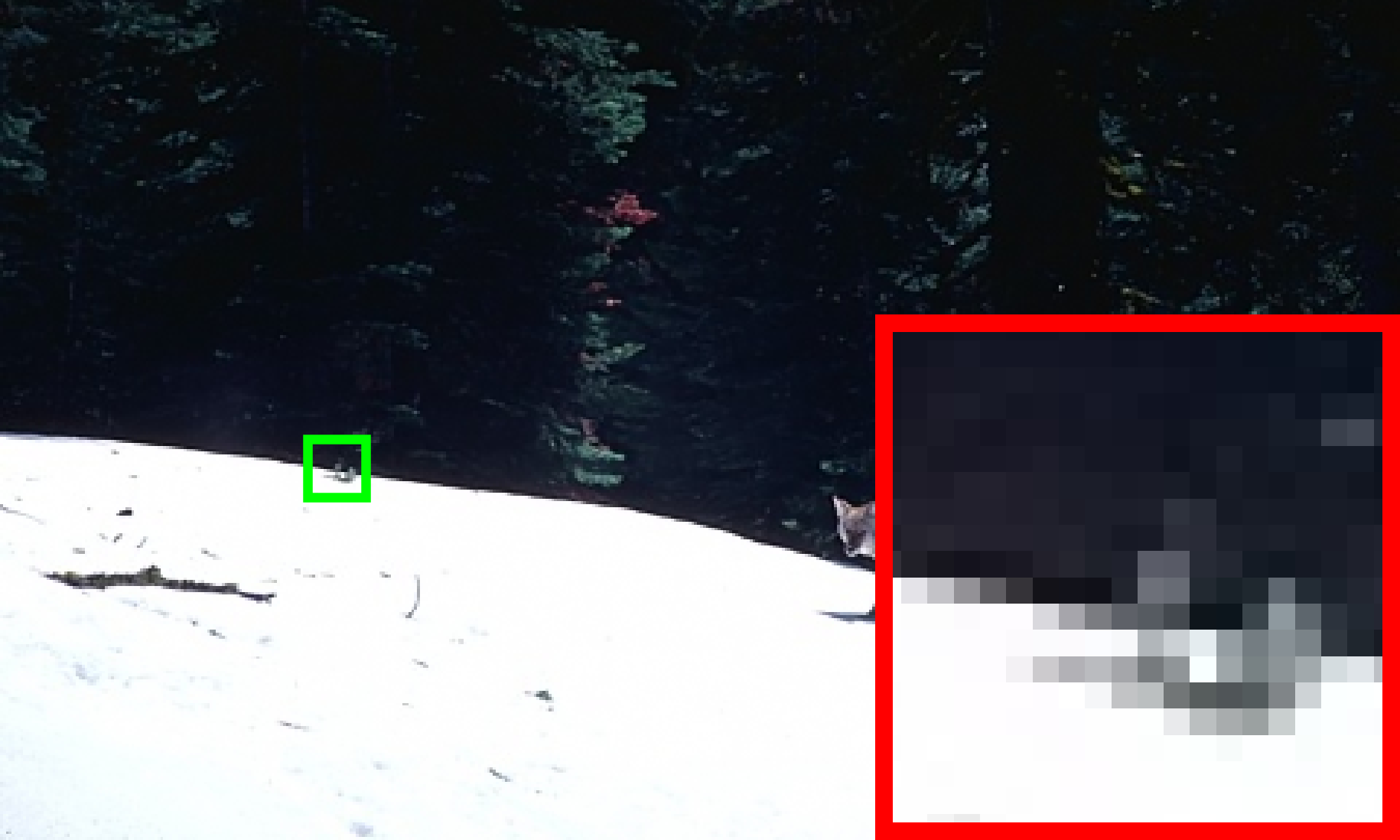}\\
    PSNR/SSIM & 34.87/0.91 & 41.82/\underline{\textcolor{blue}{0.98}} & 42.42/\underline{\textcolor{blue}{0.98}} & 42.45/0.98 & 28.58/0.88 & 40.62/0.97 & 41.81/\underline{\textcolor{blue}{0.98}} & 35.68/0.92 & 41.82/\underline{\textcolor{blue}{0.98}} & 43.01/\underline{\textcolor{blue}{0.98}} & \textbf{\textcolor{red}{43.76}}/\textbf{\textcolor{red}{0.99}} & 43.20/\textbf{\textcolor{red}{0.99}} & \underline{\textcolor{blue}{43.72}}/\textbf{\textcolor{red}{0.99}}\\
    \includegraphics[width=0.08\textwidth]{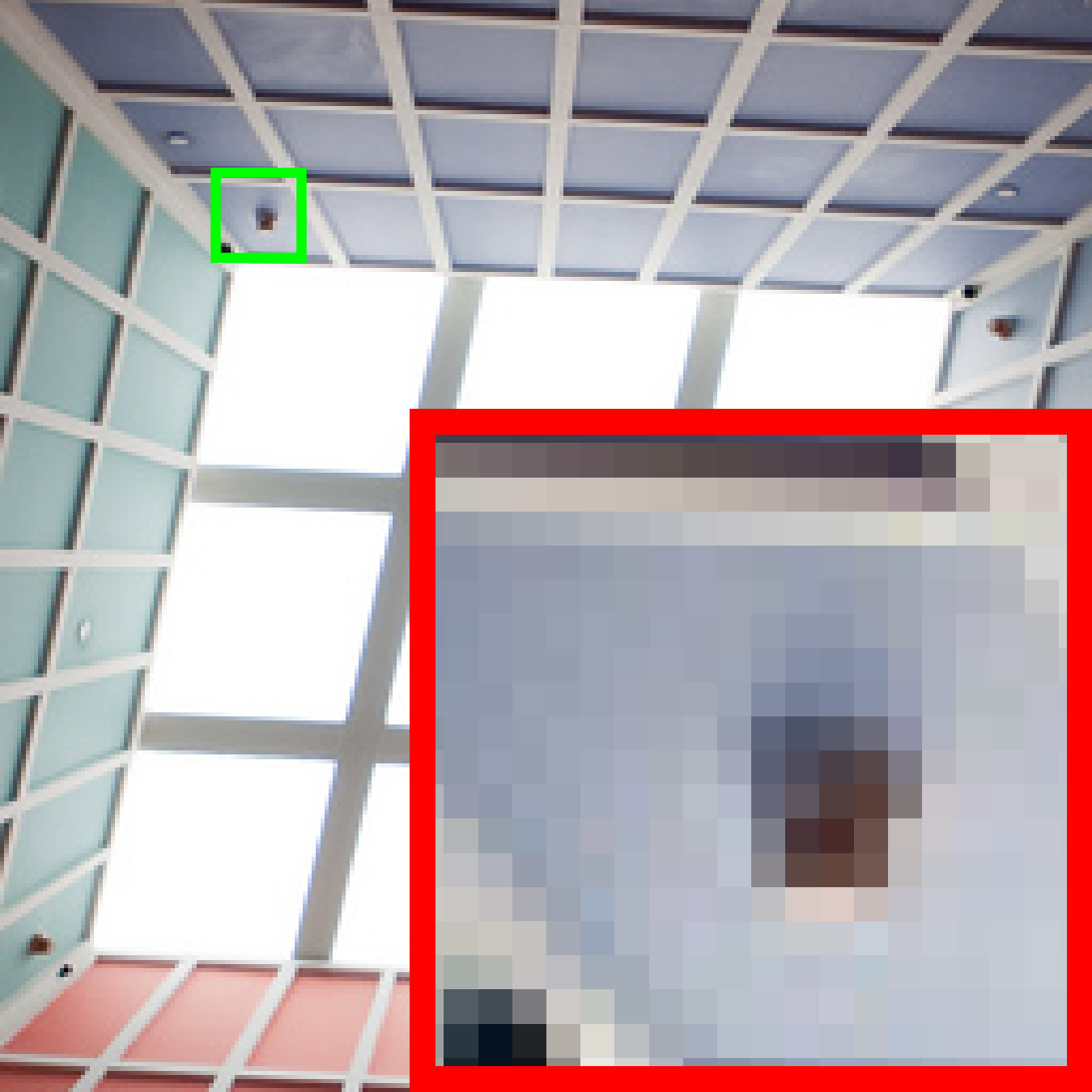}
    &\includegraphics[width=0.08\textwidth]{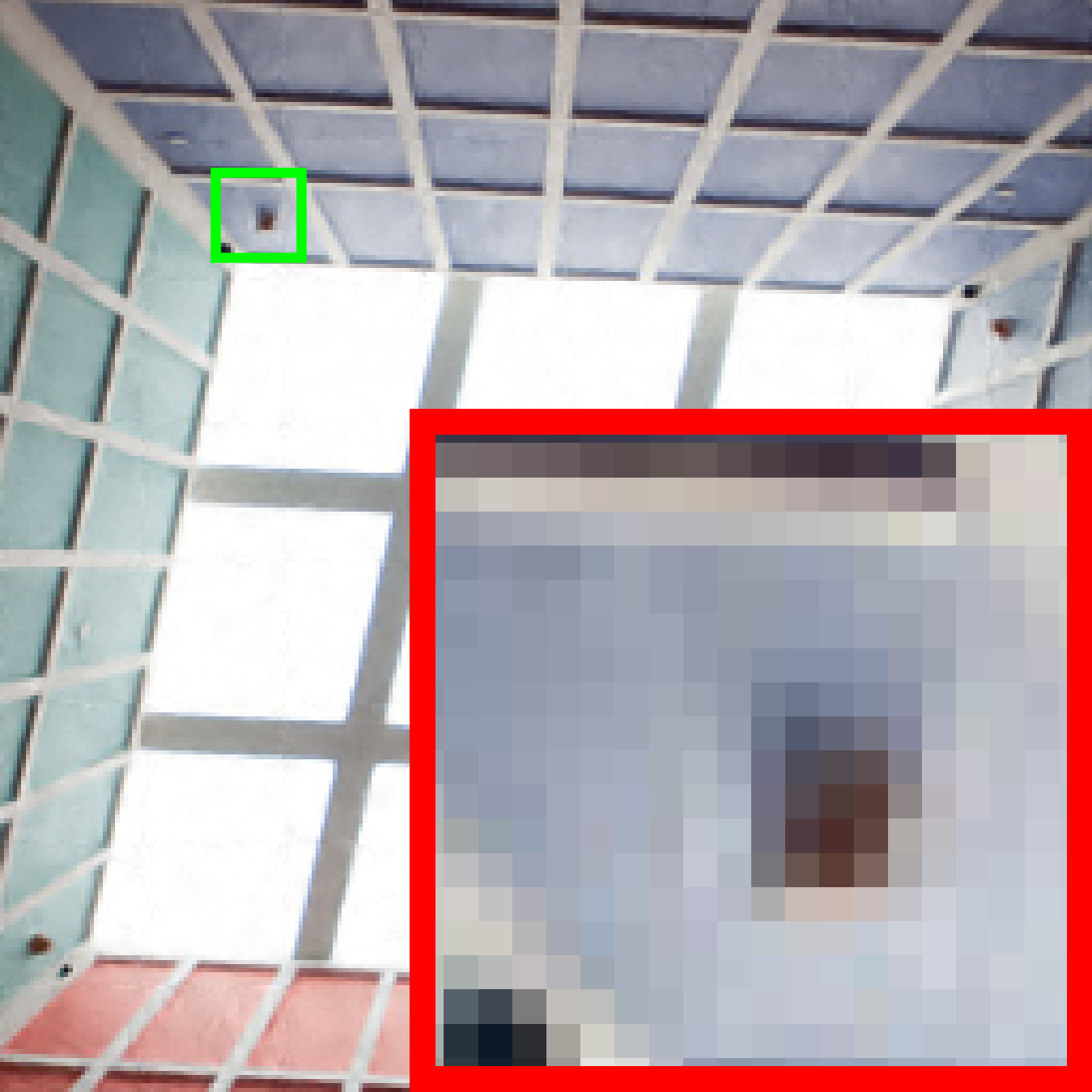}
    &\includegraphics[width=0.08\textwidth]{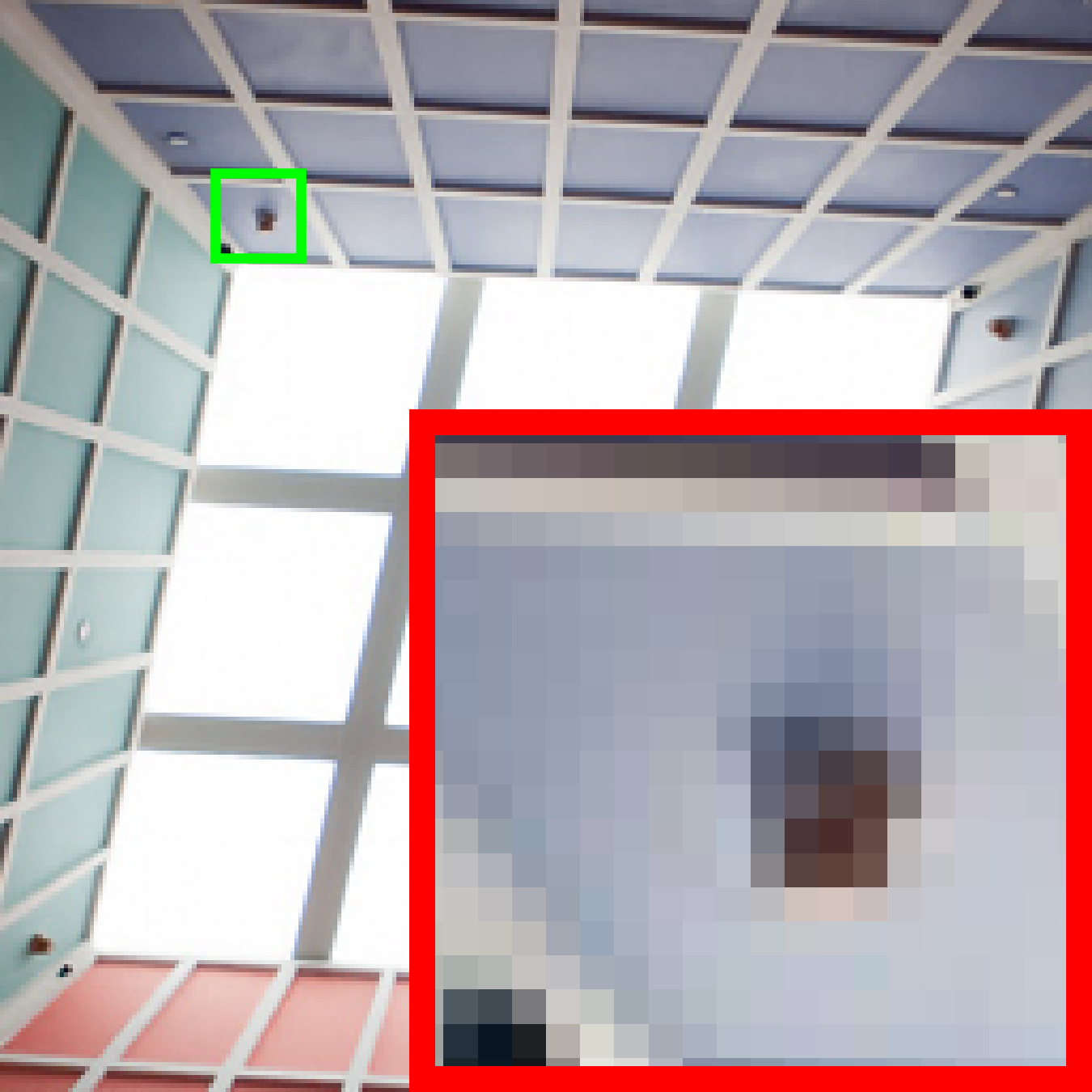}
    &\includegraphics[width=0.08\textwidth]{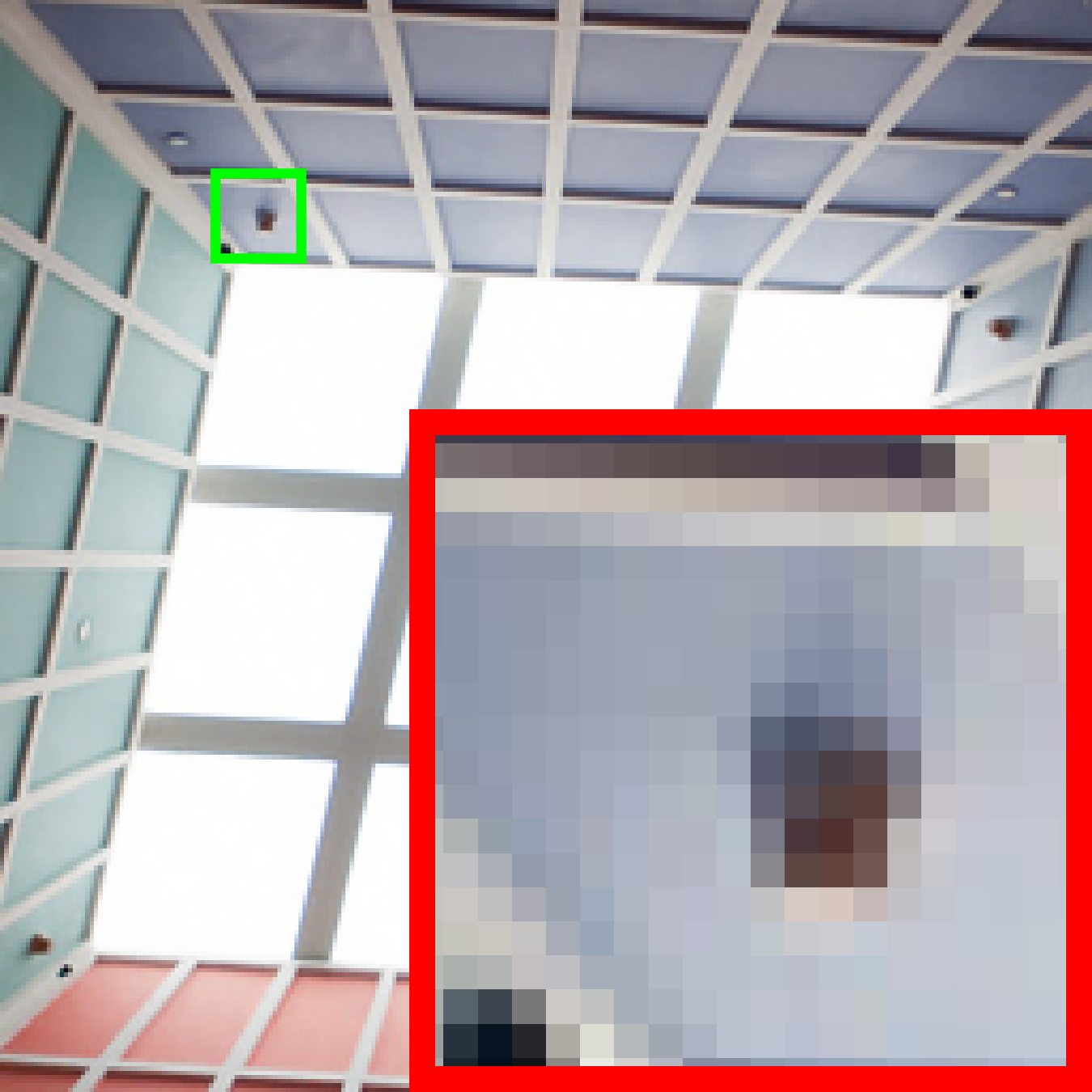}
    &\includegraphics[width=0.08\textwidth]{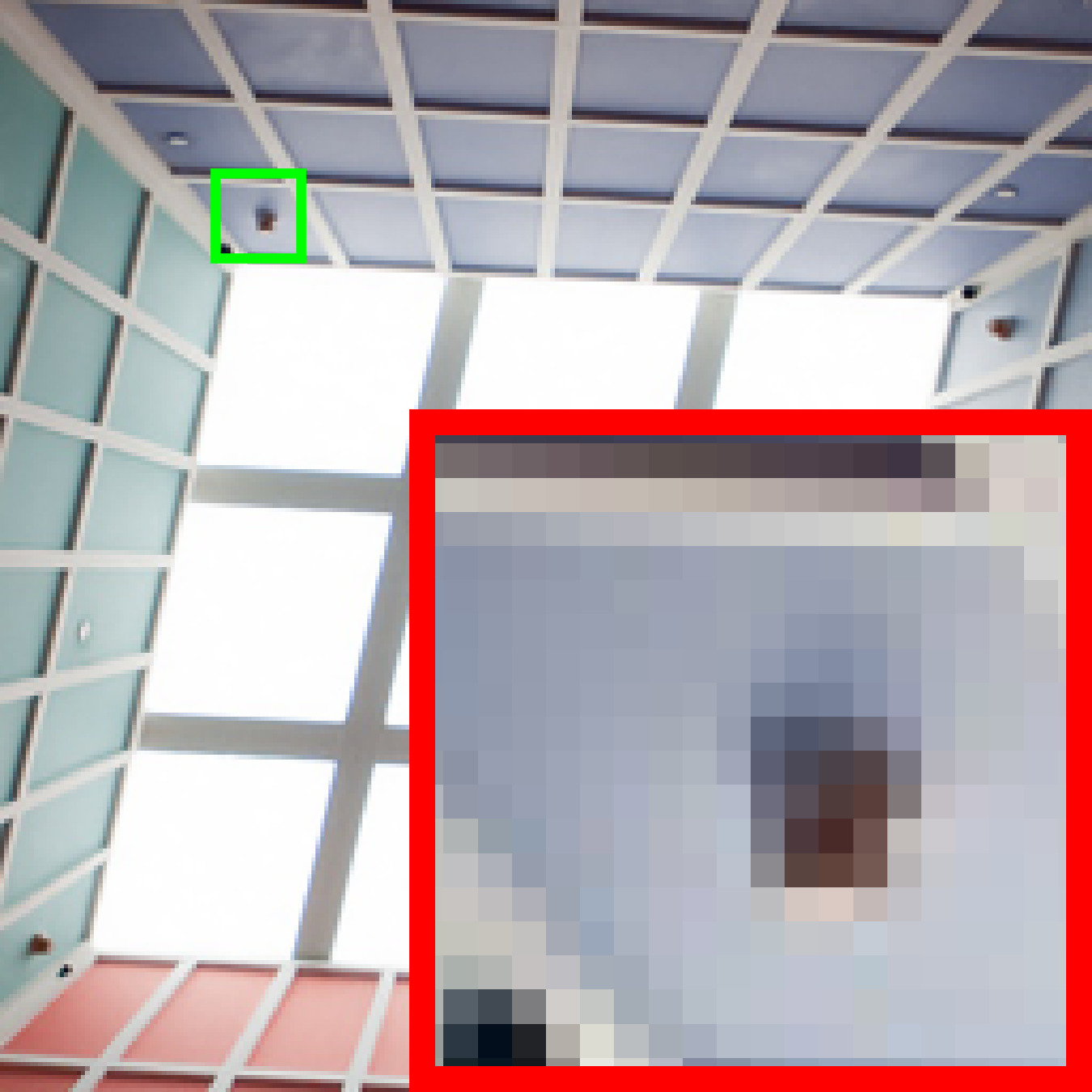}
    &\includegraphics[width=0.08\textwidth]{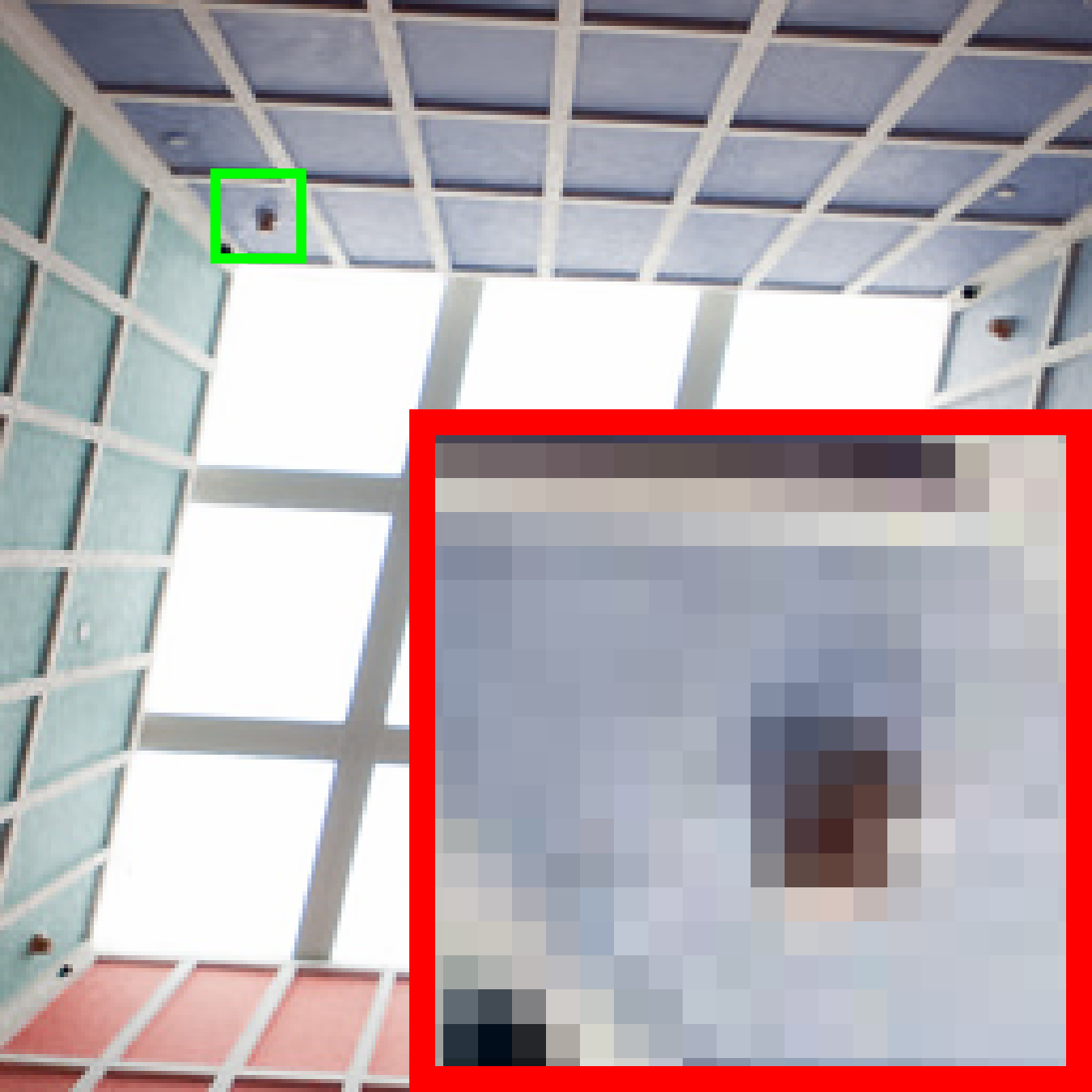}
    &\includegraphics[width=0.08\textwidth]{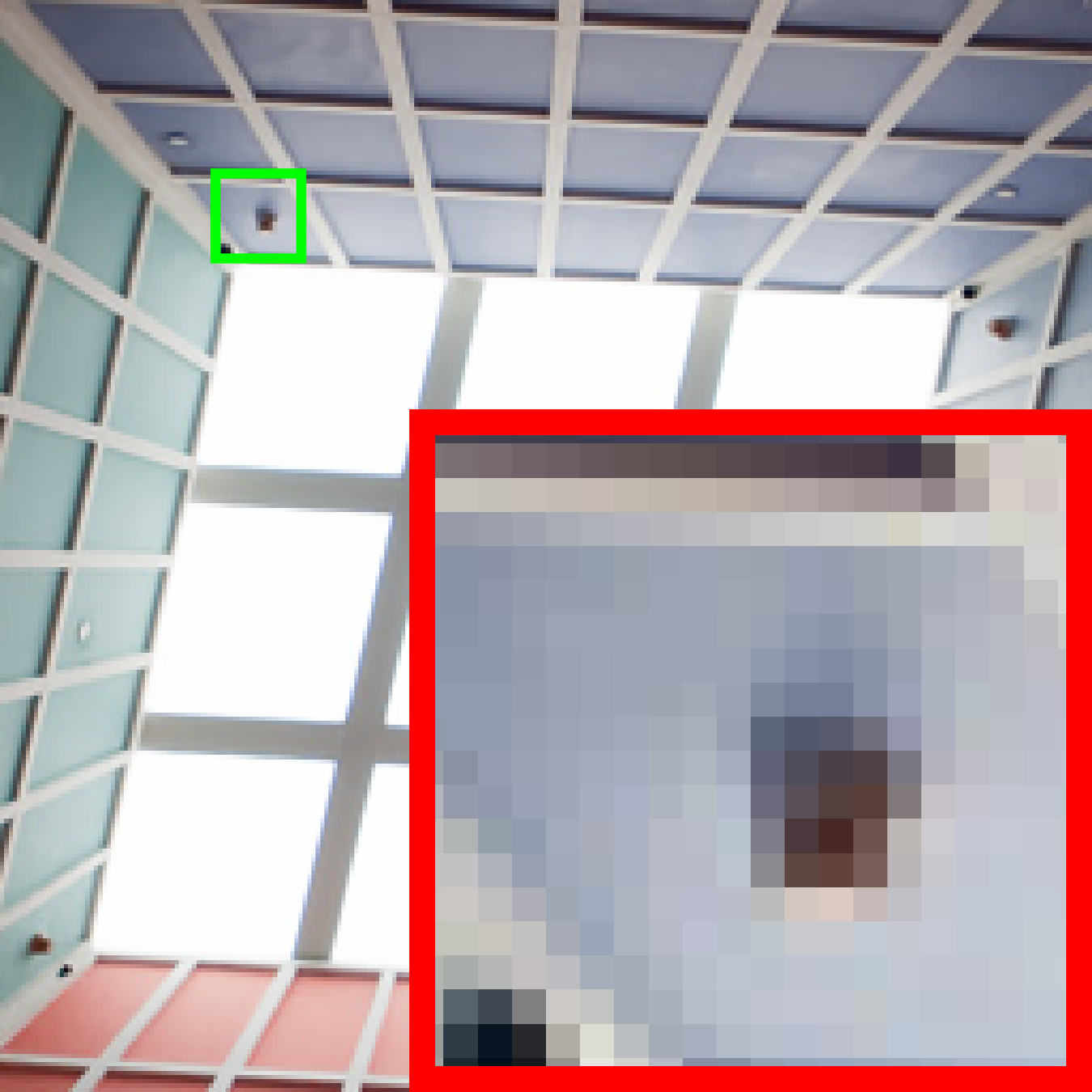}
    &\includegraphics[width=0.08\textwidth]{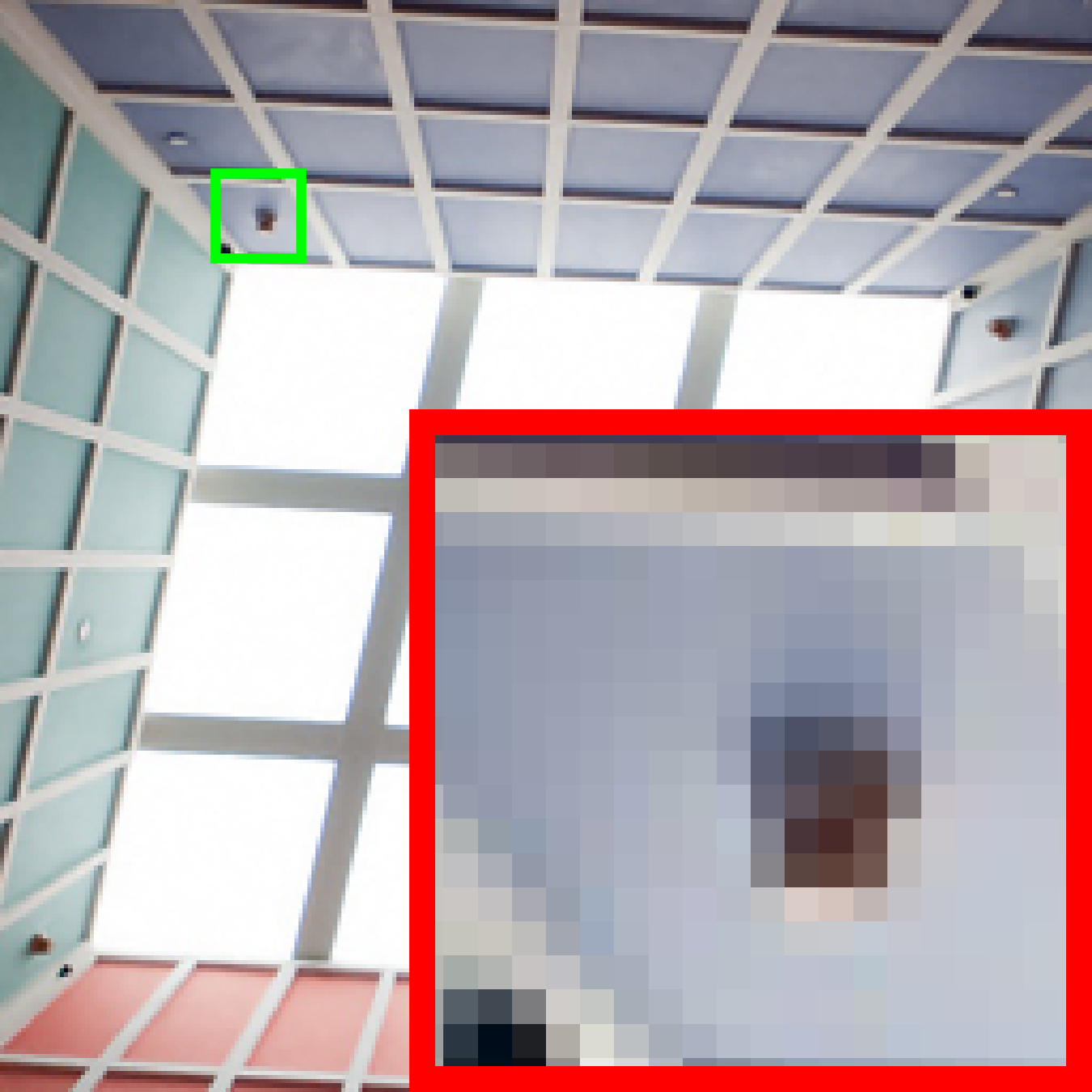}
    &\includegraphics[width=0.08\textwidth]{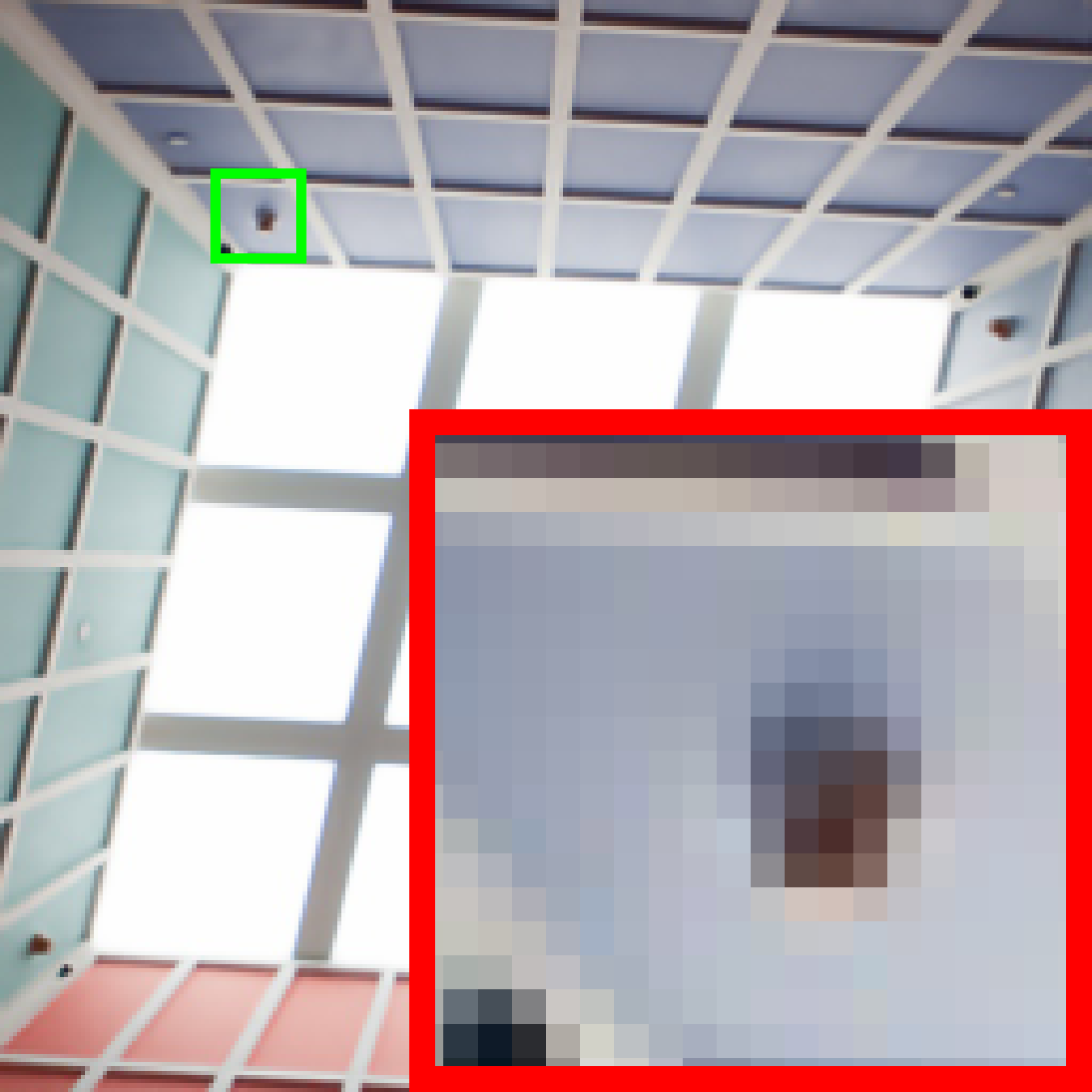}
    &\includegraphics[width=0.08\textwidth]{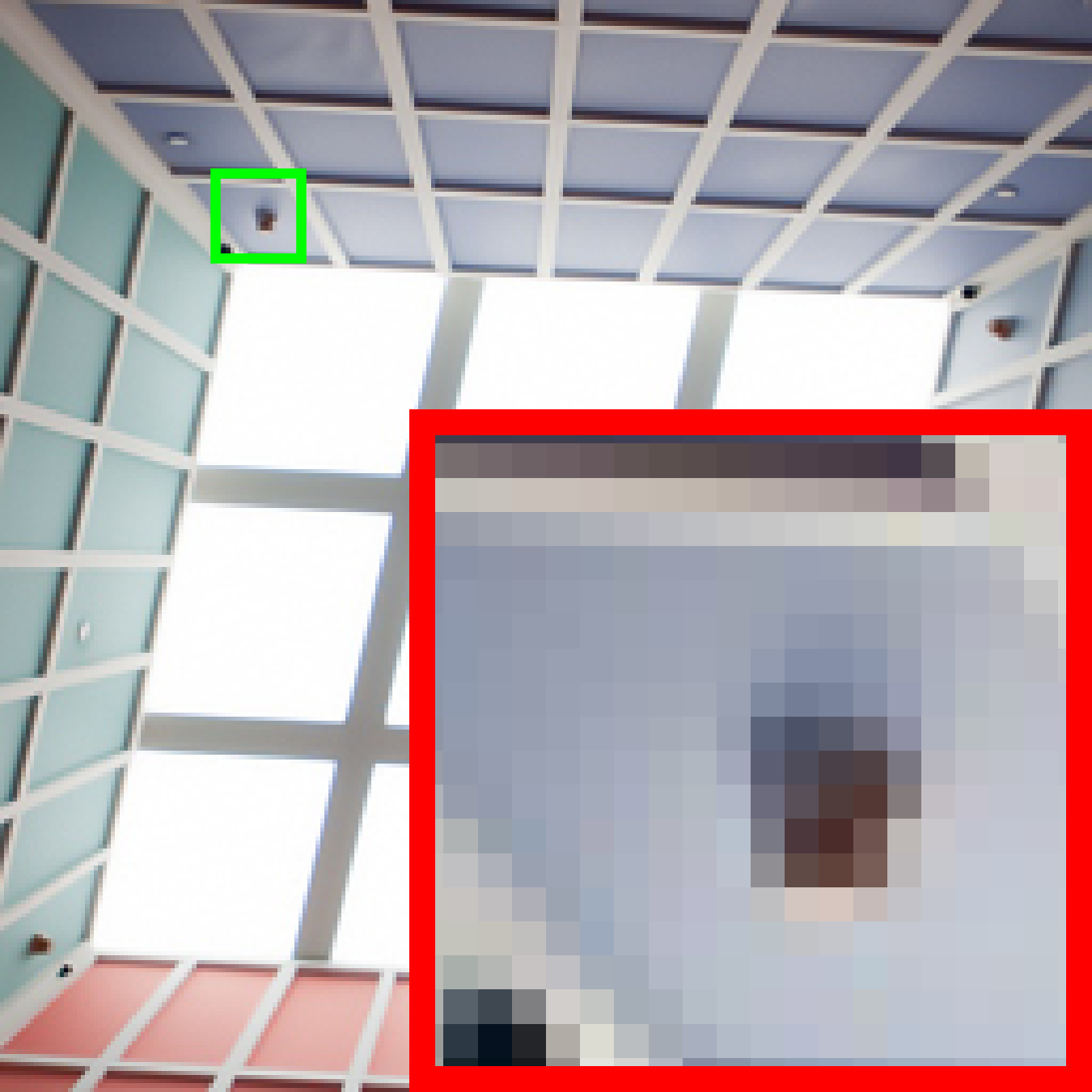}
    &\includegraphics[width=0.08\textwidth]{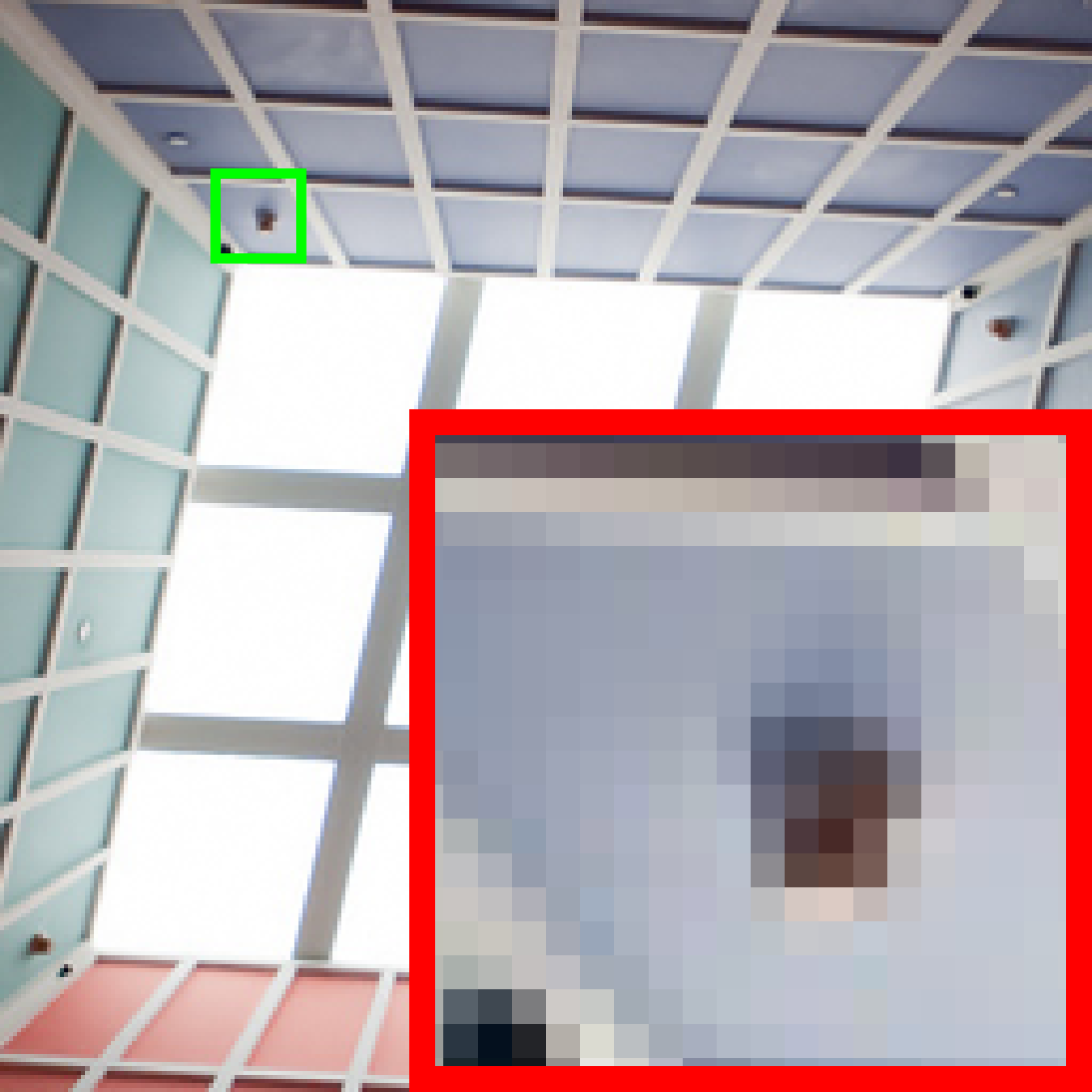}
    &\includegraphics[width=0.08\textwidth]{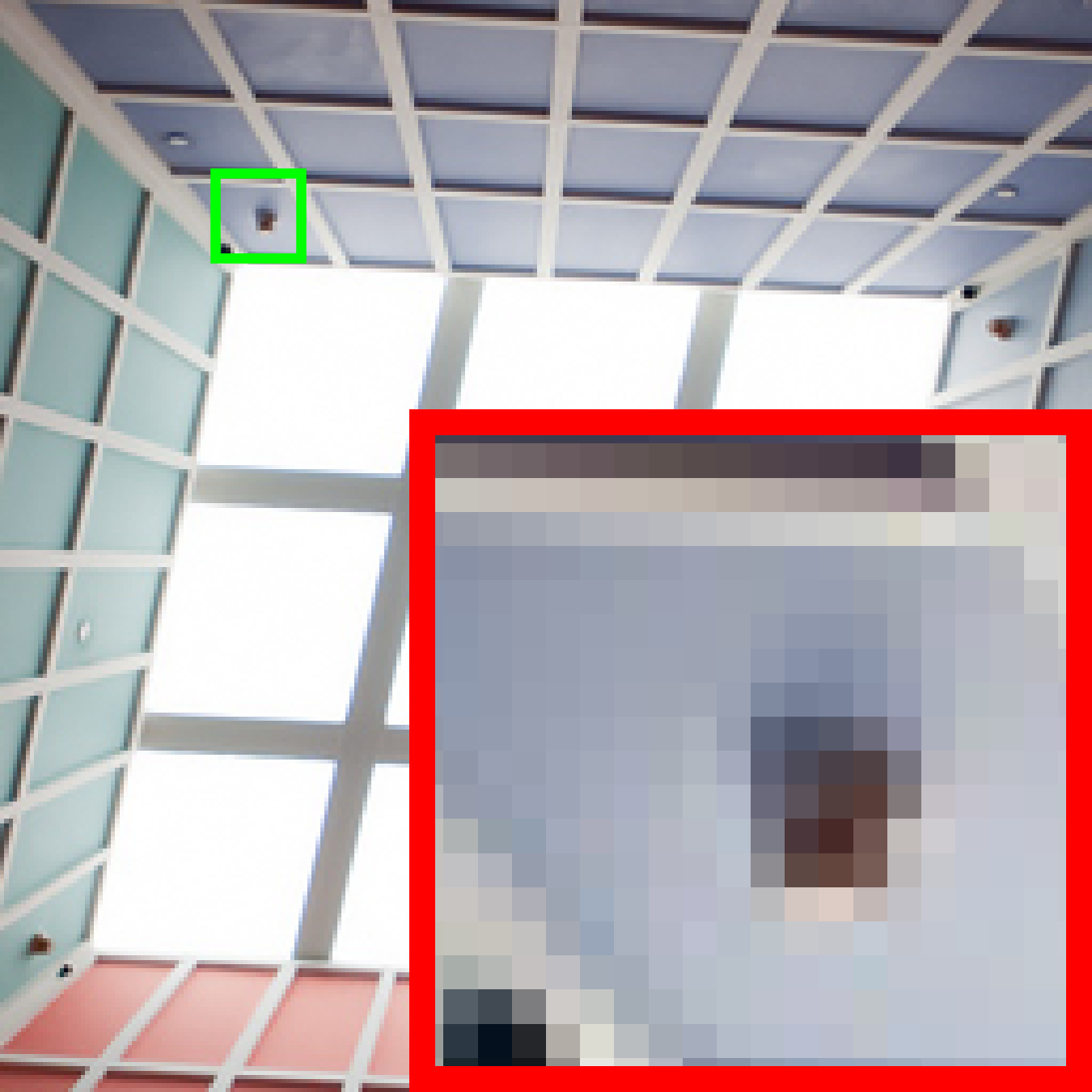}
    &\includegraphics[width=0.08\textwidth]{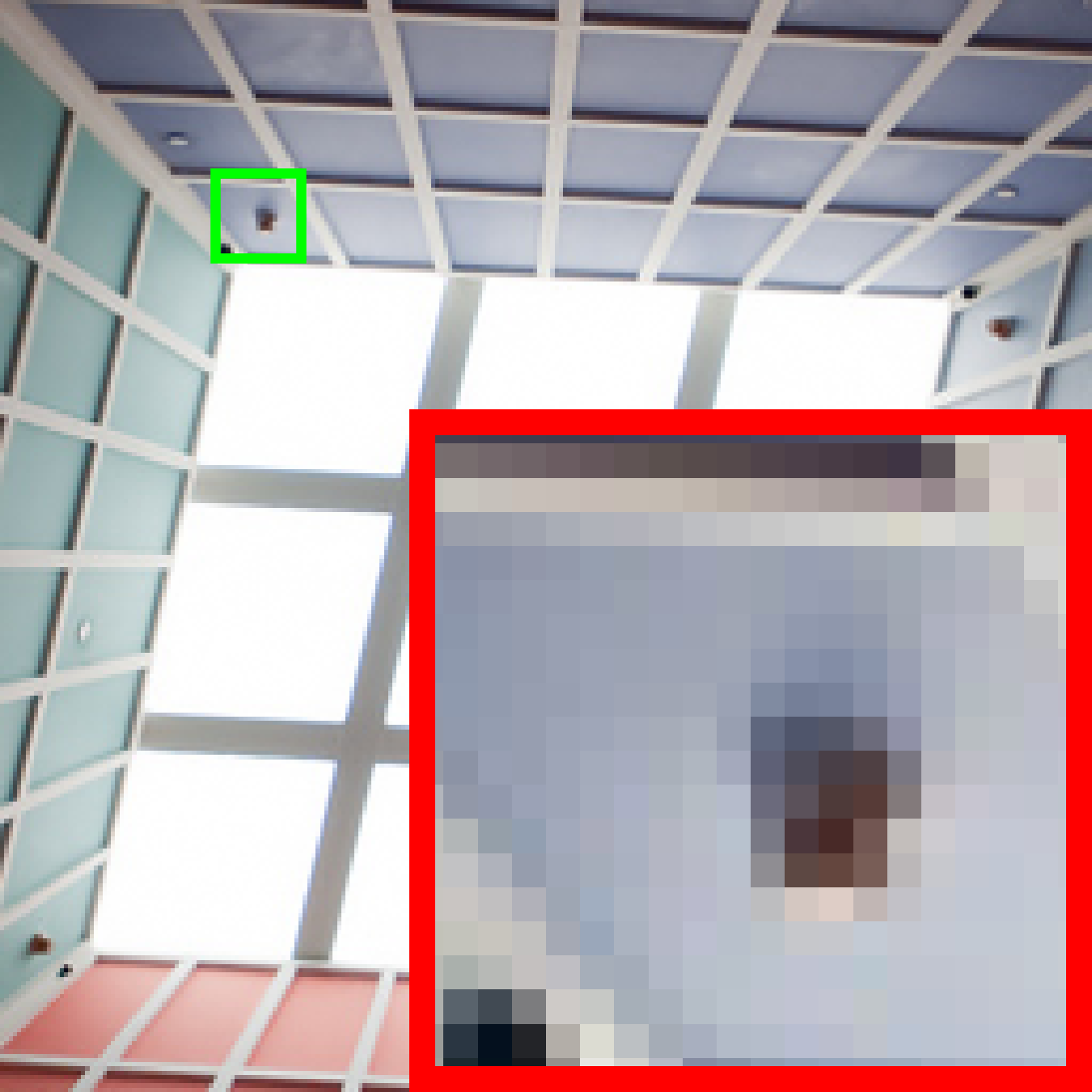}
    &\includegraphics[width=0.08\textwidth]{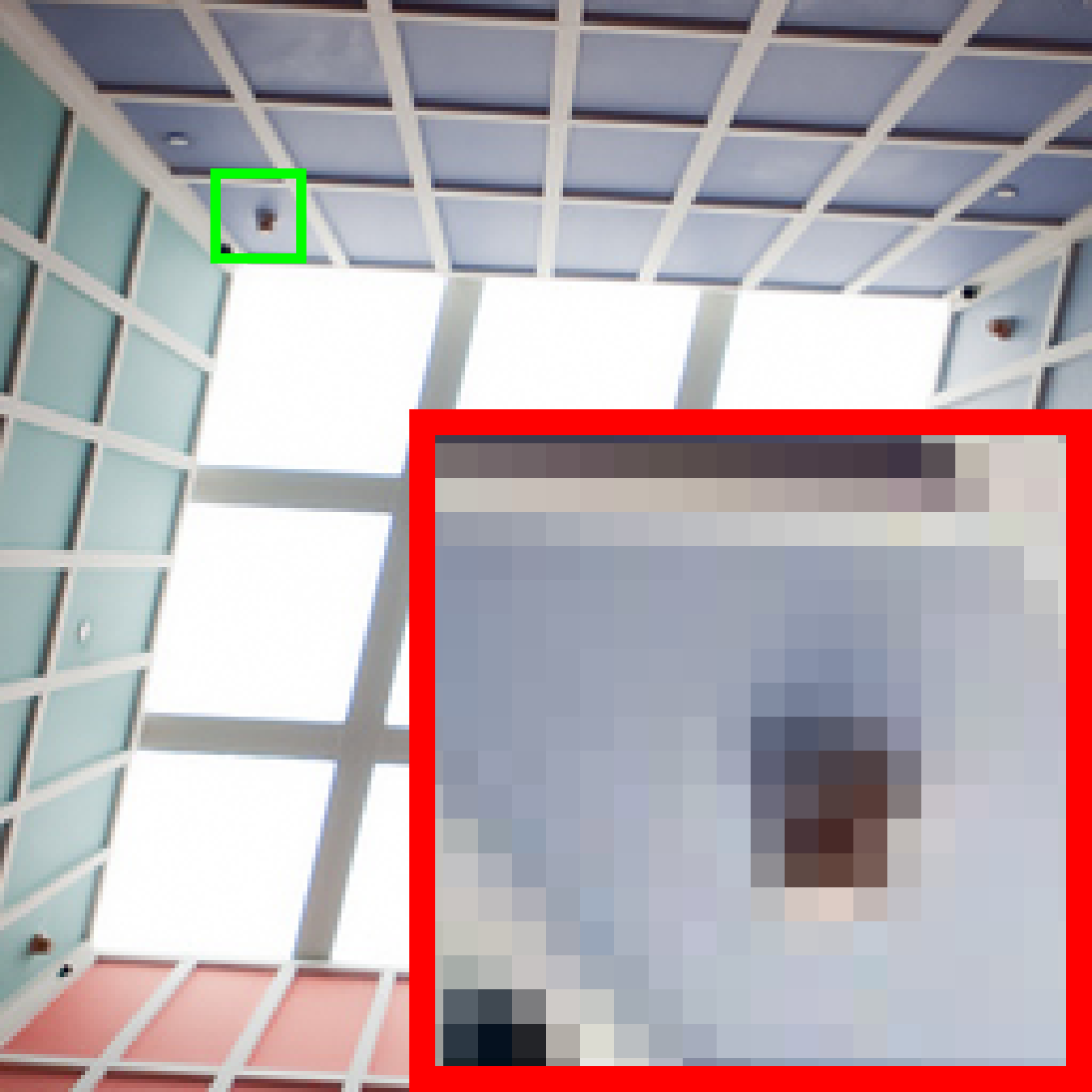}\\
    PSNR/SSIM & 34.76/0.95 & 46.16/\textbf{\textcolor{red}{1.00}} & 46.18/\textbf{\textcolor{red}{1.00}} & 47.23/\textbf{\textcolor{red}{1.00}} & 40.24/0.98 & 45.49/\textbf{\textcolor{red}{1.00}} & 46.24/\textbf{\textcolor{red}{1.00}} & 41.45/\underline{\textcolor{blue}{0.99}} & 45.02/\underline{\textcolor{blue}{0.99}} & 48.36/\textbf{\textcolor{red}{1.00}} & \underline{\textcolor{blue}{49.27}}/\textbf{\textcolor{red}{1.00}} & 48.33/\textbf{\textcolor{red}{1.00}} & \textbf{\textcolor{red}{49.45}}/\textbf{\textcolor{red}{1.00}} \\
    \includegraphics[width=0.08\textwidth]{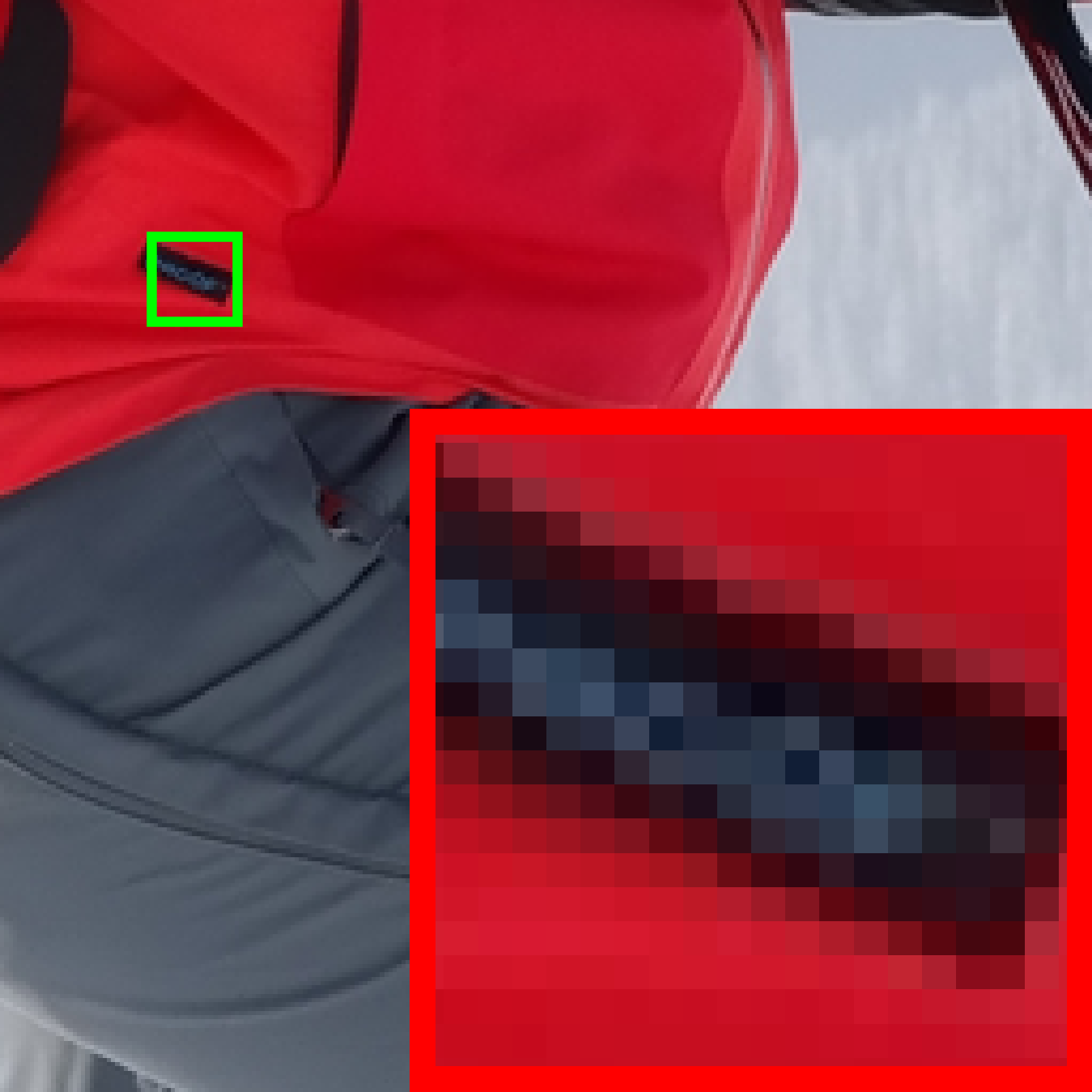}
    &\includegraphics[width=0.08\textwidth]{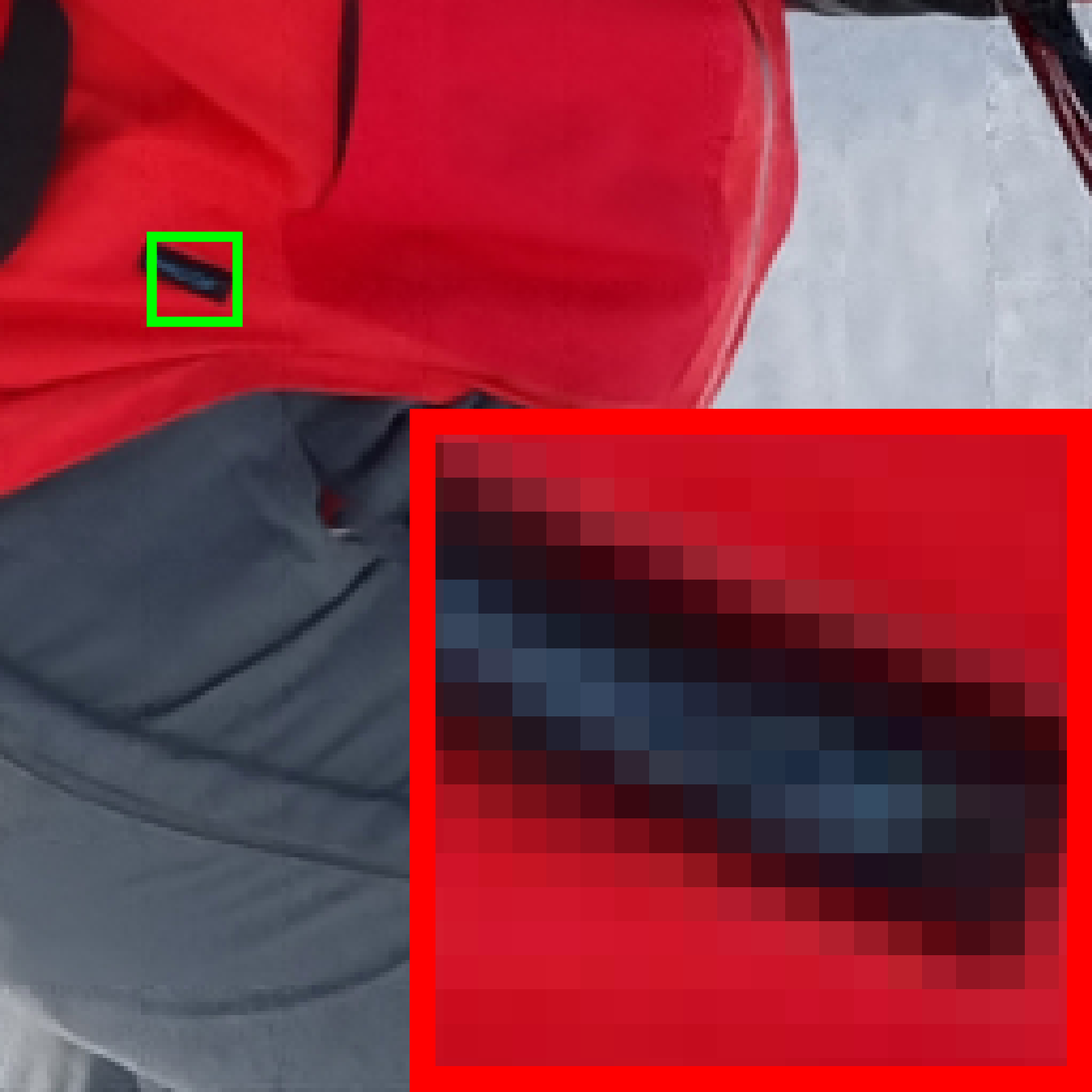}
    &\includegraphics[width=0.08\textwidth]{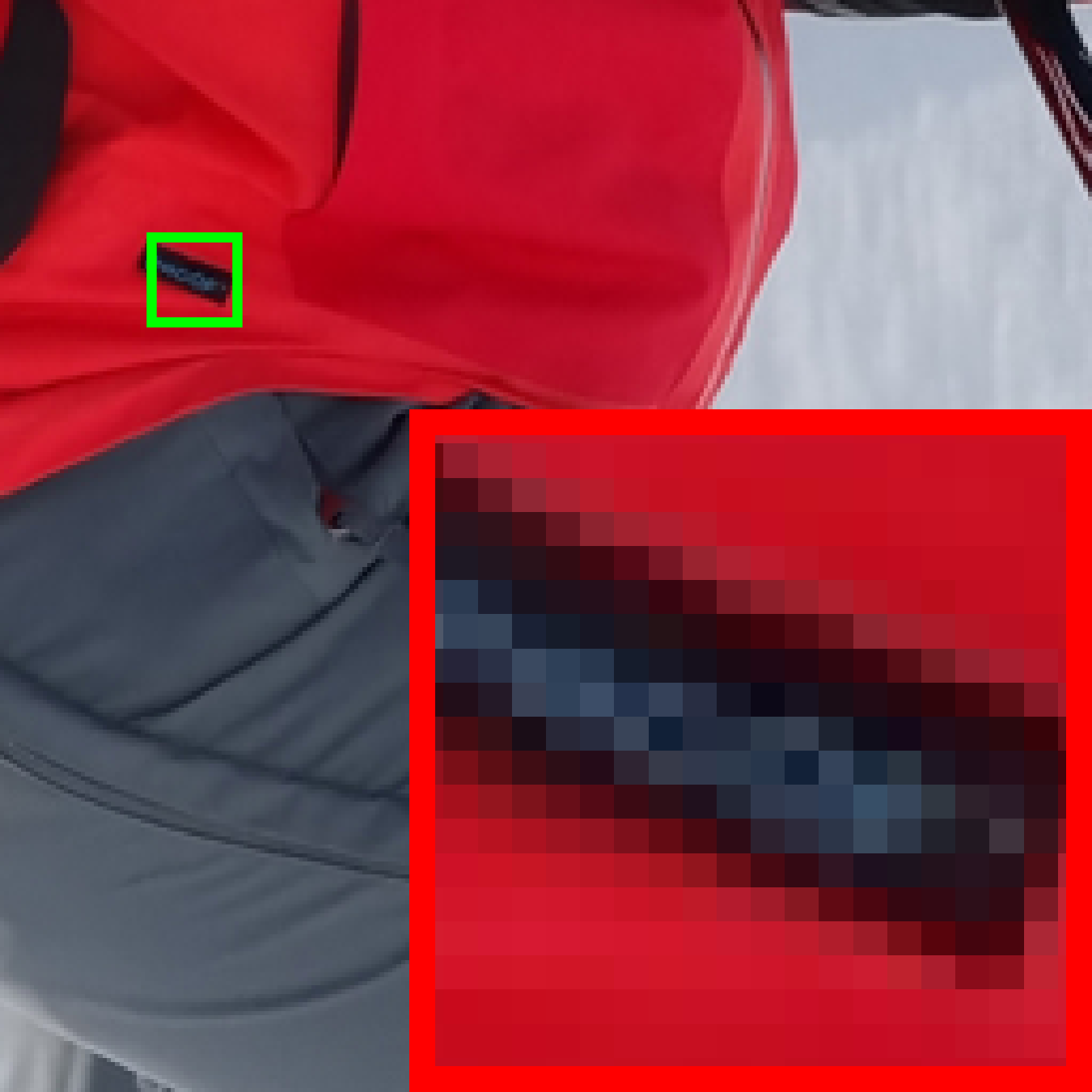}
    &\includegraphics[width=0.08\textwidth]{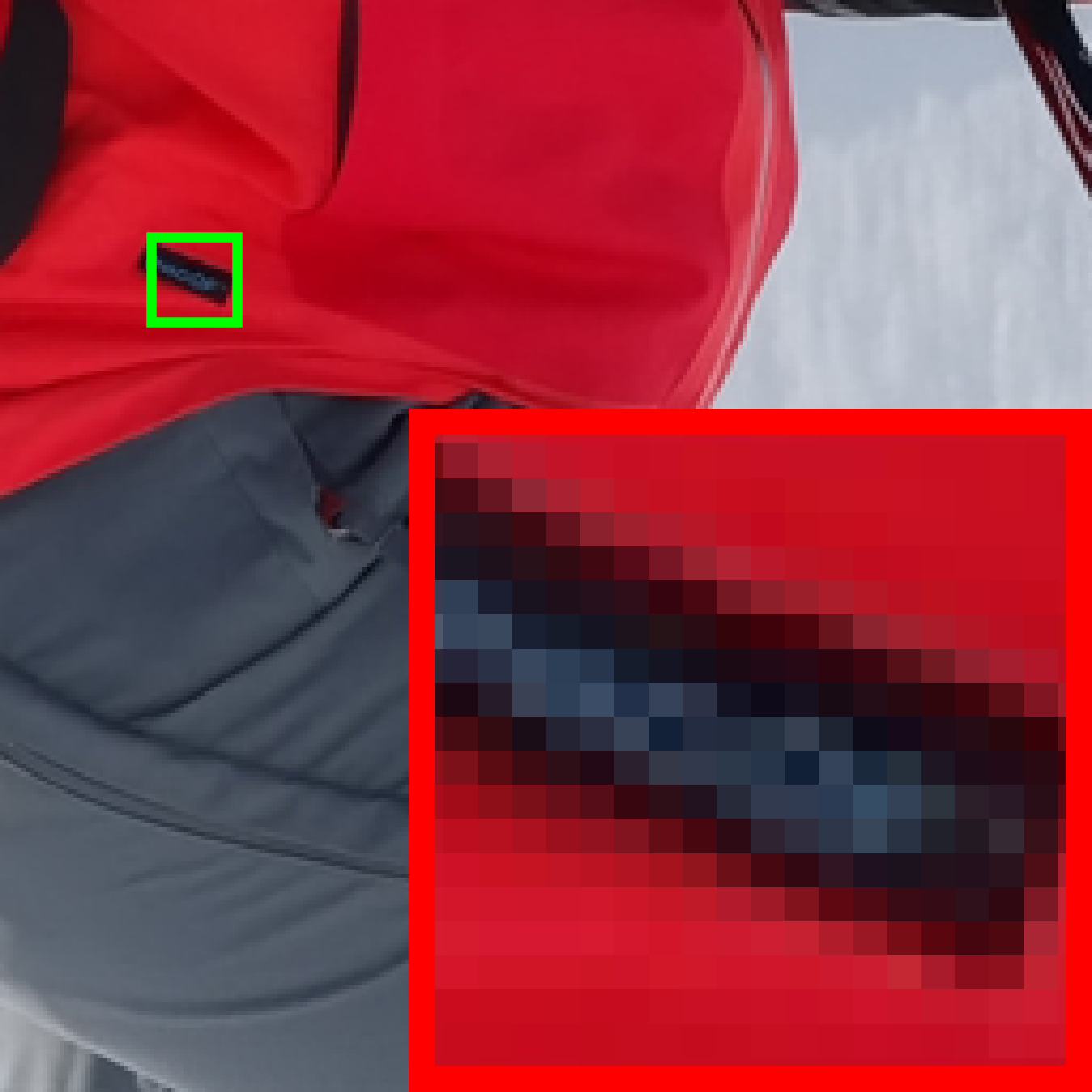}
    &\includegraphics[width=0.08\textwidth]{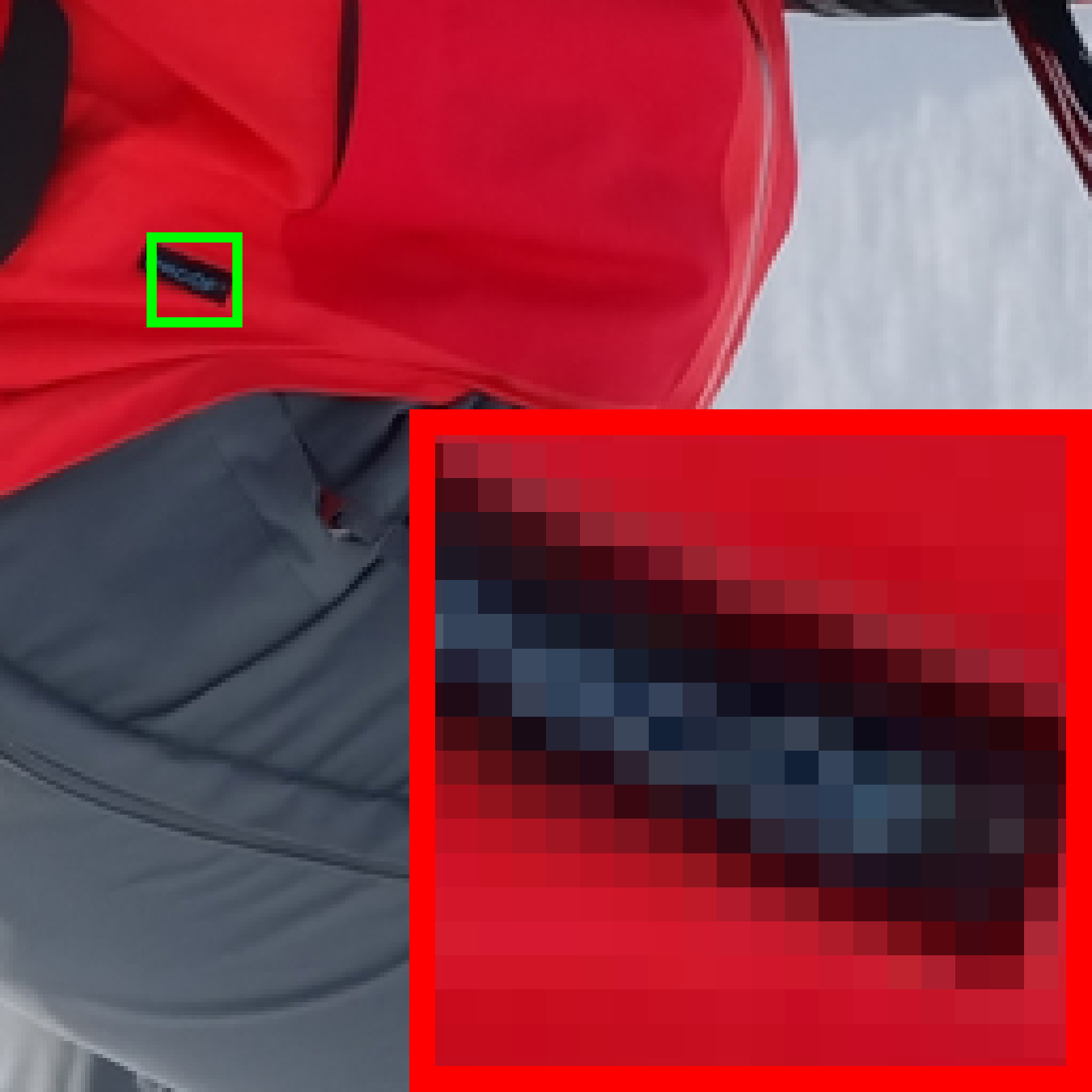}
    &\includegraphics[width=0.08\textwidth]{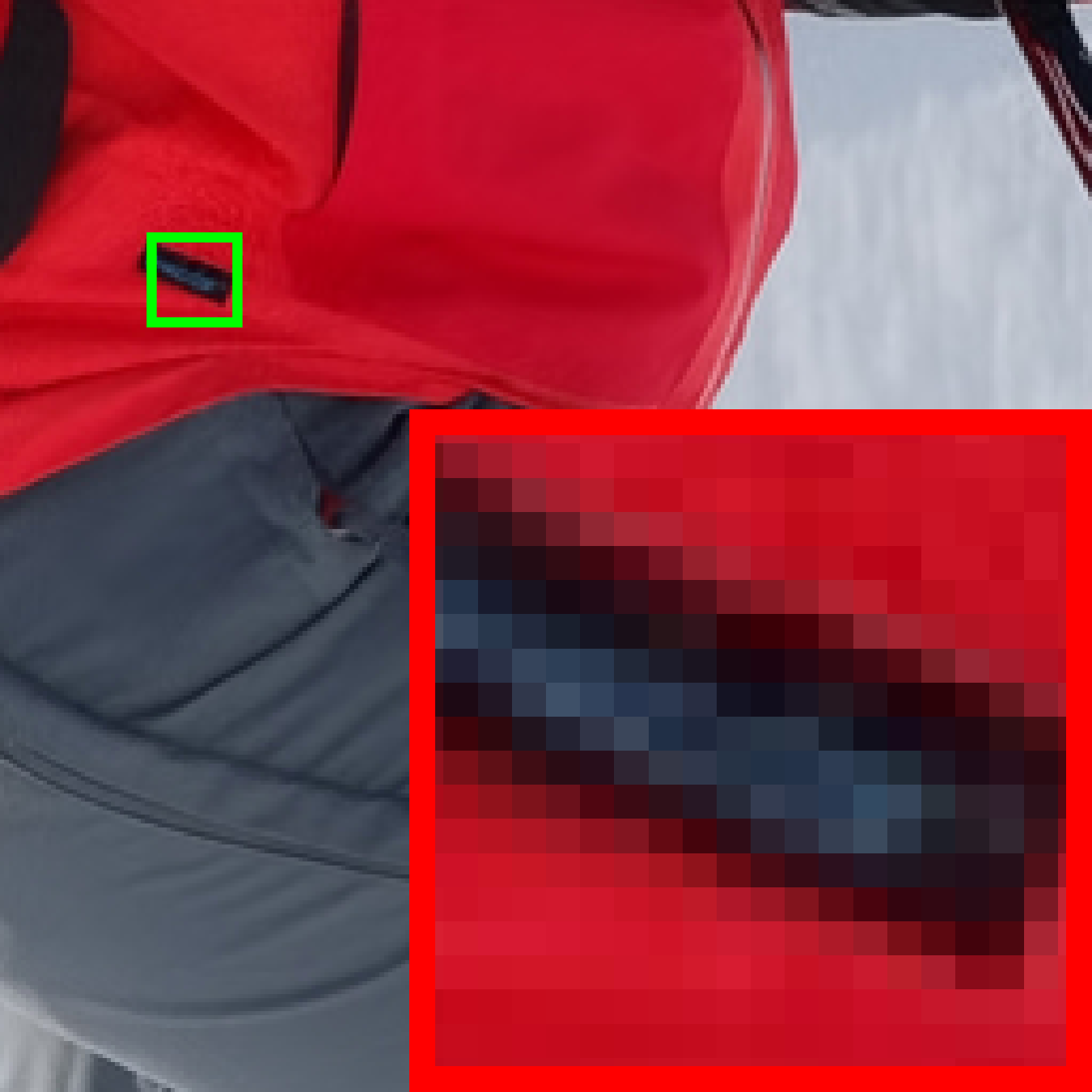}
    &\includegraphics[width=0.08\textwidth]{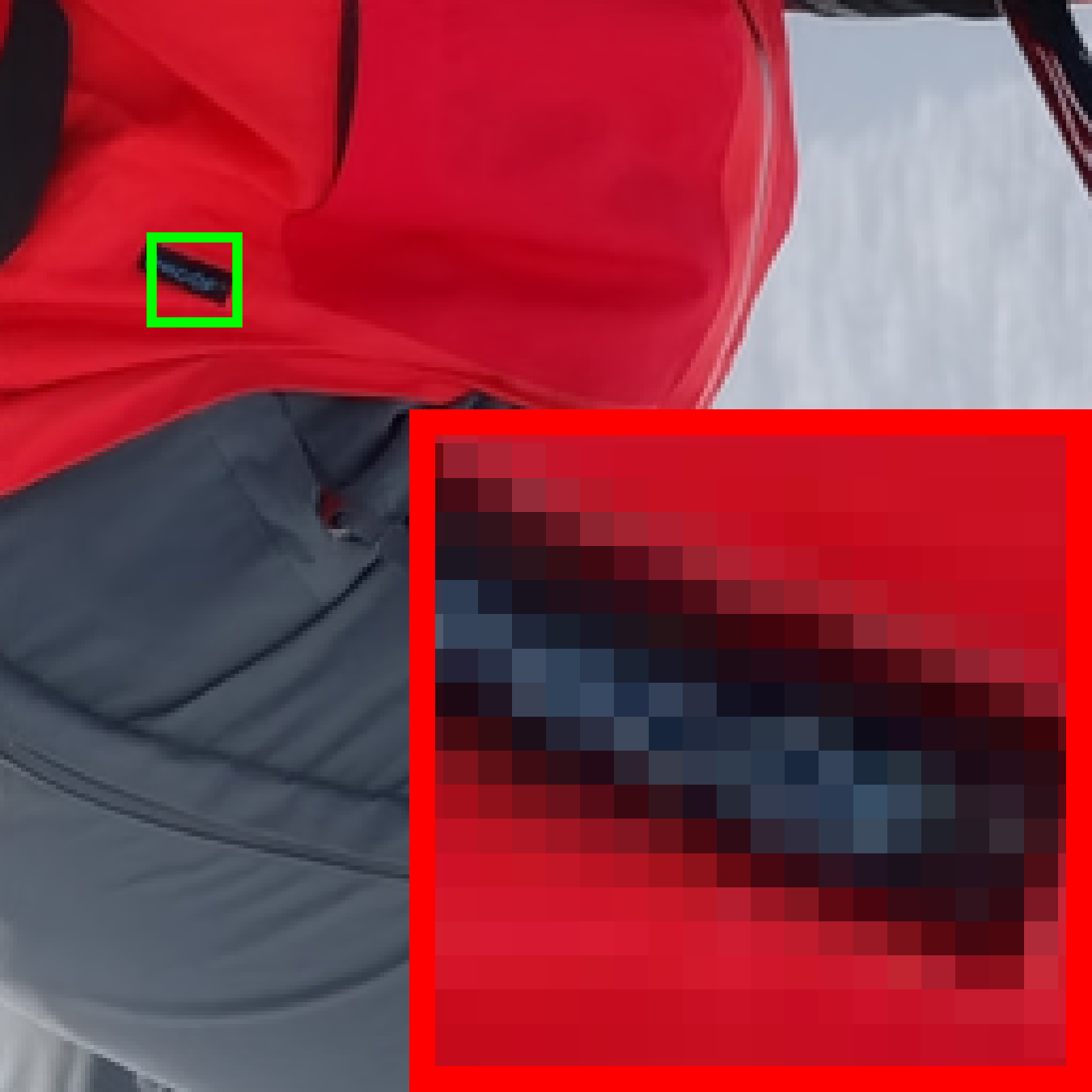}
    &\includegraphics[width=0.08\textwidth]{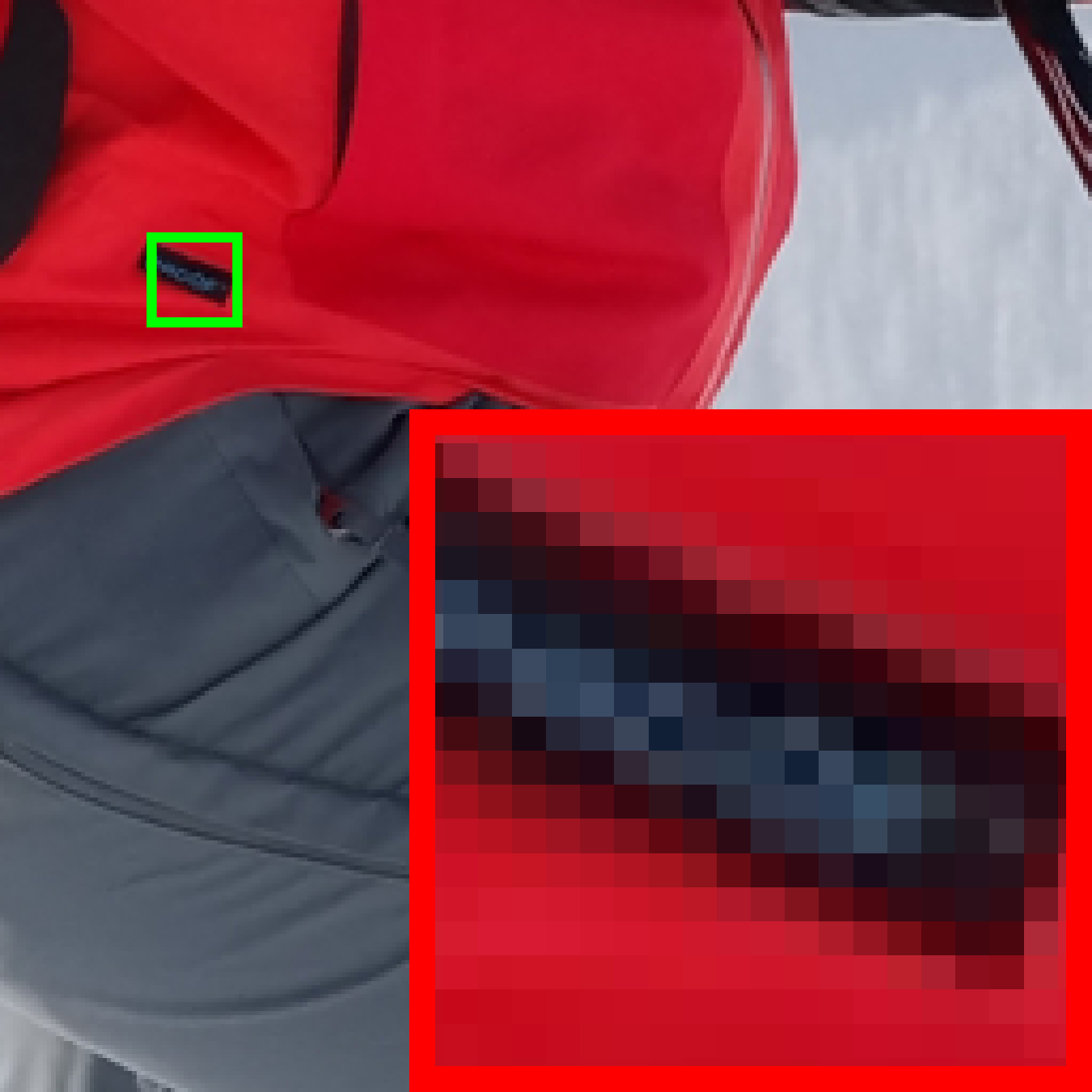}
    &\includegraphics[width=0.08\textwidth]{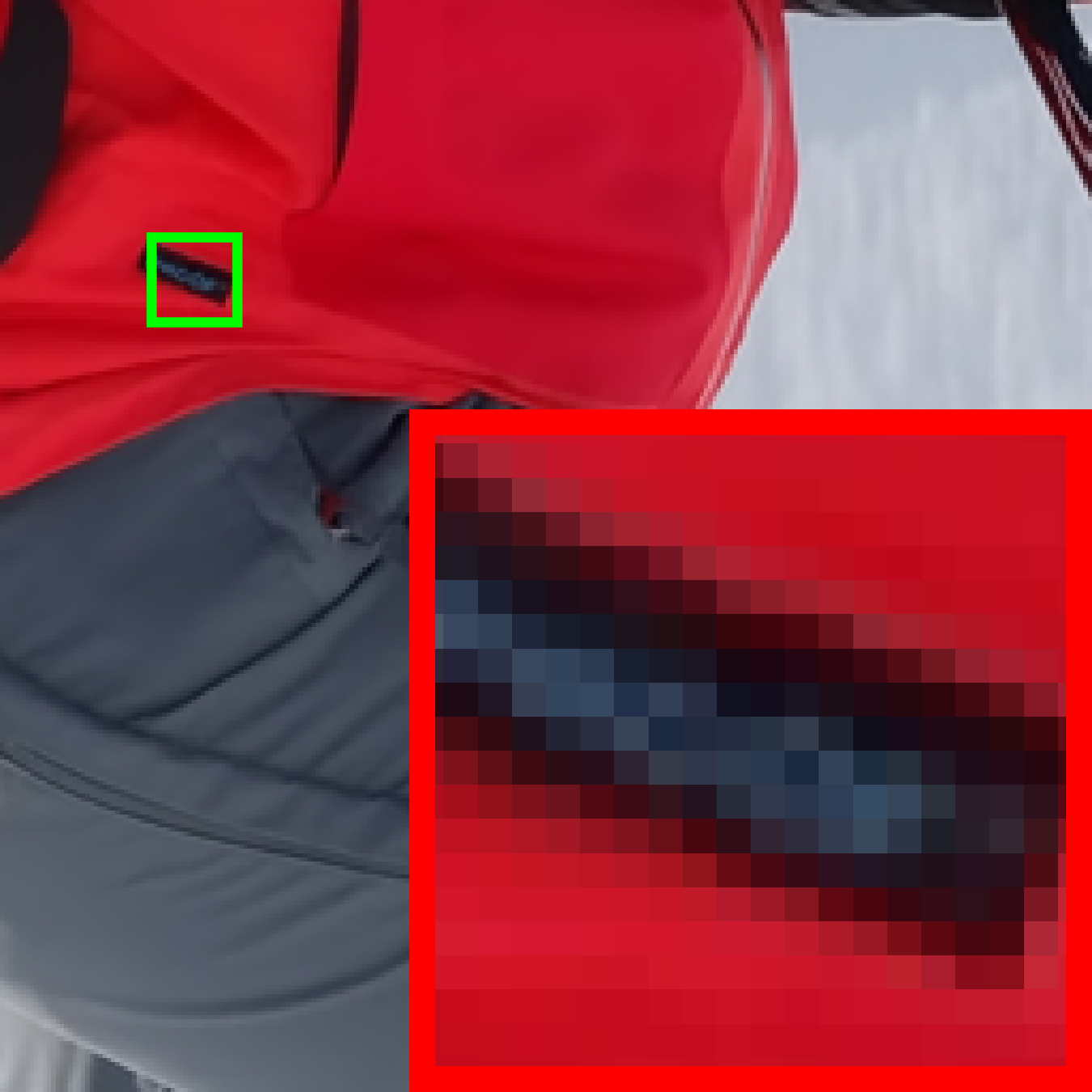}
    &\includegraphics[width=0.08\textwidth]{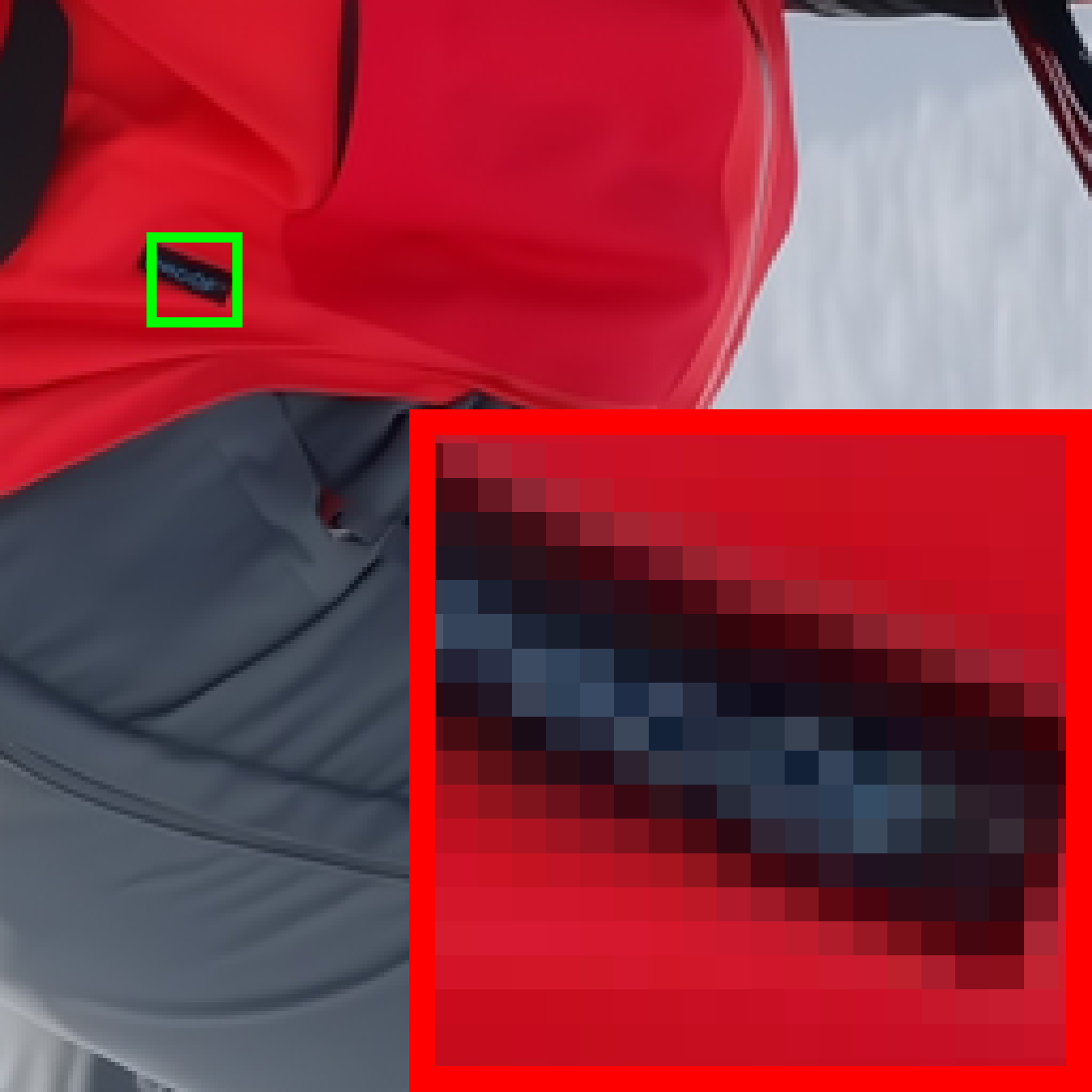}
    &\includegraphics[width=0.08\textwidth]{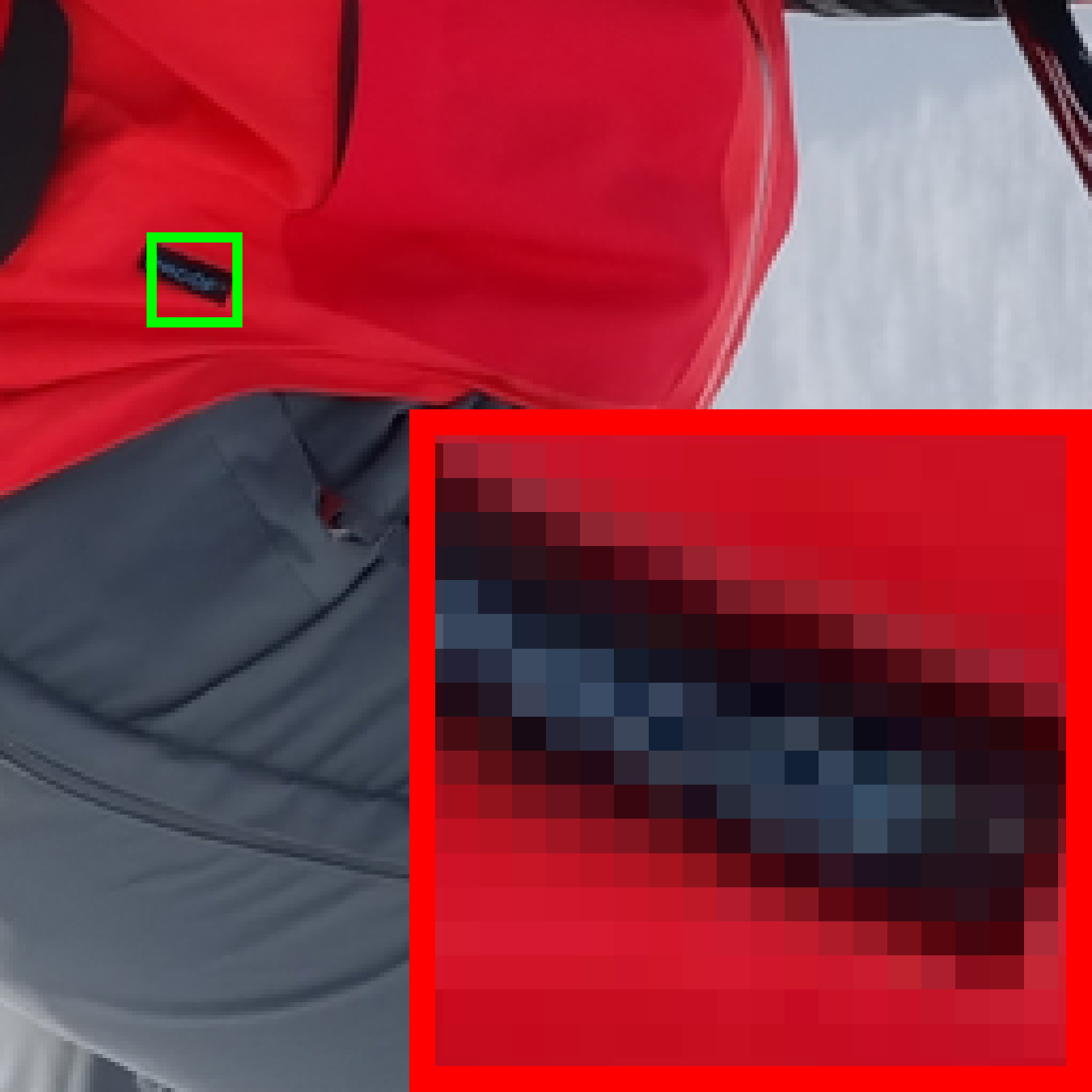}
    &\includegraphics[width=0.08\textwidth]{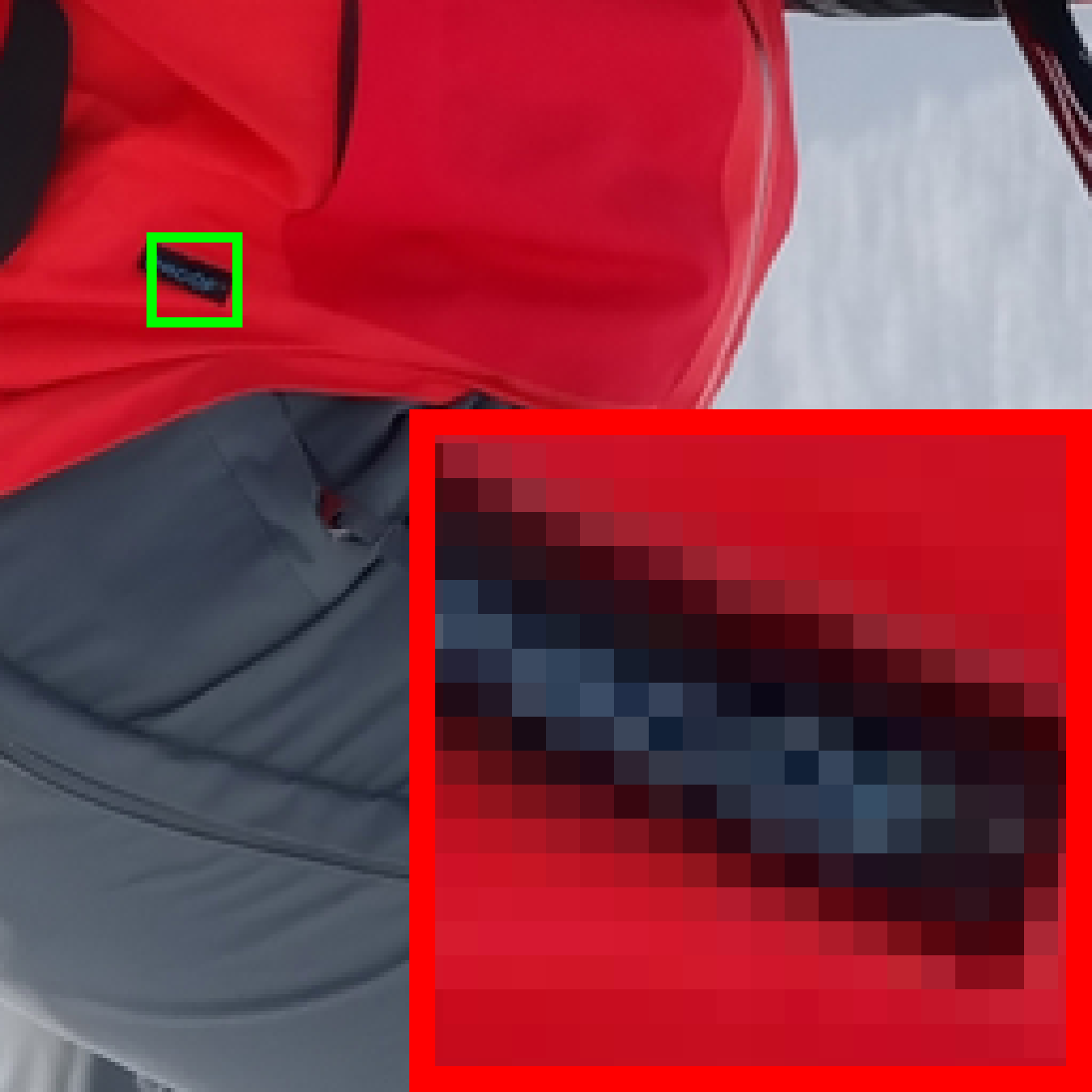}
    &\includegraphics[width=0.08\textwidth]{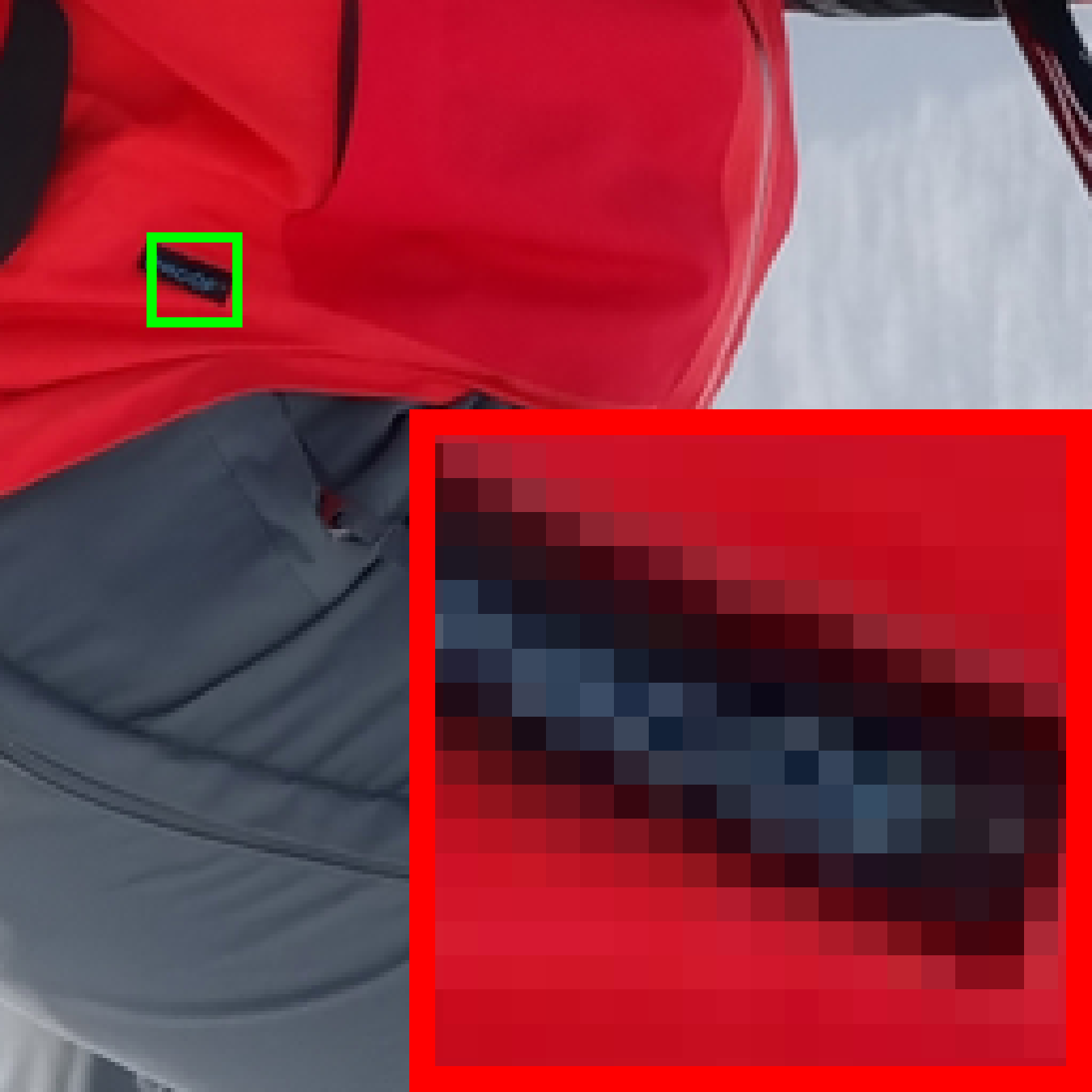}
    &\includegraphics[width=0.08\textwidth]{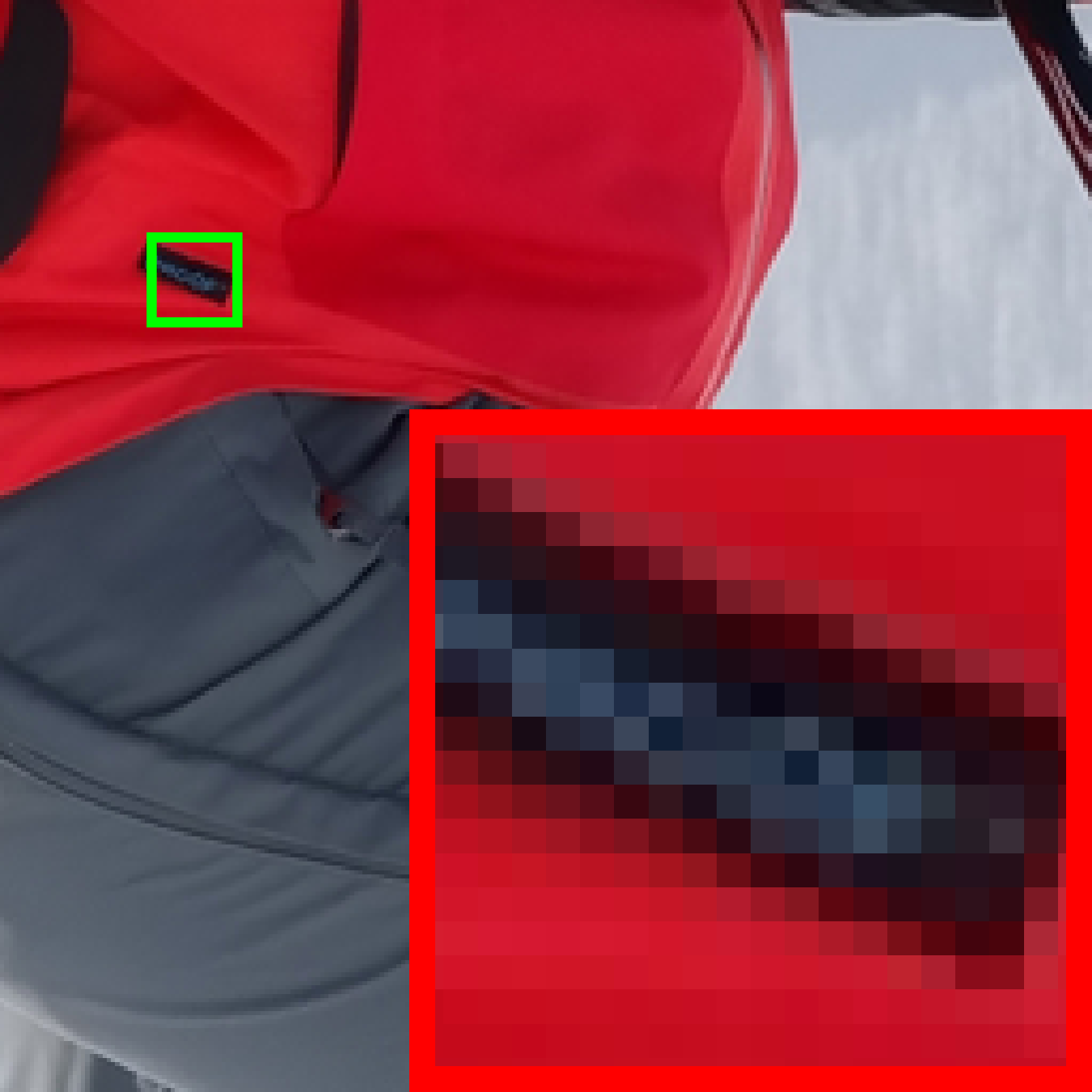}\\
    PSNR/SSIM & 38.57/0.95 & 48.11/\textbf{\textcolor{red}{0.99}} & 47.72/\textbf{\textcolor{red}{0.99}} & 48.96/\textbf{\textcolor{red}{0.99}} & 43.59/\underline{\textcolor{blue}{0.98}} & 47.18/\textbf{\textcolor{red}{0.99}} & 48.59/\textbf{\textcolor{red}{0.99}} & 47.64/\textbf{\textcolor{red}{0.99}} & 46.55/\textbf{\textcolor{red}{0.99}} & 49.74/\textbf{\textcolor{red}{0.99}} & \textbf{\textcolor{red}{50.37}}/\textbf{\textcolor{red}{0.99}} & 49.88/\textbf{\textcolor{red}{0.99}} & \underline{\textcolor{blue}{50.27}}/\textbf{\textcolor{red}{0.99}}
\end{tabular}}
\caption{Visual comparison of different methods on four natural benchmark images named ``Monarch'', ``test\_28'', ``img\_090'' and ``0894'' from Set11~\cite{kulkarni2016reconnet} \textcolor{blue}{(top)}, CBSD68~\cite{martin2001database} \textcolor{blue}{(upper middle)}, Urban100~\cite{huang2015single} \textcolor{blue}{(lower middle)}, and DIV2K~\cite{agustsson2017ntire} \textcolor{blue}{(bottom)}, respectively, with $\gamma =50\%$ and $\sigma =0$.}
\label{fig:comparison_standard_natural_images_r50}
\end{figure*}

\subsection{CS Reconstruction for 2D Natural Images}
\subsubsection{More Implementation Details}

\textbf{Block-based CS Sampling.} The fixed random sampling matrix $\A\in\Rbb^{M\times N}$ for natural images is generated by orthogonalizing all the rows of an i.i.d. Gaussian matrix \cite{adler2016deep,zhang2018ista}, \textit{i.e.}, $\A \A^\top =\mathbf{I}_M$ (with $\A^\top =\A^\dagger$). To sample a high-resolution image $\x\in\Rbb^{1\times H\times W}$, we perform a non-overlapping block division of size $N=B\times B$ and acquire the measurements $\y$ block-by-block. This approach is called block-based (or block-diagonal) CS \cite{gan2007block}, where $B=33$ and $N=1089$ \cite{kulkarni2016reconnet}. In practice, we implement the block-based operations of $\A (\cdot)$ and $\A^\top (\cdot)$ or $\A^\dagger (\cdot)$ using a convolution and a transposed convolution, respectively. The kernel weights are fixed and shared to be the reshaped $\A$ of size $M\times 1\times B\times B$, with strides being $B$ and biases being zero.

\textbf{SCL Training and Inference.} For our dual-domain loss $\Loss =\Loss_{DMC} +\alpha \Loss_{DOC}$, we set $p=1$, $\alpha =0.1$ (for $\sigma =0$), and $\alpha =0.001$ (for $\sigma =10$) by default. In external learning of stage-1, we use batch size 64 and learning rate $1\times 10^{-4}$ for 500 epochs, and learning rate $2.5\times 10^{-5}$ for fine-tuning with 100 epochs on the 88912 image blocks of size $33\times 33$ \cite{zhang2018image} from T91 \cite{dong2014learning,kulkarni2016reconnet}. For the cross-image internal learning in stage-2, we employ batch size 1 and learning rate $1\times 10^{-4}$ for $5\times 10^5$ iterations, and employ learning rate $2.5\times 10^{-5}$ for fine-tuning with $1\times 10^5$ iterations on the entire test set. During the single-image internal learning in our stage-3, we use batch size 1 and learning rate $1\times 10^{-4}$ for $5\times 10^3$ iterations, and learning rate $2.5\times 10^{-5}$ for fine-tuning with $1\times 10^3$ iterations and positional embedding (PE) size $h=w\equiv 8$ on each test image. For the self-ensemble inference in stage-4, we set the number of Monte Carlo sampling to \cite{gal2016dropout} $D=80$ and the masking probability of each matrix row to $r=0.015$. In each learning batch, different combinations of $M_1$, $M_2$, $\A_1$, $\A_2$, $\Tilde{M}_1$, $\Tilde{M}_2$, $\Tilde{\A}_1$, $\Tilde{\A}_2$, $\Tilde{\n}_1$, and $\Tilde{\n}_2$ are randomly and independently generated for different $\y$ sample instances to improve the CS task diversity and the stability of NN optimization processes.
\begin{table*}[!t]
\caption{Comparison of average PSNR (dB) among self-supervised deep CS methods on two natural image benchmarks \cite{kulkarni2016reconnet,martin2001database} with ratio $\gamma \in \{10\%, 30\%, 50\%\}$ and noise level $\sigma =10$. The best and second-best results of each case are highlighted in bold red and underlined blue, respectively.}
\label{tab:compare_sota_psnr_noisy}
\centering
\resizebox{0.75\textwidth}{!}{
\begin{tabular}{lc|ccc|ccc}
\shline
\rowcolor[HTML]{EFEFEF} 
\multicolumn{1}{l|}{\cellcolor[HTML]{EFEFEF}} &
  Test Set &
  \multicolumn{3}{c|}{\cellcolor[HTML]{EFEFEF}Set11} &
  \multicolumn{3}{c}{\cellcolor[HTML]{EFEFEF}CBSD68} \\ \hhline{>{\arrayrulecolor[HTML]{EFEFEF}}->{\arrayrulecolor{black}}|-------} 
\rowcolor[HTML]{EFEFEF} 
\multicolumn{1}{l|}{\multirow{-2}{*}{\cellcolor[HTML]{EFEFEF}Method}} &
  CS Ratio $\gamma$~~($\sigma =10$) &
  10\% &
  30\% &
  50\% &
  10\% &
  30\% &
  50\% \\ \hline \hline
\multicolumn{2}{l|}{DIP (CVPR 2018)}   & 23.96 & 27.83 & 29.48 & 22.25 & 25.36 & 27.19 \\
\multicolumn{2}{l|}{BCNN (ECCV 2020)}   & 25.31 & 29.34 & 31.09 & 23.80 & 26.98 & 28.93        \\
\multicolumn{2}{l|}{REI (CVPR 2022)}   & 21.53 & 29.02 & 29.65 & 22.14 & 26.93 & 27.84       \\
\multicolumn{2}{l|}{ASGLD (CVPR 2022)}   & 24.42 & 26.81 & 28.26 & 23.07 & 25.87 & 28.21      \\
\multicolumn{2}{l|}{DDSSL (ECCV 2022)}   & 26.12 & 30.38 & 32.15 & 24.66 & 28.26 & 30.25       \\ \hline \hline
\multicolumn{2}{l|}{\textbf{SC-CNN (Ours)}}   & 26.15 & 30.47 & 32.31 & 24.86 & 28.20 & 30.14        \\
\multicolumn{2}{l|}{\textbf{SC-CNN$^+$ (Ours)}}   & \textcolor{red}{\textbf{27.61}} & \textcolor{red}{\textbf{31.75}} & \textcolor{red}{\textbf{33.31}} & \textcolor{red}{\textbf{25.01}} & \textcolor{red}{\textbf{28.47}} & \textcolor{red}{\textbf{30.42}} \\
\multicolumn{2}{l|}{\textbf{SCT (Ours)}}   & 26.39 & 30.70 & 32.51 & \textcolor{blue}{\underline{25.00}} & 28.32 & 30.25 \\
\multicolumn{2}{l|}{\textbf{SCT$^+$ (Ours)}}   & \textcolor{blue}{\underline{27.58}} & \textcolor{blue}{\underline{31.48}} & \textcolor{blue}{\underline{33.13}} & 24.93 & \textcolor{blue}{\underline{28.46}} & \textcolor{blue}{\underline{30.30}} \\
\shline
\end{tabular}}
\end{table*}

\begin{figure*}[!t]
\setlength{\tabcolsep}{0.5pt}
\hspace{-4pt}
\resizebox{1.0\textwidth}{!}{
\tiny
\begin{tabular}{cccccccccc}
    GT & DIP & BCNN & EI & ASGLD & DDSSL & \textbf{SC-CNN} & \textbf{SC-CNN$^\text{+}$} & \textbf{SCT} & \textbf{SCT$^\text{+}$}\\
    \includegraphics[width=0.08\textwidth]{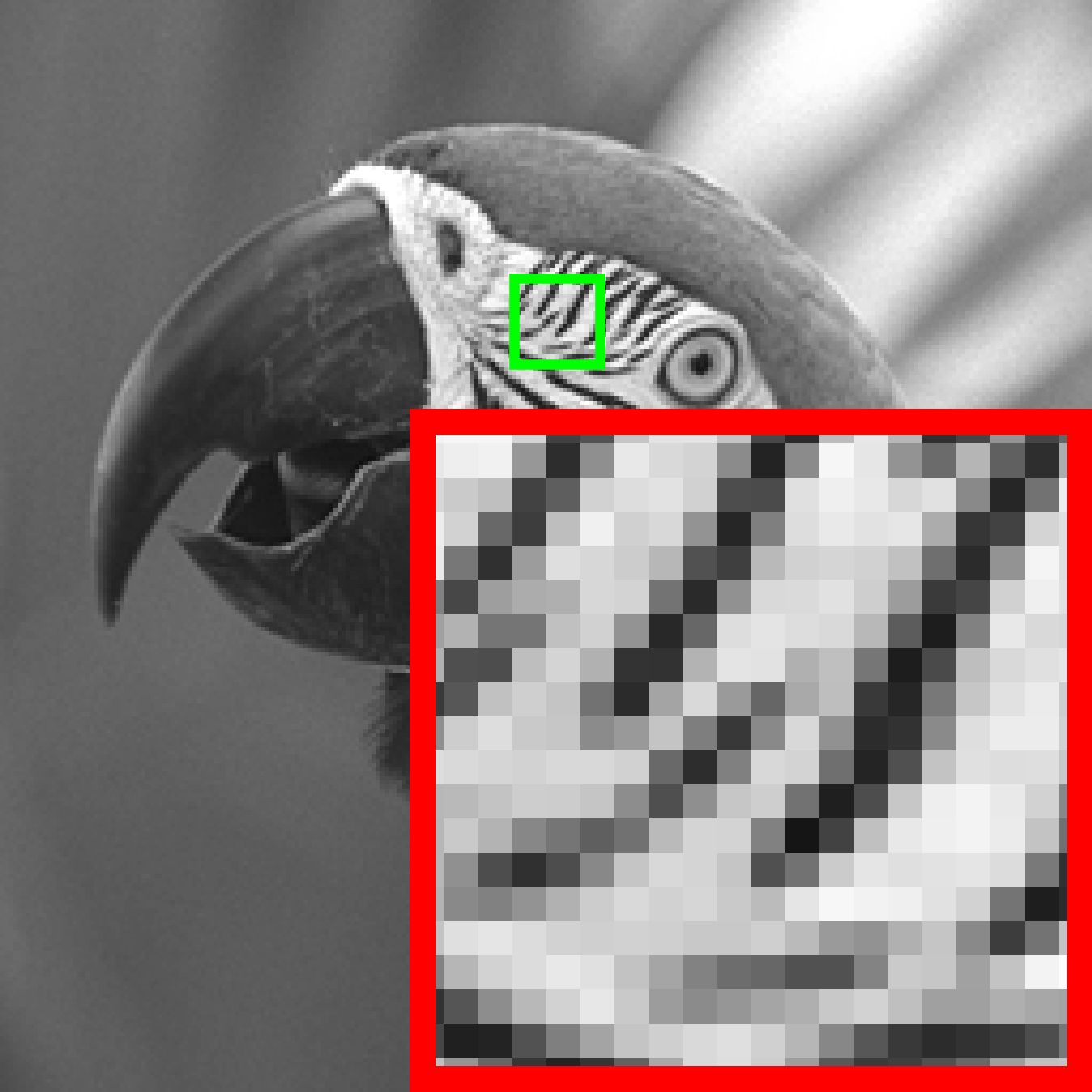}
    &\includegraphics[width=0.08\textwidth]{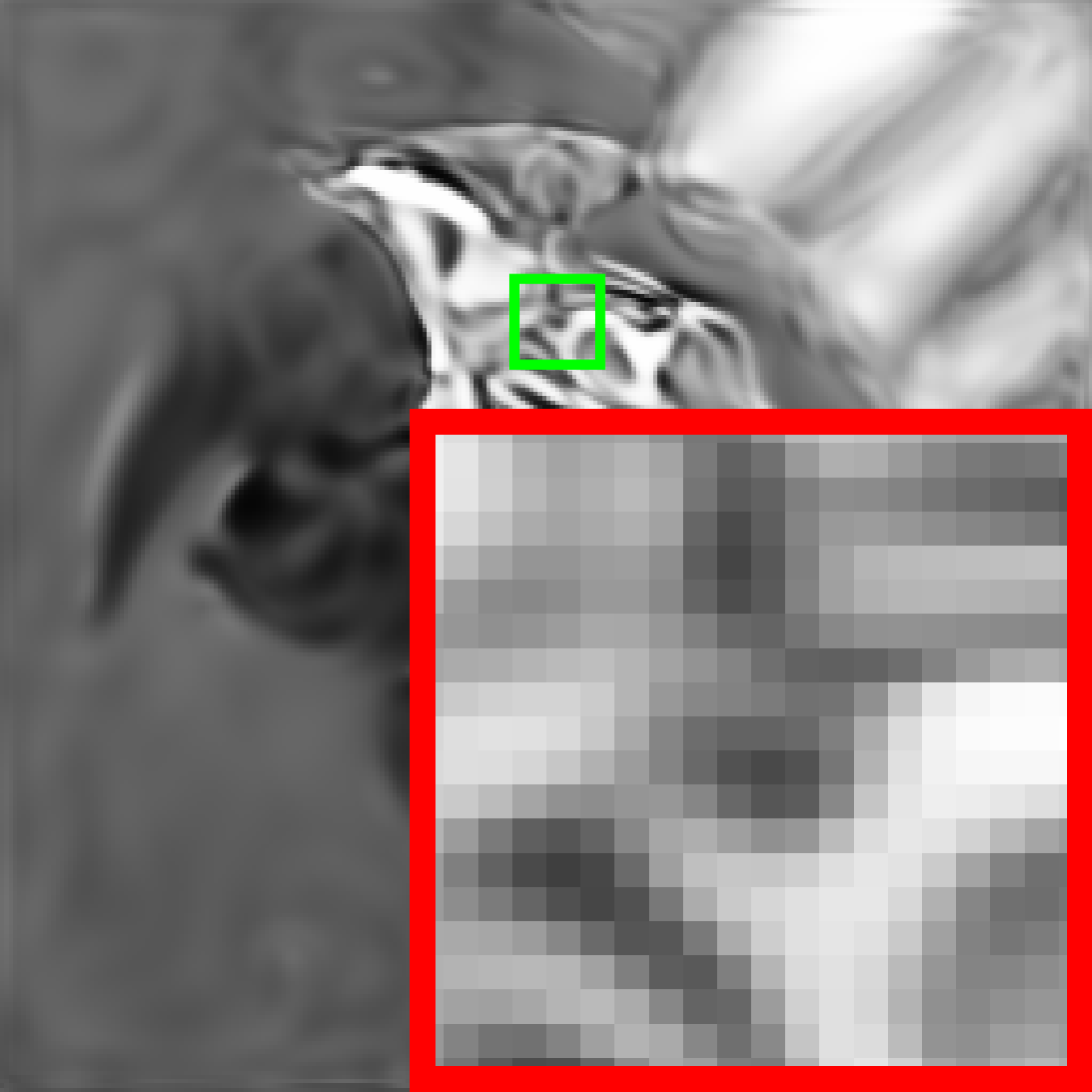}
    &\includegraphics[width=0.08\textwidth]{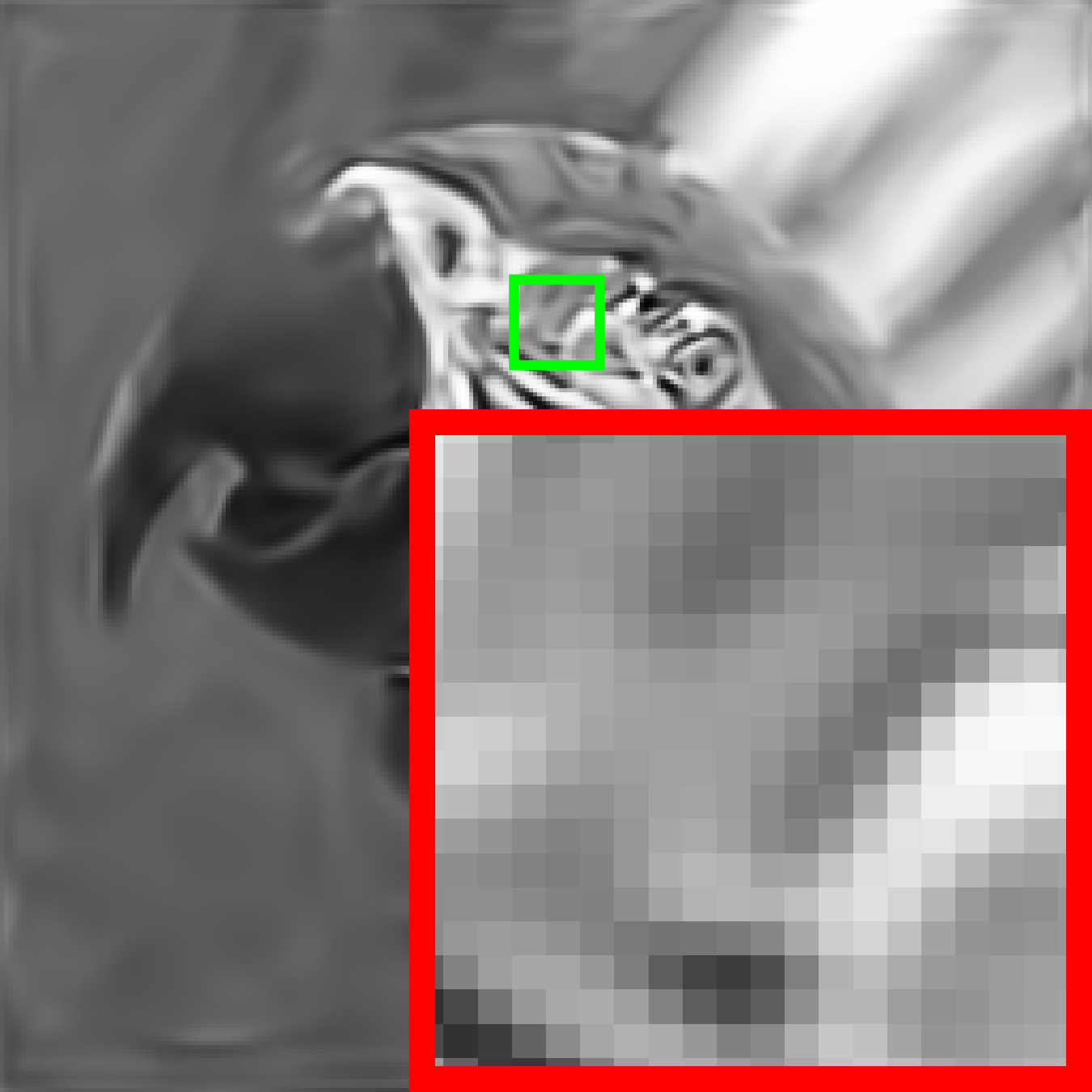}
    &\includegraphics[width=0.08\textwidth]{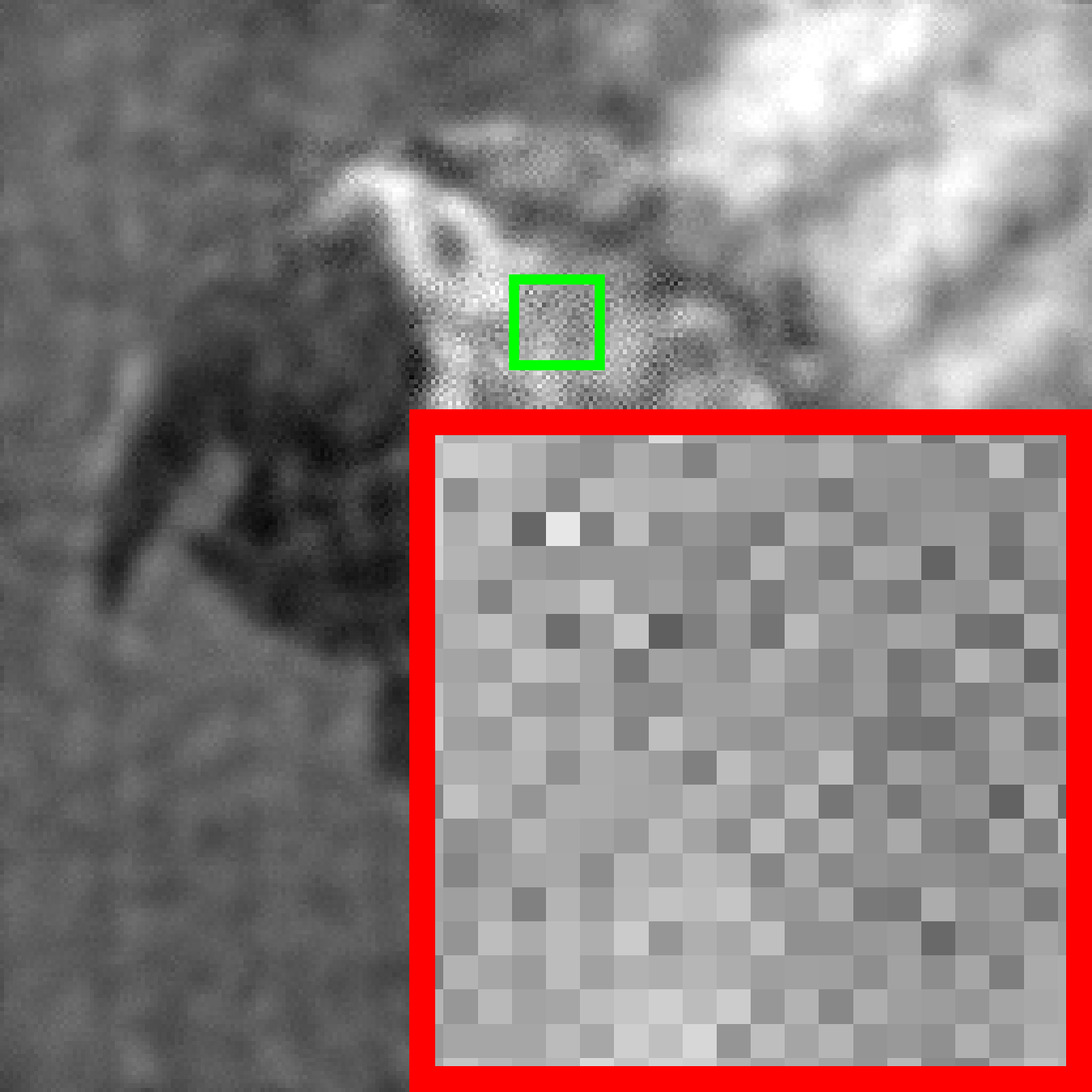}
    &\includegraphics[width=0.08\textwidth]{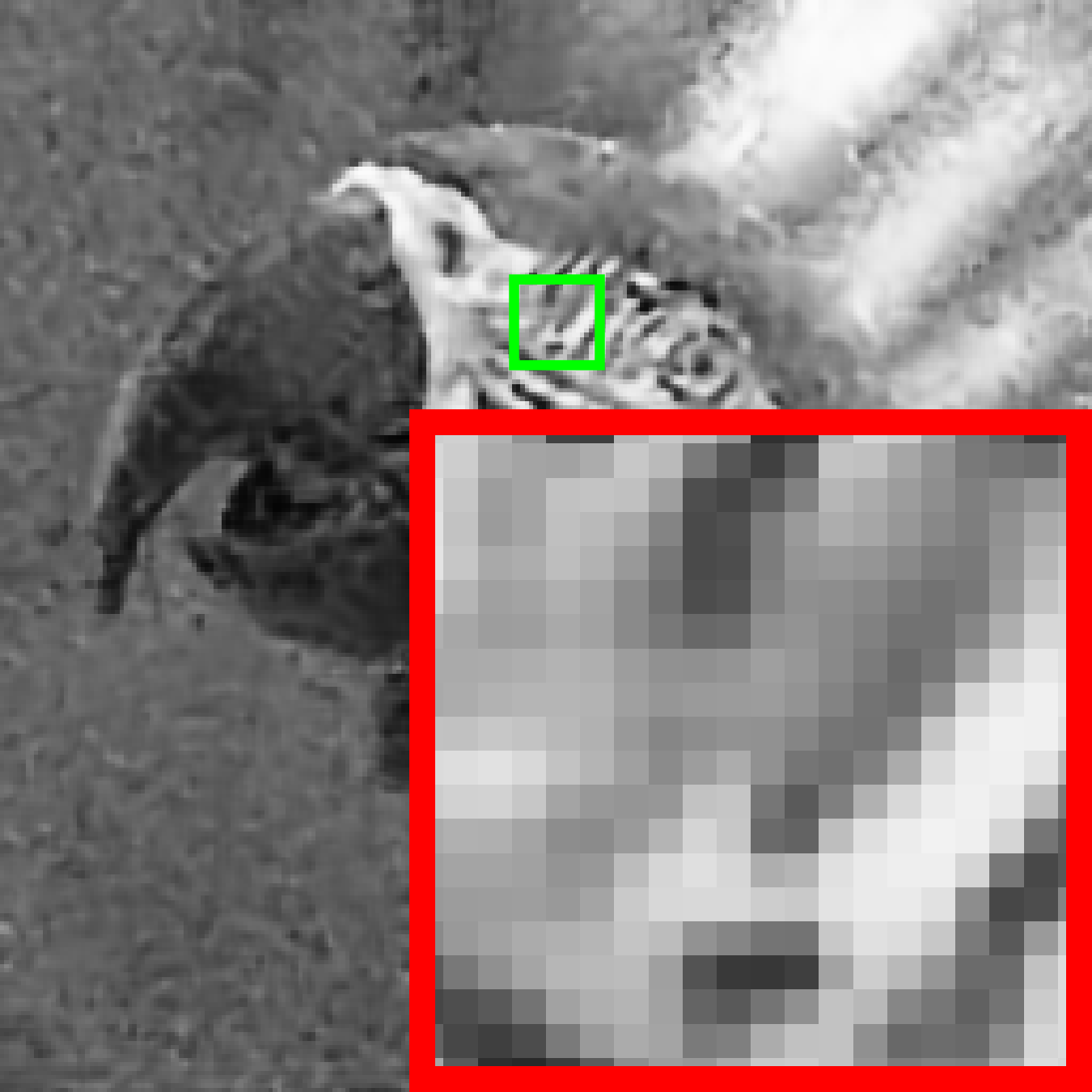}
    &\includegraphics[width=0.08\textwidth]{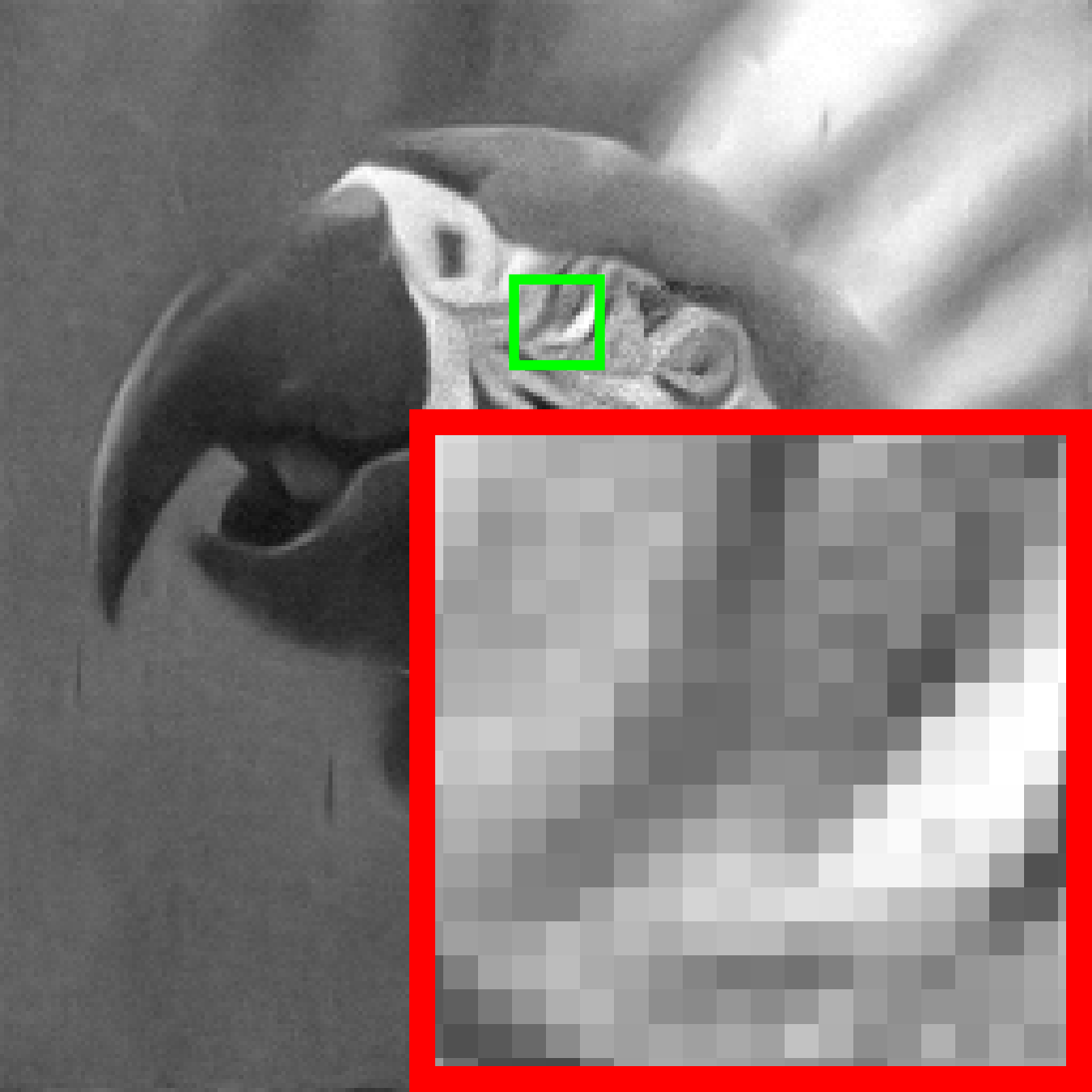}
    &\includegraphics[width=0.08\textwidth]{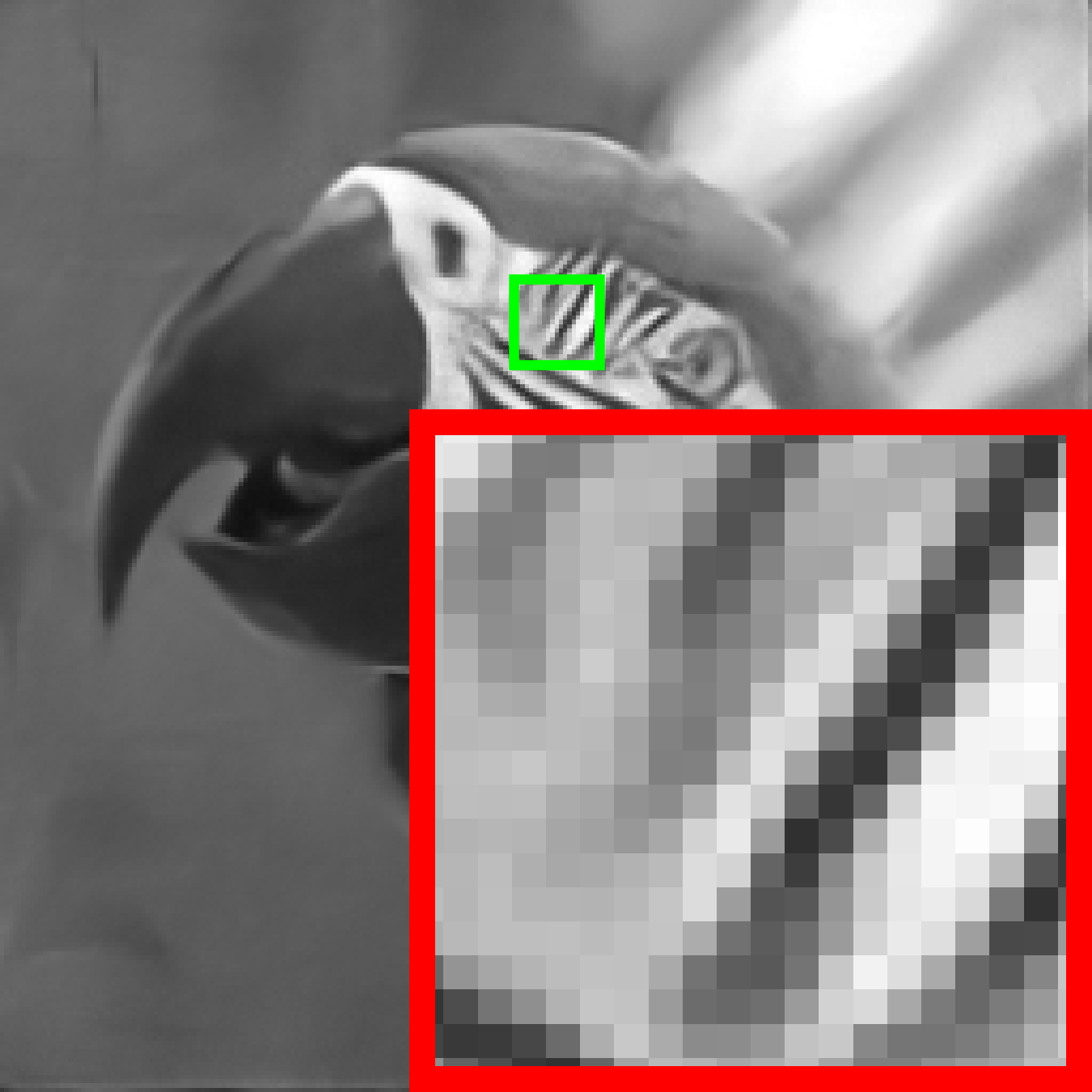}
    &\includegraphics[width=0.08\textwidth]{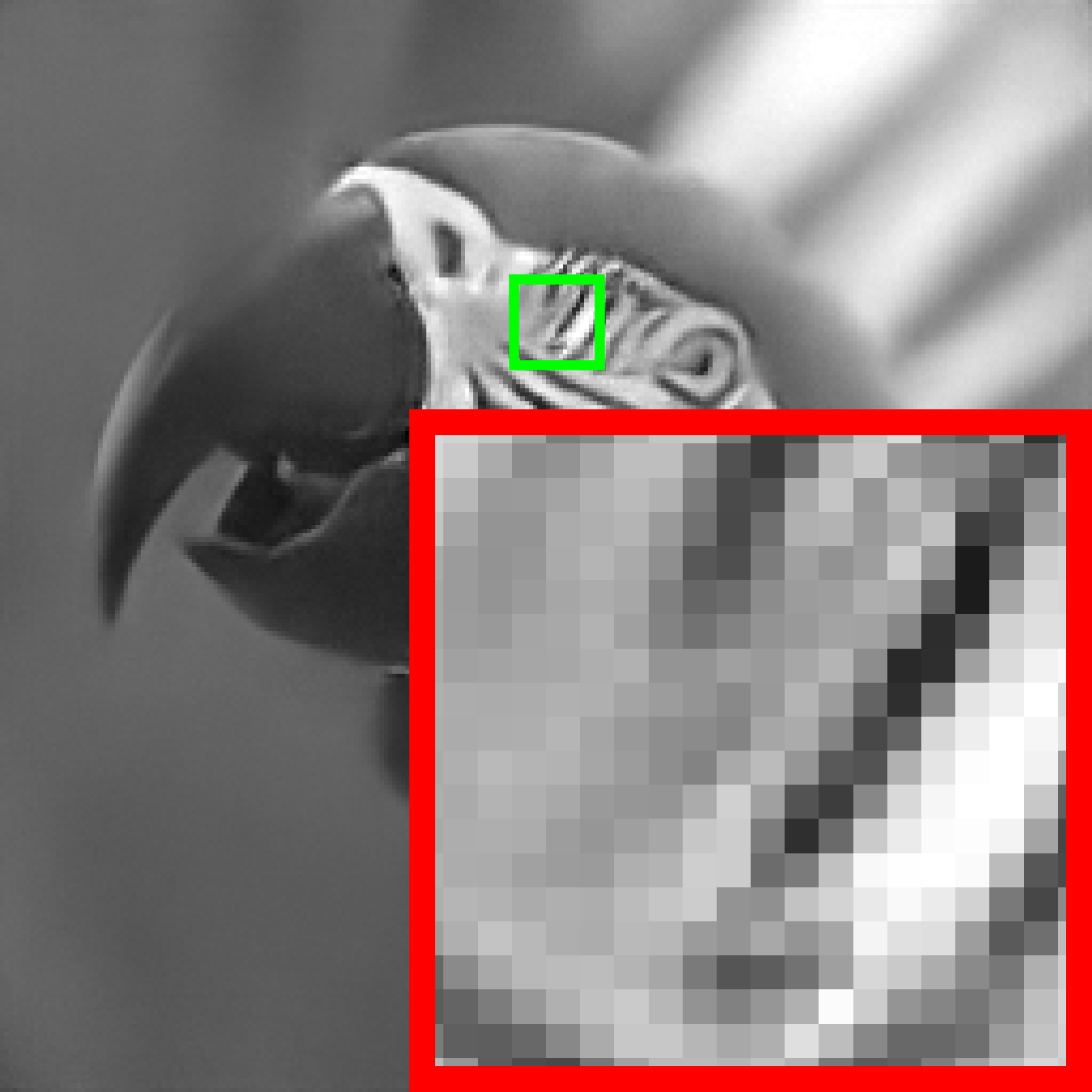}
    &\includegraphics[width=0.08\textwidth]{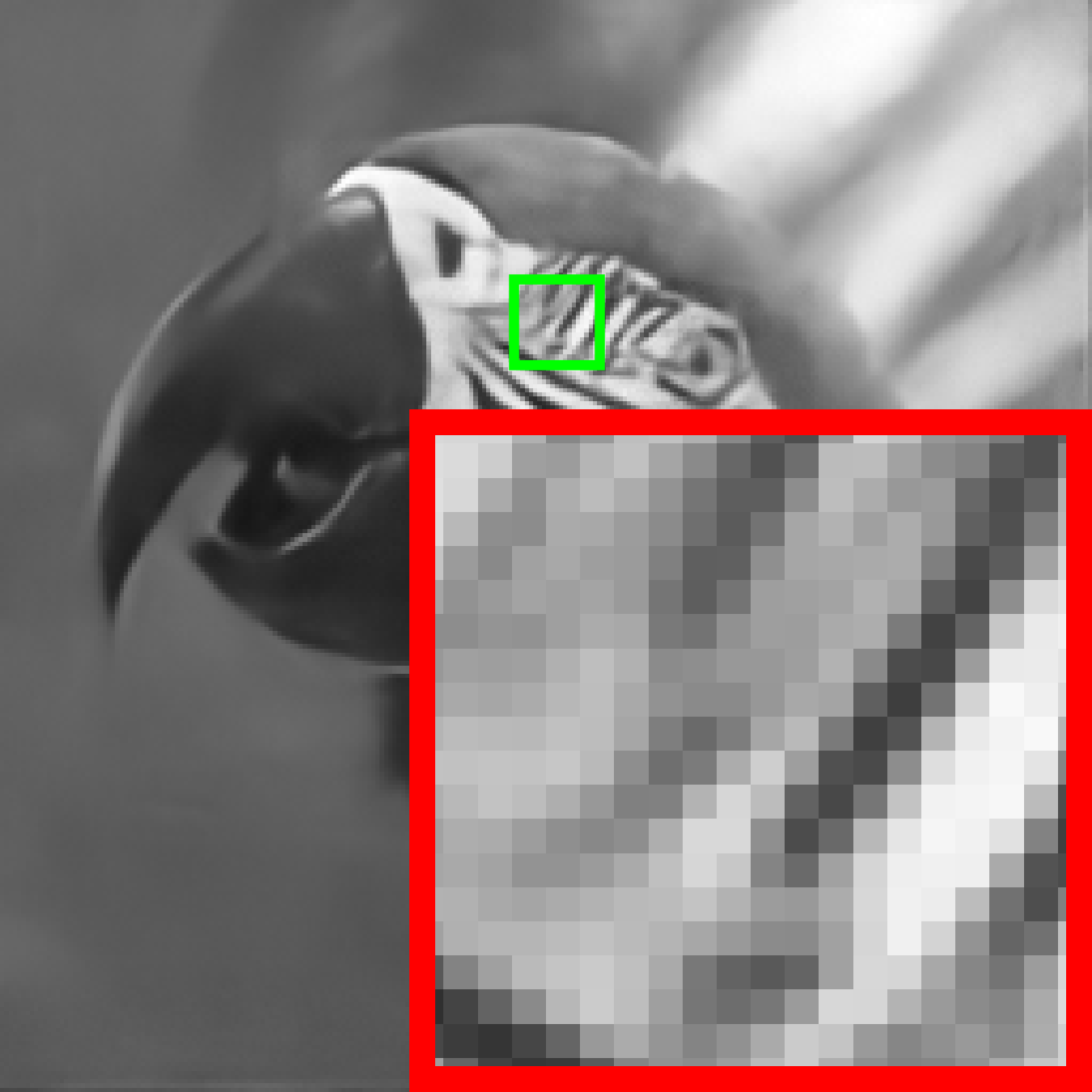}
    &\includegraphics[width=0.08\textwidth]{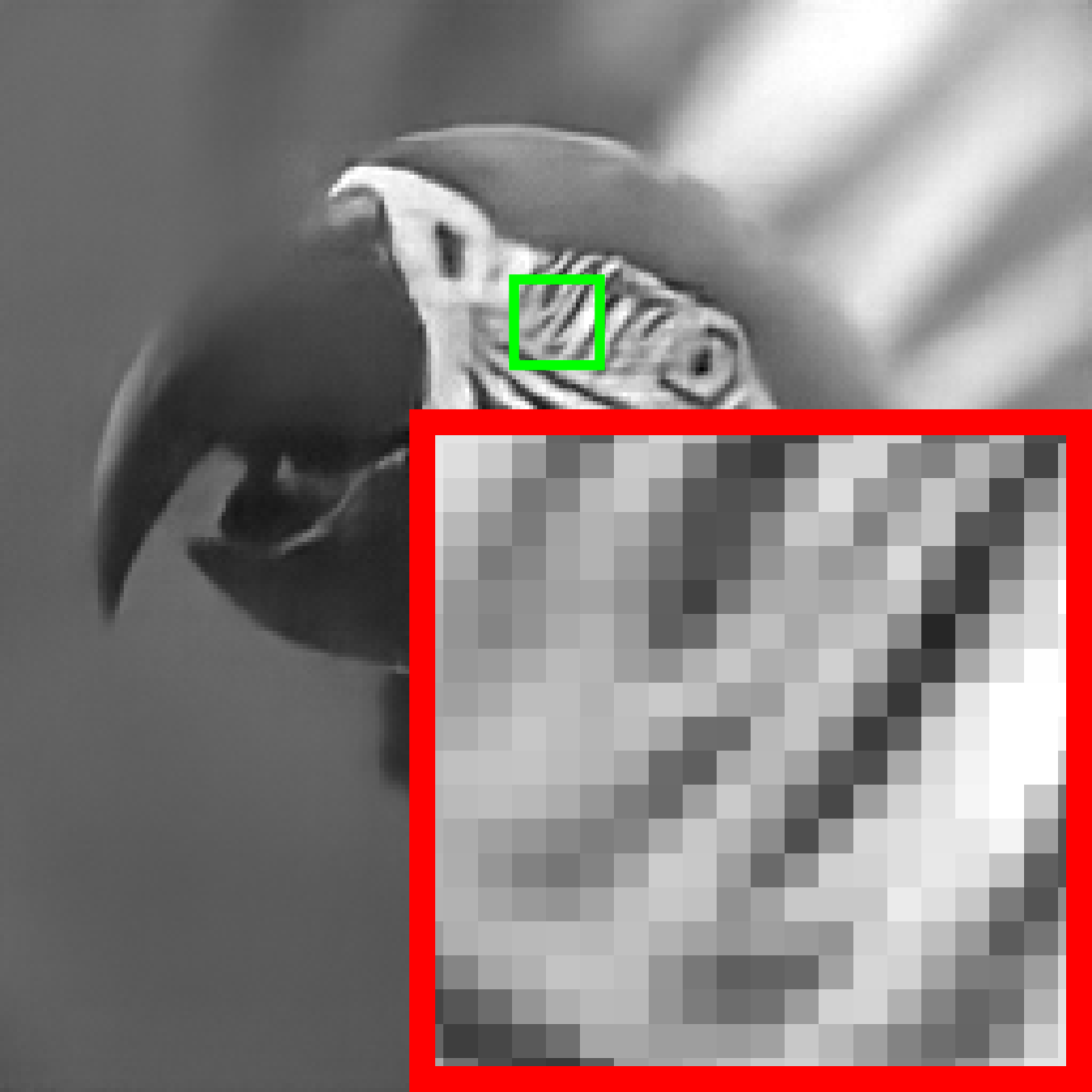}\\
    PSNR/SSIM & 23.86/0.76 & 25.31/0.80 & 21.99/0.62 & 24.51/0.62 & 25.92/0.78 & 26.90/\underline{\textcolor{blue}{0.84}} & \underline{\textcolor{blue}{27.66}}/\textbf{\textcolor{red}{0.85}} & 27.22/\underline{\textcolor{blue}{0.84}} & \textbf{\textcolor{red}{27.76}}/\textbf{\textcolor{red}{0.85}}\\
    \includegraphics[width=0.08\textwidth]{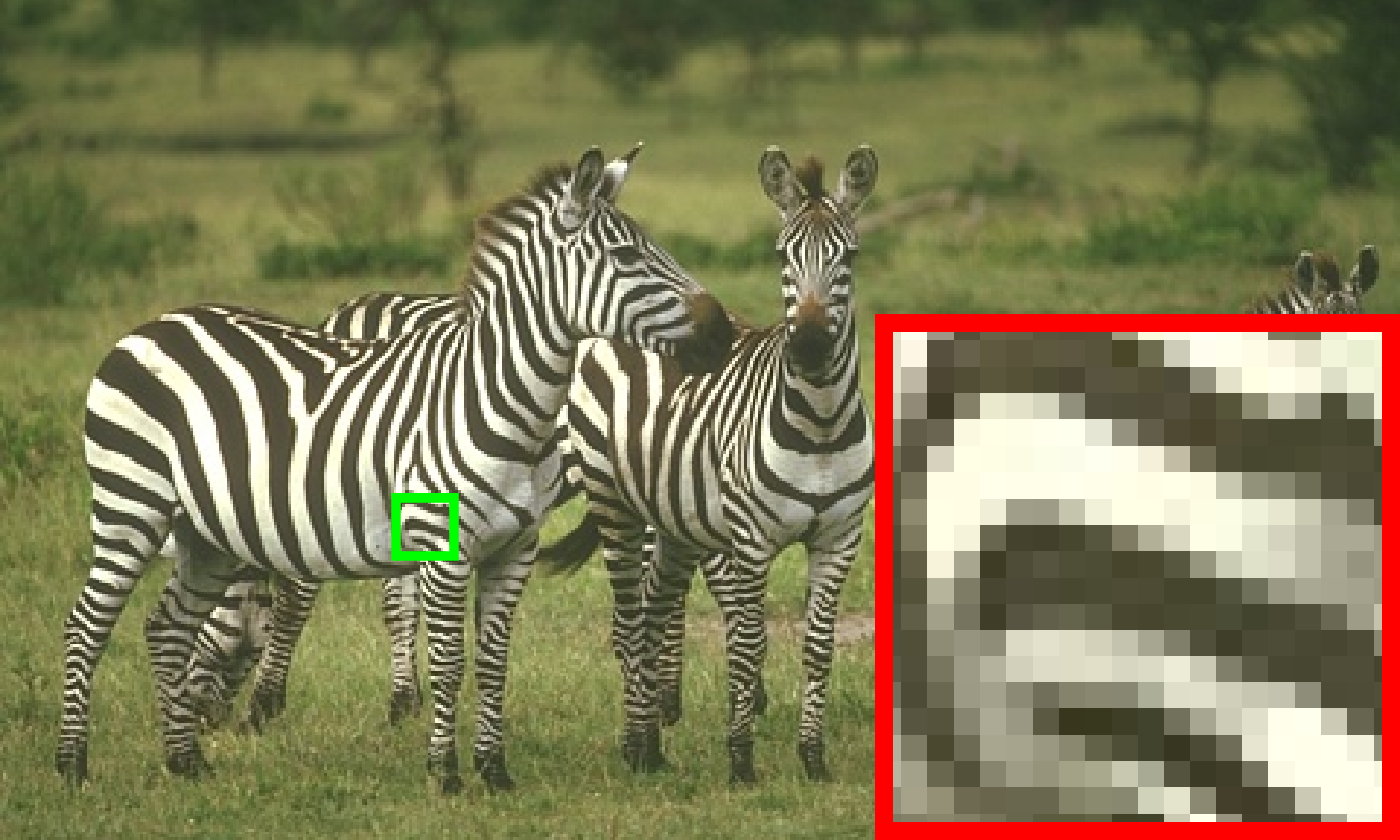}
    &\includegraphics[width=0.08\textwidth]{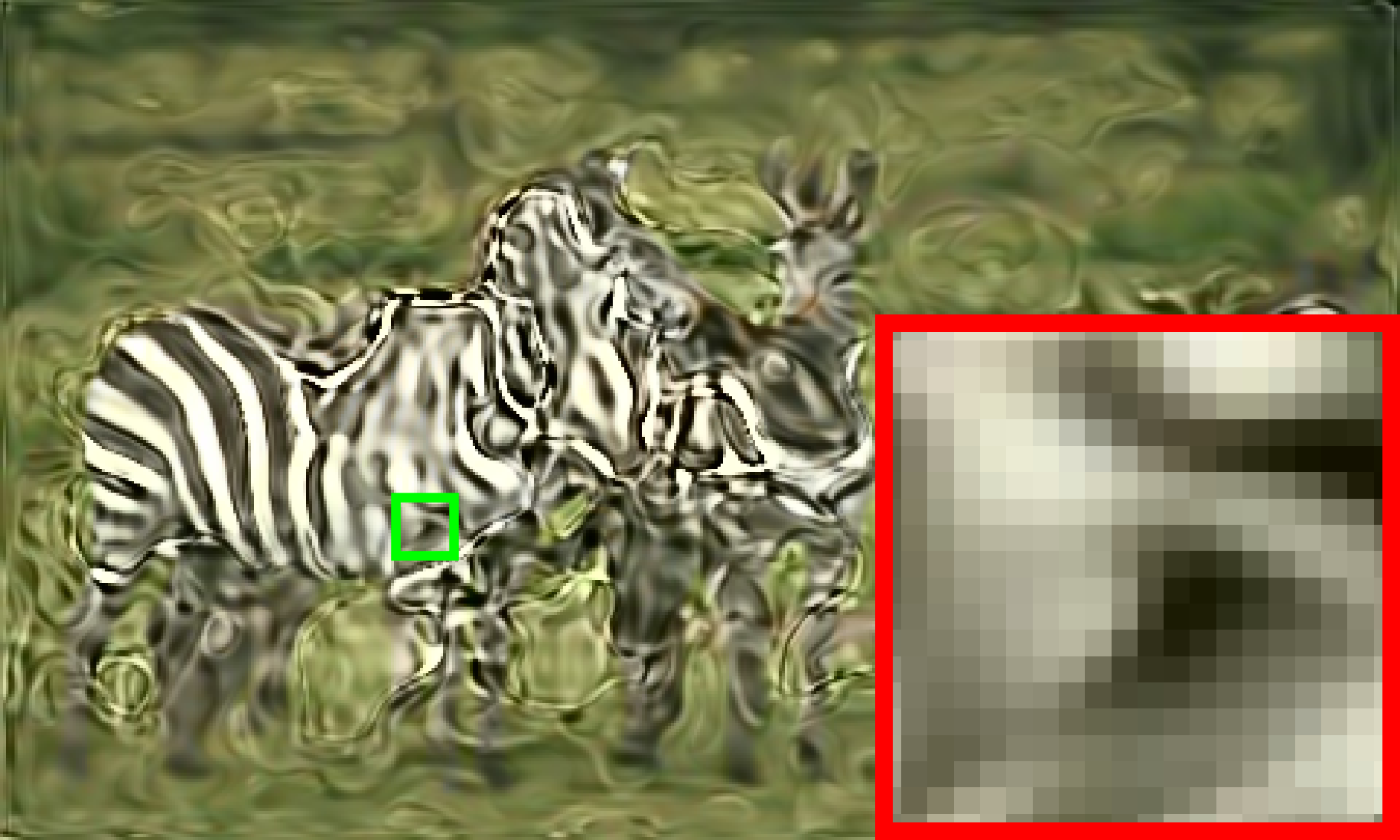}
    &\includegraphics[width=0.08\textwidth]{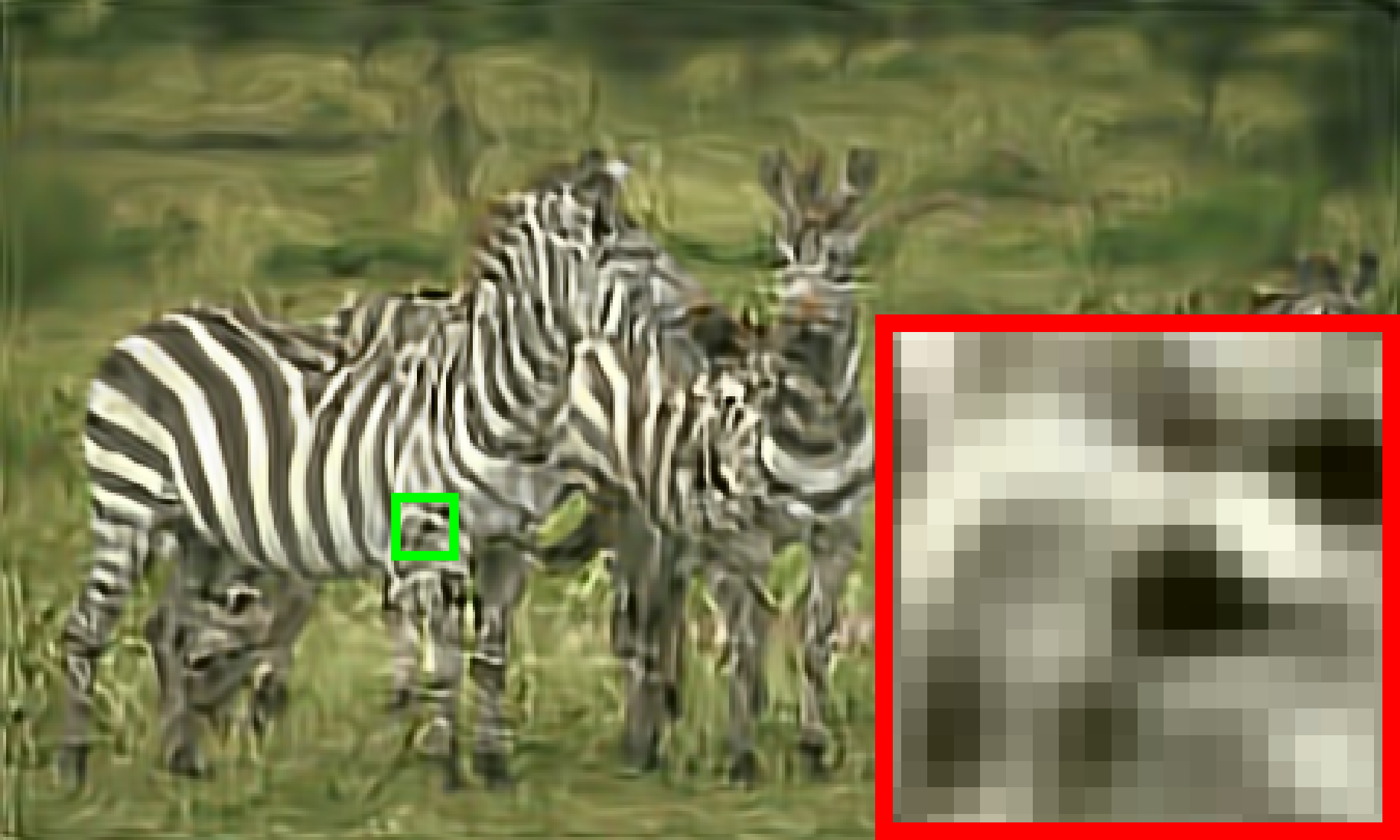}
    &\includegraphics[width=0.08\textwidth]{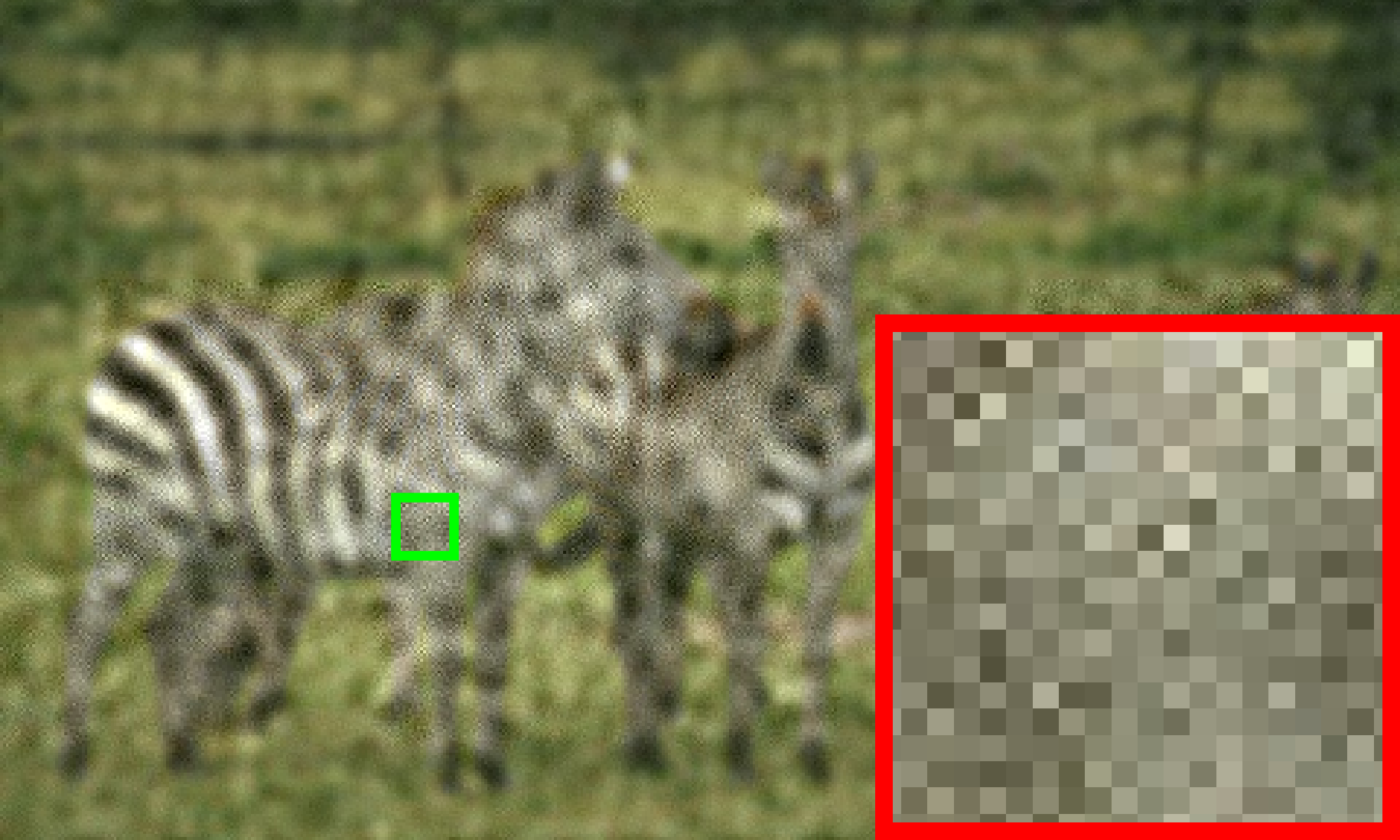}
    &\includegraphics[width=0.08\textwidth]{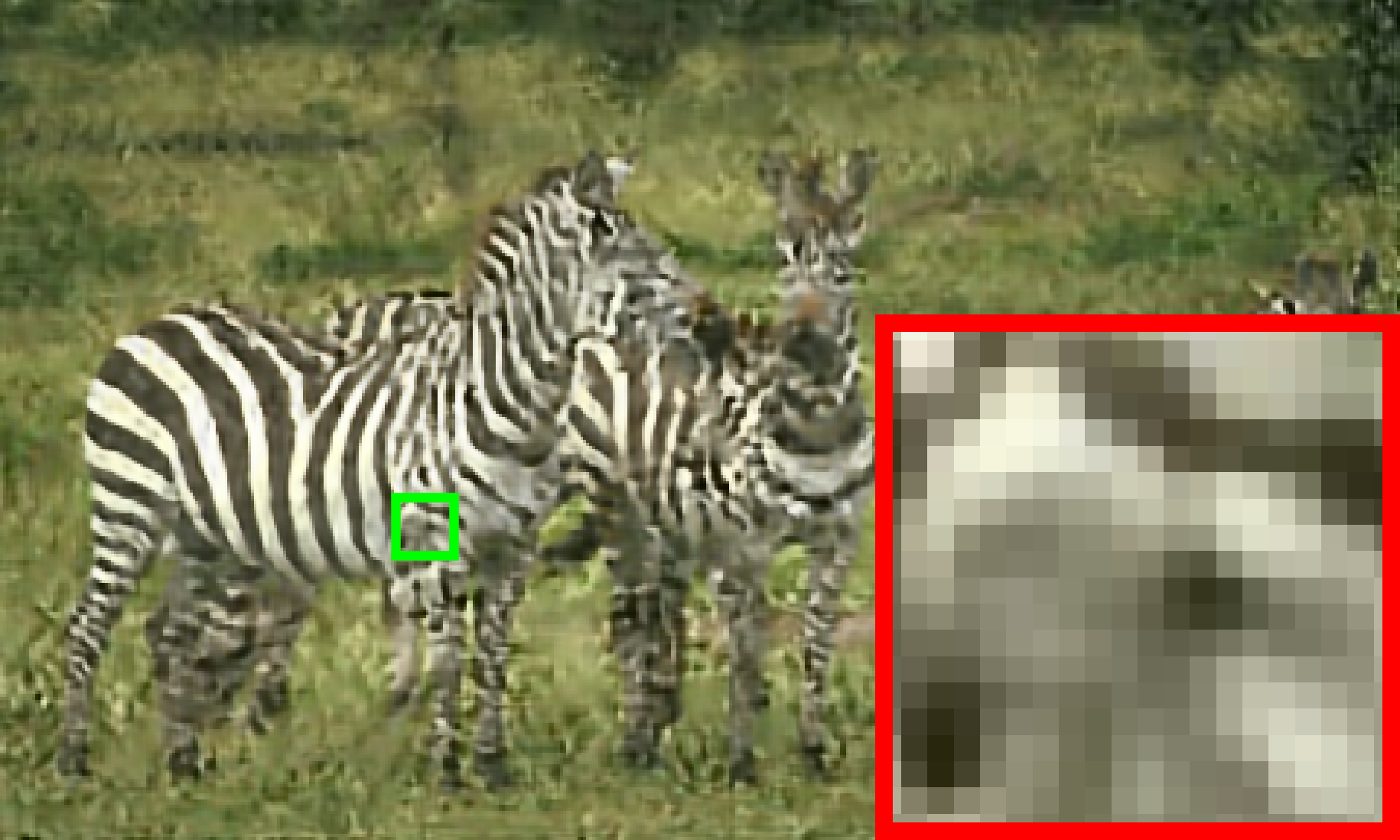}
    &\includegraphics[width=0.08\textwidth]{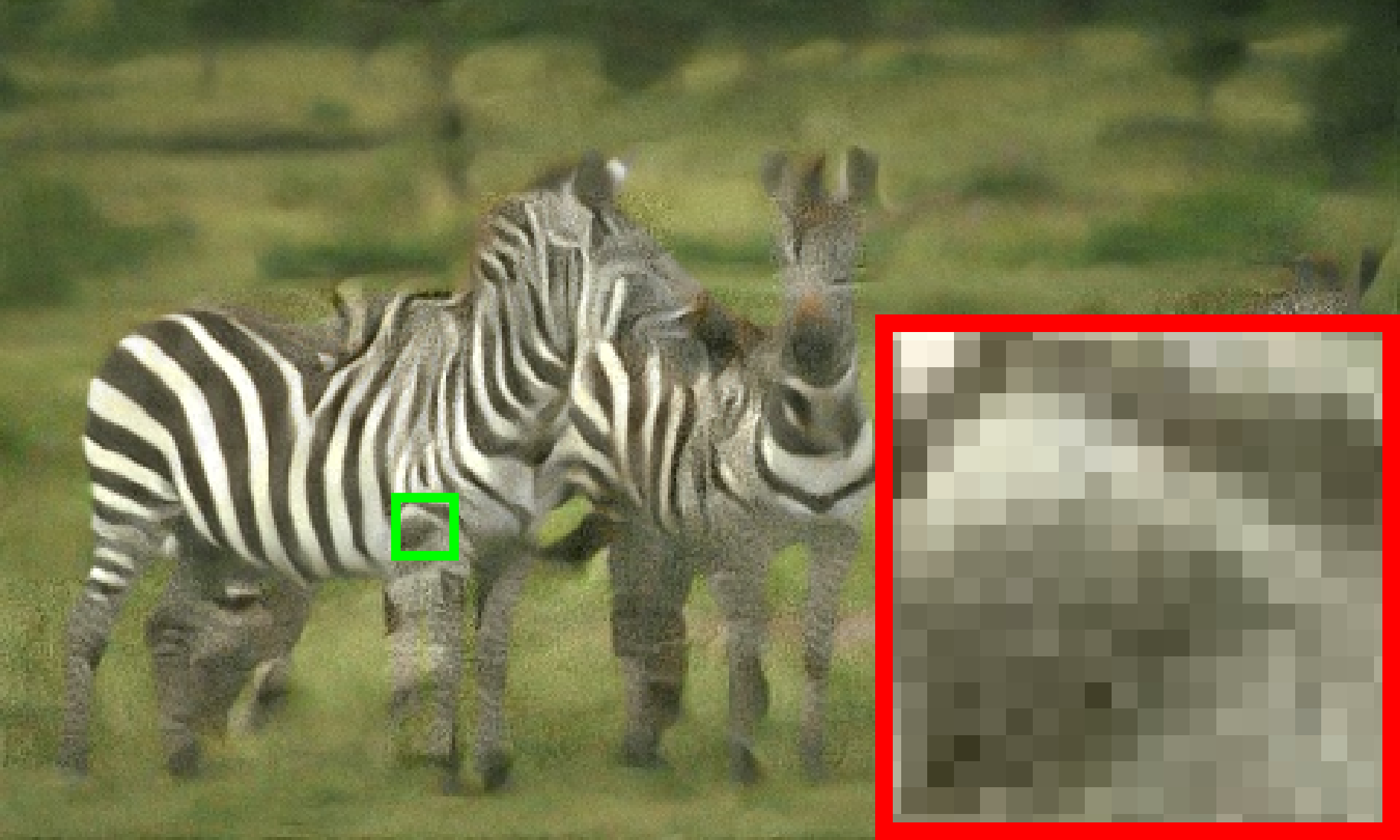}
    &\includegraphics[width=0.08\textwidth]{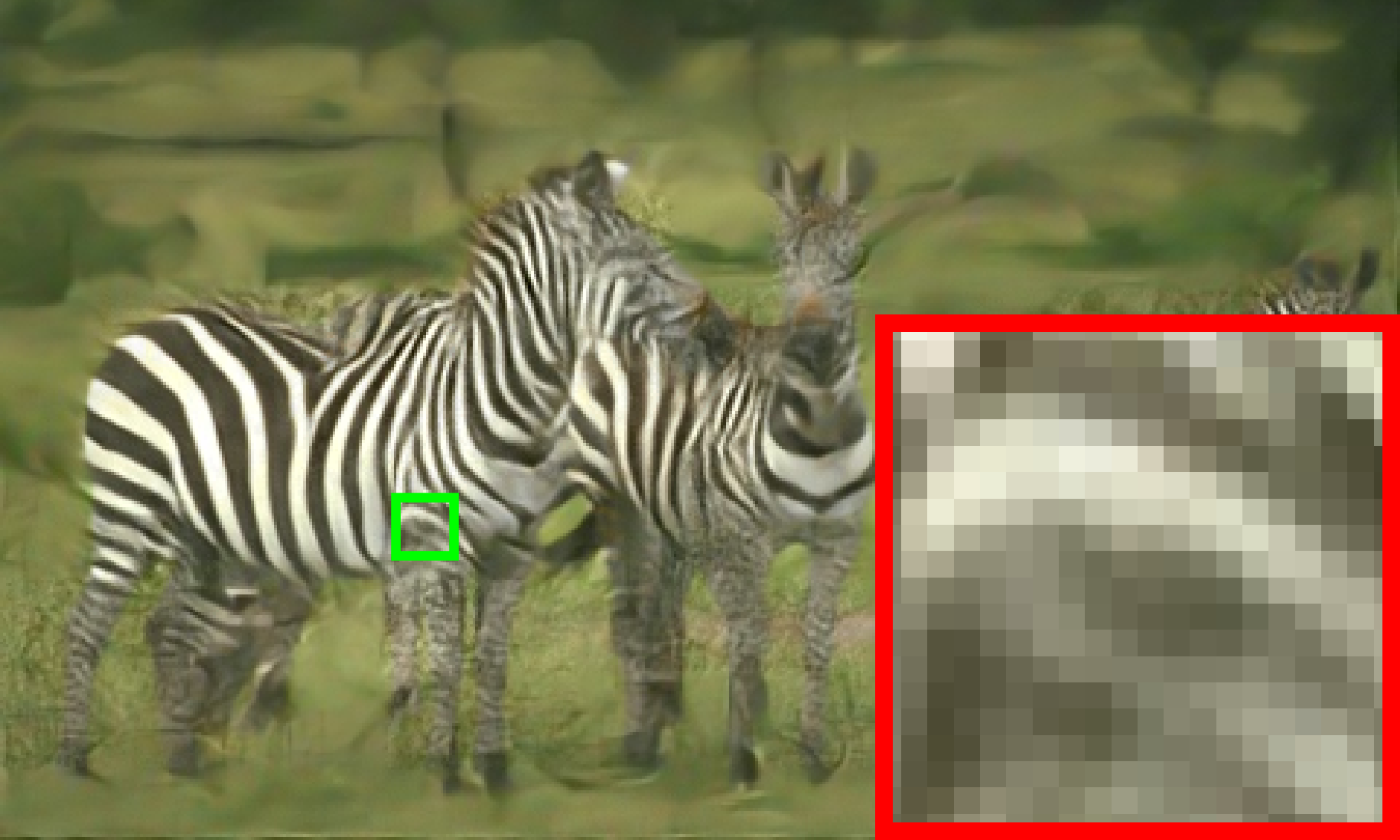}
    &\includegraphics[width=0.08\textwidth]{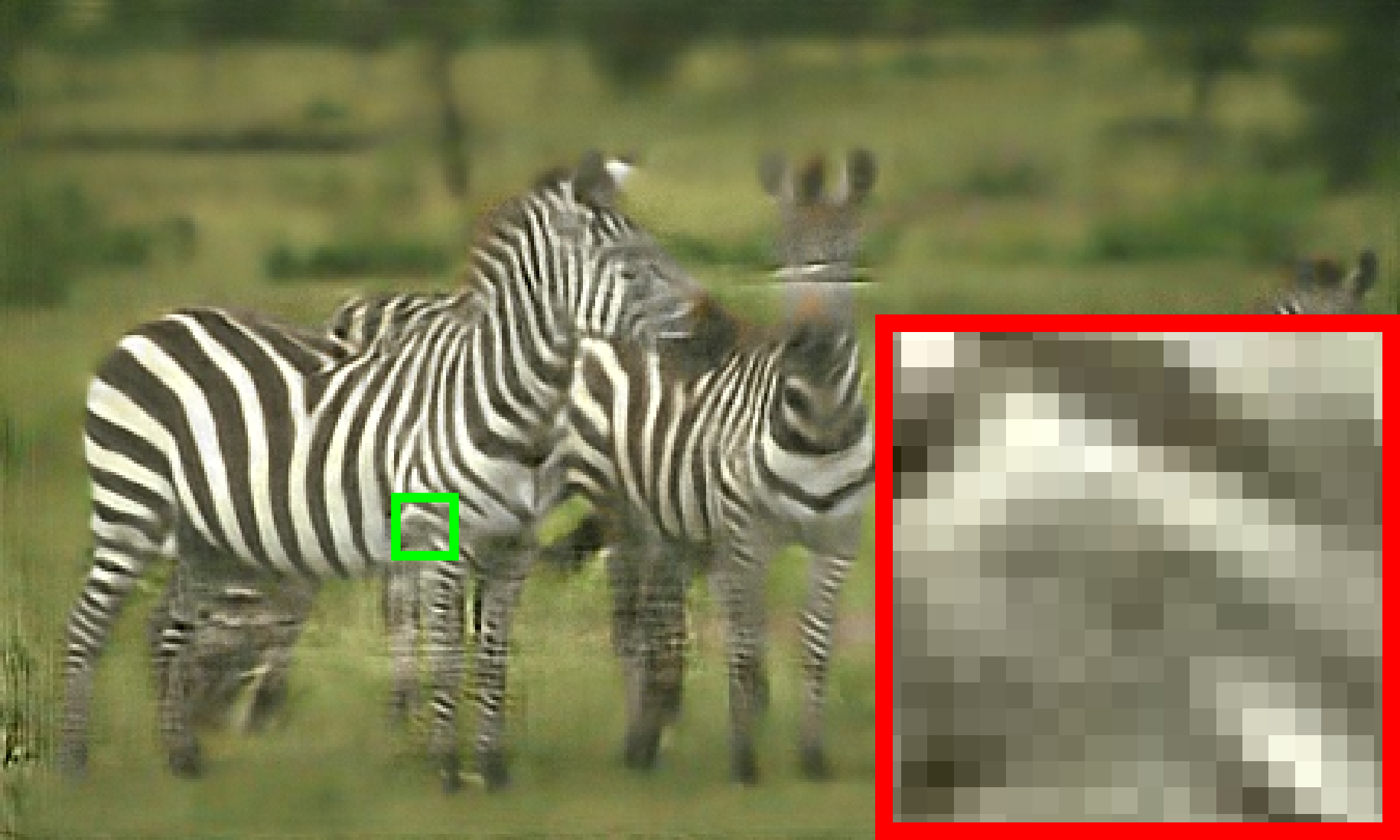}
    &\includegraphics[width=0.08\textwidth]{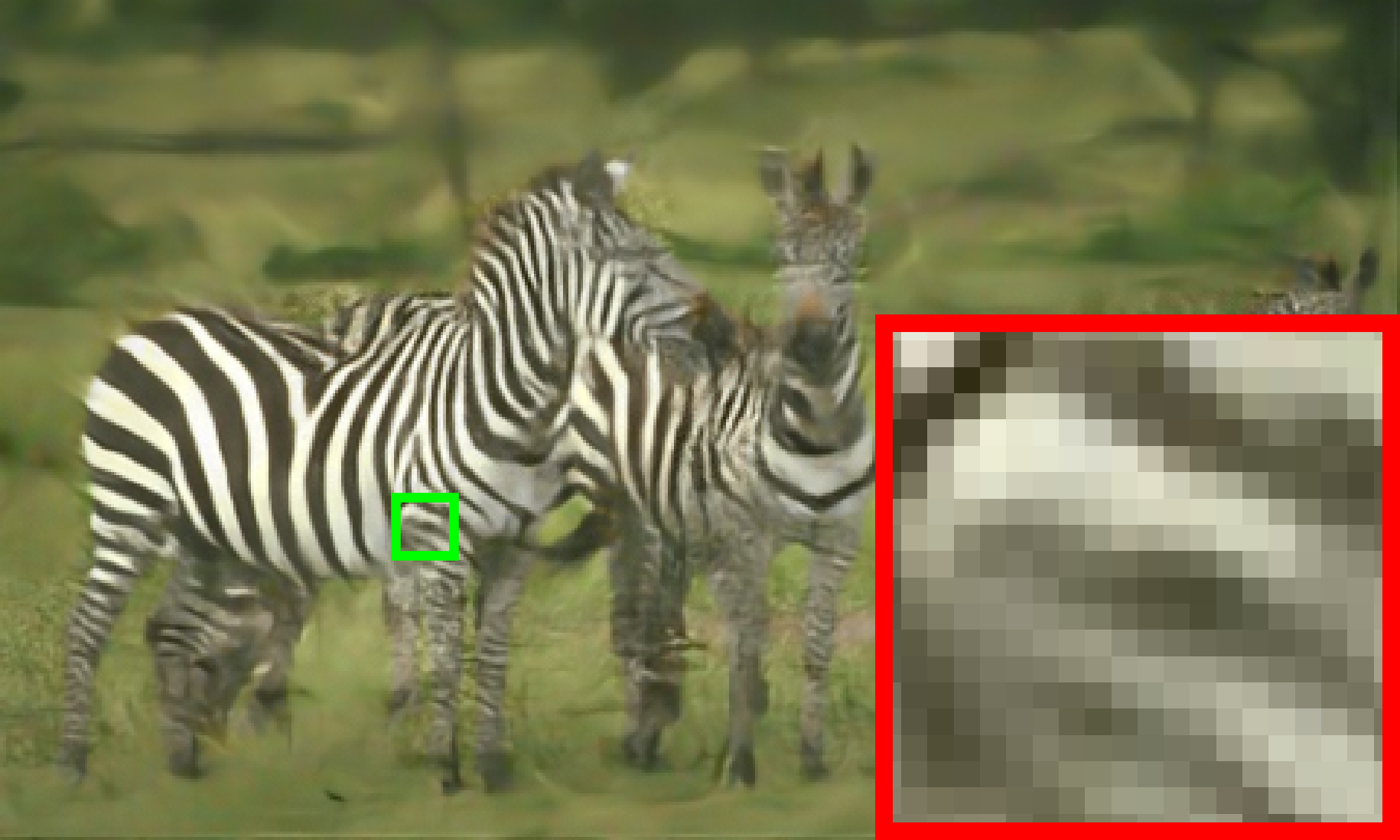}
    &\includegraphics[width=0.08\textwidth]{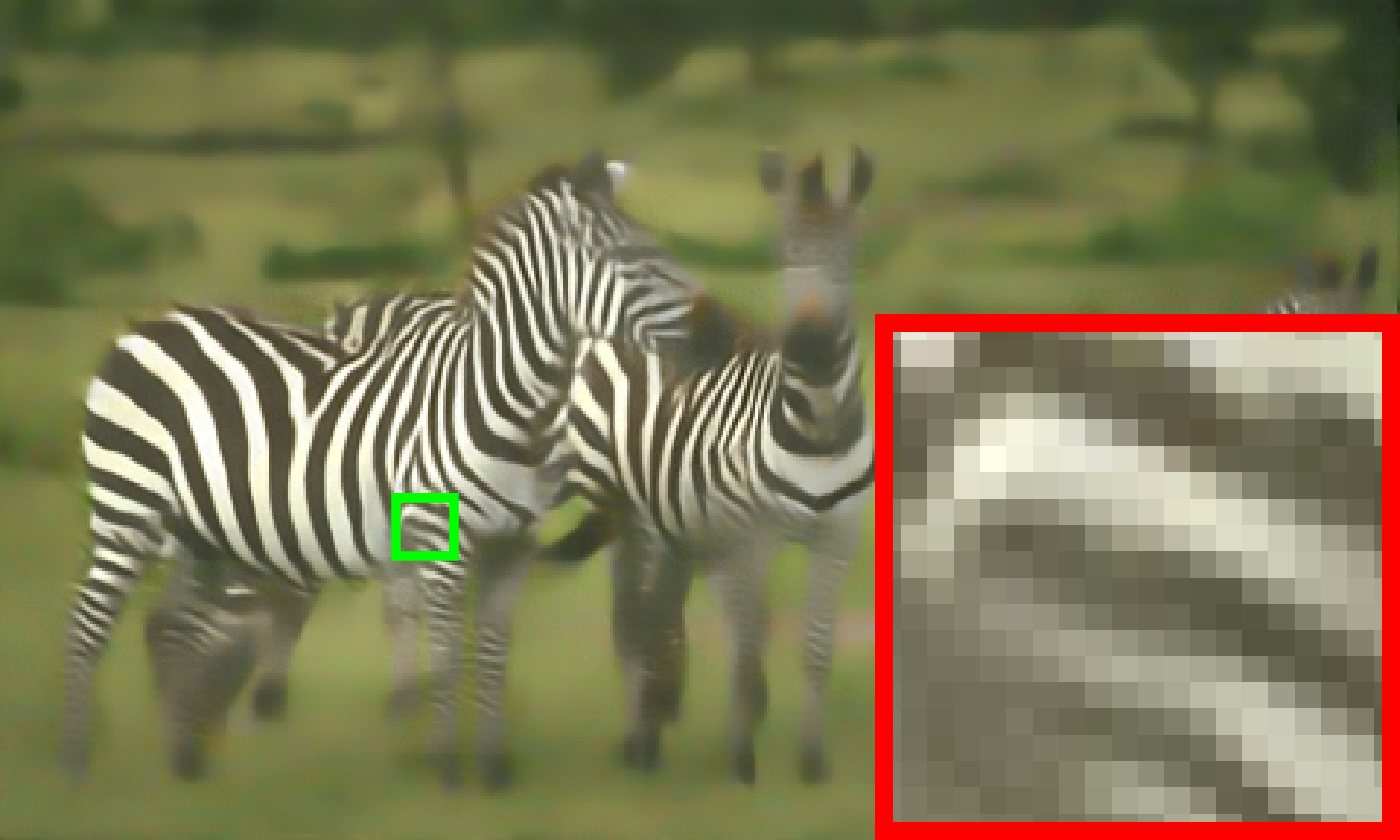}\\
    PSNR/SSIM & 16.69/0.45 & 19.47/0.56 & 17.83/0.46 & 19.73/0.52 & 20.42/0.58 & 21.45/\textbf{\textcolor{red}{0.63}} & \underline{\textcolor{blue}{22.05}}/\textbf{\textcolor{red}{0.63}} & 21.79/\textbf{\textcolor{red}{0.63}} & \textbf{\textcolor{red}{22.38}}/\underline{\textcolor{blue}{0.61}}
\end{tabular}}
\caption{Visual comparison of self-supervised methods on two natural benchmark images named ``Parrots'' and ``test\_51'' from Set11~\cite{kulkarni2016reconnet} \textcolor{blue}{(top)} and CBSD68~\cite{martin2001database} \textcolor{blue}{(bottom)}, with $\gamma =10\%$ and $\sigma =10$.}
\label{fig:comparison_standard_natural_images_r10_s10}
\end{figure*}

\begin{figure*}[!t]
\setlength{\tabcolsep}{0.5pt}
\hspace{-4pt}
\resizebox{1.0\textwidth}{!}{
\tiny
\begin{tabular}{cccccccccc}
    GT & DIP & BCNN & EI & ASGLD & DDSSL & \textbf{SC-CNN} & \textbf{SC-CNN$^\text{+}$} & \textbf{SCT} & \textbf{SCT$^\text{+}$}\\
    \includegraphics[width=0.08\textwidth]{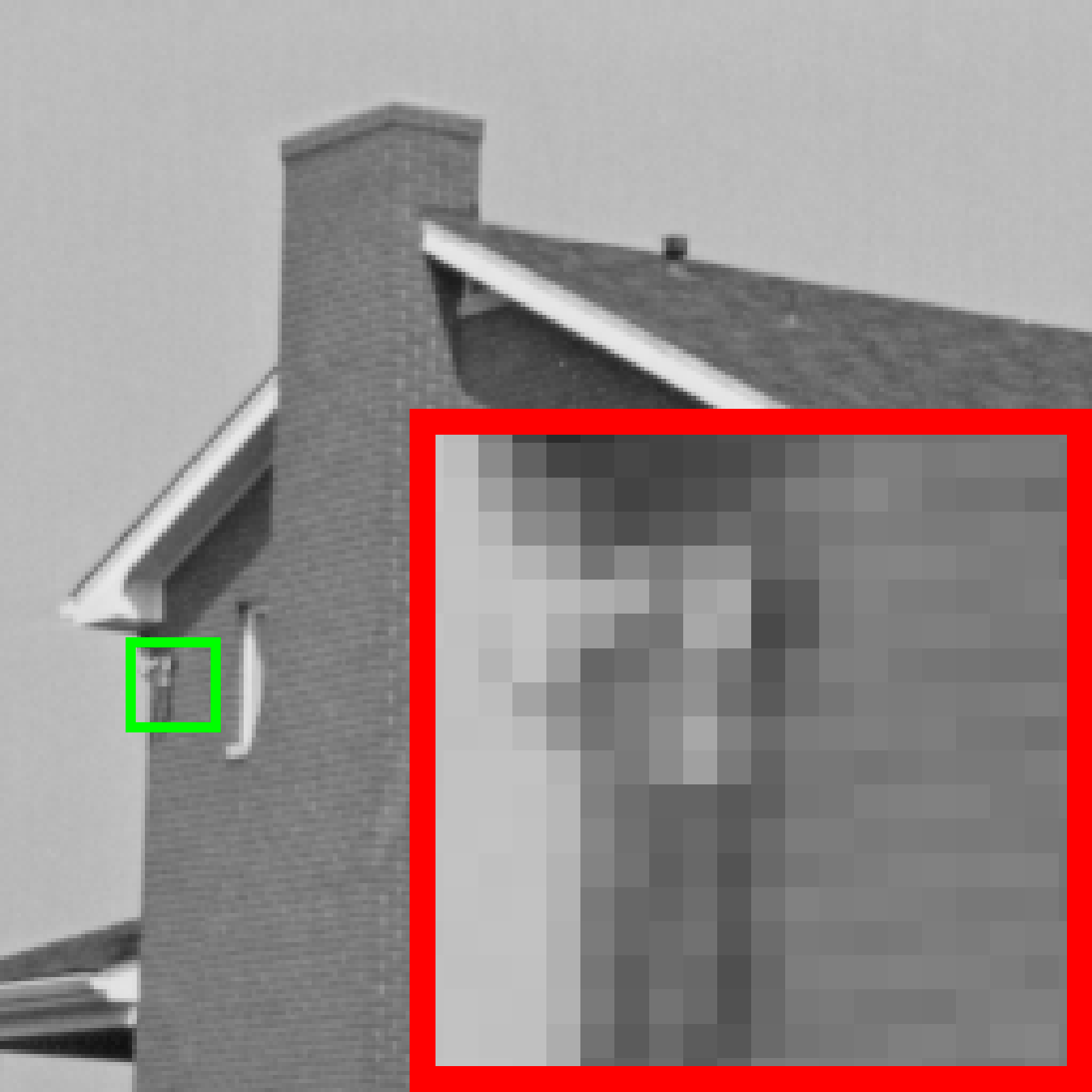}
    &\includegraphics[width=0.08\textwidth]{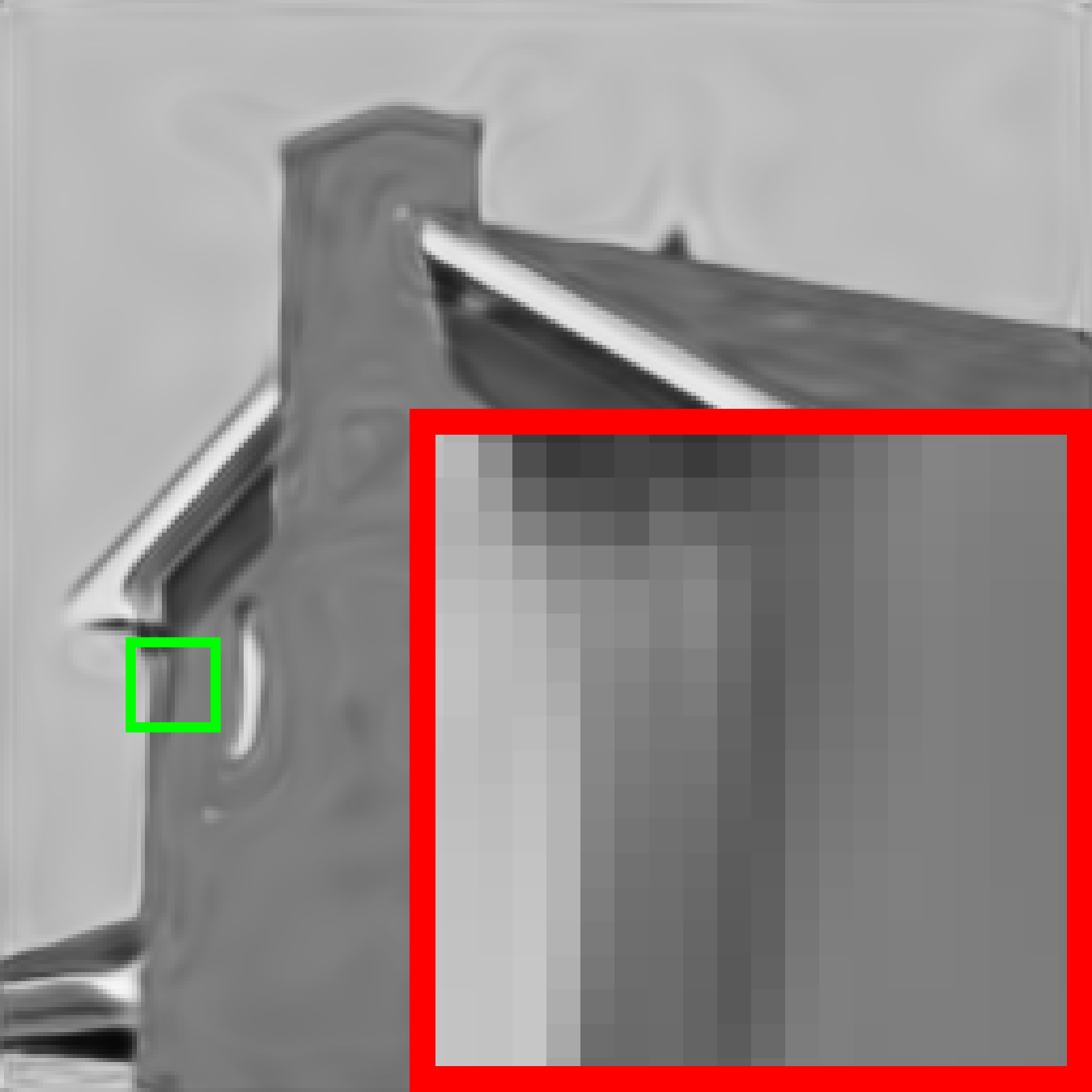}
    &\includegraphics[width=0.08\textwidth]{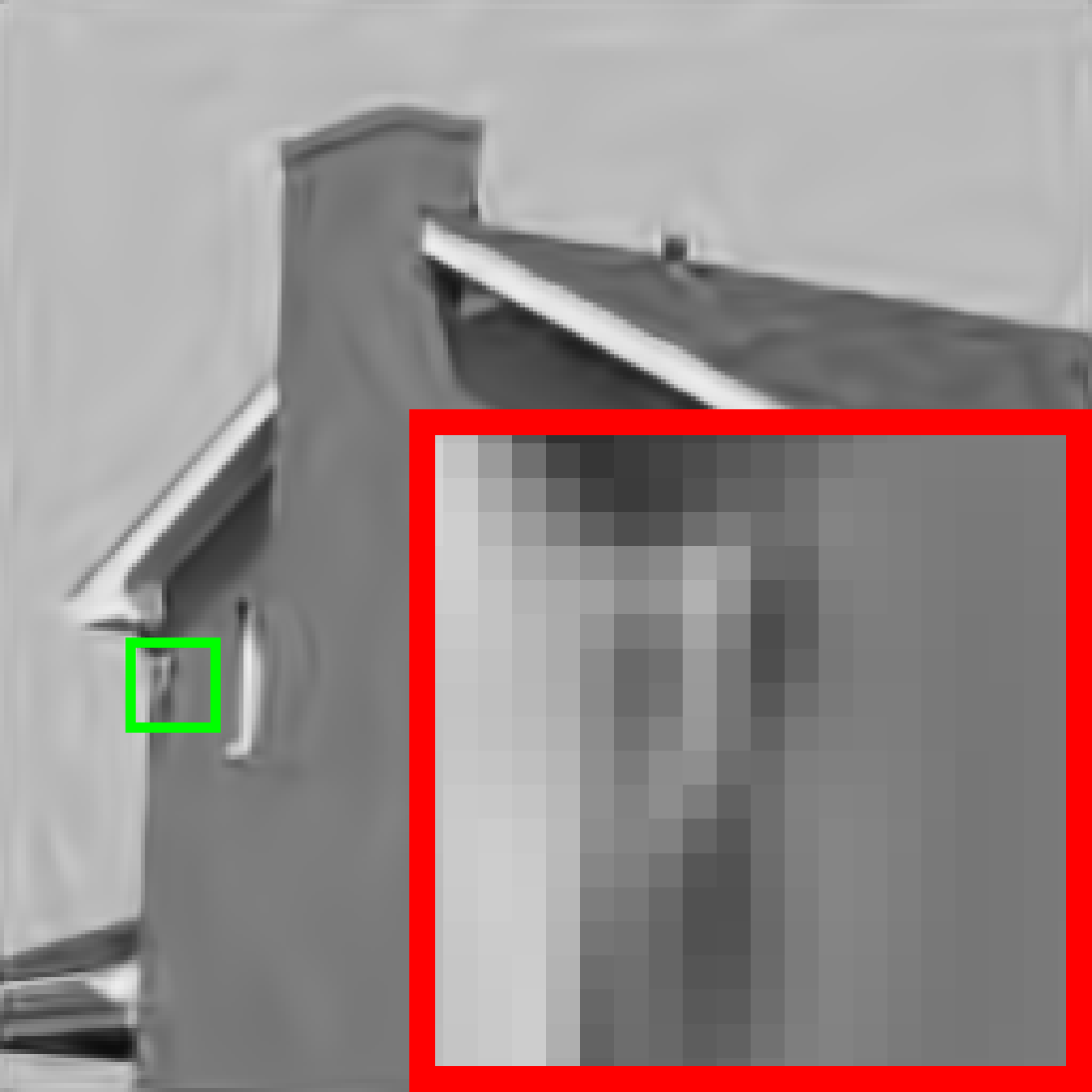}
    &\includegraphics[width=0.08\textwidth]{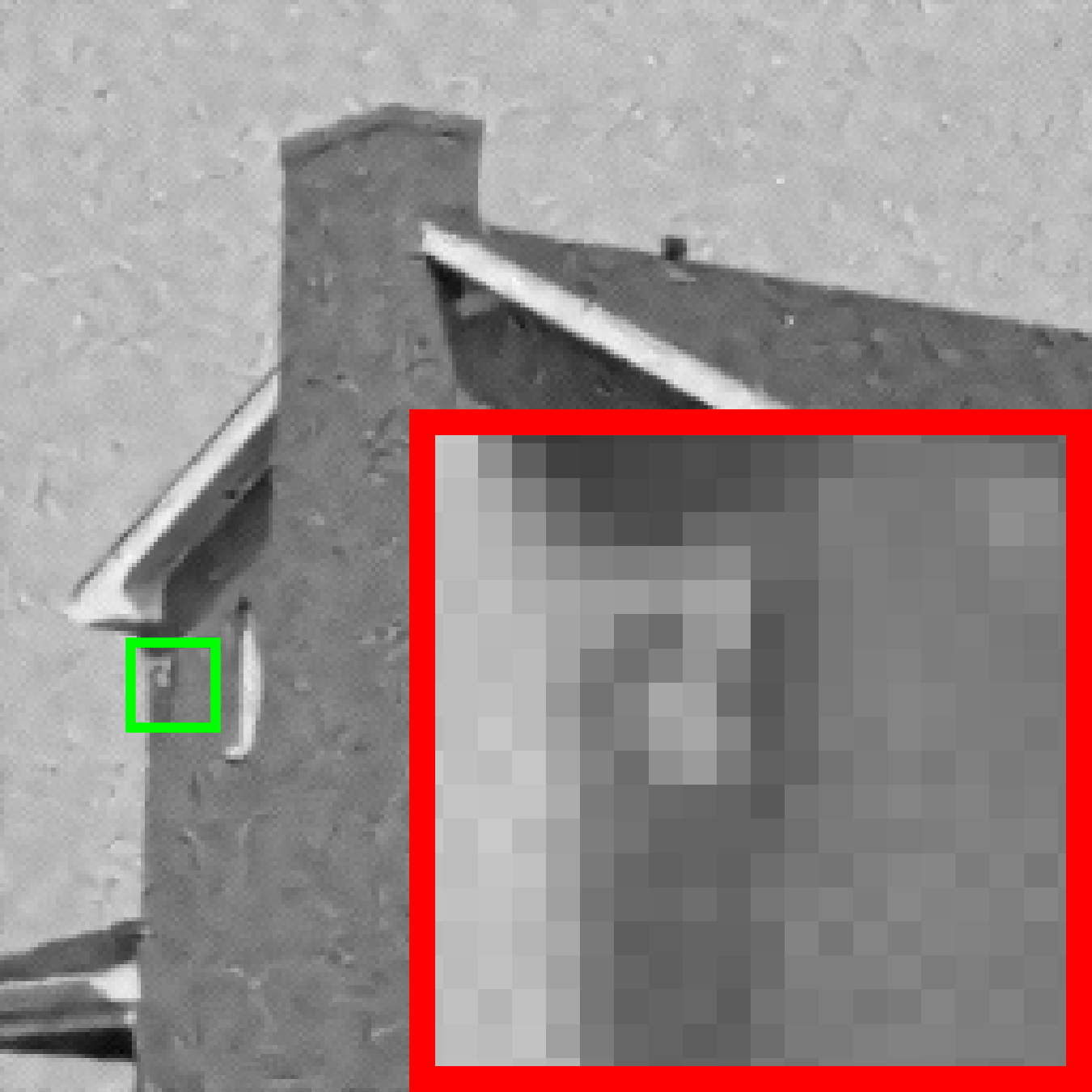}
    &\includegraphics[width=0.08\textwidth]{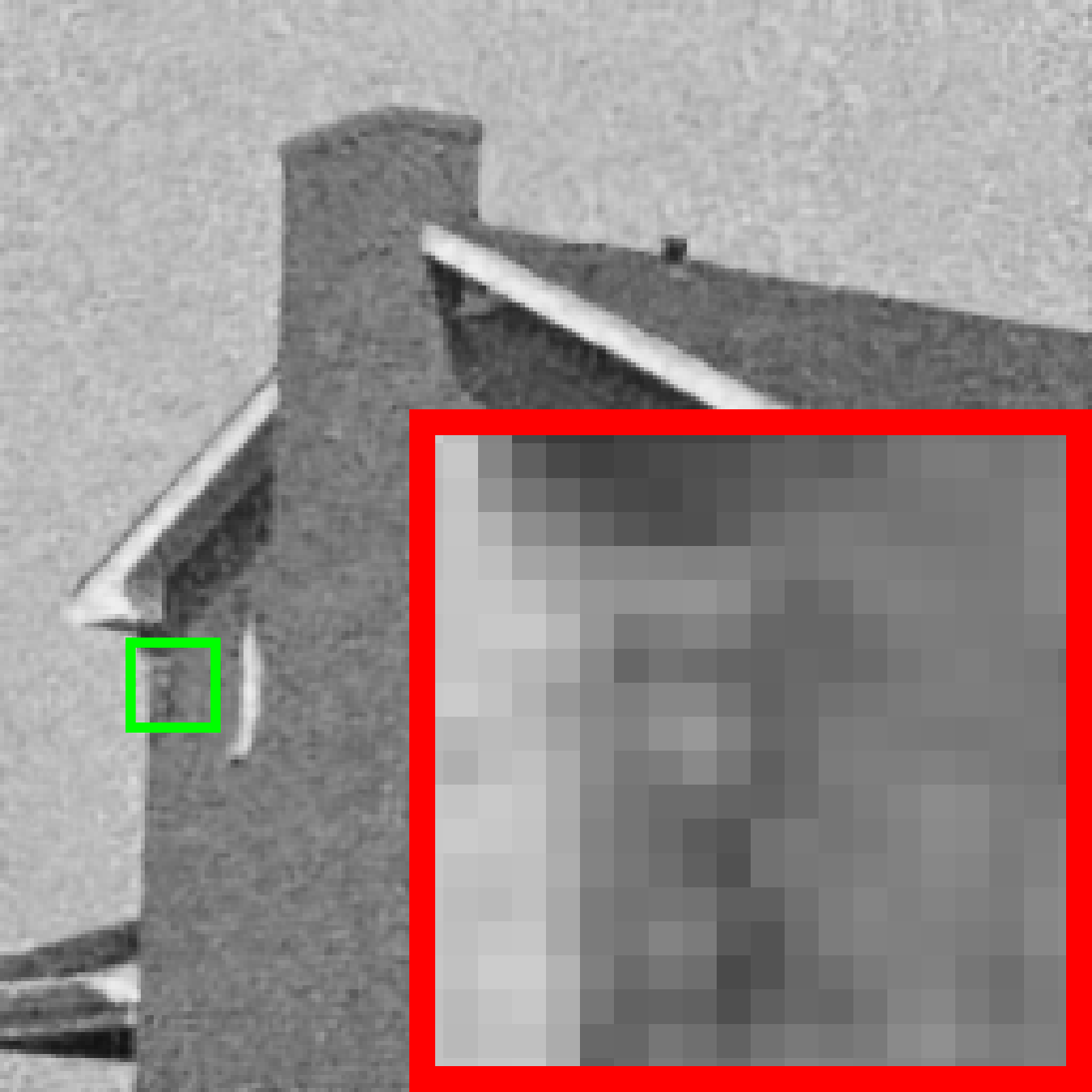}
    &\includegraphics[width=0.08\textwidth]{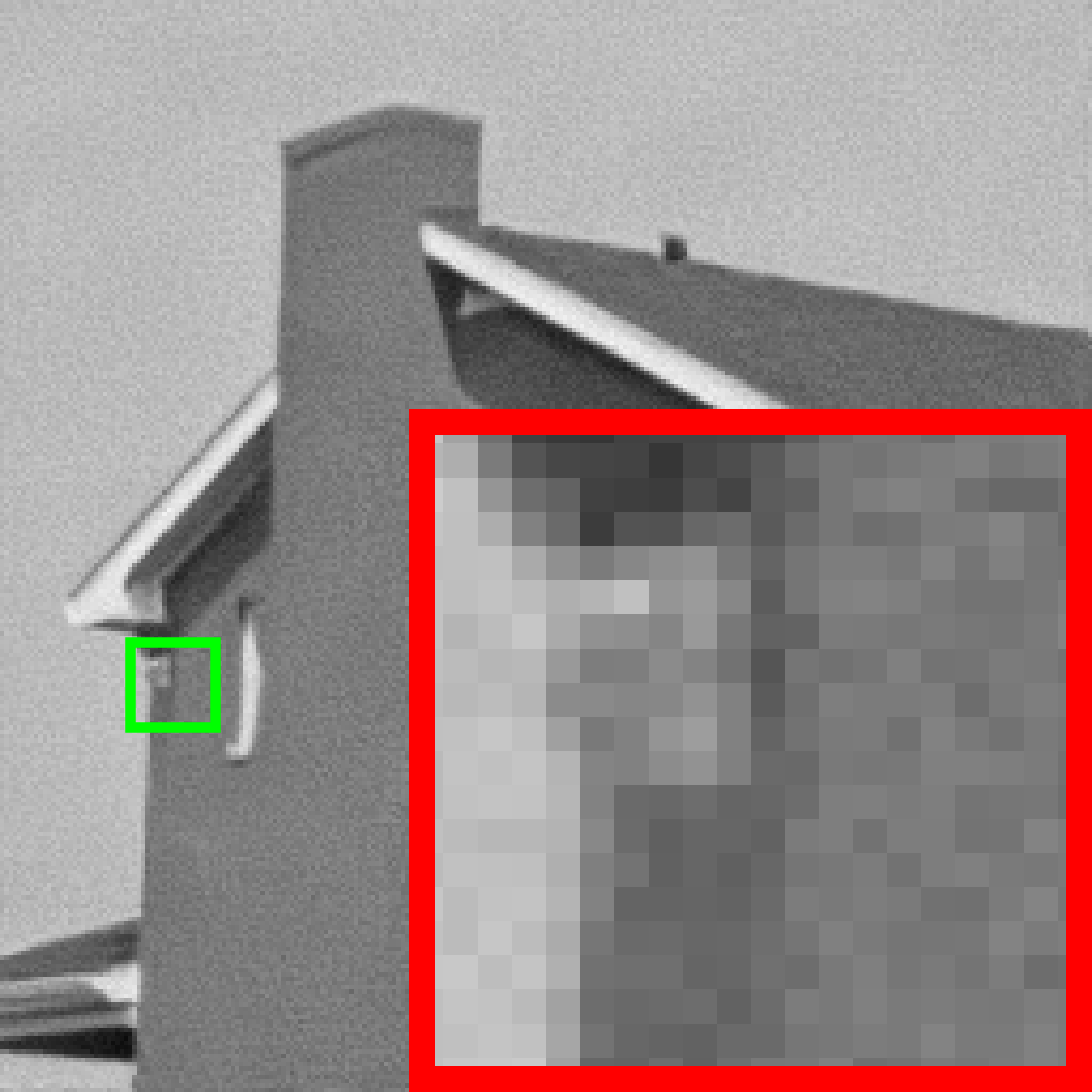}
    &\includegraphics[width=0.08\textwidth]{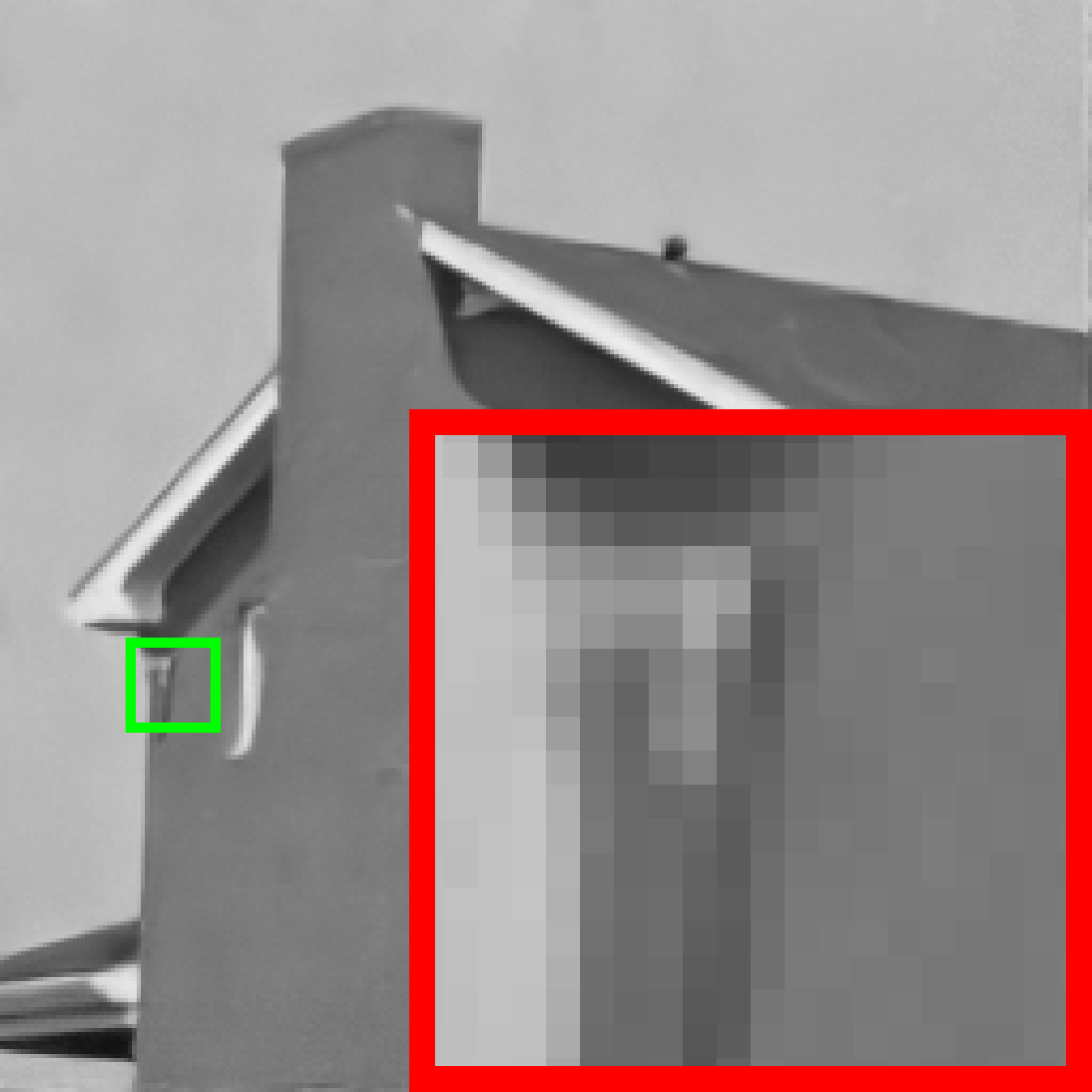}
    &\includegraphics[width=0.08\textwidth]{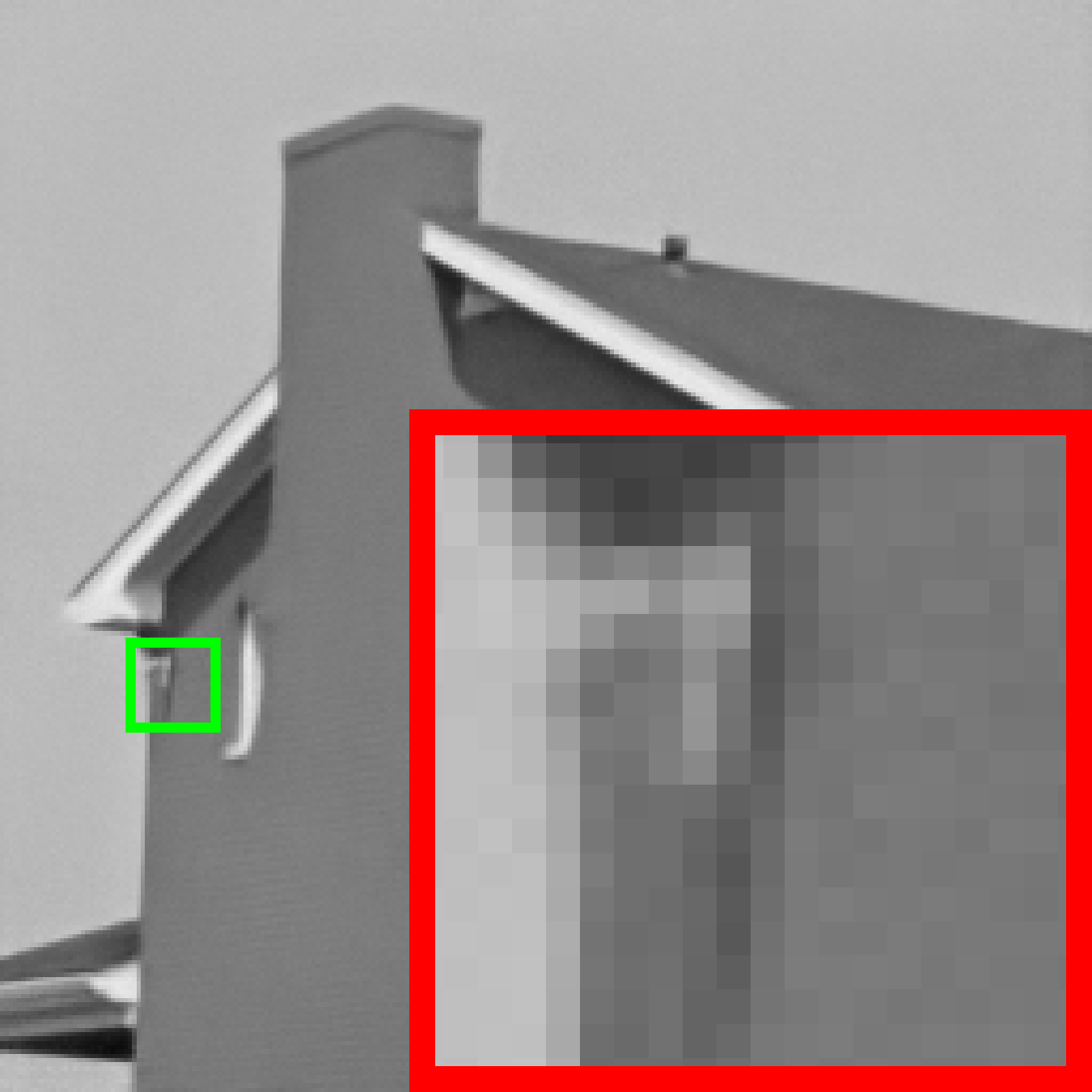}
    &\includegraphics[width=0.08\textwidth]{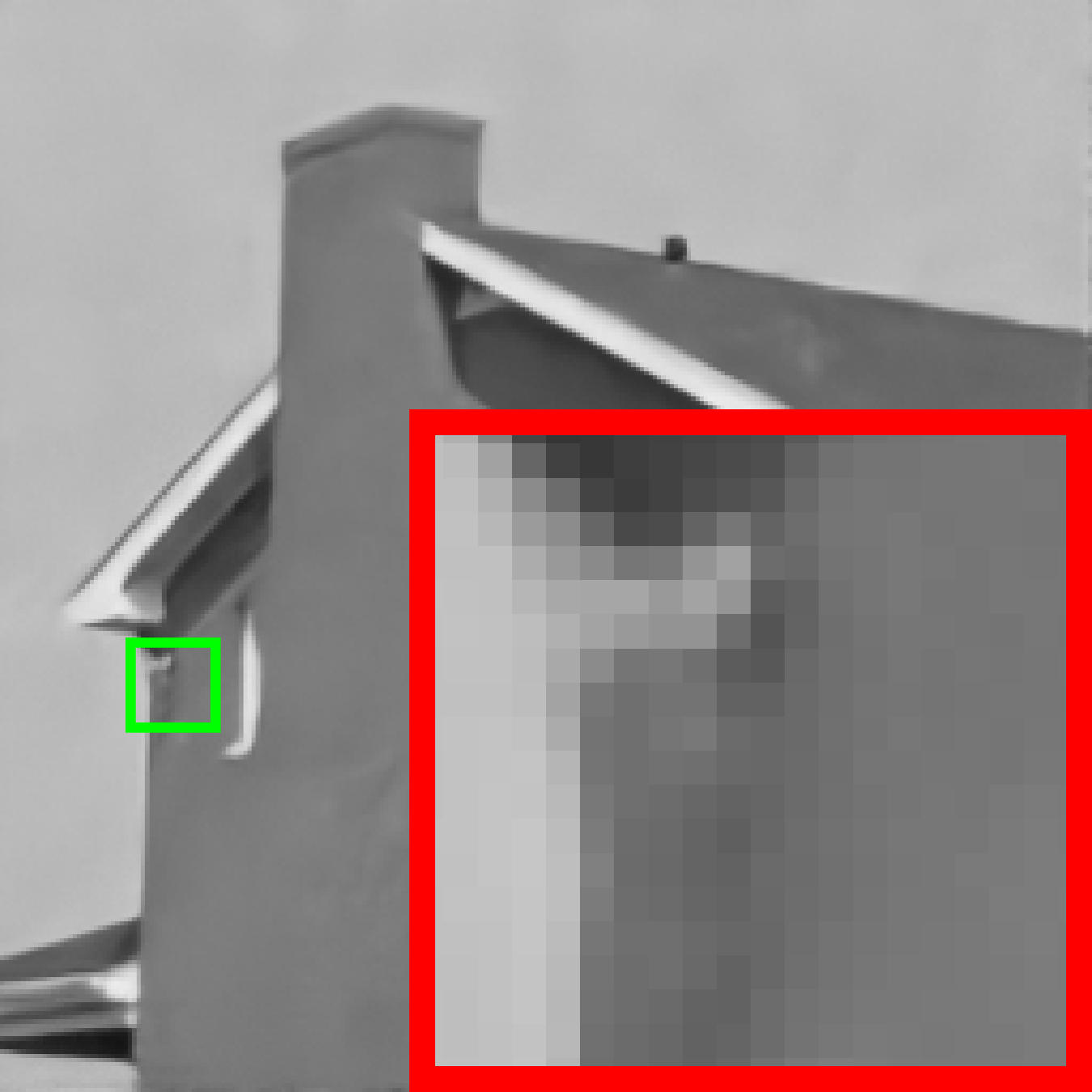}
    &\includegraphics[width=0.08\textwidth]{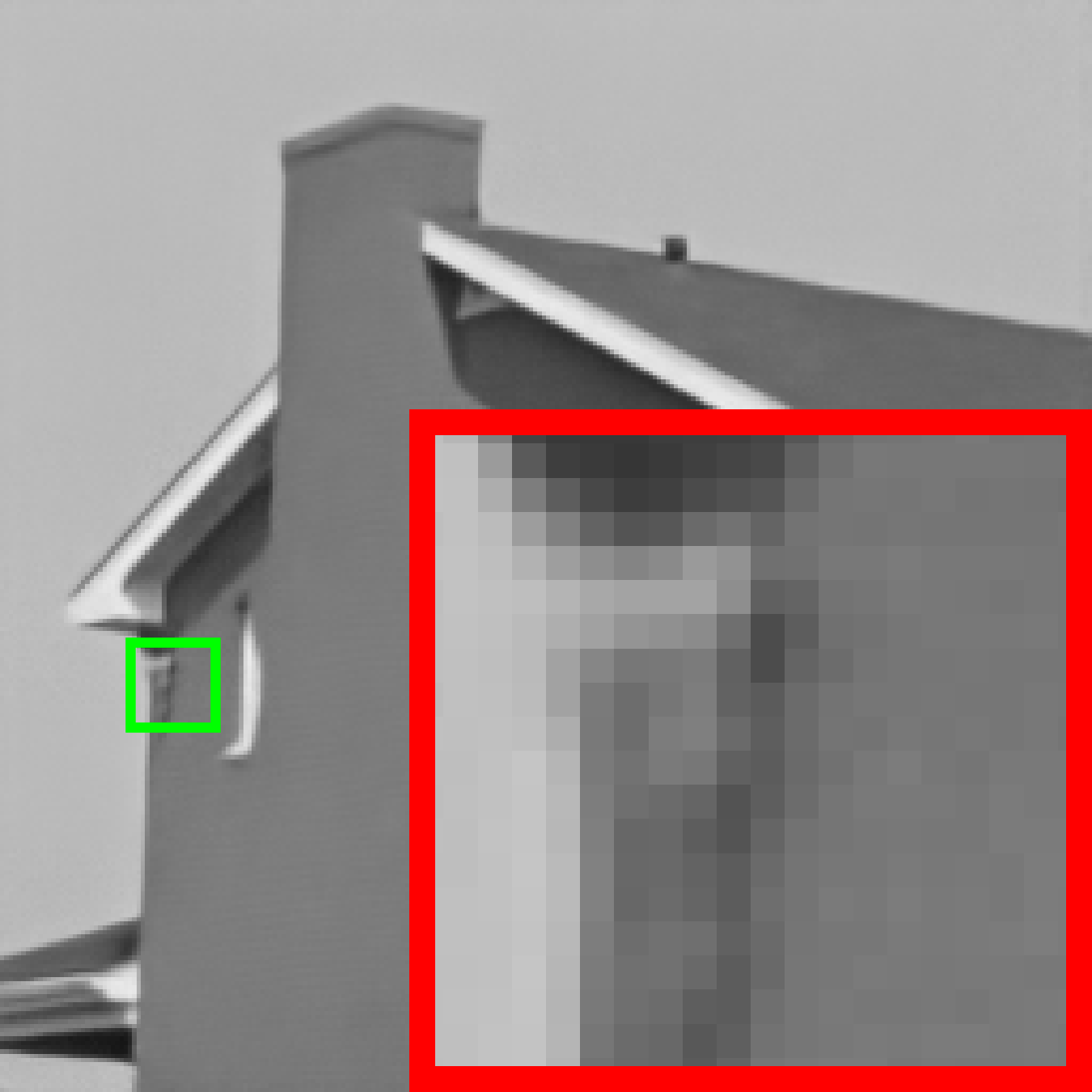}\\
    PSNR/SSIM & 31.63/0.83 & 32.56/0.84 & 30.71/0.74 & 29.53/0.68 & 31.18/0.73 & 33.55/\underline{\textcolor{blue}{0.86}} & \textbf{\textcolor{red}{34.66}}/\textbf{\textcolor{red}{0.87}} & 33.47/\underline{\textcolor{blue}{0.86}} & \underline{\textcolor{blue}{34.46}}/\textbf{\textcolor{red}{0.87}}\\
    \includegraphics[width=0.08\textwidth]{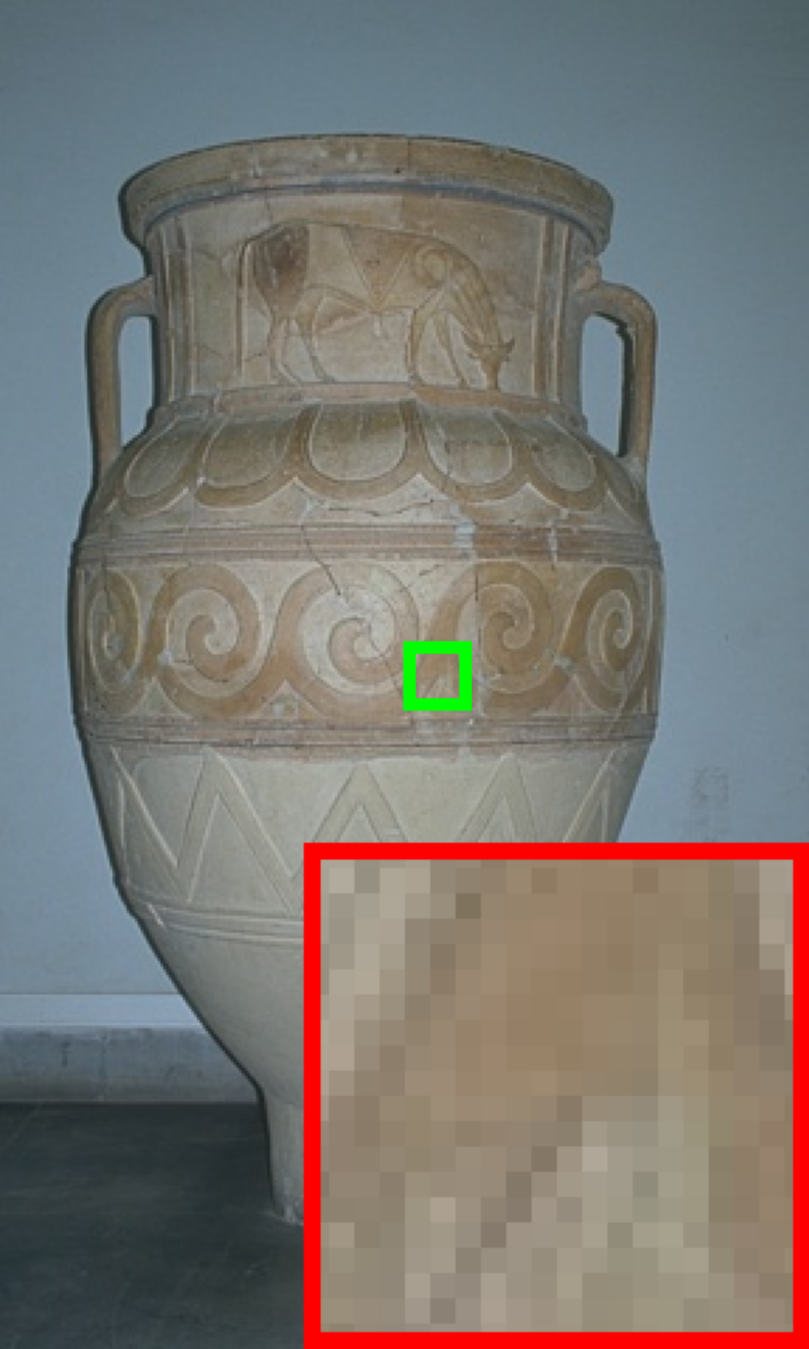}
    &\includegraphics[width=0.08\textwidth]{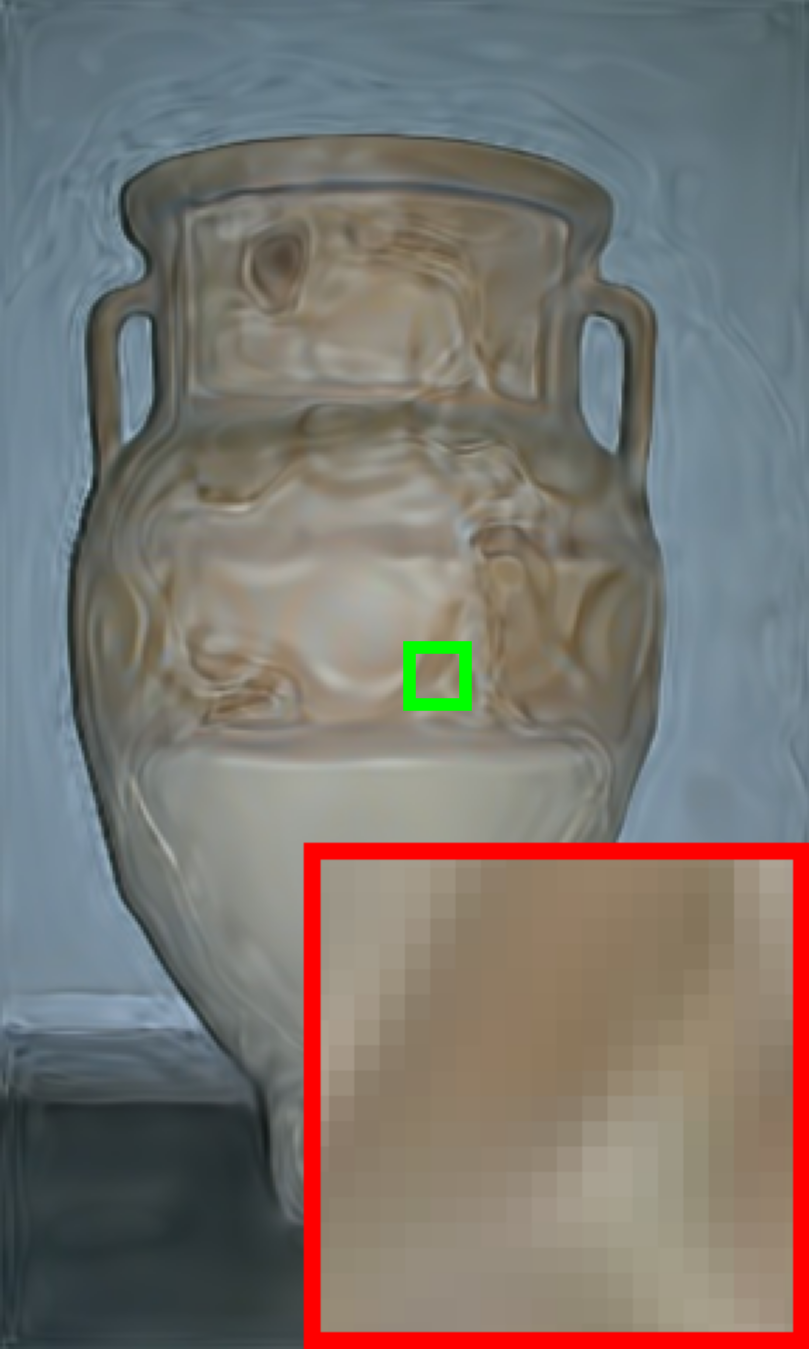}
    &\includegraphics[width=0.08\textwidth]{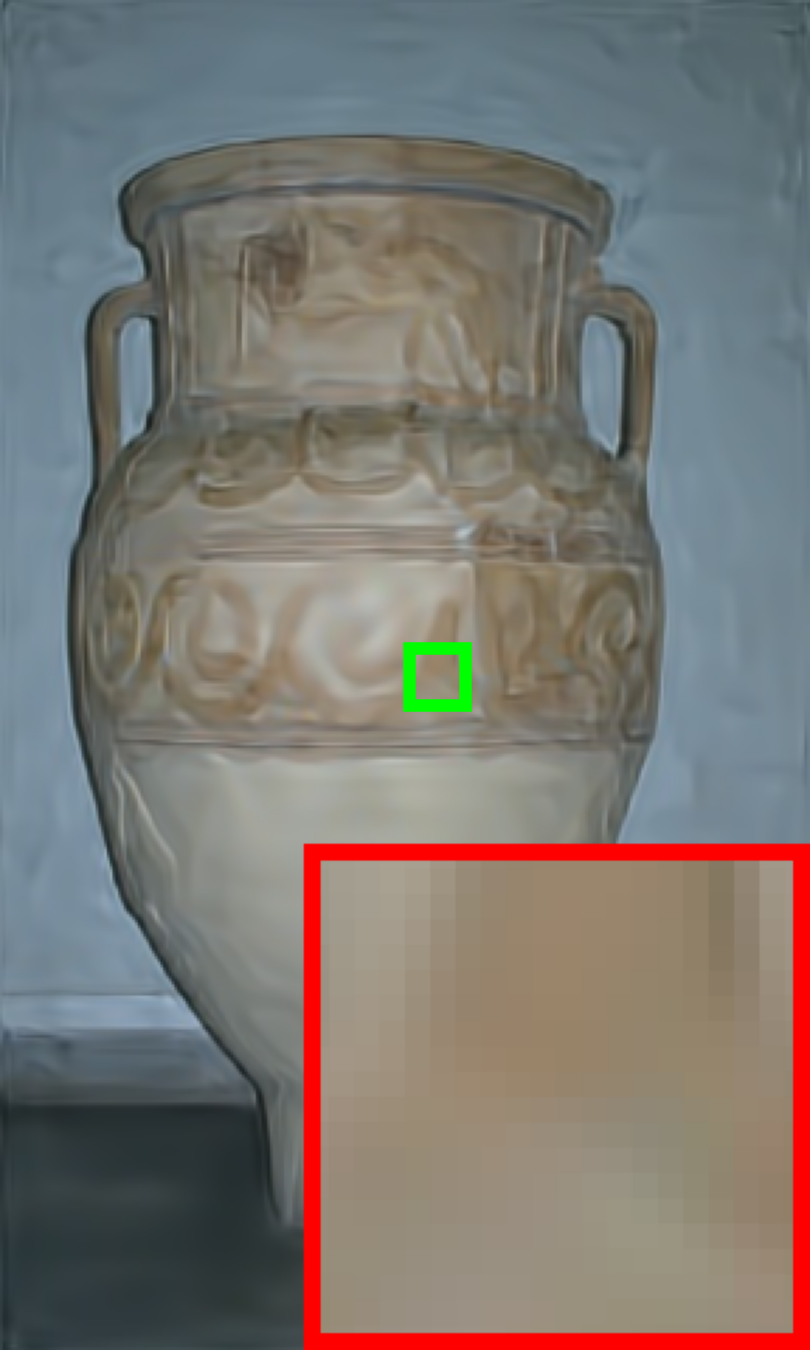}
    &\includegraphics[width=0.08\textwidth]{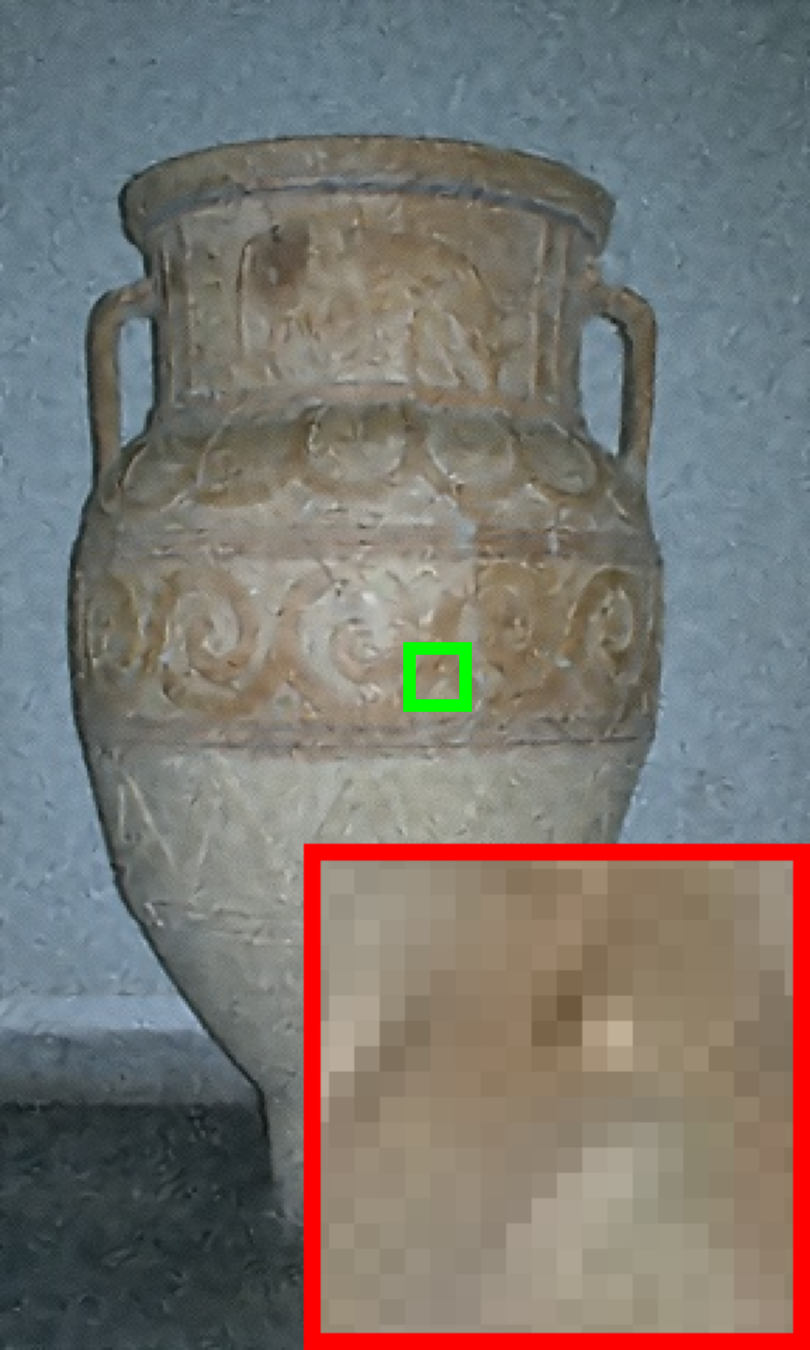}
    &\includegraphics[width=0.08\textwidth]{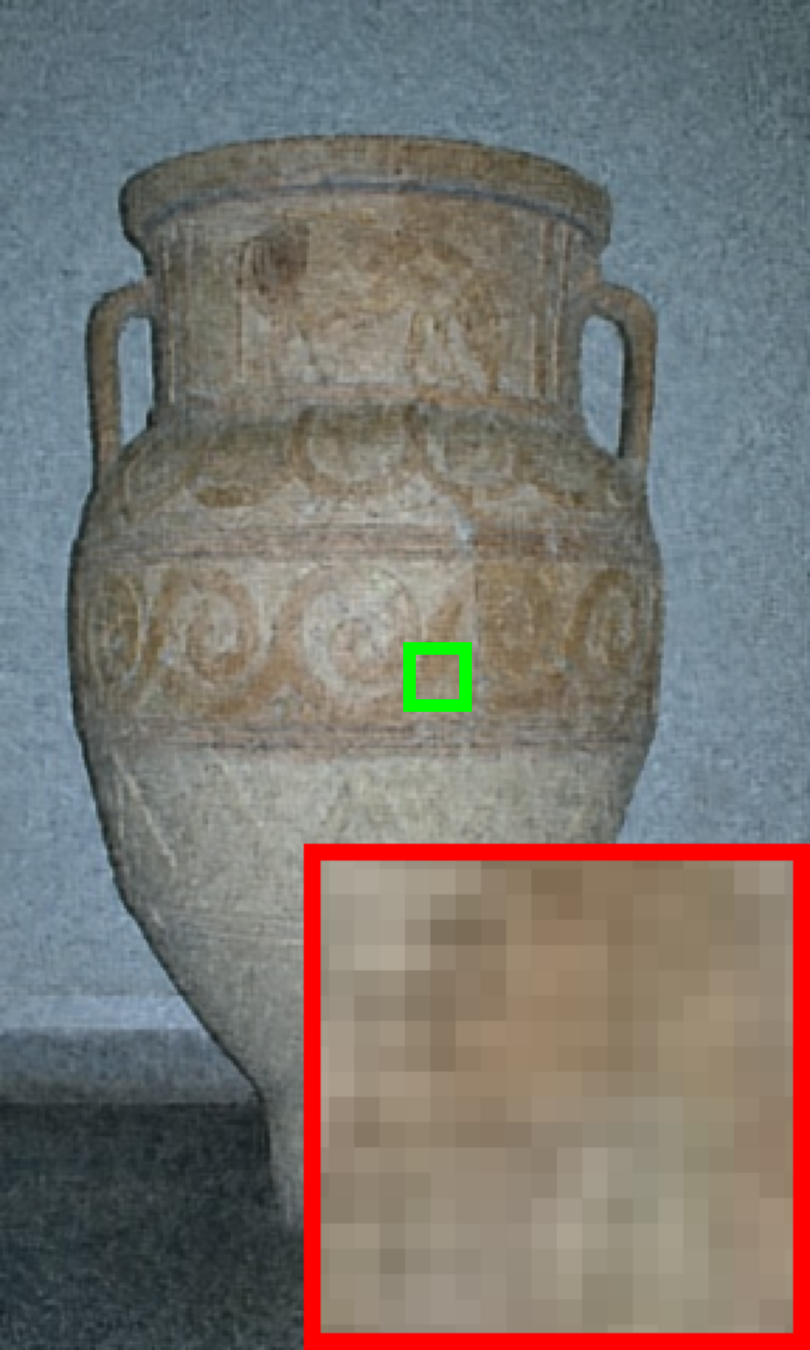}
    &\includegraphics[width=0.08\textwidth]{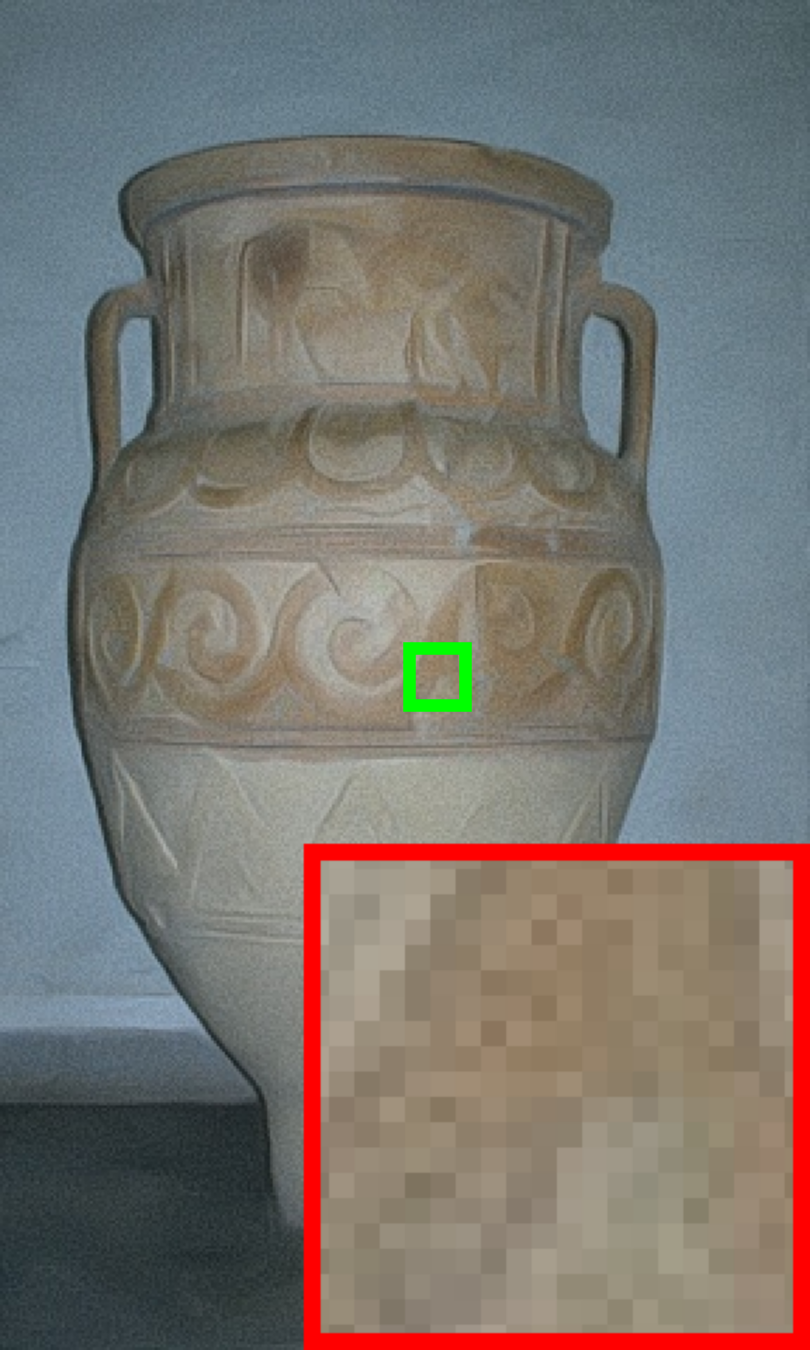}
    &\includegraphics[width=0.08\textwidth]{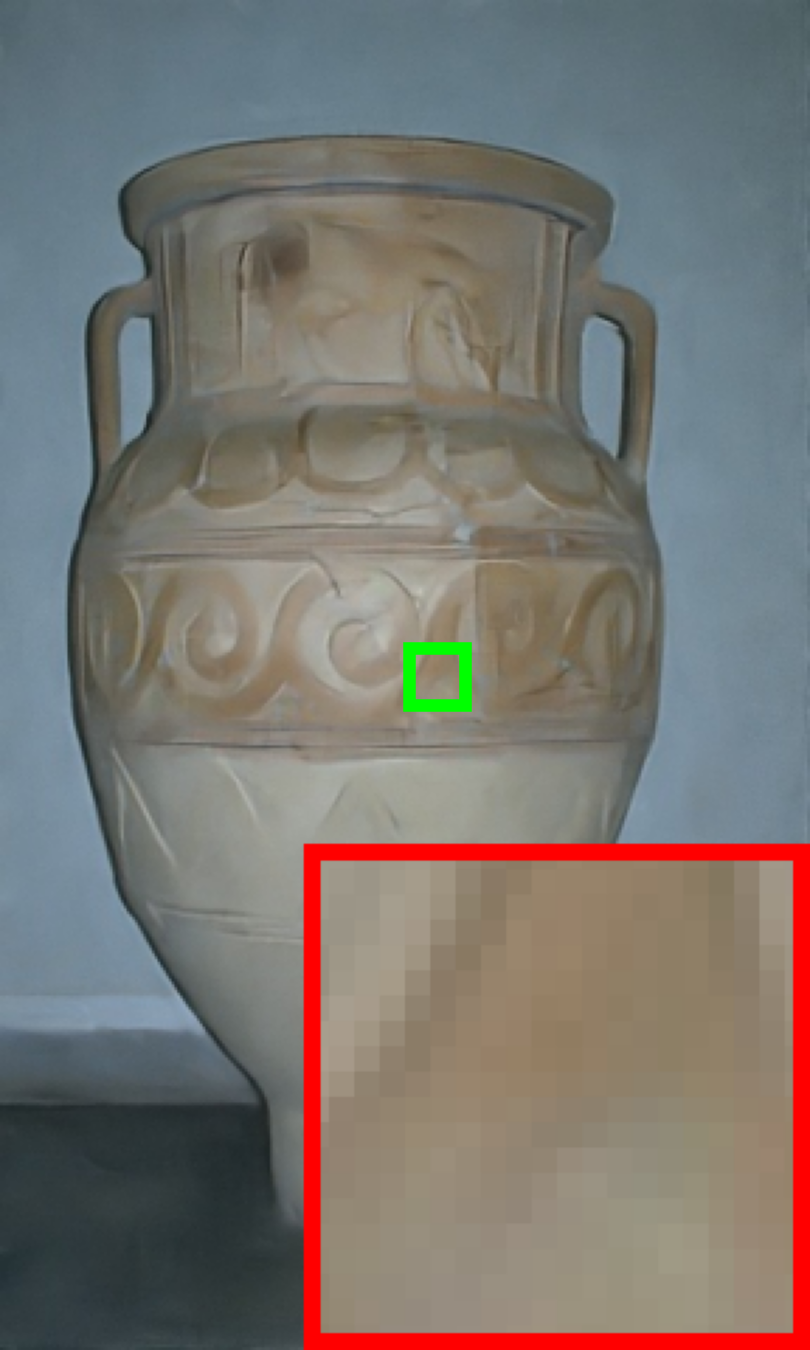}
    &\includegraphics[width=0.08\textwidth]{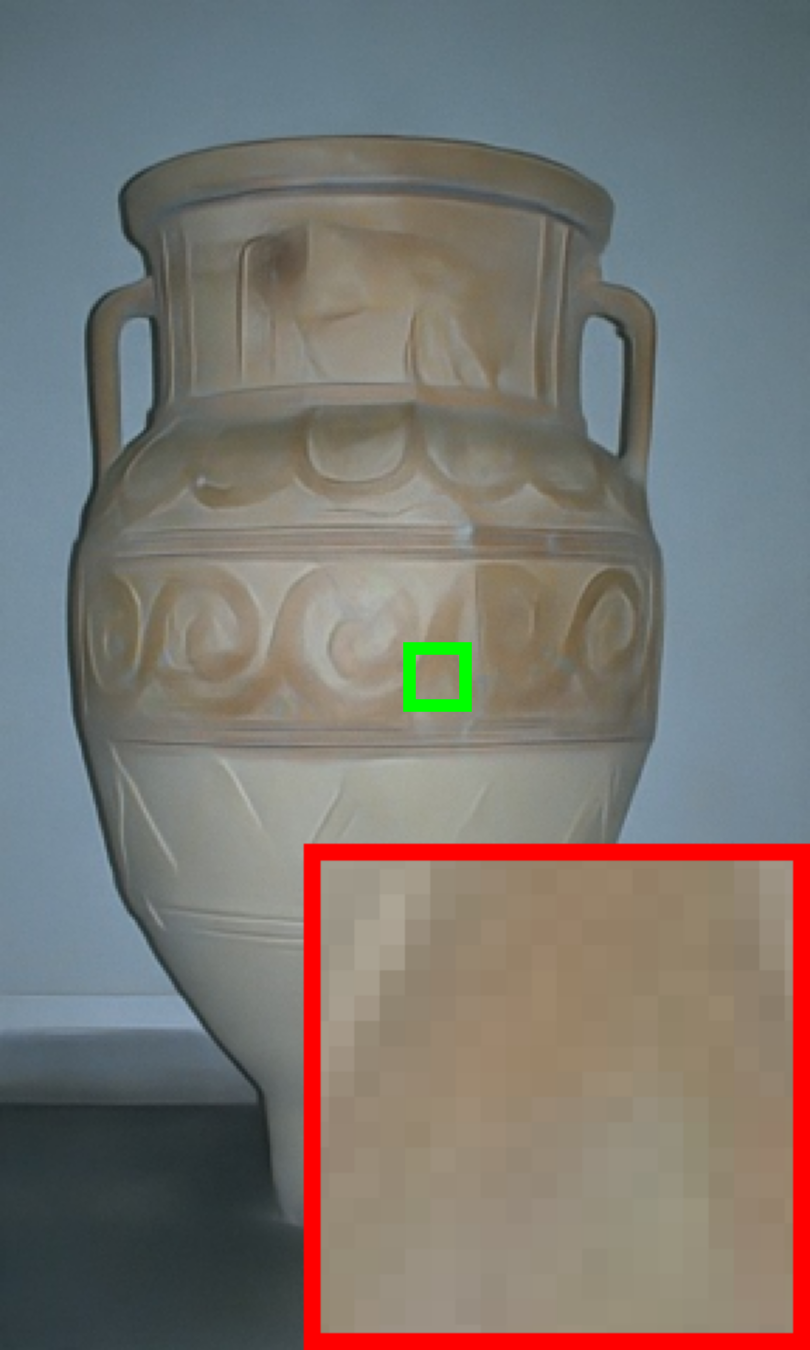}
    &\includegraphics[width=0.08\textwidth]{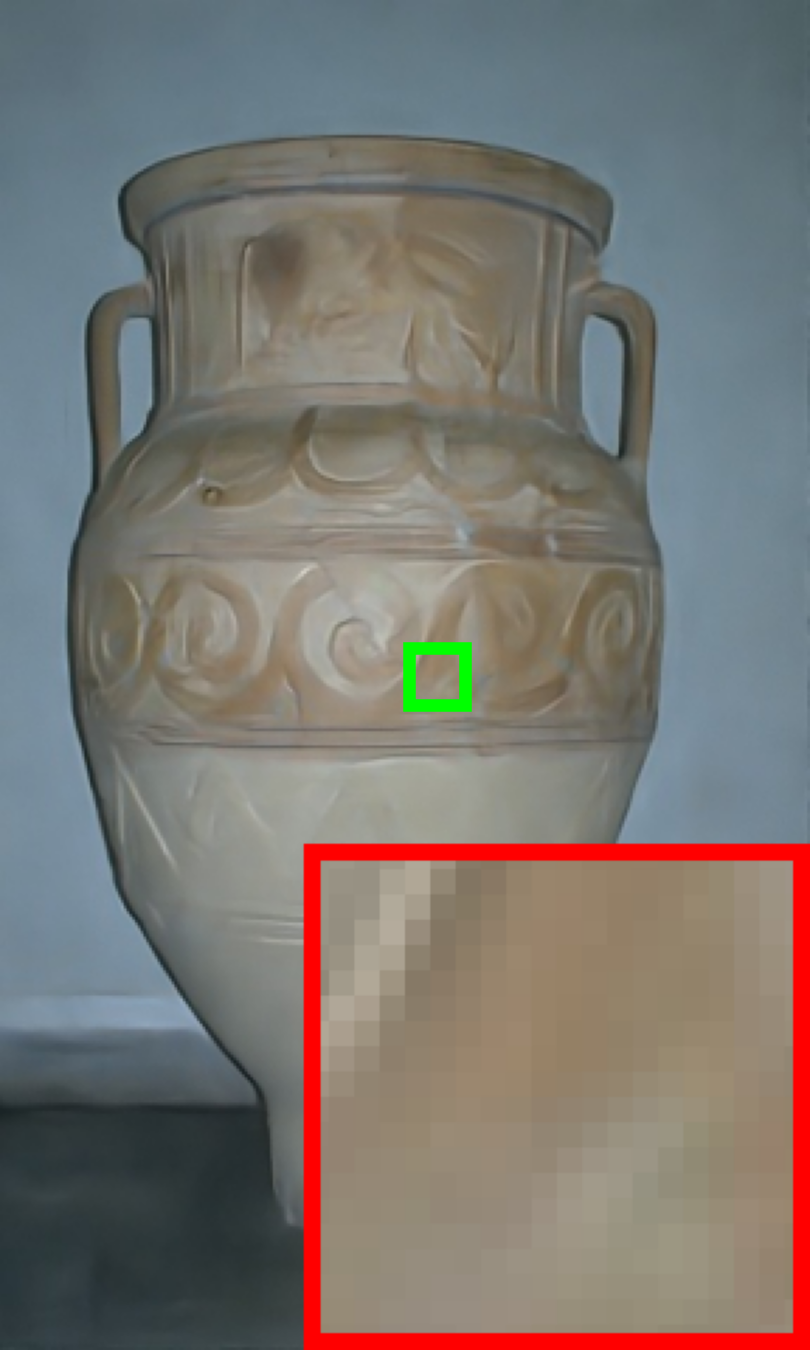}
    &\includegraphics[width=0.08\textwidth]{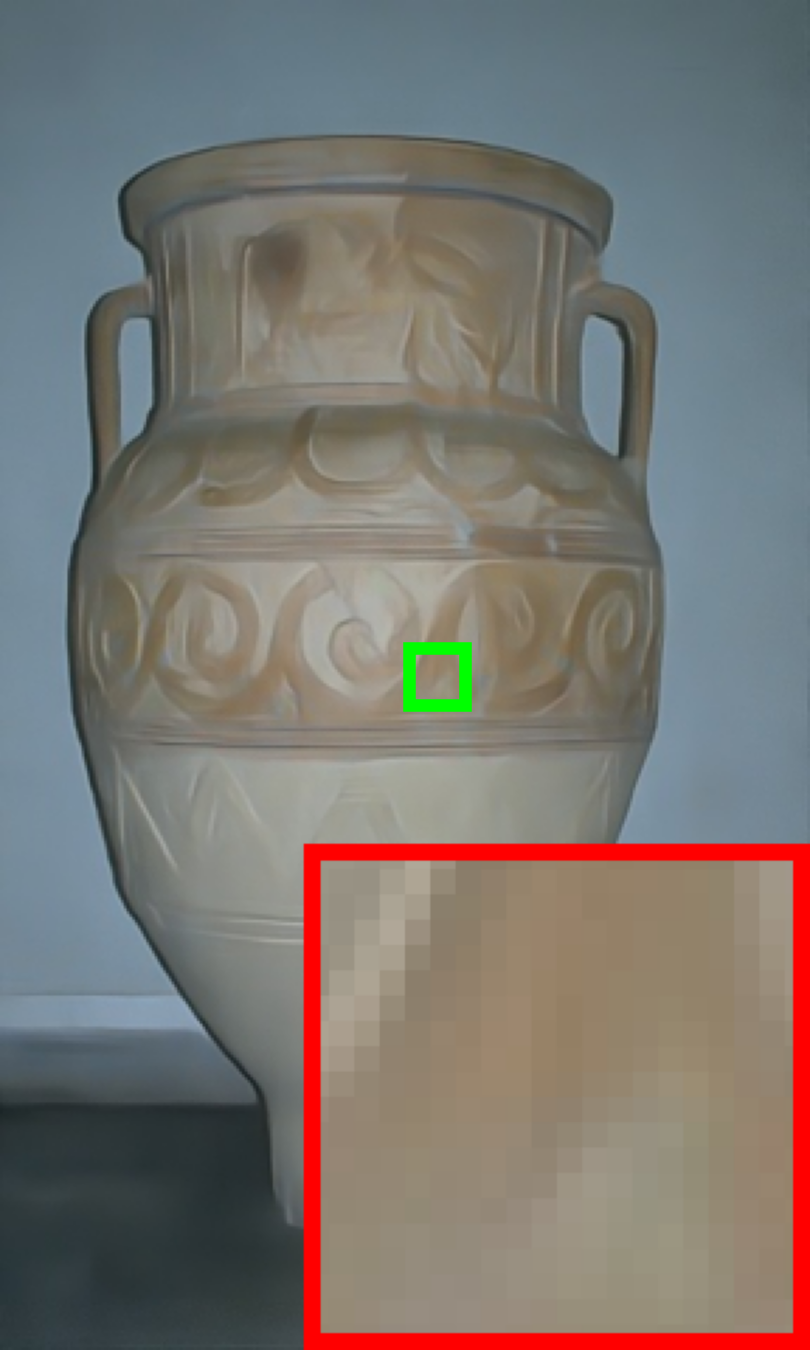}\\
    PSNR/SSIM & 31.26/0.78 & 32.28/0.83 & 30.55/0.71 & 29.79/0.65 & 31.14/0.70 & 33.58/\underline{\textcolor{blue}{0.86}} & \textbf{\textcolor{red}{34.09}}/\textbf{\textcolor{red}{0.87}} & 33.65/\textbf{\textcolor{red}{0.87}} & \underline{\textcolor{blue}{34.00}}/\textbf{\textcolor{red}{0.87}}
\end{tabular}}
\caption{Visual comparison of self-supervised methods on two natural benchmark images named ``House'' and ``test\_45'' from Set11~\cite{kulkarni2016reconnet} \textcolor{blue}{(top)} and CBSD68~\cite{martin2001database} \textcolor{blue}{(bottom)}, with $\gamma =30\%$ and $\sigma =10$.}
\label{fig:comparison_standard_natural_images_r30_s10}
\end{figure*}

\begin{figure*}[!t]
\setlength{\tabcolsep}{0.5pt}
\hspace{-4pt}
\resizebox{1.0\textwidth}{!}{
\tiny
\begin{tabular}{cccccccccc}
    GT & DIP & BCNN & EI & ASGLD & DDSSL & \textbf{SC-CNN} & \textbf{SC-CNN$^\text{+}$} & \textbf{SCT} & \textbf{SCT$^\text{+}$}\\
    \includegraphics[width=0.08\textwidth]{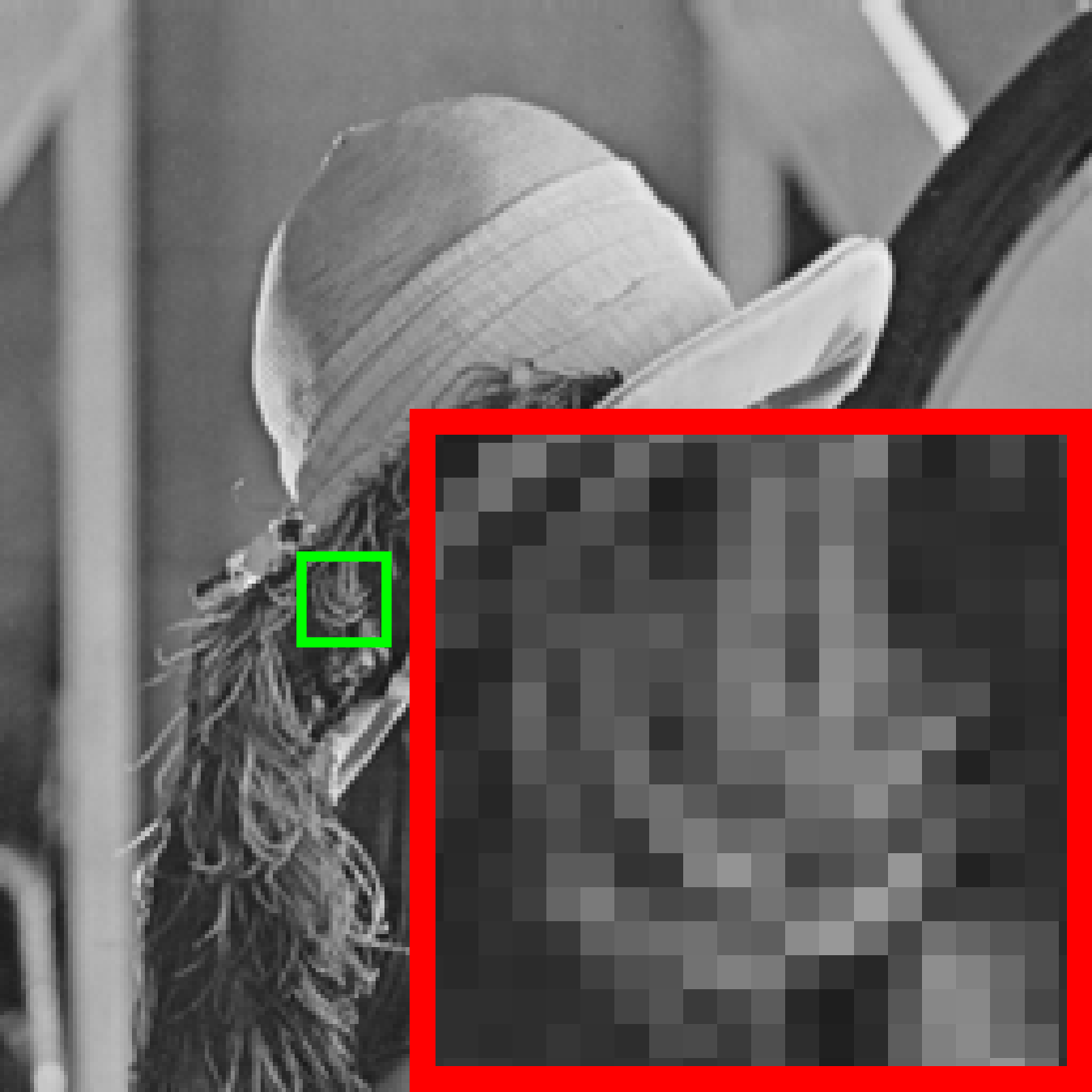}
    &\includegraphics[width=0.08\textwidth]{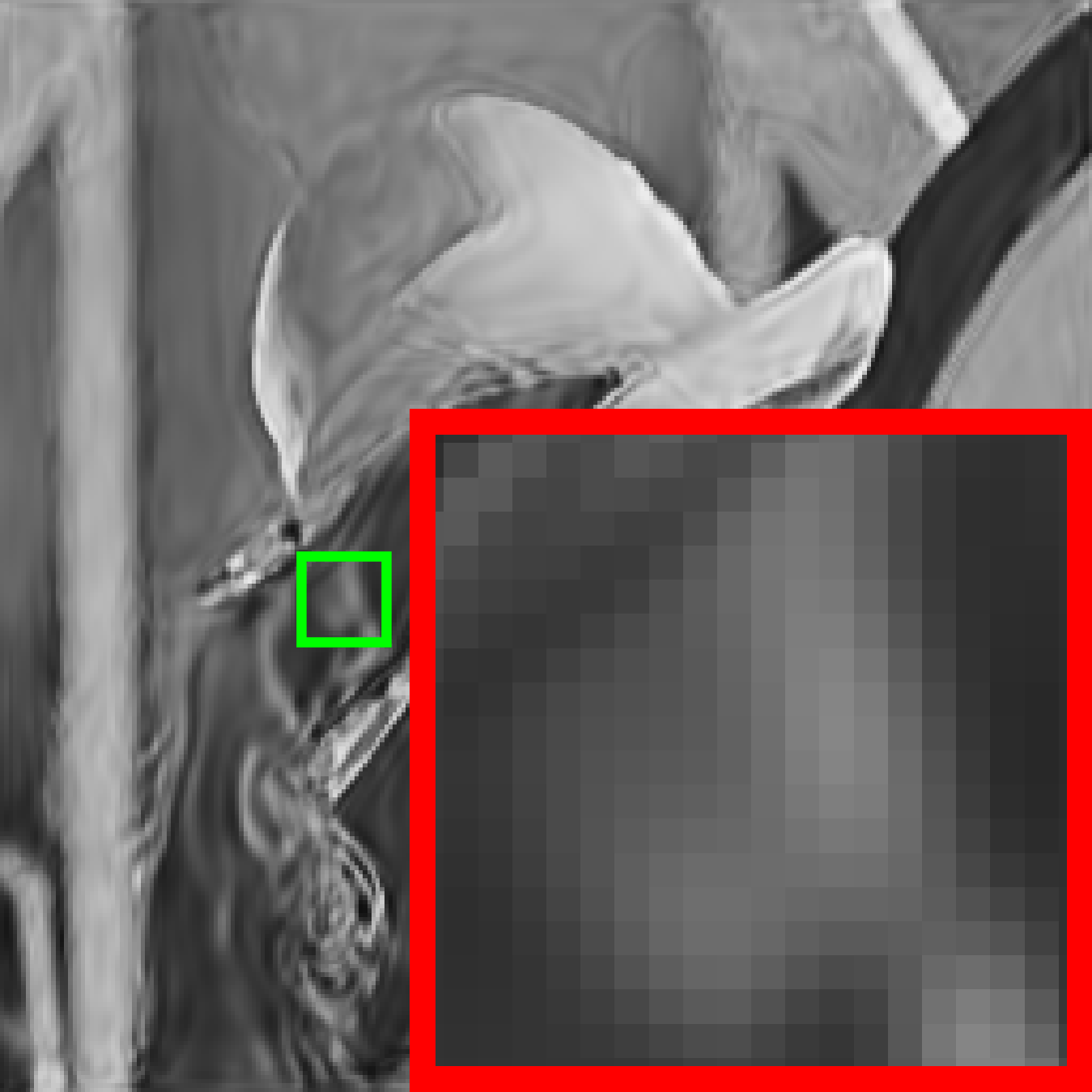}
    &\includegraphics[width=0.08\textwidth]{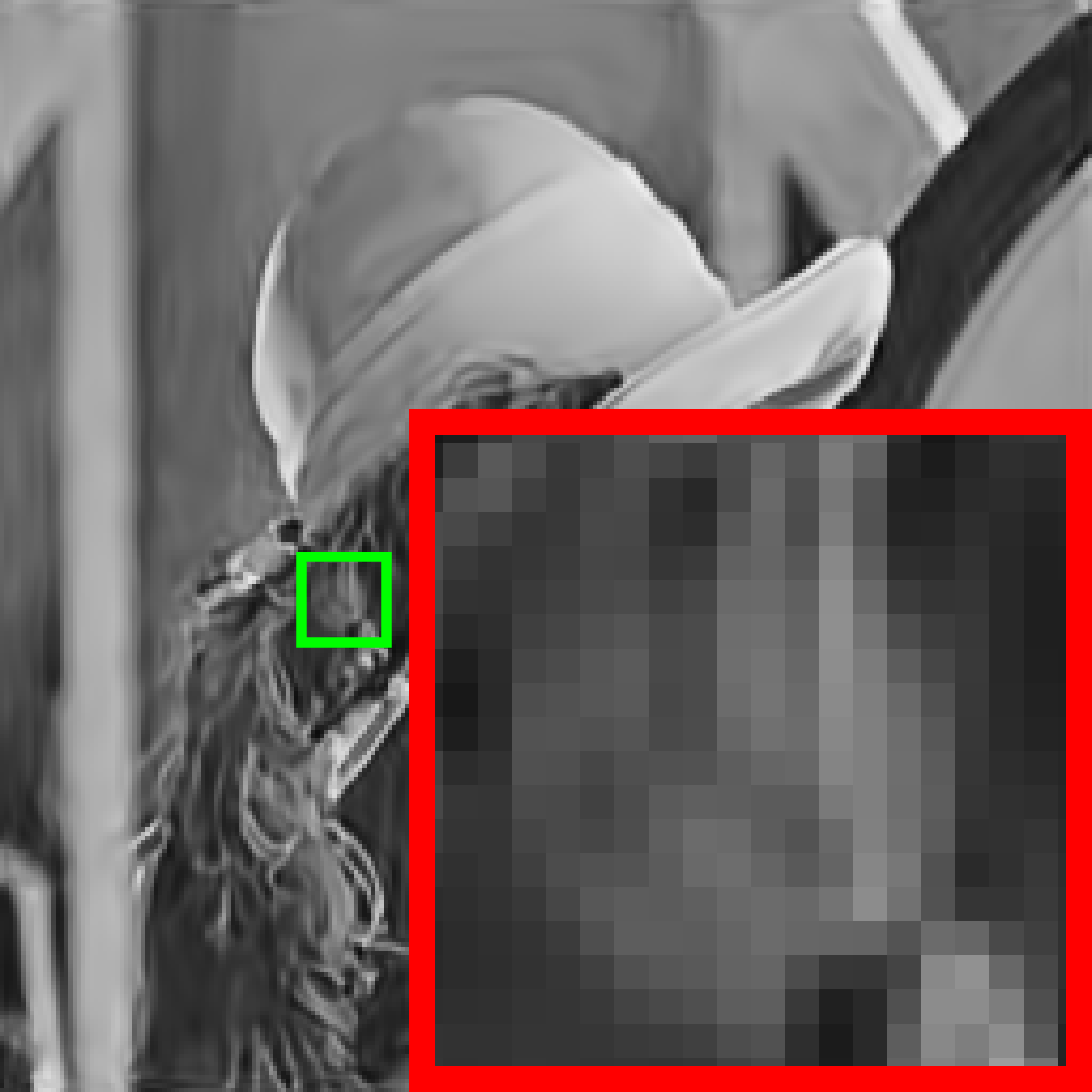}
    &\includegraphics[width=0.08\textwidth]{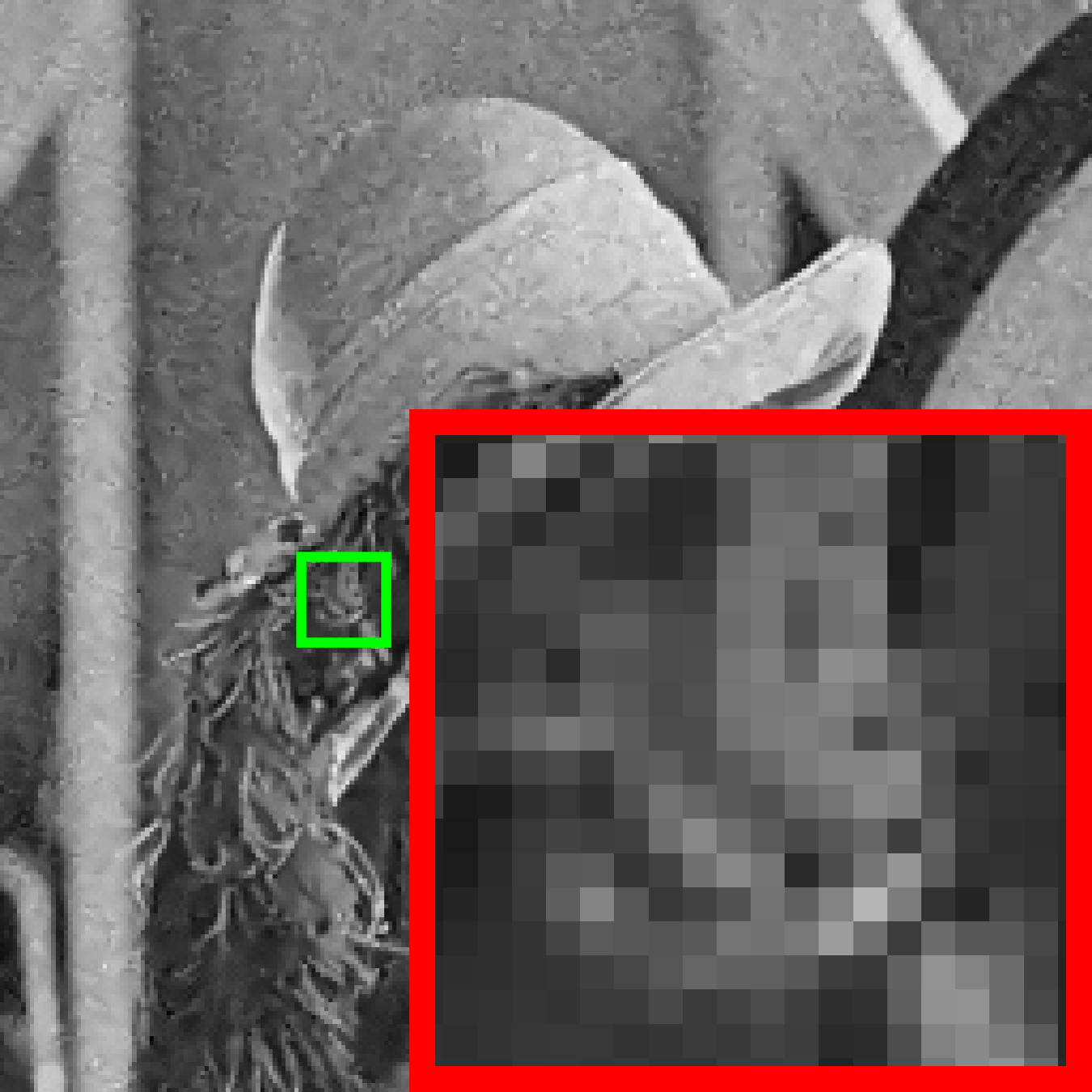}
    &\includegraphics[width=0.08\textwidth]{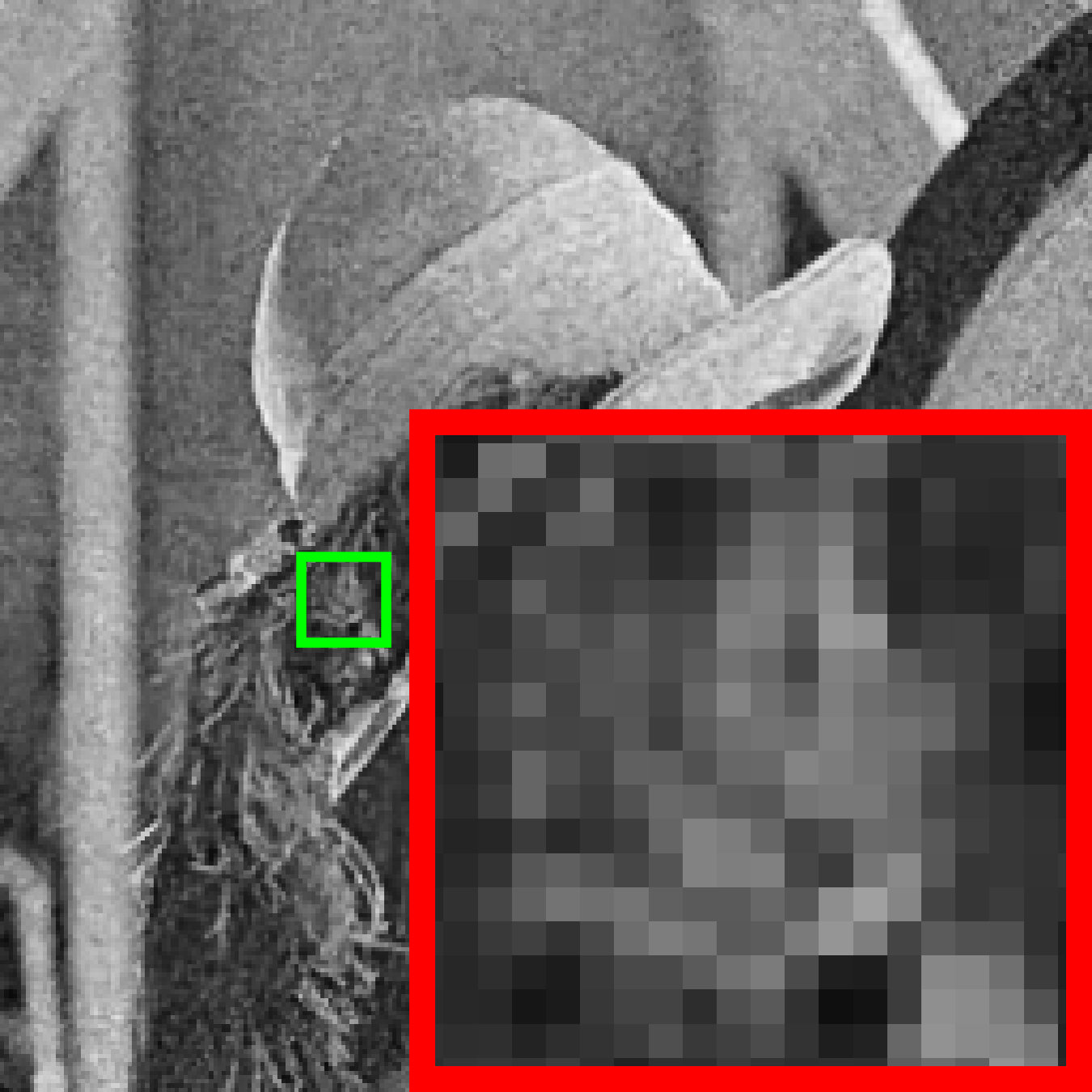}
    &\includegraphics[width=0.08\textwidth]{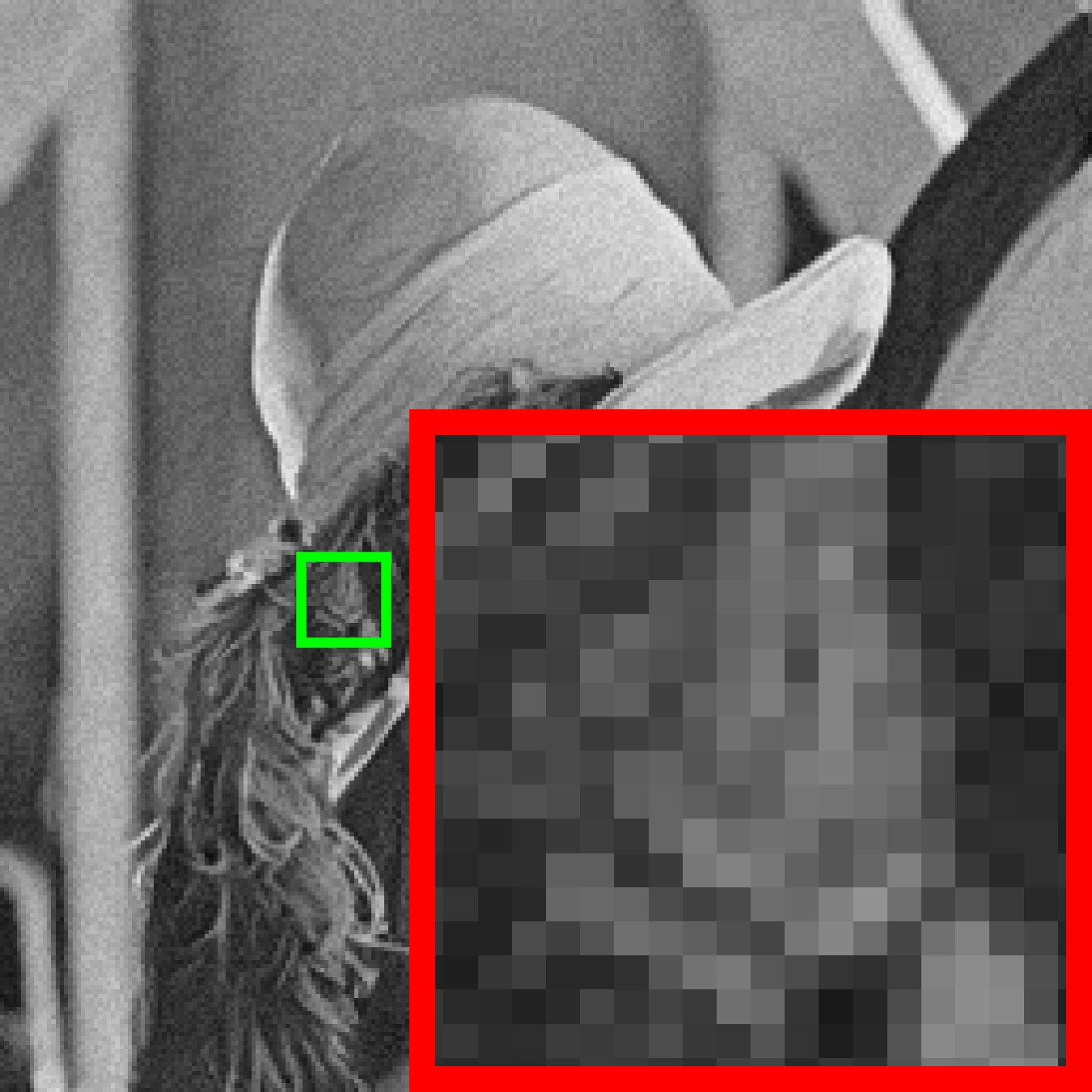}
    &\includegraphics[width=0.08\textwidth]{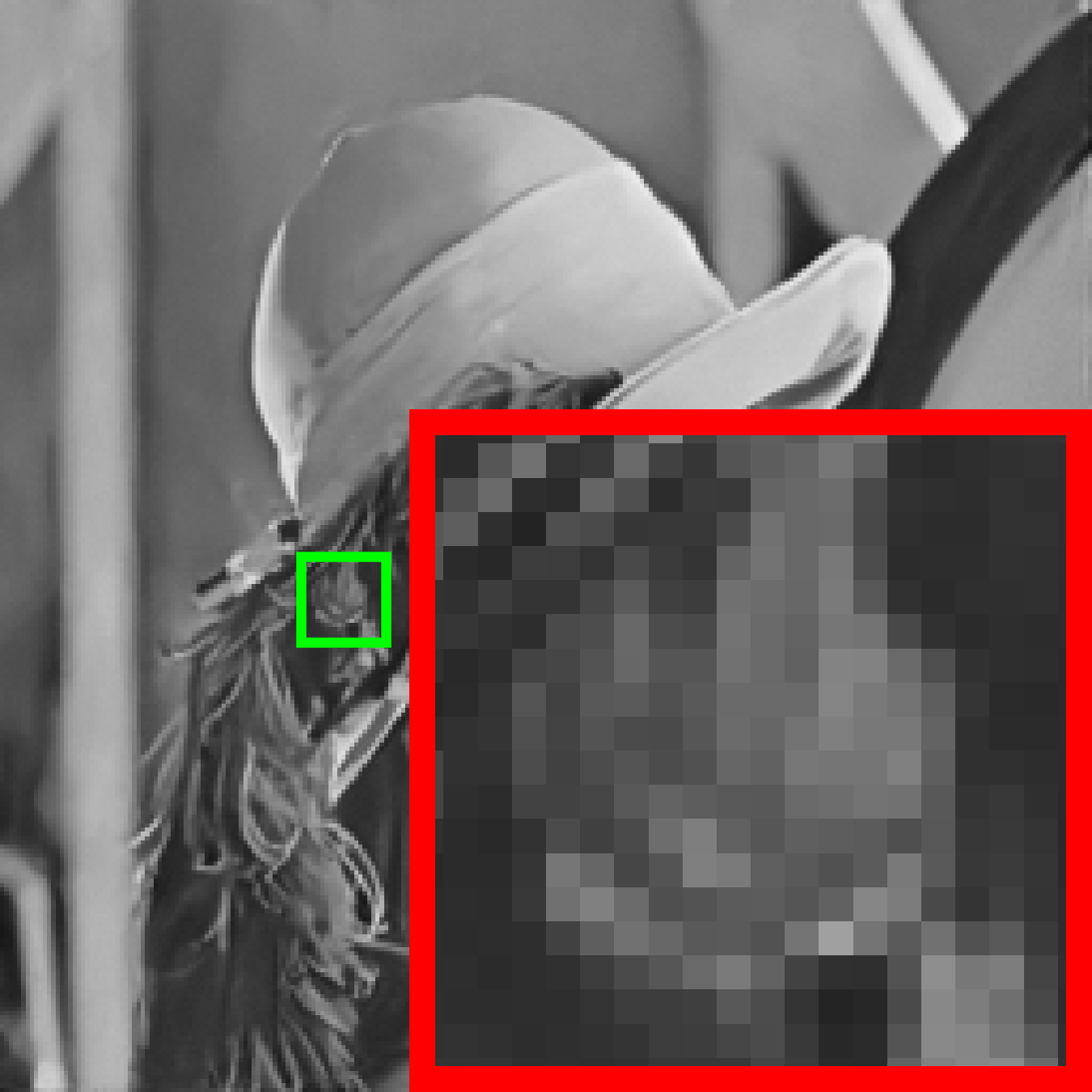}
    &\includegraphics[width=0.08\textwidth]{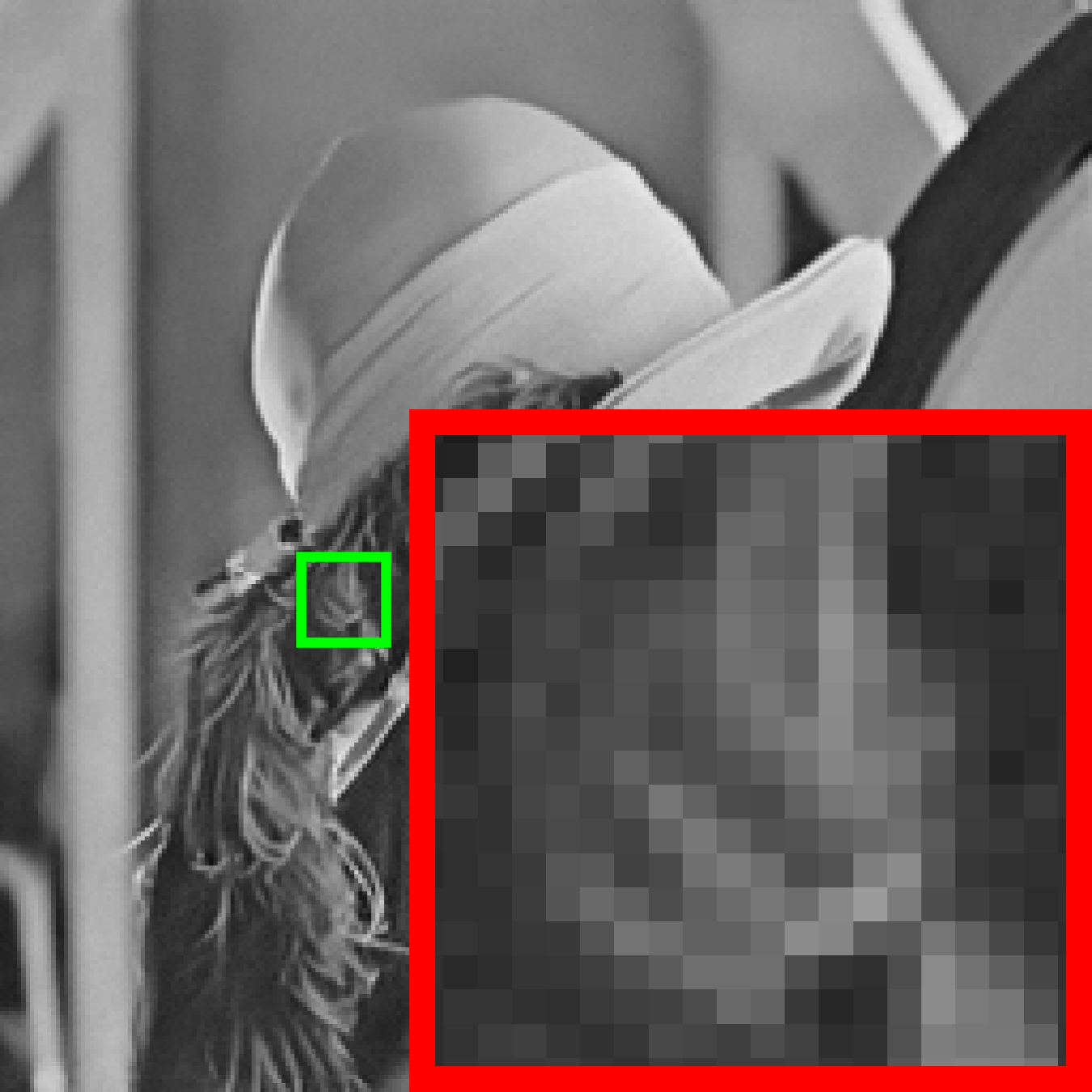}
    &\includegraphics[width=0.08\textwidth]{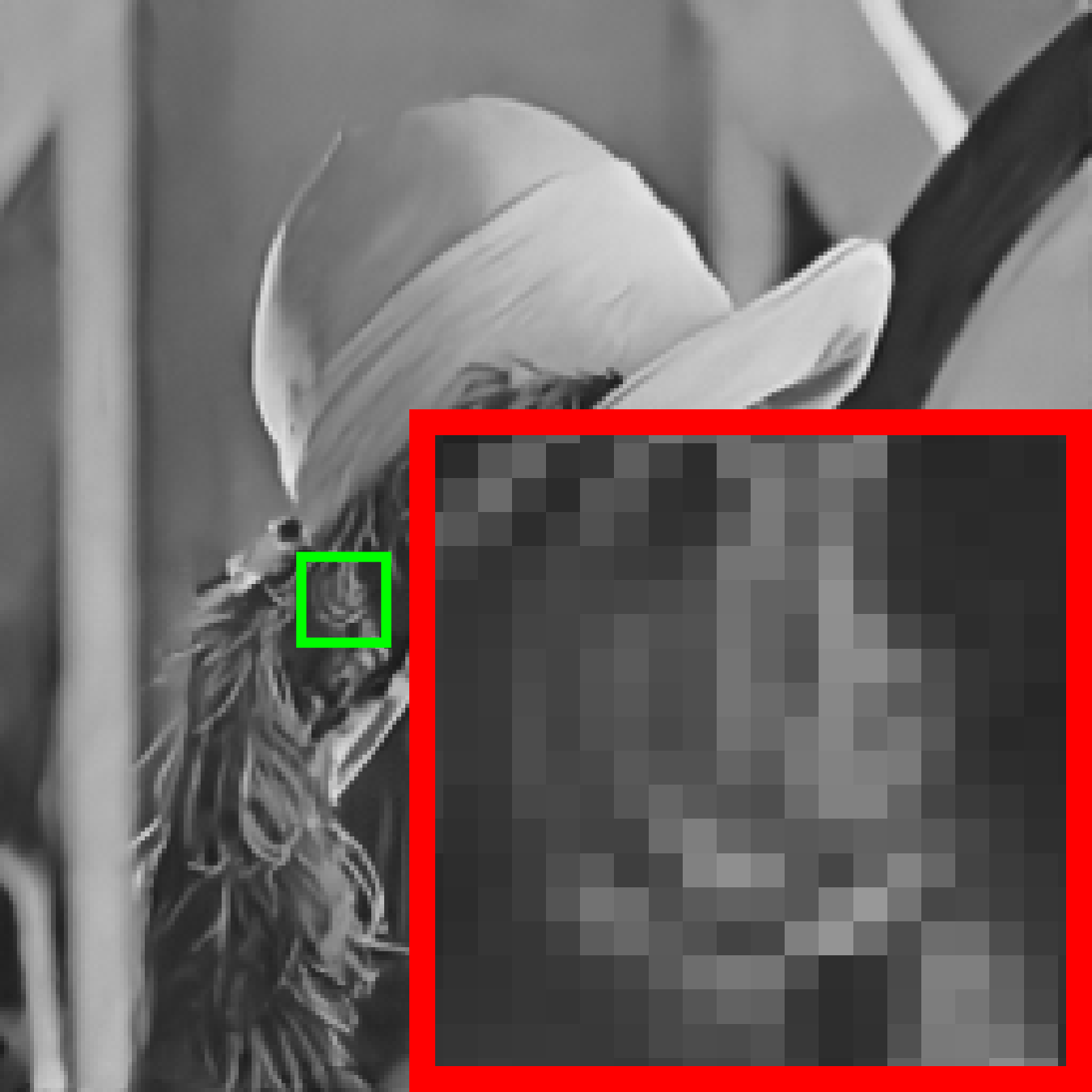}
    &\includegraphics[width=0.08\textwidth]{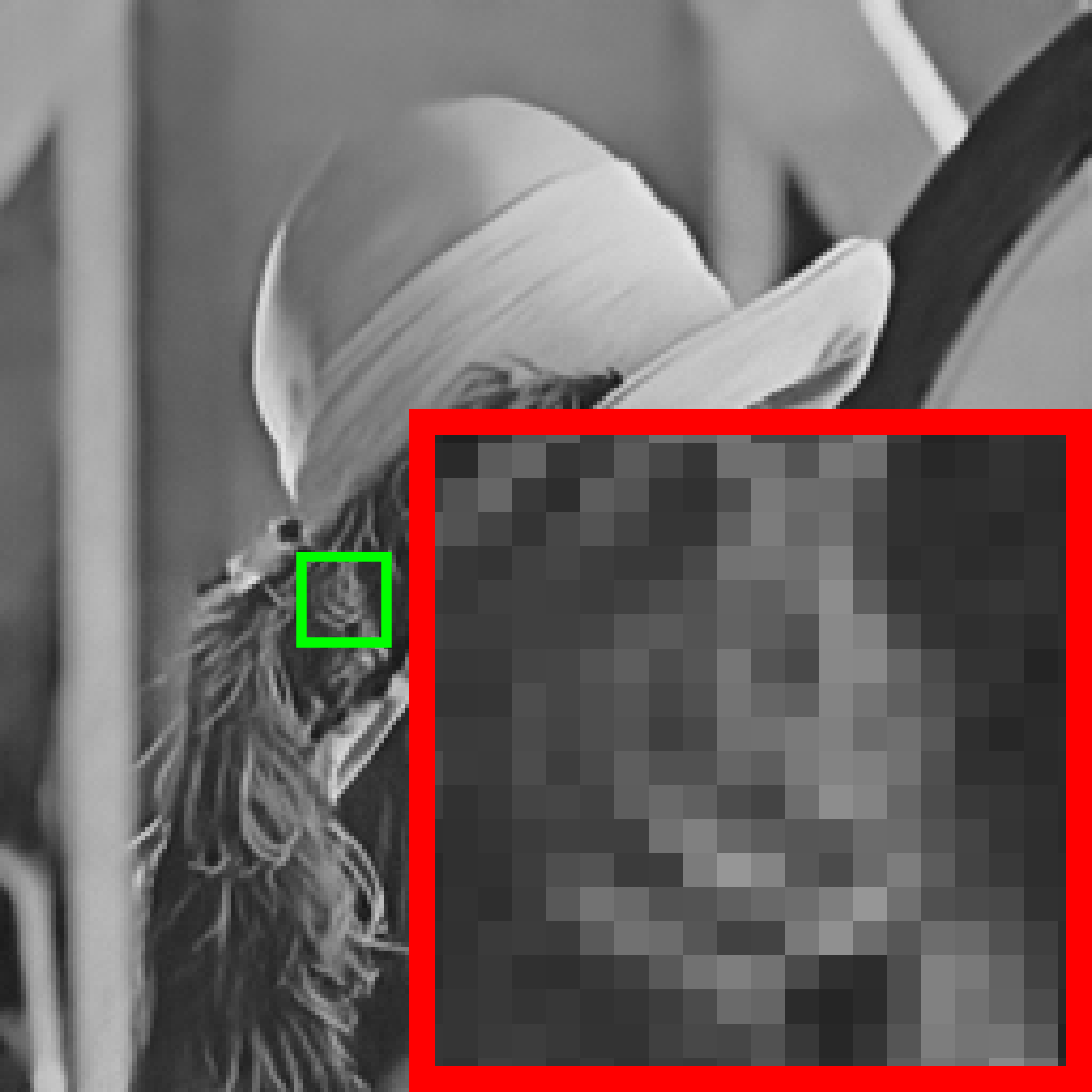}\\
    PSNR/SSIM & 29.26/0.83 & 31.12/\underline{\textcolor{blue}{0.90}} & 29.39/0.77 & 27.46/0.70 & 29.60/0.77 & 32.77/\textbf{\textcolor{red}{0.92}} & \textbf{\textcolor{red}{33.32}}/\textbf{\textcolor{red}{0.92}} & 32.88/\textbf{\textcolor{red}{0.92}} & \underline{\textcolor{blue}{33.23}}/\textbf{\textcolor{red}{0.92}}\\
    \includegraphics[width=0.08\textwidth]{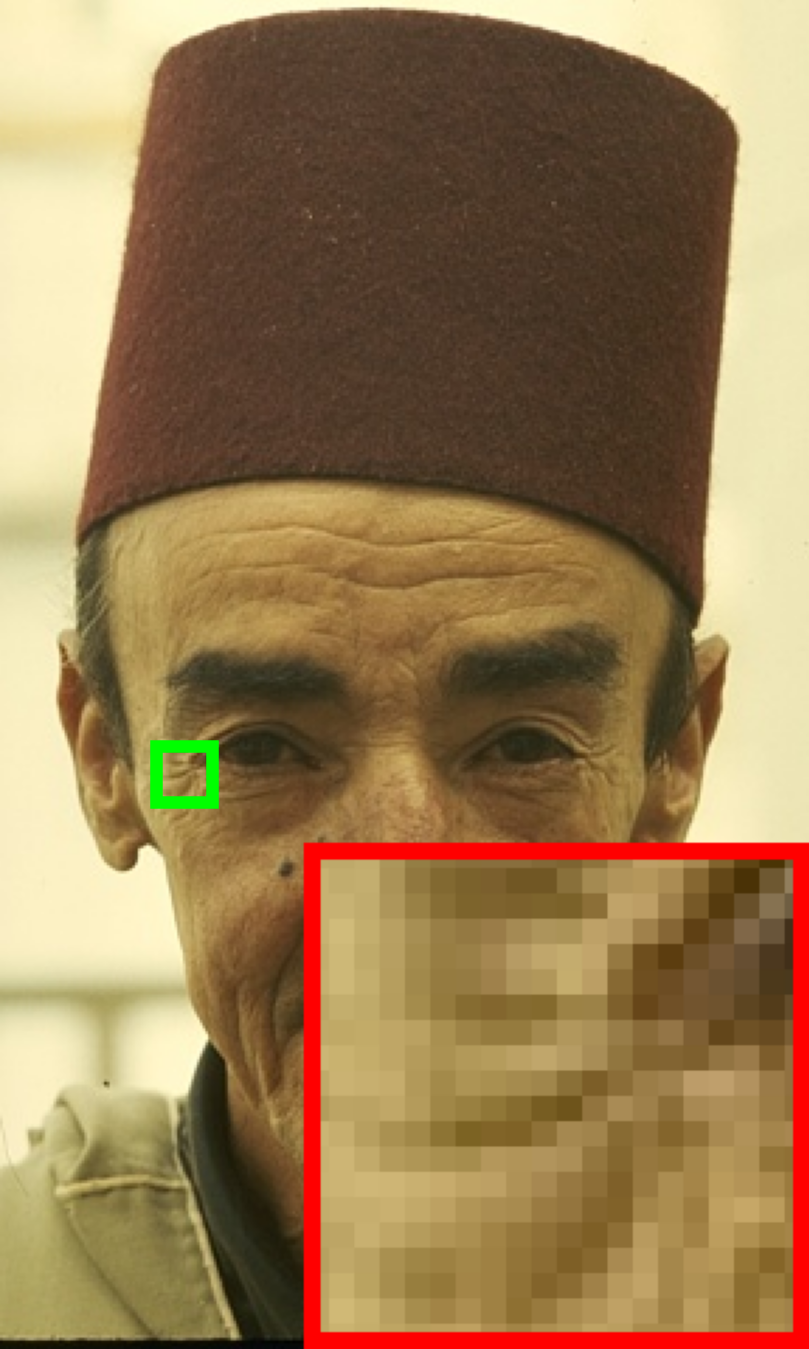}
    &\includegraphics[width=0.08\textwidth]{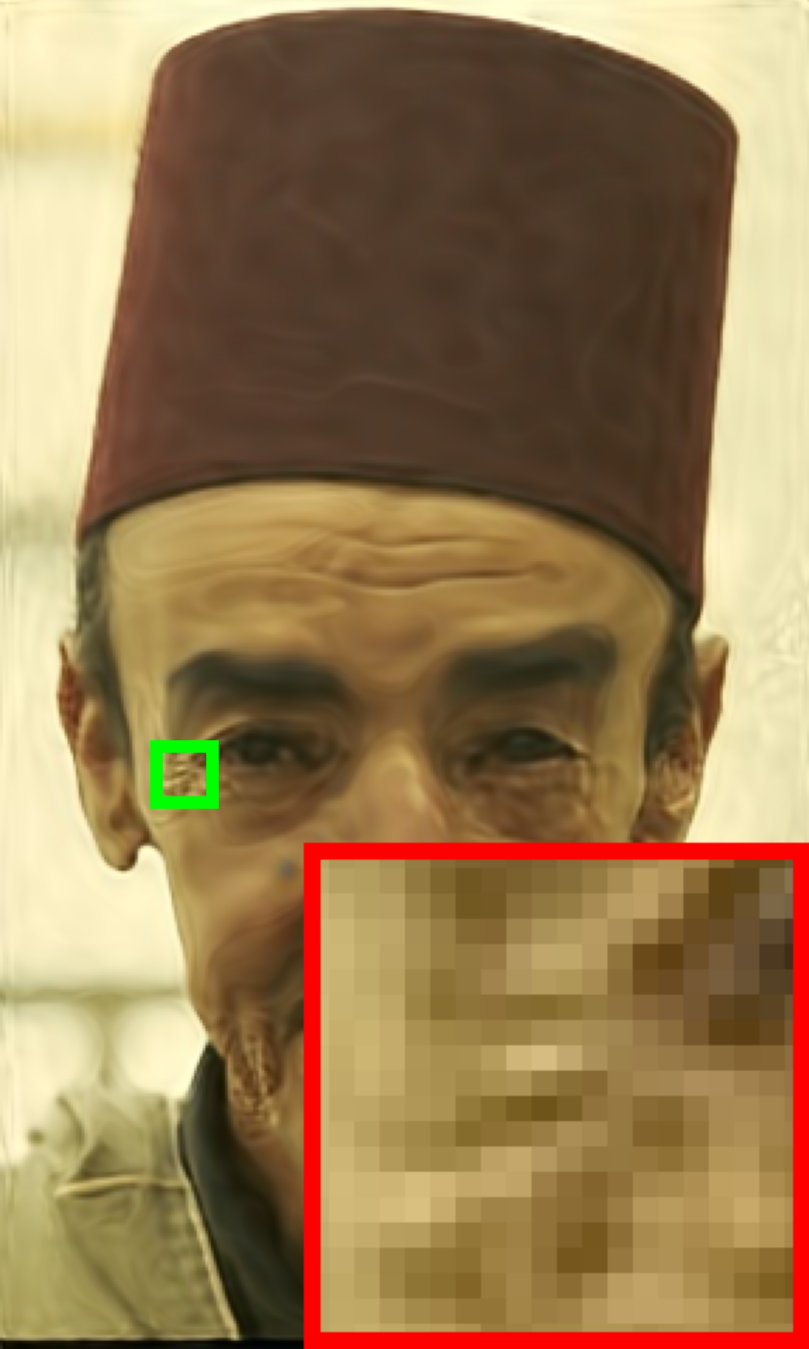}
    &\includegraphics[width=0.08\textwidth]{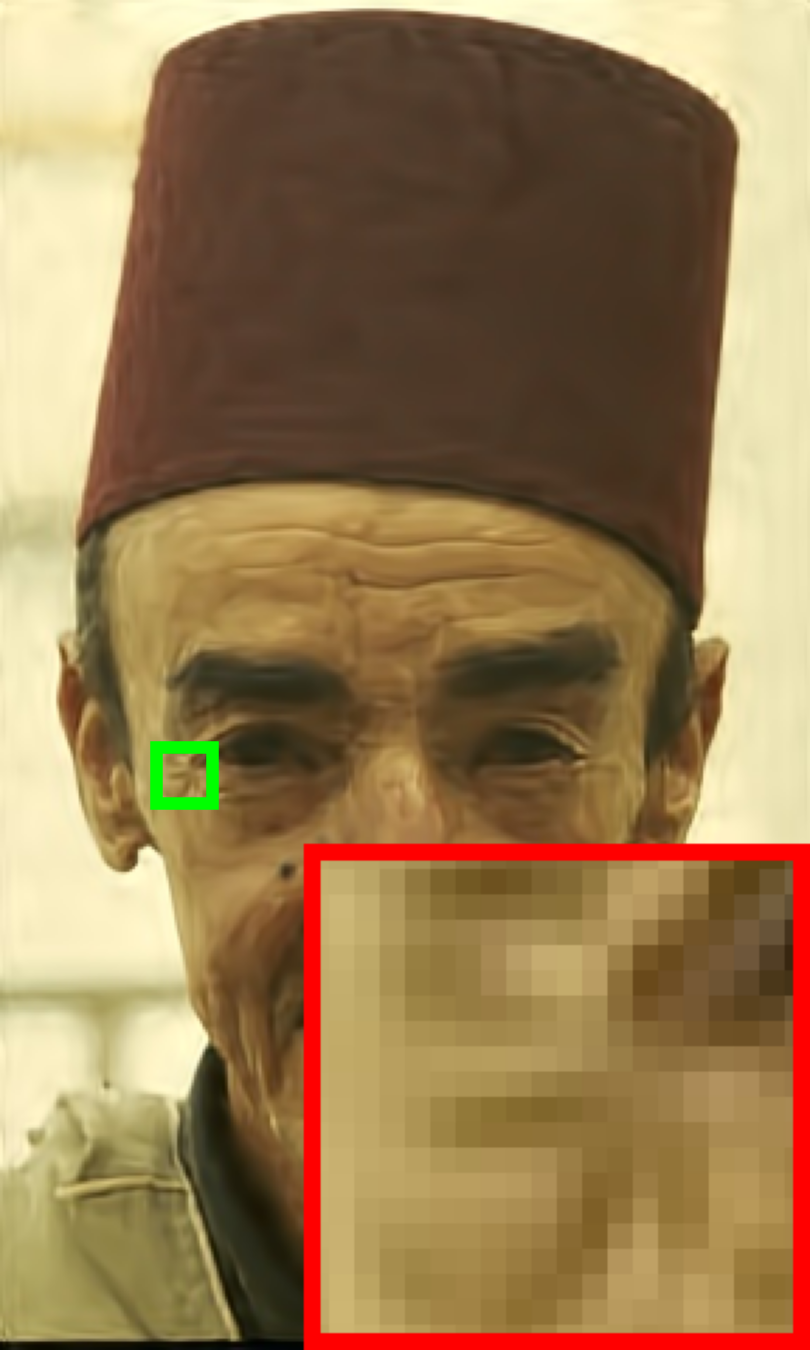}
    &\includegraphics[width=0.08\textwidth]{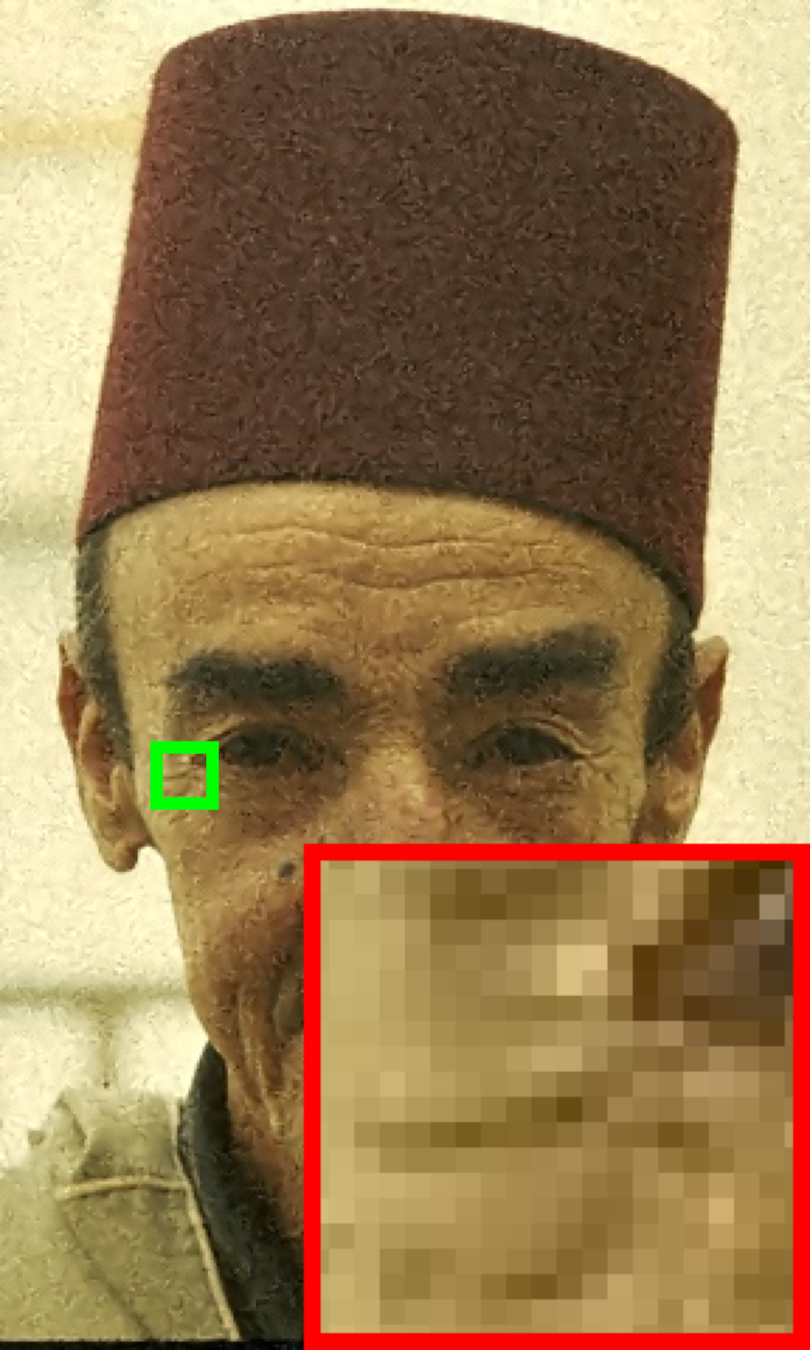}
    &\includegraphics[width=0.08\textwidth]{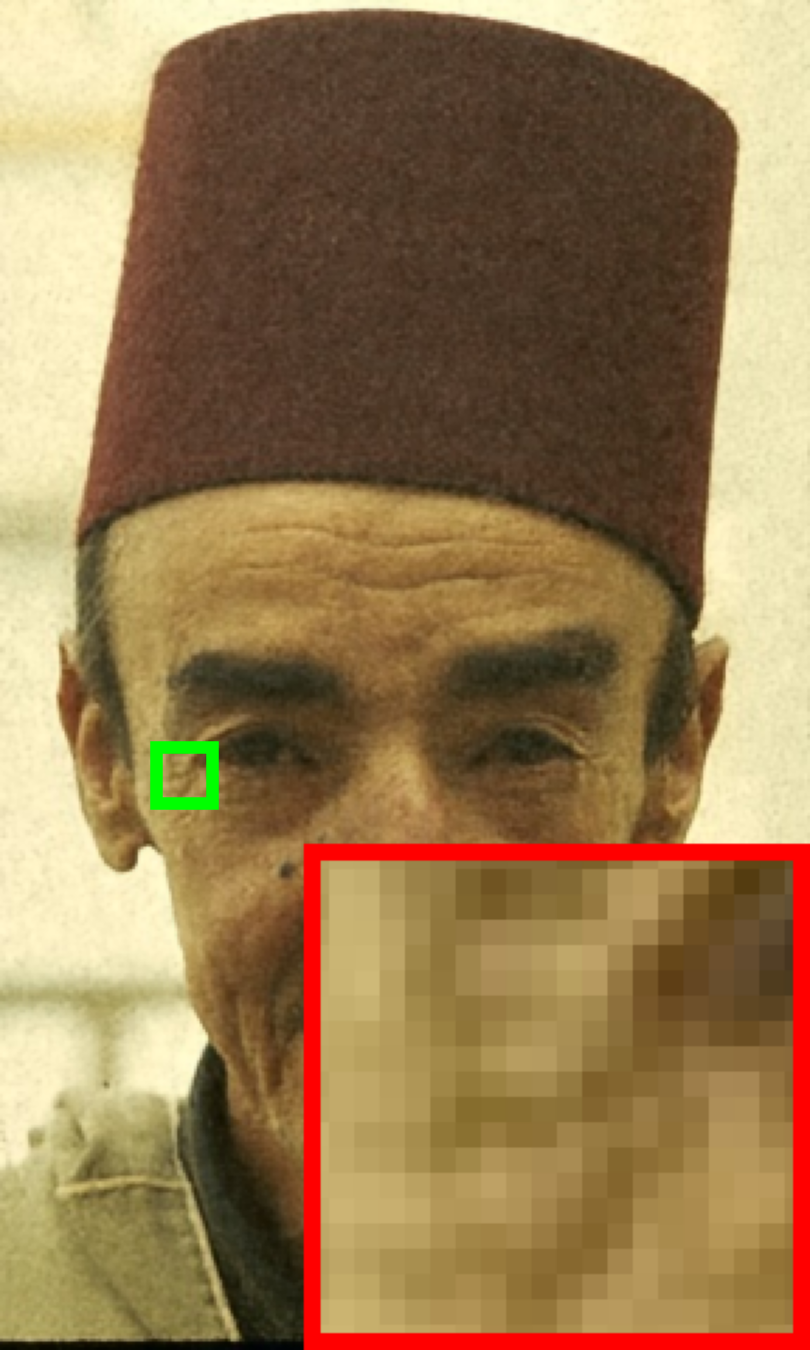}
    &\includegraphics[width=0.08\textwidth]{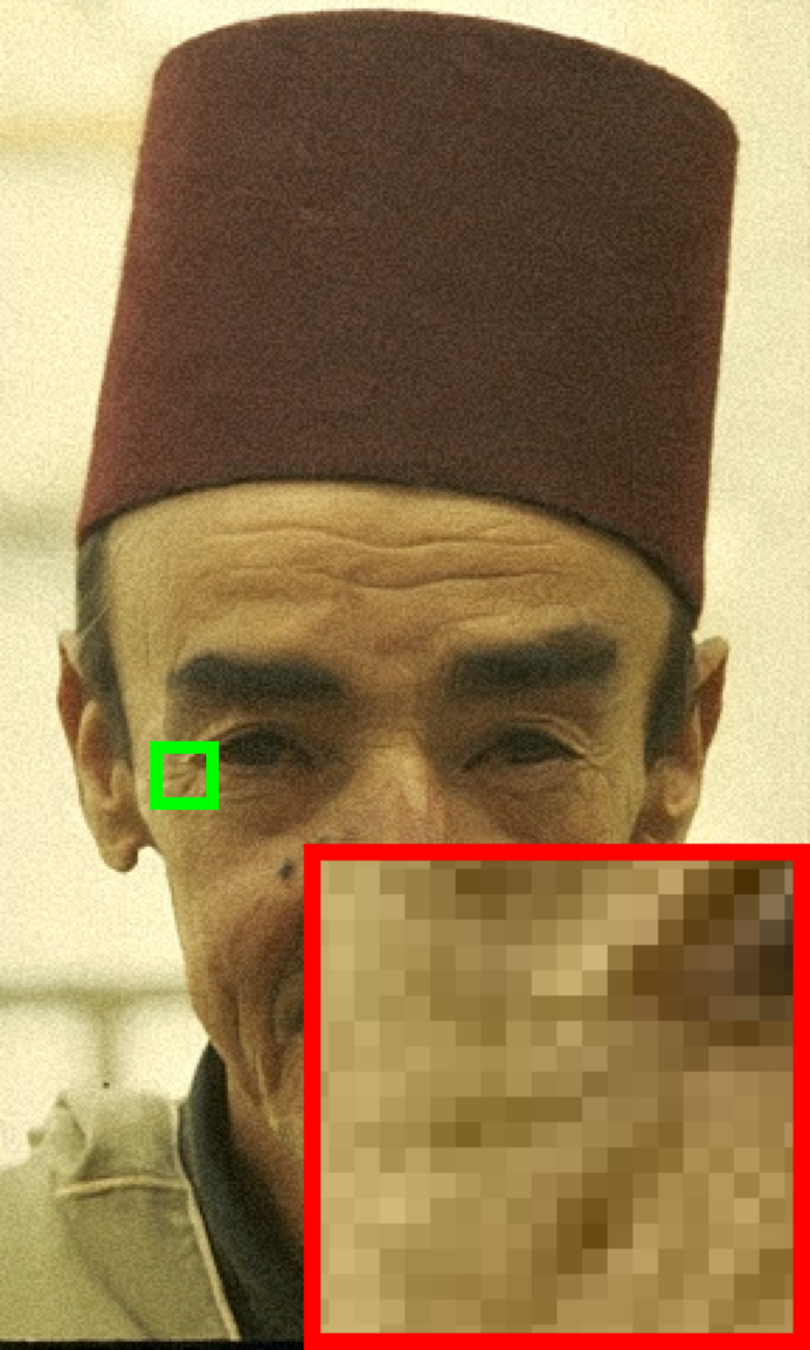}
    &\includegraphics[width=0.08\textwidth]{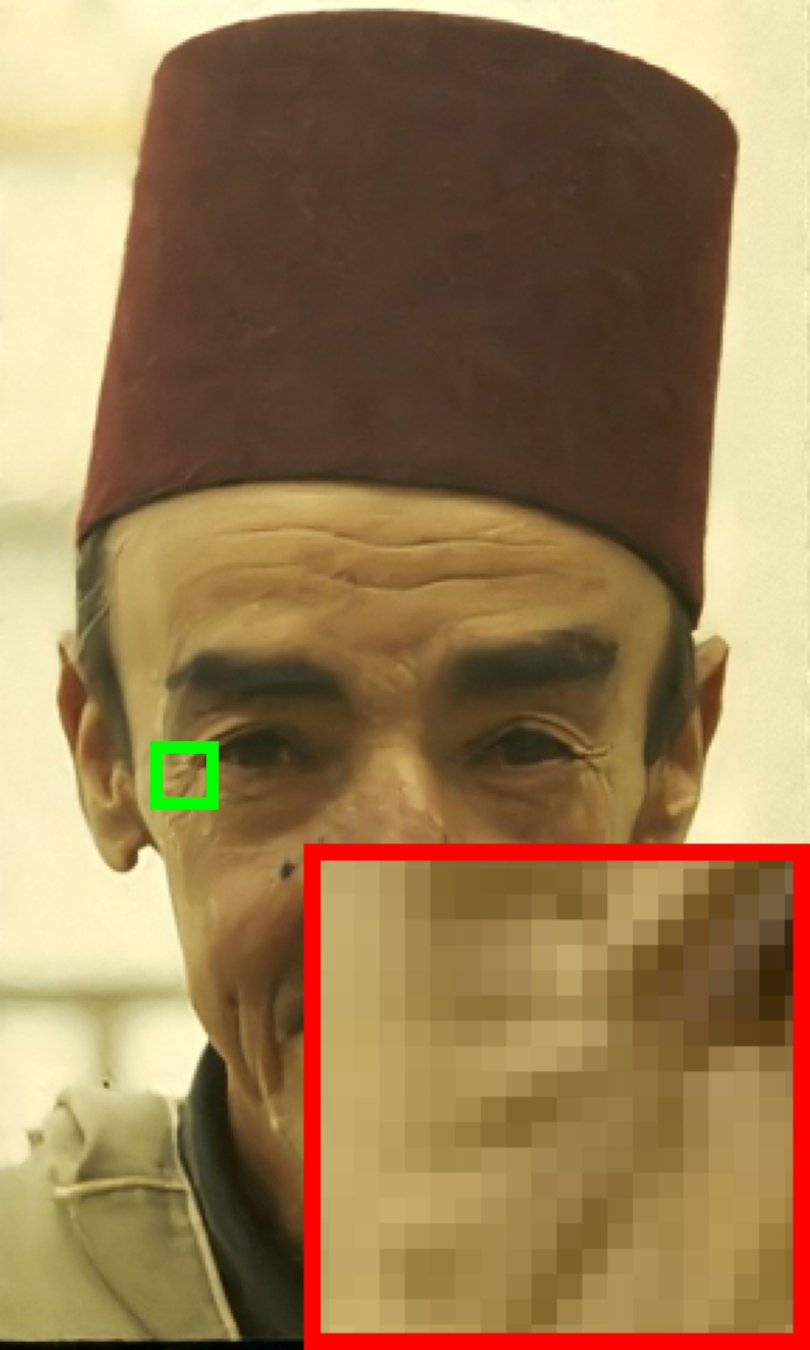}
    &\includegraphics[width=0.08\textwidth]{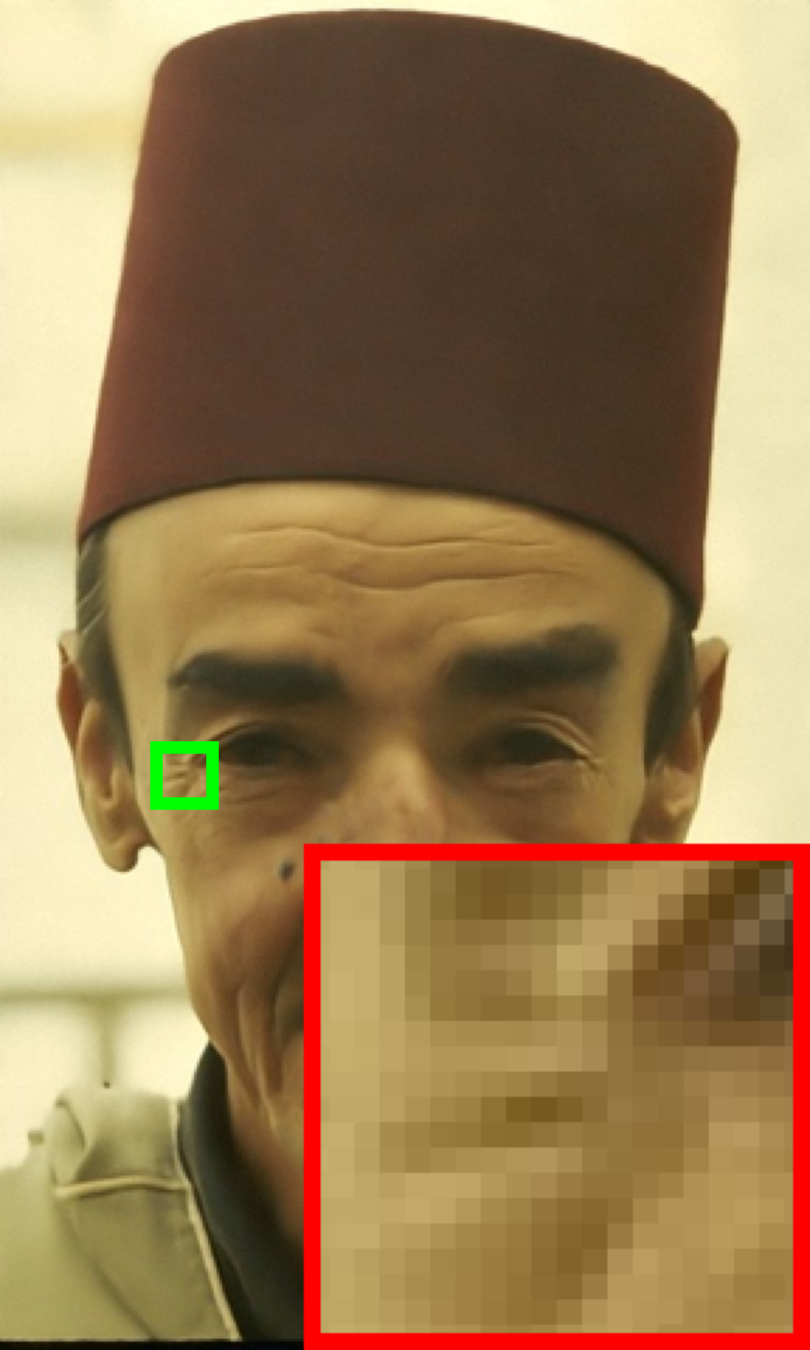}
    &\includegraphics[width=0.08\textwidth]{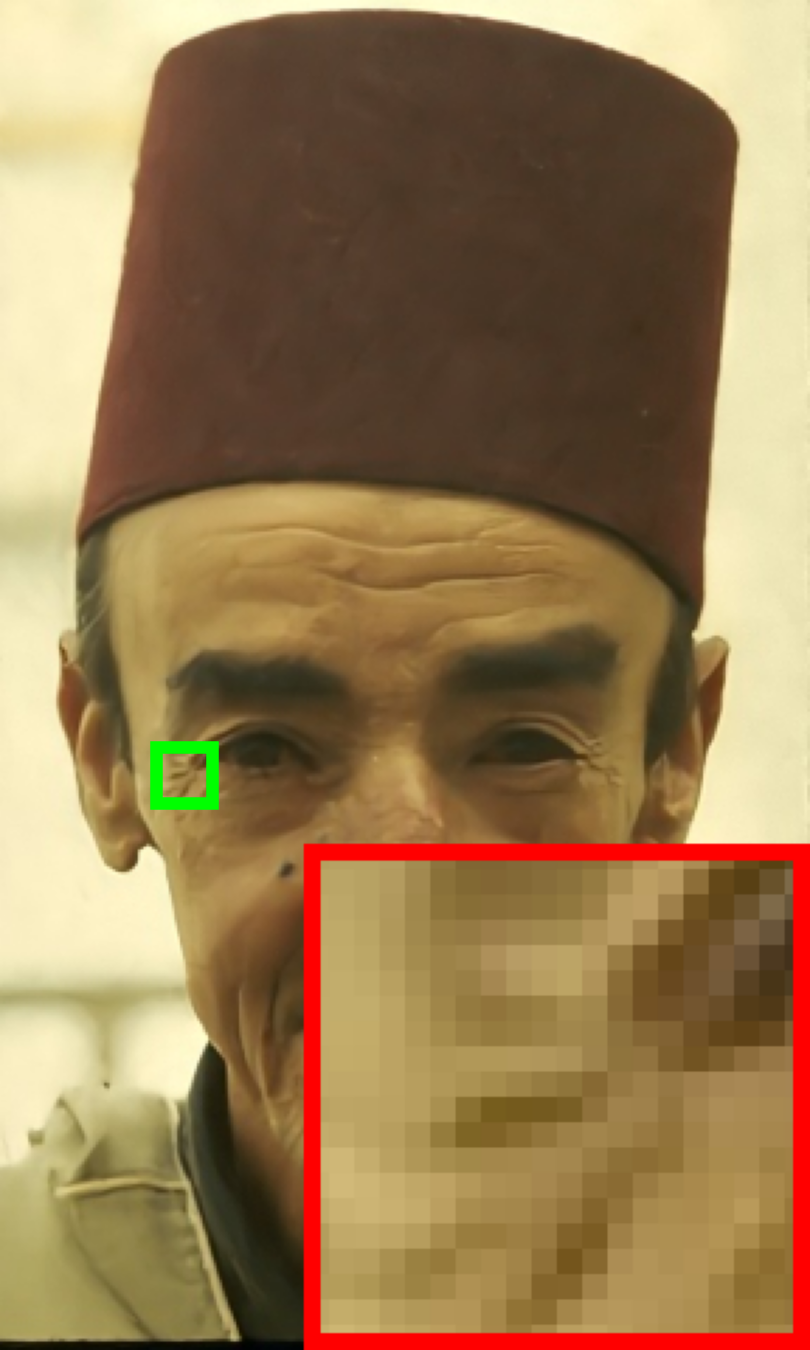}
    &\includegraphics[width=0.08\textwidth]{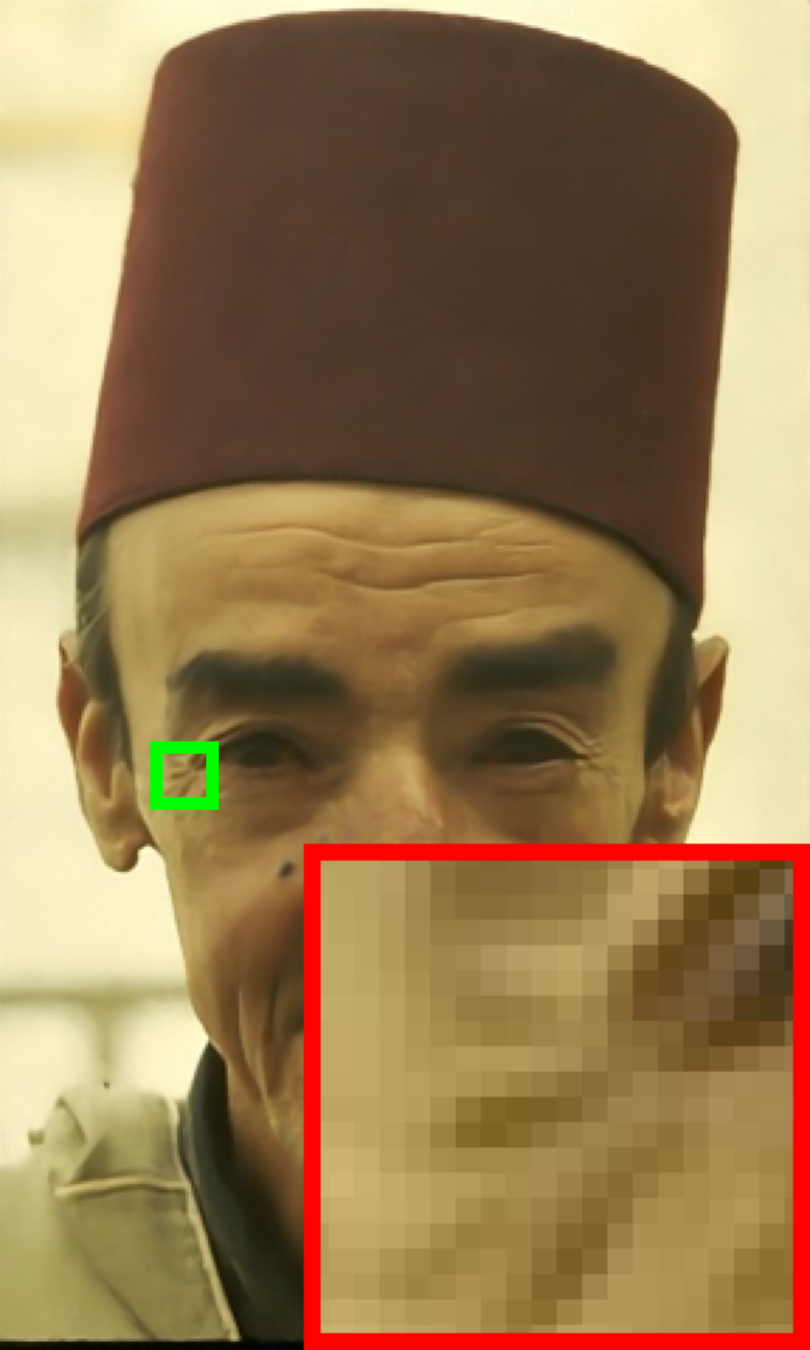}\\
    PSNR/SSIM & 32.98/0.84 & 33.59/0.86 & 29.85/0.69 & 32.78/0.81 & 30.42/0.69 & \underline{\textcolor{blue}{34.72}}/\textbf{\textcolor{red}{0.88}} & 34.60/0.86 & \textbf{\textcolor{red}{34.77}}/\textbf{\textcolor{red}{0.88}} & \textbf{\textcolor{red}{34.77}}/\underline{\textcolor{blue}{0.87}}
\end{tabular}}
\caption{Visual comparison of self-supervised methods on two natural benchmark images named ``Lena'' and ``test\_34'' from Set11~\cite{kulkarni2016reconnet} \textcolor{blue}{(top)} and CBSD68~\cite{martin2001database} \textcolor{blue}{(bottom)}, with $\gamma =50\%$ and $\sigma =10$.}
\label{fig:comparison_standard_natural_images_r50_s10}
\end{figure*}

\begin{table*}[!t]
\caption{High-level comparison of functional characteristics, reconstruction time, and parameter number of single NN in recovering an image of size $256\times 256$ from Set11 \cite{kulkarni2016reconnet} among our SCNets and twelve state-of-the-arts. \textcolor{blue}{E. L.}: external learning; \textcolor{blue}{I. L.}: internal learning; \textcolor{blue}{R. S.} \& \textcolor{blue}{M. A.}: fine-grained ratio-scalability and matrix-agnosticism of single NN once trained on a fixed dataset; \textcolor{blue}{N. R.}: noise-resistance; \textcolor{blue}{D. A.}: deblocking ability for block-based CS; \textcolor{blue}{H. T.}: high-throughput transmission without bottleneck in information flow; \textcolor{blue}{O. I.}: optimization inspiration; \textcolor{blue}{U. A.}: unrolled algorithm; \textcolor{blue}{ISTA}: iterative shrinkage-thresholding algorithm \cite{blumensath2009iterative}; \textcolor{blue}{PGD}: proximal gradient descent \cite{parikh2014proximal}.}
\label{tab:high_level_comp}
\centering
\resizebox{1.0\textwidth}{!}{
\begin{tabular}{l|c|cc|ccccccc|cc}
\shline \rowcolor[HTML]{EFEFEF}
Deep NN-based CS Method & Type & E. L. & I. L. & R. S. & M. A. & N. R. & D. A. & H. T. & O. I. & U. A. & \begin{tabular}[c]{@{}c@{}}Time\\ (s)\end{tabular} & \begin{tabular}[c]{@{}c@{}}\#Param.\\ (M)\end{tabular} \\ \hline \hline
ReconNet (CVPR 2016) & \multirow{6}{*}{\begin{tabular}[c]{@{}c@{}}Supervised\\ \\ (GT-\\ Dependent)\end{tabular}} & \multirow{6}{*}{$\checkmark$} & \multirow{6}{*}{$\times$} & \multirow{6}{*}{$\times$} & $\times$ & \multirow{6}{*}{-} & $\times$ & $\checkmark$ & $\times$ & - & \textcolor{red}{\textbf{0.0021}} & 0.62 \\
ISTA-Net$^+$ (CVPR 2018) & & & & & $\times$ & & $\times$ & $\times$ & $\checkmark$ & ISTA & \textcolor{blue}{\underline{0.0063}} & \textcolor{blue}{\underline{0.34}} \\
DPA-Net (TIP 2020) & & & & & $\times$ & & $\checkmark$ & $\checkmark$ & $\times$ & - & 0.0316 & 6.12 \\
MAC-Net (ECCV 2020) & & & & & $\times$ & & $\checkmark$ & $\checkmark$ & $\times$ & - & 0.0607 & 9.31 \\
ISTA-Net$^{++}$ (ICME 2021) & & & & & $\times$ & & $\checkmark$ & $\times$ & $\checkmark$ & PGD & 0.0167 & 0.76 \\
COAST (TIP 2021) & & & & & $\checkmark$ & & $\checkmark$ & $\times$ & $\checkmark$ & PGD & 0.0252 & 1.12 \\ \hline \hline
DIP (CVPR 2018) & \multirow{10}{*}{\begin{tabular}[c]{@{}c@{}}Self-\\ Supervised\\ \\ (GT-Free)\end{tabular}} & $\times$ & $\checkmark$ & $\times$ & $\times$ & $\checkmark$ & $\checkmark$ & $\checkmark$ & $\times$ & - & 745.3091 & 3.00 \\
BCNN (ECCV 2020) & & $\times$ & $\checkmark$ & $\times$ & $\times$ & $\checkmark$ & $\checkmark$ & $\checkmark$ & $\times$ & - & 3394.1086 & 2.55 \\
EI (ICCV 2021, with SC-CNN) & & $\checkmark$ & $\times$ & $\times$ & $\times$ & $\times$ & $\checkmark$ & $\checkmark$ & $\checkmark$ & PGD & 0.0072 & 0.37 \\
REI (CVPR 2022, with SC-CNN) & & $\checkmark$ & $\times$ & $\times$ & $\times$ & $\checkmark$ & $\checkmark$ & $\checkmark$ & $\checkmark$ & PGD & 0.0072 & 0.37 \\
ASGLD (CVPR 2022) & & $\times$ & $\checkmark$ & $\times$ & $\times$ & $\checkmark$ & $\checkmark$ & $\checkmark$ & $\times$ & - & 661.8666 & 2.20 \\
DDSSL (ECCV 2022) & & $\checkmark$ & $\checkmark$ & $\times$ & $\times$ & $\checkmark$ & $\checkmark$ & $\times$ & $\checkmark$ & PGD & 18.5351 & 0.67 \\ \hhline{-~-----------} 
\textbf{SC-CNN (Ours, Only Stage-1)} & & \multirow{4}{*}{$\checkmark$} & $\times$ & \multirow{4}{*}{$\checkmark$} & \multirow{4}{*}{$\checkmark$} & \multirow{4}{*}{$\checkmark$} & \multirow{4}{*}{$\checkmark$} & \multirow{4}{*}{$\checkmark$} & \multirow{4}{*}{$\checkmark$} & \multirow{4}{*}{PGD} & 0.0072 & 0.37 \\
\textbf{SC-CNN$^+$ (Ours, Stages 1-4)} & & & $\checkmark$ & & & & & & & & 4353.7952 & 0.37 \\
\textbf{SCT (Ours, Only Stage-1)} & & & $\times$ & & & & & & & & 0.0321 & \textcolor{red}{\textbf{0.21}} \\
\textbf{SCT$^+$ (Ours, Stages 1-4)} & & & $\checkmark$ & & & & & & & & 19202.8643 & \textcolor{red}{\textbf{0.21}} \\ \shline
\end{tabular}}
\end{table*}

\subsubsection{More Performance Comparisons in Noiseless ($\sigma =0$) Cases}

This subsection provides more comparison results among our method and other competing supervised and self-supervised deep CS approaches. In Figs.~\ref{fig:comparison_standard_natural_images_r10}, \ref{fig:comparison_standard_natural_images_r30}, and \ref{fig:comparison_standard_natural_images_r50}, we present the recovered results of twelve images from Set11 \cite{kulkarni2016reconnet}, CBSD68 \cite{martin2001database}, Urban100 \cite{huang2015single}, and DIV2K \cite{agustsson2017ntire} by different methods for $\gamma \in \{10\%,30\%,50\%\}$ and $\sigma =0$. We observe that: (1) supervised methods can achieve better results than self-supervised methods in most cases; (2) our method can not only yield basic estimations with accurate shapes and high-quality structures that compete against those given by self-supervised methods, but also leverage the coarse-to-fine internal learning schemes in stages 2-4 to recover sharper details and precise line textures, resulting in our performance leading over the supervised methods. Further comparison results regarding the noisy ($\sigma >0$) cases will be provided in the next subsection.

\begin{table*}[!t]
\caption{Ablation studies of our designs in SCL and SCNet on Set11 \cite{kulkarni2016reconnet} benchmark, started from the default SC-CNN. All NNs are trained on the same incomplete measurements sensed by a single Gaussian matrix of ratio 10\%, and evaluated on three ratios 10\%, 30\%, and 50\%. The performance impacts of each ablation in current variant compared to the original version are highlighted in purple.}
\vspace{3pt}
\label{tab:ablation_studies_from_sc-cnn}
\centering
\resizebox{1.0\textwidth}{!}{
\begin{tabular}{cl|cccc}
\shline
\rowcolor[HTML]{EFEFEF} 
\multicolumn{2}{c|}{\cellcolor[HTML]{EFEFEF}} &
  \multicolumn{4}{c}{\cellcolor[HTML]{EFEFEF}Test CS Ratio $\gamma^{test}$} \\ \hhline{>{\arrayrulecolor[HTML]{EFEFEF}}-->{\arrayrulecolor{black}}|----}
\rowcolor[HTML]{EFEFEF} 
\multicolumn{2}{c|}{\multirow{-2}{*}{\cellcolor[HTML]{EFEFEF}Experimental Setting (with Noise Level $\sigma =0$)}} &
  10\% &
  30\% &
  50\% &
  Avg. \\ \hline \hline
$<$1$>$ &
  \textbf{SC-CNN} (our default standard convolutional SCNet version) &
  27.93 &
  32.89 &
  36.25 &
  32.36 \\
$<$2$>$ &
  Add our stage-2 to $<$1$>$, \textit{i.e.}, $(1\rightarrow 2)$&
  29.16 (\textcolor{purple}{+1.23}) &
  33.57 (\textcolor{purple}{+0.68}) &
  36.55 (\textcolor{purple}{+0.30}) &
  33.09 (\textcolor{purple}{+0.73}) \\
$<$3$>$ &
  Add our stage-3 to $<$2$>$, \textit{i.e.}, $(1\rightarrow 2\rightarrow 3)$ &
  29.39 (\textcolor{purple}{+0.23}) &
  33.92 (\textcolor{purple}{+0.35}) &
  37.37 (\textcolor{purple}{+0.82}) &
  33.56 (\textcolor{purple}{+0.47}) \\
$<$4$>$ &
  Add our stage-4 to $<$3$>$, \textit{i.e.}, $(1 \rightarrow 2 \rightarrow 3 \rightarrow 4)$ &
  29.42 (\textcolor{purple}{+0.03}) &
  34.14 (\textcolor{purple}{+0.22}) &
  37.60 (\textcolor{purple}{+0.23}) &
  33.72 (\textcolor{purple}{+0.16}) \\
$<$5$>$ &
  Train our SC-CNN from scratch with only stage-2 &
  28.83 &
  33.24 &
  36.67 &
  32.91 \\
$<$6$>$ &
  Train our SC-CNN from scratch with only stage-3 &
  27.67 &
  31.79 &
  35.02 &
  31.49 \\ \hline \hline
$<$7$>$ &
  Train our SC-CNN with the supervised $\ell_2$ loss ($p=2$) &
  27.01 &
  16.81 &
  12.03 &
  18.62 \\
$<$8$>$ &
  Train our SC-CNN with the supervised $\ell_1$ loss ($p=1$) &
  27.29 &
  16.82 &
  13.51 &
  19.21 \\
$<$9$>$ &
  Train $<$7$>$ with the supervised DOC loss ($p=2$) for scalability &
  27.47 (\textcolor{purple}{+0.46}) &
  33.67 (\textcolor{purple}{+16.86}) &
  37.15 (\textcolor{purple}{+25.12}) &
  32.76 (\textcolor{purple}{+14.14}) \\
$<$10$>$ &
  Train $<$8$>$ with the supervised DOC loss ($p=1$) for scalability &
  27.73 (\textcolor{purple}{+0.44}) &
  33.90 (\textcolor{purple}{+17.08}) &
  37.49 (\textcolor{purple}{+23.98}) &
  33.04 (\textcolor{purple}{+13.83}) \\ 
$<$11$>$ &
  Remove symmetric loss terms from $<$1$>$, \textit{i.e.}, $\Loss=\Loss_{DMC_1}+0.1 \Loss_{DOC_1}$ &
  27.88 (\textcolor{purple}{-0.05}) &
  32.98 (\textcolor{purple}{+0.09}) &
  36.34 (\textcolor{purple}{+0.09}) &
  32.40 (\textcolor{purple}{+0.04}) \\
$<$12$>$ &
  Change the value of $p$ in $<$11$>$ from 1 to 2 &
  27.02 (\textcolor{purple}{-0.86}) &
  31.97 (\textcolor{purple}{-1.01}) &
  34.79 (\textcolor{purple}{-1.55}) &
  31.26 (\textcolor{purple}{-1.14}) \\
$<$13$>$ &
  Remove the geometric transformations $\T$ from $<$11$>$ &
  27.83 (\textcolor{purple}{-0.05}) &
  32.94 (\textcolor{purple}{-0.04}) &
  36.24 (\textcolor{purple}{-0.10}) &
  32.34 (\textcolor{purple}{-0.08}) \\
$<$14$>$ &
  Fix the CS ratio of our DOC loss in $<$13$>$ to $\Tilde{\gamma}= \gamma = 10\%$ &
  27.21 (\textcolor{purple}{-0.62}) &
  6.67 (\textcolor{purple}{-26.27}) &
  5.18 (\textcolor{purple}{-31.06}) &
  13.02 (\textcolor{purple}{-19.32}) \\ 
$<$15$>$ &
  Fix the sampling matrix of our DOC loss in $<$14$>$ to $\Tilde{\A}\equiv \A$ &
  27.96 (\textcolor{purple}{+0.75}) &
  30.65 (\textcolor{purple}{+23.98}) &
  8.38 (\textcolor{purple}{+3.20}) &
  22.33 (\textcolor{purple}{+9.31}) \\
$<$16$>$ &
  Remove the DOC loss from $<$15$>$, \textit{i.e.}, $\Loss=\Loss_{DMC_1}$ &
  27.94 (\textcolor{purple}{-0.02}) &
  18.27 (\textcolor{purple}{-12.38}) &
  8.82 (\textcolor{purple}{+0.44}) &
  18.34 (\textcolor{purple}{-3.99}) \\
$<$17$>$ &
  Fix the ratios of two divided parts as $\gamma_1= \gamma_2= 5\%$ in $<$16$>$ &
  17.63 (\textcolor{purple}{-10.31}) &
  7.95 (\textcolor{purple}{-10.32}) &
  5.41 (\textcolor{purple}{-3.41}) &
  10.33 (\textcolor{purple}{-8.01}) \\
$<$18$>$ &
  Remove final GD from $<$14$>$ and set $\Loss =\Loss_{MC}+ 0.1 \Loss_{DOC}$ $(p=2)$ &
  27.53 (\textcolor{purple}{+0.32}) &
  16.20 (\textcolor{purple}{+9.53}) &
  9.85 (\textcolor{purple}{+4.67}) &
  17.86 (\textcolor{purple}{+4.84}) \\
$<$19$>$ &
  Set the total number of random matrices in the DOC of $<$18$>$ to 40 &
  27.52 (\textcolor{purple}{-0.01}) &
  14.67 (\textcolor{purple}{-1.53}) &
  9.32 (\textcolor{purple}{-0.63}) &
  17.17 (\textcolor{purple}{-0.69}) \\
$<$20$>$ &
  Set the total number of random matrices in the DOC of $<$19$>$ to 10 &
  27.40 (\textcolor{purple}{-0.12}) &
  23.58 (\textcolor{purple}{+8.91}) &
  16.40 (\textcolor{purple}{+7.08}) &
  22.46 (\textcolor{purple}{+5.29}) \\ 
$<$21$>$ &
  Set the total number of random matrices in the DOC of $<$20$>$ to 1 &
  27.20 (\textcolor{purple}{-0.20}) &
  13.86 (\textcolor{purple}{-9.72}) &
  7.27 (\textcolor{purple}{-9.13}) &
  16.11 (\textcolor{purple}{-6.35}) \\ \hline \hline
$<$22$>$ &
  Make the GD step size $\rho$ in $<$11$>$ learnable and intialized to 1 &
  21.53 (\textcolor{purple}{-6.35}) &
  24.37 (\textcolor{purple}{-8.61}) &
  23.86 (\textcolor{purple}{-12.48}) &
  23.25 (\textcolor{purple}{-9.15}) \\
$<$23$>$ &
  Remove all the image embeddings (IE) from $<$2$>$ &
  28.83 (\textcolor{purple}{-0.33}) &
  33.36 (\textcolor{purple}{-0.21}) &
  36.53 (\textcolor{purple}{-0.02}) &
  32.91 (\textcolor{purple}{-0.18}) \\
$<$24$>$ &
  Remove all the positional embeddings (PE) from $<$3$>$ &
  29.23 (\textcolor{purple}{-0.16}) &
  33.71 (\textcolor{purple}{-0.21}) &
  37.09 (\textcolor{purple}{-0.28}) &
  33.34 (\textcolor{purple}{-0.22}) \\ \shline
\end{tabular}} \\ \vspace{3pt}
\resizebox{1.0\textwidth}{!}{
\begin{tabular}{cl|cccc}
\shline
\rowcolor[HTML]{EFEFEF} 
\multicolumn{2}{c|}{\cellcolor[HTML]{EFEFEF}} &
  \multicolumn{4}{c}{\cellcolor[HTML]{EFEFEF}Test CS Ratio $\gamma^{test}$} \\ \hhline{>{\arrayrulecolor[HTML]{EFEFEF}}-->{\arrayrulecolor{black}}|----}
\rowcolor[HTML]{EFEFEF} 
\multicolumn{2}{c|}{\multirow{-2}{*}{\cellcolor[HTML]{EFEFEF}Experimental Setting (with Noise Level $\sigma =10$)}} &
  10\% &
  30\% &
  50\% &
  Avg. \\ \hline \hline
$<$25$>$ &
  \textbf{SC-CNN} (our default standard convolutional SCNet version) &
  26.15 &
  29.59 &
  30.83 &
  28.86 \\
$<$26$>$ &
  Add a self-loss term to $<$25$>$, \textit{i.e.}, $\Loss =\Loss_{DMC}+ 0.1 \Loss_{DOC} + 0.01 \Loss_{self}$ &
  25.99 (\textcolor{purple}{-0.16}) &
  28.60 (\textcolor{purple}{-0.99}) &
  30.20 (\textcolor{purple}{-0.63}) &
  28.26 (\textcolor{purple}{-0.60}) \\
$<$27$>$ &
  Add a SURE-loss term to $<$25$>$, \textit{i.e.}, $\Loss =\Loss_{DMC}+ 0.1 \Loss_{DOC} + 0.01 \Loss_{SURE}$ &
  26.06 (\textcolor{purple}{-0.09}) &
  28.79 (\textcolor{purple}{-0.80}) &
  30.44 (\textcolor{purple}{-0.39}) &
  28.43 (\textcolor{purple}{-0.43}) \\
$<$28$>$ &
  Remove the noise injection of $\Tilde{\y}=\Tilde{\A}\Tilde{\x}+\Tilde{\n}$ in DOC loss from $<$25$>$ &
  26.07 (\textcolor{purple}{-0.08}) &
  28.33 (\textcolor{purple}{-1.26}) &
  28.97 (\textcolor{purple}{-1.86}) &
  27.79 (\textcolor{purple}{-1.07}) \\ \shline
\end{tabular}}
\end{table*}

\subsubsection{Performance Comparisons in Noisy ($\sigma >0$) Cases}

This subsection presents comparisons among our method and state-of-the-art self-supervised approaches with an observation noise level of $\sigma =10$. The detailed quantitative and qualitative visual comparison results are demonstrated in Tab.~\ref{tab:compare_sota_psnr_noisy}, Figs.~\ref{fig:comparison_standard_natural_images_r10_s10}, \ref{fig:comparison_standard_natural_images_r30_s10}, and \ref{fig:comparison_standard_natural_images_r50_s10}. We observe that: (1) the existing methods may suffer from over-smoothing and distortion, and exhibit undesirable noise artifacts in the recovered images; (2) our standard SCNet versions can outperform the competing approaches in almost all cases, yielding superior reconstructions with significantly higher visual quality; (3) the internal learning and self-ensemble schemes employed in our stages 2-4 are effective in improving the PSNR up to 1.46dB in the recovered results; (4) our SC-CNN$^+$ outperforms SCT$^+$ in all cases with PSNR distances of 0.01-0.27dB. This superiority can be attributed to the more appropriate regularization of locality, shift invariance, and content independence offered by convolutions and residual blocks (RB) \cite{he2016deep,lim2017enhanced} with higher constraint strength compared to the shifted window-based self-attention mechanisms in Swin-Conv blocks (SCB) \cite{zhang2022practical} without such inductive bias.

\subsubsection{Comparisons of Functional Characteristics, Test Time, and Parameter Number}

Tab.~\ref{tab:high_level_comp} provides a high-level conceptual comparison among our method and twelve other competing approaches \cite{kulkarni2016reconnet,zhang2018ista,sun2020dual,chen2020learning,you2021ista,you2021coast,ulyanov2018deep,pang2020self,chen2021equivariant,chen2022robust,wang2022self,quan2022dual}, highlighting their respective strengths and weaknesses. Our CS method can stand out from the rest for its organic integration of advantages of ratio-scalability, matrix-agnosticism, deblocking ability, high-throughput transmission, \textit{etc.}, resulting in its superior performance once trained on a fixed incomplete measurement set. In addition, our SCNets are capable of learning the common signal priors from external measurements, which makes them practical and adaptable for real-time reconstruction ($>30$ frames per second) by their standard versions (SC-CNN and SCT) or for further enhancing accuracy by exploiting the internal statistics of test samples and learned NN models in their enhanced versions (SC-CNN$^+$ and SCT$^+$). The high performance, low parameter and memory cost, scalability, and flexibility of our method make the ``SCL+SCNet'' combination an appealing choice for real-world applications with diverse requirements and deployment environments.

\begin{table*}[!t]
\caption{Quantitative comparison of PSNR (dB) among seven SC-CNNs trained with different $p$ settings at ratio 10\%.}
\label{tab:PSNR_different_p_values}
\centering
\resizebox{0.925\textwidth}{!}{
\begin{tabular}{
>{\columncolor[HTML]{EFEFEF}}c ccccccc}
\shline
Parameter $p$ & 0.8   & \textbf{1.0 (Ours)}   & 1.2   & 1.4   & 1.6 & 1.8 & 2.0 \\ \hline
Average PSNR (dB) on Set11 & - (No Convergence) & \textcolor{red}{\textbf{27.93}} & \textcolor{blue}{\underline{27.92}} & 27.86 & 27.65 & 27.34 & 27.10 \\ \shline
\end{tabular}}
\end{table*}

\begin{table*}[!t]
\caption{Quantitative comparison of PSNR (dB) among six SC-CNNs trained with six different $\alpha$ settings at ratio 10\%.}
\label{tab:PSNR_different_alpha_values}
\centering
\resizebox{0.75\textwidth}{!}{
\begin{tabular}{
>{\columncolor[HTML]{EFEFEF}}c cccccc}
\shline
Weighting Factor $\alpha$ & 10   & 1   & \textbf{0.1 (Ours)}   & 0.01   & 0.001 & 0 \\ \hline
Average PSNR (dB) on Set11 & 24.94 & 25.81 & \textcolor{blue}{\underline{27.93}} & 27.92 & 27.92 & \textcolor{red}{\textbf{27.95}} \\ \shline
\end{tabular}}
\end{table*}

\begin{figure}
\centering
\includegraphics[width=0.48\textwidth]{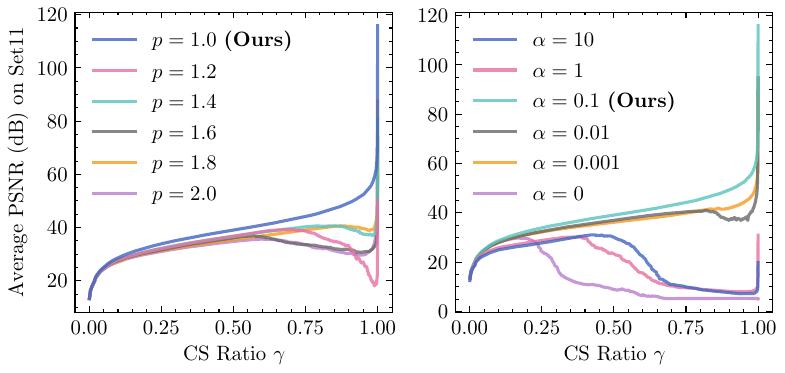}
\caption{Comparisons among SC-CNNs trained on the same measurement set of ratio 10\% with twelve different $p$ \textcolor{blue}{(left)} and $\alpha$ settings \textcolor{blue}{(right)} for the proposed self-supervised dual-domain loss function. Our default SC-CNN setting with $p=1$ and $\alpha =0.1$ achieves best results in almost all cases and can generalize well to ratio range $\gamma \in(0,1]$.}
\label{fig:scalable_curve_more_analyses_p_and_alpha}
\end{figure}

\subsubsection{Ablation Study and More Analyses}
This subsection aims to investigate the effectiveness of our proposed designs in SCL and SCNet, the influence of alternative approaches, and to offer insights into the underlying principles that contribute to our method's superior performance compared to other competing approaches.

\begin{figure}
\centering
\includegraphics[width=0.45\textwidth]{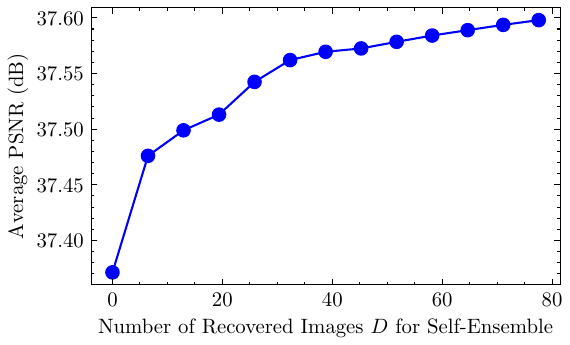}
\caption{Performance of our SC-CNN, which has been trained through stages 1-3 as indicated in Tab.~\ref{tab:ablation_studies_from_sc-cnn} $<$3$>$, and then utilized in stage 4 for the self-ensemble process with different numbers of image reconstructions $D$. The results are evaluated on Set11 \cite{kulkarni2016reconnet} at CS ratio $\gamma = 50\%$.}
\label{fig:self_ensemble_PSNR}
\end{figure}

\textbf{Effect of the Four Stages in Our Progressive Reconstruction Strategy.} As shown in Tab.~\ref{tab:ablation_studies_from_sc-cnn} $<$1-4$>$, we conduct four fine-grained experiments by sequentially training our SC-CNN with stages 1-4. The results exhibit that our cross-image and single-image internal learning, and self-ensemble schemes can effectively and respectively bring average PSNR improvements of 0.73dB, 0.47dB, and 0.16dB. In Fig.~\ref{fig:self_ensemble_PSNR}, we present experiments using our SC-CNN, which has been trained through stages 1-3 as outlined in Tab.~\ref{tab:ablation_studies_from_sc-cnn} $<$3$>$. We then implement stage 4, the self-ensemble process. This involves perturbing the NN input $\F_\Th$ $D$ times to produce $D$ distinct recovery results. These results are then combined into a single image by averaging, as described in Eq.~(\ref{eq:self_ensemble}). Our findings demonstrate that increasing the number of images $D$ in the self-ensemble process leads to a corresponding rise in recovery PSNR. Although the process demands more computational time, the substantial performance boost—a 0.23dB improvement at $\gamma = 50\%$—warrants the extra effort in situations where high recovery precision is essential. In Tab.~\ref{tab:ablation_studies_from_sc-cnn} $<$5-6$>$, we independently train two SC-CNNs from scratch on only the 11 test measurements of images from Set11 \cite{kulkarni2016reconnet} and observe that: (1) our method can achieve a superior PSNR (27.67dB at ratio 10\%) with only sufficient stage-3 training compared to most competing methods, including the supervised ReconNet \cite{kulkarni2016reconnet} (24.08dB), ISTA-Net$^+$ \cite{zhang2018ista} (26.49dB), and DPA-Net \cite{sun2020dual} (27.66dB), as well as the self-supervised DIP \cite{ulyanov2018deep} (26.09dB), BCNN \cite{pang2020self} (27.58dB), EI \cite{chen2022robust} (21.76dB), and DDSSL \cite{quan2022dual} (27.65dB); (2) by leveraging the relationships among different test samples, our SC-CNN trained only on stage-2 (28.83dB at ratio 10\%) can surpass the other methods, including the supervised MAC-Net \cite{chen2020learning} (27.92dB), ISTA-Net$^{++}$ \cite{you2021ista} (28.34dB), and COAST \cite{you2021coast} (28.78dB), as well as the self-supervised ASGLD \cite{wang2022self} (27.81dB), standard SC-CNN (27.93dB), and SCT (28.21dB) with stage-1. These results demonstrate the efficacy, superiority, and flexibility of our four stages in data utilization and their ability to benefit from the GT-free dual-domain loss and learned NNs.

\begin{figure*}[!t]
\centering
\begin{minipage}[c]{0.495\linewidth}
\centering
\includegraphics[width=\linewidth]{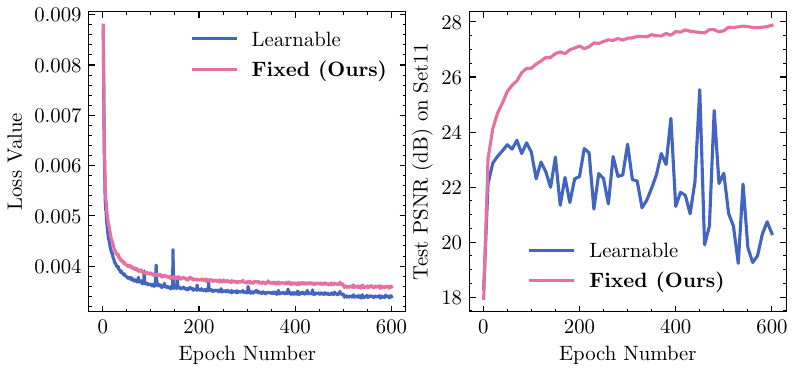}
\end{minipage}
\hfill
\begin{minipage}[c]{0.495\linewidth}
\centering
\includegraphics[width=\linewidth]{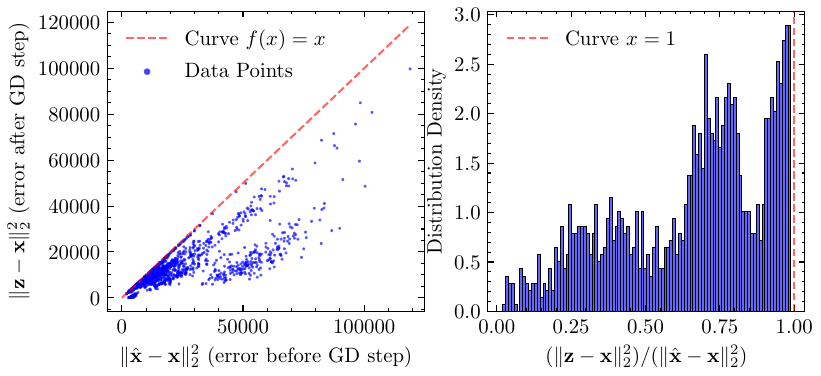}
\end{minipage}
\caption{\textcolor{blue}{(Left)} The loss and test PSNR curves of the SC-CNNs with learnable GD step size $\rho$ initialized to 1, and our default non-adaptive and fixed $\rho \equiv 1$ setting for ratio $\gamma =10\%$ and $\sigma =0$ (corresponding to Tab.~\ref{tab:ablation_studies_from_sc-cnn} $<$11$>$ and $<$22$>$). While the learnable $\rho$ induces overfitting and unstable convergence, our design effectively guides SC-CNN to learn useful signal priors and avoid trivial results. \textcolor{blue}{(Right)} Comparison between the prediction errors of images before and after GD ($\xhat$ and $\mathbf{z}$), calculated by $\lVert \xhat - \x \rVert_2^2$ and $\lVert \mathbf{z} - \x \rVert_2^2$, respectively. One can observe that GD steps effectively reduce the errors with $\lVert \mathbf{z} - \x \rVert_2^2 < \lVert \xhat - \x \rVert_2^2$. The evaluations are conducted on the images from CBSD68 \cite{martin2001database} with 21 (middle/final) estimations produced by an SC-CNN ($K=20$) at $\gamma =10\%$ and $\sigma =10$.}
\label{fig:effect_of_GD_step}
\end{figure*}

\begin{table*}[!t]
\caption{Comparison among our default SC-CNN with re-parameterized learnable GD step size $\rho =\text{Sigmoid}(\tau)$ and five variants with fixed $\rho \in \{0.6,0.7,0.8,0.9,1.0\}$ for $\gamma =10\%$ and $\sigma =10$.}
\label{tab:PSNR_different_rho_values}
\centering
\resizebox{0.85\textwidth}{!}{
\begin{tabular}{
>{\columncolor[HTML]{EFEFEF}}c cccccc}
\shline
GD Step Size $\rho$        & 0.6   & 0.7   & 0.8   & 0.9   & \textbf{0.9266 (Our Learned)} & 1.0 \\ \hline
Average PSNR (dB) on Set11 & 25.98 & 26.03 & 26.08 & \textcolor{blue}{\underline{26.13}} & \textcolor{red}{\textbf{26.15}} & 25.88 \\ \shline
\end{tabular}}
\end{table*}

\textbf{Effect of Our Dual-Domain Loss Function Designs.} In Tab.~\ref{tab:ablation_studies_from_sc-cnn} $<$7-10$>$, we train four SC-CNNs using supervised loss functions as a reference. To be concrete, the SC-CNNs in $<$7$>$ and $<$8$>$ are trained using $\Loss_{sup}=\lVert \F_\Th(\A\x,\A,\gamma) -\x \rVert_p^p$ with $p\in \{1,2\}$ and a fixed $\A$ of ratio $\gamma =10\%$, while $<$9$>$ and $<$10$>$ simulate thousands of CS tasks by using an improved supervised DOC loss $\Loss_{DOC~(sup)}=\lVert \F_\Th(\Tilde{\A}\x,\Tilde{\A},\Tilde{\gamma}) -\x \rVert_p^p$, where $\Tilde{\A}$ and $\Tilde{\gamma}$ are independently and randomly generated for the different instances in each training batch. We observe that: (1) the supervised scalable SC-CNN versions surpass the non-scalable ones in all cases, validating the feasibility and effectiveness of matrix-network disentanglement in our DOC loss for better optimization regularization and ratio-scalability achievement; (2) the NNs trained using $\ell_1$ losses ($p=1$) consistently outperform those with $\ell_2$ losses ($p=2$); (3) our self-supervised SC-CNN in $<$1$>$ can even achieve higher PSNR (27.93dB) at the original ratio 10\% than the supervised $<$9$>$ (27.47dB) and $<$10$>$ (27.73dB), but it is not superior in high-ratio ($\gamma \ge 30\%$) cases. This fact can be attributed to the sufficient utilization of limited measurements of the ratio 10\% by our SCL, and the much larger and more informative GT dataset available for supervised learning, which enjoys full supervision on the entire ratio range $(0,1]$. It is worth noting that our SC-CNN trained on the measurement set of ratio 50\% (\textit{self-supervised}, with only about half of the data size) can achieve a PSNR of 39.07dB (please refer to Tab.~1 in the \MP), which is significantly higher than the 37.15dB and 37.49dB obtained by $<$9$>$ and $<$10$>$ (\textit{supervised}, with the full data size), respectively, with PSNR distances $> 1.5$dB at $\gamma =50\%$.

\begin{table*}[!t]
\caption{Comparison among our default SC-CNN with PE size $h=w\equiv 8$ and six variants with different $(h,w)$ settings in stage-3 at $\gamma =10\%$ and $\sigma =0$. $(H,W)$ denotes the size of target image.}
\label{tab:PSNR_different_PE_sizes}
\centering
\resizebox{1.0\textwidth}{!}{
\begin{tabular}{
>{\columncolor[HTML]{EFEFEF}}c ccccccc}
\shline
Spatial Size $(h,w)$ of PE        & $(0,0)$ (w/o PE) & $(2,2)$   & $(4,4)$   & \textbf{$(8,8)$ (Ours)}   & $(16,16)$ & $(32,32)$ & $(H,W)$ \\ \hline
Average PSNR (dB) on Set11 & 29.23 & 29.28 & \textcolor{blue}{\underline{29.37}} & \textcolor{red}{\textbf{29.39}} & 29.30 & 28.99 & 28.80 \\ \hline
Extra \#Param. for Single NN & \textcolor{red}{\textbf{0}} & \textcolor{blue}{\underline{128}} & 512 & 2048 & 8192 & 32768 & 32$HW$ \\ \shline
\end{tabular}}
\end{table*}

\begin{figure}
\centering
\includegraphics[width=0.48\textwidth]{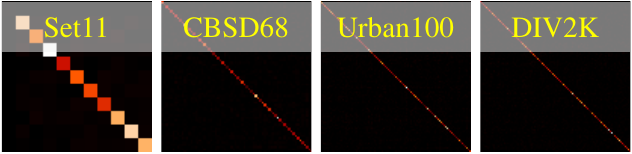}
\caption{Visualization of the correlation maps among IEs and PEs learned by our SC-CNNs in stages 2 and 3 at ratio 10\% for the different images from four natural image benchmarks \cite{kulkarni2016reconnet,martin2001database,huang2015single,agustsson2017ntire}. Specifically, we first extract the learned weights of IEs and PEs, and then concatenate them to form a matrix $\mathbf{E}\in\Rbb^{l^{test}\times (32+32\times 8\times 8)}$, where $l^{test}$ and $(32+32\times 8\times 8)$ represent the sizes of test set and stretched IE $\mathbf{E}^{img}_{i}$ and PE $\mathbf{E}^{pos}_{i}$ for the $i$-th image, corresponding to row $i$ of $\mathbf{E}$, respectively. We observe that all the four constructed embedding matrices ($\mathbf{E}$s) approximately satisfy the diagonal form of $\mathbf{EE}^\top \approx \text{diag}(\{q_1,\cdots,q_{l^{test}}\})\in \Rbb^{l^{test}\times l^{test}}$. Lighter red indicates larger absolute values.}
\label{fig:visualize_IE_and_PE_ortho}
\end{figure}

Tab.~\ref{tab:ablation_studies_from_sc-cnn} $<$11-17$>$ conduct seven break-down ablations on our $\Loss$. Specifically, in Tab.~\ref{tab:ablation_studies_from_sc-cnn} $<$11$>$, we remove the symmetric loss terms regarding $\xhat_2$, resulting in an acceptable 0.05dB PSNR drop with 0.04dB average improvement and $\sim 2\times$ learning acceleration. We also highlight that the original SC-CNN in $<$1$>$ enjoys more stable training convergence than $<$11$>$, with smaller fluctuations in loss values and test PSNRs. In Tab.~\ref{tab:ablation_studies_from_sc-cnn} $<$12$>$, we investigate the impact of two $p$ values of 1 and 2 and observe that our setting $p=1$ achieves better performance than $p=2$. Tab.~\ref{tab:ablation_studies_from_sc-cnn} $<$13$>$ shows an average PSNR drop of 0.08dB due to the removal of our introduced data augmentation with eight geometric transformations in the DOC loss, which is inspired by the group invariance of natural image sets and can effectively provide useful information beyond the range space of $\A^\dagger$ \cite{chen2021equivariant}. In Tab.~\ref{tab:ablation_studies_from_sc-cnn} $<$14-15$>$, we sequentially close the ratio and matrix augmentations in the DOC loss, and observe a weakening of NN generalization ability to the unseen ratios 30\% and 50\%. In Tab.~\ref{tab:ablation_studies_from_sc-cnn} $<$16-17$>$, we investigate the performance of SC-CNN trained by $\Loss_{DMC_1}$ and its weakened version using only random equal measurement divisions similar to the splitting operations in \cite{yaman2020self,tachella2022unsupervised}. The results indicate that these models fail to generalize to unseen cases with higher sampling ratios.

\begin{figure*}[!t]
\centering
\includegraphics[width=\linewidth]{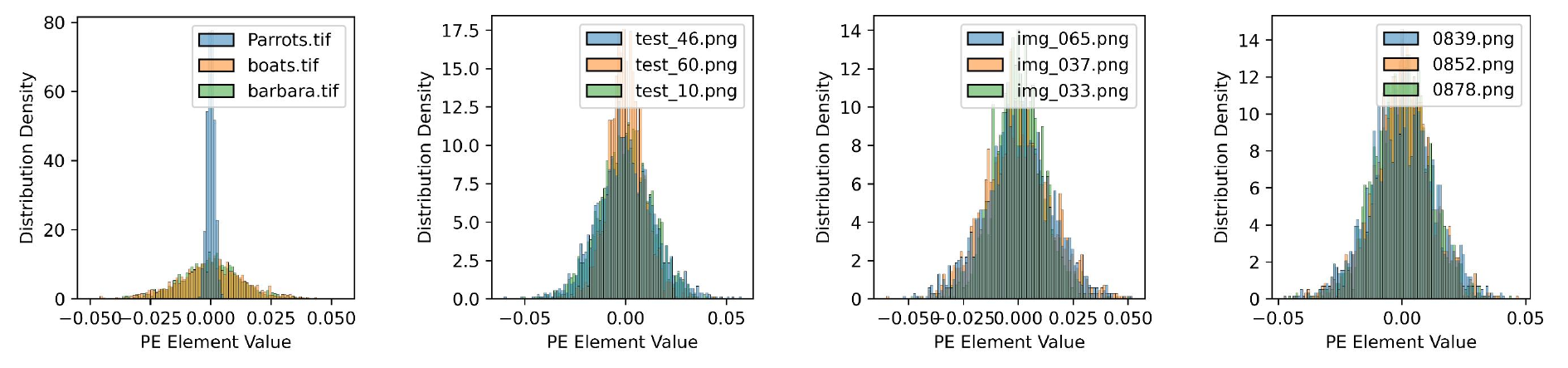}
\caption{Histogram visualizations of the element distributions of the learned PE weights corresponding to twelve different images randomly selected from Set11 \cite{kulkarni2016reconnet}, CBSD68 \cite{martin2001database}, Urban100 \cite{huang2015single}, and DIV2K \cite{agustsson2017ntire}. One can observe that the PEs for different images can exhibit diverse distributions.}
\label{fig:visualize_PE_distribution_by_histogram}
\end{figure*}

\begin{table*}[!t]
\caption{Quantitative comparison of test PSNR (dB) and parameter number (M) of single NN among our four SCNet versions with PGD-unrolled module number $K\in\{10,20,30\}$. All performance evaluations are conducted on the Set11 \cite{kulkarni2016reconnet} benchmark with ratio $\gamma =50\%$ and noise level $\sigma =0$.}
\label{tab:stage_number}
\centering
\resizebox{0.9\textwidth}{!}{
\begin{tabular}{
>{\columncolor[HTML]{EFEFEF}}c cccccccc}
\shline
Unrolled Module & \multicolumn{2}{c}{SC-CNN} & \multicolumn{2}{c}{SC-CNN$^+$} & \multicolumn{2}{c}{SCT} & \multicolumn{2}{c}{SCT$^+$} \\ \hhline{~--------}
Number $K$      & PSNR       & \#Param.      & PSNR         & \#Param.        & PSNR      & \#Param.    & PSNR        & \#Param.      \\ \hline \hline
10              & 38.38      & 0.19          & 39.68        & 0.19            & 38.95     & 0.11        & 39.95       & 0.11          \\
20              & 39.07      & 0.37          & 40.16        & 0.37            & 39.50     & 0.21        & 40.41       & 0.21          \\
30              & 39.29      & 0.56          & 40.19        & 0.56            & 39.53     & 0.32        & 40.54       & 0.32          \\ \shline
\end{tabular}}
\end{table*}

\begin{figure*}[!t]
\setlength{\tabcolsep}{0.5pt}
\hspace{-4pt}
\resizebox{1.0\textwidth}{!}{
\tiny
\begin{tabular}{ccccccccccc}
    Estimation $\xhat$ & Channel-0 & Channel-1 & Channel-2 & Channel-3 & Channel-4 & Channel-5 & Channel-6 & Channel-7 & Channel-8 & Channel-9\\
    \includegraphics[width=0.09\textwidth]{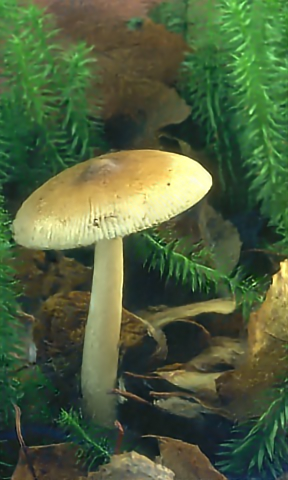}
    &\includegraphics[width=0.09\textwidth]{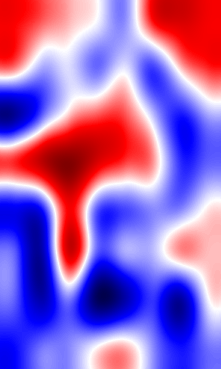}
    &\includegraphics[width=0.09\textwidth]{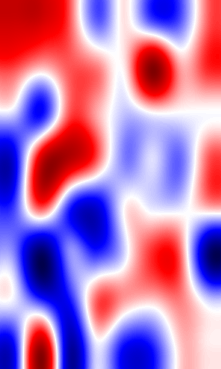}
    &\includegraphics[width=0.09\textwidth]{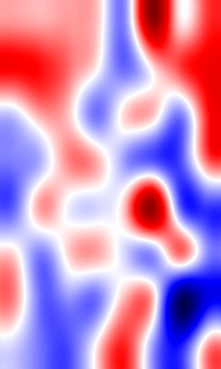}
    &\includegraphics[width=0.09\textwidth]{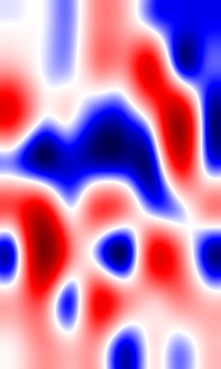}
    &\includegraphics[width=0.09\textwidth]{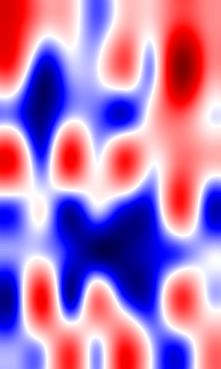}
    &\includegraphics[width=0.09\textwidth]{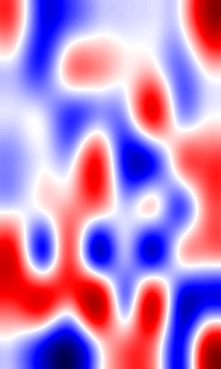}
    &\includegraphics[width=0.09\textwidth]{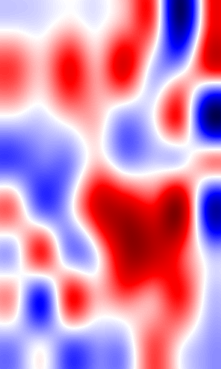}
    &\includegraphics[width=0.09\textwidth]{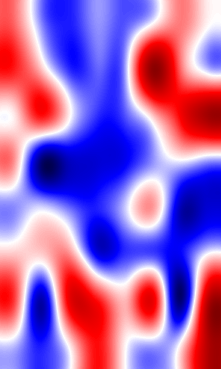}
    &\includegraphics[width=0.09\textwidth]{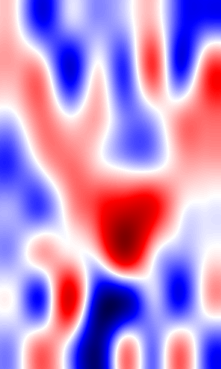}
    &\includegraphics[width=0.09\textwidth]{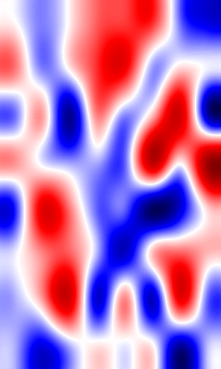}\\
    Channel-10 & Channel-11 & Channel-12 & Channel-13 & Channel-14 & Channel-15 & Channel-16 & Channel-17 & Channel-18 & Channel-19 & Channel-20\\
    \includegraphics[width=0.09\textwidth]{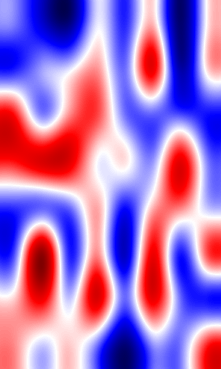}
    &\includegraphics[width=0.09\textwidth]{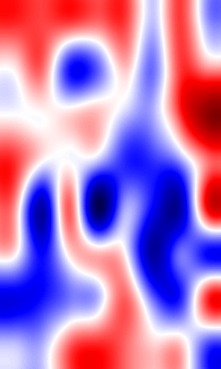}
    &\includegraphics[width=0.09\textwidth]{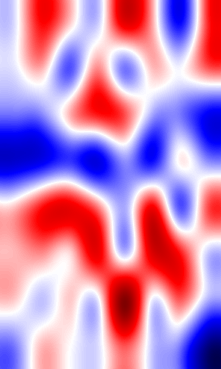}
    &\includegraphics[width=0.09\textwidth]{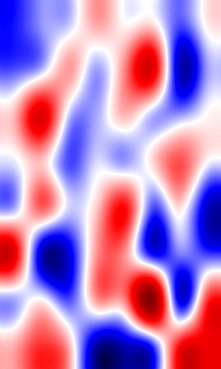}
    &\includegraphics[width=0.09\textwidth]{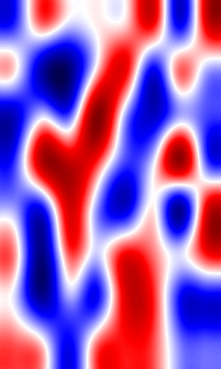}
    &\includegraphics[width=0.09\textwidth]{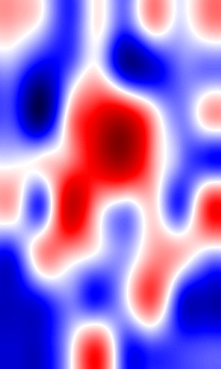}
    &\includegraphics[width=0.09\textwidth]{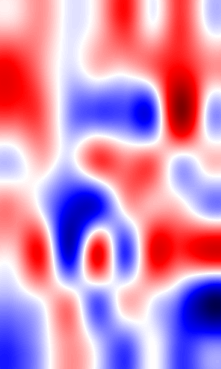}
    &\includegraphics[width=0.09\textwidth]{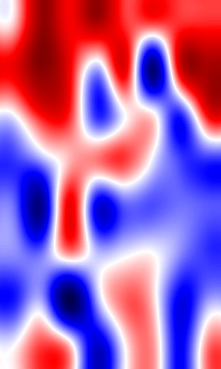}
    &\includegraphics[width=0.09\textwidth]{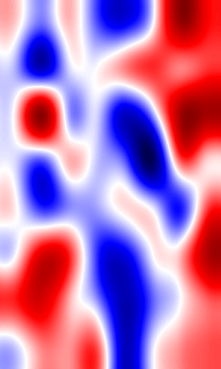}
    &\includegraphics[width=0.09\textwidth]{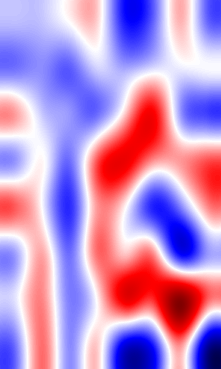}
    &\includegraphics[width=0.09\textwidth]{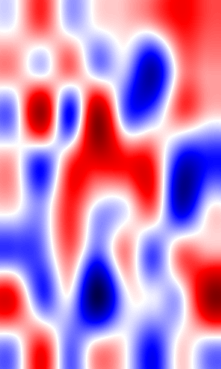}\\
    Channel-21 & Channel-22 & Channel-23 & Channel-24 & Channel-25 & Channel-26 & Channel-27 & Channel-28 & Channel-29 & Channel-30 & Channel-31\\
    \includegraphics[width=0.09\textwidth]{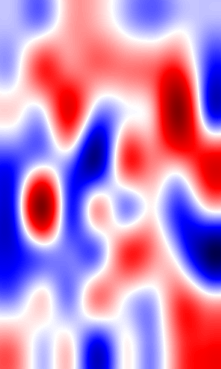}
    &\includegraphics[width=0.09\textwidth]{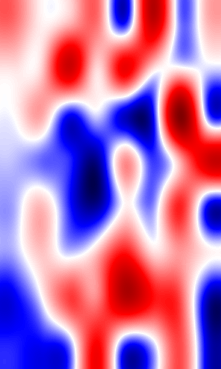}
    &\includegraphics[width=0.09\textwidth]{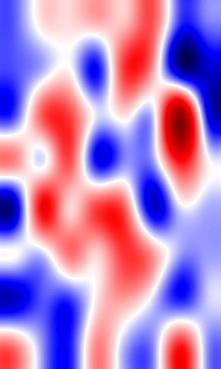}
    &\includegraphics[width=0.09\textwidth]{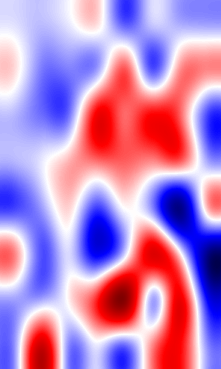}
    &\includegraphics[width=0.09\textwidth]{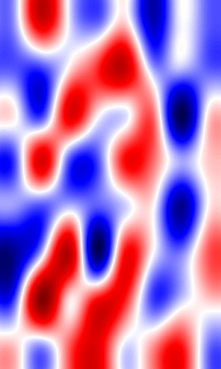}
    &\includegraphics[width=0.09\textwidth]{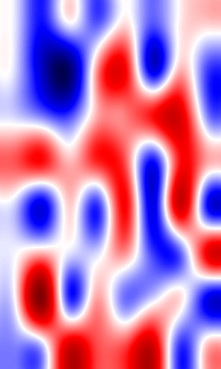}
    &\includegraphics[width=0.09\textwidth]{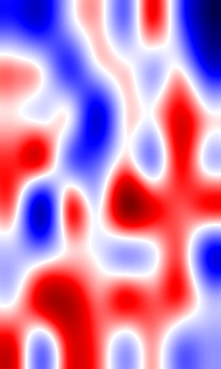}
    &\includegraphics[width=0.09\textwidth]{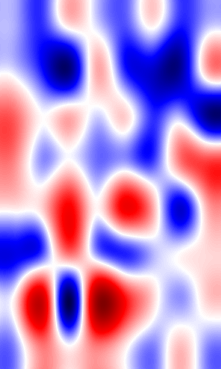}
    &\includegraphics[width=0.09\textwidth]{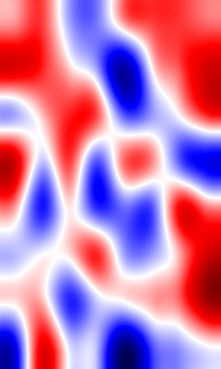}
    &\includegraphics[width=0.09\textwidth]{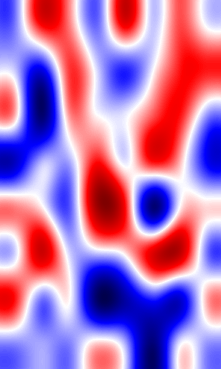}
    &\includegraphics[width=0.09\textwidth]{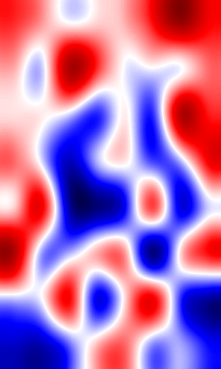}\\~\\
    Estimation $\xhat$ & Channel-0 & Channel-1 & Channel-2 & Channel-3 & Channel-4 & Channel-5 & Channel-6 & Channel-7 & Channel-8 & Channel-9\\
    \includegraphics[width=0.09\textwidth]{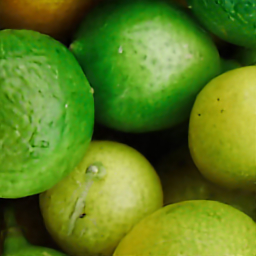}
    &\includegraphics[width=0.09\textwidth]{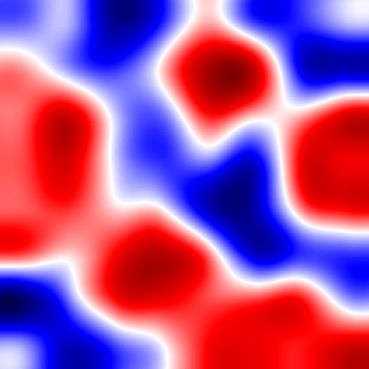}
    &\includegraphics[width=0.09\textwidth]{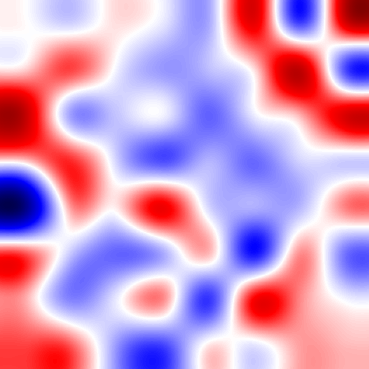}
    &\includegraphics[width=0.09\textwidth]{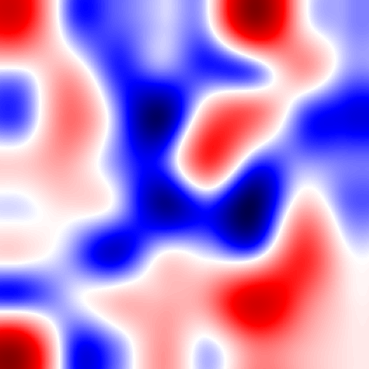}
    &\includegraphics[width=0.09\textwidth]{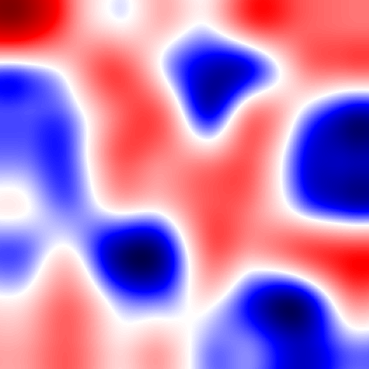}
    &\includegraphics[width=0.09\textwidth]{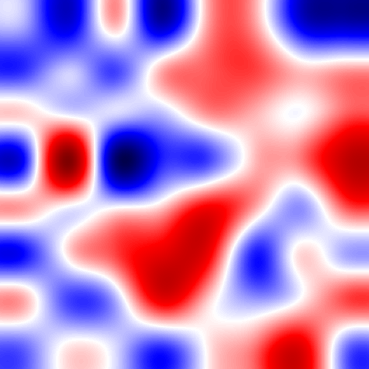}
    &\includegraphics[width=0.09\textwidth]{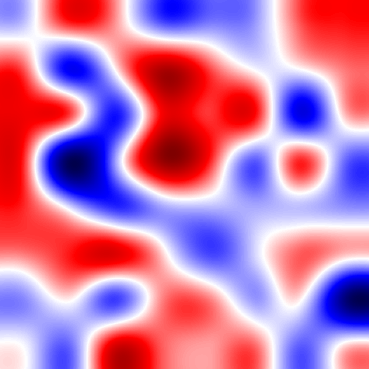}
    &\includegraphics[width=0.09\textwidth]{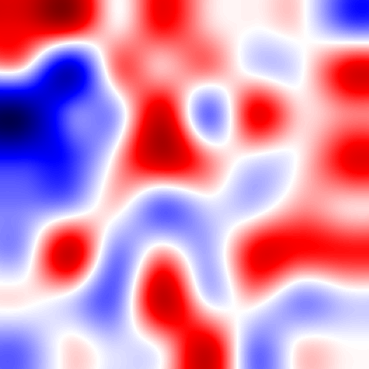}
    &\includegraphics[width=0.09\textwidth]{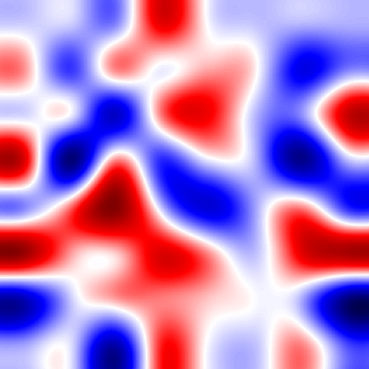}
    &\includegraphics[width=0.09\textwidth]{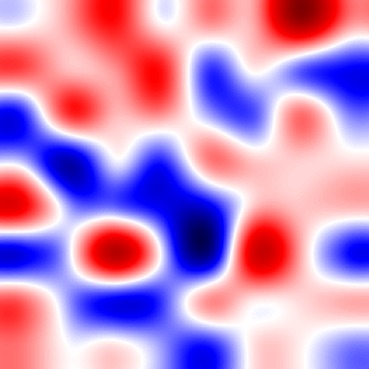}
    &\includegraphics[width=0.09\textwidth]{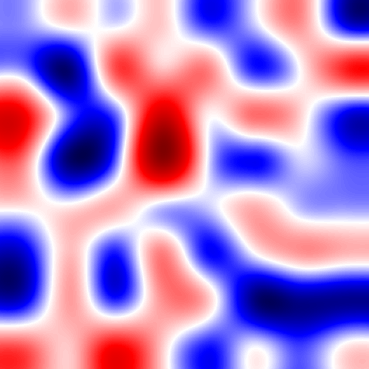}\\
    Channel-10 & Channel-11 & Channel-12 & Channel-13 & Channel-14 & Channel-15 & Channel-16 & Channel-17 & Channel-18 & Channel-19 & Channel-20\\
    \includegraphics[width=0.09\textwidth]{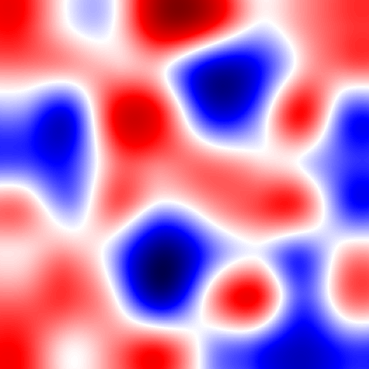}
    &\includegraphics[width=0.09\textwidth]{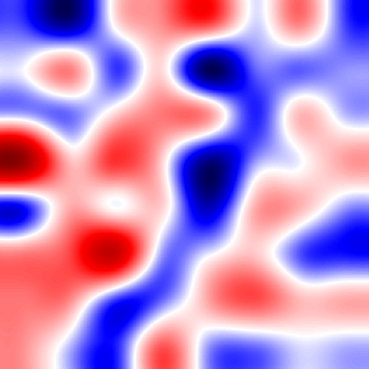}
    &\includegraphics[width=0.09\textwidth]{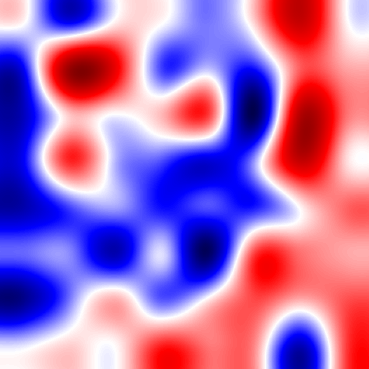}
    &\includegraphics[width=0.09\textwidth]{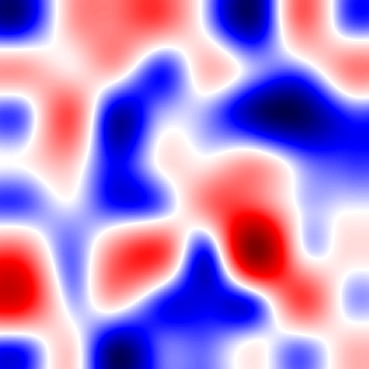}
    &\includegraphics[width=0.09\textwidth]{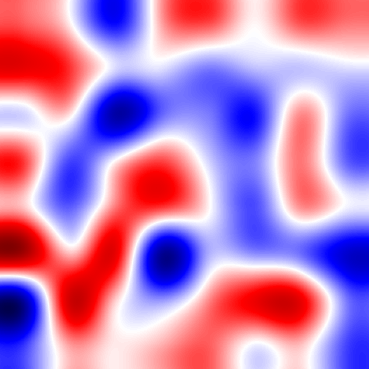}
    &\includegraphics[width=0.09\textwidth]{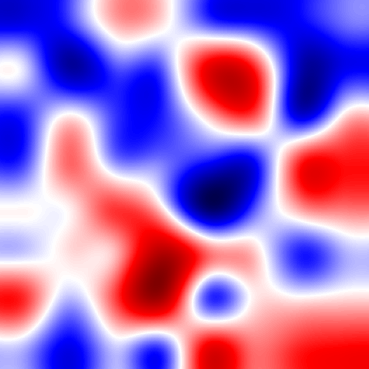}
    &\includegraphics[width=0.09\textwidth]{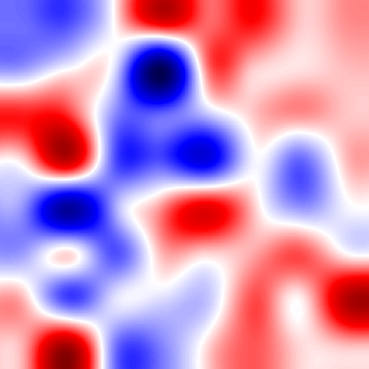}
    &\includegraphics[width=0.09\textwidth]{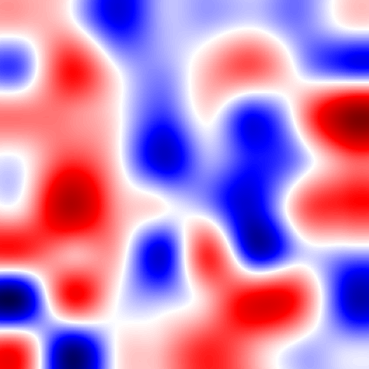}
    &\includegraphics[width=0.09\textwidth]{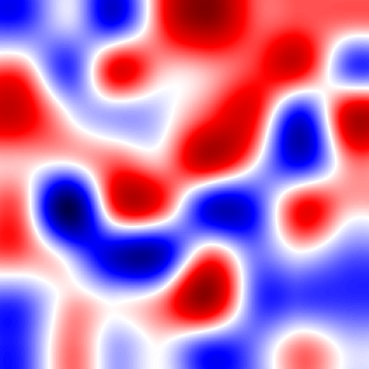}
    &\includegraphics[width=0.09\textwidth]{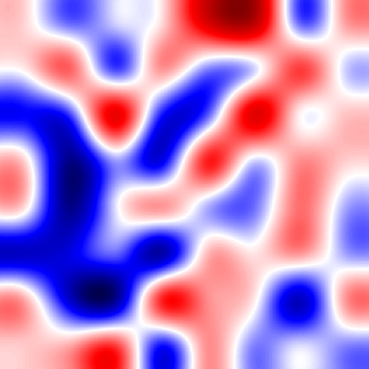}
    &\includegraphics[width=0.09\textwidth]{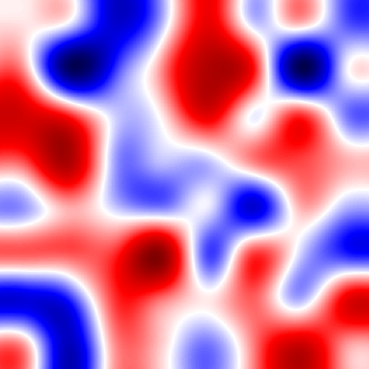}\\
    Channel-21 & Channel-22 & Channel-23 & Channel-24 & Channel-25 & Channel-26 & Channel-27 & Channel-28 & Channel-29 & Channel-30 & Channel-31\\
    \includegraphics[width=0.09\textwidth]{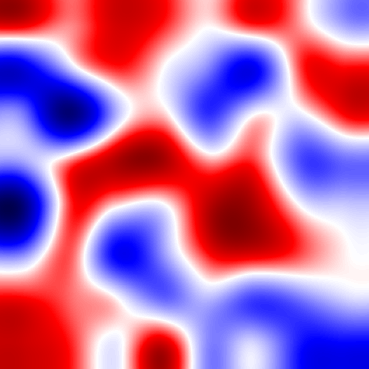}
    &\includegraphics[width=0.09\textwidth]{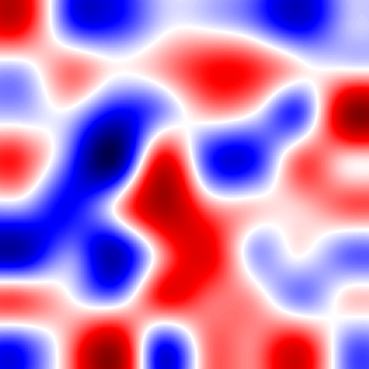}
    &\includegraphics[width=0.09\textwidth]{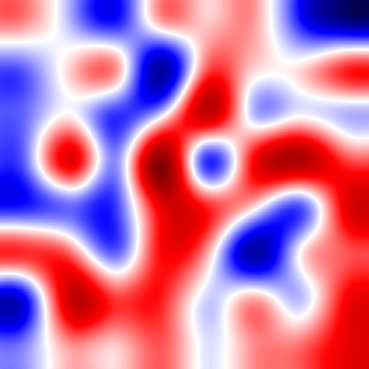}
    &\includegraphics[width=0.09\textwidth]{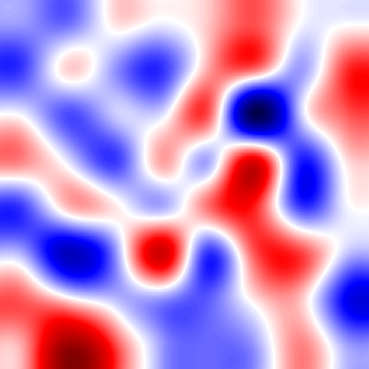}
    &\includegraphics[width=0.09\textwidth]{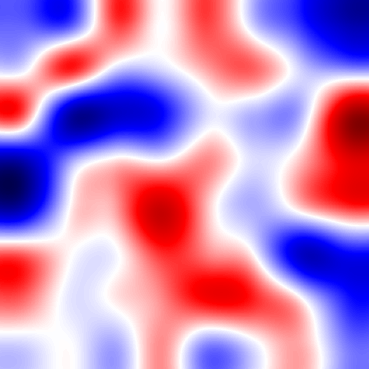}
    &\includegraphics[width=0.09\textwidth]{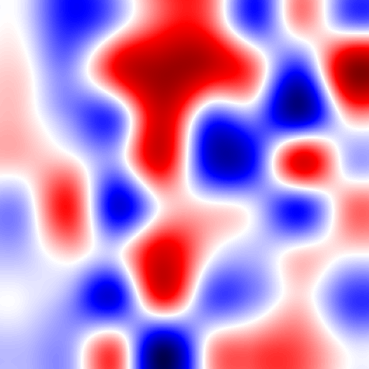}
    &\includegraphics[width=0.09\textwidth]{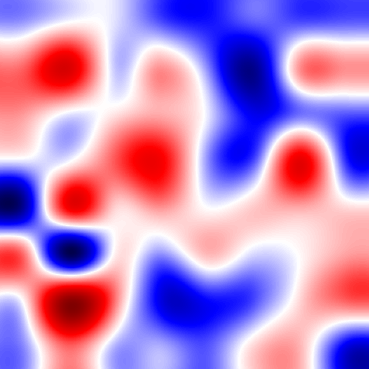}
    &\includegraphics[width=0.09\textwidth]{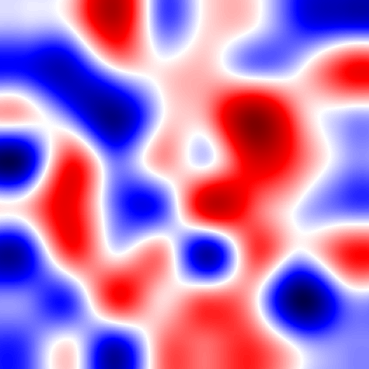}
    &\includegraphics[width=0.09\textwidth]{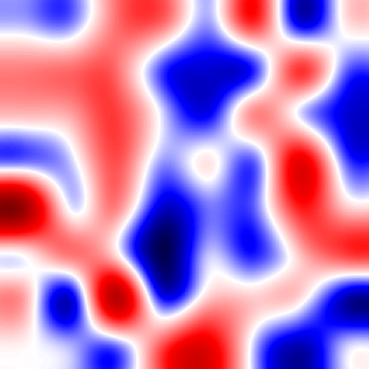}
    &\includegraphics[width=0.09\textwidth]{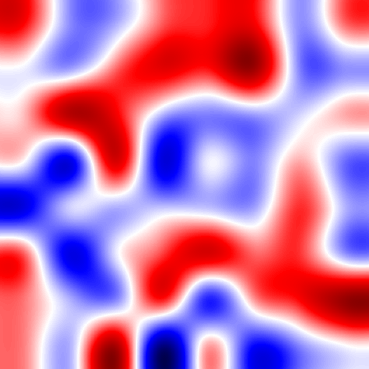}
    &\includegraphics[width=0.09\textwidth]{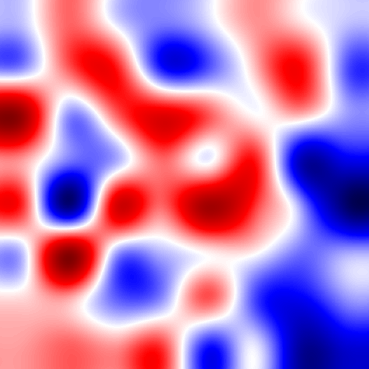}
\end{tabular}}
\caption{Visualizations of the learned 32-channel PE weights for two images named ``test\_38'' and ``0802'' from CBSD68 \cite{martin2001database} \textcolor{blue}{(top)} and DIV2K \cite{agustsson2017ntire} \textcolor{blue}{(bottom)}, respectively, at ratio $\gamma =10\%$ and noise level $\sigma =0$. PEs of different images and channels are upscaled using bicubic interpolation to match the image size as shown by Fig.~2 in the \MP. It is interesting to note that the PEs of different images and channels exhibit varied content and spatial structures. Channel-0 of the PEs is associated with the fundamental image structures, while the other channels may capture supplemental information. Darker red/blue colors indicate positive/negative elements with larger absolute values.}
\label{fig:visualize_PE_spatial}
\end{figure*}

In Tab.~\ref{tab:ablation_studies_from_sc-cnn} $<$18-21$>$, we follow the multi-operator imaging (MOI) approach proposed in \cite{tachella2022unsupervised} and simulate the MOI loss by combining $\Loss_{MC}$ and $\Loss_{DOC}$ ($p=2$) without the final GD step, for the four cases of infinite, 40, 10, and 1 available in-distribution sampling matrices in DOC constraint. We observe that: (1) our original dual-domain loss in $<$1$>$ outperforms $<$18$>$ by 0.4dB PSNR distance at ratio 10\%, confirming the effectiveness of our final GD inspired by residual learning \cite{he2016deep,zhang2017beyond} and introduced loss augmentations; (2) the inclusion of more matrices for the DOC constraint leads to better regularization and higher performance. In Tab.~\ref{tab:ablation_studies_from_sc-cnn} $<$25-28$>$, following \cite{xia2019training,chen2022robust}, we introduce a self-loss term and a SURE \cite{stein1981estimation}-loss term to our $\Loss$, remove the noise injection of sampling simulation in DOC loss, and observe the average PSNR drops of 0.60dB, 0.43dB, and 1.07dB brought by these three modifications, respectively. All the results manifest the effectiveness and competence of our loss function designs that comprehensively and sufficiently utilize available measurement data, given the fixed sampling matrix and information of distributions $q_\x$, $q_\A$, $q_\n$, and ratio range $(0,1]$.

\textbf{More Analyses about Parameters $p$ and $\alpha$ in Our Loss $\Loss$.} In Tab.~\ref{tab:PSNR_different_p_values} and Fig.~\ref{fig:scalable_curve_more_analyses_p_and_alpha} (left), we train the SC-CNNs with seven different settings of $p\in\{0.8,1.0,1.2,1.4,1.6,1.8,2.0\}$ at given measurement ratio $\gamma =10\%$ and noise level $\sigma =0$, and evaluate them on the entire ratio range $(0,1]$. We observe that: (1) when $p=0.8<1$, the NN training process becomes unstable without convergence. This phenomenon may be attributed to numerical instability brought by the non-differentiability of $\lVert \cdot \rVert_p^p~(p<1)$ with input values close to zero. Some similar results can be observed with $p\in\{0.2,0.4,0.6\}$; (2) when $p>1$, NNs will suffer significant accuracy drops in the high-ratio cases. This fact can be attributed to their unbalanced loss values and undesirable biased learning toward the low-ratio tasks with $\gamma <10\%$; (3) our default $p=1$ setting yields the best recoveries and demonstrates excellent generalization to ratios beyond the range $(0,10\%]$. In Tab.~\ref{tab:PSNR_different_alpha_values} and Fig.~\ref{fig:scalable_curve_more_analyses_p_and_alpha} (right), we evaluate six SC-CNNs trained with $\alpha \in\{10,1,0.1,0.01,0.001,0\}$, and observe that our default $\alpha =0.1$ setting outperforms the others in most cases. These results demonstrate the efficacy of our dual-domain loss designs, including DMC and DOC loss terms, and loss form generalization from $\lVert \cdot \rVert_2^2$ to $\lVert \cdot \rVert_p^p$ with $p=1$ for performance improvement and the assurance of ratio-scalability.

\textbf{Effect of GD Step Size Settings in Our SCNet.} Tab.~\ref{tab:ablation_studies_from_sc-cnn} $<$22$>$ investigates the influence of making the GD step size $\rho$ learnable and initialized to 1, similar to state-of-the-art optimization-inspired networks \cite{monga2021algorithm,quan2022dual,tachella2022unsupervised}. Fig.~\ref{fig:effect_of_GD_step} (left) presents a comparison between the learnable setting and our default $\rho \equiv 1$ configuration of the loss and test PSNR curves. We observe that our fixed $\rho$ setting effectively guides SC-CNN to learn more accurate recoveries from measurements, exhibiting a steady and continuous increase in test PSNR, with a leading advantage of 6.35dB at ratio 10\%. However, the learnable $\rho$ may lead to overfitting and unstable convergence. We attribute this fact to that our $\rho \equiv 1$ setting in Tab.~\ref{tab:ablation_studies_from_sc-cnn} $<$11$>$ can encourage NNs to focus on recovering the lost measurements (corresponding to nullspace components regarding $\A_1$ and $\A_2$). This is helpful for avoiding the undesirable overfitting to trivial results, especially in the early stage of NN training.

As shown in Fig.~\ref{fig:effect_of_GD_step} (right), we further investigate the distances between estimated middle or final images and GT before GD (calculated by $\lVert \mathbf{z} -\x \rVert_2^2$) and after GD (calculated by $\lVert \xhat - \x \rVert_2^2$), and observe that GD steps can consistently reduce the prediction error and provide more accurate recoveries. Additionally, in Tab.~\ref{tab:PSNR_different_rho_values}, we train five SC-CNNs for $\sigma =10$ with fixed $\rho \in \{0.6,0.7,0.8,0.9,1.0\}$, and find that our original re-parameterization of $\rho =\text{Sigmoid}(\tau)$ with a freely learnable $\tau$ achieves the best result, verifying the effectiveness and feasibility of our proposed design. It is worth noting that we may currently not observe more interpretable quantitative evidence from the learned $\rho \approx 0.9266$, and believe that there still exists significant room for improving the final recovery performance through better GD mechanisms.

\textbf{Effect of the IE and PE in Our SCNet.} Tab.~\ref{tab:ablation_studies_from_sc-cnn} $<$23$>$ and $<$24$>$ investigate the influence of removing our image embeddings (IE) and positional embeddings (PE) from the stages 2 and 3, and report average PSNR drops of 0.18dB and 0.22dB, respectively. In Tab.~\ref{tab:PSNR_different_PE_sizes}, we train seven SC-CNN variants based on Tab.~\ref{tab:ablation_studies_from_sc-cnn} $<$2$>$ with different PE spatial sizes. We observe that our default setting of $h=w\equiv 8$ achieves the best PSNR, striking a balance between underfitting and overfitting: SC-CNNs with smaller PEs still suffer the lack of optimization freedom, while overly larger PEs may lead to weak regularization and suboptimal reconstruction results.

To explore the insights from IE and PE, we extract their learned parameter weights from SC-CNNs in Tab.~\ref{tab:ablation_studies_from_sc-cnn} $<$2$>$ and $<$3$>$ after training in stages 2 and 3. Here we present our three interesting findings. Firstly, the learned embeddings for different test images tend to be mutually orthogonal without introducing any specific constraints, as demonstrated in Fig.~\ref{fig:visualize_IE_and_PE_ortho}. Secondly, as exhibited in Fig.~\ref{fig:visualize_PE_distribution_by_histogram}, the PEs learned for different images exhibit different distributions. Thirdly, Fig.~\ref{fig:visualize_PE_spatial} visualizes the learned PEs of two test images, displaying distinctive content and spatial structures for their different channels. All these results lead to our inference that the embeddings can adaptively learn distinctive features, which may be helpful for NNs to distinguish among different images/channels/locations and can adapt well with fewer iterations, verifying the feasibility of our NN-embedding joint learning to obtain collaborative and flexible deep signal representations.

\textbf{Effect of PGD-Unrolled Module Number of Our SCNet.} Given that each unrolled module of SCNet corresponds to one iteration in the PGD algorithm \cite{parikh2014proximal}, it is reasonable to assume that a larger module number $K$ will result in higher reconstruction performance. Tab.~\ref{tab:stage_number} examines the impact of $K$ on recovery PSNR and parameter number. We notice that there is still significant room for improvement (up to 0.22 dB at ratio 50\%) in recovery quality by increasing the NN capacity (from $K=20$ to $K=30$), while still maintaining an acceptable parameter number ($<$$0.6$M). However, the recovery time also increases linearly with $K$. Considering the trade-off between computational complexity and recovery performance, we set the default module number for SCNet to $K=20$.

\begin{table}[!t]
\caption{Comparison of average PSNR (dB) on Set11 \cite{kulkarni2016reconnet} at $\gamma =10\%$ between DDSSL \cite{quan2022dual} and SC-CNN, equipped with our proposed four-stage progressive reconstruction strategy. Notably, the original DDSSL method, when augmented with test-time adaptation, attains a PSNR of 27.65dB.}
\label{tab:application_of_four_stage_strategy_to_existing_method}
\centering
\resizebox{1.0\linewidth}{!}{
\begin{tabular}{l|cccc}
\shline
\rowcolor[HTML]{EFEFEF} 
Stage             & 1     & 2     & 3     & 4     \\ \hline \hline
DDSSL (ECCV 2022) & 26.86 & 28.28 & 28.36 & 28.49 \\ \hline
\textbf{SC-CNN (Ours)}     & 27.93 & 29.16 & 29.39 & 29.42 \\ \hline
\end{tabular}}
\end{table}

\begin{figure*}[!t]
\centering
\includegraphics[width=1.0\textwidth]{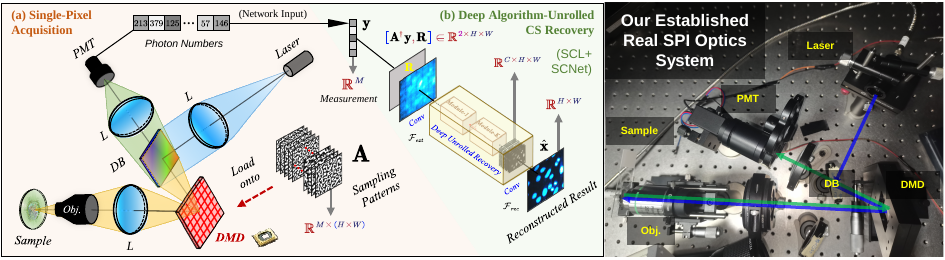}
\caption{\textcolor{blue}{(Left)} Illustration of our SPI system with more details about the reconstruction by our SCL and SC-CNN. \textcolor{blue}{(1)} A set of optics instruments and a sequence of DMD modulation patterns are used to implement CS sampling process, \textit{i.e.}, $\y= \A\x$. \textcolor{blue}{(2)} An SC-CNN (please refer to our Fig.~2 in the \MP~for structural details) variant is employed for scalable imaging once trained by SCL on a single measurement set. \textcolor{blue}{(Right)} The layout of our SPI optics system on the testbed for single-pixel CS acquisition. Blue arrows represent emission path of light from laser to sample, while green arrows correspond to excitation path of fluorescence from sample to PMT. \textcolor{blue}{Obj.}: objective; \textcolor{blue}{L}: lens; \textcolor{blue}{DMD}: digital micro-mirror device; \textcolor{blue}{DB}: dichroic beamsplitter; \textcolor{blue}{PMT}: photon-counting photomultiplier tube.}
\label{fig:real_arch}
\end{figure*}

\begin{figure*}[!t]
\setlength{\tabcolsep}{0.5pt}
\resizebox{1.0\textwidth}{!}{
\tiny
\begin{tabular}{cccccccccc}
$\A^\dagger \y$ & TVAL3 & MC & DMC & \textbf{DMC+DOC}~~~ & $\A^\dagger \y$ & TVAL3 & MC & DMC & \textbf{DMC+DOC} \\
\includegraphics[width=0.09\textwidth]{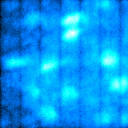} & \includegraphics[width=0.09\textwidth]{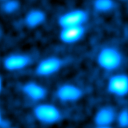} &
\includegraphics[width=0.09\textwidth]{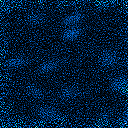} &
\includegraphics[width=0.09\textwidth]{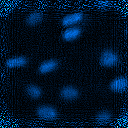} &
\includegraphics[width=0.09\textwidth]{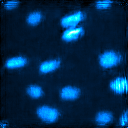}~~~ &
\includegraphics[width=0.09\textwidth]{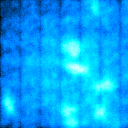} & \includegraphics[width=0.09\textwidth]{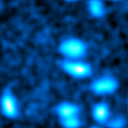} &
\includegraphics[width=0.09\textwidth]{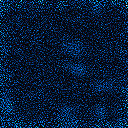} &
\includegraphics[width=0.09\textwidth]{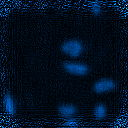} &
\includegraphics[width=0.09\textwidth]{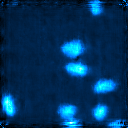}\\
\includegraphics[width=0.09\textwidth]{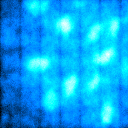} & \includegraphics[width=0.09\textwidth]{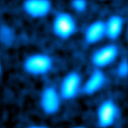} &
\includegraphics[width=0.09\textwidth]{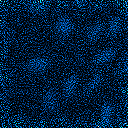} &
\includegraphics[width=0.09\textwidth]{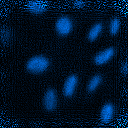} &
\includegraphics[width=0.09\textwidth]{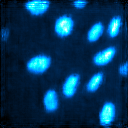}~~~ &
\includegraphics[width=0.09\textwidth]{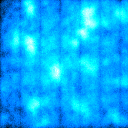} & \includegraphics[width=0.09\textwidth]{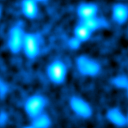} &
\includegraphics[width=0.09\textwidth]{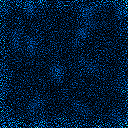} &
\includegraphics[width=0.09\textwidth]{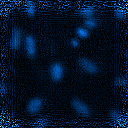} &
\includegraphics[width=0.09\textwidth]{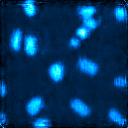}\\
\includegraphics[width=0.09\textwidth]{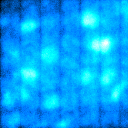} & \includegraphics[width=0.09\textwidth]{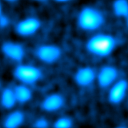} &
\includegraphics[width=0.09\textwidth]{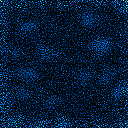} &
\includegraphics[width=0.09\textwidth]{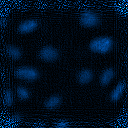} &
\includegraphics[width=0.09\textwidth]{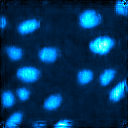}~~~ &
\includegraphics[width=0.09\textwidth]{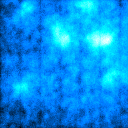} & \includegraphics[width=0.09\textwidth]{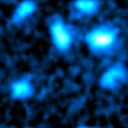} &
\includegraphics[width=0.09\textwidth]{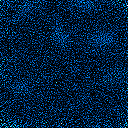} &
\includegraphics[width=0.09\textwidth]{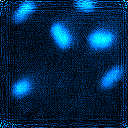} &
\includegraphics[width=0.09\textwidth]{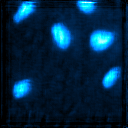}\\
\includegraphics[width=0.09\textwidth]{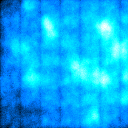} & \includegraphics[width=0.09\textwidth]{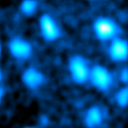} &
\includegraphics[width=0.09\textwidth]{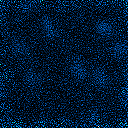} &
\includegraphics[width=0.09\textwidth]{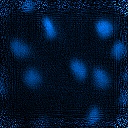} &
\includegraphics[width=0.09\textwidth]{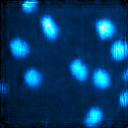}~~~ &
\includegraphics[width=0.09\textwidth]{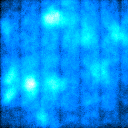} & \includegraphics[width=0.09\textwidth]{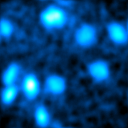} &
\includegraphics[width=0.09\textwidth]{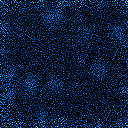} &
\includegraphics[width=0.09\textwidth]{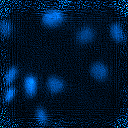} &
\includegraphics[width=0.09\textwidth]{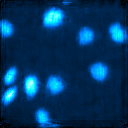}\\
\includegraphics[width=0.09\textwidth]{fc_real_Nucleus_A+y_test_8_init.png} & \includegraphics[width=0.09\textwidth]{fc_real_Nucleus_TVAL3_8.png} &
\includegraphics[width=0.09\textwidth]{fc_real_Nucleus_MC_8_q_5000.png} &
\includegraphics[width=0.09\textwidth]{fc_real_Nucleus_DMC_8_q_5000.png} &
\includegraphics[width=0.09\textwidth]{fc_real_Nucleus_DMC+DOC_8_q_5000.png}~~~ &
\includegraphics[width=0.09\textwidth]{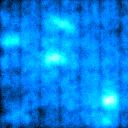} & \includegraphics[width=0.09\textwidth]{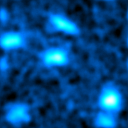} &
\includegraphics[width=0.09\textwidth]{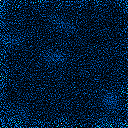} &
\includegraphics[width=0.09\textwidth]{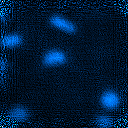} &
\includegraphics[width=0.09\textwidth]{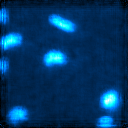}
\end{tabular}}
\caption{Visual comparison of real CS imaging among five methods on ten different nucleus SPI measurements with ratio $\gamma =5000/16384 \approx 30.5\%$, consistent as the one of training measurements.}
\label{fig:comparison_real_more_nucleus}
\end{figure*}

\begin{figure*}[!t]
\setlength{\tabcolsep}{0.5pt}
\resizebox{1.0\textwidth}{!}{
\tiny
\begin{tabular}{cccccccccc}
$\A^\dagger \y$ & TVAL3 & MC & DMC & \textbf{DMC+DOC}~~~ & $\A^\dagger \y$ & TVAL3 & MC & DMC & \textbf{DMC+DOC} \\
\includegraphics[width=0.09\textwidth]{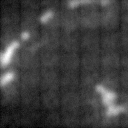} &
\includegraphics[width=0.09\textwidth]{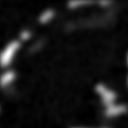} &
\includegraphics[width=0.09\textwidth]{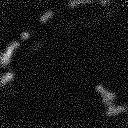} &
\includegraphics[width=0.09\textwidth]{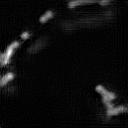} &
\includegraphics[width=0.09\textwidth]{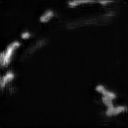}~~~ &
\includegraphics[width=0.09\textwidth]{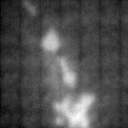} &
\includegraphics[width=0.09\textwidth]{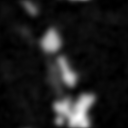} &
\includegraphics[width=0.09\textwidth]{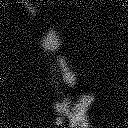} &
\includegraphics[width=0.09\textwidth]{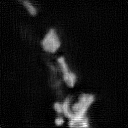} &
\includegraphics[width=0.09\textwidth]{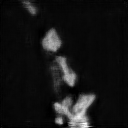}\\
\includegraphics[width=0.09\textwidth]{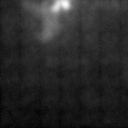} &
\includegraphics[width=0.09\textwidth]{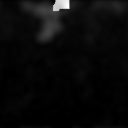} &
\includegraphics[width=0.09\textwidth]{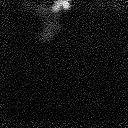} &
\includegraphics[width=0.09\textwidth]{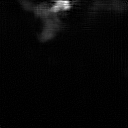} &
\includegraphics[width=0.09\textwidth]{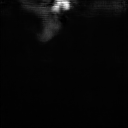}~~~ &
\includegraphics[width=0.09\textwidth]{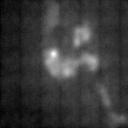} &
\includegraphics[width=0.09\textwidth]{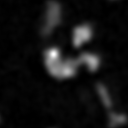} &
\includegraphics[width=0.09\textwidth]{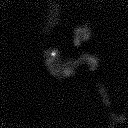} &
\includegraphics[width=0.09\textwidth]{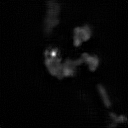} &
\includegraphics[width=0.09\textwidth]{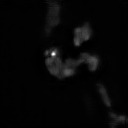}\\
\includegraphics[width=0.09\textwidth]{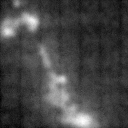} &
\includegraphics[width=0.09\textwidth]{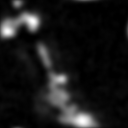} &
\includegraphics[width=0.09\textwidth]{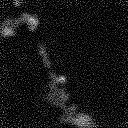} &
\includegraphics[width=0.09\textwidth]{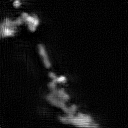} &
\includegraphics[width=0.09\textwidth]{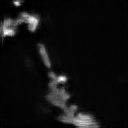}~~~ &
\includegraphics[width=0.09\textwidth]{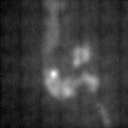} &
\includegraphics[width=0.09\textwidth]{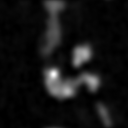} &
\includegraphics[width=0.09\textwidth]{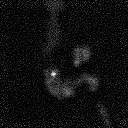} &
\includegraphics[width=0.09\textwidth]{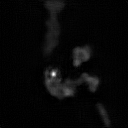} &
\includegraphics[width=0.09\textwidth]{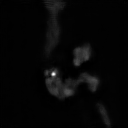}\\
\includegraphics[width=0.09\textwidth]{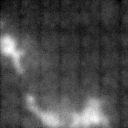} &
\includegraphics[width=0.09\textwidth]{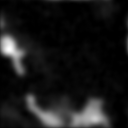} &
\includegraphics[width=0.09\textwidth]{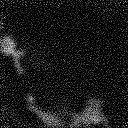} &
\includegraphics[width=0.09\textwidth]{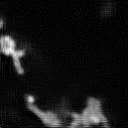} &
\includegraphics[width=0.09\textwidth]{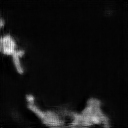}~~~ &
\includegraphics[width=0.09\textwidth]{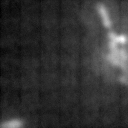} &
\includegraphics[width=0.09\textwidth]{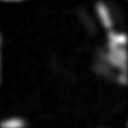} &
\includegraphics[width=0.09\textwidth]{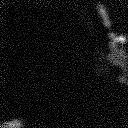} &
\includegraphics[width=0.09\textwidth]{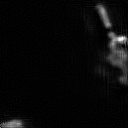} &
\includegraphics[width=0.09\textwidth]{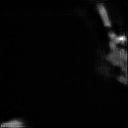}\\
\includegraphics[width=0.09\textwidth]{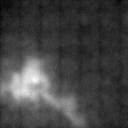} &
\includegraphics[width=0.09\textwidth]{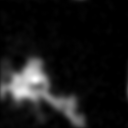} &
\includegraphics[width=0.09\textwidth]{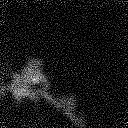} &
\includegraphics[width=0.09\textwidth]{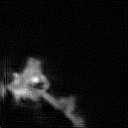} &
\includegraphics[width=0.09\textwidth]{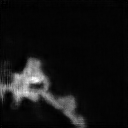}~~~ &
\includegraphics[width=0.09\textwidth]{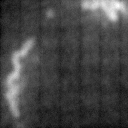} &
\includegraphics[width=0.09\textwidth]{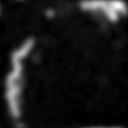} &
\includegraphics[width=0.09\textwidth]{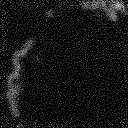} &
\includegraphics[width=0.09\textwidth]{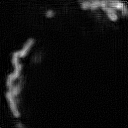} &
\includegraphics[width=0.09\textwidth]{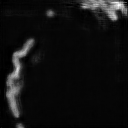}
\end{tabular}}
\caption{Visual comparison of real CS imaging among five methods on ten fluorescent microsphere (FM) SPI measurements at $\gamma =5000/16384 \approx 30.5\%$, same as ratio of training set.}
\label{fig:comparison_real_more_FM}
\end{figure*}

\begin{figure*}[!t]
\setlength{\tabcolsep}{0.5pt}
\resizebox{1.0\textwidth}{!}{
\tiny
\begin{tabular}{cccccccccc}
$\A^\dagger \y$ & TVAL3 & MC & DMC & \textbf{DMC+DOC}~~~ & $\A^\dagger \y$ & TVAL3 & MC & DMC & \textbf{DMC+DOC} \\
\includegraphics[width=0.09\textwidth]{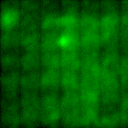} &
\includegraphics[width=0.09\textwidth]{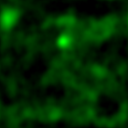} &
\includegraphics[width=0.09\textwidth]{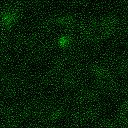} &
\includegraphics[width=0.09\textwidth]{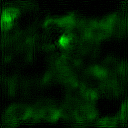} &
\includegraphics[width=0.09\textwidth]{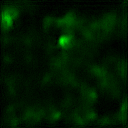}~~~ &
\includegraphics[width=0.09\textwidth]{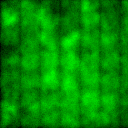} &
\includegraphics[width=0.09\textwidth]{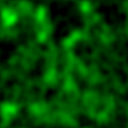} &
\includegraphics[width=0.09\textwidth]{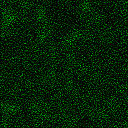} &
\includegraphics[width=0.09\textwidth]{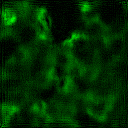} &
\includegraphics[width=0.09\textwidth]{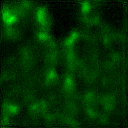}\\
\includegraphics[width=0.09\textwidth]{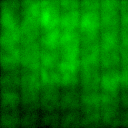} &
\includegraphics[width=0.09\textwidth]{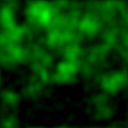} &
\includegraphics[width=0.09\textwidth]{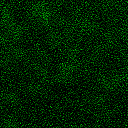} &
\includegraphics[width=0.09\textwidth]{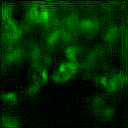} &
\includegraphics[width=0.09\textwidth]{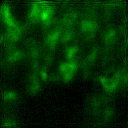}~~~ &
\includegraphics[width=0.09\textwidth]{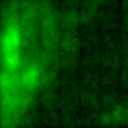} &
\includegraphics[width=0.09\textwidth]{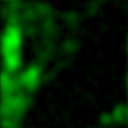} &
\includegraphics[width=0.09\textwidth]{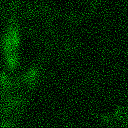} &
\includegraphics[width=0.09\textwidth]{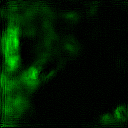} &
\includegraphics[width=0.09\textwidth]{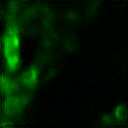}\\
\includegraphics[width=0.09\textwidth]{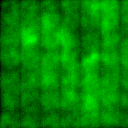} &
\includegraphics[width=0.09\textwidth]{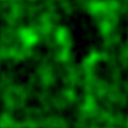} &
\includegraphics[width=0.09\textwidth]{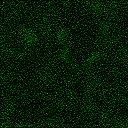} &
\includegraphics[width=0.09\textwidth]{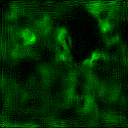} &
\includegraphics[width=0.09\textwidth]{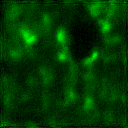}~~~ &
\includegraphics[width=0.09\textwidth]{fc_real_F-Actin-3D_A+y_test_5_0_init.png} &
\includegraphics[width=0.09\textwidth]{fc_real_F-Actin-3D_TVAL3_5_0.png} &
\includegraphics[width=0.09\textwidth]{fc_real_F-Actin-3D_MC_5_0_q_5000.png} &
\includegraphics[width=0.09\textwidth]{fc_real_F-Actin-3D_DMC_5_0_q_5000.png} &
\includegraphics[width=0.09\textwidth]{fc_real_F-Actin-3D_DMC+DOC_5_0_q_5000.png}\\
\includegraphics[width=0.09\textwidth]{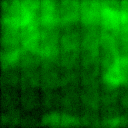} &
\includegraphics[width=0.09\textwidth]{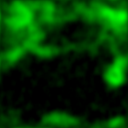} &
\includegraphics[width=0.09\textwidth]{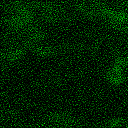} &
\includegraphics[width=0.09\textwidth]{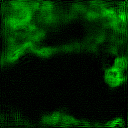} &
\includegraphics[width=0.09\textwidth]{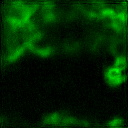}~~~ &
\includegraphics[width=0.09\textwidth]{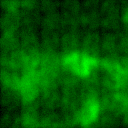} &
\includegraphics[width=0.09\textwidth]{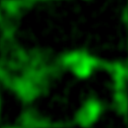} &
\includegraphics[width=0.09\textwidth]{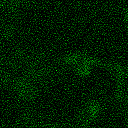} &
\includegraphics[width=0.09\textwidth]{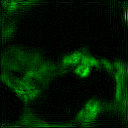} &
\includegraphics[width=0.09\textwidth]{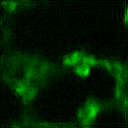}\\
\includegraphics[width=0.09\textwidth]{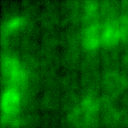} &
\includegraphics[width=0.09\textwidth]{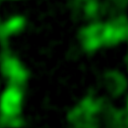} &
\includegraphics[width=0.09\textwidth]{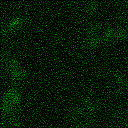} &
\includegraphics[width=0.09\textwidth]{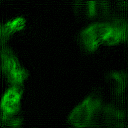} &
\includegraphics[width=0.09\textwidth]{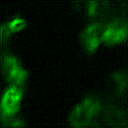}~~~ &
\includegraphics[width=0.09\textwidth]{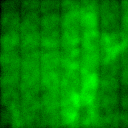} &
\includegraphics[width=0.09\textwidth]{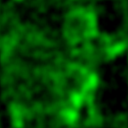} &
\includegraphics[width=0.09\textwidth]{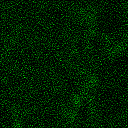} &
\includegraphics[width=0.09\textwidth]{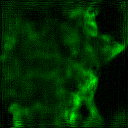} &
\includegraphics[width=0.09\textwidth]{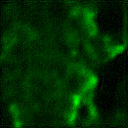}
\end{tabular}}
\caption{Visual comparison of real CS imaging among five methods on ten filamentous actin (F-Actin) SPI measurements with ratio $\gamma =5000/16384 \approx 30.5\%$, consistent as the one of training set. For 3D reconstructions, we only provide the first 2D slice if there are no specific explanations.}
\label{fig:comparison_real_more_F-Actin}
\end{figure*}

\begin{figure*}[!t]
\setlength{\tabcolsep}{0.5pt}
\resizebox{1.0\textwidth}{!}{
\tiny
\begin{tabular}{ccccccccccccccccccccc}
1 & 2 & 3 & 4 & 5 & 6 & 7 & 8 & 9 & 10 & 11 & 12 & 13 & 14 & 15 & 16 & 17 & 18 & 19 & 20 & 21 \\
\includegraphics[width=0.04\textwidth]{fc_real_F-Actin-3D_DMC+DOC_0_0_q_5000.png} &
\includegraphics[width=0.04\textwidth]{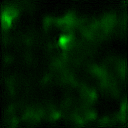} &
\includegraphics[width=0.04\textwidth]{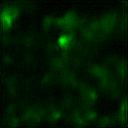} &
\includegraphics[width=0.04\textwidth]{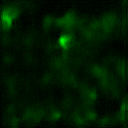} &
\includegraphics[width=0.04\textwidth]{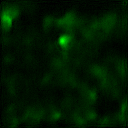} &
\includegraphics[width=0.04\textwidth]{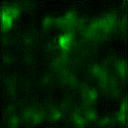} &
\includegraphics[width=0.04\textwidth]{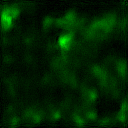} &
\includegraphics[width=0.04\textwidth]{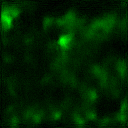} &
\includegraphics[width=0.04\textwidth]{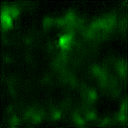} &
\includegraphics[width=0.04\textwidth]{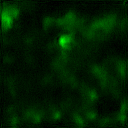} &
\includegraphics[width=0.04\textwidth]{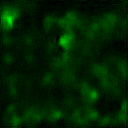} &
\includegraphics[width=0.04\textwidth]{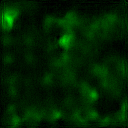} &
\includegraphics[width=0.04\textwidth]{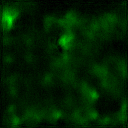} &
\includegraphics[width=0.04\textwidth]{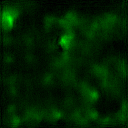} &
\includegraphics[width=0.04\textwidth]{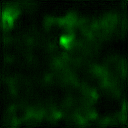} &
\includegraphics[width=0.04\textwidth]{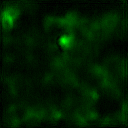} &
\includegraphics[width=0.04\textwidth]{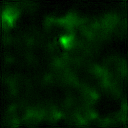} &
\includegraphics[width=0.04\textwidth]{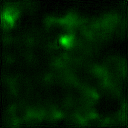} &
\includegraphics[width=0.04\textwidth]{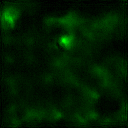} &
\includegraphics[width=0.04\textwidth]{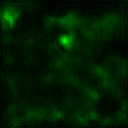} &
\includegraphics[width=0.04\textwidth]{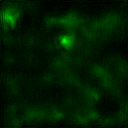}\\
\includegraphics[width=0.04\textwidth]{fc_real_F-Actin-3D_DMC+DOC_1_0_q_5000.png} &
\includegraphics[width=0.04\textwidth]{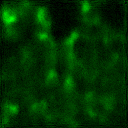} &
\includegraphics[width=0.04\textwidth]{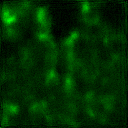} &
\includegraphics[width=0.04\textwidth]{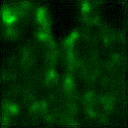} &
\includegraphics[width=0.04\textwidth]{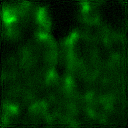} &
\includegraphics[width=0.04\textwidth]{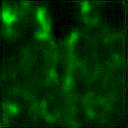} &
\includegraphics[width=0.04\textwidth]{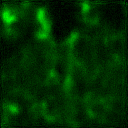} &
\includegraphics[width=0.04\textwidth]{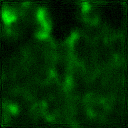} &
\includegraphics[width=0.04\textwidth]{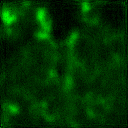} &
\includegraphics[width=0.04\textwidth]{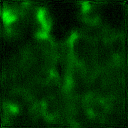} &
\includegraphics[width=0.04\textwidth]{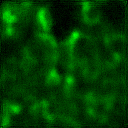} &
\includegraphics[width=0.04\textwidth]{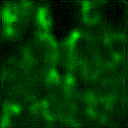} &
\includegraphics[width=0.04\textwidth]{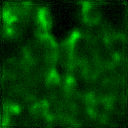} &
\includegraphics[width=0.04\textwidth]{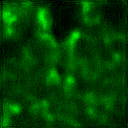} &
\includegraphics[width=0.04\textwidth]{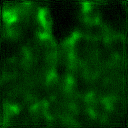} &
\includegraphics[width=0.04\textwidth]{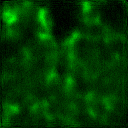} &
\includegraphics[width=0.04\textwidth]{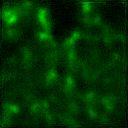} &
\includegraphics[width=0.04\textwidth]{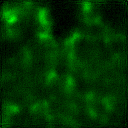} &
\includegraphics[width=0.04\textwidth]{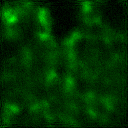} &
\includegraphics[width=0.04\textwidth]{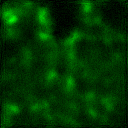} &
\includegraphics[width=0.04\textwidth]{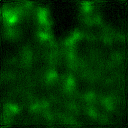}\\
\includegraphics[width=0.04\textwidth]{fc_real_F-Actin-3D_DMC+DOC_2_0_q_5000.png} &
\includegraphics[width=0.04\textwidth]{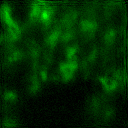} &
\includegraphics[width=0.04\textwidth]{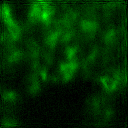} &
\includegraphics[width=0.04\textwidth]{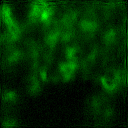} &
\includegraphics[width=0.04\textwidth]{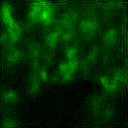} &
\includegraphics[width=0.04\textwidth]{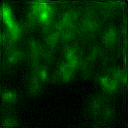} &
\includegraphics[width=0.04\textwidth]{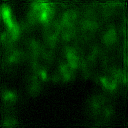} &
\includegraphics[width=0.04\textwidth]{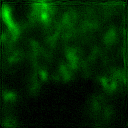} &
\includegraphics[width=0.04\textwidth]{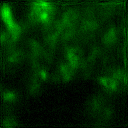} &
\includegraphics[width=0.04\textwidth]{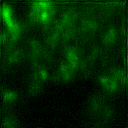} &
\includegraphics[width=0.04\textwidth]{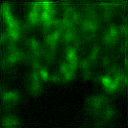} &
\includegraphics[width=0.04\textwidth]{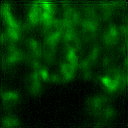} &
\includegraphics[width=0.04\textwidth]{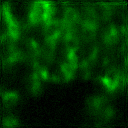} &
\includegraphics[width=0.04\textwidth]{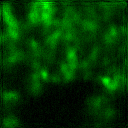} &
\includegraphics[width=0.04\textwidth]{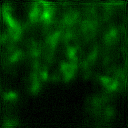} &
\includegraphics[width=0.04\textwidth]{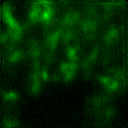} &
\includegraphics[width=0.04\textwidth]{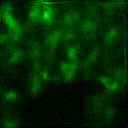} &
\includegraphics[width=0.04\textwidth]{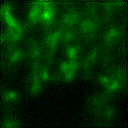} &
\includegraphics[width=0.04\textwidth]{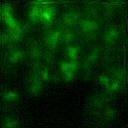} &
\includegraphics[width=0.04\textwidth]{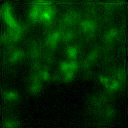} &
\includegraphics[width=0.04\textwidth]{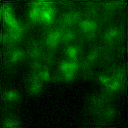}\\
\includegraphics[width=0.04\textwidth]{fc_real_F-Actin-3D_DMC+DOC_3_0_q_5000.png} &
\includegraphics[width=0.04\textwidth]{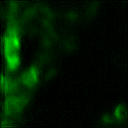} &
\includegraphics[width=0.04\textwidth]{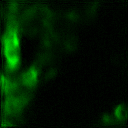} &
\includegraphics[width=0.04\textwidth]{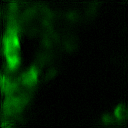} &
\includegraphics[width=0.04\textwidth]{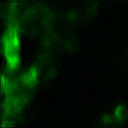} &
\includegraphics[width=0.04\textwidth]{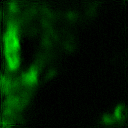} &
\includegraphics[width=0.04\textwidth]{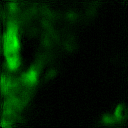} &
\includegraphics[width=0.04\textwidth]{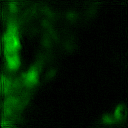} &
\includegraphics[width=0.04\textwidth]{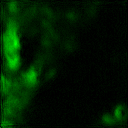} &
\includegraphics[width=0.04\textwidth]{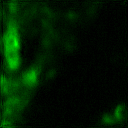} &
\includegraphics[width=0.04\textwidth]{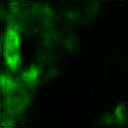} &
\includegraphics[width=0.04\textwidth]{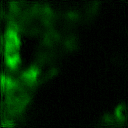} &
\includegraphics[width=0.04\textwidth]{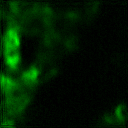} &
\includegraphics[width=0.04\textwidth]{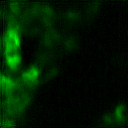} &
\includegraphics[width=0.04\textwidth]{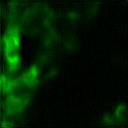} &
\includegraphics[width=0.04\textwidth]{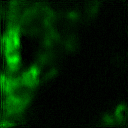} &
\includegraphics[width=0.04\textwidth]{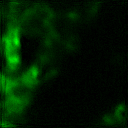} &
\includegraphics[width=0.04\textwidth]{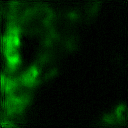} &
\includegraphics[width=0.04\textwidth]{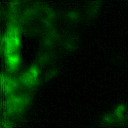} &
\includegraphics[width=0.04\textwidth]{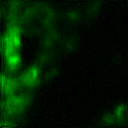} &
\includegraphics[width=0.04\textwidth]{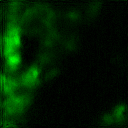}\\
\includegraphics[width=0.04\textwidth]{fc_real_F-Actin-3D_DMC+DOC_4_0_q_5000.png} &
\includegraphics[width=0.04\textwidth]{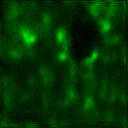} &
\includegraphics[width=0.04\textwidth]{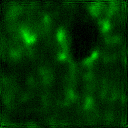} &
\includegraphics[width=0.04\textwidth]{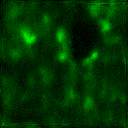} &
\includegraphics[width=0.04\textwidth]{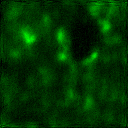} &
\includegraphics[width=0.04\textwidth]{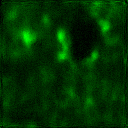} &
\includegraphics[width=0.04\textwidth]{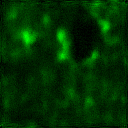} &
\includegraphics[width=0.04\textwidth]{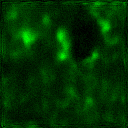} &
\includegraphics[width=0.04\textwidth]{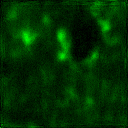} &
\includegraphics[width=0.04\textwidth]{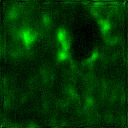} &
\includegraphics[width=0.04\textwidth]{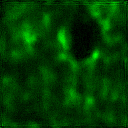} &
\includegraphics[width=0.04\textwidth]{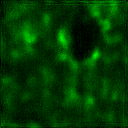} &
\includegraphics[width=0.04\textwidth]{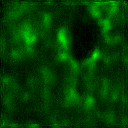} &
\includegraphics[width=0.04\textwidth]{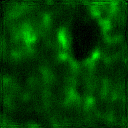} &
\includegraphics[width=0.04\textwidth]{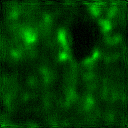} &
\includegraphics[width=0.04\textwidth]{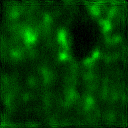} &
\includegraphics[width=0.04\textwidth]{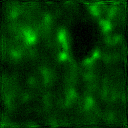} &
\includegraphics[width=0.04\textwidth]{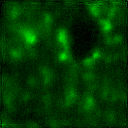} &
\includegraphics[width=0.04\textwidth]{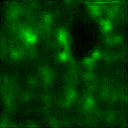} &
\includegraphics[width=0.04\textwidth]{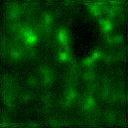} &
\includegraphics[width=0.04\textwidth]{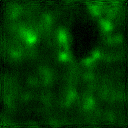}
\end{tabular}}
\caption{Visual comparison of real 3D CS imaging of our method on five groups of F-actin SPI measurements at $\gamma =5000/16384 \approx 30.5\%$, with each group having 21 2D slices along the z-axis.}
\label{fig:our_results_more_FM_slice}
\end{figure*}

\begin{figure*}[!t]
\setlength{\tabcolsep}{0.5pt}
\resizebox{1.0\textwidth}{!}{
\tiny
\begin{tabular}{ccccccccccccc}
\multirow{2}{*}{\rotatebox[origin=c]{90}{Mean $\xhat^*$~~~}}~~~~~ & $M=20$ & $50$ & $100$ & $150$ & $300$ & $400$ & $500$ & $1000$ & $2000$ & $3000$ & $4000$ & $5000$ \\
\multirow{2}{*}{\rotatebox[origin=c]{90}{Std. Dev.~~~}}~~~~~ & \includegraphics[width=0.07\textwidth]{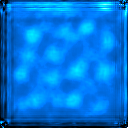} &
\includegraphics[width=0.07\textwidth]{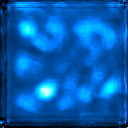} &
\includegraphics[width=0.07\textwidth]{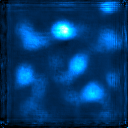} &
\includegraphics[width=0.07\textwidth]{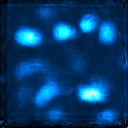} &
\includegraphics[width=0.07\textwidth]{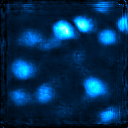} &
\includegraphics[width=0.07\textwidth]{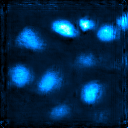} &
\includegraphics[width=0.07\textwidth]{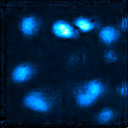} &
\includegraphics[width=0.07\textwidth]{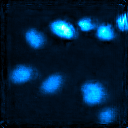} &
\includegraphics[width=0.07\textwidth]{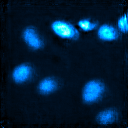} &
\includegraphics[width=0.07\textwidth]{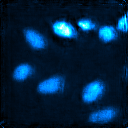} &
\includegraphics[width=0.07\textwidth]{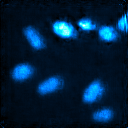} &
\includegraphics[width=0.07\textwidth]{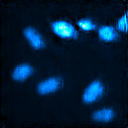} \\
\multirow{2}{*}{\rotatebox[origin=c]{90}{Mean $\xhat^*$~~~}}~~~~~ & \includegraphics[width=0.07\textwidth]{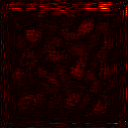} &
\includegraphics[width=0.07\textwidth]{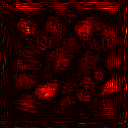} &
\includegraphics[width=0.07\textwidth]{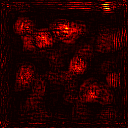} &
\includegraphics[width=0.07\textwidth]{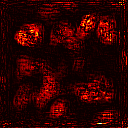} &
\includegraphics[width=0.07\textwidth]{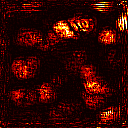} &
\includegraphics[width=0.07\textwidth]{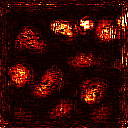} &
\includegraphics[width=0.07\textwidth]{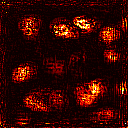} &
\includegraphics[width=0.07\textwidth]{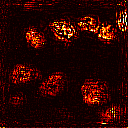} &
\includegraphics[width=0.07\textwidth]{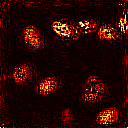} &
\includegraphics[width=0.07\textwidth]{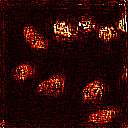} &
\includegraphics[width=0.07\textwidth]{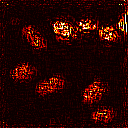} &
\includegraphics[width=0.07\textwidth]{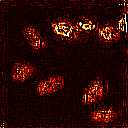} \\
\multirow{2}{*}{\rotatebox[origin=c]{90}{Std. Dev.~~~}}~~~~~ & \includegraphics[width=0.07\textwidth]{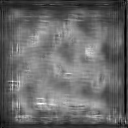} &
\includegraphics[width=0.07\textwidth]{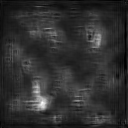} &
\includegraphics[width=0.07\textwidth]{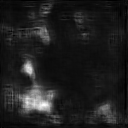} &
\includegraphics[width=0.07\textwidth]{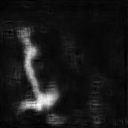} &
\includegraphics[width=0.07\textwidth]{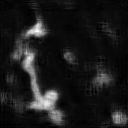} &
\includegraphics[width=0.07\textwidth]{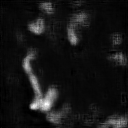} &
\includegraphics[width=0.07\textwidth]{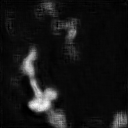} &
\includegraphics[width=0.07\textwidth]{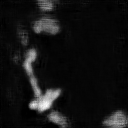} &
\includegraphics[width=0.07\textwidth]{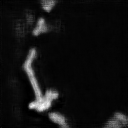} &
\includegraphics[width=0.07\textwidth]{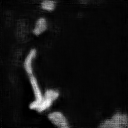} &
\includegraphics[width=0.07\textwidth]{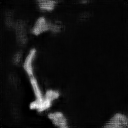} &
\includegraphics[width=0.07\textwidth]{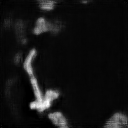} \\
\multirow{2}{*}{\rotatebox[origin=c]{90}{Mean $\xhat^*$~~~}}~~~~~ & \includegraphics[width=0.07\textwidth]{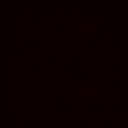} &
\includegraphics[width=0.07\textwidth]{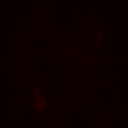} &
\includegraphics[width=0.07\textwidth]{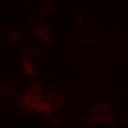} &
\includegraphics[width=0.07\textwidth]{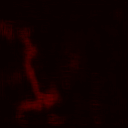} &
\includegraphics[width=0.07\textwidth]{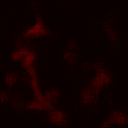} &
\includegraphics[width=0.07\textwidth]{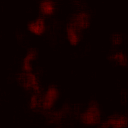} &
\includegraphics[width=0.07\textwidth]{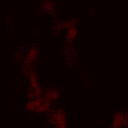} &
\includegraphics[width=0.07\textwidth]{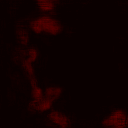} &
\includegraphics[width=0.07\textwidth]{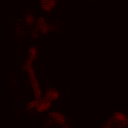} &
\includegraphics[width=0.07\textwidth]{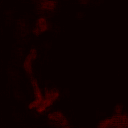} &
\includegraphics[width=0.07\textwidth]{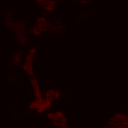} &
\includegraphics[width=0.07\textwidth]{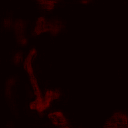} \\
\multirow{2}{*}{\rotatebox[origin=c]{90}{Std. Dev.~~~}}~~~~~ & \includegraphics[width=0.07\textwidth]{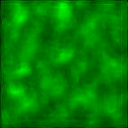} &
\includegraphics[width=0.07\textwidth]{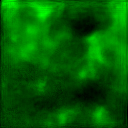} &
\includegraphics[width=0.07\textwidth]{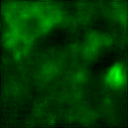} &
\includegraphics[width=0.07\textwidth]{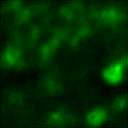} &
\includegraphics[width=0.07\textwidth]{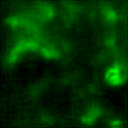} &
\includegraphics[width=0.07\textwidth]{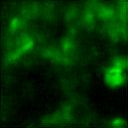} &
\includegraphics[width=0.07\textwidth]{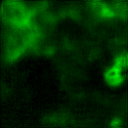} &
\includegraphics[width=0.07\textwidth]{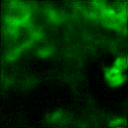} &
\includegraphics[width=0.07\textwidth]{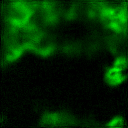} &
\includegraphics[width=0.07\textwidth]{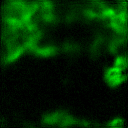} &
\includegraphics[width=0.07\textwidth]{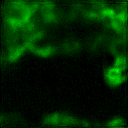} &
\includegraphics[width=0.07\textwidth]{fc_real_F-Actin-3D_DMC+DOC_6_0_q_5000.png} \\
 & \includegraphics[width=0.07\textwidth]{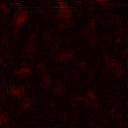} &
\includegraphics[width=0.07\textwidth]{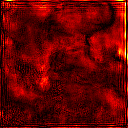} &
\includegraphics[width=0.07\textwidth]{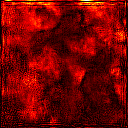} &
\includegraphics[width=0.07\textwidth]{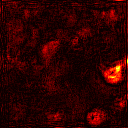} &
\includegraphics[width=0.07\textwidth]{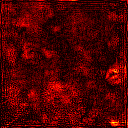} &
\includegraphics[width=0.07\textwidth]{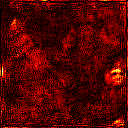} &
\includegraphics[width=0.07\textwidth]{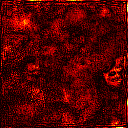} &
\includegraphics[width=0.07\textwidth]{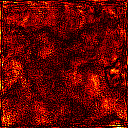} &
\includegraphics[width=0.07\textwidth]{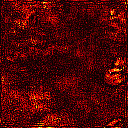} &
\includegraphics[width=0.07\textwidth]{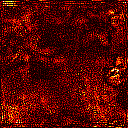} &
\includegraphics[width=0.07\textwidth]{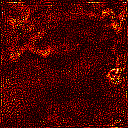} &
\includegraphics[width=0.07\textwidth]{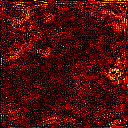}
\end{tabular}}
\caption{Visualization of the recovered intensity values and uncertainty quantifications from our method on three samples of nucleus \textcolor{blue}{(top)}, fluorescent microsphere (FM, \textcolor{blue}{middle}), and filamentous actin (F-Actin, \textcolor{blue}{bottom}). The final estimation $\xhat^*$ is obtained by averaging the $D=80$ predictions in stage-4 as Eq.~(6) \textcolor{blue}{(upper)}, while the uncertainty \cite{sun2021deep,feng2023score} is quantified by calculating the standard deviation (Std. Dev.) among all the 80 instances of each image pixel \textcolor{blue}{(lower)}. The intensity means are normalized for visualization, whereas standard deviations of nucleus, FM, and F-actin are distributed in [0,0.25], [0,2], and [0,0.2], respectively. Lighter shades of red represent higher uncertainty values.}
\label{fig:scalability_real}
\end{figure*}

\subsubsection{Extension of Our Four-Stage Progressive Reconstruction Strategy to Existing Method}
In this subsection, we extend our four-stage reconstruction strategy to the existing self-supervised method DDSSL \cite{quan2022dual} for natural image CS reconstruction. We compare its performance with our SC-CNN, as shown in Tab.~\ref{tab:application_of_four_stage_strategy_to_existing_method}. Implementing our strategy improves the DDSSL method by 0.84dB in PSNR. However, this enhanced DDSSL performance of 28.49dB is still 0.93dB lower than the 29.42dB of our SC-CNN. We conjecture that this difference may result from the less effective NN design of DDSSL compared to our proposed SC-CNN, which uses fewer parameters. These findings demonstrate the broad applicability of the four-stage strategy and the effectiveness of our NN design.

\subsection{CS Reconstruction for 2D and 3D Fluorescence Microscopy with Real SPI Optics System}
\label{sec:more_experimental_details_real}

\subsubsection{More Implementation Details}

\textbf{Additional SPI Optics, Network, and Training Details.} To verify the efficacy of our method in practical CS imaging, following the examples in \cite{duarte2008single,kulkarni2016reconnet,li2022dual}, we set up an SPI optics system for fluorescence microscopy, which is a pivotal tool for monitoring the cell physiology and tackling various biological problems \cite{lichtman2005fluorescence}. As depicted in Figs.~\ref{fig:real_arch} (left, a) and (right), our system employs two power-tunable lasers operating at $405nm$ and $488nm$, respectively, as the light source. The laser beam is reflected by a dichroic beamsplitter (DB) to a digital micro-mirror device (DMD) (V7001 DLP7000\&DLPC410), which modulates the laser and then illuminates the biological sample through an objective (Nikon 20x/0.75). The biological sample is excited and emits fluorescence at the focal plane of the objective in accordance with DMD modulation patterns. The fluorescence then returns to DMD along the excitation path and is focused onto a photon-counting photomultiplier tube (PMT, H0682-210 Hamamatsu), allowing weak signal detection. Each DMD pattern or target image corresponds to a sample area of size $87.552 \mu m \times 87.552 \mu m$. For F-actin, the adjacent optical slices (sections) are $0.5\mu m$ apart from each other. As Fig.~\ref{fig:real_arch} (left, b) exhibits, we employ our standard SC-CNN ($K=20$, $C=32$, and trained by only stage-1 for 1000 epochs with $\alpha =0.001$ and a learnable GD step size $\rho$) without IE, PE, the final GD step, and the noise injection in $\Loss_{DOC}$\footnote{In the ``+1/-1'' SPI sampling mode, the measurement noise follows mixed Skellam-Gaussian distributions \cite{jg1946frequency,mur2022deep}. Although they can be approximated by Gaussian distributions under certain conditions \cite{huang2009introduction}, accurately estimating the noise parameters and addressing the signal dependence of measurement noise pose significant challenges. In this paper, we have made a simple assumption (\textbf{Assumption} in \MP) on the noise and not explicitly accounted for the precise modeling of SPI noise, which is our future work \cite{wei2021physics,mur2022deep}.}. For 3D imaging on F-actin, we set $C=64$ and concatenate the 21 slices channel-wise to reconstruct them jointly. In other words, the input is 22-channel, containing 21 image initializations $\A^\top \y$ and a CS ratio map $\mathbf{R}\in\Rbb^{1\times H\times W}$, and the output is 21-channel, consisting of the 21 predictions $\xhat$. We have empirically observed that this joint reconstruction strategy can improve imaging quality and efficiency. In this case, the matrix elements are i.i.d. Bernoulli random variables, with each element having an equal probability of 0.5 for being +1 or -1. In the complementary CS sampling mode \cite{yu2014complementary}, the $+12^{\circ}$ and $-12^{\circ}$ direction states of each digital micro-reflector in DMD implement the matrix elements of value 1 and 0 for each pattern (matrix row), respectively. The acquisition rate can reach 7500 measurements per second, based on the DMD refresh rate of 15kHz, thereby providing an opportunity to observe transient life activities. It is worth noting that by incorporating time-correlated single photon-counting (TCSPC) \cite{becker2004fluorescence} and a spectroscopic prism, the system equipped with our proposed method can be further extended to achieve deep CS-based scalable fluorescence lifetime and spectral imaging, enabling a broader range of multimodal biomicroscopic studies.

\subsubsection{More Performance Comparisons}

Figs.~\ref{fig:comparison_real_more_nucleus}, \ref{fig:comparison_real_more_FM}, and \ref{fig:comparison_real_more_F-Actin} present the additional reconstructed results of five distinct CS methods on thirty real captured SPI measurements of nucleus, FM, and F-actin, respectively, with ratio $\gamma =5000/16384\approx 30.5\%$. Notably, our default ``DMC+DOC'' scheme can consistently outperform other ones in terms of image quality, exhibiting superior noise suppression and information preservation. This is particularly evident in the critical and informative regions, where the balance between noise reduction and fidelity is effectively achieved, mitigating artifacts and distortions that may be prevalent in other methods.

\subsubsection{Results of 3D CS Imaging, Scalable Recovery, and Uncertainty Quantification}

\textbf{3D Imaging Results.} Fig.~\ref{fig:our_results_more_FM_slice} provides the reconstruction results of our SC-CNN variant from five groups of measurements, each of which corresponds to a distinct 3D F-actin instance with 21 optical slices in z-direction. We observe that our method exhibits exceptional capabilities in delivering high-fidelity 3D imaging results by jointly reconstructing the stacks of 2D slices. Moreover, it effectively captures the subtle variations along the z-axis, thereby confirming the efficacy and extensibility of our method in handling data of diverse dimensionalities, covering 1-/2-/3-D forms.

\textbf{Scalable CS Reconstruction and Uncertainty Quantification Results.} Fig.~\ref{fig:scalability_real} evaluates our trained SC-CNN at 12 CS ratios. For each case, we generate $D=80$ estimations as Eq.~(\ref{eq:self_ensemble}) in stage-4 to further enhance imaging quality and quantify the uncertainty \cite{sun2021deep,feng2023score,qin2023ground} through multiple predicted $\xhat$ instances. Specifically, we calculate the mean and standard deviation of $\T^{-1}\F_{\Th}(\y,\M\A\T,\gamma)$ over multiple instances of $\T$ and $\mathbf{M}$ as the final estimation $\xhat^*$ and uncertainty map, respectively. We observe that: (1) our method can consistently and stably produce clear results with $M\ge 1000$. This fact implies that users have the flexibility to adjust CS ratio from original 30.5\% to approximately 6.1\%, enabling $\sim 5\times$ sampling acceleration while reducing the phototoxicity for live cells. Alternatively, researchers can also increase the CS ratio by capturing more measurements to enhance final quality without the need for measurement re-collections, set re-constructions, or NN re-trainings; (2) the uncertainty associated with informative edge and texture features is higher compared to those of non-salient smooth regions. This finding underscores the ability of our method to capture and represent uncertainties through matrix perturbations, thereby providing valuable insights for scientific research \cite{sun2021deep,feng2023score,qin2023ground}.

To summarize, the above experimental results validate the effectiveness of our method in handling real-world data with complex noise, supporting scientific researches, and demonstrating its superiority in terms of recovery performance, generalization capabilities, flexibility, and scalability compared to existing approaches. Notably, our proposed ``SCL+SCNet'' method can achieve these remarkable results once trained on a single pre-collected measurement set sensed with fixed ratio and matrix.

\section{Discussions}
\label{sec:discussions}
\subsection{Insights and Explanations behind Our Method}

Although we have demonstrated the successful application of our proposed method, which combines an SCL scheme and an SCNet family for self-supervised scalable CS, through our experiments, it is important to understand the underlying factors contributing to its performance. While a comprehensive interpretation of NNs and their optimization dynamics remains a challenging task, we present analyses and some empirical inferences and conjectures to shed light on the effectiveness of our ``SCL+SCNet'' combination. Specifically, we attribute the success of our method to the following four critical factors:

(1) The incorporation of appropriate self-supervision and random measurement and matrix division mechanisms through DMC loss ($\Loss_{DMC}$), inspired by N2N \cite{lehtinen2018noise2noise, xia2019training} and our theorem, which is further empirically augmented. Our utilization of the power set of measurement elements and their corresponding matrix rows can significantly improve data diversity to alleviate overfitting, with a $\sim 2^M\times$ expansion on the size of measurement set. Another intuition is that our cross-supervision prevents $\F_\Th$ from fitting trivial mappings like $(\y,\A)\mapsto \A^\dagger \y$; instead, $\F_\Th$ is encouraged to learn the latent common signal priors from measurements that are easier to capture by well-designed NNs in the early training stage. It is also worth noting that as Fig.~\ref{fig:scalable_curve_more_analyses_p_and_alpha} (right, $\alpha=0$) exhibits, an SC-CNN trained only by our $\Loss_{DMC}$ can already be competent to reconstruct well for ratio range $(0,\gamma^{train}]$.

(2) The incorporation of effective regularization and redundancy reduction via DOC loss ($\Loss_{DOC}$). Our encouragement of matrix-network disentanglement proves to be highly effective in imposing regularization. The combination of $\Loss_{DMC}$ and $\Loss_{DOC}$ ensures that NNs do not overfit to specific matrices and ratios but instead learn the general recovery knowledge regarding $q_\x$, $q_\A$, and $q_\n$. This may be interpreted as an operation of ``redundancy reduction'' - learning a single NN rather than multiple ones for diverse tasks itself can serve as a regularization for optimization. This concept is supported by the similarities (or redundancies) observed in learned NN weights across different tasks \cite{he2019modulating,he2020interactive}, as exhibited in our toy experiments and ablation studies. We would also highlight that the cooperation of $\Loss_{DMC}$ and $\Loss_{DOC}$ is significant: on one hand, the predictions $\xhat$ from DMC loss calculation allows NN to learn beyond range $(0,\gamma^{train}]$ and matrix $\A$ using $\Loss_{DOC}$ to its full potential; on the other hand, DOC loss effectively imposes $\xhat$ (generated by the previous $\Loss_{DMC}$ calculation) and NNs to jointly satisfy the sampling-reconstruction cycle-consistency. These mutual promotions and bidirectional positive feedback enhance training stability and performance.

(3) The incorporation of appropriate implicit structural regularizations facilitated by SCNet framework and NN blocks. The inductive bias of locality, shift invariance, continuity, and other properties offered by convolutions, RBs, and SCBs play a significant role in ensuring the robust performance. It has been validated that such NNs can naturally tend to learn low-frequency harmonious signal components \cite{ulyanov2018deep,qayyum2022untrained}. Our adoption of the inspiration from PGD, and the learnable embeddings (IE and PE) also provide effective and crucial regularization, facilitating training process and avoiding under-/over-fitting to inharmonious results. We emphasize that advanced NN blocks do not always guarantee superior performance; instead, appropriate constraints prove to be more critical for SCNets. For instance, SCT$^+$ does not always outperform SC-CNN$^+$ in the noisy cases, as shown in Tab.~\ref{tab:compare_sota_psnr_noisy}.

(4) The incorporation of appropriate explicit and implicit regularizations brought by the knowledge of common image priors shared across multiple images. One fundamental assumption of this work is that the underlying images $\{\x_i\}$ are drawn from the same distribution $q_\x$ and possess common and low-dimensional structures. The utilization of external measurements for offline training (stage-1) and cross-image internal learning (stage-2) proves to be effective for performance improvement. Another assumption is that natural image sets satisfy the invariance to a certain group of transformations $\T$ like rotations and flippings \cite{chen2021equivariant,chen2022robust,chen2023imaging,chen2022content}. Our combination of explicit and implicit priors exploits the collective structure inherent in $\{\x_i\}$, while the corresponding assumptions we make in this work can often be satisfied in many applications where there is no access to abundant clean images \cite{leong2023ill}.

To summarize, by integrating our designs and leveraging available data and information to their fullest extent, the proposed method has demonstrated its efficacy and superiority, supported by theoretical foundations, experimental validations, and intuitive explanations elucidating its underlying working principles.

\subsection{Limitations}

The main limitation of this work is that the performance improvements of our method come at the costs of (1) reduced speed, (2) increased information requirement and assumptions, and (3) partial loss of interpretability brought by the ``black-box'' property of our NNs and augmentations.

Firstly, while our standard SCNets enjoy inference speeds higher than 30 frames per second, their enhanced versions, equipped with the fully activated stages 1-4, require approximately 1-5 hours to reconstruct a $256\times 256$ test image, as reported in Tab.~\ref{tab:high_level_comp}, to utilize data and NN to their full extent and achieve the optimal performance. This may pose a challenge for real-time CS imaging applications. We will study how to improve adaption speed without increasing model complexity or sacrificing recovery accuracy.

Secondly, our work assumes access to the statistical characteristics of natural signal sets, observation noises, and sampling matrices. However, their modelings or characterizations in real-world scenarios are non-trivial. As real CS tasks usually involve more complex signals, noises, and matrix forms, it becomes necessary to customize the approach more accurately for performance improvement in specific cases. For instance, in the cases of SPI with a ``+1/-1'' sampling mode, the measurement noises are signal-dependent and follow a mixed Skellam-Gaussian model \cite{jg1946frequency,mur2022deep}. In other CS-based imaging tasks like MRI and SPI with structured sampling matrices such as FFT basis, Hadamard matrix, and learned matrices \cite{shi2019image,chen2022content}, the importance of different measurement elements and matrix rows can vary significantly. Consequently, treating them indiscriminately for performance assurance may not be sufficient, and it would be beneficial to develop more appropriate method implementations for measurement division for these cases \cite{yaman2020self}. It is worth mentioning that our SPI experiments did not incorporate such precise customizations, leaving room for further improvement in imaging quality.

Thirdly, despite the successful application of our loss and high-throughput unrolled SCNets with simple GD steps, inspired by the insights from our theorem and PGD generalization, which yields superior recovery results compared to other methods, a comprehensive understanding of the recovery mechanisms and optimization dynamics of NNs has not been thoroughly addressed within our work, and represents an interesting problem for further investigation.

\subsection{Broader Impacts}
The proposed method extends the realm of CS to real-world applications and can serve as a catalyst for the development of learning schemes and unrolled NNs for inverse problems related to natural signals. Thus far, CS techniques have not exhibited detrimental societal repercussions. Our method, in line with this trend, is not expected to yield negative foreseeable consequences.

Leveraging its merits of reduced data requirements, exceptional performance, adaptability, and generalization capability, our method possesses the potential to foster deeper integration of artificial intelligence within the domains of life sciences and cytology. By unlocking the power of self-supervised scalable CS, it creates new opportunities for enhanced scientific exploration, biomedical advancements, and further strides in understanding cellular structures and functions.

\bibliography{sn-bibliography}

\end{document}